\documentclass[a4paper,fleqn,usenatbib]{mnras}

\usepackage{mathptmx}

\usepackage[T1]{fontenc}
\usepackage{ae,aecompl}

\usepackage{graphicx}	
\usepackage{amsmath}	
\usepackage{amssymb}	
\usepackage{ifxetex}
\usepackage{multicol}
\usepackage{morefloats}

\def\br{\hfill\break}
\def\RP{R$^{\prime}$}
\def\S_AB{S$\underline{\rm A}$B}
\def\SA_B{SA$\underline{\rm B}$}
\def\SB_a{SB$_a$}
\def\ans{$_a$}
\def\^+{$^+$}
\def\^o{$^o$}
\def\_rs{$\underline{\rm r}$s}
\def\Rone{R$_1$}
\def\rRone{rR$_1$}
\def\Rtwo{R$_2$}
\def\RoneP{R$_1^{\prime}$}
\def\RoneL{R$_1$L}
\def\RtwoP{R$_2^{\prime}$}
\def\rpl{r$^{\prime}$l}
\def\rcr{$r_{CR}$}
\def\rolr{$r_{OLR}$}
\def\ro4r{$r_{O4R}$}
\def\ri4r{$r_{I4R}$}
\def\a_b{a$\underline{\rm b}$}

\def\aj{{AJ}}

\def\apj{{ApJ}}
\def\apjs{{ApJS}}
\def\mnras{{MNRAS}}

\def\pasp{{PASP}}

\def\s4g{{S$^4$G}}

\title[Galactic Rings Revisited. II]{Galactic Rings Revisited. II. Dark
Gaps and the Locations of Resonances in Early-to-Intermediate Type Disk Galaxies}

\author[Ronald J. Buta]{
Ronald J. Buta\thanks{E-mail: rbuta@ua.edu}
\\
Department of Physics \& Astronomy,University of Alabama, Box
870324, Tuscaloosa, AL 35487
}
\date{Accepted XXX. Received YYY; in original form ZZZ}

\pubyear{2016}

\begin{document}
\label{firstpage}
\pagerange{\pageref{firstpage}--\pageref{lastpage}}
\maketitle

\begin{abstract}
Dark gaps are commonly seen in early-to-intermediate type barred
galaxies having inner and outer rings or related features. In this
paper, the morphologies of 54 barred and oval ringed galaxies have been
examined with the goal of determining what the dark gaps are telling us
about the structure and evolution of barred galaxies. The analysis is
based mainly on galaxies selected from the Galaxy Zoo 2 database and the
Catalogue of Southern Ringed Galaxies. The dark gaps between inner and
outer rings are of interest because of their likely association with
the $L_4$ and $L_5$ Lagrangian points that would be present in the
gravitational potential of a bar or oval. Since the points are
theoretically expected to lie very close to the corotation resonance
(CR) of the bar pattern, the gaps provide the possibility of
locating corotation in some galaxies simply by measuring the radius
$r_{gp}$ of the gap region and setting $r_{CR}$=$r_{gp}$.  With the
additional assumption of generally flat rotation curves, the locations
of other resonances can be predicted and compared with observed
morphological features. It is shown that this ``gap method" provides
remarkably consistent interpretations of the morphology of
early-to-intermediate type barred galaxies. The paper also brings
attention to cases where the dark gaps lie inside an inner ring, rather
than between inner and outer rings. These may have a different origin 
compared to the inner/outer ring gaps.

\end{abstract}




\section{Introduction}

Barred galaxies are well-known not only for their ubiquity, but also
for the distinctive ring morphologies they often display. The
characteristic rings have been described by Buta \& Combes (1996=BC96) and
include nuclear (or circumnuclear) rings, inner rings, and outer rings
in order of increasing linear and relative size. The rings are often
made of tightly-wrapped spiral arms (``pseudorings"), and each type has
its own unique set of morphological attributes.

Rings continue to be of interest in galactic studies because of their
strong sensitivity to internal galaxy dynamics and because the features
are likely to be intimately connected with secular evolutionary
processes in galactic disks (e.g., Kormendy 1979, 2012; BC96; Knapen
2010; Buta 2012, 2013).  The most popular view is that the features are
related to specific orbital resonances with the pattern speed of a bar
or oval. Buta (1995) used distributions of {\it apparent} ring axis
ratios and relative bar-ring position angles for a large sample of
southern galaxies to infer the {\it intrinsic} axis ratios and
orientations of the rings, and determined that these were consistent
with the predictions of barred galaxy models such as those of Schwarz
(1981, 1984a), Simkin, Su, \& Schwarz (1980) and Byrd et al. (1994).
Rautiainen \& Salo (2000) further examined the link using more
sophisticated $n$-body bar models. Comer\'on et al. (2014) carried out
a similar study to Buta (1995) using deprojected mid-infrared images
from the Spitzer Survey of Stellar Structure in Galaxies (S$^4$G, Sheth
et al. 2010). Although Comer\'on et al. confirmed that inner SB rings
are generally aligned parallel to bars, these authors also identified a
population of inner rings in late-type galaxies having random
alignments with respect to bars.

The possible connection between galactic rings and orbital resonances
in galaxies perturbed by a bar, oval, or a strong global spiral has
been considered since Lindblad (1941) first proposed a density wave
view of the nature of spiral structure. This view, later developed
further by Lindblad (1955) and Lindblad \& Lindblad (1958), highlighted
the importance of the appropriately-named Lindblad resonances. The
expanded density wave theory of Lin \& Shu (1964) confined a
self-consistent spiral between the inner and outer Lindblad resonances
(ILR and OLR, respectively) of the wave, and indeed expected that the
spiral could not pass through the ILR. Lin (1970) considered rings to
be among the general features of galaxies to be explained by the
quasi-stationary spiral structure (QSSS) hypothesis. Lin, Yuan, and Shu
(1969) linked the ``3 kpc" arm in the Milky Way to the ILR, and argued
that ``According to the theory, the pattern ends here as a ring." This
would likely be a ring of star formation, and indeed Lin (1970) pointed
to NGC 5364 as an example of a spiral with a bright inner ring of star
formation where no HII regions lie inside the ring.

Although the morphology of galactic rings has long suggested a link to
resonances, studies of individual cases have been sparse, and there is
clearly more to learn about these rings than can be gleaned from
classification and morphology alone. The Sloan Digital Sky Survey
(SDSS, Gunn et al. 1998; York et al. 2000) provides the opportunity to
examine the photometric structure of large numbers of ringed galaxies
with unprecedented homogeneity and effectiveness, and is the basis for
most of the analysis presented in this paper.  The resonance idea is
also not the only theory that has been used to describe the types of
ring features discussed here. The manifold theory reviewed by
Athanassoula et al. (2011) and Romero-G\'omez (2012) provides a
remarkably different view that has had success in explaining many
aspects of the morphology of rings and spirals. This theory will 
be discussed further in section 10.

In the first paper in this series (Buta 2017 = Paper I), the Galaxy Zoo
2 citizen science project (Willett et al. 2013; GZ2 hereafter) was used
to extract a large sample of galactic rings within the SDSS footprint.
Zoo volunteers used a set of buttons to classify galaxies within broad,
but somewhat more finely detailed categories than had been used in
Galaxy Zoo 1 (Lintott et al. 2008). One button asked if there was
``anything odd" about a galaxy, and if answered ``yes", then ``Is the
odd feature a ring?" Using the sophisticated weighting scheme described
by Willett et al (2013), an average probability of having a ring could
be derived for the 300,000 galaxies in the GZ2 sample. Restricting to
those relatively nearby cases most likely to have a ring yielded 3962
galaxies, a sample comparable in size to that compiled for the
Catalogue of Southern Ringed Galaxies (CSRG, Buta 1995).

The analysis in this paper is based in part on interpretations of a
subset of these 3962 GZ2 galaxies in the Comprehensive de Vaucouleurs
revised Hubble-Sandage (CVRHS) galaxy classification system (Buta et
al. 2015). CVRHS classifications are ideally suited to studies of ring
phenomena because the system recognizes in a consistent manner all
normal inner, outer, and nuclear rings, pseudorings, and lenses, as
well as cataclysmic rings such as those seen in ring galaxies, polar
ring galaxies, inclined ring galaxies, and accretion ring galaxies.
Paper I provides CVRHS classifications based on the same SDSS colour
images used by the Galaxy Zoo 2 participants. One of the main results
from paper I is that the types of rings that caused GZ2 participants to
select "yes" for "Is the odd feature a ring" are basically identical to
those selected for the CSRG. It is thus fair to intermingle the two
catalogues and combine an analysis of the same kinds of galaxies in
both catalogues.

This paper is focussed mainly on 50 galaxies having outer rings and
pseudorings showing the distinctive morphologies thought to be
characteristic of the outer Lindblad resonance, based on early barred
galaxy models (Schwarz 1981; Byrd et al. 1994). These are classified as
types (R$_1$), (R$_1^{\prime}$), (R$_2^{\prime}$), and
(R$_1$R$_2^{\prime}$) by Buta (1995) and are known as ``OLR
subclasses." The primes refer to pseudorings.  In paper I,
additional subclasses are recognized including (R$_2$), (R$_1$R$_2$),
and (R$_1^{\prime}$R$_2^{\prime}$). Of particular interest are the
characteristic dark spaces or gaps seen in the morphology of OLR
subclass galaxies.  It is argued here that the location of these dark
gaps/zones and their shapes favour the idea that the gaps are linked to
the region of the $L_4$, $L_5$ Lagrangian points in the gravitational
potential of a bar or oval. It is suggested that the surface brightness
minima within the dark gaps can be used to trace the radius of the
corotation resonance (CR), which can be viewed as the most important
resonance governing the morphology and dynamics of a disk-shaped
galaxy.

The objective of this paper is to examine the implications of using
dark gaps to define the location of corotation in a sample of
early-to-intermediate type barred galaxies. The theoretical basis for
applying this ``gap method" is described more fully in section 3. With
the location of the CR determined, the radii of other resonances are
predicted assuming flat rotation curves. It is shown that with this
scaling, the morphology of OLR subclass galaxies is remarkably
consistent and favours the idea that these are largely single pattern
speed systems. The analysis also suggests that the outer 4:1 resonance
plays a more important role in galaxy morphology than previously
thought, which may necessitate a renaming of the ``OLR subclasses" to
``outer resonant subclasses."

The paper also considers rarer cases where the dark spaces lie
{\it inside an inner ring} rather than between inner and outer rings.
If the dark spaces in these galaxies are interpreted in the same manner
as for the inner/outer ring gap galaxies, then either the existence of
extremely high pattern speed (``superfast") barred galaxies would have
to be acknowledged, or another mechanism for forming dark gaps that has
nothing to do with Lagrangian points would have to be hypothesized.
The possible role of internal extinction in making dark gaps is
examined in Section 9 but considered unlikely.

A recent study by Kim et al. (2016) has independently touched on the
issues of barred galaxy dark spaces, which are called ``inner disc
light deficits" by these authors. This paper uses numerical simulations
and bar major and minor axis light profiles to examine the evolution of
the inner regions of barred galaxies that might lead to these light
deficits. Their Figure 1 also highlights the same two types of dark
space morphologies as are discussed here, and their Figure 5 shows the
same kinds of profiles as are used in Figures 5 and 10 of the present
paper. The implications of Kim et al.'s work on the present study are
examined in section 10.

Section 2 describes the OLR subclasses in more detail.  Section 3
describes the types of ``dark-spacer" morphologies recognized from the
GZ2 sample, and makes the case for interpreting the dark zones in terms
of the $L_4$ and $L_5$ Lagrangian points. Section 4 describes the
processing of SDSS and other images used for this paper. Section 5
illustrates the application of the gap method to two dissimilar cases,
UGC 4596 and NGC 5335, as examples. A summary of the full procedure is
provided in section 6 and the results of its application to the
remaining galaxies in the sample is described in section 7. Section 8
presents an analysis of the derived resonance radii, while section 9
examines the colours of the dark gaps. A discussion is presented in
section 10. Conclusions are presented in section 11.

\section{Outer Resonant Morphologies}

The morphological categories R$_1$, R$_1^{\prime}$, R$_2^{\prime}$, and
R$_1$R$_2^{\prime}$ are singled out of the general ring population by
their distinctive resemblance to model outer Lindblad resonance rings
(Schwarz 1981; Buta \& Crocker 1991=BC91; Buta 1995). The subclasses
are believed to be tied to two familes of periodic orbits near the OLR
of a bar or oval (Schwarz 1981). Type R$_1^{\prime}$
(Figure~\ref{fig:figure-OLR}, top) is an outer pseudoring defined by an
$\approx$180$^o$ winding of the main spiral arms from near one end of
the bar to the other, forming a characteristic figure eight pattern
with ``dimples" (indicated by the two horizontal arrows in
Figure~\ref{fig:figure-OLR}, top right) which pinch in towards the bar
axis. This type may also appear in a more closed form, called R$_1$.

 \setcounter{figure}{0}
 \begin{figure}
 \begin{minipage}[b]{0.45\linewidth}
 \centering
\includegraphics[width=\textwidth]{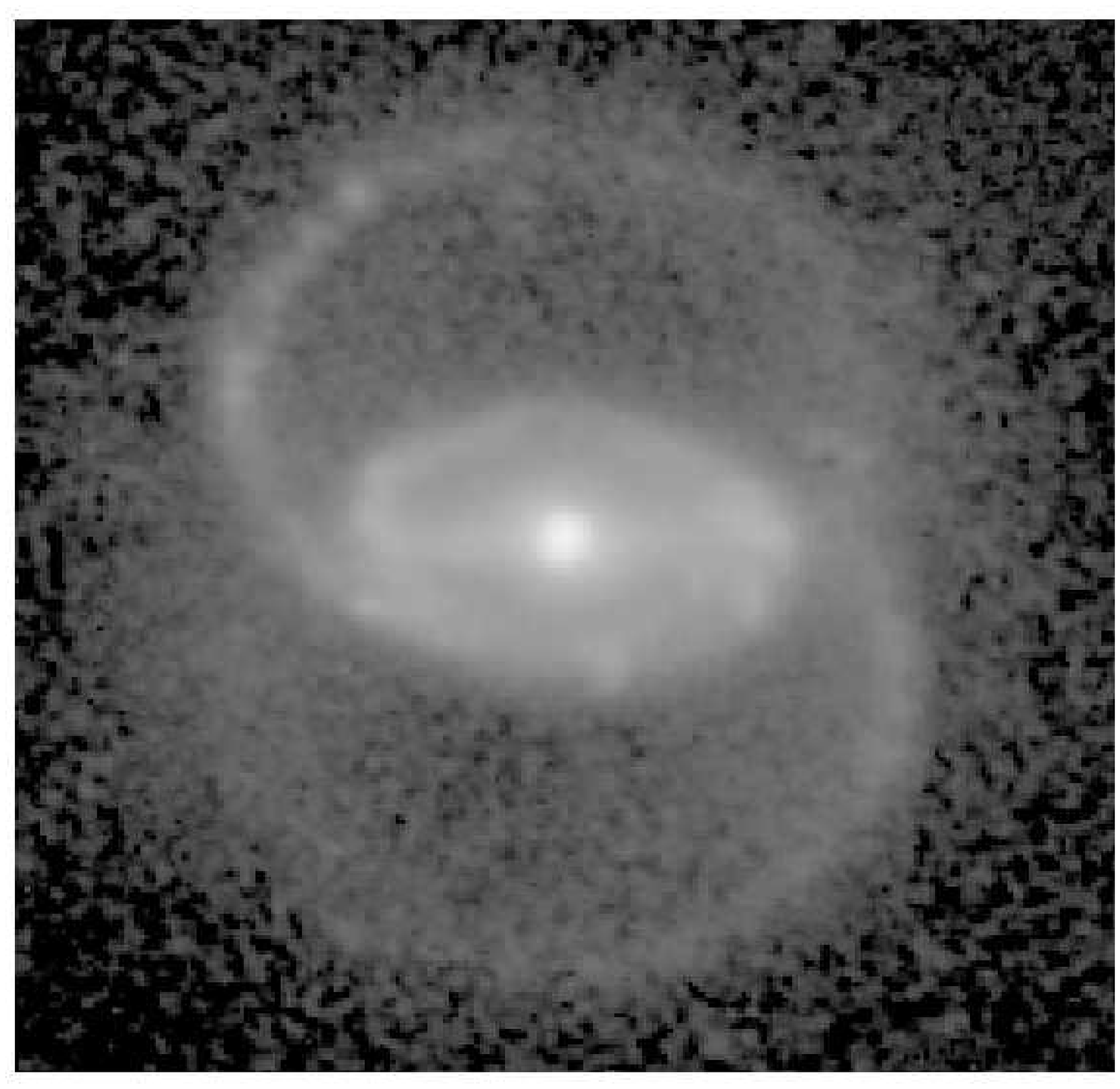}
 \hspace{0.1cm}
 \end{minipage}
 \begin{minipage}[t]{0.68\linewidth}
 \centering
\raisebox{0.5cm}{\includegraphics[width=\textwidth,trim=0 0 0 300,clip]{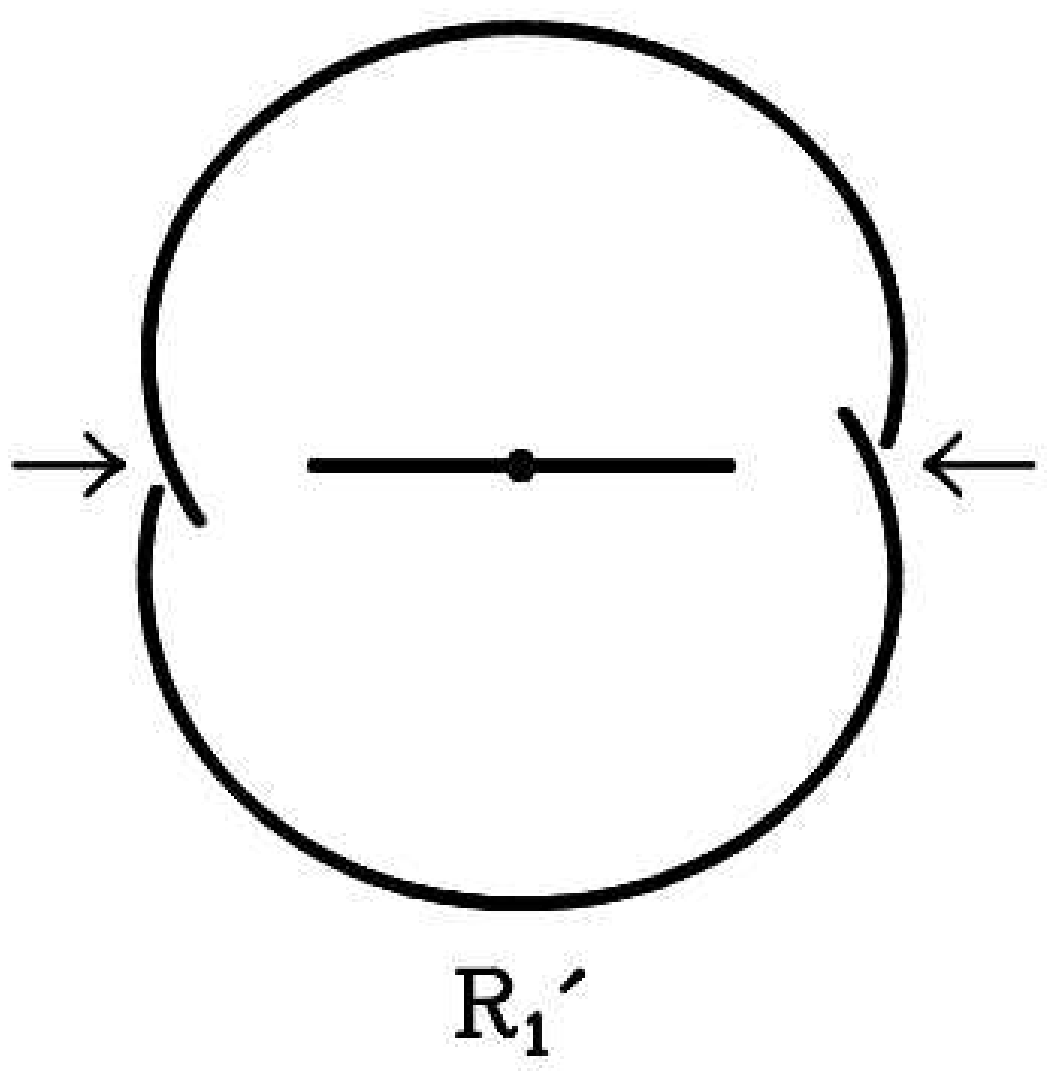}}
 \end{minipage}
 \begin{minipage}[b]{0.45\linewidth}
 \centering
\includegraphics[width=\textwidth]{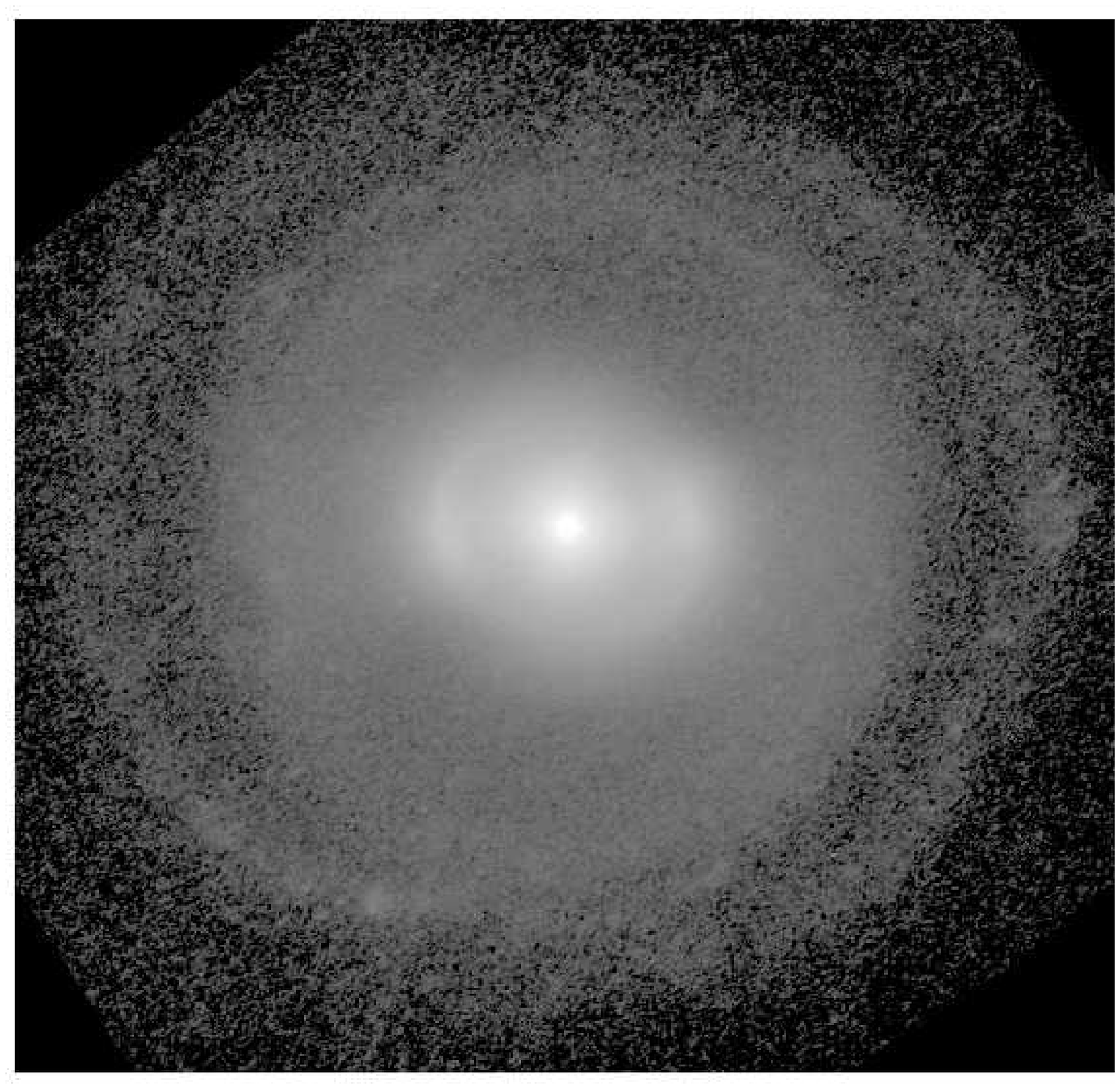}
 \hspace{0.1cm}
 \end{minipage}
 \begin{minipage}[t]{0.68\linewidth}
 \centering
\raisebox{0.5cm}{\includegraphics[width=\textwidth,trim=0 0 0 300,clip]{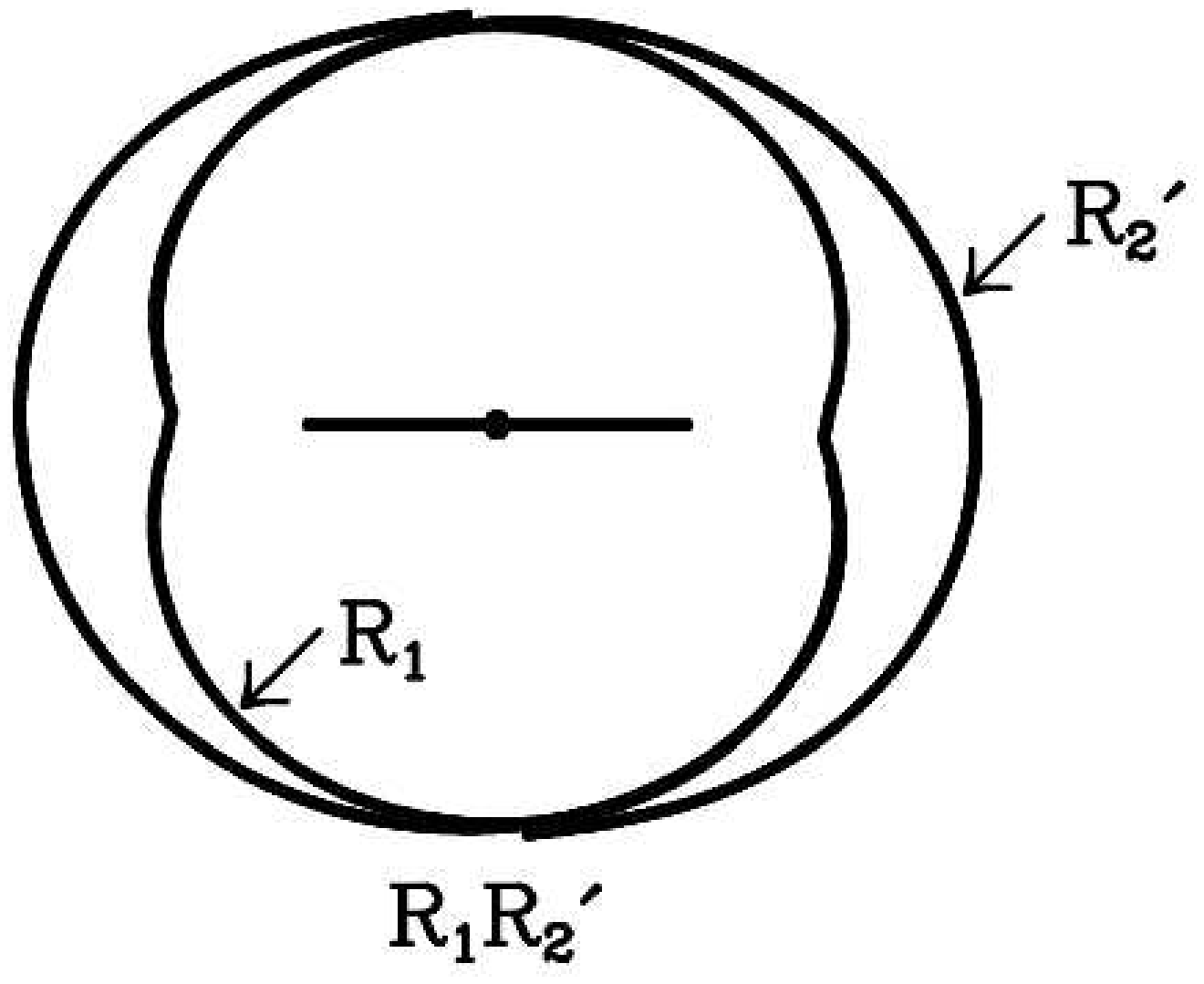}}
 \end{minipage}
 \begin{minipage}[b]{0.45\linewidth}
 \centering
\includegraphics[width=\textwidth]{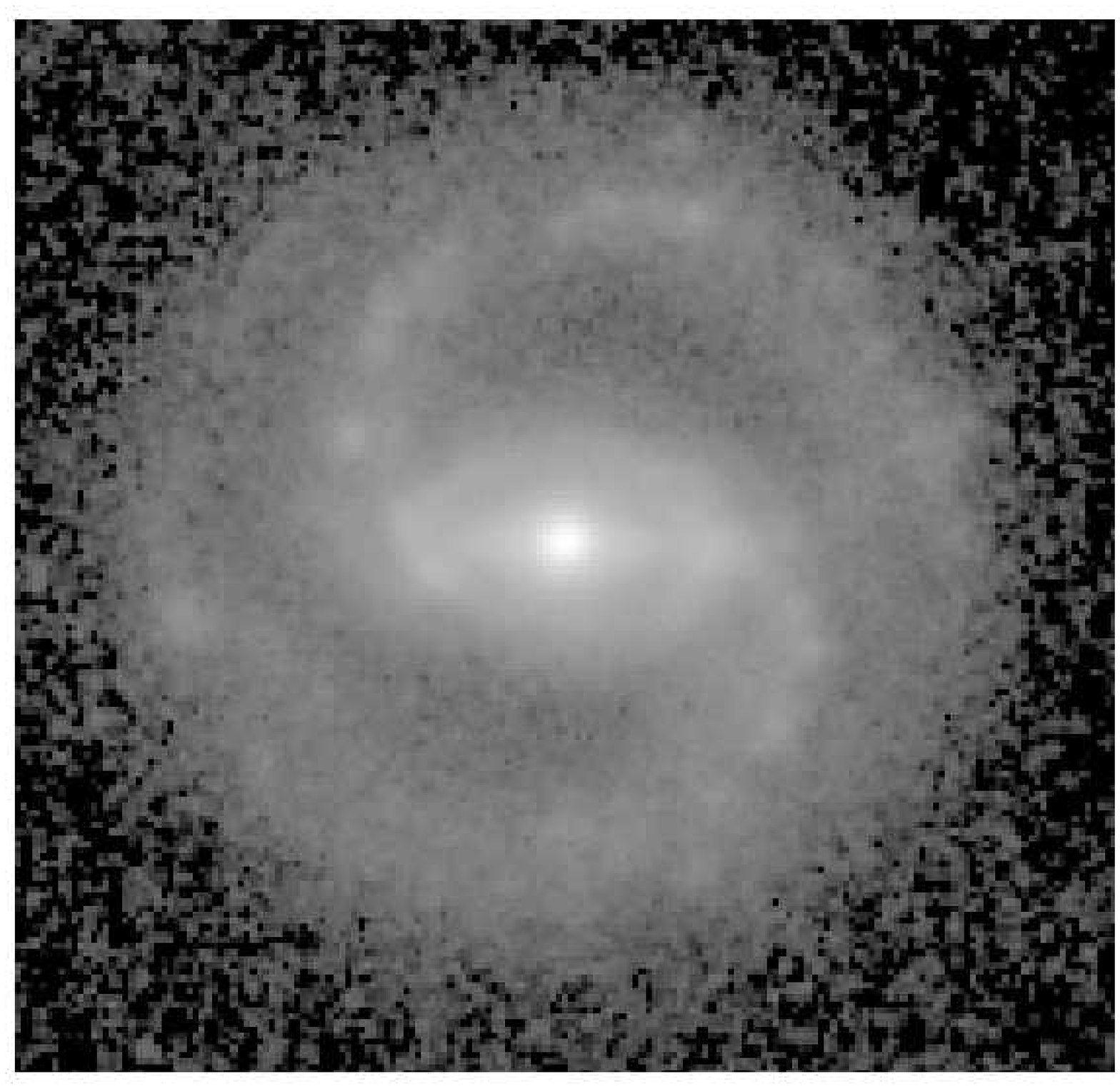}
 \hspace{0.1cm}
 \end{minipage}
 \begin{minipage}[t]{0.68\linewidth}
 \centering
\raisebox{0.5cm}{\includegraphics[width=\textwidth,trim=0 0 0 300,clip]{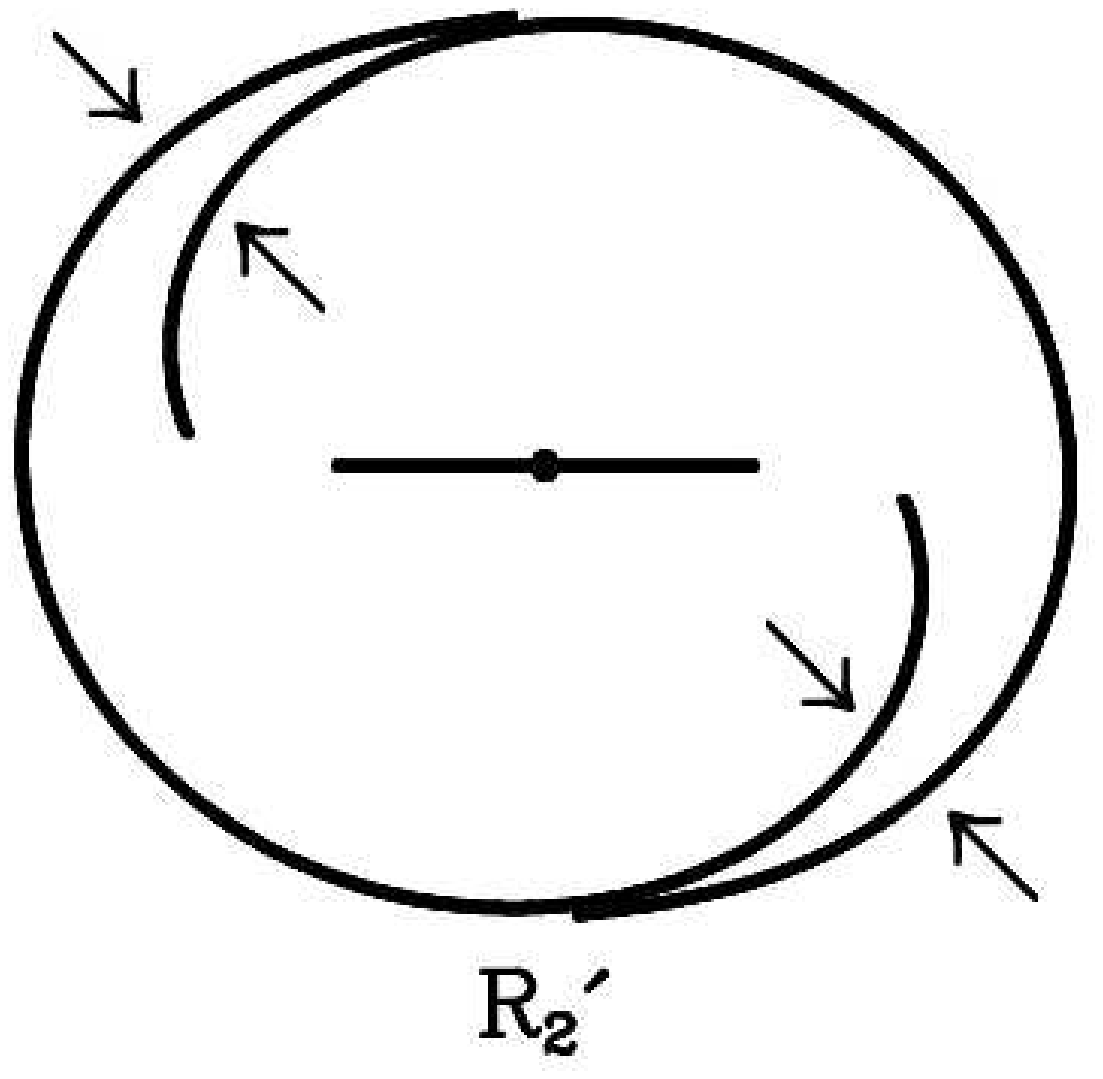}}
 \end{minipage}
\vspace{-0.5truecm}
\caption{Schematics of the outer resonant ring/pseudoring
morphologies. The arrows point to the characteristic features of each
type shown: dimples for R$_1^{\prime}$ outer pseudorings (top right);
arm-doubling in two opposing quadrants for R$_2^{\prime}$ outer
pseudorings (bottom right); and the double ring/pseudoring morphology
of type R$_1$R$_2^{\prime}$ (middle right). The left panels show an
example of each type. From top to bottom these are: UGC 12646, NGC
1079, and ESO 325-28. The structure of these morphologies can be
connected to two major families of OLR periodic orbits shown by Schwarz
(1981). The horizontal line in each case is the theoretical orientation
of the bar relative to the ring structures.
}
\label{fig:figure-OLR}
\end{figure}

\begin{table*}
\centering
\caption{CVRHS classifications. Col. 1: galaxy name; col. 2: source of
classification; col. 3: full classification. Sources: BC91=Buta and Crocker 1991; BC96=Buta and
Combes 1996; RC3=de Vaucouleurs et al. 1991, Appendix 3; deVA= Buta et al. 2007;
B17=paper I; TP= this paper}
\label{tab:types}
\begin{tabular}{llllll}
\hline
Name & Source & Type & Name & Source & Type \\
 1 & 2 & 3 & 1 & 2 &  3 \\
\hline
CGCG 8-10       & B17       & (R$_1^{\prime}$R$_2^{\prime}$)S$\underline{\rm A}$B(r)a                                                        & NGC 2665        & BC91      & (\RoneP)SA$\underline{\rm B}$(rs)a                                                                            \\
CGCG 13-75      & B17       & (R$_2^{\prime}$)S$\underline{\rm A}$B(r$^{\prime}$l)ab                                                         & NGC 2766        & B17       & (R$_1$R$_2^{\prime}$)S$\underline{\rm A}$B(r)a                                                                \\
CGCG 65-2       & B17       & (R$_1$)SB$_a$($\underline{\rm r}$s,bl)ab                                                                       & NGC 3081        & BC96,deVA & (R$_1$R$_2^{\prime}$)SAB(r,nr,nb)0/a                                                                          \\
CGCG 67-4       & B17       & (R$_1$L)SB$_a$(s,bl,nl)0/a                                                                                     & NGC 3380        & B17       & (R$_1$L)SAB(l,rs,bl)ab                                                                                        \\
CGCG 73-53      & B17       & (R$_1$R$_2$)S$\underline{\rm A}$B$_a$(l)0$^+$                                                                   & NGC 4113        & B17       & (R$_1^{\prime}$)SA$\underline{\rm B}$($\underline{\rm r}$s,nr)ab                                                 \\
CGCG 185-14     & B17       & (R$_1$R$_2^{\prime}$)SAB$_a$(rs)ab                                                                                 & NGC 4608        & B17       & (RL)SB(rl,bl)0/a                                                                                               \\
CGCG 263-22     & B17       & (R$_1^{\prime}$)S$\underline{\rm A}$B($\underline{\rm r}$s,$\underline{\rm r}$s)a                              & NGC 4736        & deVA,TP   & (R)\S_AB(\rpl,\_rs)ab                                                                                         \\
ESO 325-28      & BC91      & (\RtwoP)SB(r)b                                                                                                 & NGC 4935        & B17       & (R$_2^{\prime}$)SAB$_a$(rs)b                                                                                  \\
ESO 365-35      & BC91,TP   & (\RtwoP)\S_AB(l)0/a                                                                                            & NGC 5132        & B17       & (R$_1^{\prime}$)SB($\underline{\rm r}$s,bl,nr)a                                                               \\
ESO 426-2       & BC91      & (\Rone\RtwoP)SB(r)0/a                                                                                          & NGC 5211        & B17       & (R$_2^{\prime}$)SA(rs)b                                                                                       \\
ESO 437-33      & BC91      & (\RoneP\RtwoP)SAB(rs,nr)ab                                                                                     & NGC 5335        & B17       & SB($\underline{\rm r}$s,bl)ab                                                                                 \\
ESO 437-67      & BC91      & (\RoneP)SB(\_rs,nr)\a_b                                                                                        & NGC 5370        & B17       & (R$_1$L)SB$_a$($\underline{\rm r}$s,bl)0/a                                                                    \\
ESO 566-24      & deVA      & (R)SB(r)b                                                                                                      & NGC 5686        & B17       & SB$_a$(r,bl)0$^o$                                                                                             \\
ESO 575-47      & BC91      & (\RoneP)SB(\_rs)ab                                                                                             & NGC 5701        & RC3,TP    & (\Rone\RtwoP)SB(\rpl,bl)a                                                                                     \\
IC 1223         & B17       & (R$^{\prime}$)SB(rs,rs,bl)a                                                                   & NGC 6782        & deVA,BC96 & (\RoneP)SB(r,nr,nb)a                                                                                          \\
IC 1438         & RC3,TP    & (\Rone\RtwoP)SAB(r,nr)a                                                                                        & NGC 7098        & deVA,TP   & (\Rone\RtwoP)SAB\ans(\_rs,nb)ab                                                                               \\
IC 2473         & B17       & (R$_1^{\prime}$)SB(r,bl)ab                                                                                     & PGC 54897       & B17       & (R$_1^{\prime}$)SA$\underline{\rm B}$$_a$(r$^{\prime}$l)ab                                                    \\
IC 2628         & B17       & (R$_1$R$_2^{\prime}$)SAB(l)a                                                                                   & PGC 1857116     & B17       & (R$_1^{\prime}$)SAB$_a$($\underline{\rm r}$s)a                                                                \\
IC 4214         & RC3       & (\Rone)SA(\_rs,nr)a                                                                                            & PGC 2570478     & B17       & (R$^{\prime}$)SB(s)a                                                                                          \\
MCG 6-32-24     & B17       & (R$_1$)SAB(r)0/a                                                                                               & UGC 4596        & B17       & (R$_1$R$_2^{\prime}$)SA(rr)b                                                                                  \\
MCG 7-18-40     & B17       & (R$_2^{\prime}$)SB$_a$(l)b                                                                                     & UGC 4771        & B17       & (R$_2^{\prime}$)SAB$_a$(l)b                                                                                   \\
NGC 210         & deVA,TP   & (\RoneP\RtwoP)SAB(l,nr)b                                                                                       & UGC 5380        & B17       & (L)SB(r,bl)0/a                                                                                                \\
NGC 1079        & RC3,TP    & (\Rone\RtwoP)SAB\ans(\rpl,bl)a                                                                                 & UGC 5885        & B17       & (R$_1^{\prime}$)SAB$_a$($\underline{\rm r}$s,nr)ab                                                            \\
NGC 1291        & RC3,deVA  & (\Rone\RtwoP)SAB(l,nb)0/a                                                                                      & UGC 9418        & B17       & (R$_2^{\prime}$)SAB$_a$(r$^{\prime}$l)b                                                                       \\
NGC 1326        & BC96      & (\Rone)SAB(r,nr)0/a                                                                                            & UGC 10168       & B17       & (R$_1$R$_2^{\prime}$)SAB$_a$(r$^{\prime}$l)a                                                                  \\
NGC 1398        & B95,TP    & (R$^{\prime}$,R$_1$)SB($\underline{\rm r}$s)ab                                                                 & UGC 10712       & B17       & (R$_1$R$_2^{\prime}$)SAB(rl)b                                                                                 \\
NGC 1433        & BC96,TP   & (\RoneP)SB(p,$\underline{\rm r}s$,nr,nb)ab                                                                       & UGC 12646       & RC3       & (\RoneP)SB(r)ab                                                                                               \\
\hline
\end{tabular}
\end{table*}

In contrast to R$_1^{\prime}$ outer pseudorings, R$_2^{\prime}$
pseudorings are defined by an $\approx$270$^o$ winding of the spiral
arms relative to the bar (Figure~\ref{fig:figure-OLR}, bottom). The
defining characteristic in this case is a doubling of the spiral arms
in two opposing quadrants (indicated by the pairs of arrows in
Figure~\ref{fig:figure-OLR}, bottom right). Although closed R$_2$ rings
are not ruled out, they would be harder to recognize than R$_1$ without
the clear arm-doubling.

Most interesting is how some galaxies show a combination of these types
called R$_1$R$_2^{\prime}$, R$_1^{\prime}$R$_2^{\prime}$, or
R$_1$R$_2$. In these cases, both families of OLR periodic orbits
appear to be manifested in the morphology (Figure~\ref{fig:figure-OLR},
middle). This combined ring/pseudoring morphology, which had not been
predicted by Schwarz (1981), strongly supports the outer resonant
interpretation of the features.

Based on a detailed analysis of CSRG galaxies, Buta (1995) concluded
that R$_1$ and R$_1^{\prime}$ rings are more abundant than
R$_2^{\prime}$ rings, in the relative proportion 0.64:0.36. In the GZ2
catalogue, R$_1^{\prime}$ and R$_2^{\prime}$ are present in
approximately equal numbers, with R$_1$R$_2^{\prime}$ being less
abundant than either (paper I).

The CVRHS types of the 54 sample galaxies examined in this paper are
listed in Table~\ref{tab:types}. The GZ2 sample types are from the
catalogue in paper I, while those from the CSRG and a few others are
from different sources, mainly BC91, the de Vaucouleurs Atlas of
Galaxies (Buta et al. 2007=deVA), BC96, and a special supplementary table in
RC3 (de Vaucouleurs et al. 1991). OLR subclassifications of outer
rings and pseudorings are also recognized in the SDSS-based
morphological catalogues of Nair \& Abraham (2010) and Baillard et al.
(2011), and in the mid-infrared classifications of Buta et al. (2015).

\section{Dark Gaps in the Luminosity Distribution and the Dynamics of
Barred Galaxies}

A dark-spacer is a galaxy having well-defined and organized low surface
brightness regions that appear darker than their surroundings. In GZ2
galaxies, there are two main types of dark-spacer morphologies: (1) the
radial zone (gap) between an inner and outer ring appears dark, either
continuously or in 2-4 distinct sections [``(rR) dark-spacer"], and (2)
the interior of an inner ring appears darker than outside the ring
[``(r) dark-spacer"]. Only four examples of the latter type are
included in the sample (NGC 4608, NGC 5335, NGC 5686, and UGC 5380).
Section 5 describes both types, which have been recognized or at least
alluded to in previous studies.  For example, Gadotti \& de Souza
(2003) noted the existence of ``empty regions" around some bars that
stand out in a map of the residuals of the intensity distribution after
subtraction of a two-dimensional decomposition model. Two examples
Gadotti \& de Souza noted, NGC 4608 and NGC 5701, also independently
made it into our sample here. These authors consider that the bars in
these galaxies may have destroyed their disks through a secular
evolutionary process. Based on mid-infrared images from the 
S$^4$G, Laurikainen et al. (2013) note ``Sometimes an inner ring is
seen mainly because of dark space around the sides of the bar." 
In some cases the dark regions have a distinct banana shape, especially
if an average background is subtracted (as in the disk-subtracted image
for NGC 3081 from Buta \& Purcell 1998). The critical issue with
dark-spacer morphology is that the dark-looking regions are {\it not} a
result of dust extinction but are in many cases plateaus or genuine
minima in surface brightness. This is examined further in section 9.

The consistency in the appearance of the dark regions in (rR)
dark-spacers suggests a link to dynamics rather than dust extinction.
Here we make the following hypothesis: the dark gaps between the inner
and outer rings of OLR subclass galaxies are tied to the $L_4$ and
$L_5$ Lagrangian points in the gravitational potential of a bar or
bar-like oval. These points lie on a line perpendicular to the
perturbation major axis and, unlike the $L_1$, $L_2$ points, may be
stable or unstable (Binney \& Tremaine 1987=BT87, 2008=BT08). $L_1$ and
$L_2$ may lie at a different radius from $L_4$ and $L_5$, and a circle
through each pair of points defines what BT87 call the ``region of
corotation."

A useful potential for analytically deriving the location of these
points is the logarithmic potential $\Phi_L$ (equation 3.103 of BT08),
which is defined by an axis ratio $q$, a core radius $R_c$, and which
has a rotation curve which rises quickly near $R_c$ and then asymptotes
to $v = v_0$ = constant at radii $r >> R_c$.  The bar pattern speed is
$\Omega_p$ such that the radius of corotation $r_{CR} = v_0/\Omega_p$.
If the bar is oriented horizontally, then the $L_4$ and $L_5$ points
lie at $y_L$ = $\pm$$r_{CR}\sqrt{1-q^2 \big({R_c \over r_{CR}}\big)^2}$
(BT08, equation 3.131). This equation shows that if $q^2 \big({R_c
\over r_{CR}}\big)^2$ $<<$1, then $r_{CR}$ = $|y_L|$. Such a situation
would occur if the rotation curves of OLR subclass galaxies are
generally flat, and if $r_{CR}$ $>>$ $R_c$.  From known rotation curves
of several outer resonant subclass cases [NGC 1433 (Buta et al. 2001);
NGC 3081 (Buta \& Purcell 1998); NGC 6782 (Lin et al. 2008); IC 4214
(Salo et al. 1999); ESO 566$-$24 (Rautiainen et al. 2004)], and from
the analysis in this paper, these conditions are likely to be
satisfied.

The logarithmic potential is mostly an ovally-distorted disk (meaning
there is a flattening in the planar potential which affects the entire
disk, not just a part of the disk). Athanassoula (1992) examined the
locations of $L_1$, $L_2$, $L_4$, and $L_5$ for a more realistic model
barred galaxy having a bulge, a disk, and a Ferrers ellipsoid bar. She
found the radius of the $L_4$ and $L_5$ points to be slightly inside
the radius of the $L_1$ and $L_2$ points, and that \rcr\ was between
these radii.

An important question is, on what basis should we identify the dark
zones in (rR) dark-spacers as being linked to Lagrangian points? Why
should the area of such points be dark? The reason could be linked to
the stability of orbits around these points. According to Pfenniger
(1990; see also Contopoulos and Papayannopoulos 1980), ``if the $L_4$
and $L_5$ points are stable, they must trap high-energy orbits ...
while if unstable, they generate chaotic motion around them, which
might tend to depopulate these regions." Presumably, depopulation could
lead to a lower surface mass density around these points, and therefore
a lower surface brightness.

In the case of stability, the high energy orbits are the long period
orbits (LPOs; Contopoulos \& Grosbol 1989), which tend to be
banana-shaped orbits representing ``a slow tangential wallowing in the
weak non-axisymmetric component of $\Phi_L$" (BT87). Stars would be
trapped in these orbits circulating around the $L_4$ and $L_5$ points
as seen from the reference frame rotating with the bar.  Additionally,
short-period orbits (SPOs; Contopoulos \& Grosbol 1989) would also be
circulating around these points with smaller excursions (BT87). 

Numerical simulations have shown that the $L_4$ and $L_5$ points can be
unstable in the presence of a strong bisymmetric perturbation. In his
early test-particle models of barred galaxies, Schwarz (1981) noted how
the region between CR and OLR became increasingly depopulated as bar
strength increased. He states: ``As the bar strength is increased, the
region becomes larger and takes on the same appearance as the holes
which are common in almost all SB galaxies - the spaces in the
`$\theta$' shape." Contopoulos (1981) noted that this happens because
all the periodic orbits between CR and OLR become unstable, and therefore
are unable to trap any non-periodic orbits around them.

\begin{figure}
\includegraphics[width=\columnwidth]{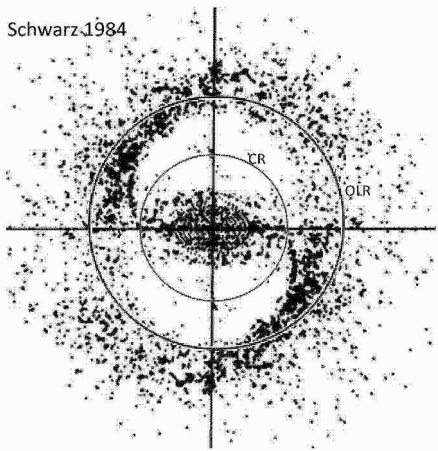}
\caption{A test-particle frame from a simulation (Schwarz 1984a, Figure 6) of a
strongly-barred disk galaxy showing the depopulation of the broad zones
around the $L_4$ and $L_5$ Lagrangian points. The CR and OLR circles
for the model are superposed. Reprinted with the permission
of the author.
}
\label{fig:schwarz1984}
\end{figure}

Schwarz (1984a) shows a strong bar model where the regions around $L_4$
and $L_5$ are unstable for the cases of a collisionless disk and a disk
with test-particle clouds allowed to collide. In both cases, the area
around these points is strongly depleted. Figure~\ref{fig:schwarz1984}
selects the frame of the collisional case after 10 bar rotations.
Superposed on the particle plot are circles having the radius of CR and
OLR. This shows how CR lies roughly in the middle of the gap between
the inner structure and an R$_1^{\prime}$ outer pseudoring. Similar
results were found by Byrd et al. (1994) and Salo et al. (1999).

{\it We interpret the dark spaces in (rR) dark-spacer galaxies to be
the locations of such unstable $L_4$ and $L_5$ points, and that the
average radius $r_{gp}$ of the minima of surface brightness or residual
intensity in these regions, based on parabolic fits to bar minor axis
profiles, traces the location of corotation, i. e., $r_{gp}$ =
$r_{CR}$.} We also expect that the more unstable the orbits around
$L_4$ and $L_5$ are, the darker and better-defined the gaps will be.
This can be seen comparing the models of Schwarz (1981) with those of
Schwarz (1984a), where the bar strength for the former models was only
half that used in Figures 4 and 6 of Schwarz (1984a).

\section{Surface Photometry}

For each GZ2 galaxy selected for photometry, the SDSS DAS Query form
was used to download Data Release 7 (DR7; Abazajian et al. 2009) fpC
images in the filters $u$, $g$, $r$, and $i$.  Although SDSS $u$-band
images are not as well-exposed as the $gri$ images, $u$ is still an
extremely useful filter for examining star formation in a galaxy, which
is very relevant to galactic rings. From these images, a smaller area
(512x512 or 256x256 pixels) centred on the galaxy was extracted. Each
filter image was then cleaned of contaminating foreground and
background objects using IRAF routine IMEDIT. The background was
subtracted by fitting a plane to either pixels in a specified border
region or pixels in specified squares in the corners of the images. The
fitted plane was then subtracted from both the cleaned and uncleaned
images. Filter alignments were made using IRAF routine IMALIGN.

The above procedure was applied mainly to objects that had only a
single field to download. In some cases, 2-3 fields were available,
allowing an improved signal-to-noise ratio for surface photometry at
fainter light levels. To combine multiple images after background
subtractions, IRAF routines IMALIGN, IMSHIFT and, where needed,
IMCOMBINE, were used to remove small shifts and rotations.

Image calibrations were taken from the tsfield fits tables
that accompany the fpC images. The zero points were derived as

$$zp = -(a+kx)+2.5log(A_{pix}t)$$

\noindent
where $a$ is the zero point for 1 second, $k$ is the extinction
coefficient, $x$ is the airmass, and $A_{pix}$ is the area of a single
pixel in square arcsec, where 1 pix = 0\rlap{.}$^{\prime\prime}$396.
The uniform exposure time for all SDSS images is $t$ = 53.907 seconds.

The IRAF routine ELLIPSE was used to fit ellipses to the isophotes of
each galaxy. The main goal of these fits was an estimate of the
orientation parameters of the disks, meaning the inclination through
the mean projected disk axis ratio $q_d$ and the line of nodes through
the photometric major axis position angle $\phi_d$. These are compiled
in Table~\ref{tab:orient}. To insure that the faintest outer isophotes
are used for this purpose, the ellipse fits were generally performed on
2x2 or 4x4 block-averaged images. With estimates of the orientation
parameters of the galaxies, the IRAF routine IMLINTRAN was used to
deproject the galaxies. All images illustrated in this paper are
deprojected and largely cleaned of foreground stars except where
noted.

For each GZ2 galaxy, we derived (a) $ugri$ azimuthally-averaged
luminosity profiles; (b) the deprojected radius and position angle of
the bar or oval; (c) luminosity profiles parallel and perpendicular to
the bar axis; and (d) relative Fourier amplitudes and phases.  No
photometric decompositions of the luminosity profiles are presented in
this paper. The emphasis instead is on profile shapes, feature
contrasts, and morphology.

The GZ2 subsample accounts for 33 of the sample of 54 galaxies. The
remaining 21 cases are mainly from the CSRG and involve Johnson-Cousins
$BVI$ CCD images from the author's private image library. Many of these
images were illustrated in the study of BC91.
Although most of these images are of high quality, they lack the
homogeneity of the SDSS images. Zero points were determined either from
published $BVI$ photoelectric photometry (e.g., Buta and Crocker 1992;
Buta et al. 1995; Buta and Williams 1995), or standard star
transformations, with image scales ranging from
0\rlap{.}$^{\prime\prime}$19 to 0\rlap{.}$^{\prime\prime}$6
pix$^{-1}$. The final cleaned and background-subtracted $BVI$ 
images were used to derive the same kinds of profiles and parameters
as were derived for the GZ2 galaxies.

We focus first on two GZ2 examples with exceptionally dark but very
regular spaces flanking a bar or oval, to show the photometric
information we have derived for each galaxy, and to illustrate our
procedure for deriving $r_{gp}$.

\section{Two Examples}

\subsection{UGC 4596}

\begin{figure}
\includegraphics[width=\columnwidth]{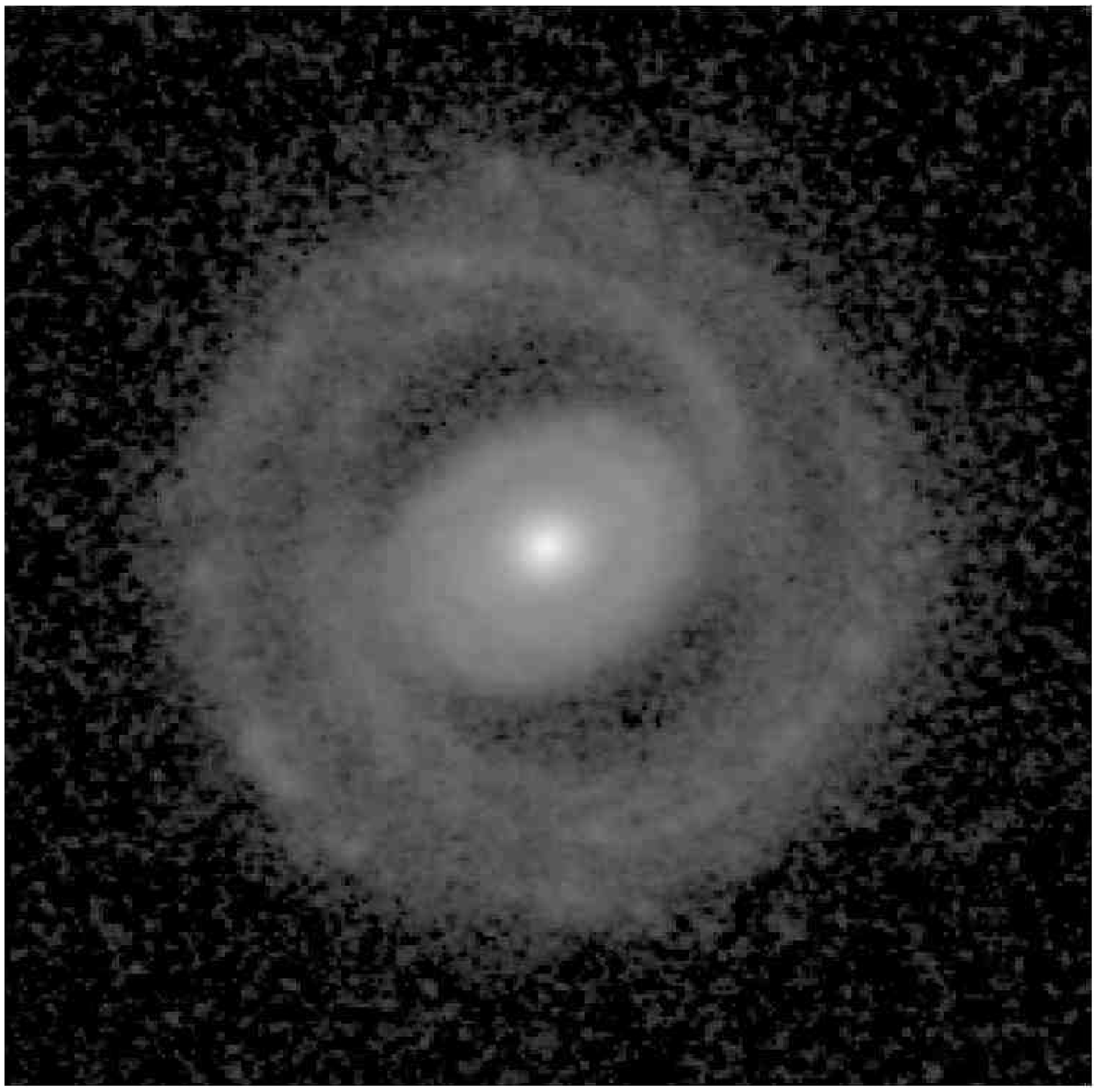}
\includegraphics[width=\columnwidth]{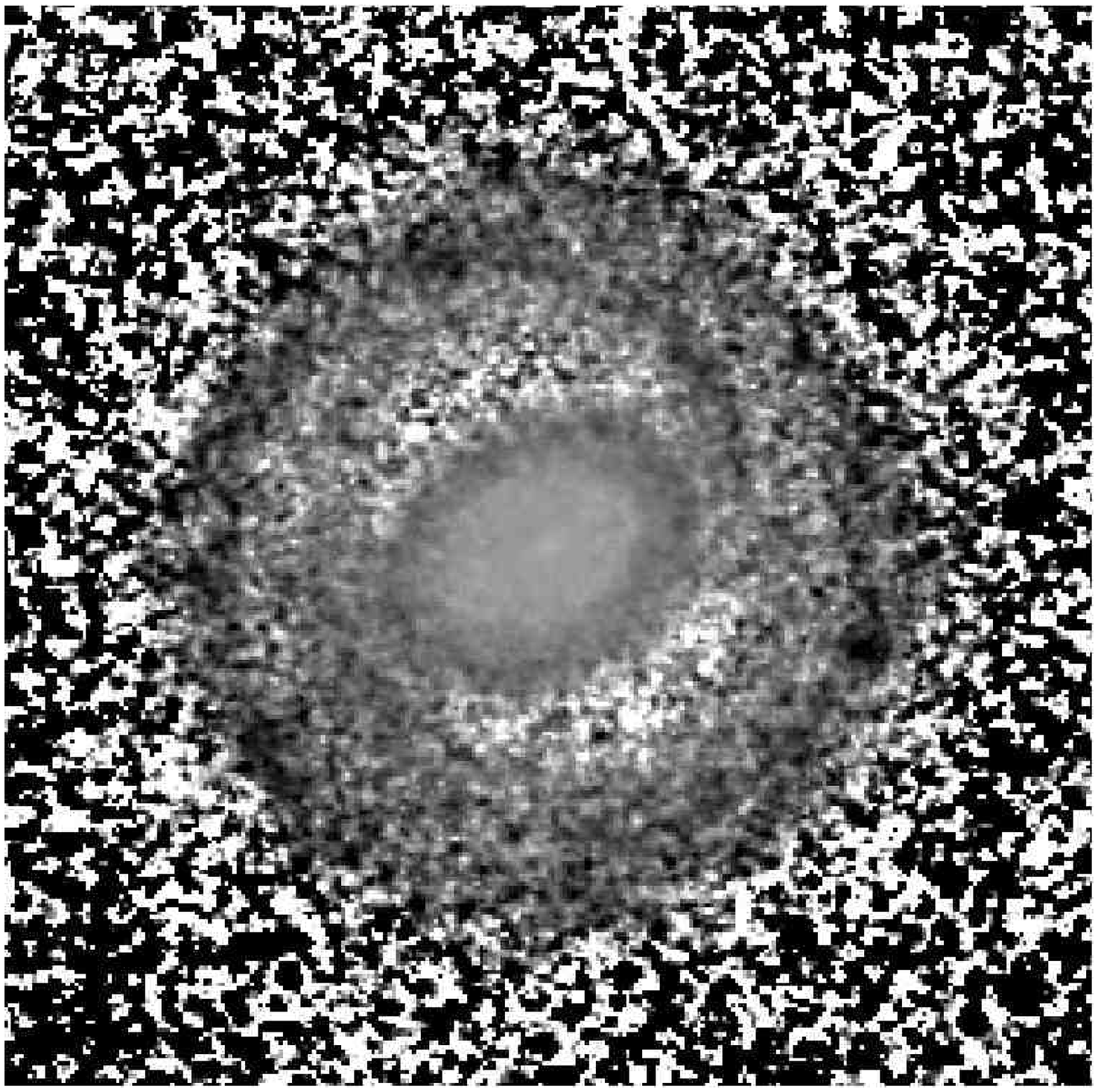}
\caption{Deprojected $g$-band image (top) and $g-i$ colour index map of UGC
4596, type (R$_1$R$_2^{\prime}$)SA(rr)b.  Both images are in units of
mag arcsec$^{-2}$. The $g-i$ colour index map is coded such that redder
features are light and bluer features are dark. The field shown is
1\rlap{.}$^{\prime}$69 arcminutes square.} 
\label{fig:ugc4596-images}
\end{figure}

UGC 4596 (Figure~\ref{fig:ugc4596-images}) is a strong (rR) dark-spacer
having four well-defined ring or pseudoring features. Its CVRHS
classification is (R$_1$R$_2^{\prime}$)SA(rr)b.  Although there is no
conventional bar, the two inner rings lie in the outer regions of a
massive oval, which must act as the bar of the system. Based on ellipse
fits to $gri$ isophotes in the radius range
31$^{\prime}$$^{\prime}$--48$^{\prime}$$^{\prime}$, the average disk
axis ratio and major axis position angle are 0.855$\pm$0.014 and
130\rlap{.}$^{\circ}$1$\pm$3\rlap{.}$^{\circ}$8, respectively. These
parameters were used to derive azimuthally-averaged luminosity profiles
and to deproject the $ugri$ images.

\begin{figure}
\includegraphics[width=\columnwidth,bb=14 14 600 700]{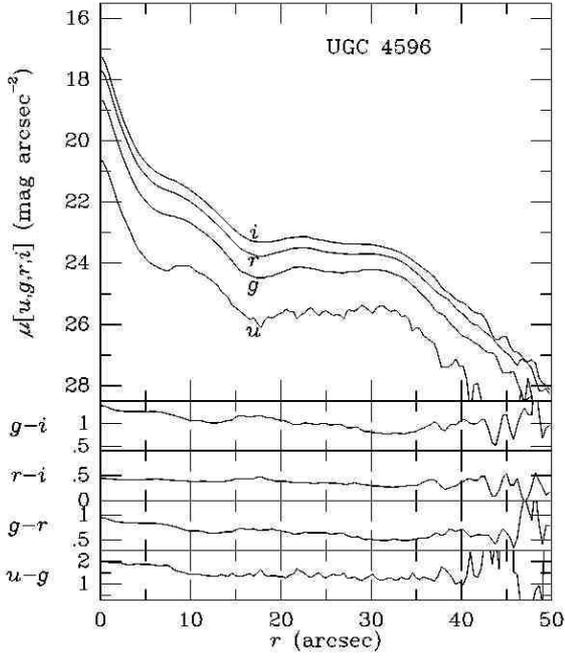}
\caption{Azimuthally-averaged $ugri$ surface brightness and colour index profiles of UGC 4596}
\label{fig:ugc4596a}
\end{figure}

\begin{figure}
\includegraphics[width=\columnwidth,bb=14 14 600 730]{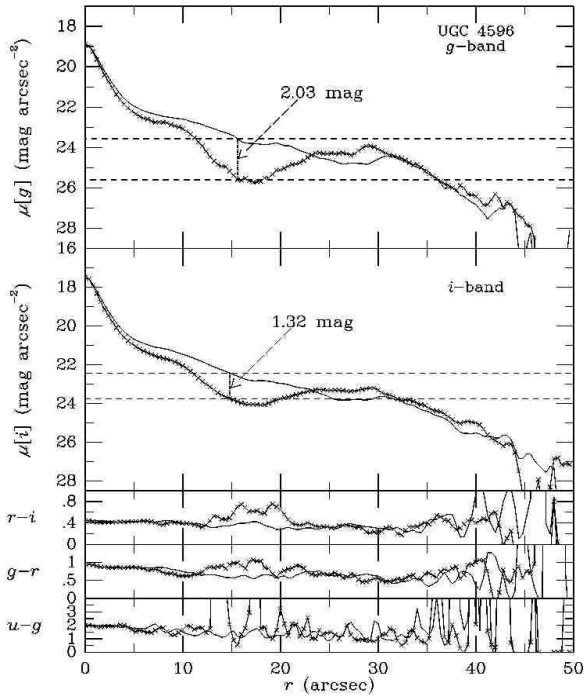}
\caption{Profiles of UGC 4596 parallel (solid
curve) and perpendicular (solid curve marked with crosses) to the
deprojected massive oval}
\label{fig:ugc4596b}
\end{figure}

Figure~\ref{fig:ugc4596a} shows the $ugri$ azimuthally-averaged surface
brightness and colour index profiles of UGC 4596. The galaxy is an
exceptional example where a strong R$_1$R$_2^{\prime}$ morphology
causes two clear ``bumps" in the outer plateau. The profiles in all
four filters show a steep decline beyond the second outer bump.
Figure~\ref{fig:ugc4596b} compares $g$- and $i$-band profiles along
axes parallel and perpendicular to the deprojected oval. The vertical
lines show the radial positions where the difference between the two
profiles is maximum (see also section 10). In the $g$-band, this
difference amounts to 2.03 mag, but in the $i$-band, the difference
amounts to only 1.32 mag. The parallel and perpendicular profiles show
that the gaps in UGC 4596 are redder than much of the rest of the disk.
The explanation for this is likely to be that recent star formation
extends over much of the disk (as seen in both the $g-i$ colour index
map and in the azimuthally-averaged $u$-band surface brightness
profile), except for the gaps which are red most likely because their
light is dominated by older stars (section 9). The redness of the gaps
is still present but subdued in the azimuthally-averaged colour index
profiles.

\begin{figure}
\includegraphics[width=\columnwidth,bb=14 14 600 280]{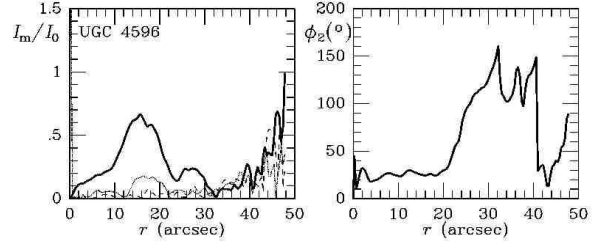}
\caption{Fourier intensity analysis of UGC 4596:  (left) $i$-band
relative Fourier amplitudes as a function of radius, for $m$ = 2 (solid
curve), 4 (dotted curve), and 6 (dashed curve); (right) Phase of $m$ =
2 Fourier component. This phase is in the arbitary units of the
deprojected images.}
\label{fig:ugc4596c}
\end{figure}

Figure~\ref{fig:ugc4596c} shows relative $m$ = 2, 4, and 6 Fourier
intensity amplitudes and the phase of the $m$ = 2 component as a
function of radius, based on the deprojected $i$-band image. In spite
of the lack of an apparent bar, the maximum $m$ = 2 relative amplitude
is $A_2$ = 0.66, with much lower $m$ = 4 and 6 amplitudes.

\begin{figure}
\includegraphics[width=\columnwidth,bb=14 14 600 500]{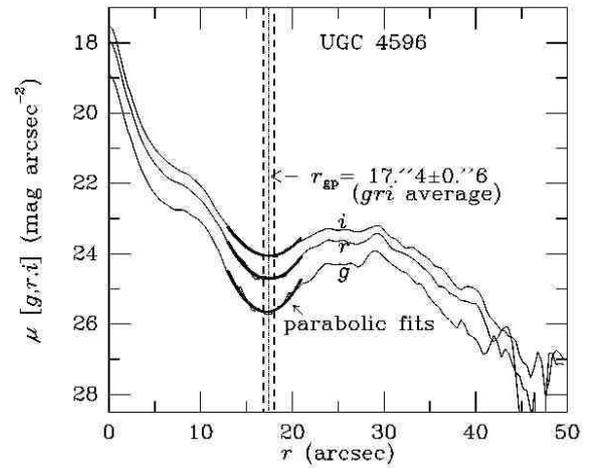}
\caption{Illustration of the method used to determine $r_{gp}$, the
radius in the gap region where the surface intensity along the bar
minor axis is a minimum, for UGC 4596 and other similar examples of
(rR) dark-spacers where the gaps are especially well-defined. The
parabolic fits to the $\mu$ values in these regions are indicated by
the darker points.}
\label{fig:ugc4596-proc}
\end{figure}

Figure~\ref{fig:ugc4596-proc} shows how $r_{gp}$ is derived.  In this
case, for each filter we fitted a parabola to the surface brightnesses
in the gap region. An uncertainty was assigned to the derived radius
using the standard deviation of the data points around the fitted
parabola. This gave $r_{gp}$ =
17\rlap{.}$^{\prime\prime}$2$\pm$1\rlap{.}$^{\prime\prime}$2,
17\rlap{.}$^{\prime\prime}$5$\pm$1\rlap{.}$^{\prime\prime}$1, and
17\rlap{.}$^{\prime\prime}$5$\pm$0\rlap{.}$^{\prime\prime}$8, for the
$g$, $r$, and $i$ filters, respectively.  The final value, $<r_{gp}>$ =
17\rlap{.}$^{\prime\prime}$4$\pm$0\rlap{.}$^{\prime\prime}$6, is a
weighted average over the three filters using the estimated
uncertainties $\sigma_g$ = 1\rlap{.}$^{\prime\prime}$2, $\sigma_r$ =
1\rlap{.}$^{\prime\prime}$1, and $\sigma_i$ =
0\rlap{.}$^{\prime\prime}$8 as weights $w$=1/$\sigma^2$.

The above procedure to get $r_{gp}$ works well for UGC 4596 because its
oval minor axis gaps are dark and well-defined. In some cases, however,
the apparent gaps are subtle, such that in profiles the gap minima are
diffcult to measure reliably in a regular surface brightness profile.
In such cases, we subtracted a heavily median-smoothed background from
the intensity images and fitted parabolas to the intensity residuals in
the enhanced gap region. This is illustrated and discussed further in
the Appendix.

\subsection{NGC 5335}

NGC 5335 provides a remarkable contrast to UGC 4596. The dark zones in
UGC 4596 are located in the transition region between the galaxy's
multiple rings, while the dark zones in NGC 5335 lie inside a
conspicuous inner ring, the only ring in that system. NGC 5335 is the
strongest (r) dark-spacer in the GZ2 catalog and it suggests that dark
spaces inside inner rings are not merely contrast effects but are
actually dynamical in nature.

\begin{figure}
\includegraphics[width=\columnwidth]{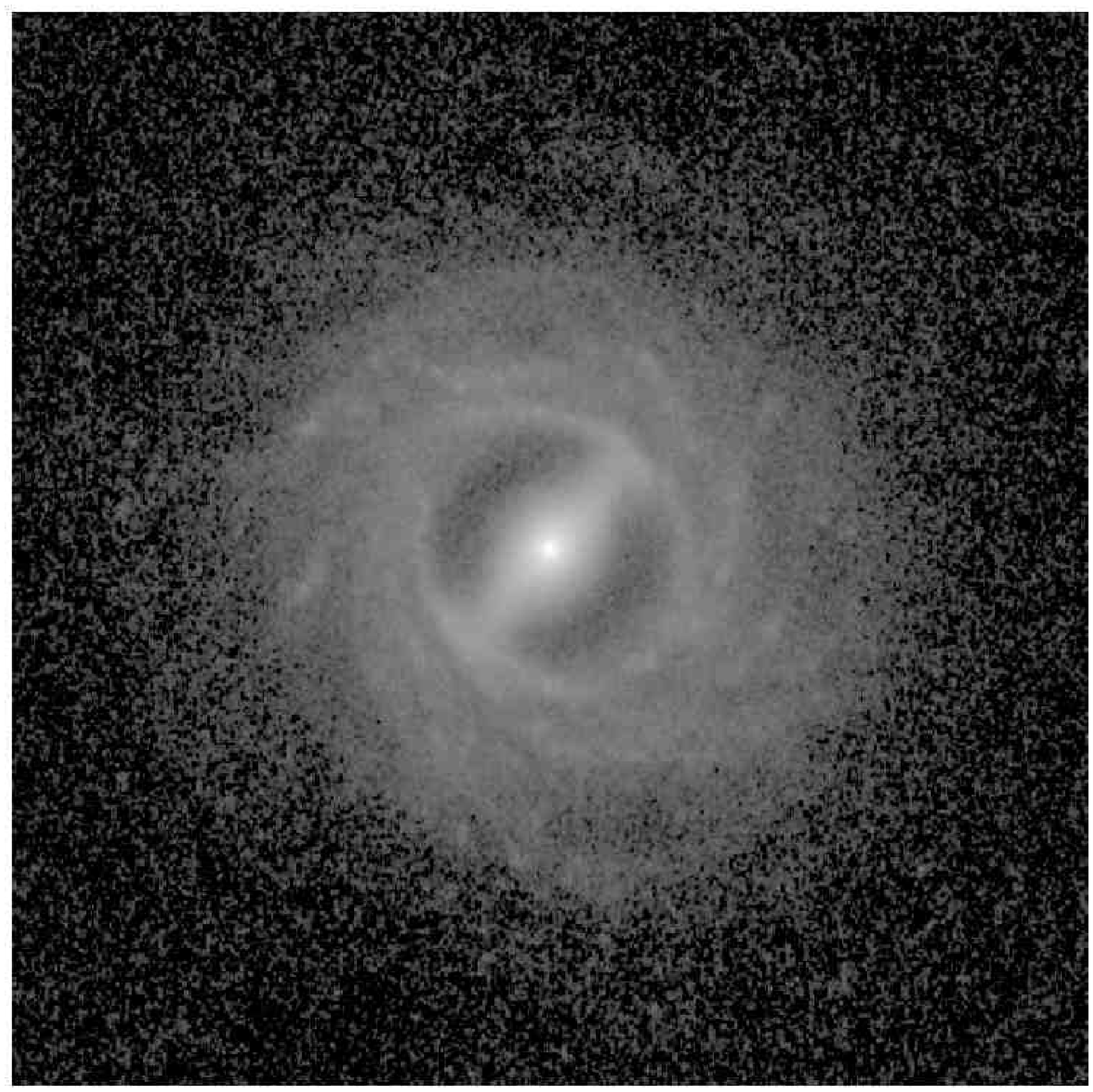}
\includegraphics[width=\columnwidth]{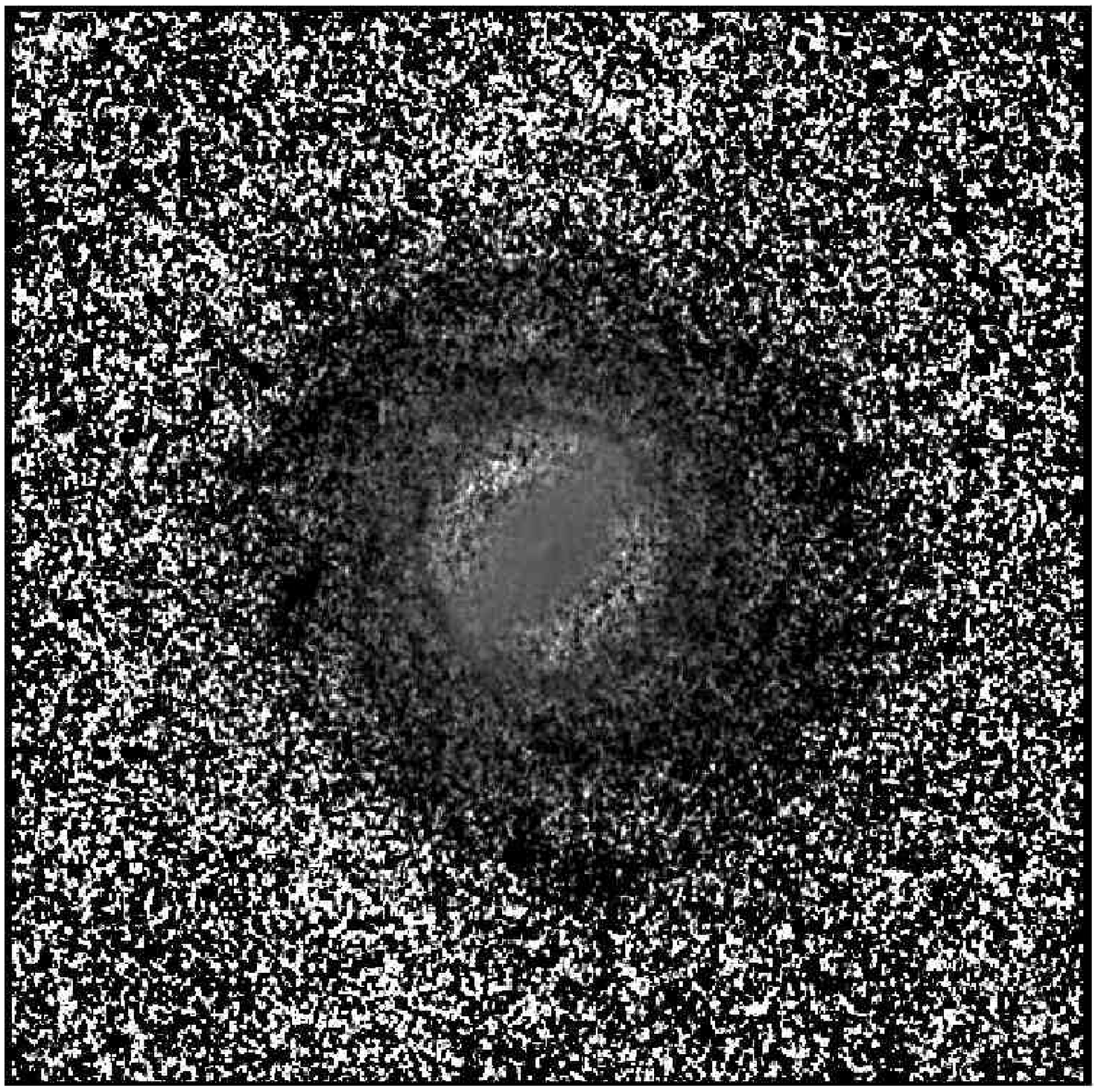}
\caption{Deprojected $g$-band image (top) and $g-i$ colour index map of NGC
5335, type SB($\underline{\rm r}$s,bl)ab.  The field shown is
3\rlap{.}$^{\prime}$38 arcminutes square.}
\label{fig:NGC5335-images}
\end{figure}

\begin{figure}
\includegraphics[width=\columnwidth,bb=14 14 600 800]{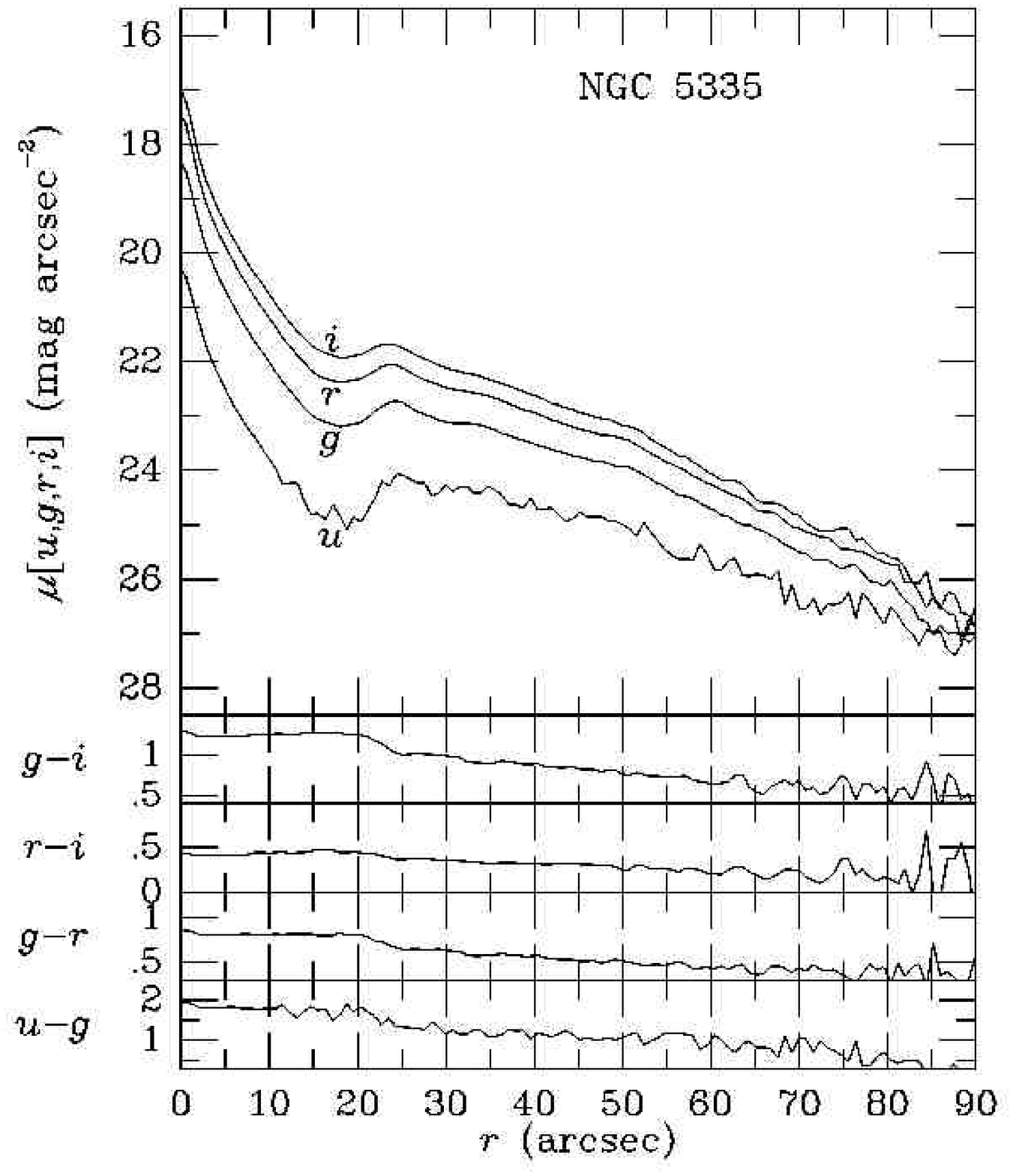}
\caption{Azimuthally-averaged $ugri$ surface brightness and colour index profiles of NGC 5335}
\label{fig:NGC5335a}
\end{figure}

\begin{figure}
\includegraphics[width=\columnwidth,bb=14 14 600 830]{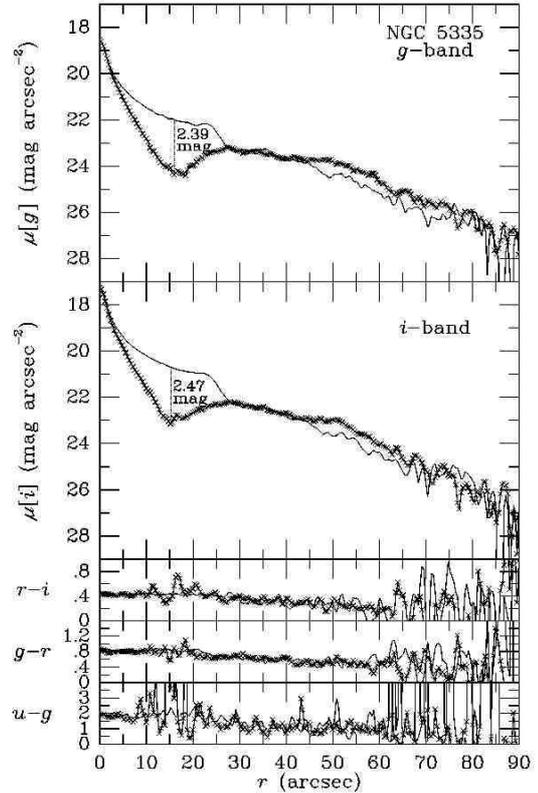}
\caption{Profiles of NGC 5335 parallel (solid
curve) and perpendicular (solid curve marked with crosses) to the
deprojected bar axis}
\label{fig:NGC5335b}
\end{figure}

\begin{figure}
\includegraphics[width=\columnwidth,bb=14 14 600 280]{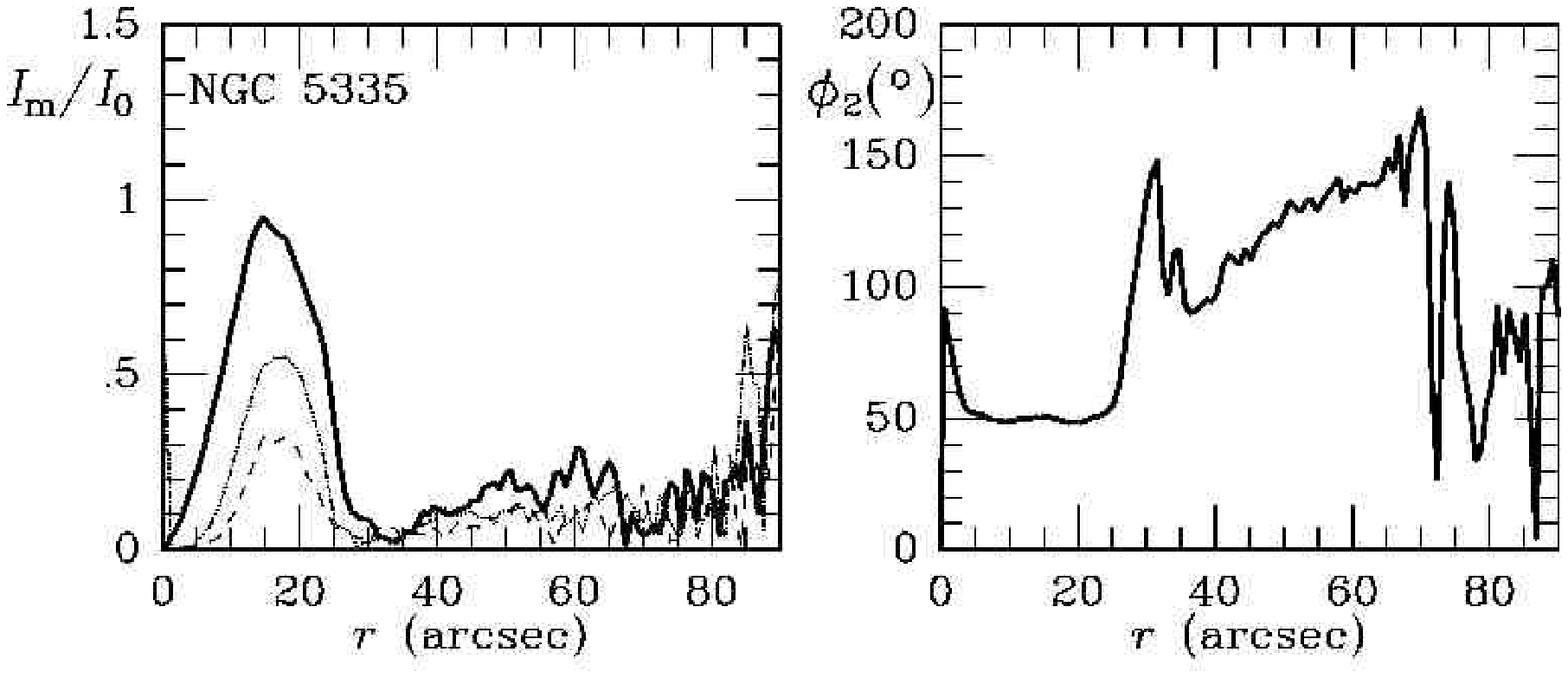}
\caption{Fourier intensity analysis of NGC 5335: (left) $i$-band relative Fourier amplitudes as a function of radius, for $m$ =
2 (solid curve), 4 (dotted curve), and 6 (dashed curve); (right) Phase of $m$ = 2 Fourier component }
\label{fig:NGC5335c}
\end{figure}

Figures~\ref{fig:NGC5335-images}--~\ref{fig:NGC5335c} show deprojected
images, azimuthally-averaged profiles, bar major and minor axis
profiles, and relative Fourier intensity profiles for NGC 5335. The
CVRHS classification of the galaxy is SB($\underline{\rm r}$s,bl)ab,
and the deprojection is based on a disk axis ratio of 0.856 and a disk
major axis position angle of 109\rlap{.}$^o$4. Both the $g$-band image
and the $g-i$ colour index map show a considerable amount of noise in
the dark spaces. The colour index maps are in fact so noisy in this
region that it is difficult to decide what the colours of the dark
spaces are. This is definitively clarified by the azimuthally-averaged
colour index profiles, all of which are nearly constant across the bar
region. {\it The colours of the dark spaces are nearly identical to
those of the bar.} This suggests that a dynamical effect associated
with the bar has caused the dark spaces.

The deprojected bar major and minor axis profiles in
Figure~\ref{fig:NGC5335b} show that both the $g$- and $i$-band surface
brightnesses of NGC 5335 are about 2.4 mag fainter at radii of
15$^{\prime\prime}$--16$^{\prime\prime}$ compared to the bar axis, and
that the surface brightness inside the ring is not achieved again until
a radius of nearly 59$^{\prime\prime}$ in $g$ and 53$^{\prime\prime}$
in $i$. The shapes of the $u-g$, $g-r$, and $r-i$ colour index profiles
is nearly identical along the two axes, except for the increased noise
level between 10$^{\prime\prime}$ and 20$^{\prime\prime}$ due to the
drastic difference in surface brightness between the bar major and
minor axis profiles in this range.

The inner ring of NGC 5335 is also somewhat unusual. The average inner
ring in an SB galaxy is intrinsically oval (mean axis ratio =
0.81$\pm$0.06) and is aligned parallel to the bar (Buta 1995). The
visually-mapped deprojected inner ring in NGC 5335 has an axis ratio of
0.893 and is oriented 82$^o$ to the deprojected bar axis.
Thus, the inner ring of NGC 5335 is not a typical SB inner ring. When
combined with a multi-armed outer spiral pattern as opposed to an outer
ring or pseudoring, then NGC 5335 becomes even more atypical.
The relative Fourier intensity profiles of NGC 5335 also show that $A_2$ is
0.95, making the bar one of the strongest in our sample. 

\begin{figure}
\includegraphics[width=\columnwidth,bb=14 14 600 500]{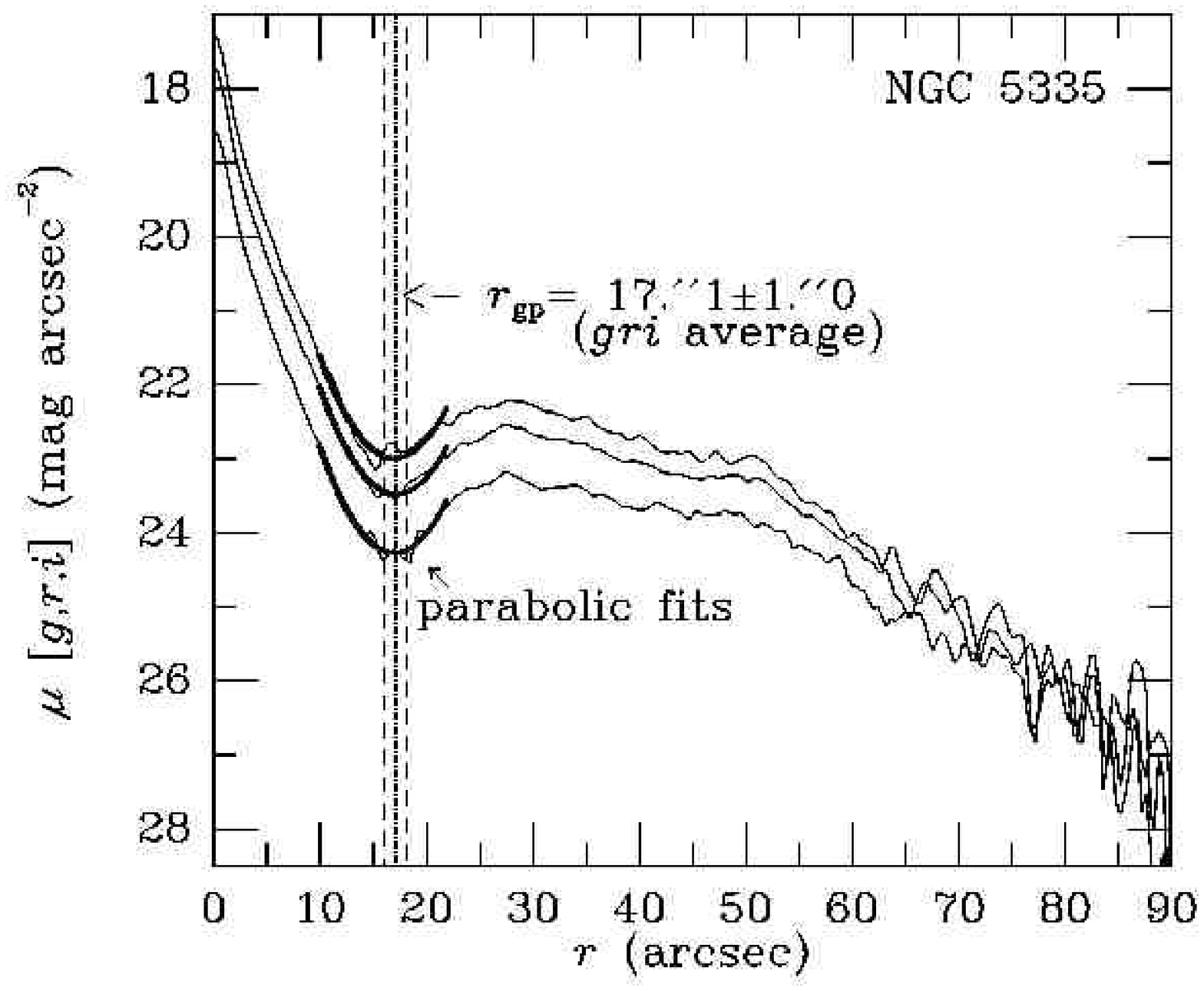}
\caption{Illustration of the method used to determine $r_{gp}$
for NGC 5335 and other similar examples of (r) dark-spacers.
}
\label{fig:NGC5335-proc}
\end{figure}

Figure~\ref{fig:NGC5335-proc} shows the derivation of $r_{gp}$ for NGC 5335.
As for UGC 4596, we fitted a parabola to
the surface brightnesses in the dark gap region in each filter and then
computed a weighted average.  The resulting value, $r_{gp}$ =
17\rlap{.}$^{\prime\prime}$1$\pm$1\rlap{.}$^{\prime\prime}$0, 
has a larger relative error than UGC 4596 because of the high noise
level in the gap zone.

\section{Application of the Gap Method to the Sample}

The gap method is a photometric/morphological approach to locating the
corotation resonance. Other more sophisticated methods have been
proposed and applied for the same purpose, including the Tremaine \&
Weinberg (1984) photometric/kinematic method, the Canzian (1993)
geometric phase method, the potential-density phase shift method (Zhang
\& Buta 2007), and most recently the streaming phase reversal
(kinematic) method (Font et al. 2011, 2014).  In the remainder of this
paper, the method is applied to a total sample of 50 (rR) dark-spacer
galaxies and 4 (r) dark-spacer cases. The steps are:

\noindent
1. Clean images of foreground and background objects, and then subtract off
the sky level as a plane using IRAF routine IMSURFIT.

\noindent
2. Using IRAF routine ELLIPSE, fit ellipses to faint outer isophotes to
get reliable orientation parameters, including the mean disk axis ratio
$q_d$ and the mean disk position angle $\phi_d$. The adopted values are
listed in Table~\ref{tab:orient}, including for the CSRG sample. The
typical uncertainty in $q_d$ is $\pm$0.010 or less, while that for
$\phi_d$ can be 2$^o$-20$^o$ depending on how round the isophotes are.

\begin{table*}
\centering
\caption{Orientation parameters and filters used for deriving $r_{gp}$.
Col. 1: galaxy name; cols. 2 and 3: mean disk axis ratio and major axis
position angle (degrees) based on ellipse fits to faint outer
isophotes; col. 4: filters and method used to derive $r_{gp}$. ``gri"
means that $r_{gp}$ is based on parabolic fits to surface brightnesses
(mag arcsec$^{-2}$) in SDSS filters $gri$. ``intgri" means that $r_{gp}$
is based on parabolic fits to residual intensities in $gri$ after
subtraction of a heavily median-smoothed background.}
\label{tab:orient}
\begin{tabular}{lcrllcrl}
\hline
Name & $q_d$ & $\phi_d$ & filters & Name & $q_d$ & $\phi_d$ & filters \\
 1 & 2 & 3 & 4 & 1 & 2 &  3 & 4 \\ 
\hline
CGCG 8-10  & 0.783 & 159.2 & gri      & NGC 2665      & 0.767 & 121.1 & BVI      \\
CGCG 13-75  & 0.870 & 108.6 & gri      & NGC 2766      & 0.403 & 127.0 & intgri   \\
CGCG 65-2  & 0.602 &  74.2 & intgri   & NGC 3081      & 0.850 &  89.0 & BI       \\
CGCG 67-4  & 0.792 &  39.7 & intgri   & NGC 3380      & 0.963 &  62.9 & intgri   \\
CGCG 73-53  & 0.825 &  98.3 & intgri   & NGC 4113      & 0.830 & 133.1 & gri      \\
CGCG 185-14  & 0.478 &  28.2 & intgri   & NGC 4608      & 0.873 & 110.5 & intgri   \\
CGCG 263-22  & 0.927 &  75.4 & intgri   & NGC 4736      & 0.865 & 116.0 & gi       \\
ESO 325-28   & 0.974 & 128.3 & intBVI   & NGC 4935      & 0.949 &  84.9 & intgri   \\
ESO 365-35   & 0.751 & 155.2 & BVI      & NGC 5132      & 0.800 & 102.8 & gri      \\
ESO 426-2   & 0.906 &  64.7 & BVI      & NGC 5211      & 0.772 & 175.0 & gri      \\
ESO 437-33   & 0.718 &   0.9 & BVI      & NGC 5335      & 0.856 & 109.4 & gri      \\
ESO 437-67   & 0.883 &   5.0 & BVI      & NGC 5370      & 0.885 &  76.1 & intgri   \\
ESO 566-24   & 0.749 &  69.1 & BVI      & NGC 5686      & 0.922 &  29.4 & intgri   \\
ESO 575-47   & 0.811 & 111.5 & BVI      & NGC 5701      & 0.960 &  49.2 & gri      \\
IC 1223       & 0.761 &   0.4 & intgri   & NGC 6782      & 0.894 &  34.9 & BVI      \\
IC 1438       & 0.898 & 143.1 & intBVI   & NGC 7098      & 0.557 &  70.1 & BVI      \\
IC 2473       & 0.739 &  32.3 & gri      & PGC 54897   & 0.847 &  84.2 & gri      \\
IC 2628       & 0.911 & 151.8 & intgri   & PGC 1857116   & 0.549 &  36.4 & intgri   \\
IC 4214       & 0.616 & 170.5 & BVI      & PGC 2570478   & 0.701 &  91.3 & intgri   \\
MCG 6-32-24  & 0.743 & 141.0 & intgri   & UGC 4596     & 0.855 & 130.1 & gri      \\
MCG 7-18-40  & 0.938 & 179.9 & intgri   & UGC 4771     & 0.634 &  30.2 & gri      \\
NGC 210      & 0.690 & 159.0 & intR    & UGC 5380     & 0.668 & 147.9 & intgri   \\
NGC 1079      & 0.592 &  81.3 & intBVI   & UGC 5885     & 0.332 & 136.5 & gri      \\
NGC 1291      & 1.000 &   0.0 & BVRI  & UGC 9418     & 0.972 &  54.4 & gri      \\
NGC 1326      & 0.750 &  73.4 & intBVI & UGC 10168     & 0.725 & 167.5 & gri      \\
NGC 1398      & 0.702 &  95.9 & intBVI   & UGC 10712     & 0.960 &  84.2 & gri      \\
NGC 1433      & 0.850 &  17.0 & BI       & UGC 12646     & 0.909 &  28.0 & BV       \\
\hline
\end{tabular}
\end{table*}

\noindent
3. For each galaxy, derive azimuthally-averaged luminosity and colour
index profiles within the fixed ellipses defined by $q_d$ and $\phi_d$,
in order to highlight the homogeneity of the the average structure of
each galaxy. These profiles are shown in Figures~\ref{fig:ugc4596a} and
~\ref{fig:NGC5335a} for UGC 4596 and NGC 5335, respectively, and for
the remaining 52 sample galaxies in Figure~\ref{fig:azim} (see
Appendix). Exponential extrapolations of the surface brightness
profiles were used to estimate total magnitudes and colour indices for
the galaxies. These are provided in Table~\ref{tab:totmags} for the GZ2
galaxies and Table~\ref{tab:totmagsBVI} for the CSRG and other sample
galaxies.

The last three lines in Table~\ref{tab:totmags} compare the derived
total $g$-band magnitudes, $g-i$, $g-r$, and $r-i$ colours with the SDSS
pipeline ``modelMag" values described by Stoughton et al. (2002). In
general, the Table~\ref{tab:totmags} values are brighter than the
modelMag values by an amount that increases with decreasing
wavelength. The last three lines in Table~\ref{tab:totmagsBVI}
similarly compare the derived $BVI$ parameters with published values
based on mean growth curve fits to photoelectric multi-aperture
photometry. Since this was the same photometry used to calibrate most
of the non-GZ2 images, the good agreement on the colours is not
surprising. However, the $B_T$ magnitudes in Table~\ref{tab:totmagsBVI}
are 0.08 mag fainter than the published values, most likely because
of the different extrapolations used.

Also listed in both tables are the absolute magnitudes of the galaxies,
ranging from $-$18.7 to $-$21.5 for the $g$-band and $-$19.2 to $-$21.3
for the $B$-band, in each case typical of intermediate to high mass
disk galaxies. Distance moduli are taken from the NASA/IPAC
Extragalactic Database (NED). The colours of the galaxies are further
examined in section 9.

\begin{table*}
\centering
\caption{Integrated magnitudes and colours of 33 GZ2 sample galaxies plus 1 non-GZ2 case.
Col.  1: galaxy name; cols. 2-6: integrated magnitude and colours,
obtained by extrapolation of an exponential beyond the outer regions;
col. 7:  distance modulus with respect to the galactic standard of rest
(GSR), from the NASA/IPAC Extragalactic Database (NED); col. 8:
absolute $g$-band magnitude corrected for galactic extinction from NED.
The last three lines compare the derived magnitudes and colours with
SDSS DR9 ``modelMag" values (Stoughton et al. 2002) for the same
galaxies, extracted through NED.}
\label{tab:totmags}
\begin{tabular}{lrrrrrrr}
\hline
Name  & $g_T$ & $(u-g)_T$ & $(g-i)_T$ & $(g-r)_T$ & $(r-i)_T$ & $m-M$ & $M_g^o$ \\
 1 & 2 & 3 & 4 & 5 & 6 & 7 & 8 \\  
\hline
CGCG  8-10       &  15.24 &   1.74 &   1.25 &   0.80 &   0.45 &  35.91 & $-$20.8\\
CGCG 13-75       &  14.62 &   1.47 &   1.09 &   0.70 &   0.40 &  35.07 & $-$20.6\\
CGCG 65-2        &  15.05 &   1.31 &   1.01 &   0.66 &   0.35 &  34.36 & $-$19.5\\
CGCG 67-4        &  15.43 &   1.36 &   1.04 &   0.65 &   0.39 &  36.20 & $-$20.8\\
CGCG 73-53       &  14.96 &   .... &   1.14 &   0.75 &   0.40 &  34.90 & $-$20.0\\
CGCG 185-14      &  14.87 &   1.47 &   1.05 &   0.68 &   0.37 &  35.13 & $-$20.3\\
CGCG 263-22      &  14.96 &   1.58 &   1.31 &   0.87 &   0.45 &  35.61 & $-$20.8\\
MCG 6-32-24      &  15.21 &   1.38 &   1.03 &   0.67 &   0.36 &  35.37 & $-$20.3\\
MCG 7-18-40      &  15.86 &   1.41 &   1.10 &   0.71 &   0.40 &  37.24 & $-$21.5\\
IC 1223          &  14.66 &   1.66 &   1.25 &   0.83 &   0.42 &  35.48 & $-$20.9\\
IC 2628          &  15.20 &   1.46 &   1.08 &   0.70 &   0.38 &  36.18 & $-$21.1\\
IC 2473          &  13.96 &   1.37 &   1.01 &   0.65 &   0.36 &  35.21 & $-$21.3\\
NGC 2766         &  14.10 &   1.44 &   1.11 &   0.71 &   0.39 &  33.77 & $-$19.8\\
NGC 3380         &  13.02 &   1.31 &   0.99 &   0.62 &   0.37 &  31.66 & $-$18.7\\
NGC 4113         &  14.35 &   1.15 &   0.80 &   0.51 &   0.29 &  35.19 & $-$20.9\\
NGC 4608         &  11.46 &   1.75 &   1.15 &   0.74 &   0.41 &  31.95 & $-$20.5\\
NGC 4736$^a$     &   8.32 &   .... &   1.04 &   .... &   .... &  28.47 & $-$20.2\\
NGC 4935         &  13.69 &   1.28 &   0.95 &   0.60 &   0.35 &  34.71 & $-$21.1\\
NGC 5132         &  13.52 &   1.56 &   1.21 &   0.79 &   0.42 &  35.00 & $-$21.6\\
NGC 5211         &  13.01 &   1.42 &   1.10 &   0.73 &   0.37 &  33.50 & $-$20.6\\
NGC 5335         &  13.09 &   1.27 &   1.00 &   0.64 &   0.36 &  34.00 & $-$21.0\\
NGC 5370         &  13.49 &   1.66 &   1.13 &   0.73 &   0.40 &  33.20 & $-$19.8\\
NGC 5686         &  14.62 &   1.64 &   1.12 &   0.72 &   0.40 &  33.83 & $-$19.2\\
NGC 5701         &  11.43 &   1.55 &   1.12 &   0.69 &   0.43 &  31.58 & $-$20.3\\
PGC 54897        &  14.59 &   1.31 &   1.00 &   0.64 &   0.36 &  34.94 & $-$20.5\\
PGC 1857116      &  15.65 &   1.33 &   1.09 &   0.70 &   0.39 &  ..... & ......\\
PGC 2570478      &  15.57 &   1.83 &   1.29 &   0.88 &   0.41 &  ..... & ......\\
UGC 4596         &  14.67 &   1.54 &   1.10 &   0.72 &   0.39 &  35.54 & $-$21.0\\
UGC 4771         &  14.67 &   1.61 &   1.13 &   0.74 &   0.39 &  35.88 & $-$21.3\\
UGC 5380         &  14.39 &   .... &   1.19 &   0.78 &   0.41 &  34.60 & $-$20.4\\
UGC 5885         &  14.91 &   1.35 &   1.12 &   0.71 &   0.41 &  35.59 & $-$20.8\\
UGC 9418         &  14.45 &   1.54 &   1.05 &   0.68 &   0.37 &  35.66 & $-$21.3\\
UGC 10168        &  13.71 &   1.74 &   1.22 &   0.80 &   0.42 &  34.64 & $-$21.0\\
UGC 10712        &  15.01 &   1.49 &   1.05 &   0.68 &   0.37 &  35.89 & $-$21.0\\
                 &        &        &        &        &        &        &         \\
$<$mag,col-modelMag$>$&  $-$0.24  & $-$0.27 &  $-$0.11 & $-$0.10  & $-$0.01 &  ......      & ...... \\
$\sigma_1$       &     0.18  &    0.16 &   0.08   &   0.05   &    0.04 &  ......      & ...... \\
$N$              &       32  &      30 &     32   &     32   &      32 &        &  \\     
\hline
$^a$not part of GZ2 subsample
\end{tabular}
\end{table*}

\begin{table*}
\centering
\caption{Integrated magnitudes and colours of 19 CSRG and other
galaxies. Col. 1: galaxy name; cols. 2-5: integrated magnitude and
colours, obtained by extrapolation of an exponential beyond the outer
regions; col. 6:  distance modulus with respect to the galactic
standard of rest (GSR), from NED; col. 7: absolute $B$-band magnitude
corrected for galactic extinction from NED. The last three lines compare
the derived magnitudes and colours with published values from Buta \&
Crocker (1992) and RC3.
}
\label{tab:totmagsBVI}
\begin{tabular}{lcccccc}
\hline
Name  & $B_T$ & $(B-V)_T$ & $(V-I)_T$ & $(B-I)_T$ & $m-M$ & $M_B^o$ \\
 1 & 2 & 3 & 4 & 5 & 6 & 7 \\  
\hline
ESO 325-28      &  14.17 &   0.77 &   1.06 &   1.83 &  34.95 &  $-$21.1 \\
ESO 365-35      &  14.95 &   1.06 &   1.25 &   2.31 &  35.05 &  $-$20.4 \\
ESO 426-2       &  14.42 &   0.95 &   1.21 &   2.16 &  34.76 &  $-$20.5 \\
ESO 437-33      &  13.77 &   0.77 &   1.11 &   1.88 &  33.78 &  $-$20.2 \\
ESO 437-67      &  13.54 &   0.83 &   1.18 &   2.01 &  33.06 &  $-$19.8 \\
ESO 566-24      &  13.65 &   0.69 &   1.08 &   1.77 &  33.28 &  $-$19.8 \\
ESO 575-47      &  13.78 &   0.88 &   1.21 &   2.09 &  34.07 &  $-$20.6 \\
IC 1438         &  13.02 &   0.82 &   1.15 &   1.97 &  32.82 &  $-$19.9 \\
IC 4214         &  12.30 &   0.88 &   1.15 &   2.03 &  32.36 &  $-$20.3 \\
NGC 1079        &  12.26 &   0.91 &   1.10 &   2.01 &  31.38 &  $-$19.2 \\
NGC 1291        &   9.51 &   0.88 &   1.09 &   1.97 &  29.95 &  $-$20.5 \\
NGC 1326        &  11.47 &   0.86 &   1.14 &   2.00 &  31.15 &  $-$19.7 \\
NGC 1398        &  10.61 &   0.92 &   1.22 &   2.14 &  31.24 &  $-$20.7 \\
NGC 1433        &  10.79 &   .... &   .... &   1.82 &  30.52 &  $-$19.8 \\
NGC 2665        &  13.02 &   0.70 &   1.11 &   1.81 &  31.61 &  $-$19.0 \\
NGC 3081        &  12.88 &   .... &   .... &   2.06 &  32.38 &  $-$19.7 \\
NGC 6782        &  12.83 &   0.89 &   1.22 &   2.12 &  33.61 &  $-$21.0 \\
NGC 7098        &  12.44 &   1.02 &   1.18 &   2.20 &  32.45 &  $-$20.3 \\
UGC 12646       &  14.15 &   0.83 &   .... &   .... &  35.26 &  $-$21.3 \\
                &        &        &        &        &        &          \\
$<$mag,col-pub$>$& 0.08  &   0.01 &   0.00 &   0.01 &  ..... &  .....   \\
$\sigma_1$      &  0.09  &   0.03 &   0.04 &   0.06 &  ..... &  .....   \\
$N$             &    17  &     15 &     12 &     13 &  ..... &  .....   \\     
\hline
\end{tabular}
\end{table*}

\noindent
4. Deproject the galaxies using $q_d$ and $\phi_d$ with IRAF routine
IMLINTRAN. From the deprojected $i$-band image of each GZ2 galaxy, and
the deprojected $I$-band image of each CSRG galaxy and other non-GZ2
cases, derive relative Fourier intensity amplitudes $I_m/I_0$ as a
function of radius. These profiles are shown in
Figures~\ref{fig:ugc4596c} and ~\ref{fig:NGC5335c} for UGC 4596 and NGC
5335, respectively, and for the remaining 52 galaxies in
Figure~\ref{fig:allfours} (see Appendix). Relative maximum amplitudes $A_m =
(I_m/I_0)_{max}$ for $m$ = 2, 4, and 6 are compiled in
Table~\ref{tab:fourier}. The perturbation strengths range from
relatively weak ($A_2$ = 0.25, CGCG 185-14) to very strong ($A_2$ =
1.39, NGC 4113).

\begin{table*}
\centering
\caption{
Maximum relative Fourier amplitudes for terms $m$ = 2, 4, and 6 for all 54
sample galaxies.  Col.
1: name of galaxy; cols. 2-4: maximum relative Fourier intensity amplitude
for $m$ = 2, 4, and 6, respectively. For the GZ2 galaxies, the amplitudes are 
based on the $i$-band, while for the non-GZ2 galaxies, the amplitudes are based on the $I$-band
except for UGC 12646, where the $V$-band was used instead.
}
\label{tab:fourier}
\begin{tabular}{lccclccclccc}
\hline
Galaxy & $A_2$ & $A_4$ & $A_6$ & Galaxy & $A_2$ & $A_4$ & $A_6$ & Galaxy & $A_2$ & $A_4$ & $A_6$ \\
 1 & 2 & 3 & 4 & 1 & 2 & 3 & 4 & 1 & 2 & 3 & 4 \\
\hline
CGCG 8-10   & 0.70 & 0.29 & 0.14 & IC 4214     & 0.63 & 0.23 & 0.08 & NGC 5211    & 0.41 & 0.15 & 0.05\\
CGCG 13-75  & 0.54 & 0.18 & 0.08 & MCG 6-32-24 & 0.61 & 0.16 & 0.05 & NGC 5335    & 0.95 & 0.55 & 0.33\\
CGCG 65-2   & 0.62 & 0.34 & 0.18 & MCG 7-18-40 & 0.48 & 0.16 & 0.07 & NGC 5370    & 0.64 & 0.33 & 0.17\\
CGCG 67-4   & 0.98 & 0.56 & 0.31 & NGC 210     & 0.53 & 0.14 & 0.06 & NGC 5686    & 0.31 & 0.17 & 0.10\\
CGCG 73-53  & 0.42 & 0.11 & 0.04 & NGC 1079    & 0.65 & 0.26 & 0.12 & NGC 5701    & 0.49 & 0.20 & 0.09\\
CGCG 185-14 & 0.25 & 0.07 & 0.04 & NGC 1291    & 0.48 & 0.16 & 0.07 & NGC 6782    & 0.64 & 0.28 & 0.13\\
CGCG 263-22 & 0.56 & 0.10 & 0.07 & NGC 1326    & 0.62 & 0.21 & 0.09 & NGC 7098    & 0.68 & 0.23 & 0.11\\
ESO 325-28  & 0.78 & 0.31 & 0.17 & NGC 1398    & 0.46 & 0.23 & 0.16 & PGC 54897   & 0.96 & 0.49 & 0.22\\
ESO 365-35  & 0.51 & 0.12 & 0.10 & NGC 1433    & 0.97 & 0.40 & 0.21 & PGC 1857116 & 0.99 & 0.46 & 0.26\\
ESO 426-2   & 0.74 & 0.36 & 0.23 & NGC 2665    & 1.06 & 0.62 & 0.45 & PGC 2570478 & 0.47 & 0.20 & 0.06\\
ESO 437-33  & 0.60 & 0.19 & 0.07 & NGC 2766    & 0.40 & 0.18 & 0.11 & UGC 4596    & 0.66 & 0.18 & 0.07\\
ESO 437-67  & 1.17 & 0.53 & 0.24 & NGC 3081    & 0.69 & 0.24 & 0.11 & UGC 4771    & 0.68 & 0.32 & 0.15\\
ESO 566-24  & 0.61 & 0.36 & 0.22 & NGC 3380    & 0.81 & 0.34 & 0.15 & UGC 5380    & 0.74 & 0.43 & 0.25\\
ESO 575-47  & 0.91 & 0.40 & 0.27 & NGC 4113    & 1.39 & 0.83 & 0.52 & UGC 5885    & 1.13 & 0.60 & 0.37\\
IC 1223     & 0.48 & 0.20 & 0.10 & NGC 4608    & 0.71 & 0.43 & 0.27 & UGC 9418    & 0.72 & 0.30 & 0.15\\
IC 1438     & 0.67 & 0.29 & 0.13 & NGC 4736    & 0.44 & 0.09 & 0.03 & UGC 10168   & 0.49 & 0.15 & 0.06\\
IC 2473     & 1.04 & 0.49 & 0.26 & NGC 4935    & 0.57 & 0.21 & 0.09 & UGC 10712   & 0.71 & 0.28 & 0.10\\
IC 2628     & 0.36 & 0.11 & 0.05 & NGC 5132    & 1.07 & 0.48 & 0.21 & UGC 12646   & 1.16 & 0.61 & 0.29\\
\hline
\end{tabular}
\end{table*}

\noindent
5. Convert deprojected images into units of mag/arcsec$^2$, allowing
the morphology of each galaxy to be displayed as well as for making
colour index maps. 

\noindent
6. Derive the deprojected radius $a_{bar}$ and position angle
$\theta_{bar}$ of the bar/oval in the $i$-band (GZ2 galaxies) or the
$I$-band (non-GZ2 galaxies, if available), using IRAF/STSDAS routine
ELLIPSE. In most cases the radius of the bar isophote of maximum
ellipticity provided a reasonable estimate of this parameter.

\noindent
7. Visually map the ring, pseudoring, or spiral features in the
$g$-band (GZ2 galaxies) or $B$-band (CSRG and other galaxies). The
mappings (Figure~\ref{fig:results}) were made using IRAF routine
TVMARK, and usually more than one mapping was needed to get the best
representation of a feature. The major axis radius $a$, the minor axis
radius $b$, and the major axis position angle $\theta_f$ of each
feature were derived from least squares ellipse fits to the mapped
points.

The derived values of $a$, $b$, and $\theta_{bf} = \theta_{bar} -
\theta_f$ for each galaxy and morphological feature $f$ are listed in
Tables~\ref{tab:R1gals}-~\ref{tab:inner-features}.

\noindent
8. Derive luminosity profiles along and perpendicular to the bar/oval
axis. Depending on how strong the dark gaps are, fit a parabola either
to gap surface  brightnesses $\mu$ or to residual intensities $\Delta$
after subtraction of a heavily median-smoothed version of the image, to
get $r_{gp}(j)$ in filter $j$. Do this for as many filters available as
possible, and then take the weighted average of $r_{gp}(j)$ from all
the applicable filters. For GZ2 galaxies, all $r_{gp}$ are based on
$gri$ images. For non-GZ2 galaxies, most are based on $BVI$ images, with a
few having only $BV$ (UGC 12646), $BI$ (NGC 1433, NGC 3081), or a red
continuum image (NGC 210).  Table~\ref{tab:orient} lists the filters
used for each sample galaxy and the method (surface brightness versus
residual intensities) used to get $r_{gp}(j)$.

\noindent
9. set $r_{CR}$ = $<r_{gp}>$

The determination of $r_{CR}$ allows us to locate other important
resonances if we can make a reasonable assumption about the shapes of
the rotation curves of typical OLR subclass galaxies. Based on detailed
studies of five well-observed examples:  NGC 1433  (Buta et al. 2001),
NGC 3081 (Buta \& Purcell 1998; Buta et al.  2004), IC 4214 (Buta et
al.  1999; Salo et al. 1999), ESO 566$-$24 (Rautiainen et al. (2004),
and NGC 6782 (Lin et al. 2008), the most reasonable choice is that of a
flat rotation curve. In this case, we can predict the locations of the
likely major ring-forming resonances as (Athanassoula et al. 1982):

\noindent
$r_{OLR}$ = 1.71 $r_{CR}$\br
\noindent
$r_{ILR}$ = $r_{OLR}$/5.83\br
\noindent
$r_{O4R}$ = $r_{OLR}$/1.26\br
\noindent
$r_{I4R}$ = $r_{OLR}$/2.64\br

In the vicinity of the inner Lindblad resonance (ILR), the flat
rotation curve assumption is unlikely to be correct for most of our
sample galaxies, and for this reason the discussion will focus on the
other major resonances. Table~\ref{tab:resrads} summarizes the results
for the 50 (rR) dark-spacers in the GZ2/CSRG sample, while
Table~\ref{tab:rdkresrads} does the same for the 4 (r) dark-spacers in
the sample.  In both tables, the gap CR radii are in col. 2, the
predicted radii of the inner and outer 4:1 resonances are in cols. 3
and 4, and the predicted radii of the inner and outer Lindblad
resonances are in cols. 5 and 6.

\begin{table*}
\centering
\caption{
Gap Corotation, Predicted Resonance, and Relative Bar Radii for
(rR) dark-spacers.  Col. 1:  galaxy name; col. 2: radius of corotation
resonance based on gap method; cols. 3-6: inferred radii of other
resonances, assuming a flat rotation curve; col. 7: deprojected radius
of bar based on ellipse of maximum ellipticity; cols. 8-9: relative
radius of bar to corotation and inner 4:1 resonance radii,
respectively.}
\label{tab:resrads}
\begin{tabular}{lrrrrrrrr}
\hline
Name & $r_{CR}$ & $r_{I4R}$ & $r_{O4R}$ & $r_{ILR}$ & $r_{OLR}$ & $a_{bar}$ & $a_{bar}/r_{CR}$  & $a_{bar}/r_{I4R}$ \\
 1 & 2 & 3 & 4 & 5 & 6 & 7 & 8 & 9 \\
\hline
CGCG 8-10    &    21.9 $\pm$    1.6 &    14.2 &    29.7 &     6.4 &    37.4 &    12.7 &    0.58 &    0.90 \\
CGCG 13-75   &    12.6 $\pm$    0.6 &     8.2 &    17.1 &     3.7 &    21.5 &     7.9 &    0.63 &    0.97 \\
CGCG 65-2    &    12.9 $\pm$    0.7 &     8.3 &    17.5 &     3.8 &    22.0 &     9.5 &    0.74 &    1.14 \\
CGCG 67-4    &    13.5 $\pm$    0.9 &     8.7 &    18.3 &     4.0 &    23.1 &     8.9 &    0.66 &    1.02 \\
CGCG 73-53   &    10.3 $\pm$    0.3 &     6.7 &    14.0 &     3.0 &    17.6 &     5.9 &    0.57 &    0.88 \\
CGCG 185-14  &    11.4 $\pm$    0.2 &     7.4 &    15.5 &     3.3 &    19.5 &     7.1 &    0.62 &    0.96 \\
CGCG 263-22  &    22.0 $\pm$    1.2 &    14.2 &    29.7 &     6.4 &    37.5 &    23.0 &    1.05 &    1.62 \\
ESO 325-28   &    17.6 $\pm$    1.3 &    11.4 &    23.8 &     5.1 &    30.0 &    13.1 &    0.75 &    1.15 \\
ESO 365-35   &    20.8 $\pm$    1.2 &    13.4 &    28.1 &     6.1 &    35.5 &    12.6 &    0.61 &    0.94 \\
ESO 426-2    &    22.1 $\pm$    1.1 &    14.3 &    30.0 &     6.5 &    37.8 &    18.1 &    0.82 &    1.27 \\
ESO 437-33   &    23.7 $\pm$    1.0 &    15.3 &    32.1 &     6.9 &    40.4 &    13.7 &    0.58 &    0.90 \\
ESO 437-67   &    54.0 $\pm$    3.7 &    34.9 &    73.1 &    15.8 &    92.1 &    37.2 &    0.69 &    1.07 \\
ESO 566-24   &    26.6 $\pm$    0.9 &    17.2 &    36.0 &     7.8 &    45.3 &    19.4 &    0.73 &    1.13 \\
ESO 575-47   &    36.8 $\pm$    2.3 &    23.8 &    49.8 &    10.8 &    62.7 &    31.2 &    0.85 &    1.31 \\
IC 1223      &    22.5 $\pm$    1.0 &    14.5 &    30.5 &     6.6 &    38.4 &    12.7 &    0.56 &    0.87 \\
IC 1438      &    33.3 $\pm$    2.2 &    21.6 &    45.2 &     9.8 &    56.9 &    21.8 &    0.65 &    1.01 \\
IC 2473      &    35.2 $\pm$    2.0 &    22.8 &    47.7 &    10.3 &    60.1 &    24.1 &    0.68 &    1.06 \\
IC 2628      &    10.0 $\pm$    0.4 &     6.5 &    13.6 &     2.9 &    17.1 &     5.5 &    0.55 &    0.85 \\
IC 4214      &    46.2 $\pm$    2.0 &    29.9 &    62.6 &    13.5 &    78.8 &    44.4 &    0.96 &    1.49 \\
MCG 6-32-24  &    11.7 $\pm$    0.7 &     7.5 &    15.8 &     3.4 &    19.9 &     5.0 &    0.43 &    0.66 \\
MCG 7-18-40  &     7.6 $\pm$    0.5 &     4.9 &    10.3 &     2.2 &    12.9 &     5.2 &    0.69 &    1.06 \\
NGC 210      &    60.2 $\pm$    5.5 &    38.9 &    81.6 &    17.6 &   102.8 &    33.8 &    0.56 &    0.87 \\
NGC 1079     &    83.5 $\pm$    6.5 &    54.0 &   113.1 &    24.4 &   142.5 &    42.0 &    0.50 &    0.78 \\
NGC 1291     &   175.0 $\pm$    6.3 &   113.1 &   237.0 &    51.2 &   298.7 &    90.5 &    0.52 &    0.80 \\
NGC 1326     &    62.5 $\pm$    4.9 &    40.4 &    84.7 &    18.3 &   106.7 &    39.9 &    0.64 &    0.99 \\
NGC 1398     &    88.7 $\pm$    3.8 &    57.3 &   120.1 &    26.0 &   151.4 &    54.8 &    0.62 &    0.96 \\
NGC 1433     &   132.8 $\pm$    8.4 &    85.9 &   180.0 &    38.9 &   226.7 &    85.3 &    0.64 &    0.99 \\
NGC 2665     &    42.4 $\pm$    2.7 &    27.4 &    57.5 &    12.4 &    72.4 &    33.4 &    0.79 &    1.22 \\
NGC 2766     &    21.3 $\pm$    0.4 &    13.8 &    28.8 &     6.2 &    36.3 &    18.2 &    0.85 &    1.32 \\
NGC 3081     &    59.1 $\pm$    2.7 &    38.2 &    80.1 &    17.3 &   100.9 &    25.7 &    0.43 &    0.67 \\
NGC 3380     &    31.6 $\pm$    2.3 &    20.4 &    42.8 &     9.2 &    53.9 &    20.0 &    0.63 &    0.98 \\
NGC 4113     &    24.1 $\pm$    2.1 &    15.6 &    32.6 &     7.1 &    41.1 &    18.4 &    0.76 &    1.18 \\
NGC 4736     &   246.1 $\pm$    8.6 &   159.1 &   333.4 &    72.1 &   420.1 &   141.8 &    0.58 &    0.89 \\
NGC 4935     &    17.0 $\pm$    0.9 &    11.0 &    23.1 &     5.0 &    29.1 &     9.5 &    0.56 &    0.86 \\
NGC 5132     &    42.0 $\pm$    2.3 &    27.2 &    57.0 &    12.3 &    71.8 &    32.5 &    0.77 &    1.20 \\
NGC 5211     &    34.9 $\pm$    1.8 &    22.6 &    47.3 &    10.2 &    59.6 &    18.8 &    0.54 &    0.83 \\
NGC 5370     &    31.9 $\pm$    1.6 &    20.6 &    43.2 &     9.3 &    54.5 &    18.2 &    0.57 &    0.88 \\
NGC 5701     &    69.7 $\pm$    3.0 &    45.1 &    94.4 &    20.4 &   118.9 &    39.0 &    0.56 &    0.87 \\
NGC 6782     &    44.5 $\pm$    2.7 &    28.7 &    60.2 &    13.0 &    75.9 &    26.1 &    0.59 &    0.91 \\
NGC 7098     &    86.8 $\pm$    3.2 &    56.1 &   117.6 &    25.4 &   148.2 &    48.3 &    0.56 &    0.86 \\
PGC 54897    &    17.7 $\pm$    1.2 &    11.4 &    24.0 &     5.2 &    30.2 &     9.9 &    0.56 &    0.87 \\
PGC1857116   &    15.3 $\pm$    0.7 &     9.9 &    20.8 &     4.5 &    26.2 &    10.7 &    0.70 &    1.08 \\
PGC2570478   &    13.6 $\pm$    0.9 &     8.8 &    18.5 &     4.0 &    23.3 &     8.5 &    0.62 &    0.96 \\
UGC 4596     &    17.4 $\pm$    0.6 &    11.3 &    23.6 &     5.1 &    29.8 &     9.5 &    0.54 &    0.84 \\
UGC 4771     &    18.2 $\pm$    1.0 &    11.8 &    24.6 &     5.3 &    31.1 &    11.1 &    0.61 &    0.94 \\
UGC 5885     &    25.7 $\pm$    2.0 &    16.6 &    34.9 &     7.5 &    43.9 &    22.2 &    0.86 &    1.33 \\
UGC 9418     &    14.7 $\pm$    0.9 &     9.5 &    19.9 &     4.3 &    25.1 &     9.9 &    0.67 &    1.04 \\
UGC10168     &    27.4 $\pm$    1.3 &    17.7 &    37.2 &     8.0 &    46.9 &    16.0 &    0.58 &    0.90 \\
UGC10712     &    11.5 $\pm$    0.6 &     7.4 &    15.6 &     3.4 &    19.6 &     7.5 &    0.65 &    1.01 \\
UGC12646     &    32.2 $\pm$    3.9 &    20.8 &    43.6 &     9.4 &    55.0 &    23.4 &    0.73 &    1.12 \\
             &                      &         &         &         &         &         &         &         \\
means        &    ................. &   ..... &   ..... &   ....  & ....    & ....    &    0.65 &    1.01 \\
mean error   &    ................. &   ..... &   ..... &   ....  & ....    & ....    &    0.02 &    0.03 \\
$\sigma_1$   &    ................. &   ..... &   ..... &   ....  & ....    & ....    &    0.12 &    0.19 \\
$N$          &    ................. &   ..... &   ..... &   ....  & ....    & ....    &      50 &      50 \\
\hline
\end{tabular}
\end{table*}

\begin{table*}
\centering
\caption{
Gap Corotation, Predicted Resonance, and Relative Bar Radii for
(r) dark-spacers.  Col. 1:  galaxy name; col. 2: radius of corotation
resonance based on gap method; cols. 3-6: inferred radii of other
resonances, assuming a flat rotation curve; col. 7: deprojected radius
of bar based on ellipse of maximum ellipticity; cols. 8-9: relative
radius of bar to corotation and inner 4:1 resonance radii,
respectively.
}
\label{tab:rdkresrads}
\begin{tabular}{lrrrrrrrr}
\hline
Name & $r_{CR}$ & $r_{I4R}$ & $r_{O4R}$ & $r_{ILR}$ & $r_{OLR}$ & $a_{bar}$ & $a_{bar}/r_{CR}$  & $a_{bar}/r_{I4R}$ \\
 1 & 2 & 3 & 4 & 5 & 6 & 7 & 8 & 9 \\
\hline
NGC 4608     &    36.8 $\pm$    1.4 &    23.8 &    49.9 &    10.8 &    62.9 &    50.3 &    1.37 &    2.11 \\
NGC 5335     &    17.1 $\pm$    1.0 &    11.1 &    23.2 &     5.0 &    29.2 &    24.2 &    1.41 &    2.19 \\
NGC 5686     &     4.7 $\pm$    0.1 &     3.0 &     6.4 &     1.4 &     8.0 &     5.9 &    1.25 &    1.94 \\
UGC 5380     &     8.5 $\pm$    0.3 &     5.5 &    11.5 &     2.5 &    14.5 &    14.7 &    1.73 &    2.67 \\
             &                      &         &         &         &         &         &         &         \\
means        &    ................. &   ..... &   ..... &   ....  & ....    & ....    &    1.44 &    2.23 \\
mean error   &    ................. &   ..... &   ..... &   ....  & ....    & ....    &    0.10 &    0.16 \\
$\sigma_1$   &    ................. &   ..... &   ..... &   ....  & ....    & ....    &    0.20 &    0.31 \\
$N$          &    ................. &   ..... &   ..... &   ....  & ....    & ....    &       4 &       4 \\
\hline
\end{tabular}
\end{table*}

\section{Description of Individual Galaxies}

The way the morphology of the sample galaxies is compared with the
predicted locations of resonances from the gap method is in the form of
schematic graphs where details of the ring and pseudoring features have
been visually mapped (Figure~\ref{fig:results}). These schematics are useful
because they highlight the spiral and dimpling characteristics of many
outer rings; the spiral, cusping, and arc-ansae characteristics of many
inner rings/lenses; and the limited extent of bars inside inner
rings/lenses. In these schematics, partial ellipse fits were used to
map shapes where possible. For the outer arms that make up
R$_2^{\prime}$ components, the radius $r$ versus position angle
$\theta$ along each arm was fitted with a polynomial usually of order
5-7. Also, for each schematic, the bar has been rotated to a horizontal
position for ease of comparison with other galaxies in the sample. 

Figure~\ref{fig:results} shows two frames per galaxy: The left frame is
the $g$ or $B$-band image (in units of mag arcsec$^{-2}$) with the
location of the CR (white circle) superposed, while the right frame is
the schematic showing the visually-mapped features which include all of
the ring, lens, and bar features recognized in the CVRHS
classfication.  The assumed locations of the $L_4$ and $L_5$ points are
indicated. In a few cases, larger filled circles or ovals indicate the
locations of strong bar ansae (as, for example, in the schematic of NGC
7098). Strong bars (mostly type SB in Table~\ref{tab:types}) are
indicated by the horizontal solid lines, while weaker bars (mostly type
SAB in Table~\ref{tab:types}) are indicated by light or dashed lines. No
horizontal lines are used for the apparently weakest bars (mostly types
SA and S$\underline{\rm A}$B in Table~\ref{tab:types}). Ovals are
indicated in some cases by arc ansae. Also, bar-ring misalignment is
highlighted in a few cases, such as NGC 5132 and IC 2473. In these
cases, the bar line is not horizontal.

\noindent
CGCG 8-10 - The galaxy is of type
(\RoneP\RtwoP)\S_AB(r)a, where the \S_AB\ refers to arc ansae around
the major axis of the inner ring, which are prominent in both $g$ and
$i$. This galaxy also displays a dichotomy first noted by Buta (1995)
for IC 1438: the \Rone\ component is a prominent dimpled feature in the
$i$-band, while the \RtwoP\ component is most prominent in the
$g$-band. This suggests that the \Rone\ component formed first and left
behind a mostly stellar remnant. Colour index maps show that most of
the recent star formation in the galaxy is in the bright ansae arcs of
the inner ring and the bright opposing quadrants of the
\RtwoP\ pseudoring. The gap corotation method places the
\Rone\ component between \rcr\ and \rolr, and in fact the feature
straddles \ro4r. The \RtwoP\ reaches but still is largely inside \rolr,
while the major axis of the inner ring and its ansae are very close to
$r_{I4R}$. The small inner oval in the schematic refers to a possible
nuclear lens at the resolution of the SDSS image.

 \setcounter{figure}{12}
 \begin{figure}
\vspace{-1.27cm}
 \begin{minipage}[b]{0.45\linewidth}
 \centering
\includegraphics[width=\textwidth]{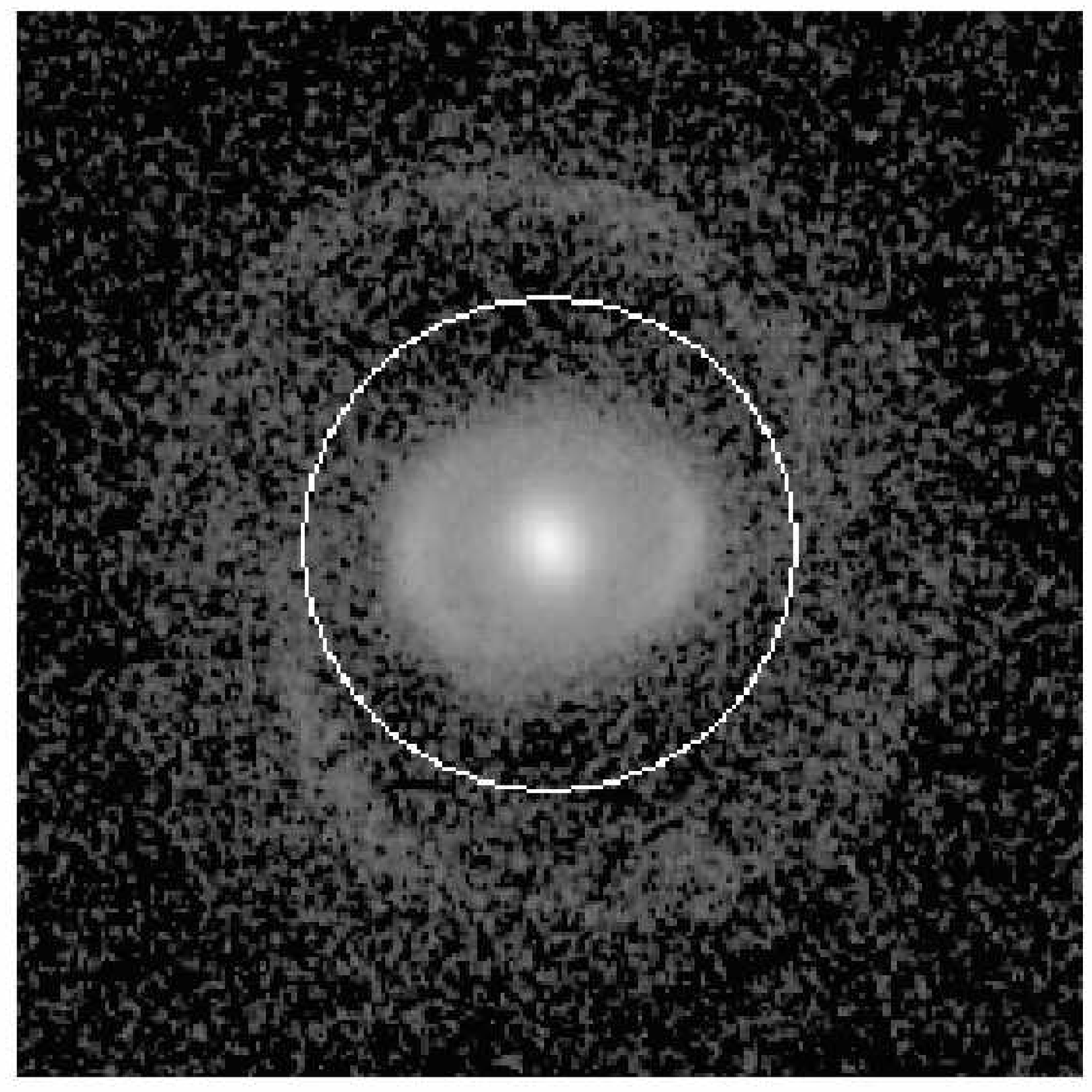}
 \hspace{0.1cm}
 \end{minipage}
 \begin{minipage}[t]{0.68\linewidth}
 \centering
\raisebox{0.5cm}{\includegraphics[width=\textwidth,trim=0 0 0 250,clip]{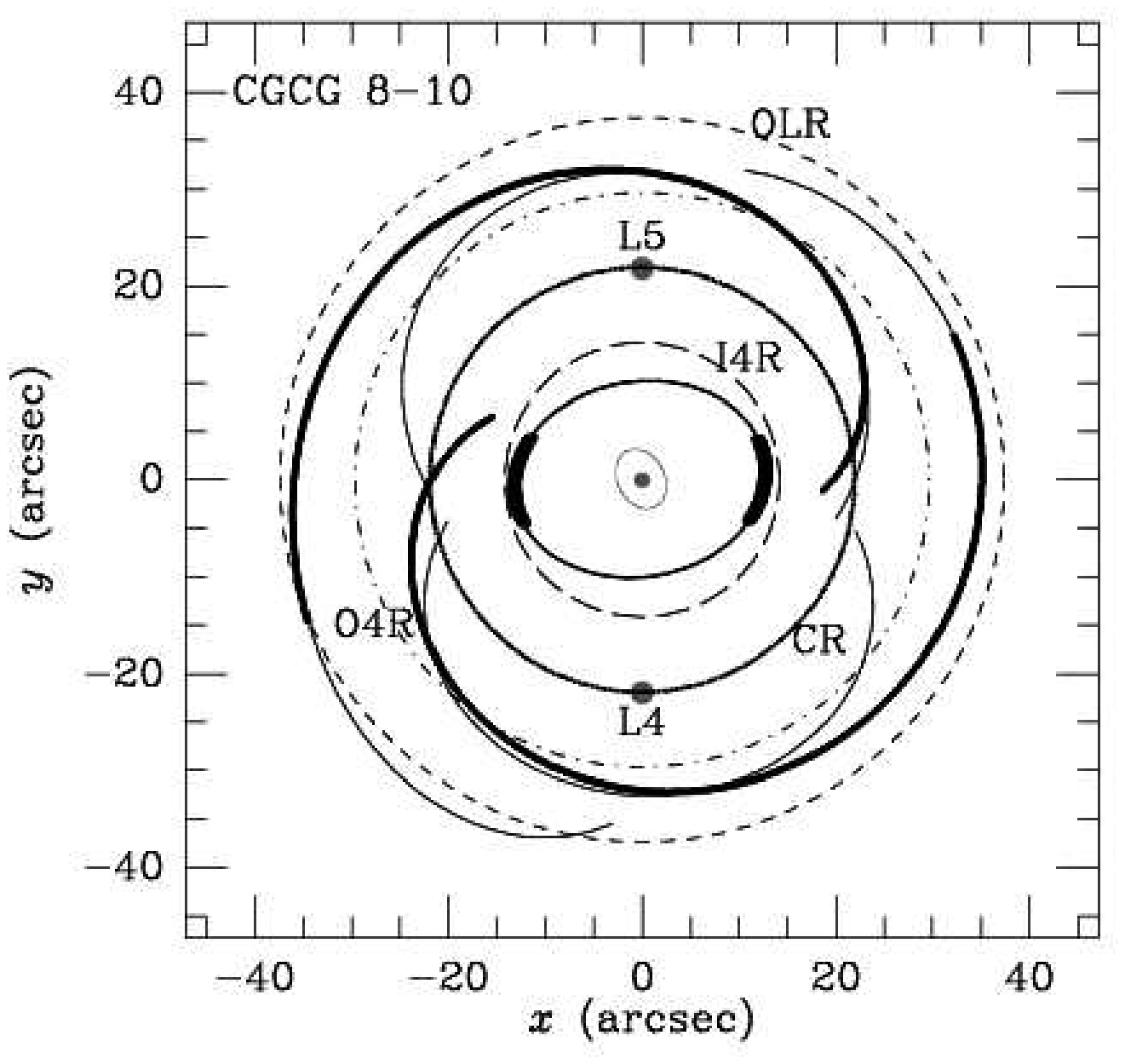}}
 \end{minipage}
 \begin{minipage}[b]{0.45\linewidth}
 \centering
\includegraphics[width=\textwidth]{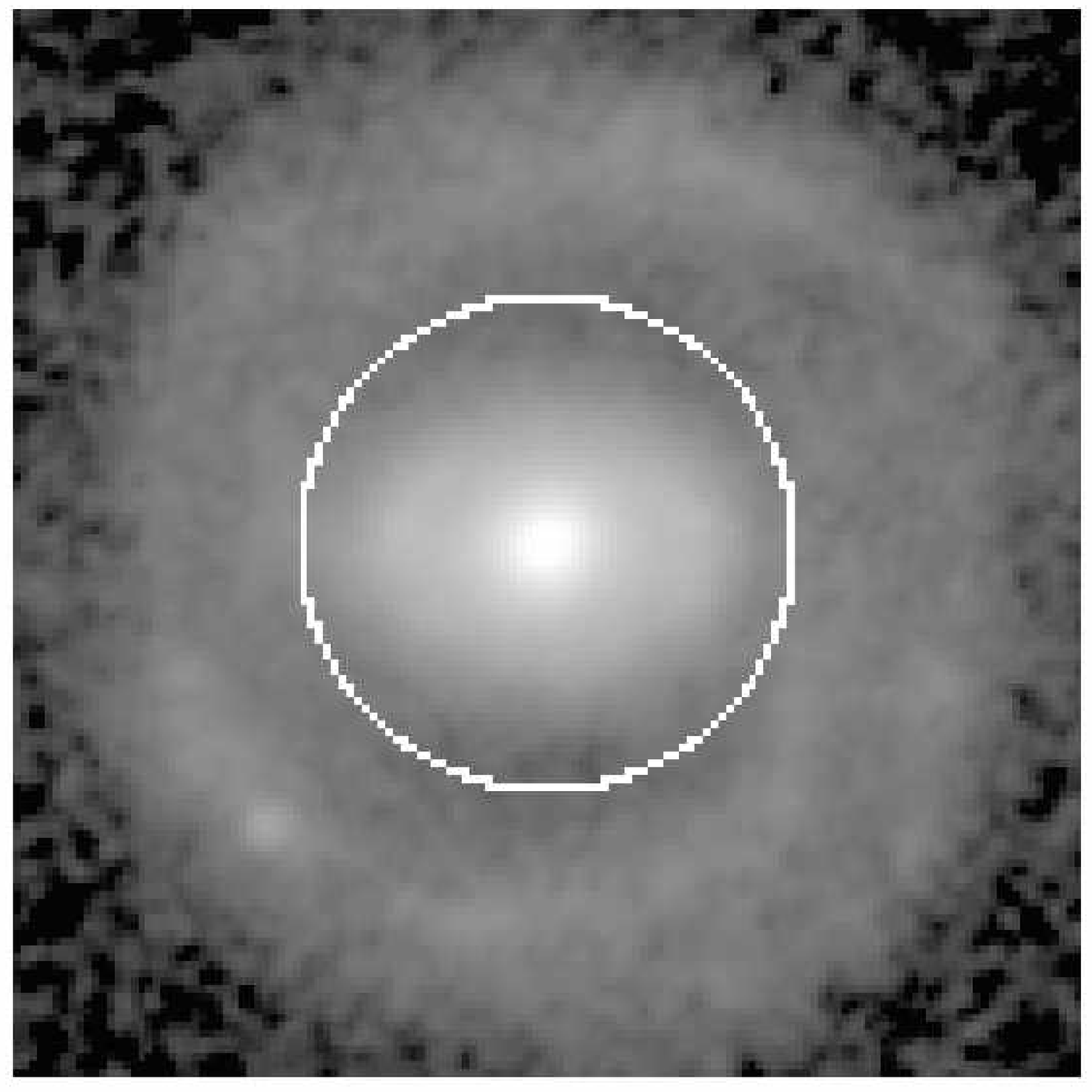}
 \hspace{0.1cm}
 \end{minipage}
 \begin{minipage}[t]{0.68\linewidth}
 \centering
\raisebox{0.5cm}{\includegraphics[width=\textwidth,trim=0 0 0 250,clip]{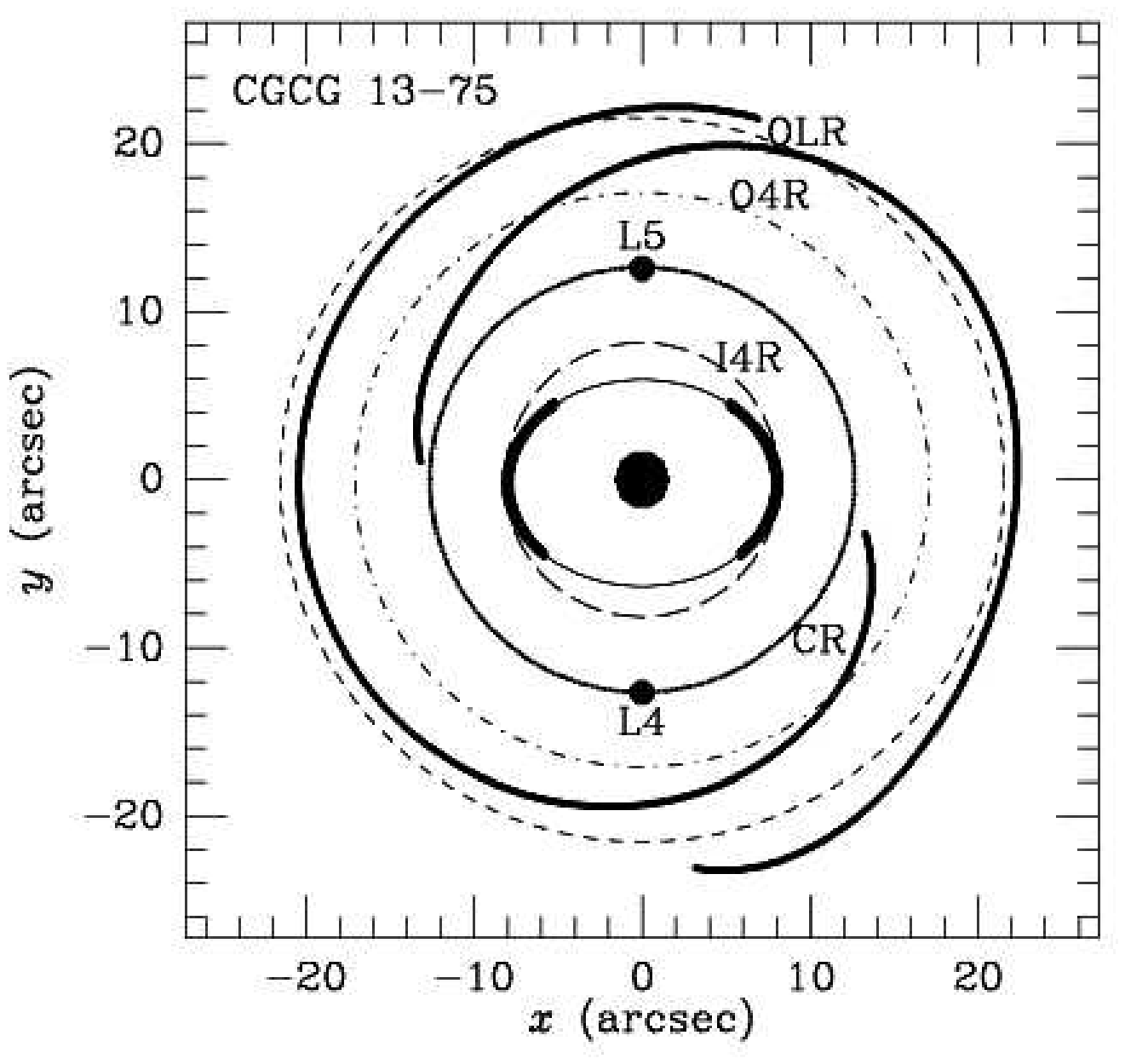}}
 \end{minipage}
 \begin{minipage}[b]{0.45\linewidth}
 \centering
\includegraphics[width=\textwidth]{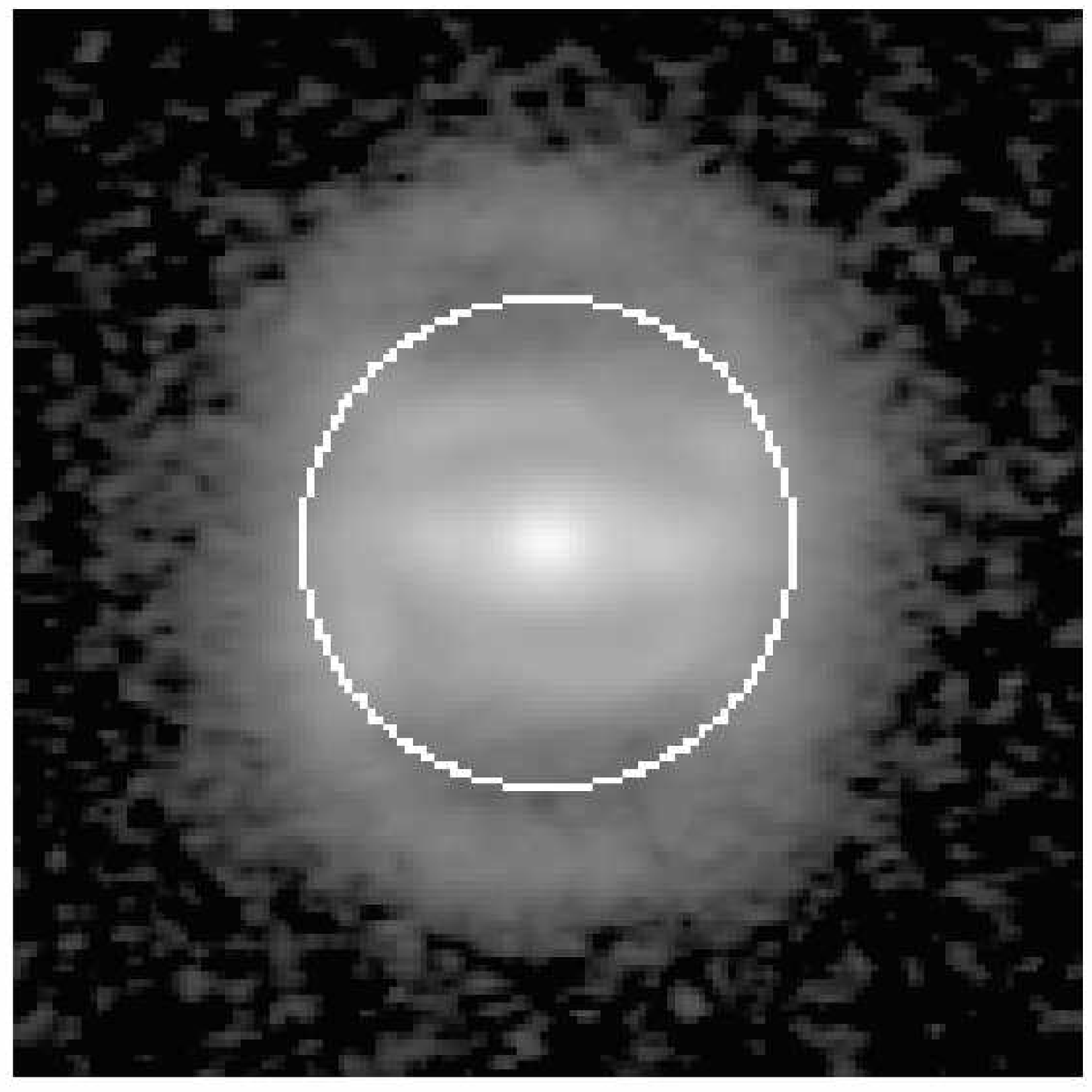}
 \hspace{0.1cm}
 \end{minipage}
 \begin{minipage}[t]{0.68\linewidth}
 \centering
\raisebox{0.5cm}{\includegraphics[width=\textwidth,trim=0 0 0 250,clip]{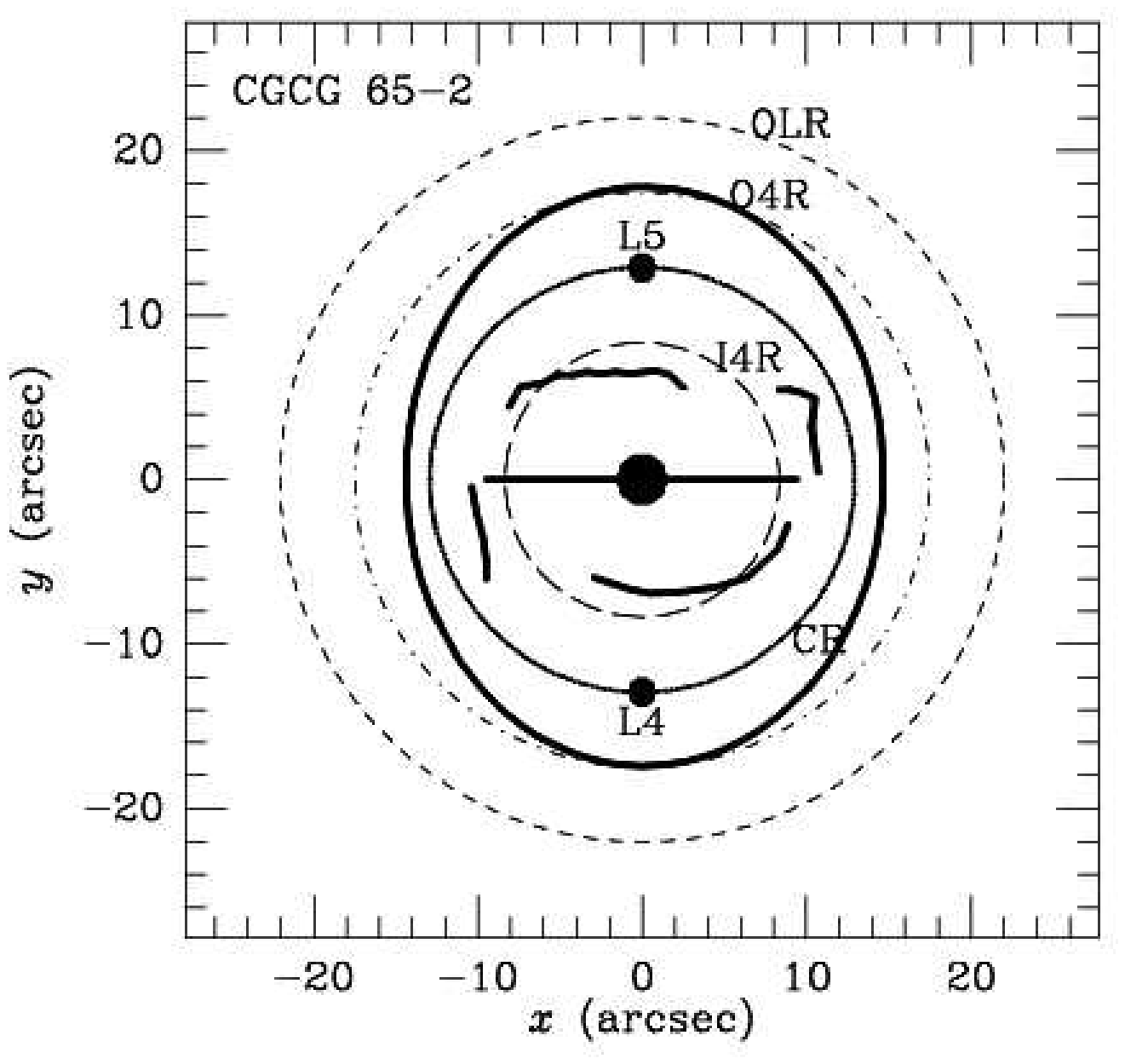}}
 \end{minipage}
 \begin{minipage}[b]{0.45\linewidth}
 \centering
\includegraphics[width=\textwidth]{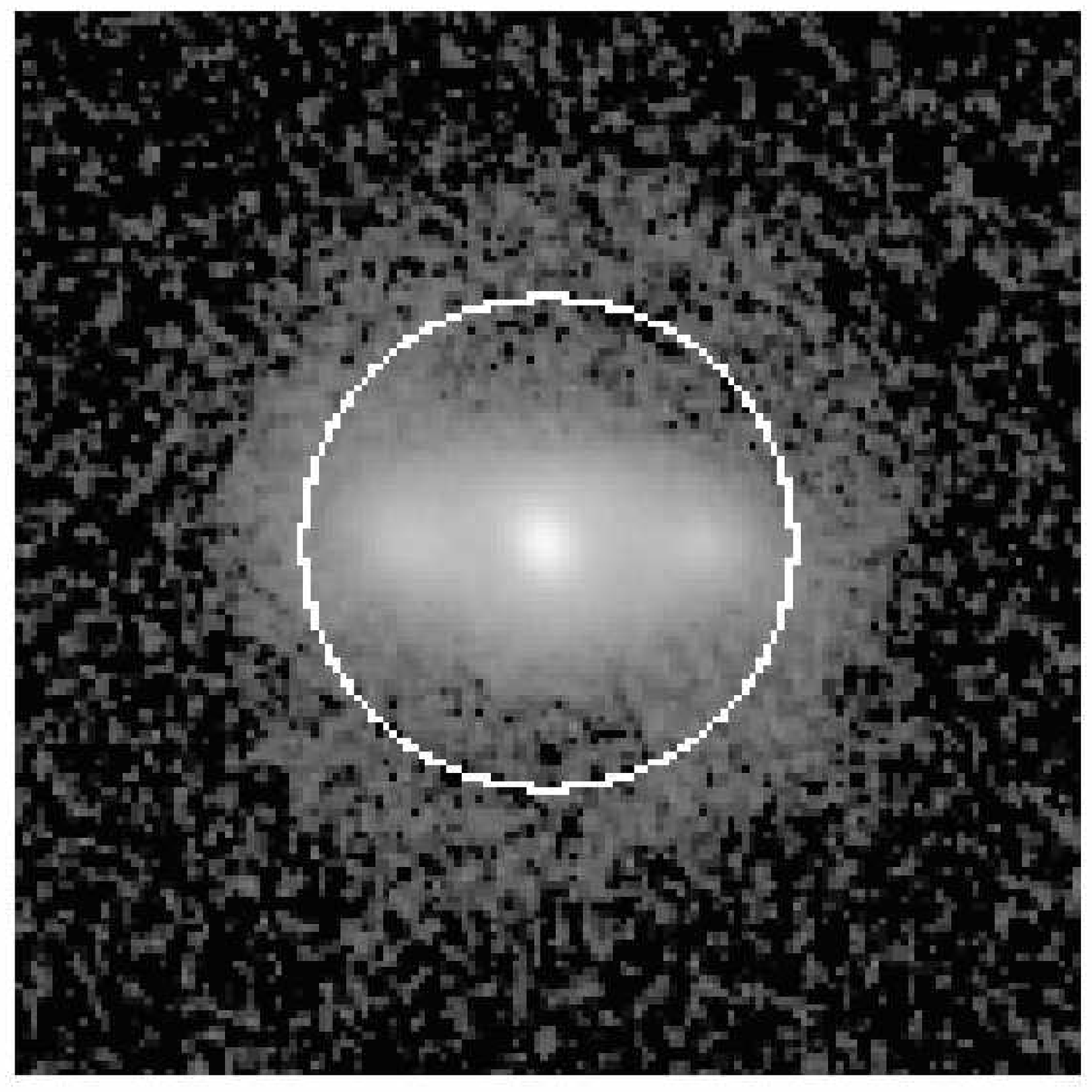}
 \hspace{0.1cm}
 \end{minipage}
 \begin{minipage}[t]{0.68\linewidth}
 \centering
\raisebox{0.5cm}{\includegraphics[width=\textwidth,trim=0 0 0 250,clip]{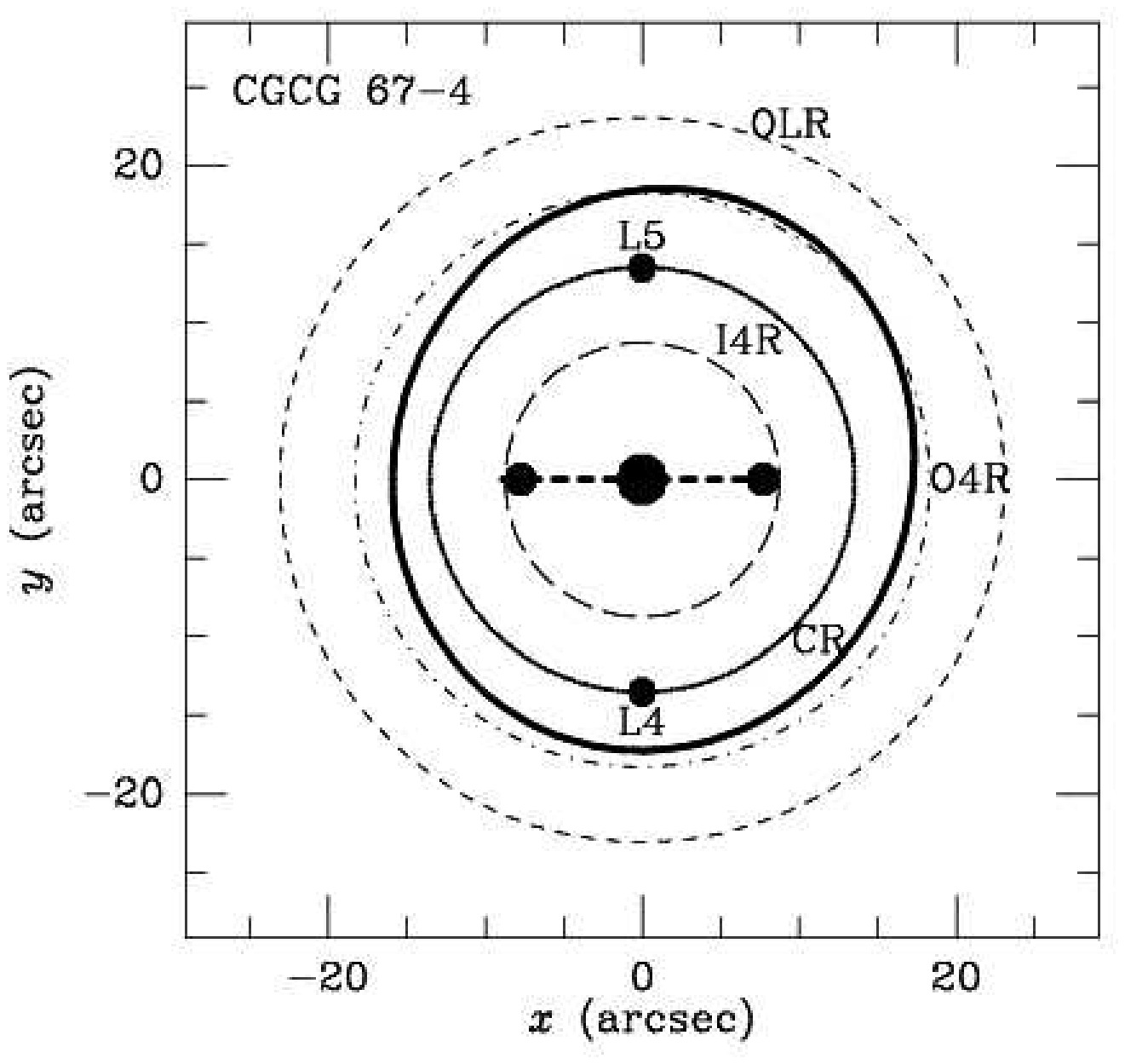}}
 \end{minipage}
 \begin{minipage}[b]{0.45\linewidth}
 \centering
\includegraphics[width=\textwidth]{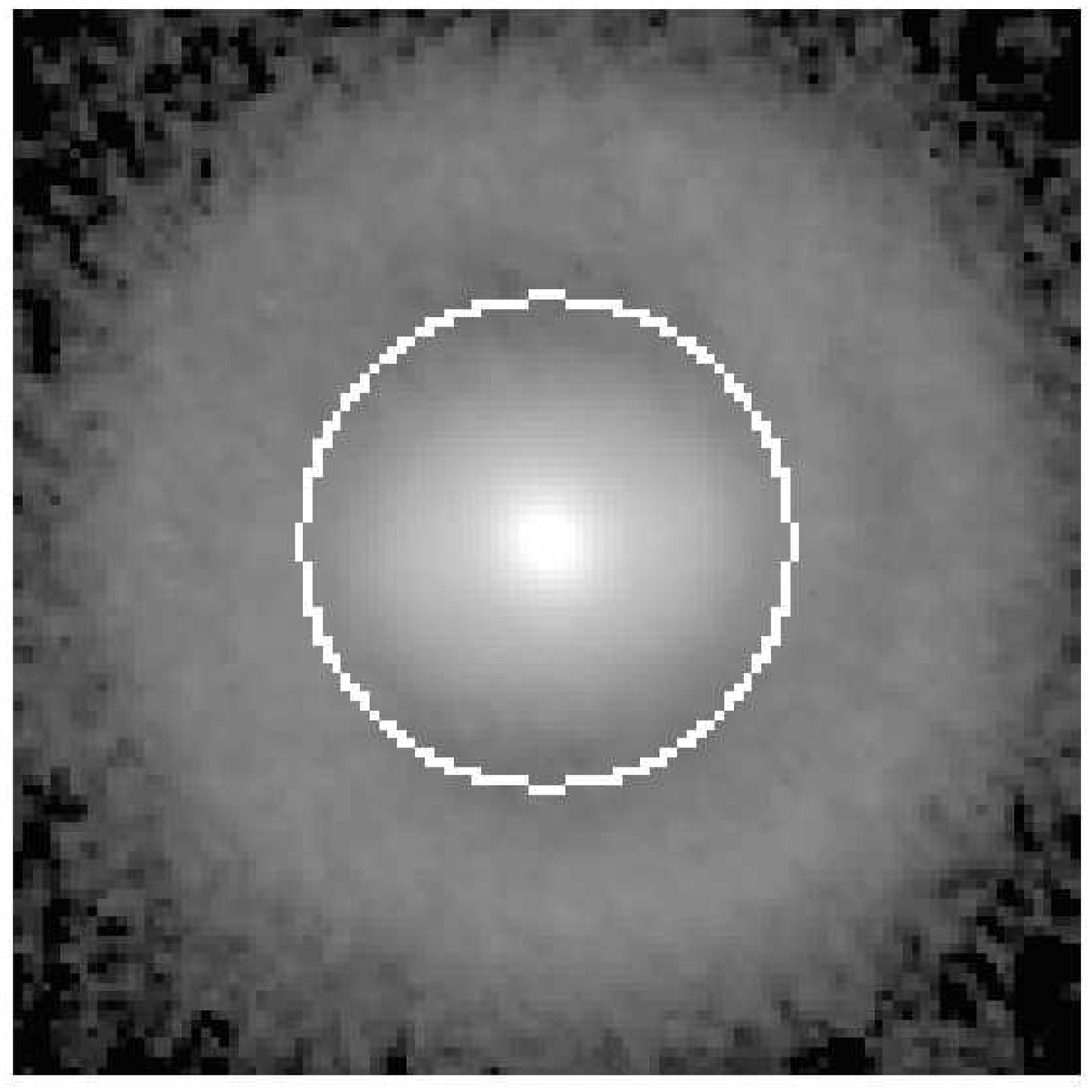}
 \hspace{0.1cm}
 \end{minipage}
 \begin{minipage}[t]{0.68\linewidth}
 \centering
\raisebox{0.5cm}{\includegraphics[width=\textwidth,trim=0 0 0 250,clip]{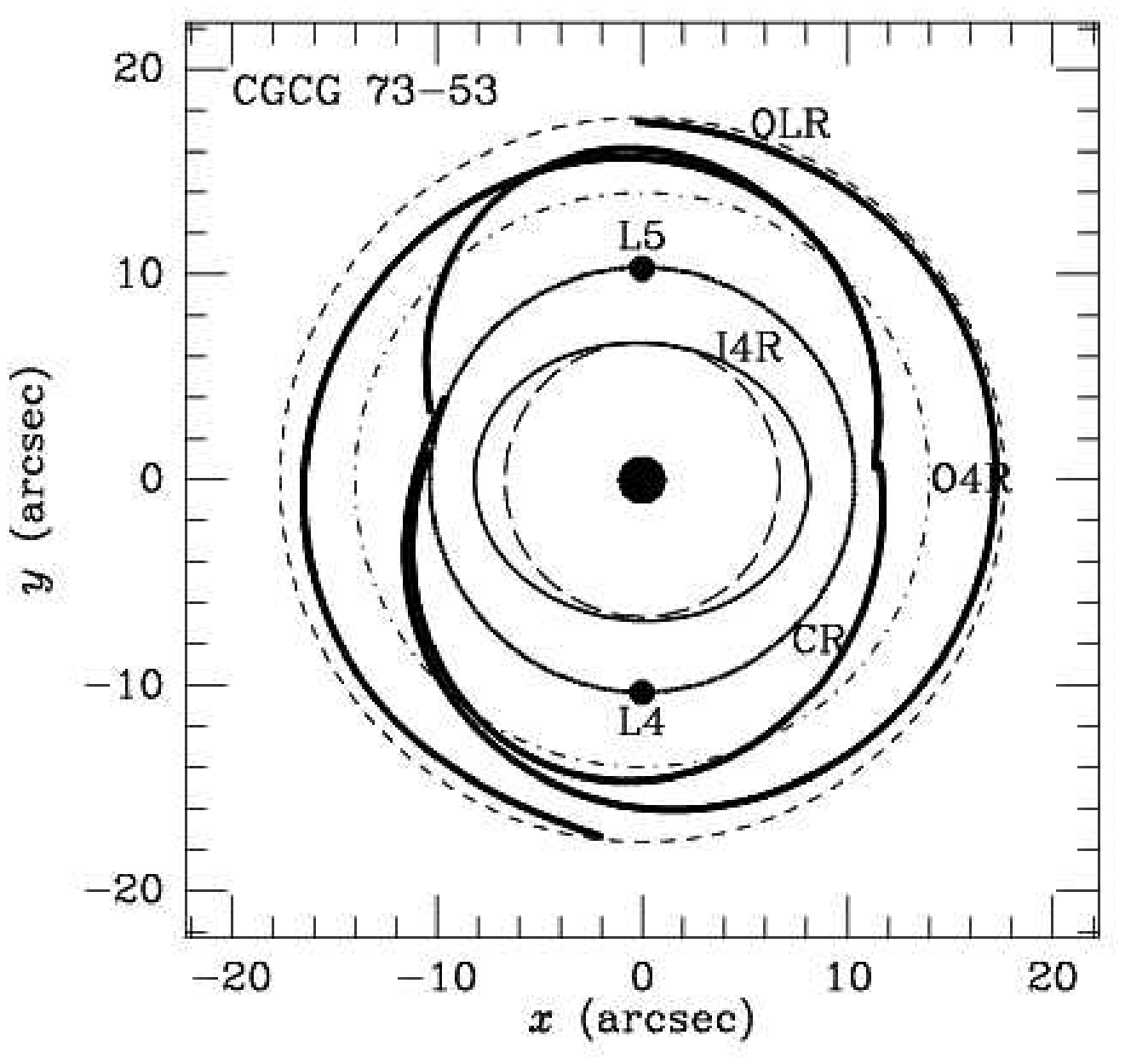}}
 \end{minipage}
\vspace{-1.0truecm}
\caption{}
\label{fig:results}
 \end{figure}
 \setcounter{figure}{12}
 \begin{figure}
\vspace{-1.27cm}
 \begin{minipage}[b]{0.45\linewidth}
 \centering
\includegraphics[width=\textwidth]{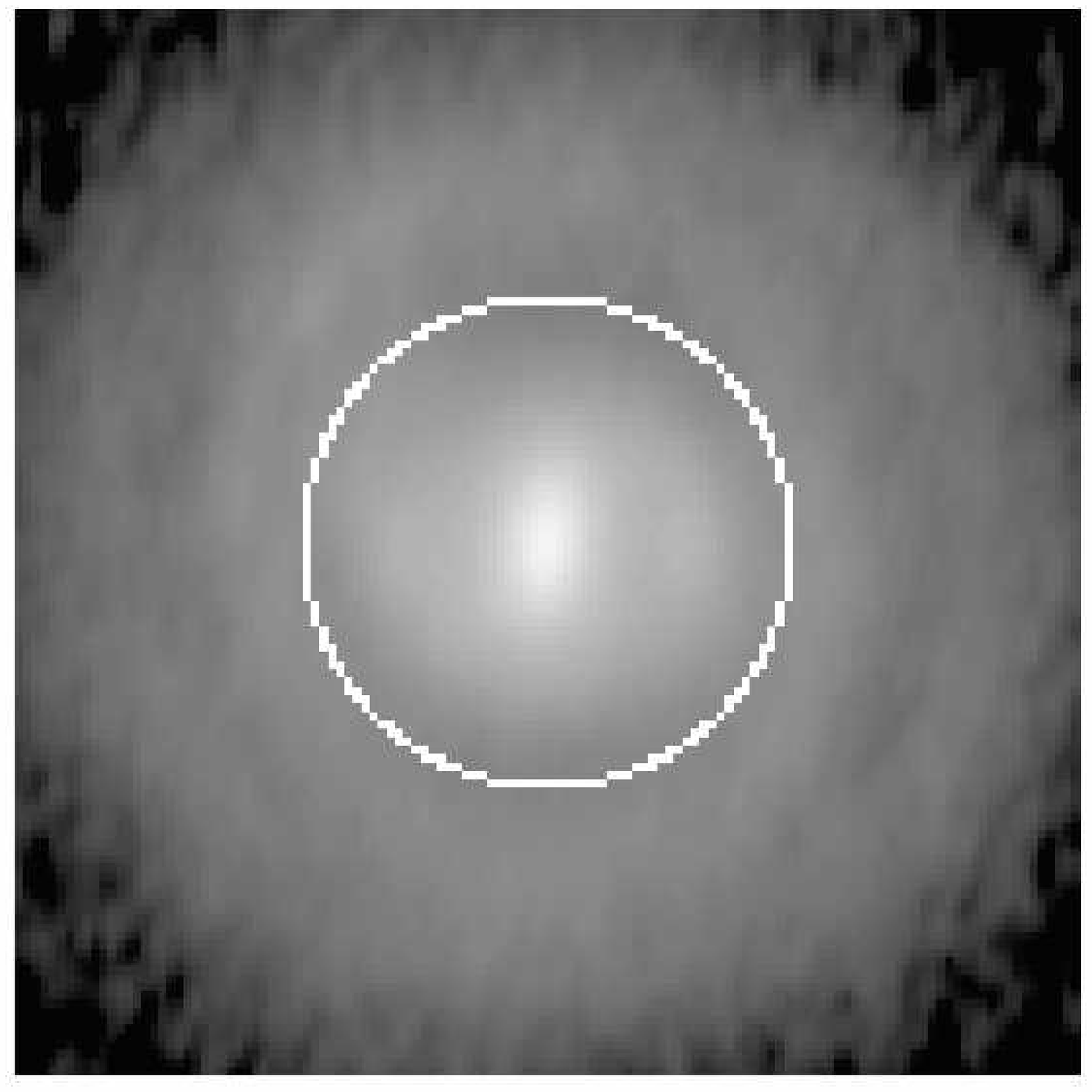}
 \hspace{0.1cm}
 \end{minipage}
 \begin{minipage}[t]{0.68\linewidth}
 \centering
\raisebox{0.5cm}{\includegraphics[width=\textwidth,trim=0 0 0 250,clip]{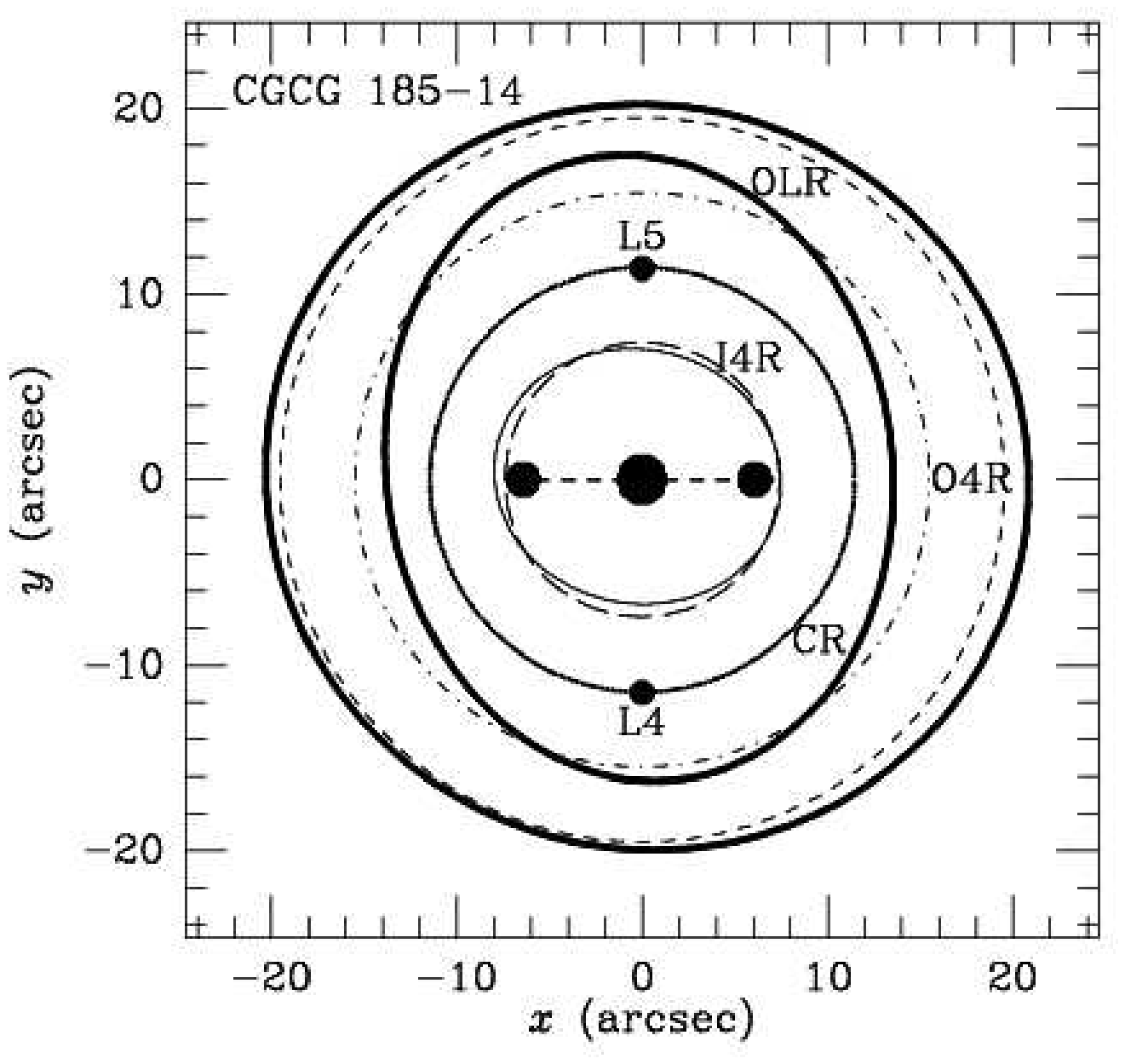}}
 \end{minipage}
 \begin{minipage}[b]{0.45\linewidth}
 \centering
\includegraphics[width=\textwidth]{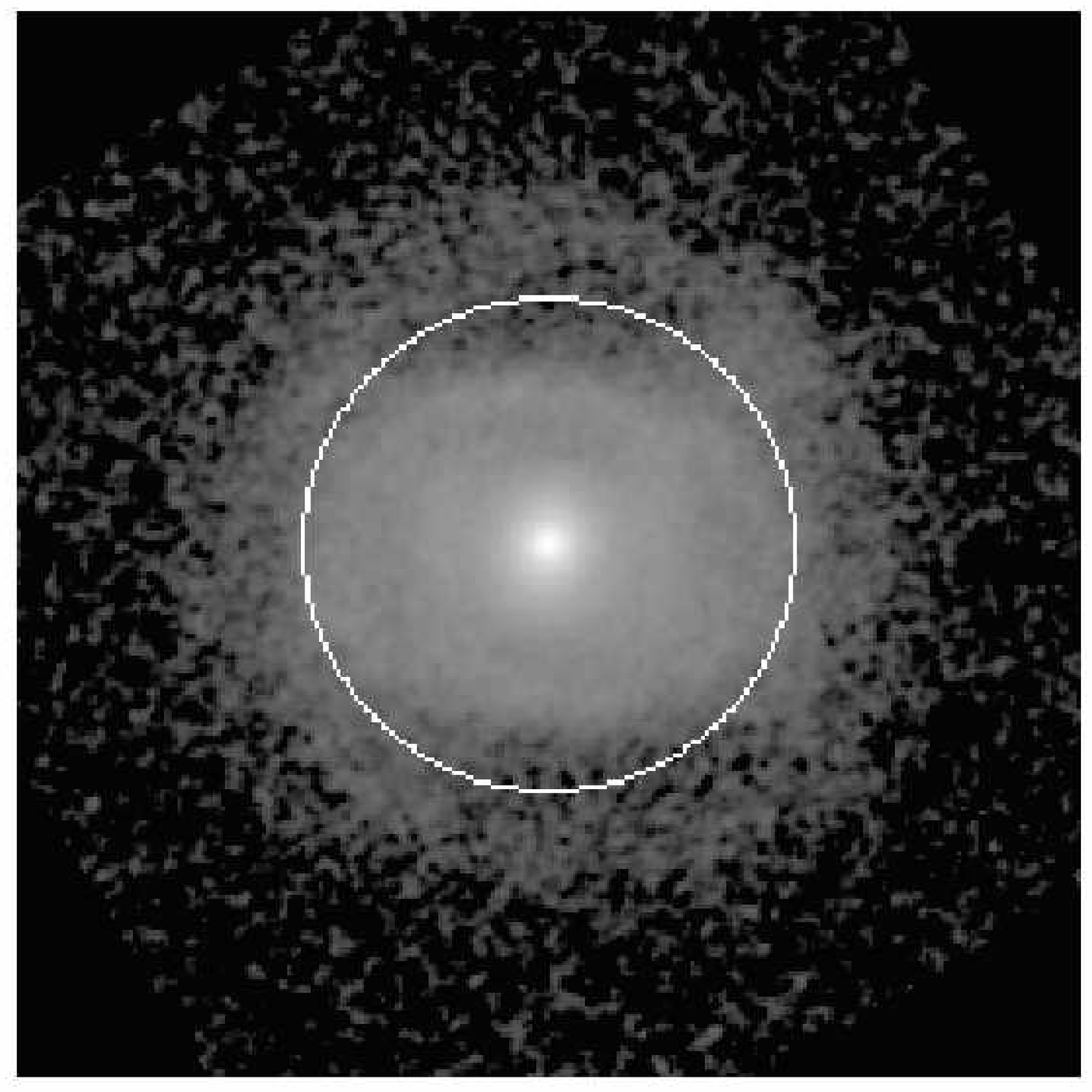}
 \hspace{0.1cm}
 \end{minipage}
 \begin{minipage}[t]{0.68\linewidth}
 \centering
\raisebox{0.5cm}{\includegraphics[width=\textwidth,trim=0 0 0 250,clip]{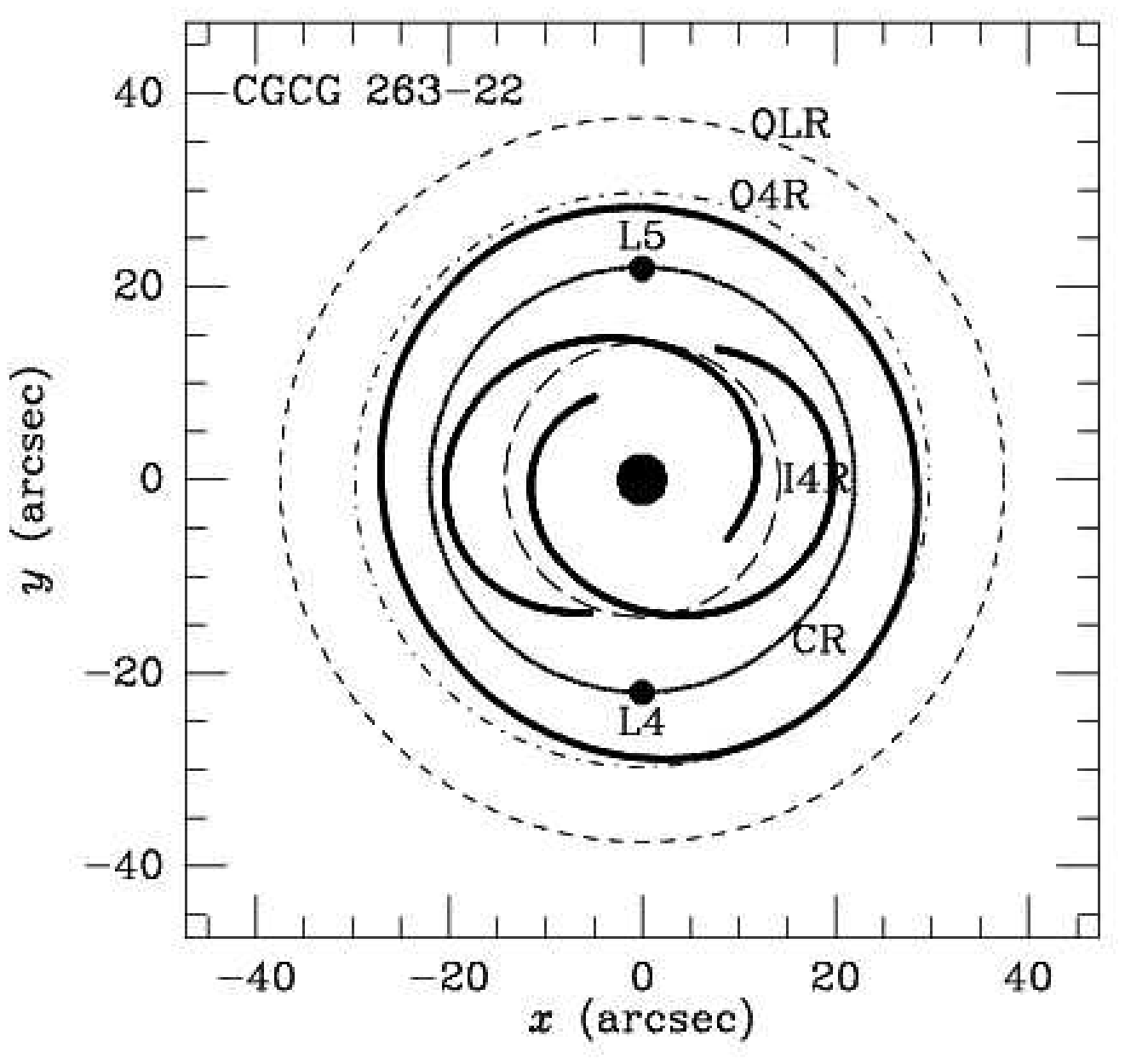}}
 \end{minipage}
 \begin{minipage}[b]{0.45\linewidth}
 \centering
\includegraphics[width=\textwidth]{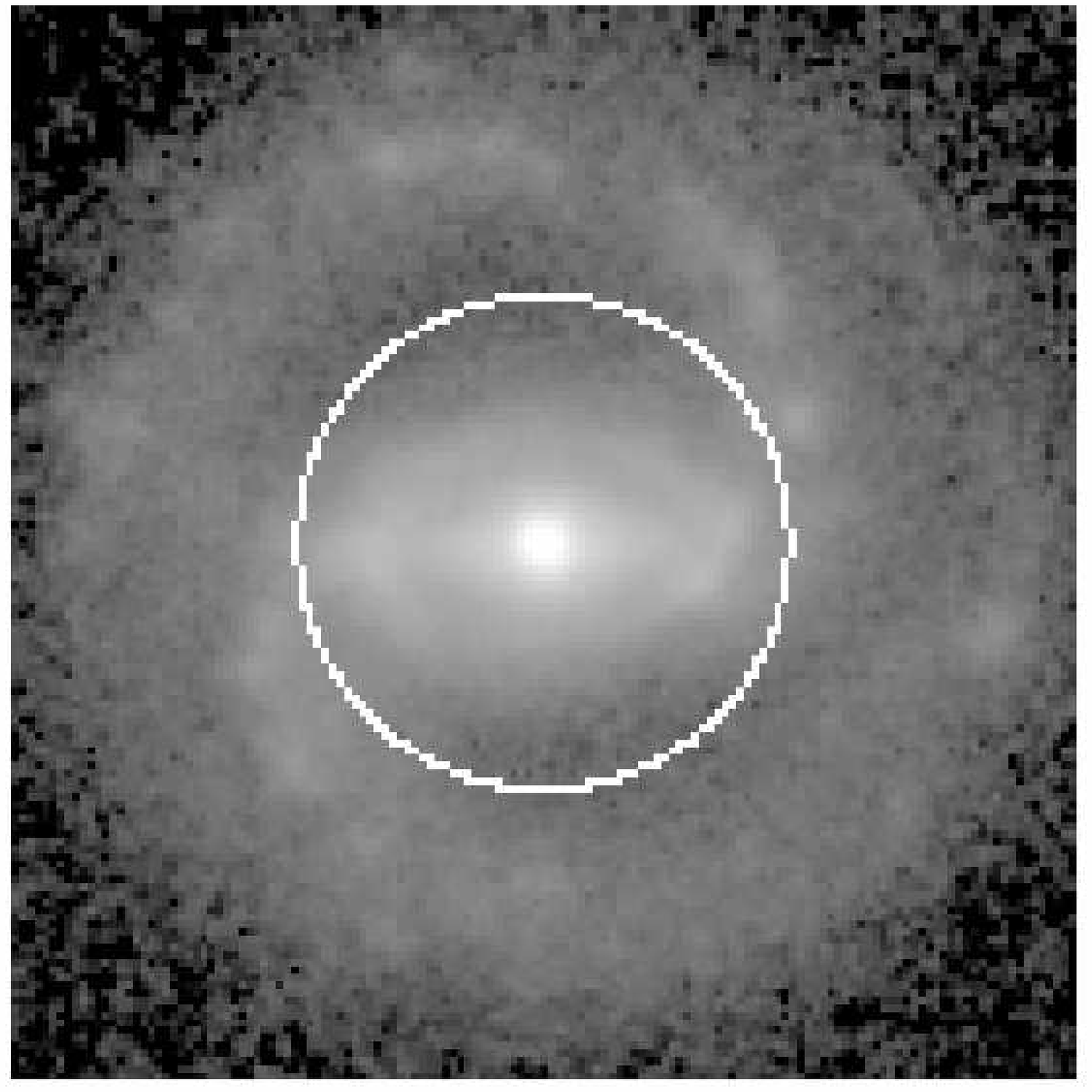}
 \hspace{0.1cm}
 \end{minipage}
 \begin{minipage}[t]{0.68\linewidth}
 \centering
\raisebox{0.5cm}{\includegraphics[width=\textwidth,trim=0 0 0 250,clip]{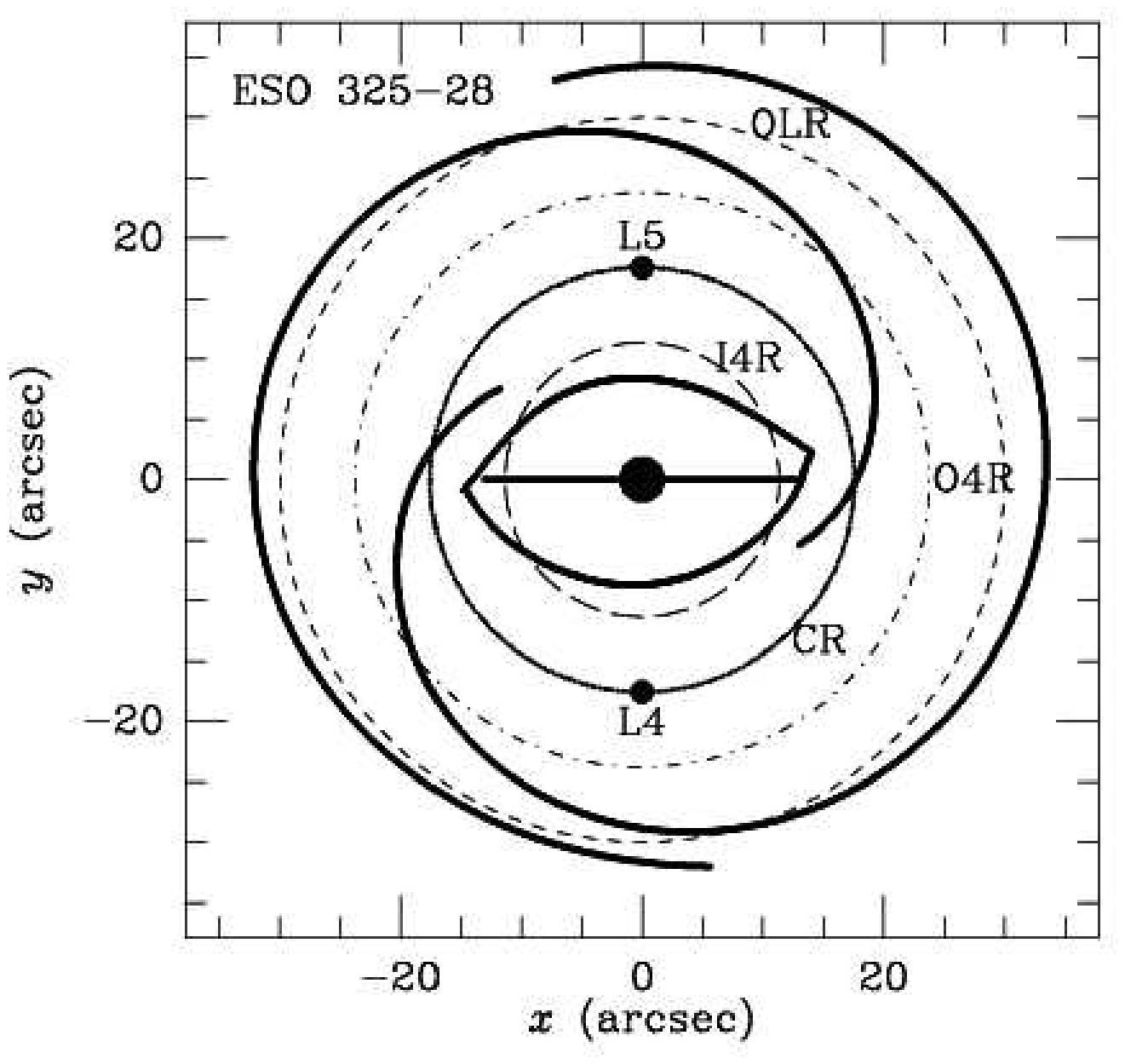}}
 \end{minipage}
 \begin{minipage}[b]{0.45\linewidth}
 \centering
\includegraphics[width=\textwidth]{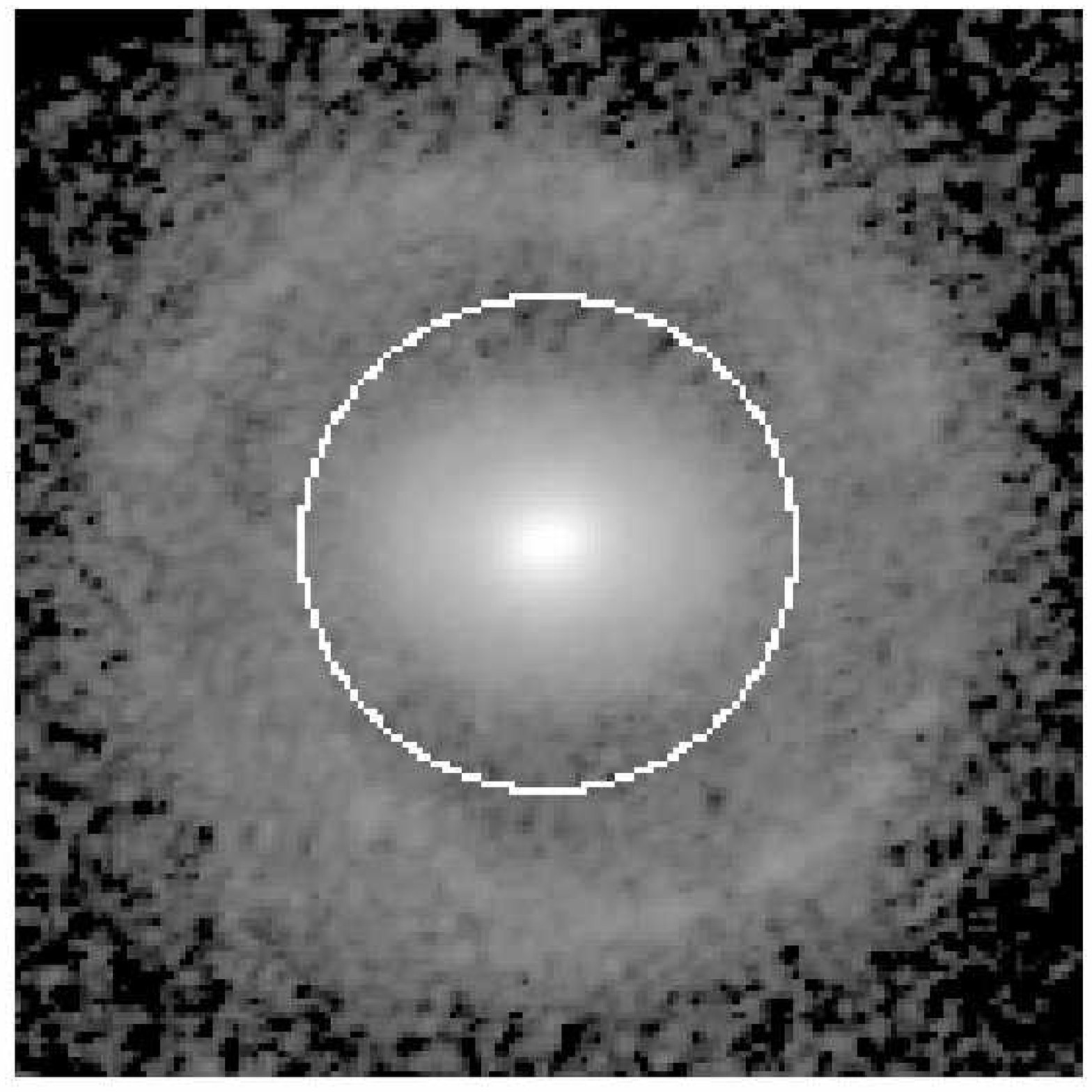}
 \hspace{0.1cm}
 \end{minipage}
 \begin{minipage}[t]{0.68\linewidth}
 \centering
\raisebox{0.5cm}{\includegraphics[width=\textwidth,trim=0 0 0 250,clip]{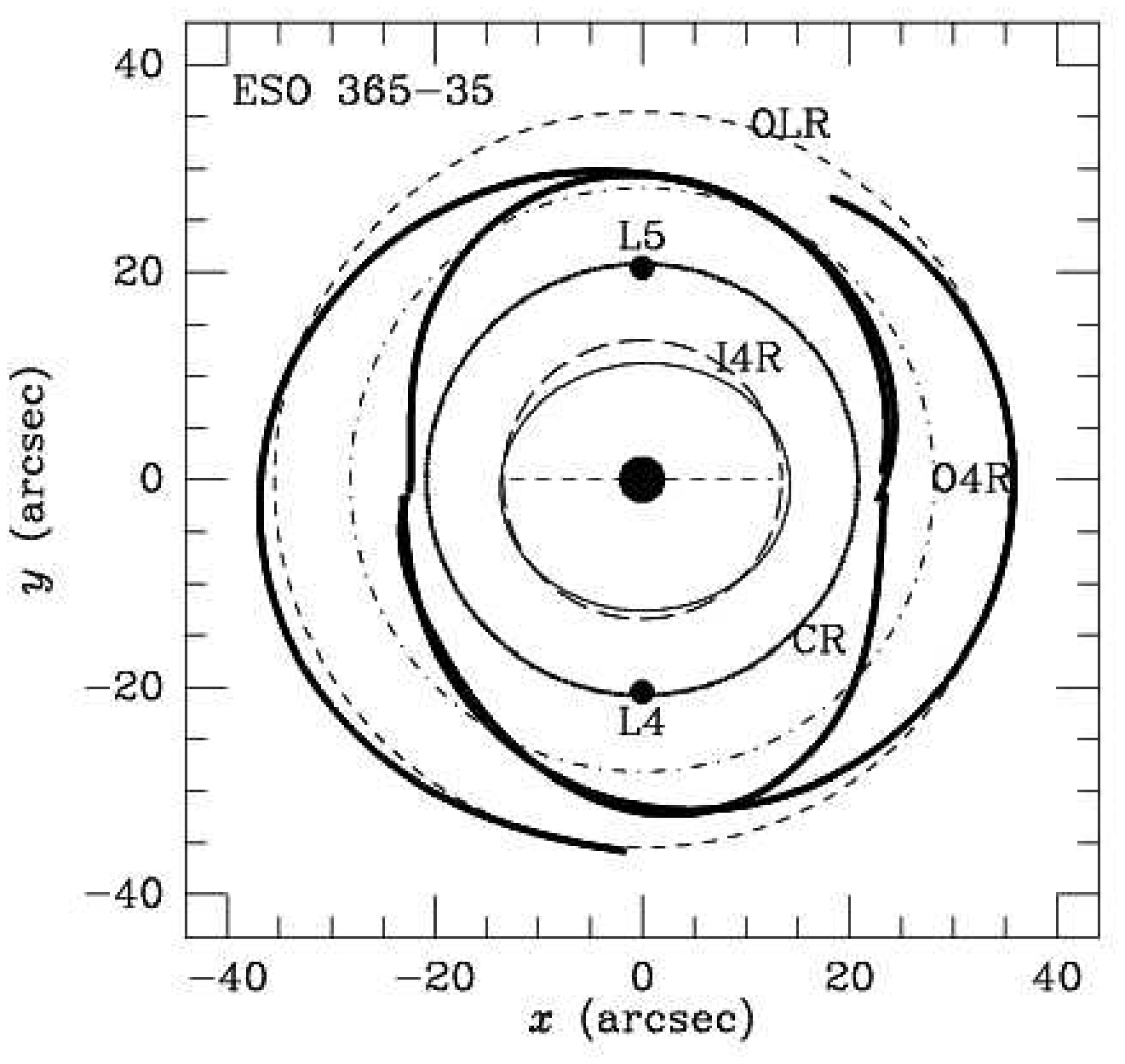}}
 \end{minipage}
 \begin{minipage}[b]{0.45\linewidth}
 \centering
\includegraphics[width=\textwidth]{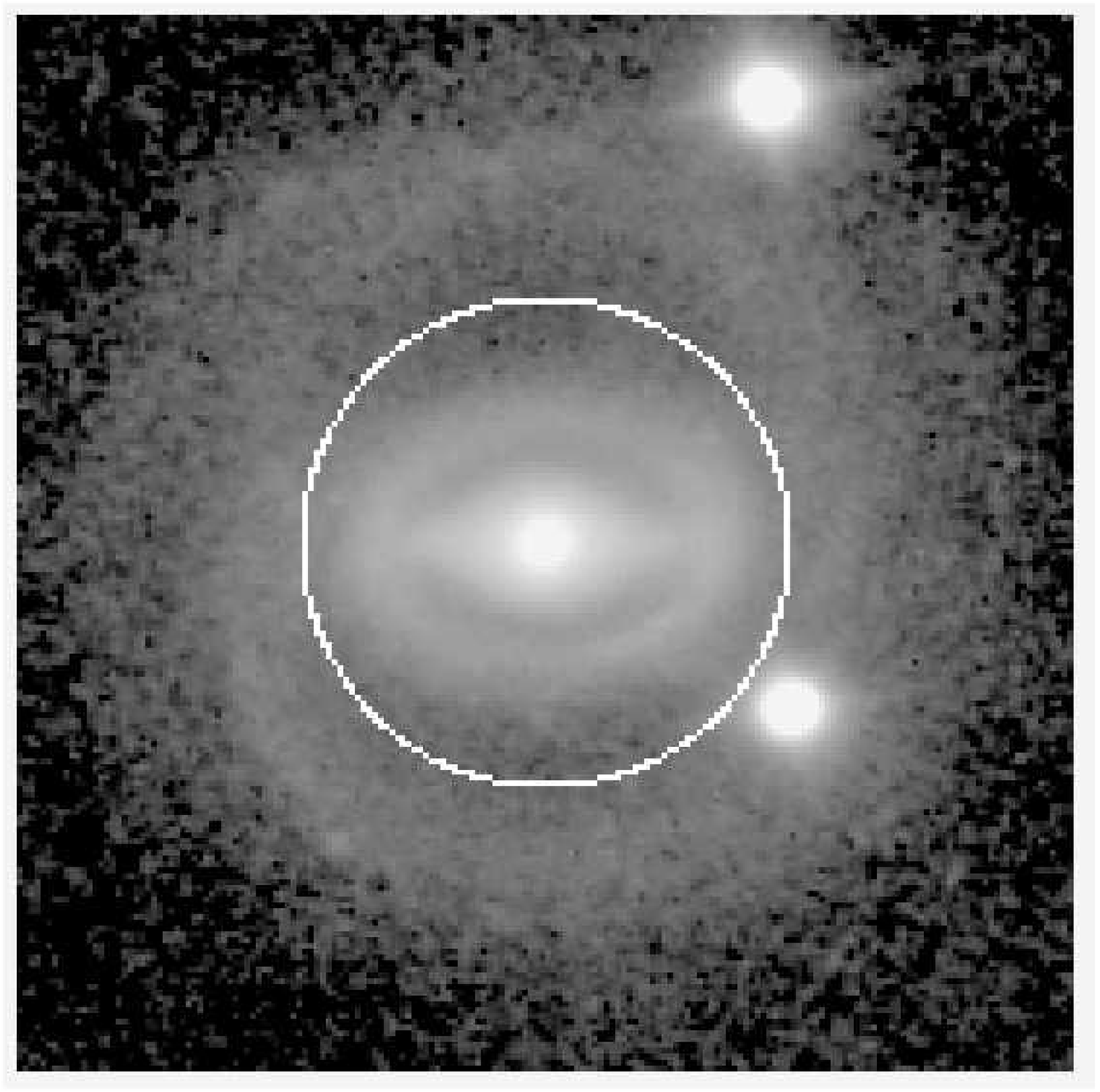}
 \hspace{0.1cm}
 \end{minipage}
 \begin{minipage}[t]{0.68\linewidth}
 \centering
\raisebox{0.5cm}{\includegraphics[width=\textwidth,trim=0 0 0 250,clip]{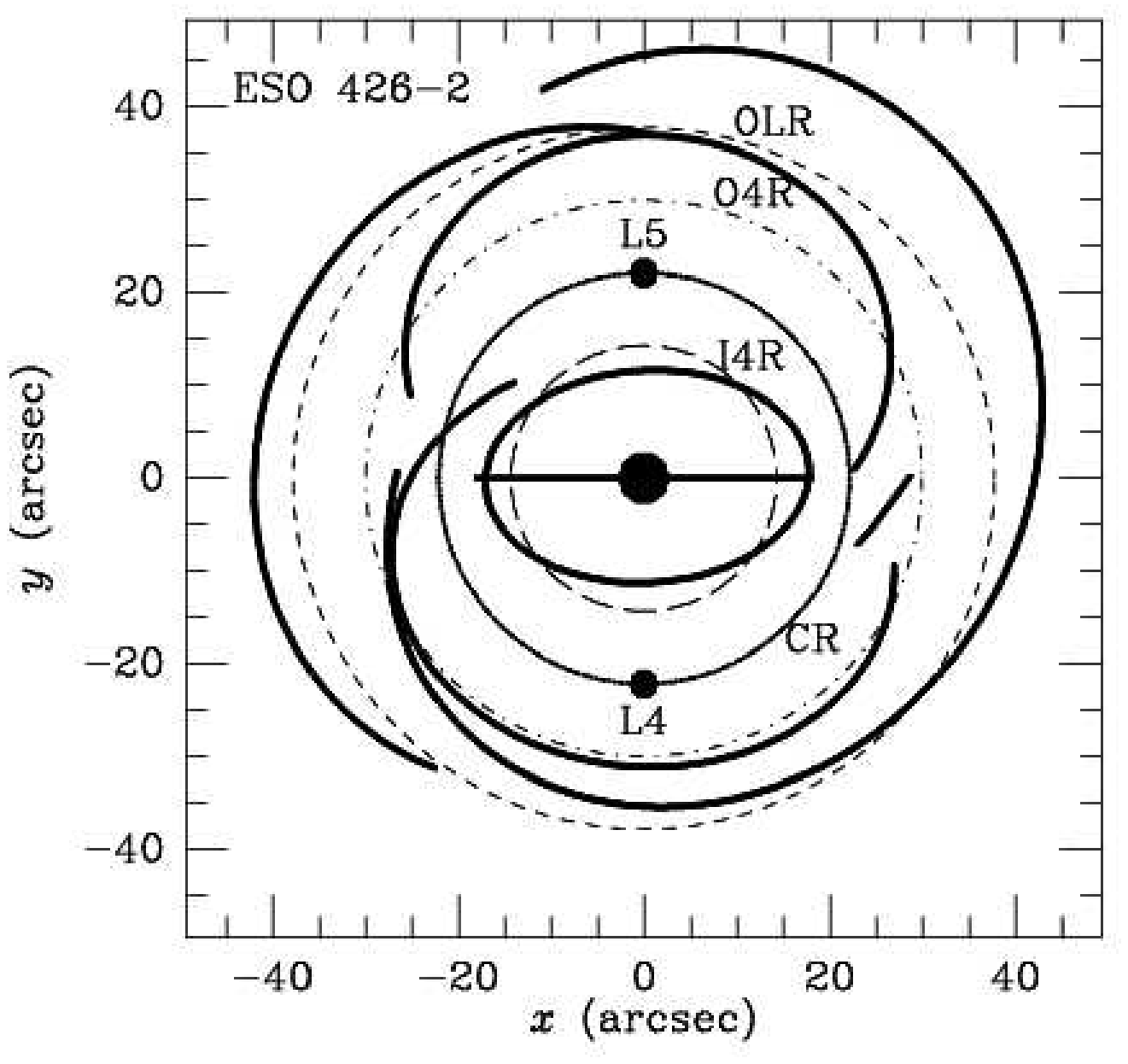}}
 \end{minipage}
\vspace{-1.0truecm}
\caption{(cont.)}
 \end{figure}
 \setcounter{figure}{12}
 \begin{figure}
\vspace{-1.27cm}
 \begin{minipage}[b]{0.45\linewidth}
 \centering
\includegraphics[width=\textwidth]{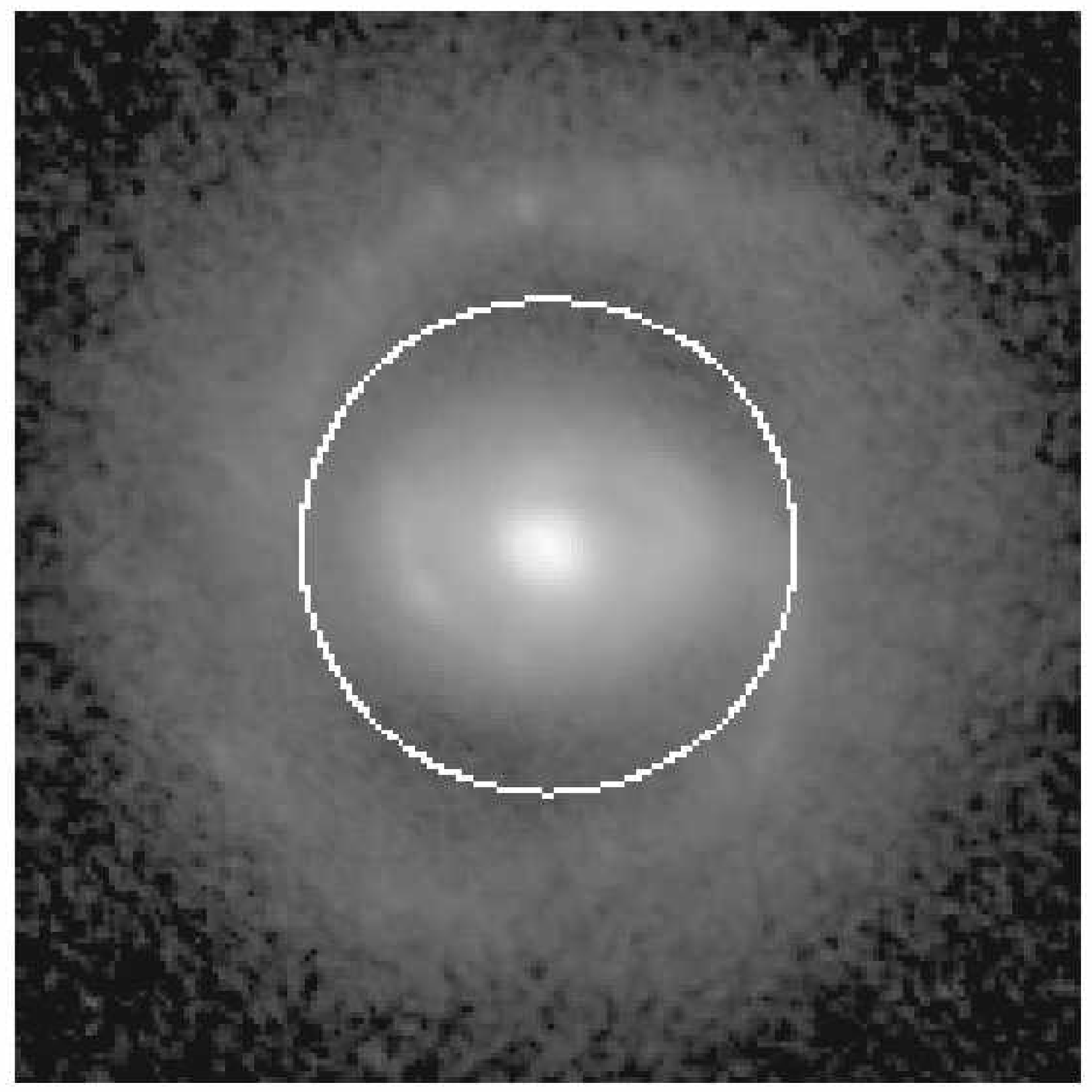}
 \hspace{0.1cm}
 \end{minipage}
 \begin{minipage}[t]{0.68\linewidth}
 \centering
\raisebox{0.5cm}{\includegraphics[width=\textwidth,trim=0 0 0 250,clip]{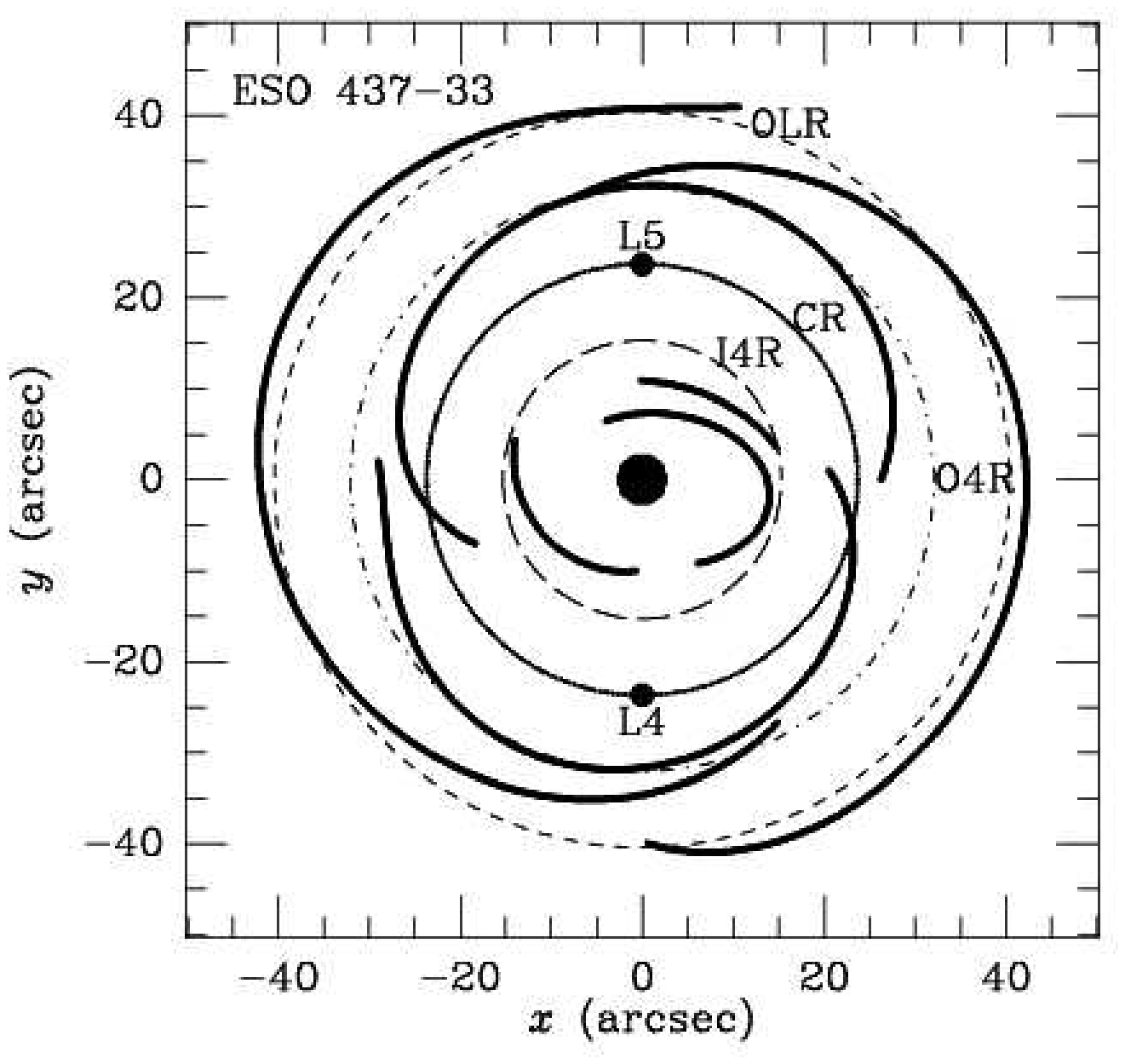}}
 \end{minipage}
 \begin{minipage}[b]{0.45\linewidth}
 \centering
\includegraphics[width=\textwidth]{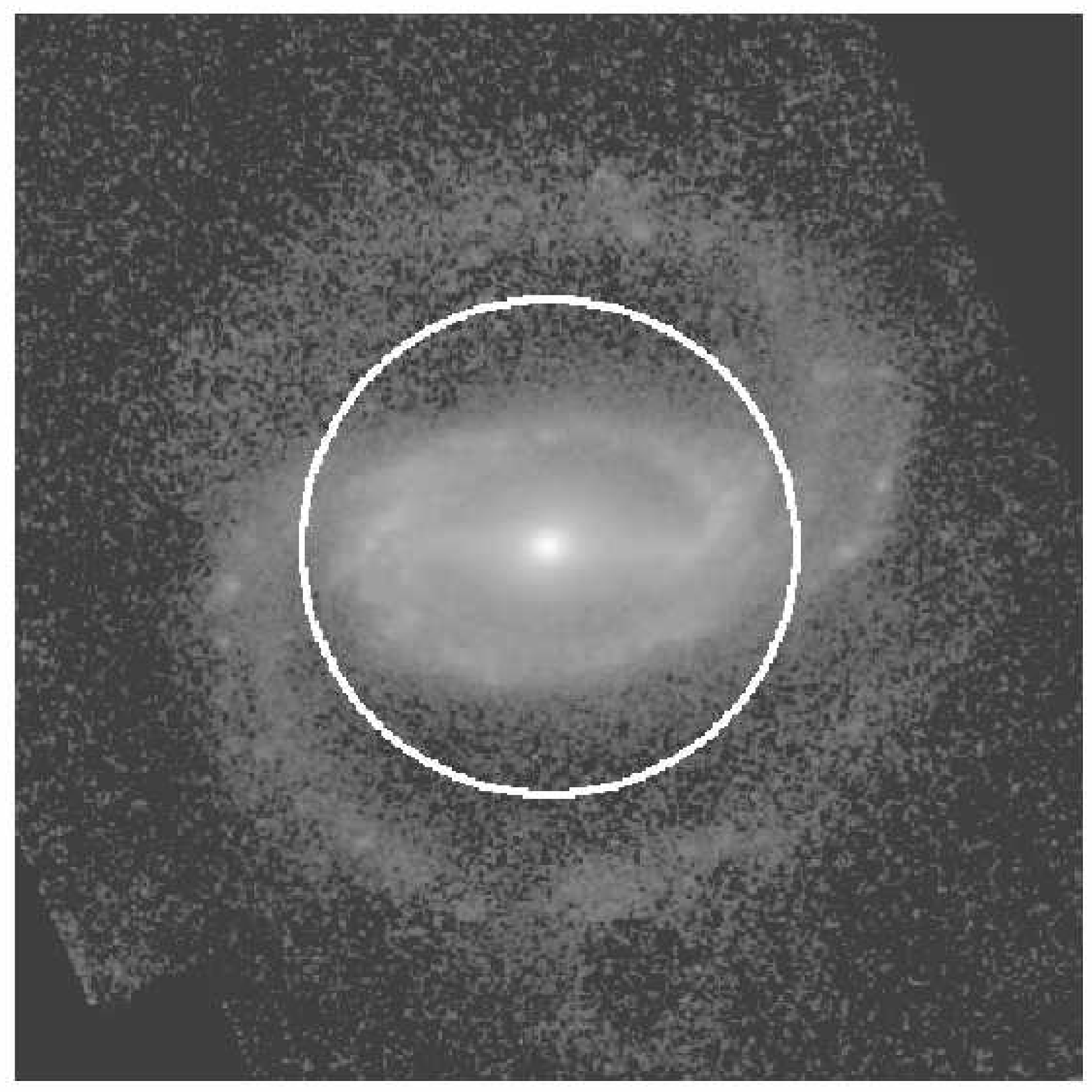}
 \hspace{0.1cm}
 \end{minipage}
 \begin{minipage}[t]{0.68\linewidth}
 \centering
\raisebox{0.5cm}{\includegraphics[width=\textwidth,trim=0 0 0 250,clip]{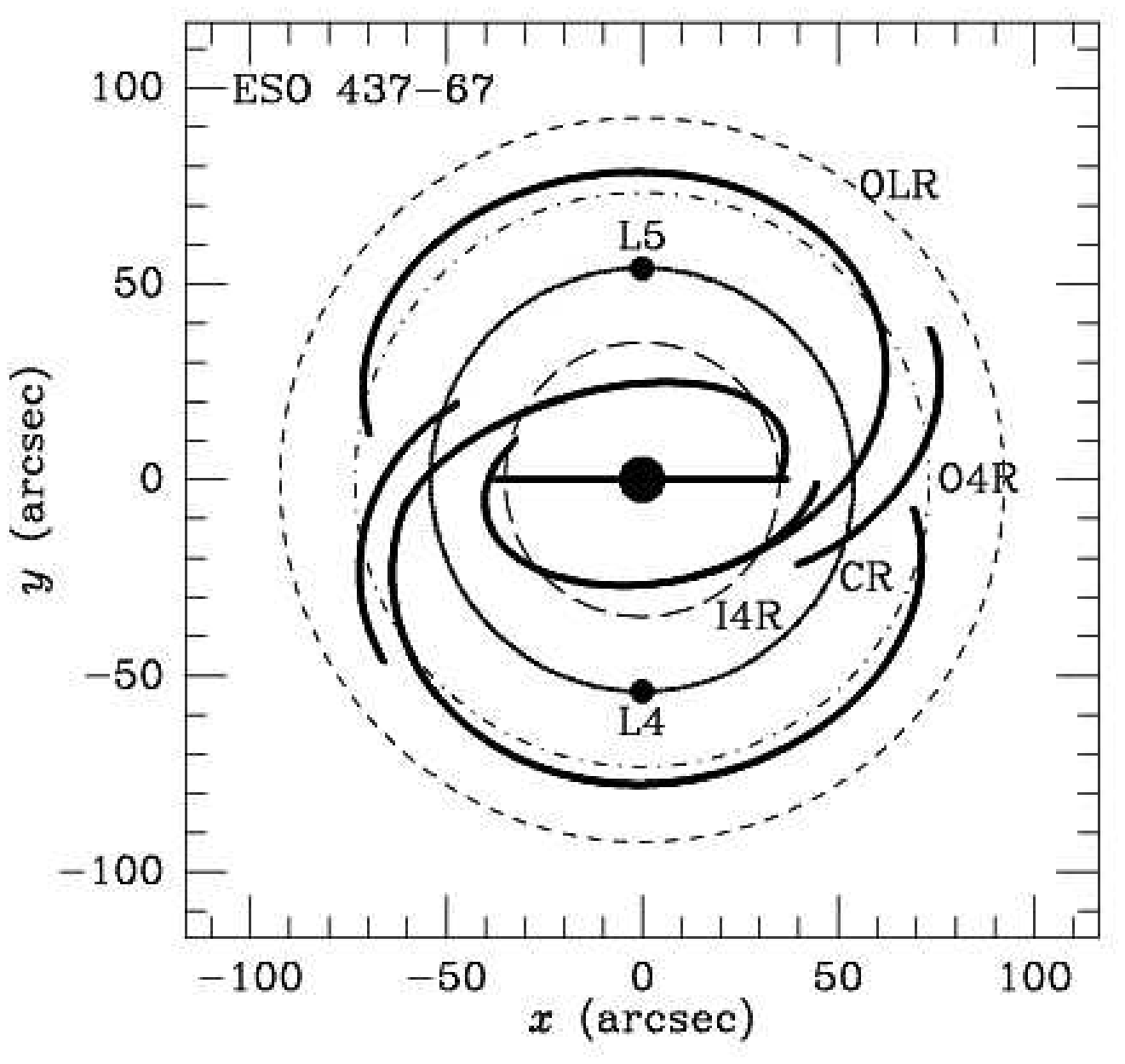}}
 \end{minipage}
 \begin{minipage}[b]{0.45\linewidth}
 \centering
\includegraphics[width=\textwidth]{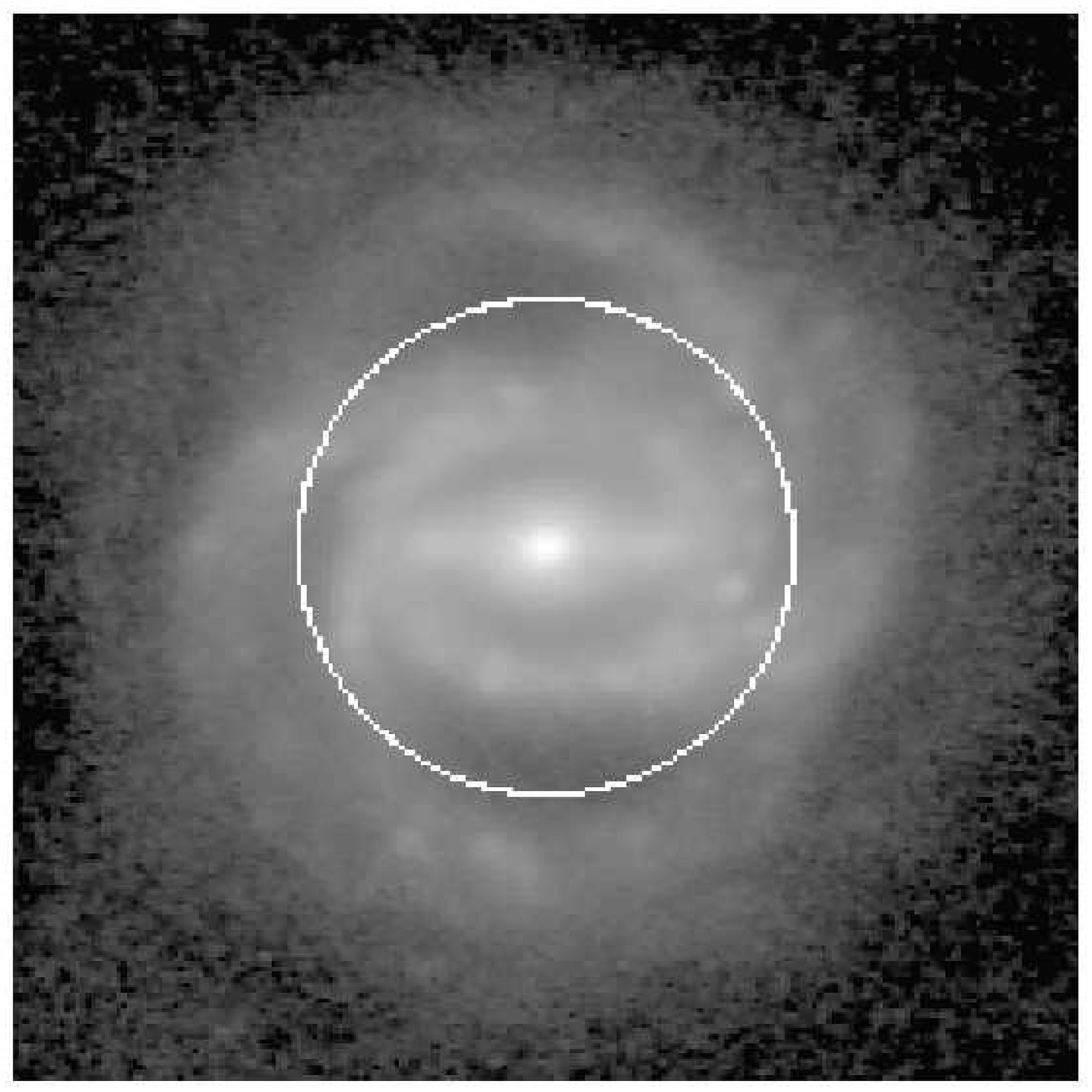}
 \hspace{0.1cm}
 \end{minipage}
 \begin{minipage}[t]{0.68\linewidth}
 \centering
\raisebox{0.5cm}{\includegraphics[width=\textwidth,trim=0 0 0 250,clip]{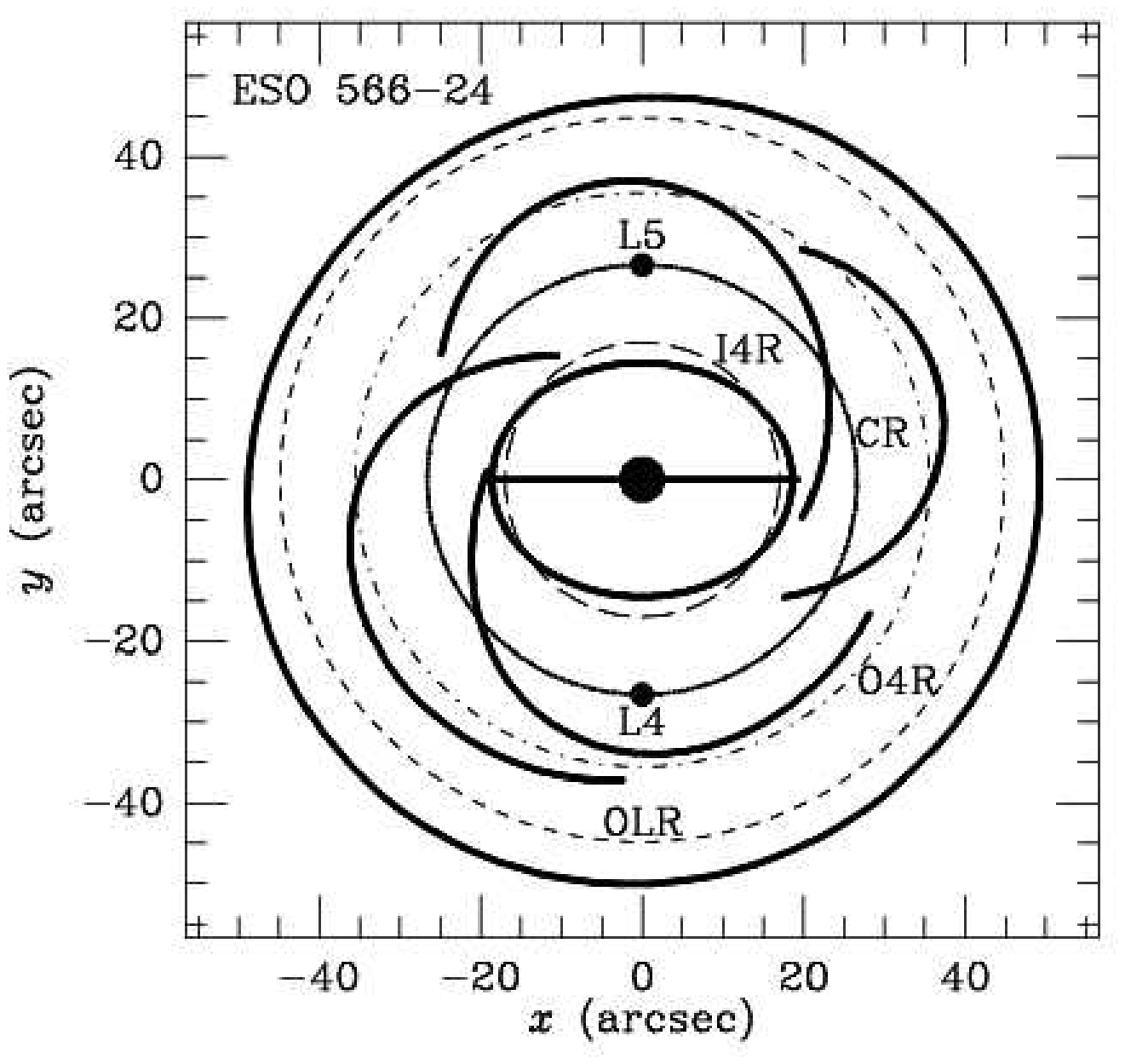}}
 \end{minipage}
 \begin{minipage}[b]{0.45\linewidth}
 \centering
\includegraphics[width=\textwidth]{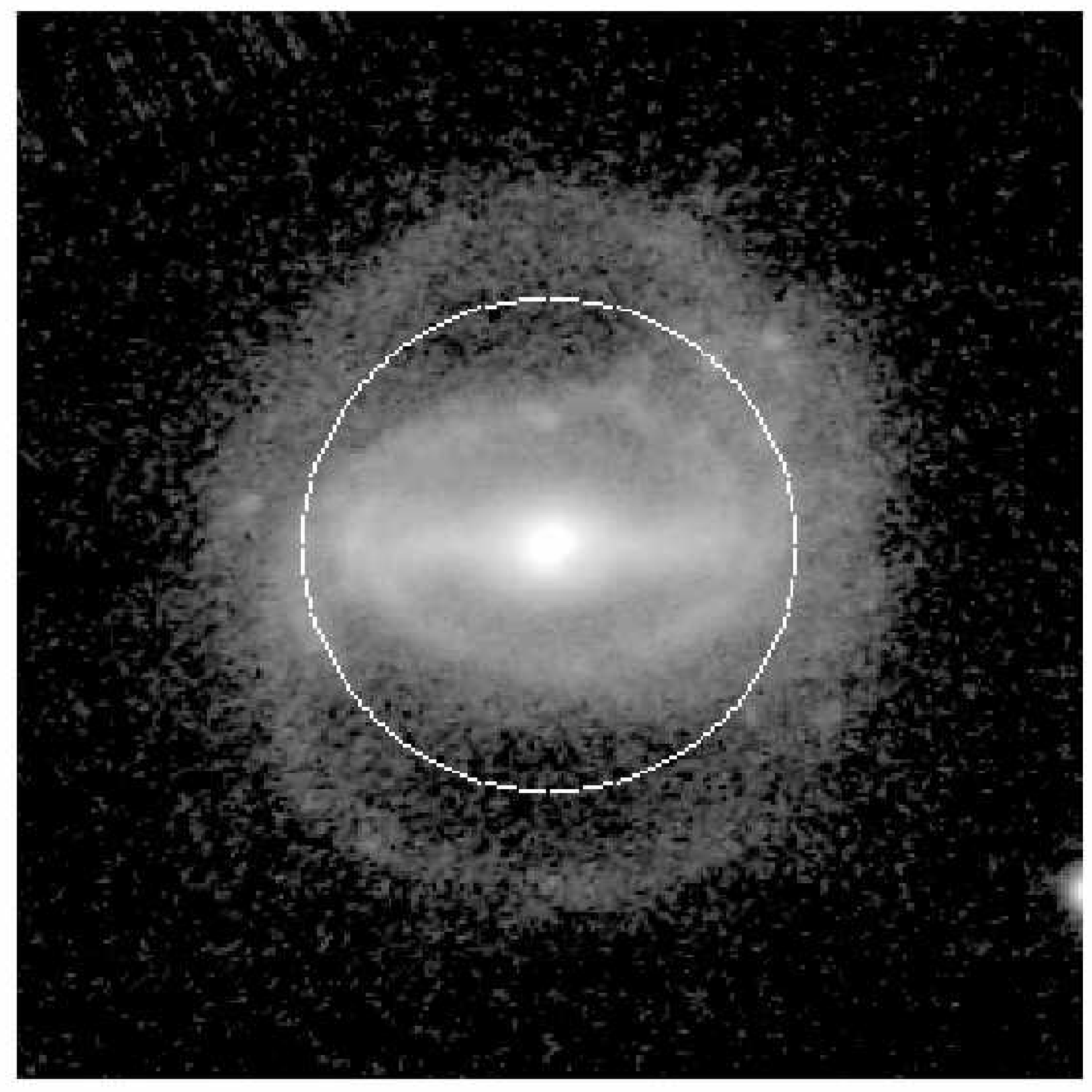}
 \hspace{0.1cm}
 \end{minipage}
 \begin{minipage}[t]{0.68\linewidth}
 \centering
\raisebox{0.5cm}{\includegraphics[width=\textwidth,trim=0 0 0 250,clip]{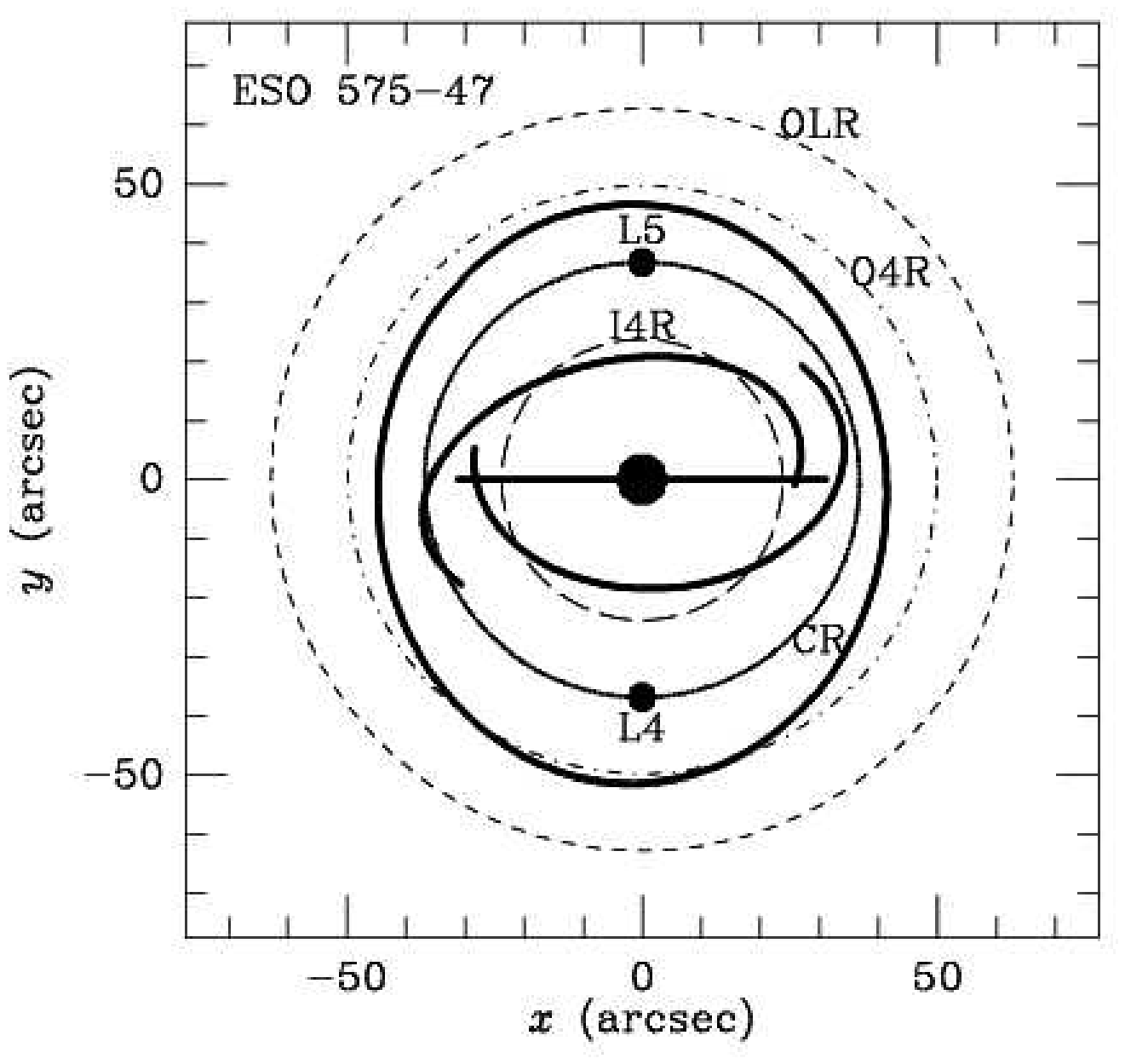}}
 \end{minipage}
 \begin{minipage}[b]{0.45\linewidth}
 \centering
\includegraphics[width=\textwidth]{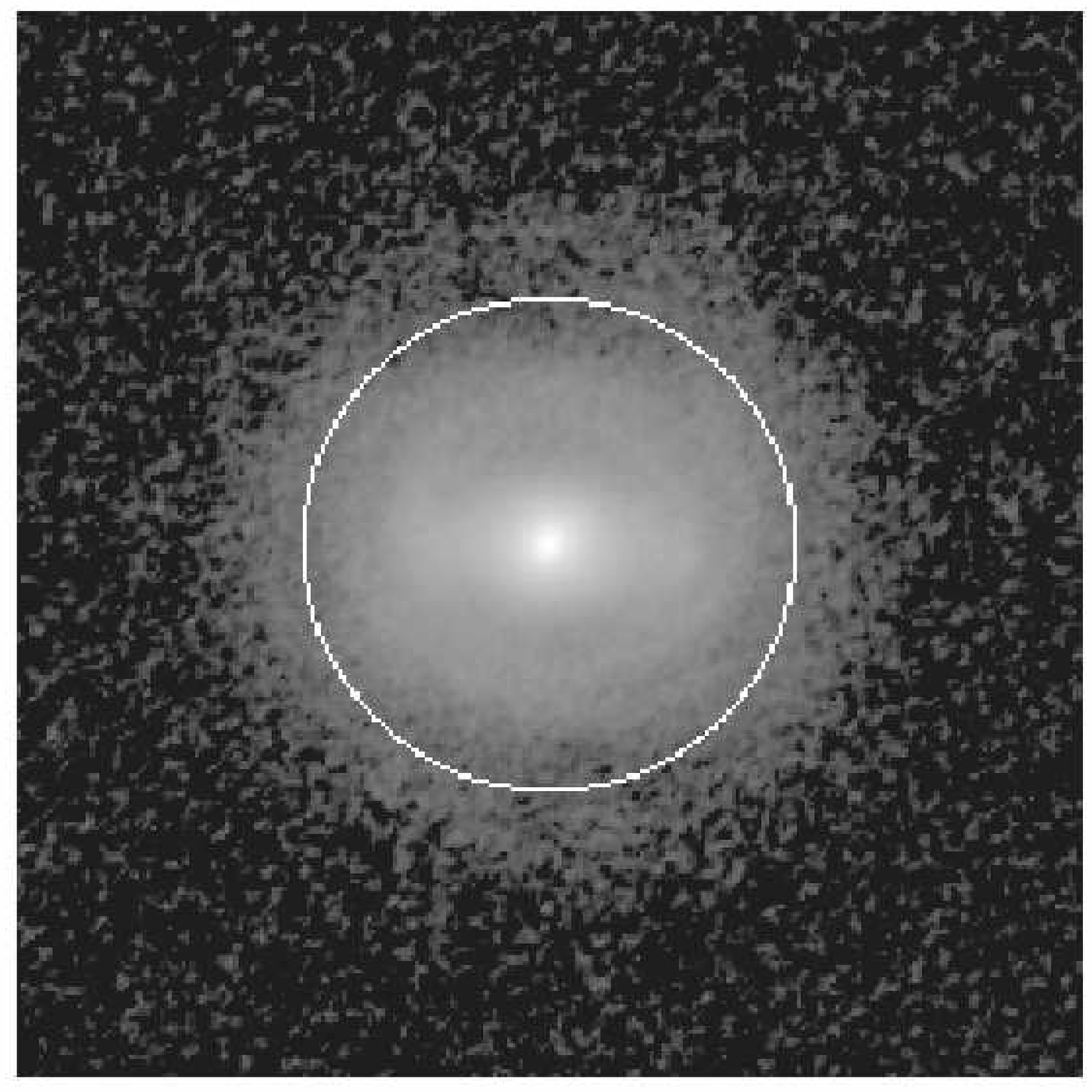}
 \hspace{0.1cm}
 \end{minipage}
 \begin{minipage}[t]{0.68\linewidth}
 \centering
\raisebox{0.5cm}{\includegraphics[width=\textwidth,trim=0 0 0 250,clip]{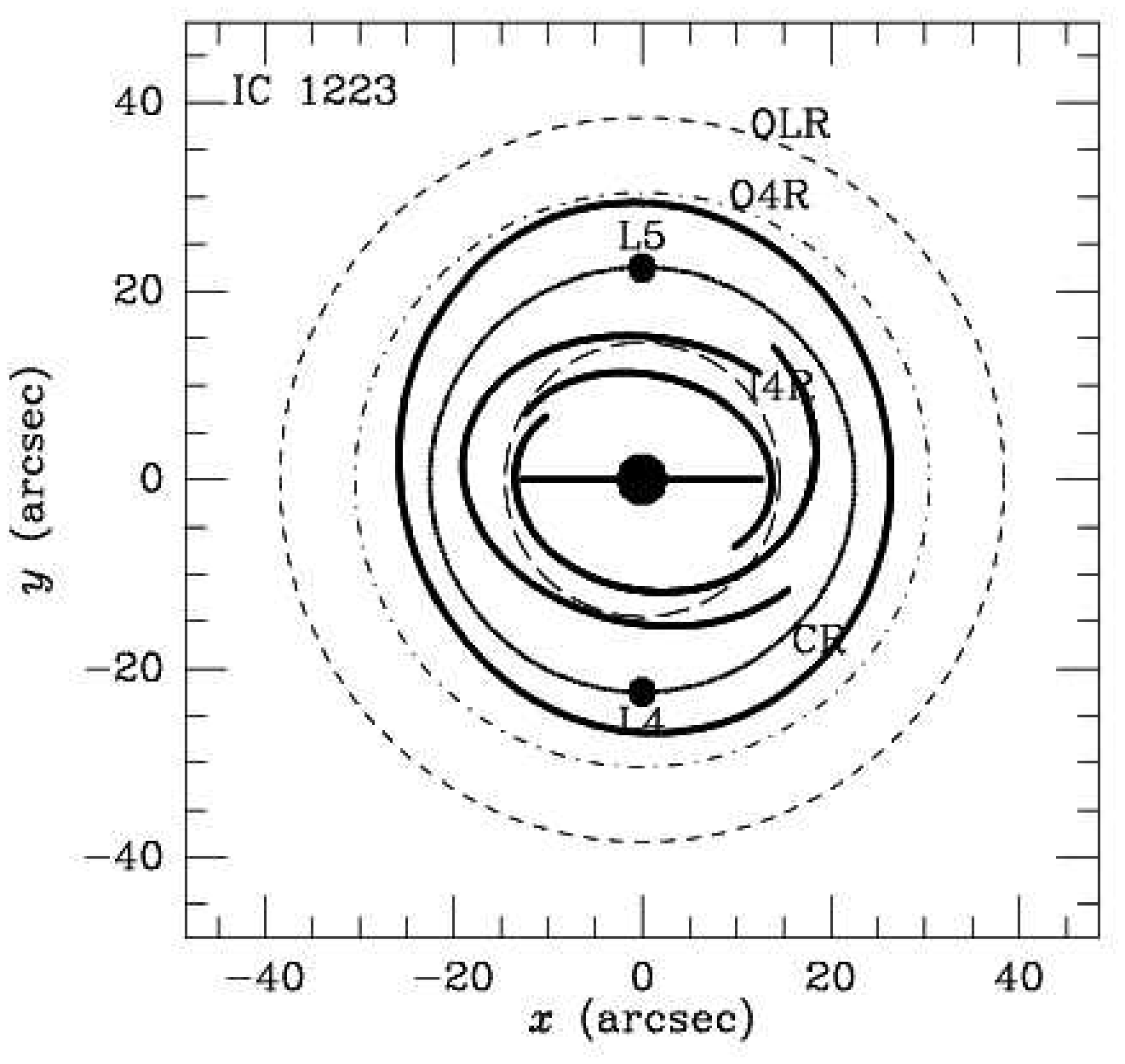}}
 \end{minipage}
\vspace{-1.0truecm}
\caption{(cont.)}
 \end{figure}
 \setcounter{figure}{12}
 \begin{figure}
\vspace{-1.27cm}
 \begin{minipage}[b]{0.45\linewidth}
 \centering
\includegraphics[width=\textwidth]{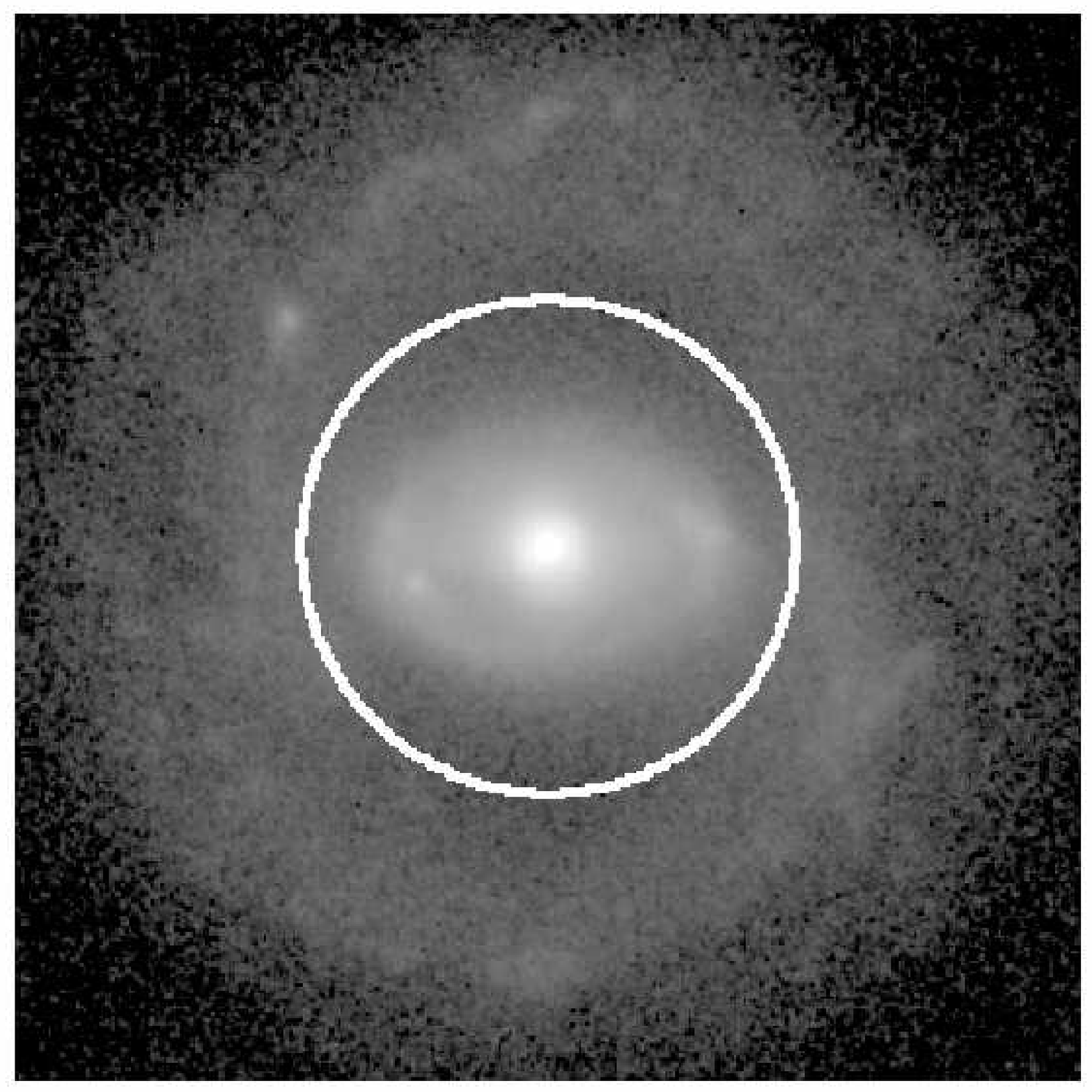}
 \hspace{0.1cm}
 \end{minipage}
 \begin{minipage}[t]{0.68\linewidth}
 \centering
\raisebox{0.5cm}{\includegraphics[width=\textwidth,trim=0 0 0 250,clip]{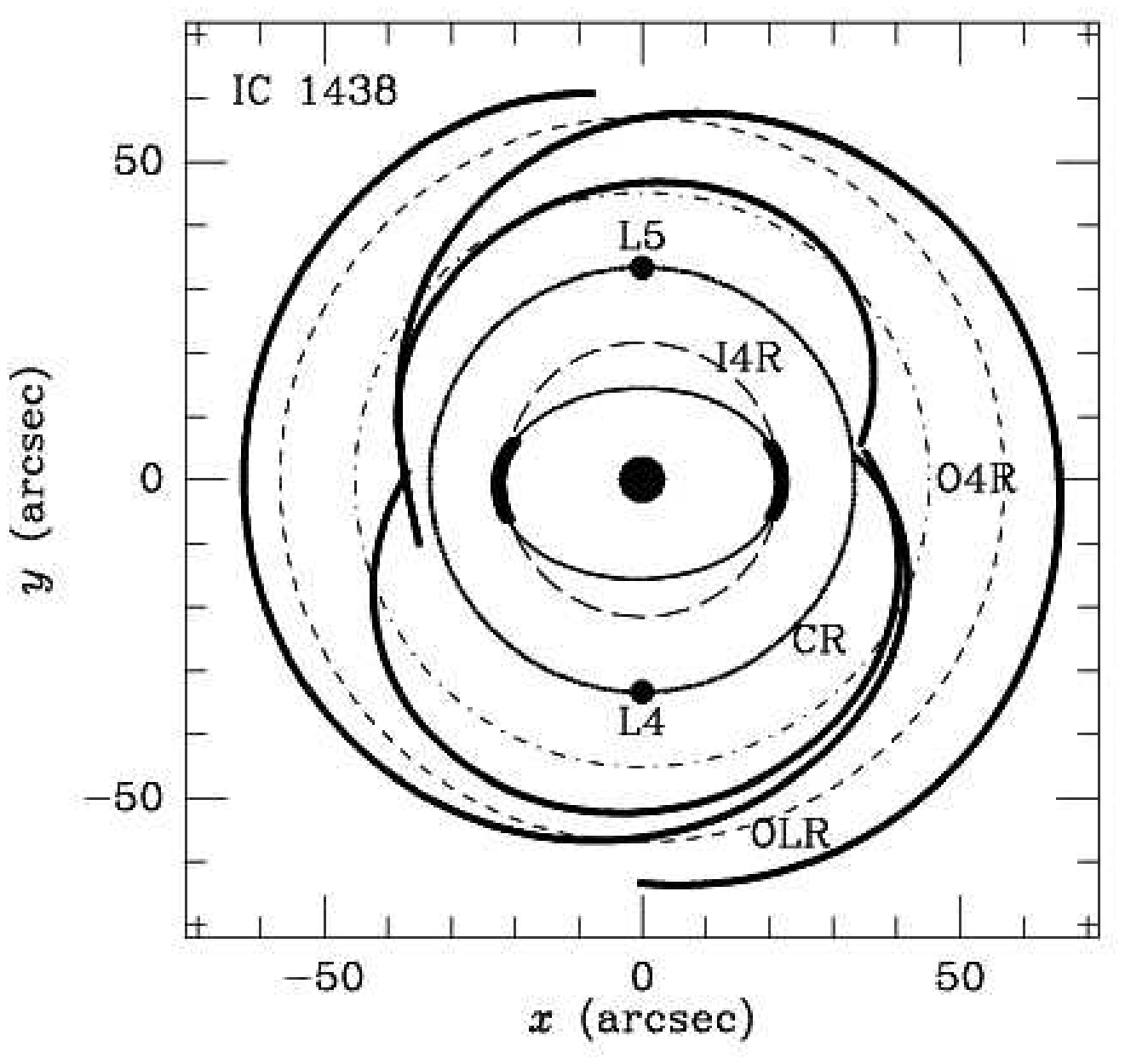}}
 \end{minipage}
 \begin{minipage}[b]{0.45\linewidth}
 \centering
\includegraphics[width=\textwidth]{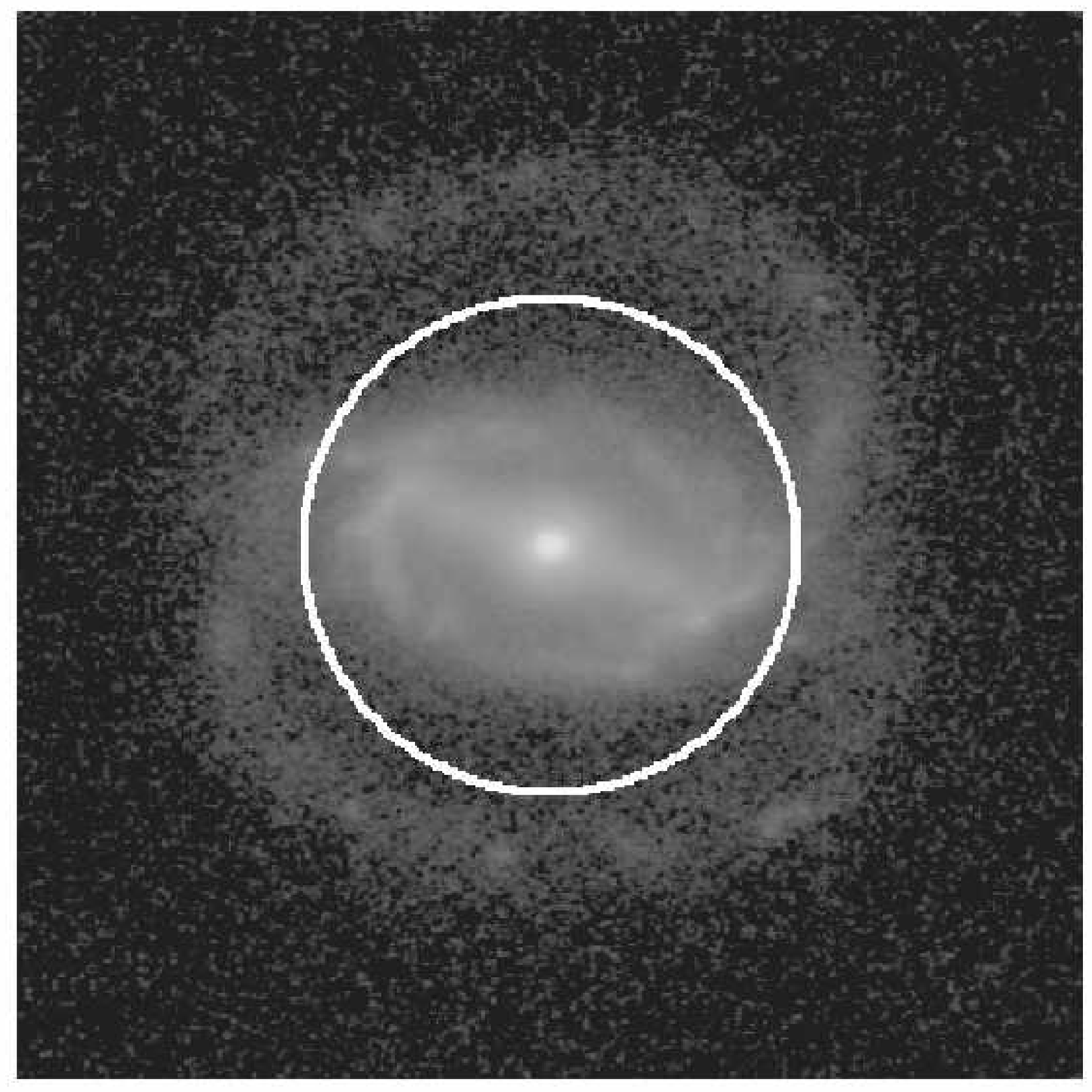}
 \hspace{0.1cm}
 \end{minipage}
 \begin{minipage}[t]{0.68\linewidth}
 \centering
\raisebox{0.5cm}{\includegraphics[width=\textwidth,trim=0 0 0 250,clip]{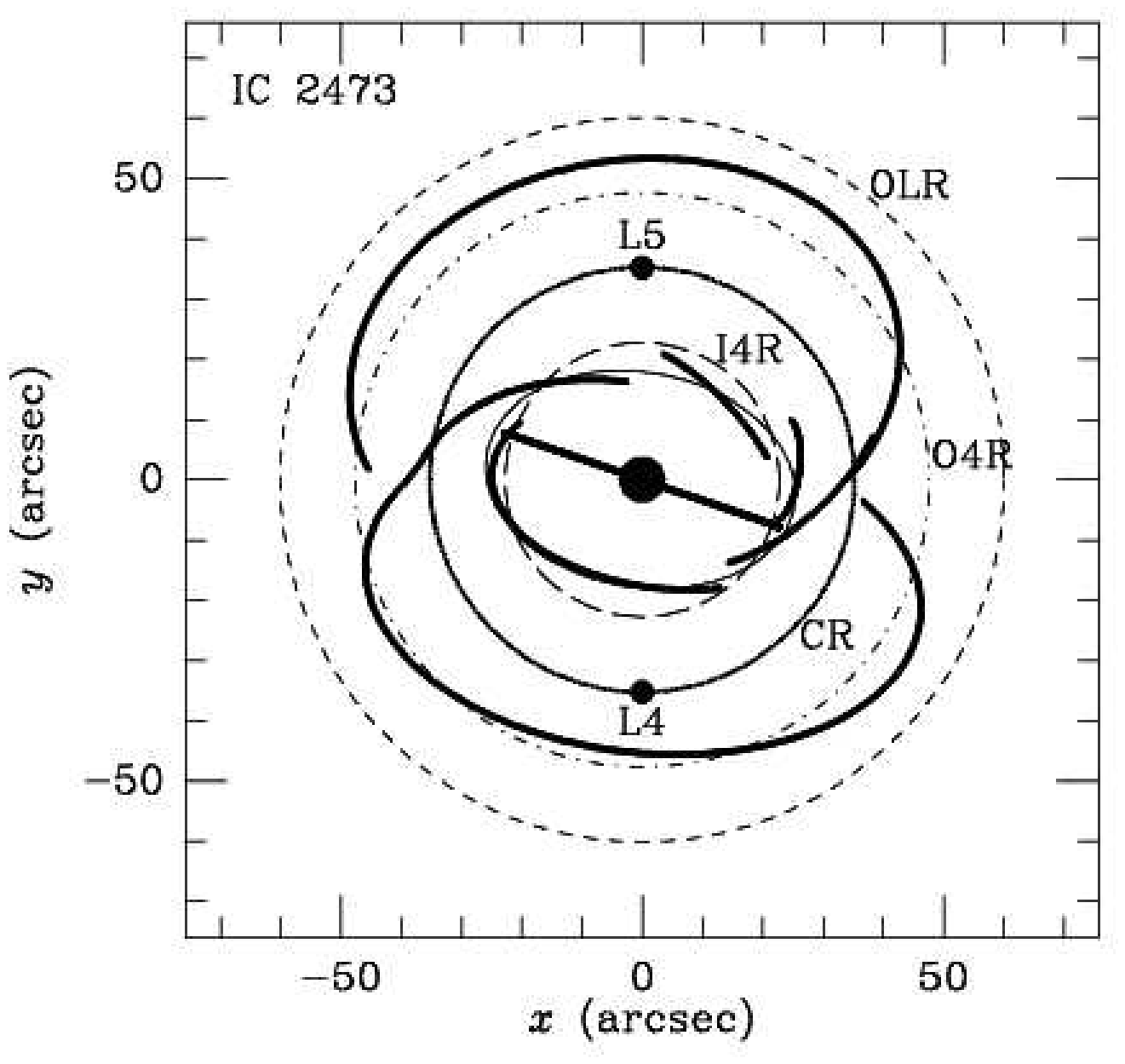}}
 \end{minipage}
 \begin{minipage}[b]{0.45\linewidth}
 \centering
\includegraphics[width=\textwidth]{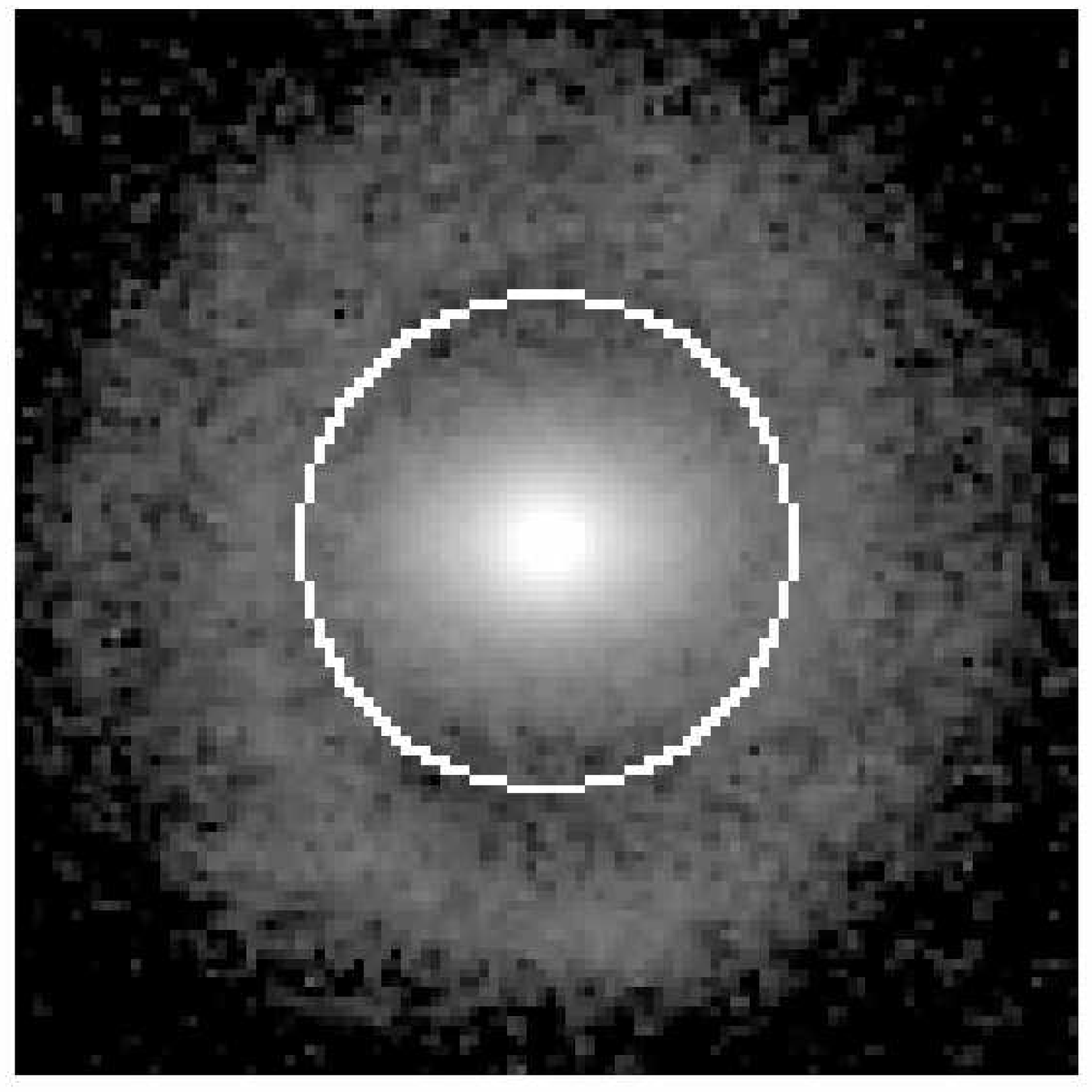}
 \hspace{0.1cm}
 \end{minipage}
 \begin{minipage}[t]{0.68\linewidth}
 \centering
\raisebox{0.5cm}{\includegraphics[width=\textwidth,trim=0 0 0 250,clip]{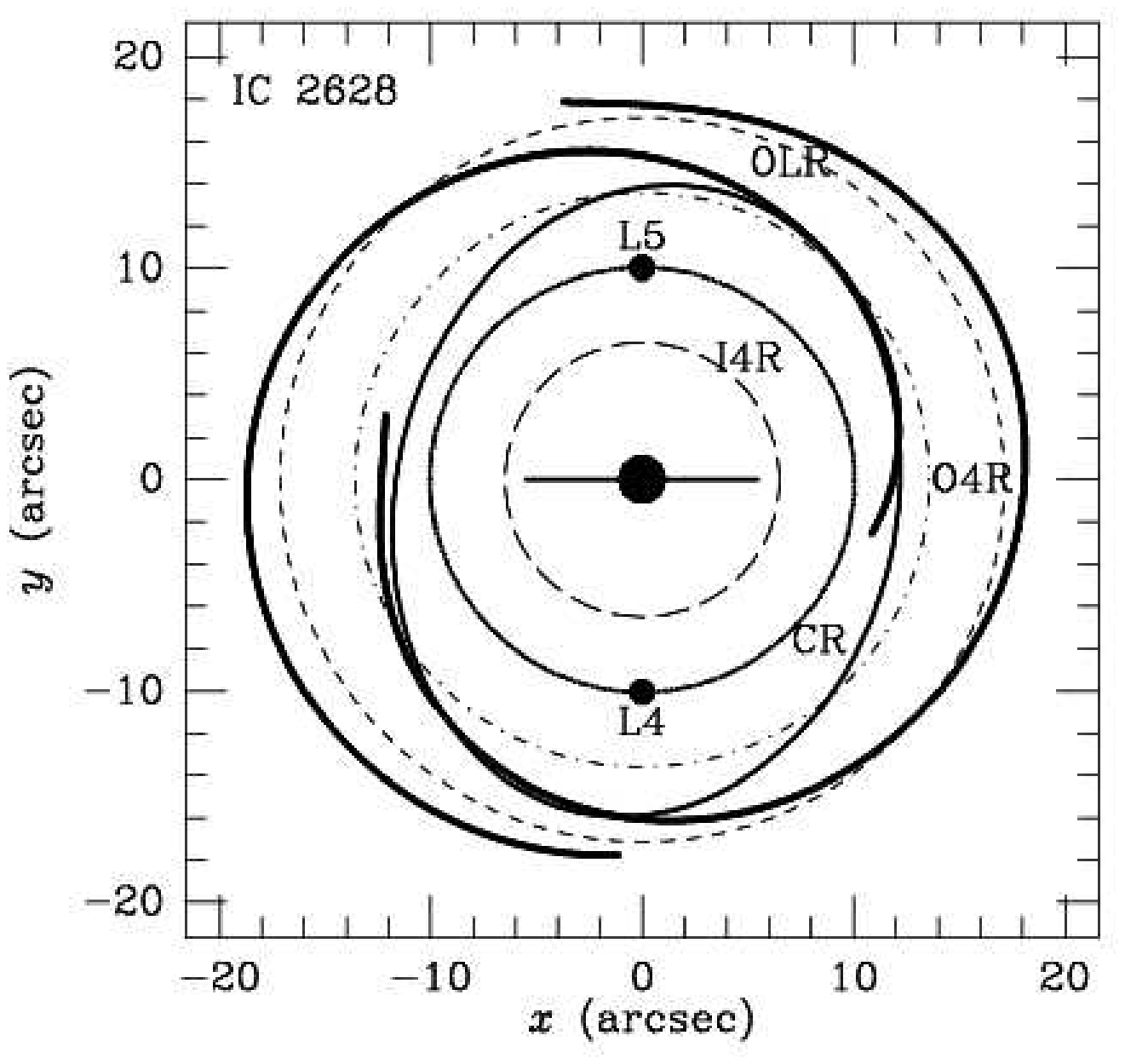}}
 \end{minipage}
 \begin{minipage}[b]{0.45\linewidth}
 \centering
\includegraphics[width=\textwidth]{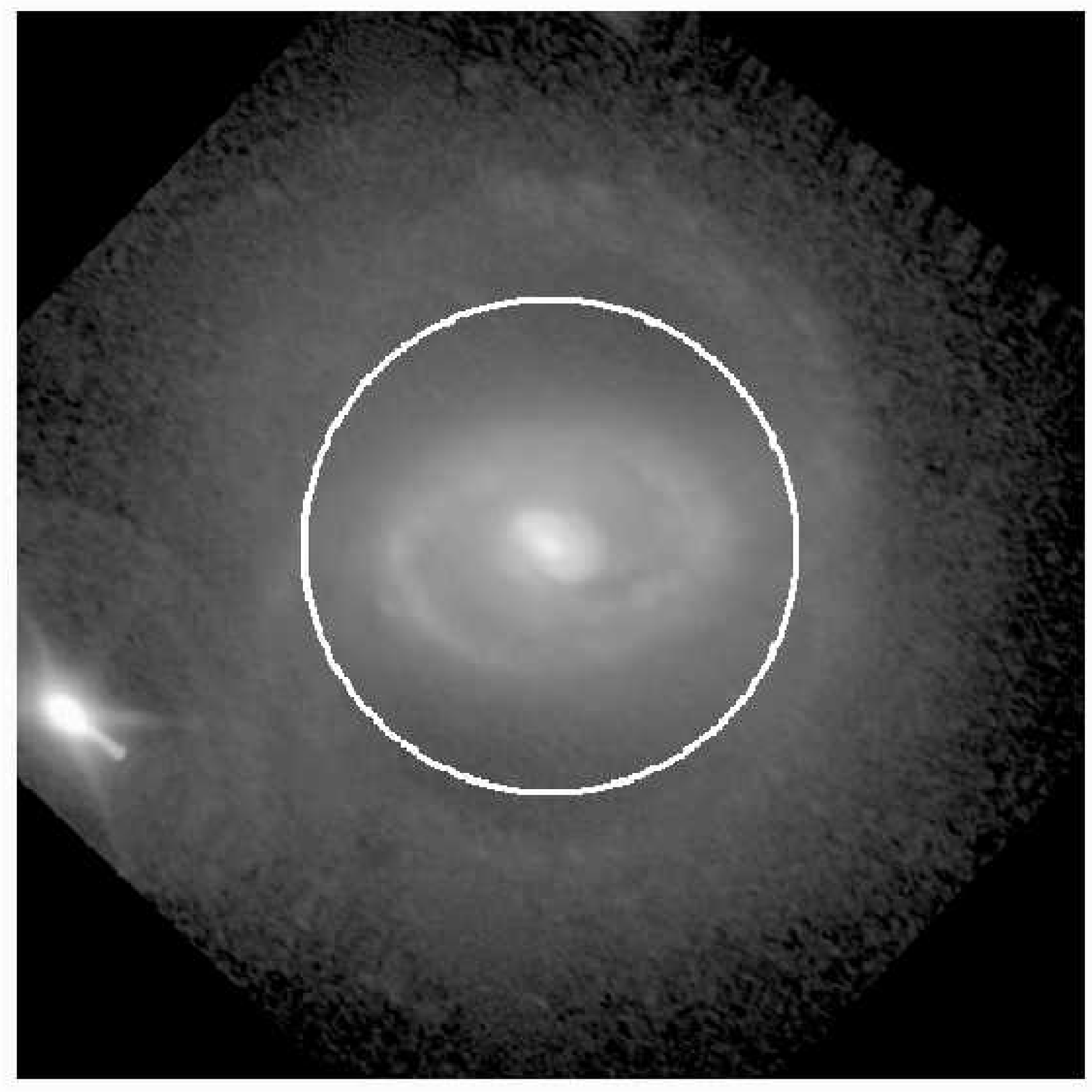}
 \hspace{0.1cm}
 \end{minipage}
 \begin{minipage}[t]{0.68\linewidth}
 \centering
\raisebox{0.5cm}{\includegraphics[width=\textwidth,trim=0 0 0 250,clip]{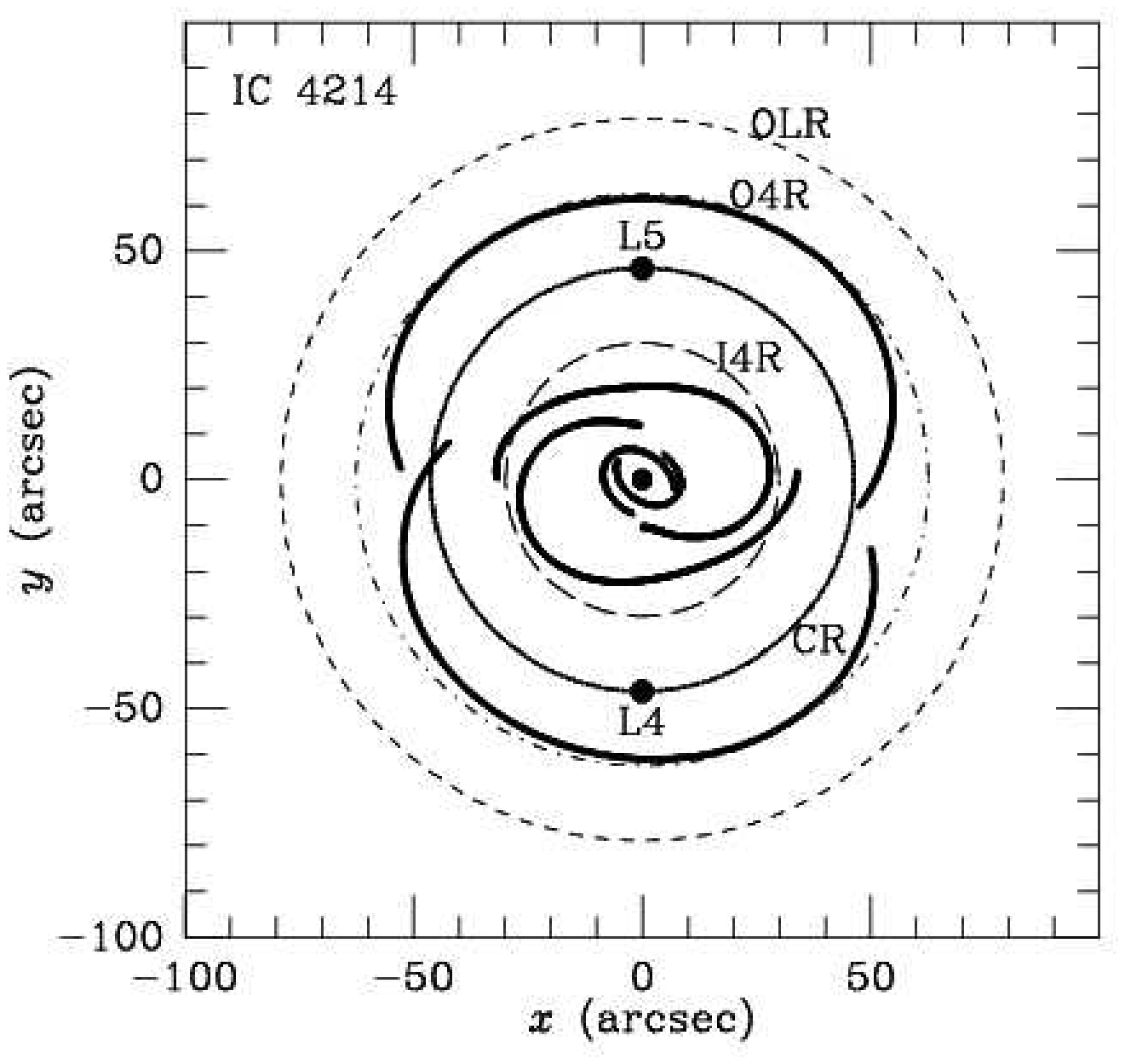}}
 \end{minipage}
 \begin{minipage}[b]{0.45\linewidth}
 \centering
\includegraphics[width=\textwidth]{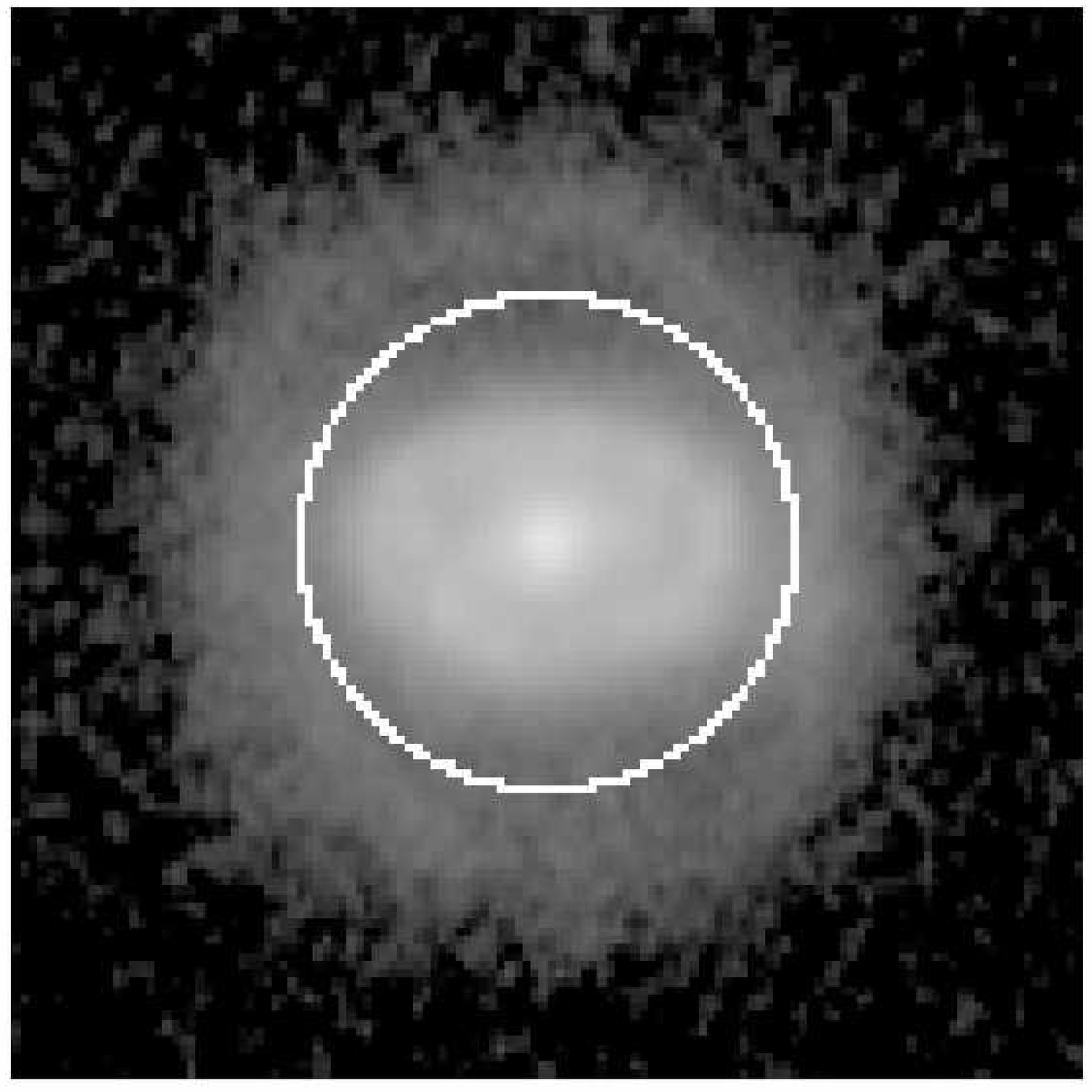}
 \hspace{0.1cm}
 \end{minipage}
 \begin{minipage}[t]{0.68\linewidth}
 \centering
\raisebox{0.5cm}{\includegraphics[width=\textwidth,trim=0 0 0 250,clip]{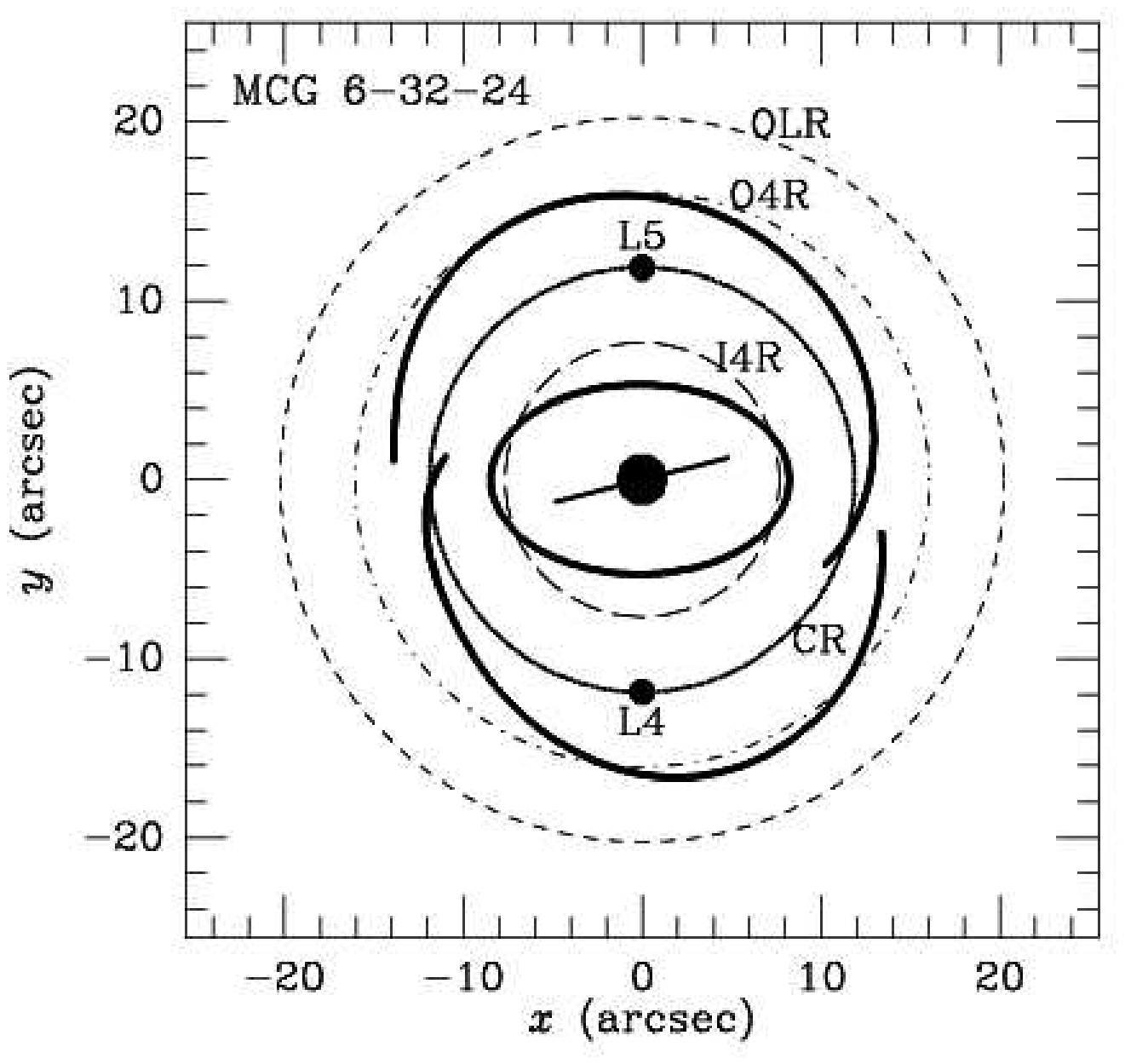}}
 \end{minipage}
\vspace{-1.0truecm}
\caption{(cont.)}
 \end{figure}
 \setcounter{figure}{12}
 \begin{figure}
\vspace{-1.27cm}
 \begin{minipage}[b]{0.45\linewidth}
 \centering
\includegraphics[width=\textwidth]{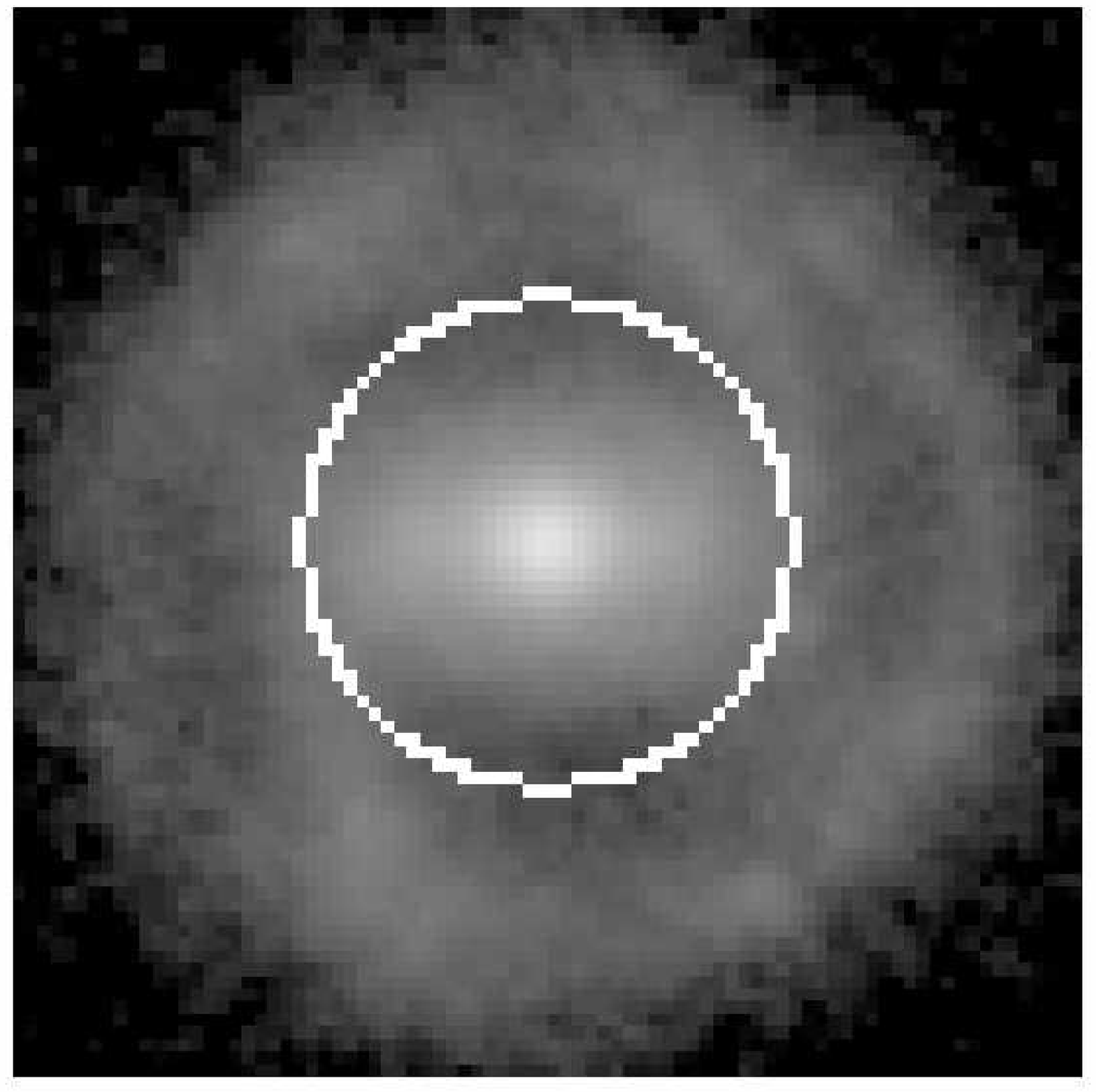}
 \hspace{0.1cm}
 \end{minipage}
 \begin{minipage}[t]{0.68\linewidth}
 \centering
\raisebox{0.5cm}{\includegraphics[width=\textwidth,trim=0 0 0 250,clip]{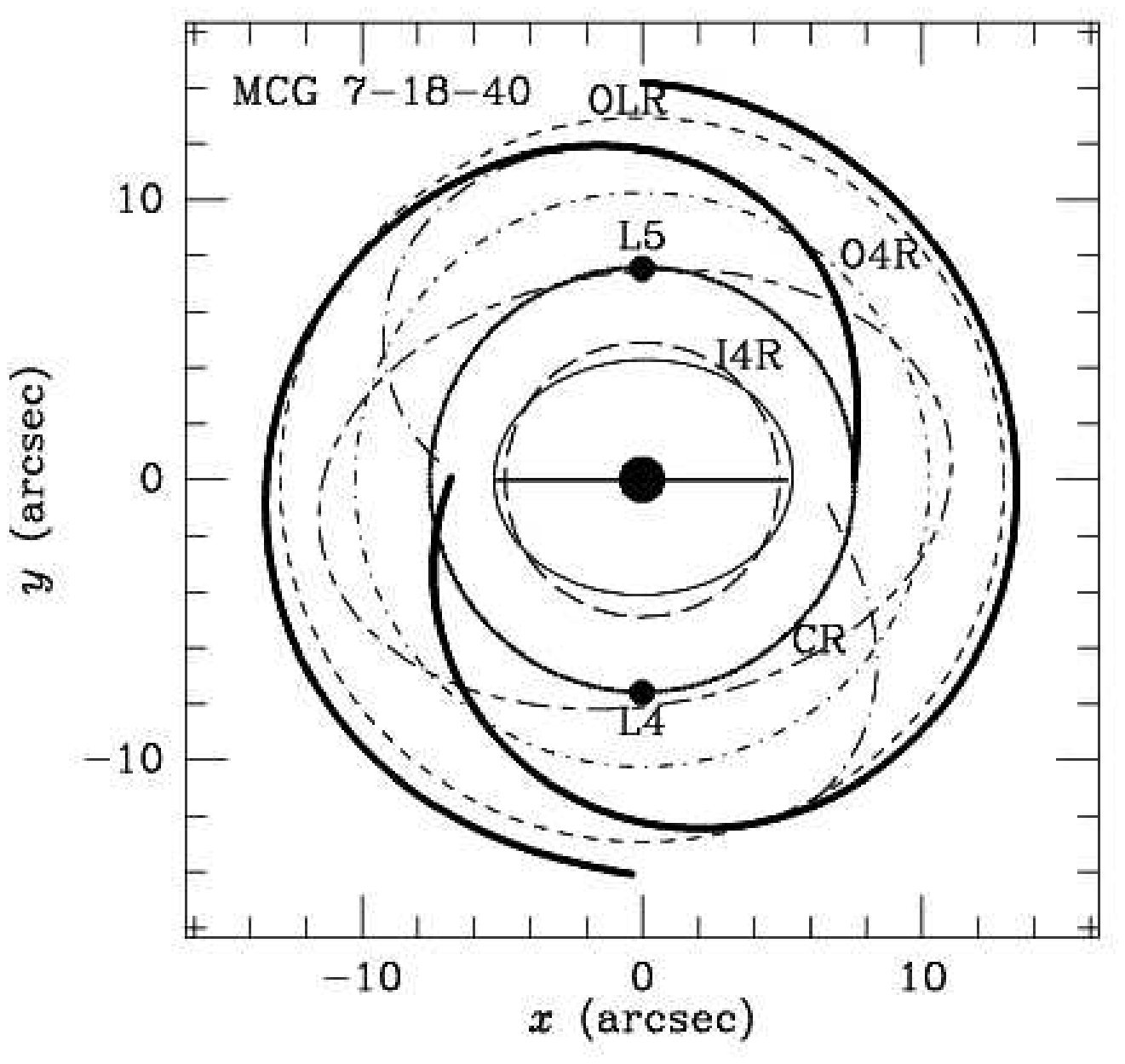}}
 \end{minipage}
 \begin{minipage}[b]{0.45\linewidth}
 \centering
\includegraphics[width=\textwidth]{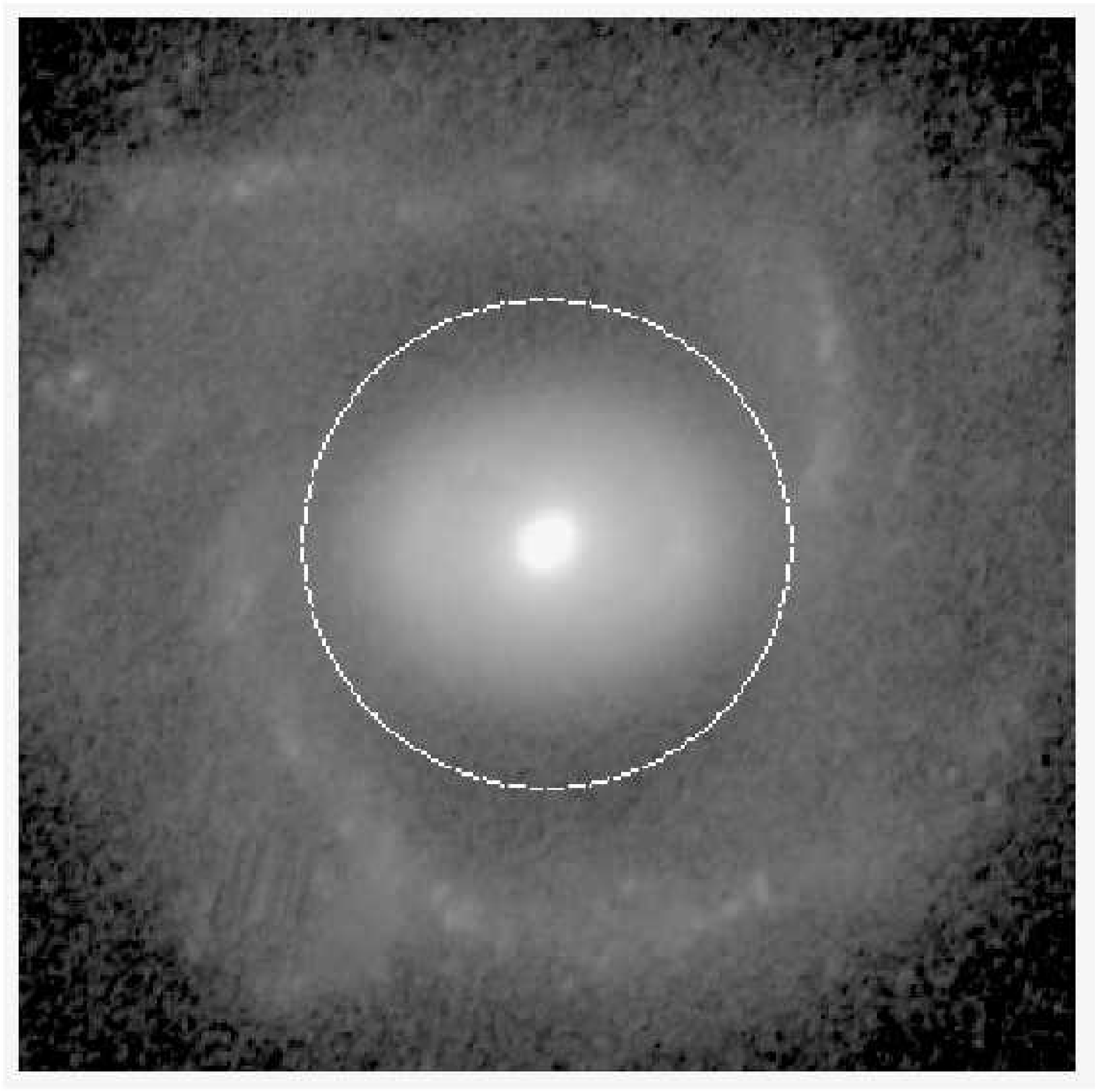}
 \hspace{0.1cm}
 \end{minipage}
 \begin{minipage}[t]{0.68\linewidth}
 \centering
\raisebox{0.5cm}{\includegraphics[width=\textwidth,trim=0 0 0 250,clip]{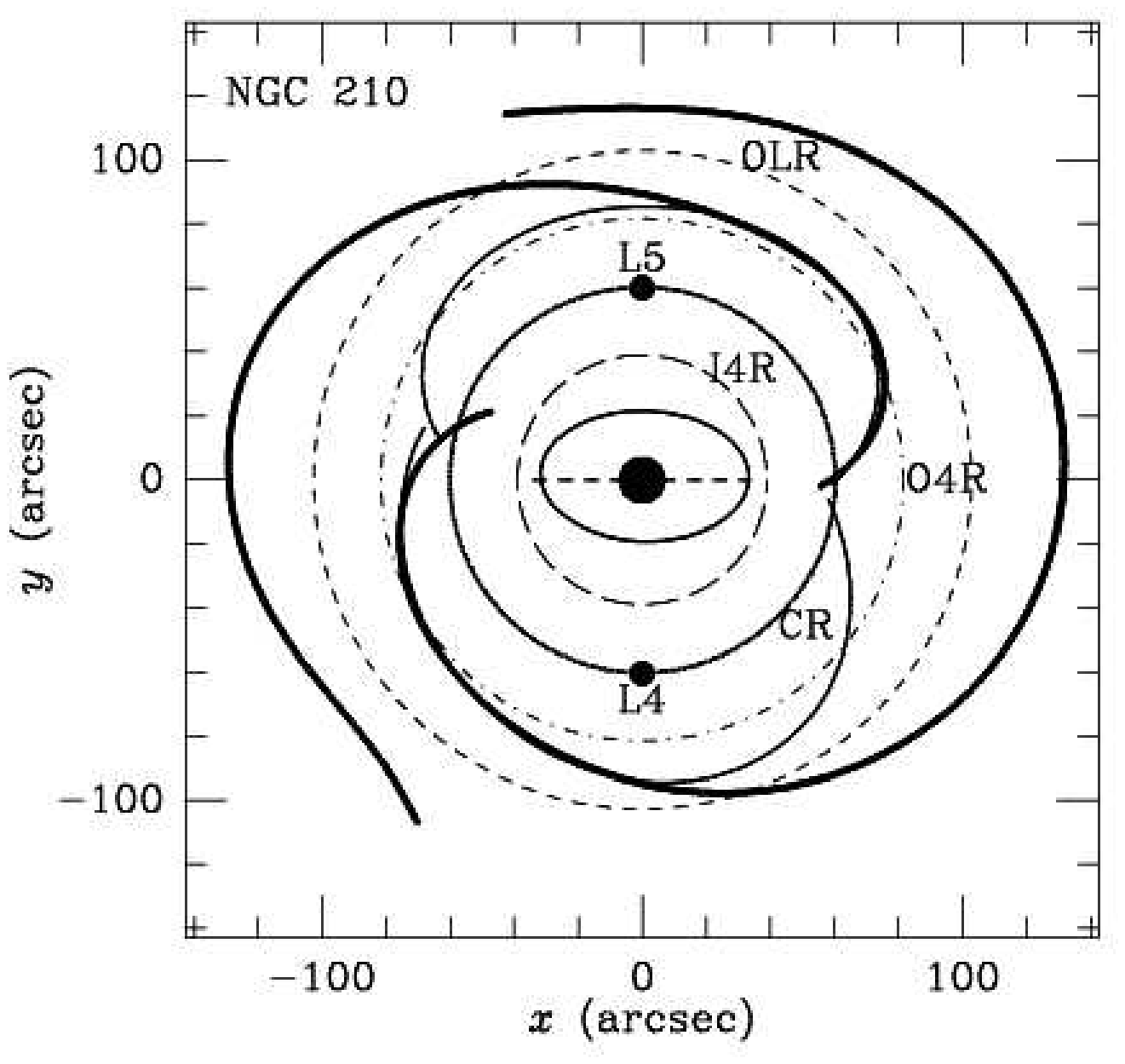}}
 \end{minipage}
 \begin{minipage}[b]{0.45\linewidth}
 \centering
\includegraphics[width=\textwidth]{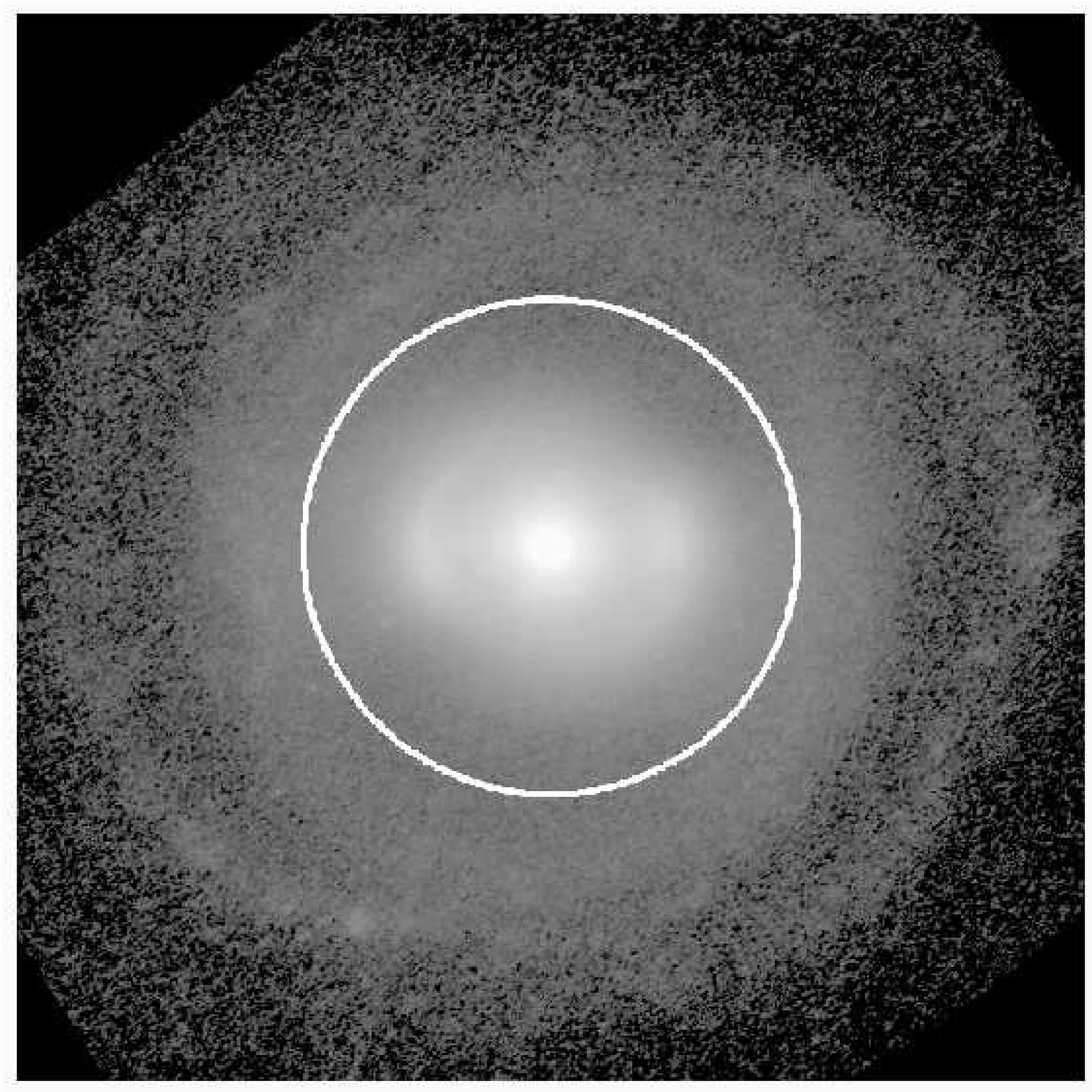}
 \hspace{0.1cm}
 \end{minipage}
 \begin{minipage}[t]{0.68\linewidth}
 \centering
\raisebox{0.5cm}{\includegraphics[width=\textwidth,trim=0 0 0 250,clip]{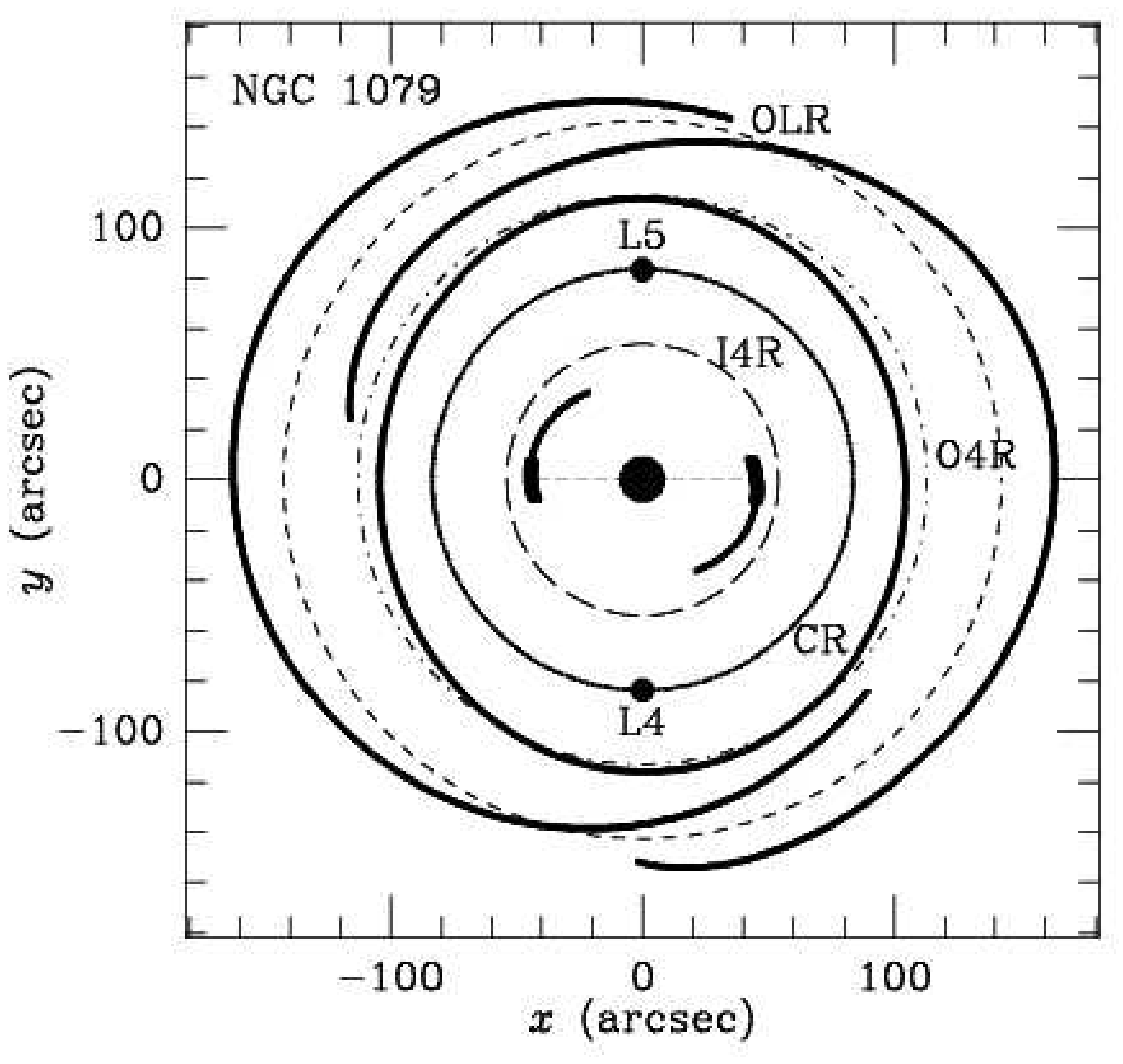}}
 \end{minipage}
 \begin{minipage}[b]{0.45\linewidth}
 \centering
\includegraphics[width=\textwidth]{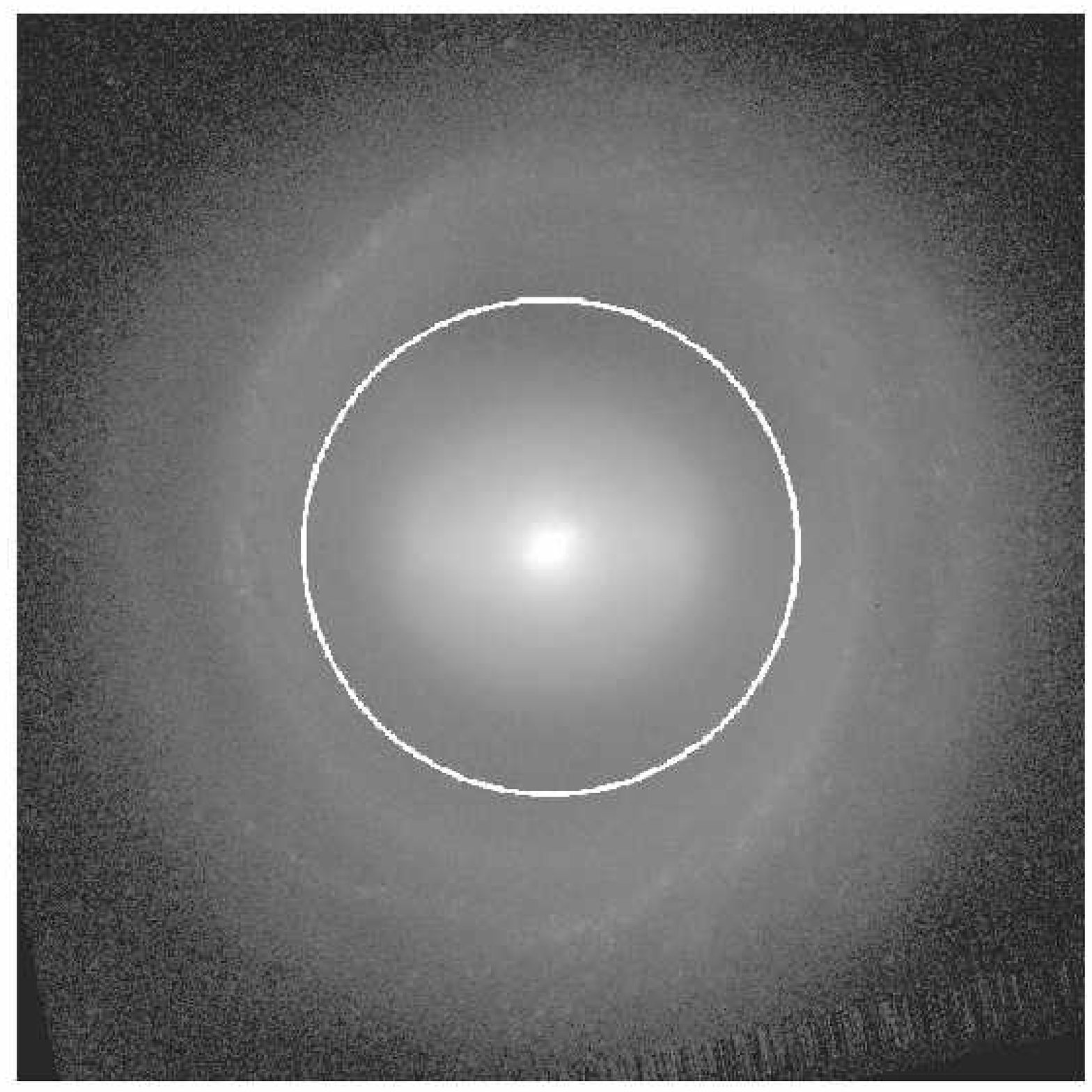}
 \hspace{0.1cm}
 \end{minipage}
 \begin{minipage}[t]{0.68\linewidth}
 \centering
\raisebox{0.5cm}{\includegraphics[width=\textwidth,trim=0 0 0 250,clip]{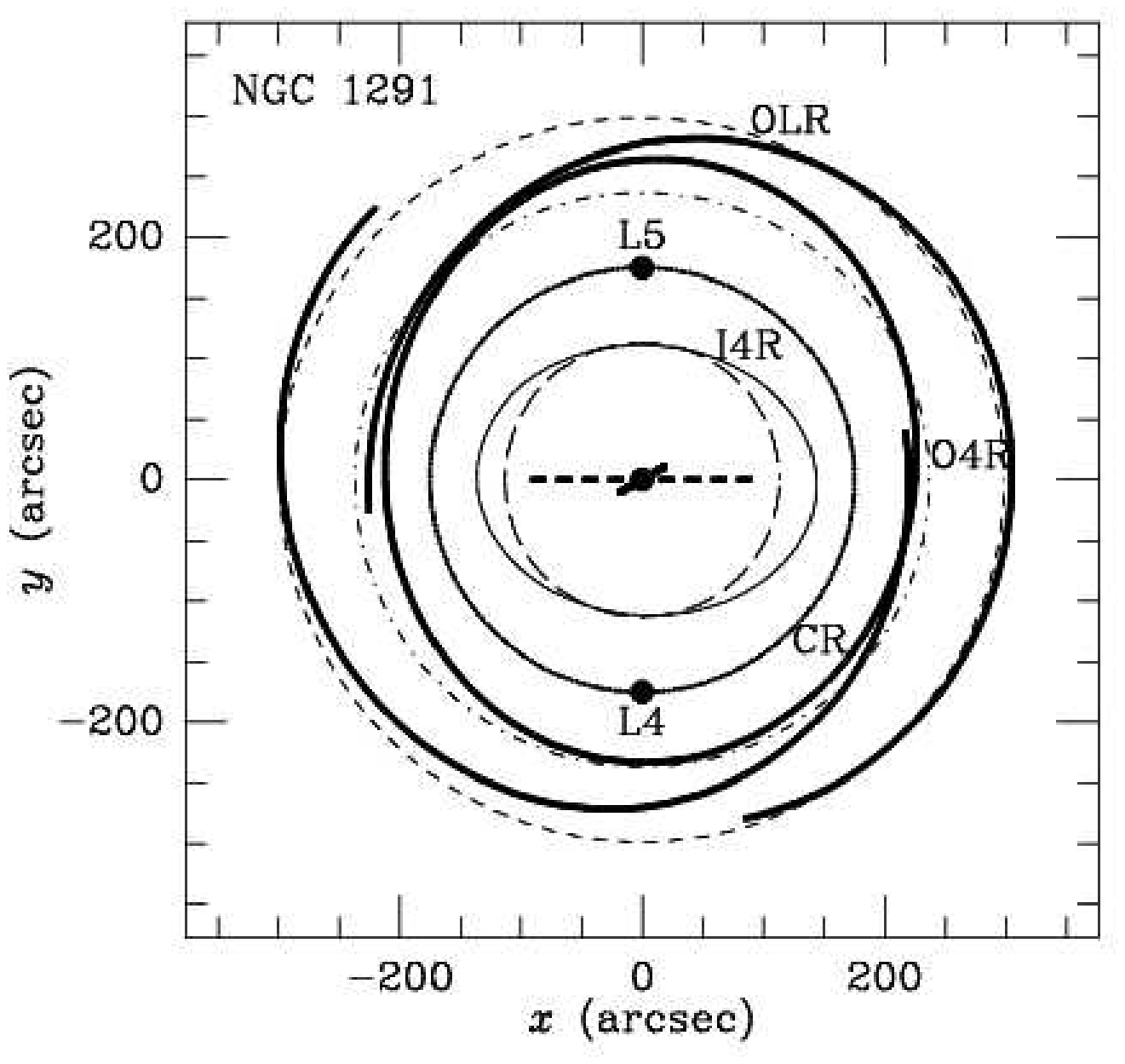}}
 \end{minipage}
 \begin{minipage}[b]{0.45\linewidth}
 \centering
\includegraphics[width=\textwidth]{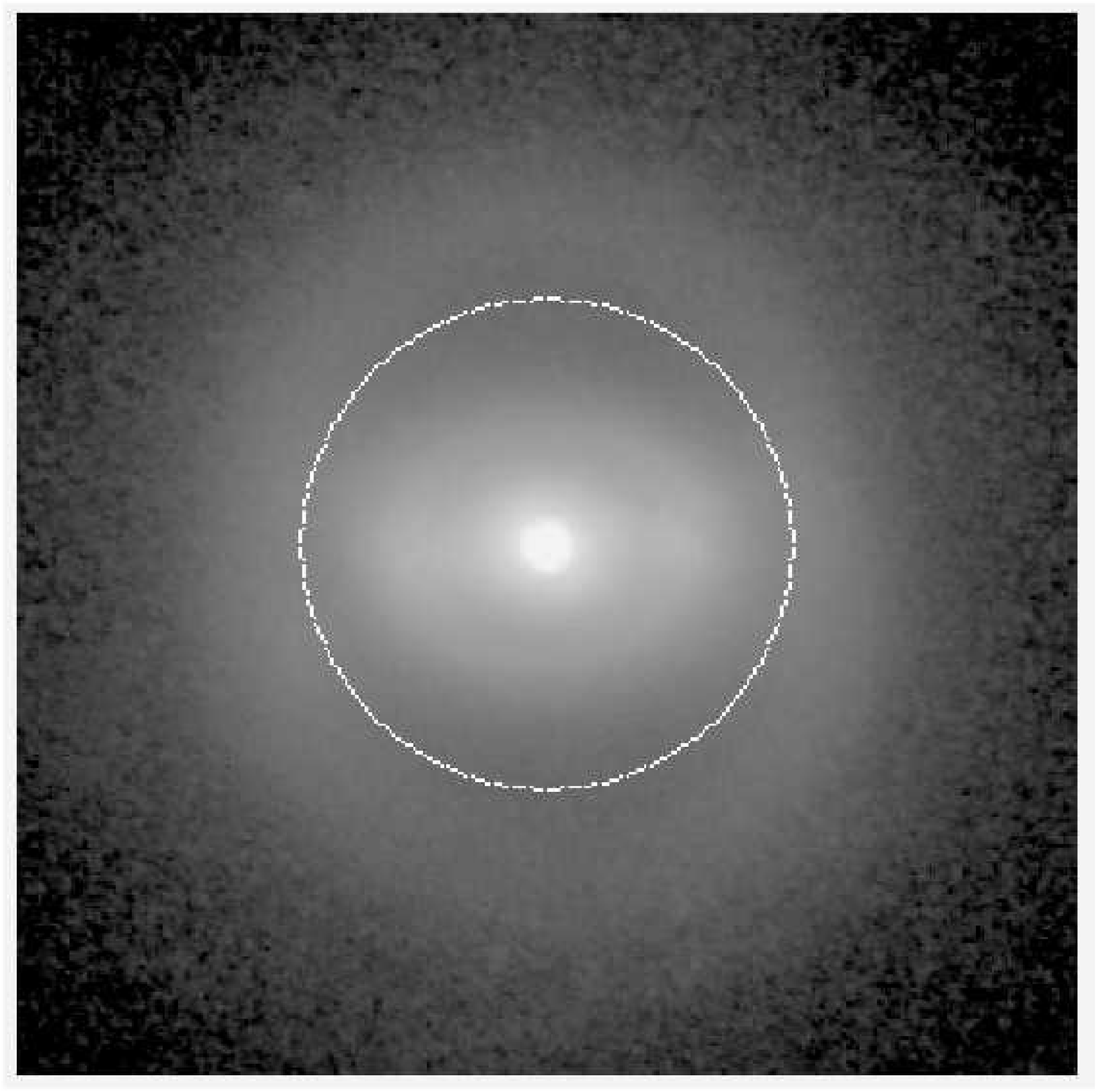}
 \hspace{0.1cm}
 \end{minipage}
 \begin{minipage}[t]{0.68\linewidth}
 \centering
\raisebox{0.5cm}{\includegraphics[width=\textwidth,trim=0 0 0 250,clip]{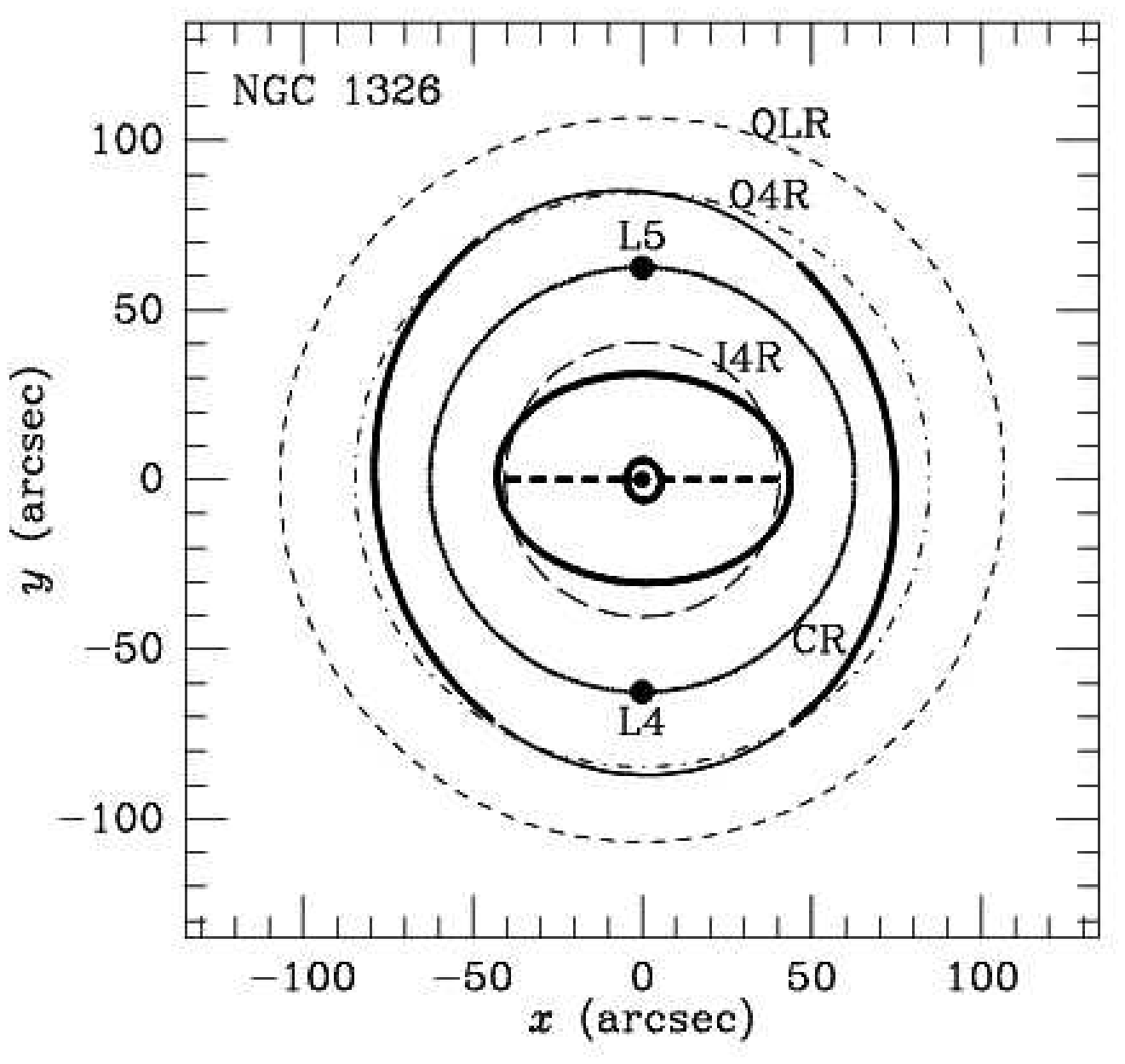}}
 \end{minipage}
\vspace{-1.0truecm}
\caption{(cont.)}
 \end{figure}
 \setcounter{figure}{12}
 \begin{figure}
\vspace{-1.27cm}
 \begin{minipage}[b]{0.45\linewidth}
 \centering
\includegraphics[width=\textwidth]{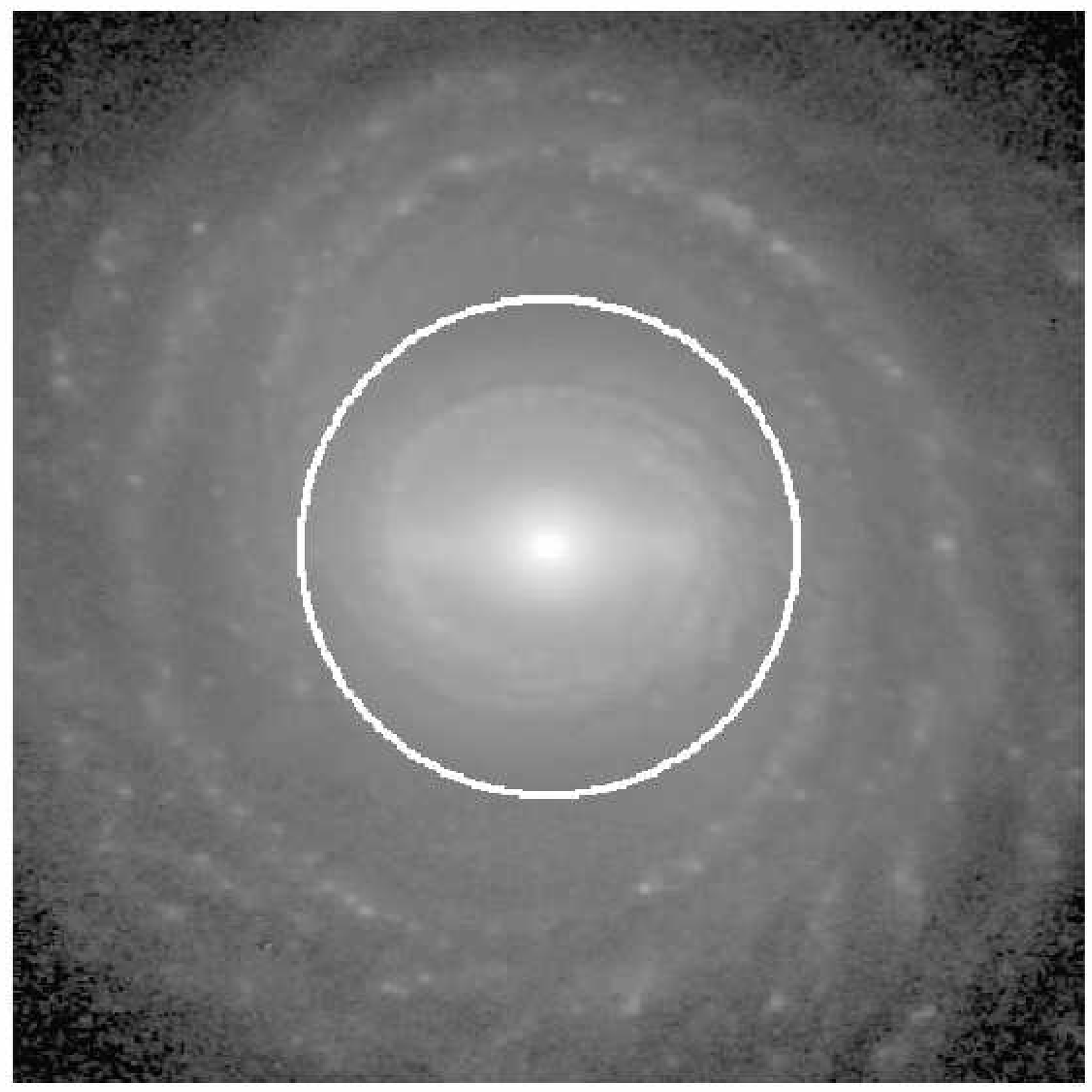}
 \hspace{0.1cm}
 \end{minipage}
 \begin{minipage}[t]{0.68\linewidth}
 \centering
\raisebox{0.5cm}{\includegraphics[width=\textwidth,trim=0 0 0 250,clip]{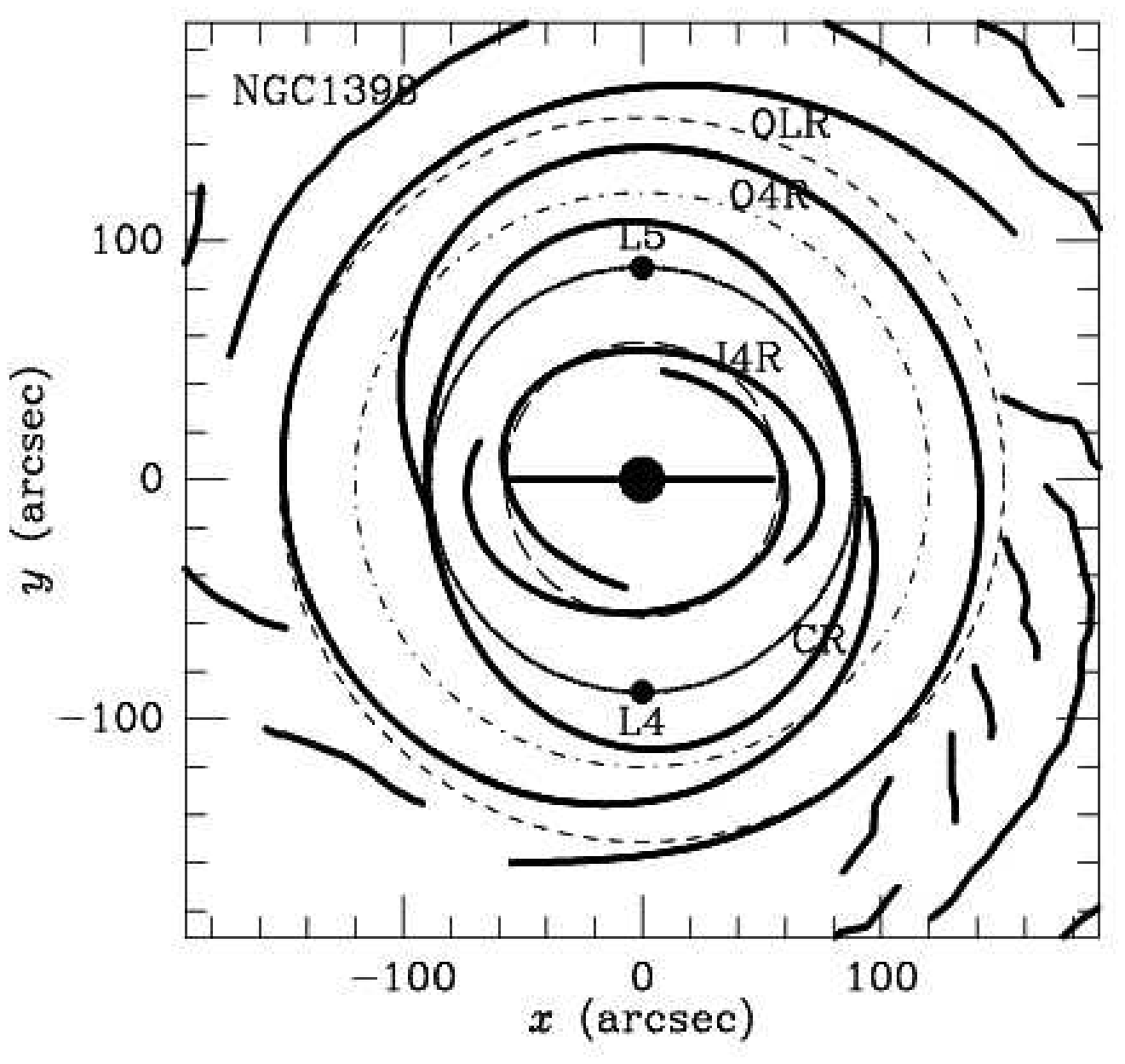}}
 \end{minipage}
 \begin{minipage}[b]{0.45\linewidth}
 \centering
\includegraphics[width=\textwidth]{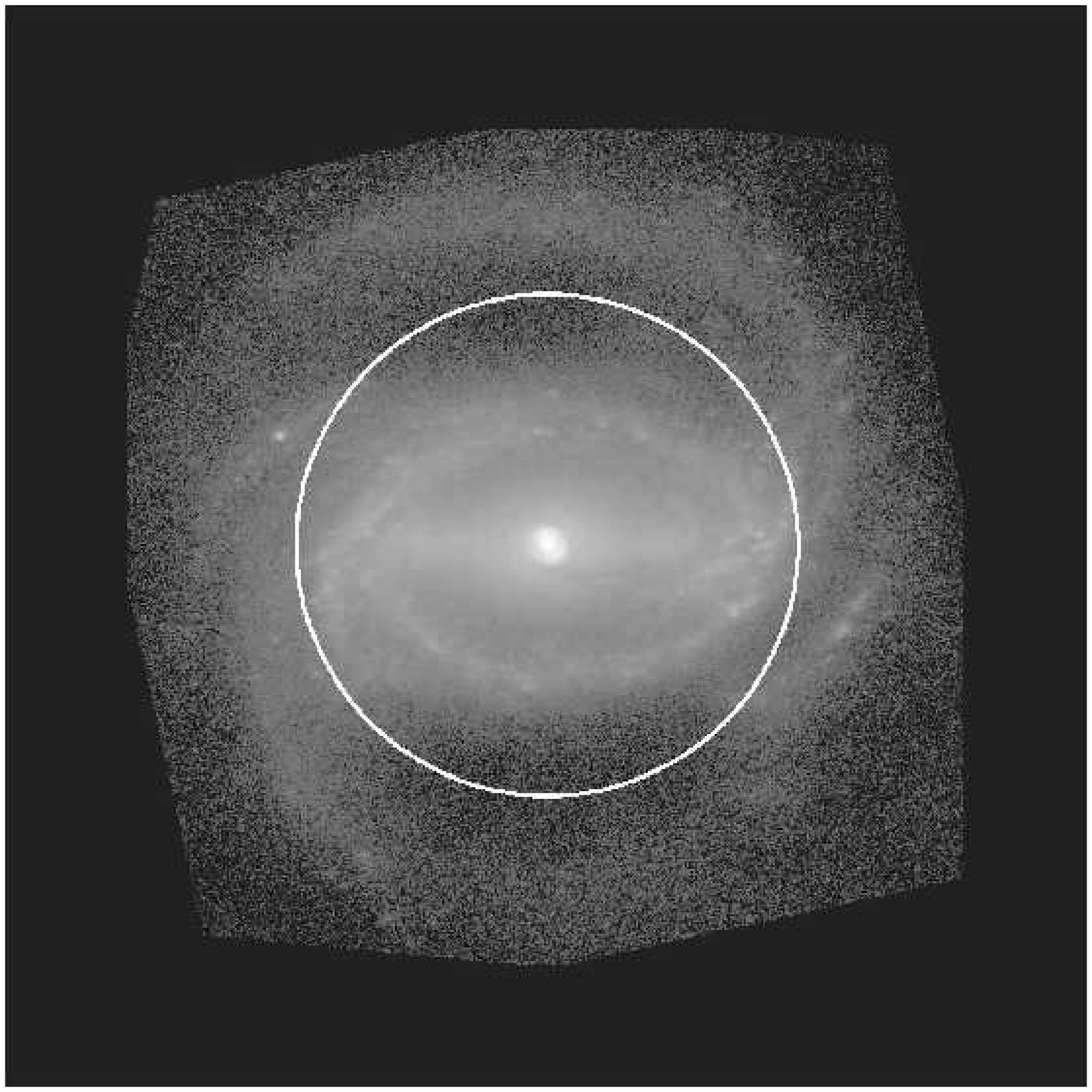}
 \hspace{0.1cm}
 \end{minipage}
 \begin{minipage}[t]{0.68\linewidth}
 \centering
\raisebox{0.5cm}{\includegraphics[width=\textwidth,trim=0 0 0 250,clip]{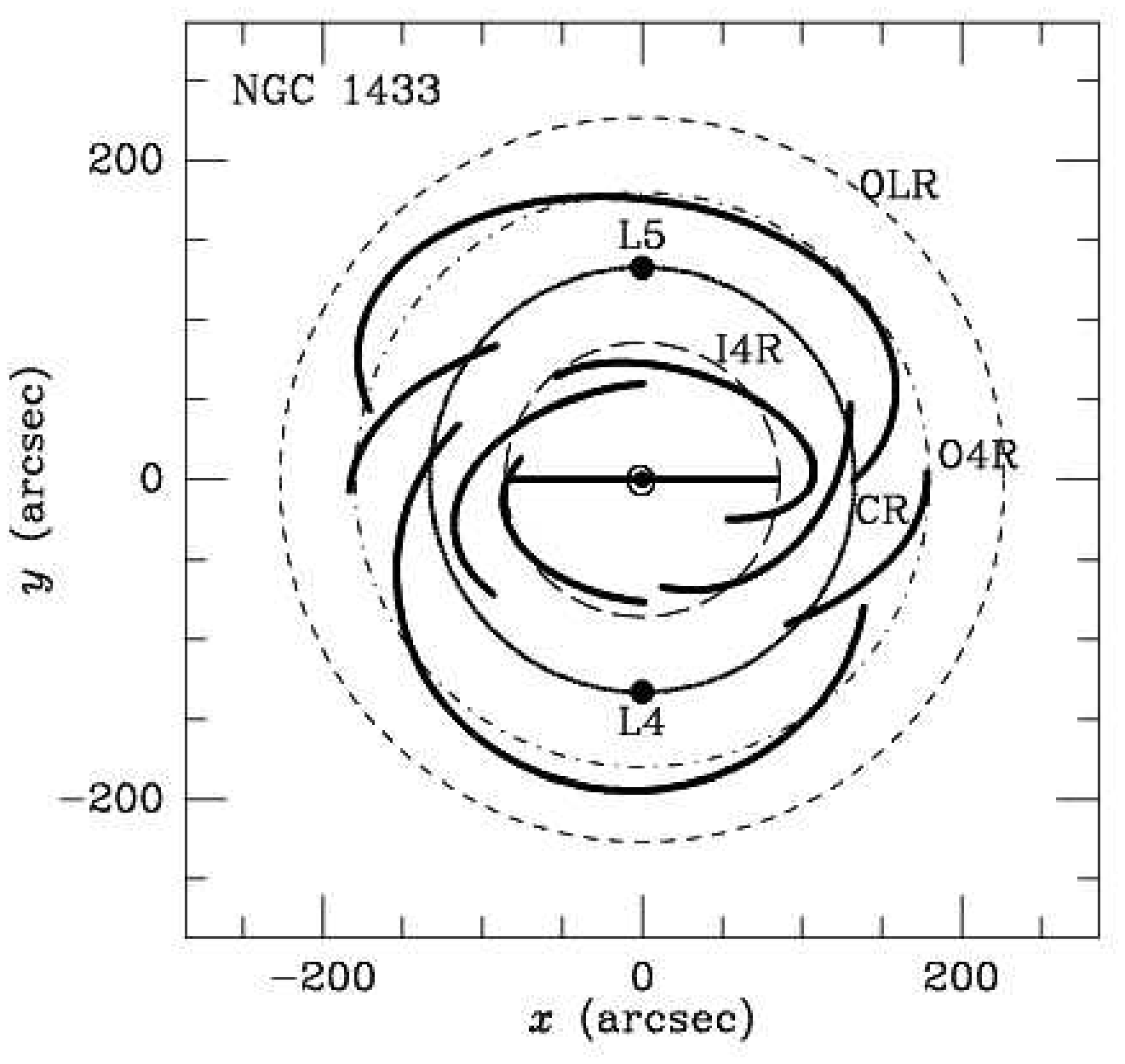}}
 \end{minipage}
 \begin{minipage}[b]{0.45\linewidth}
 \centering
\includegraphics[width=\textwidth]{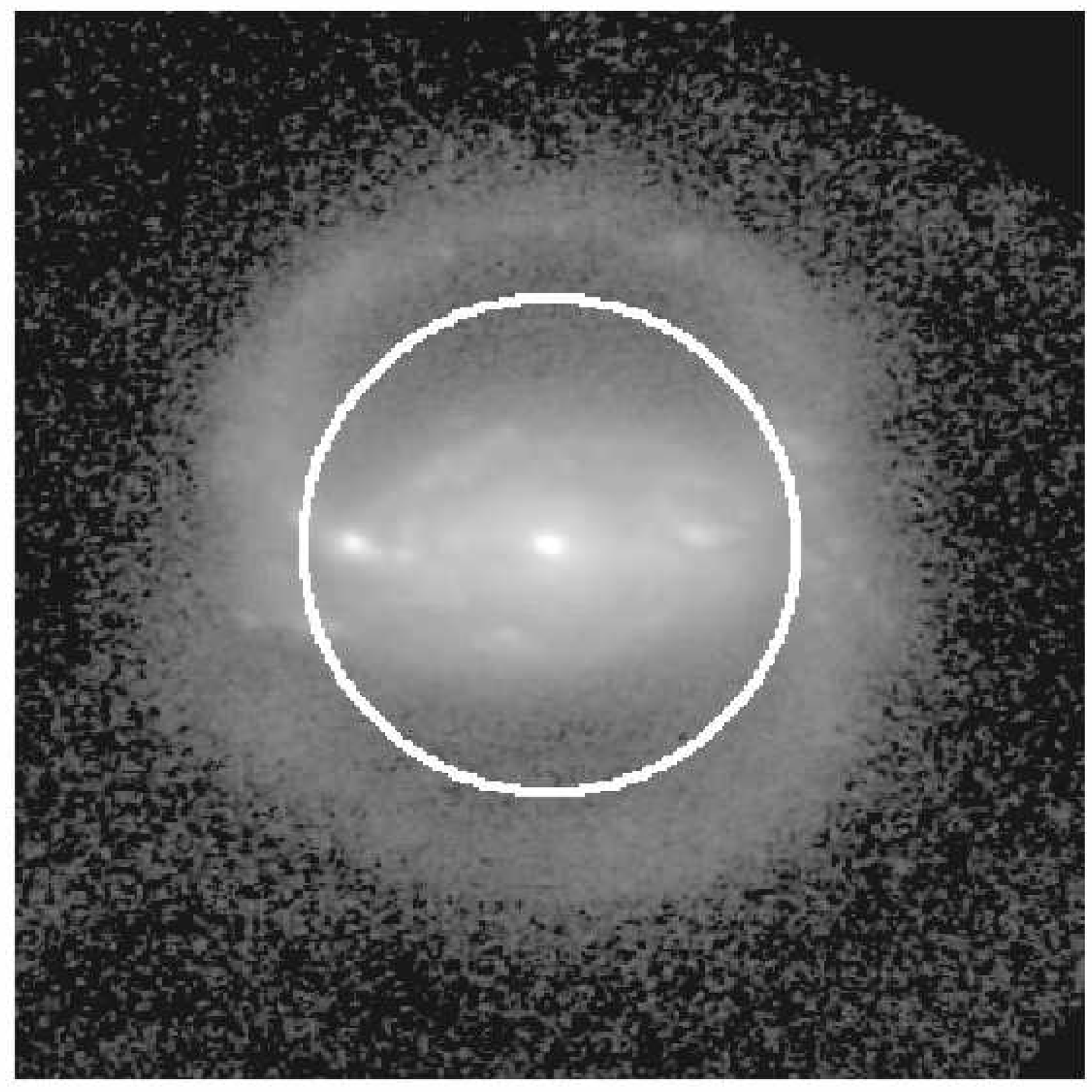}
 \hspace{0.1cm}
 \end{minipage}
 \begin{minipage}[t]{0.68\linewidth}
 \centering
\raisebox{0.5cm}{\includegraphics[width=\textwidth,trim=0 0 0 250,clip]{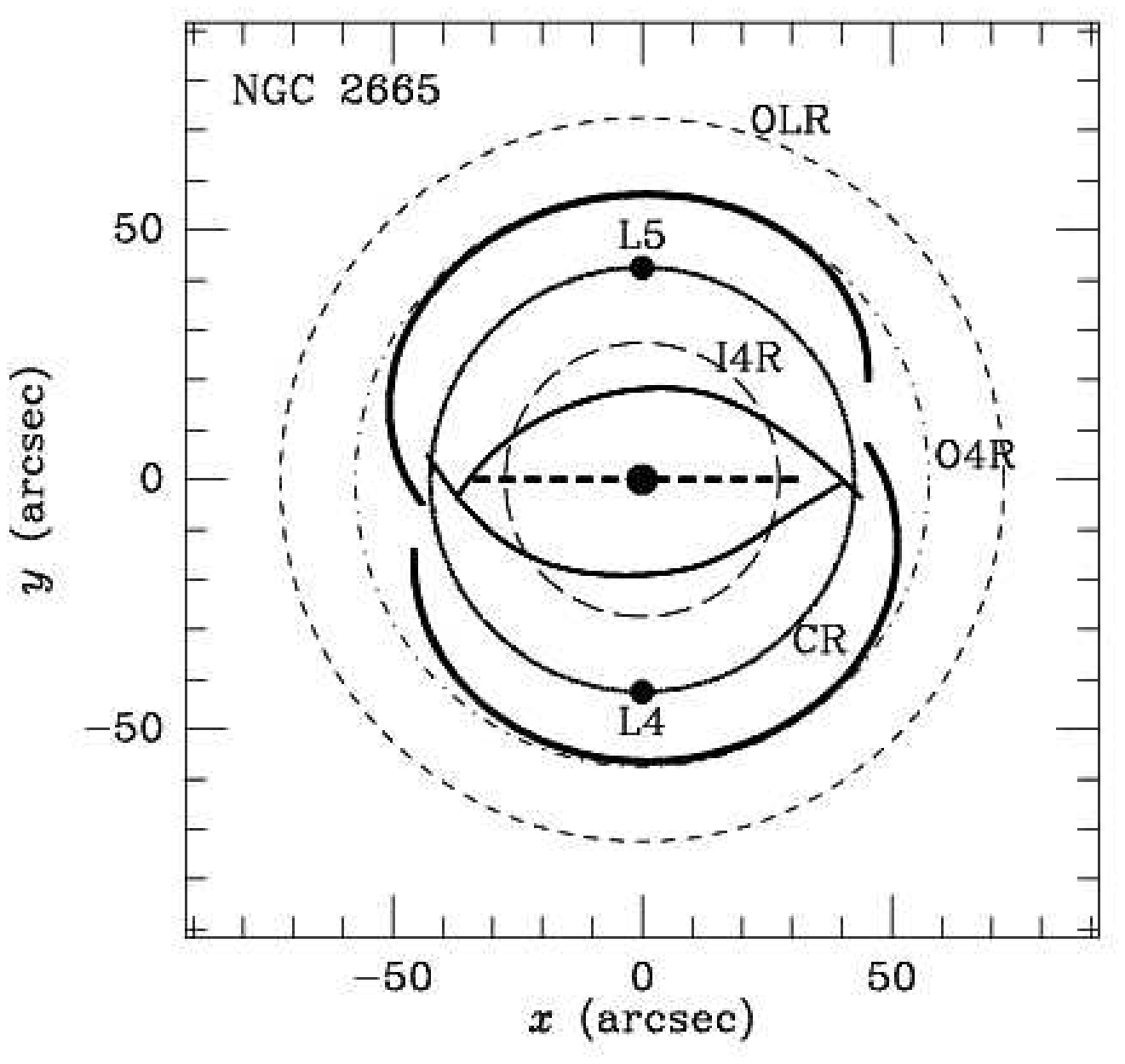}}
 \end{minipage}
 \begin{minipage}[b]{0.45\linewidth}
 \centering
\includegraphics[width=\textwidth]{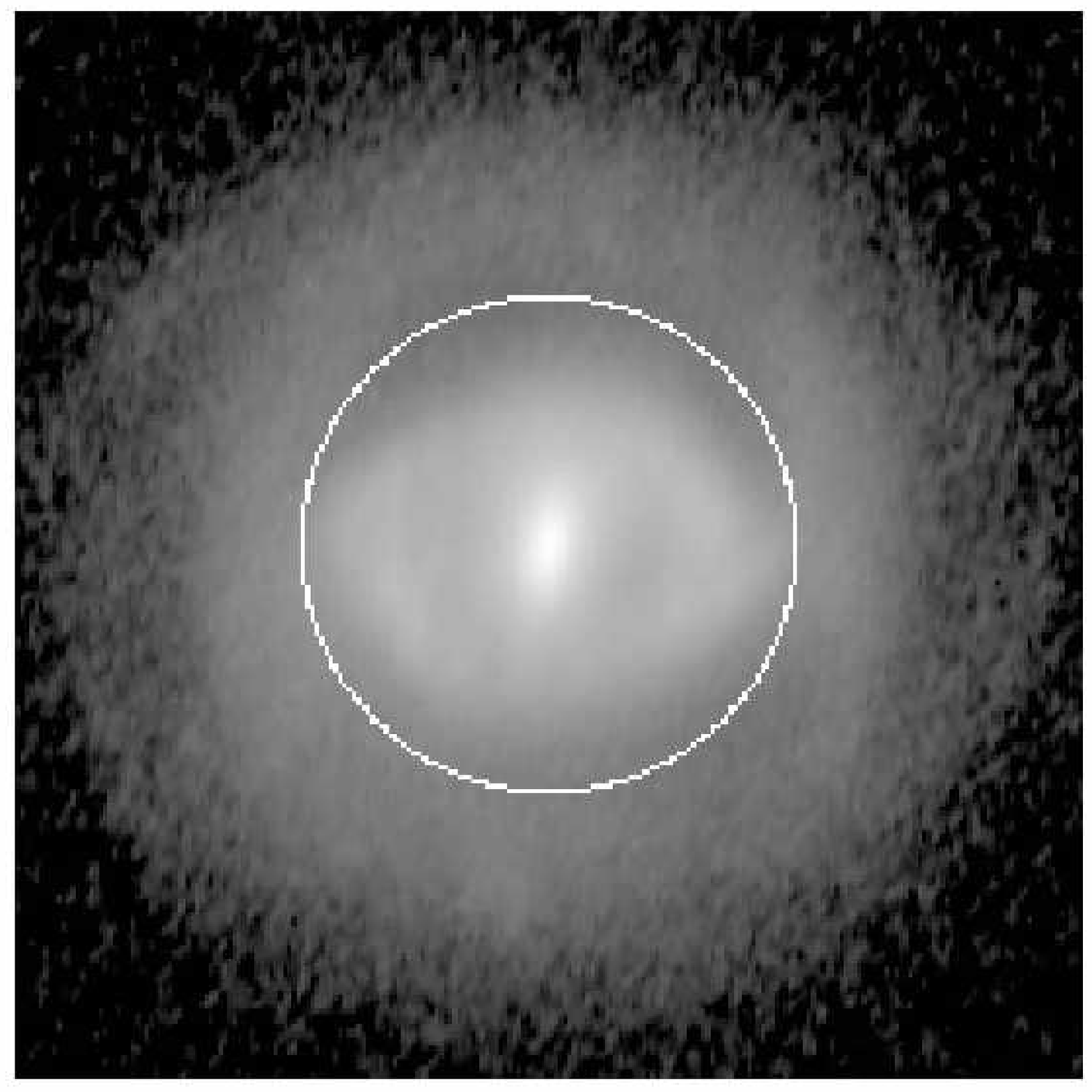}
 \hspace{0.1cm}
 \end{minipage}
 \begin{minipage}[t]{0.68\linewidth}
 \centering
\raisebox{0.5cm}{\includegraphics[width=\textwidth,trim=0 0 0 250,clip]{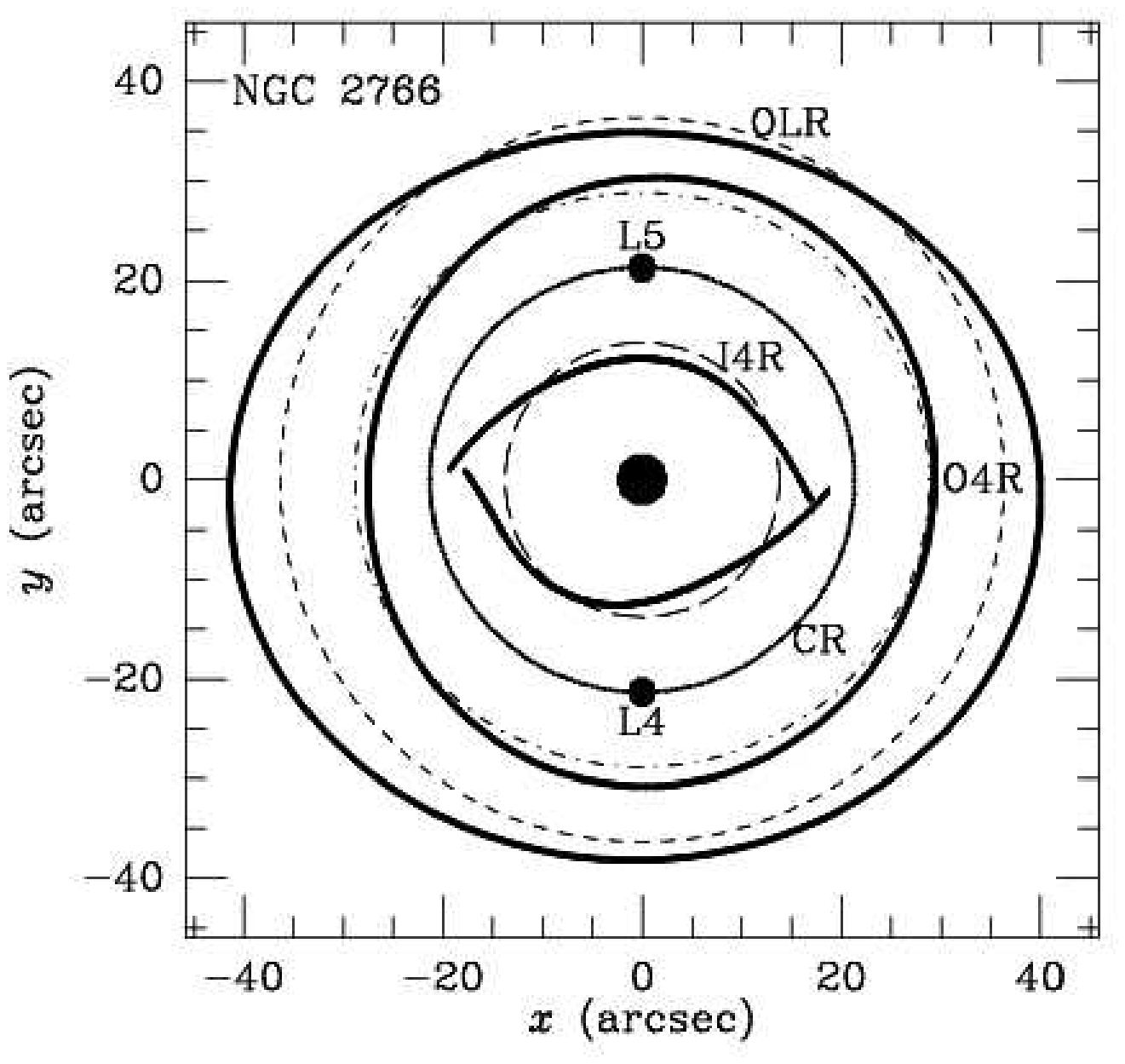}}
 \end{minipage}
 \begin{minipage}[b]{0.45\linewidth}
 \centering
\includegraphics[width=\textwidth]{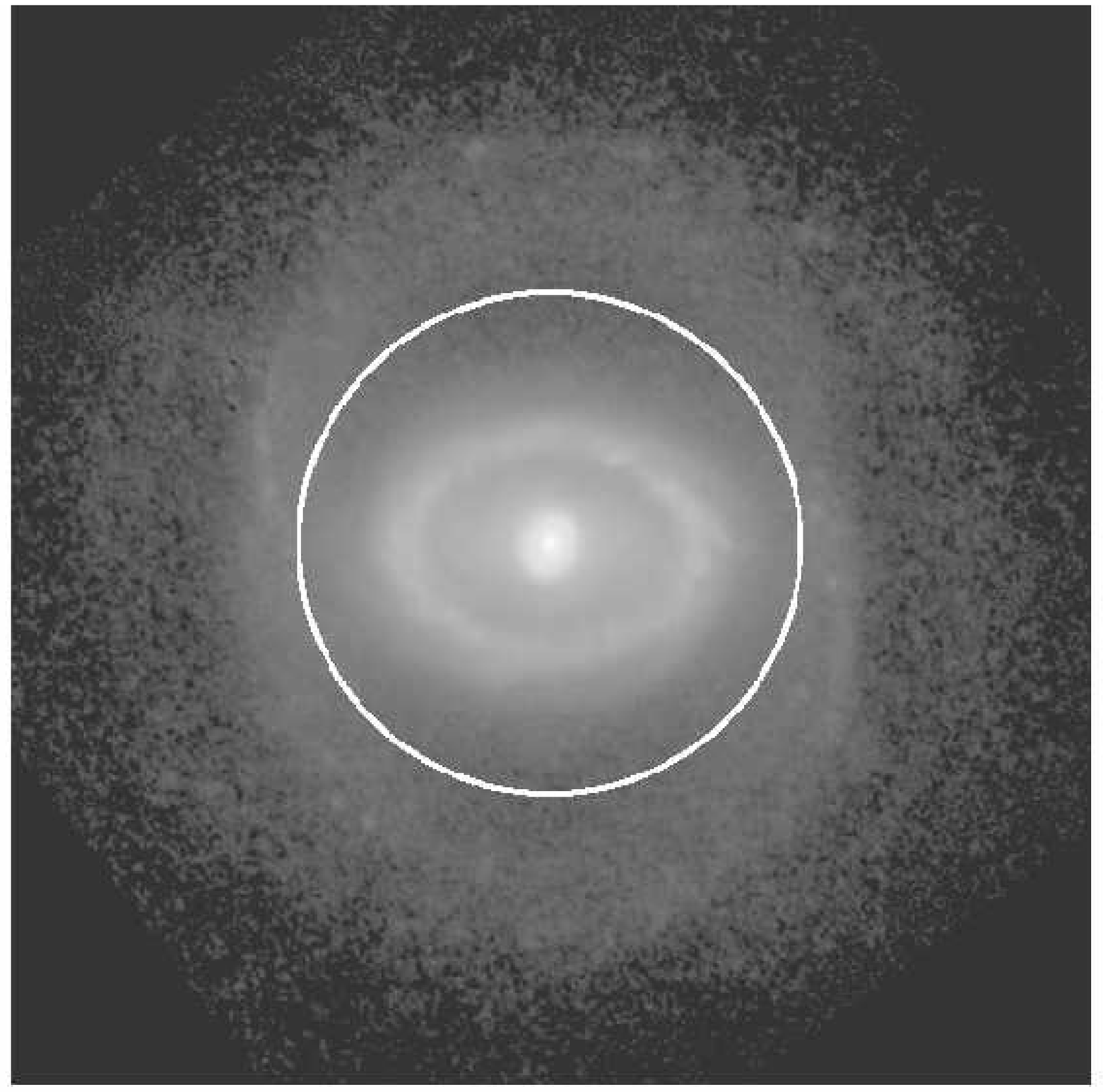}
 \hspace{0.1cm}
 \end{minipage}
 \begin{minipage}[t]{0.68\linewidth}
 \centering
\raisebox{0.5cm}{\includegraphics[width=\textwidth,trim=0 0 0 250,clip]{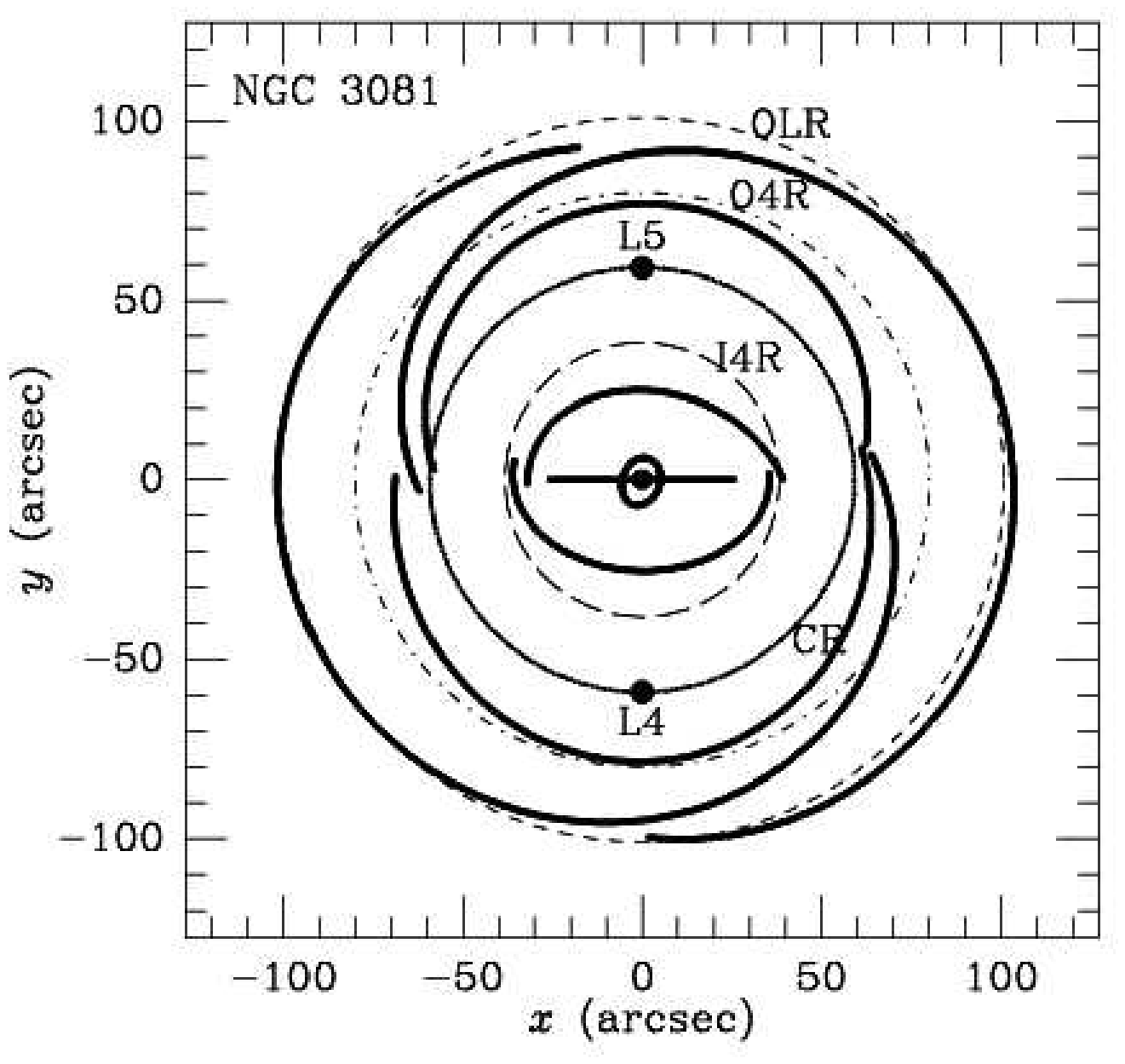}}
 \end{minipage}
\vspace{-1.0truecm}
\caption{(cont.)}
 \end{figure}
 \setcounter{figure}{12}
 \begin{figure}
\vspace{-1.27cm}
 \begin{minipage}[b]{0.45\linewidth}
 \centering
\includegraphics[width=\textwidth]{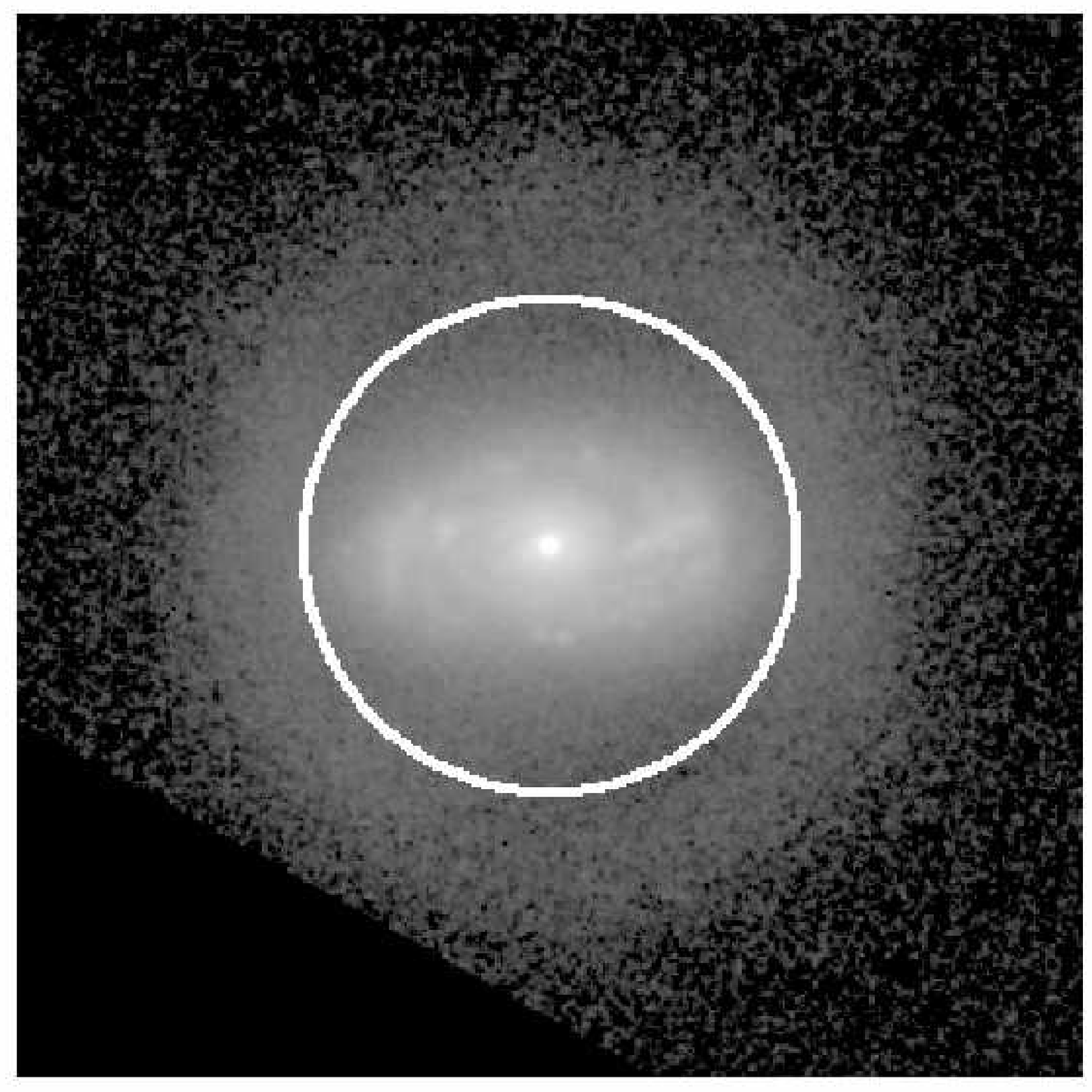}
 \hspace{0.1cm}
 \end{minipage}
 \begin{minipage}[t]{0.68\linewidth}
 \centering
\raisebox{0.5cm}{\includegraphics[width=\textwidth,trim=0 0 0 250,clip]{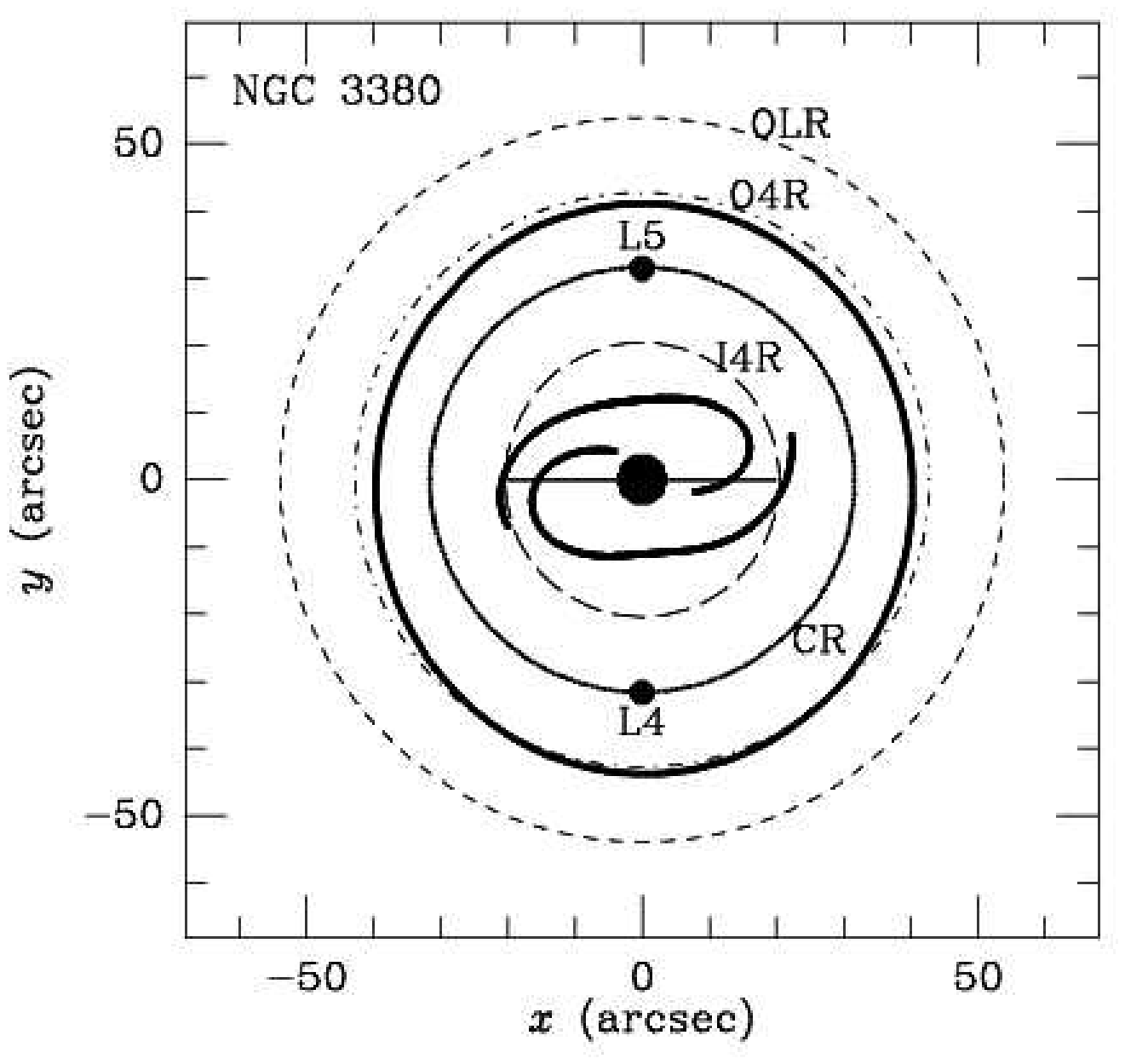}}
 \end{minipage}
 \begin{minipage}[b]{0.45\linewidth}
 \centering
\includegraphics[width=\textwidth]{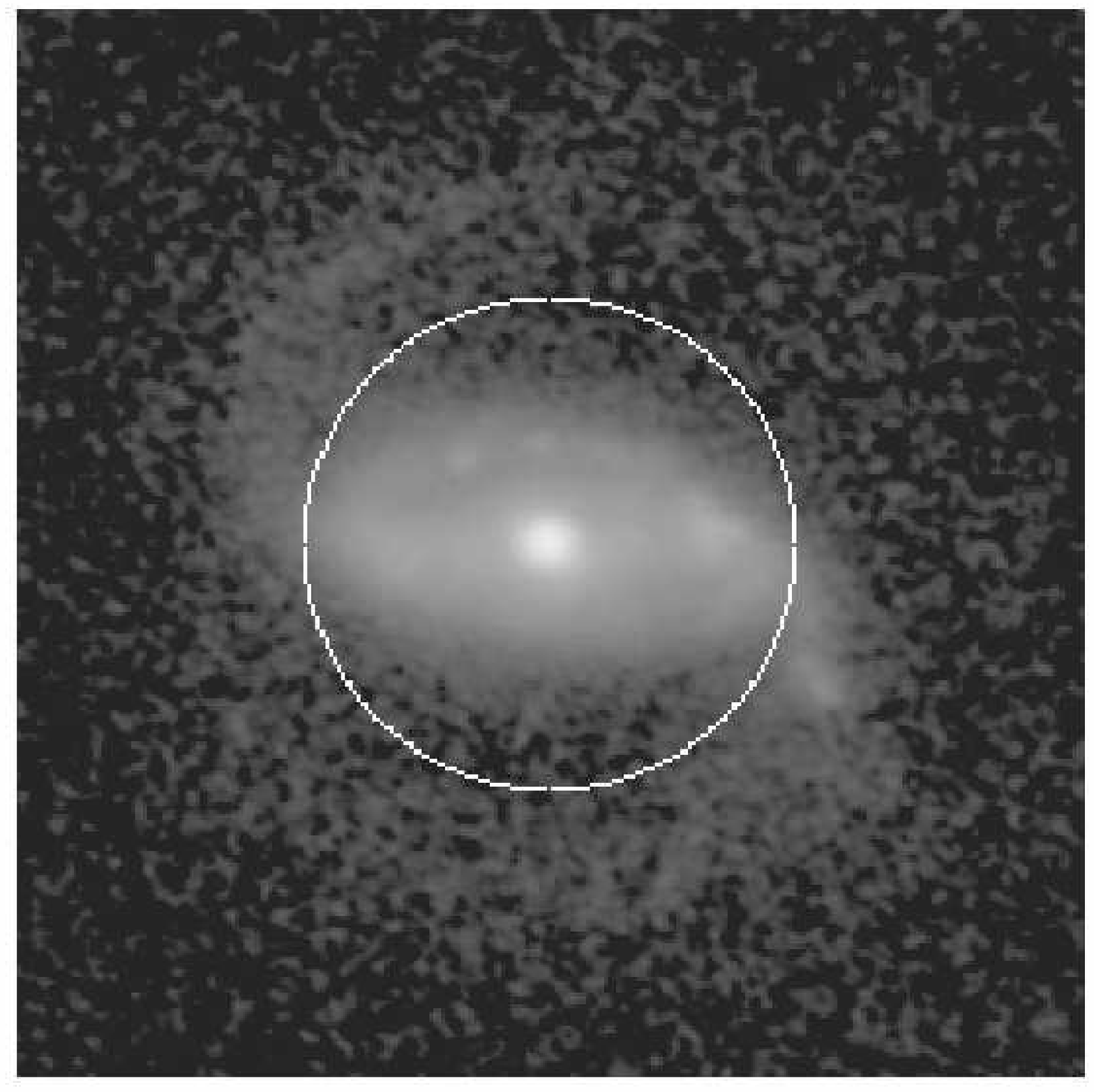}
 \hspace{0.1cm}
 \end{minipage}
 \begin{minipage}[t]{0.68\linewidth}
 \centering
\raisebox{0.5cm}{\includegraphics[width=\textwidth,trim=0 0 0 250,clip]{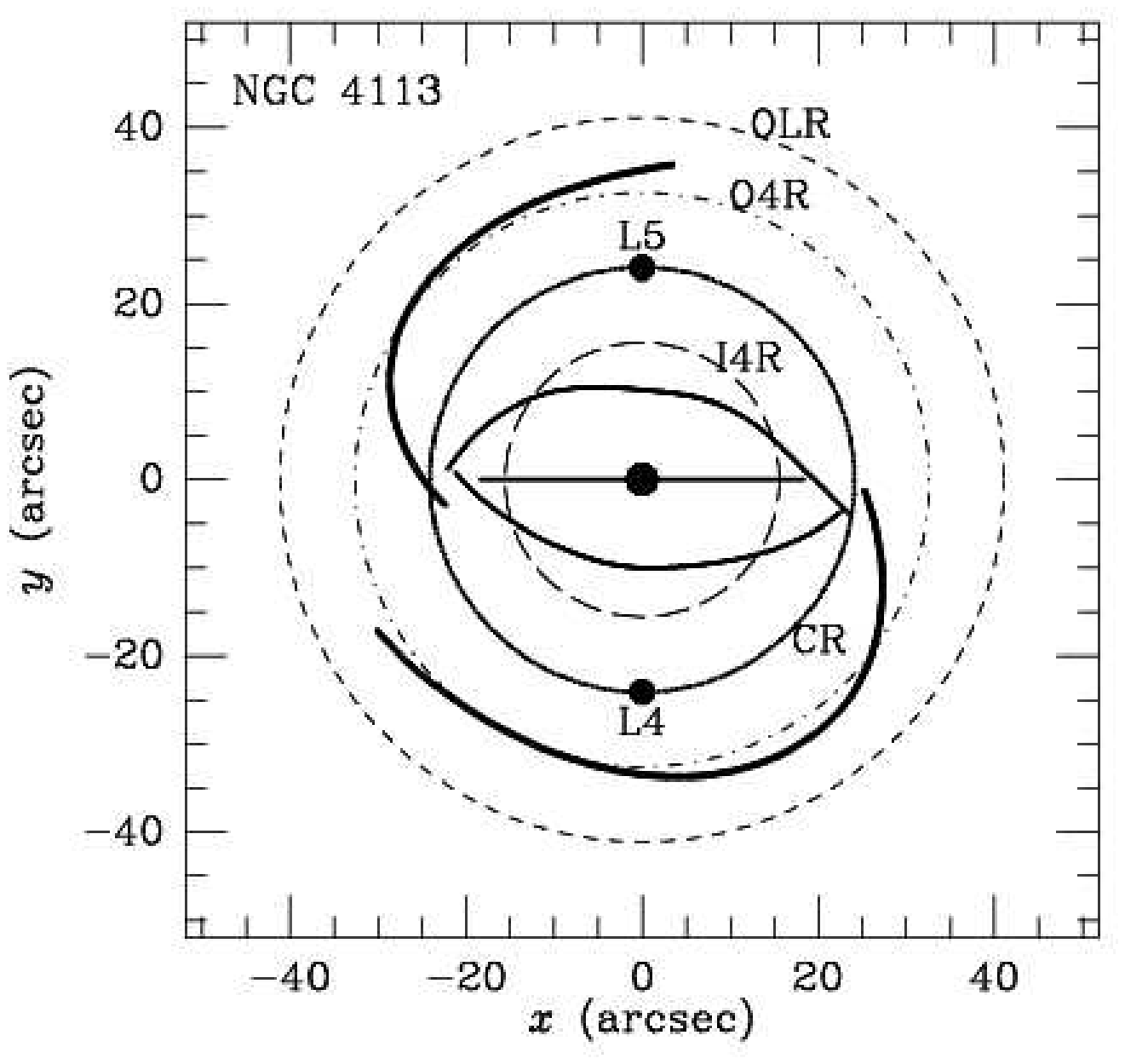}}
 \end{minipage}
 \begin{minipage}[b]{0.45\linewidth}
 \centering
\includegraphics[width=\textwidth]{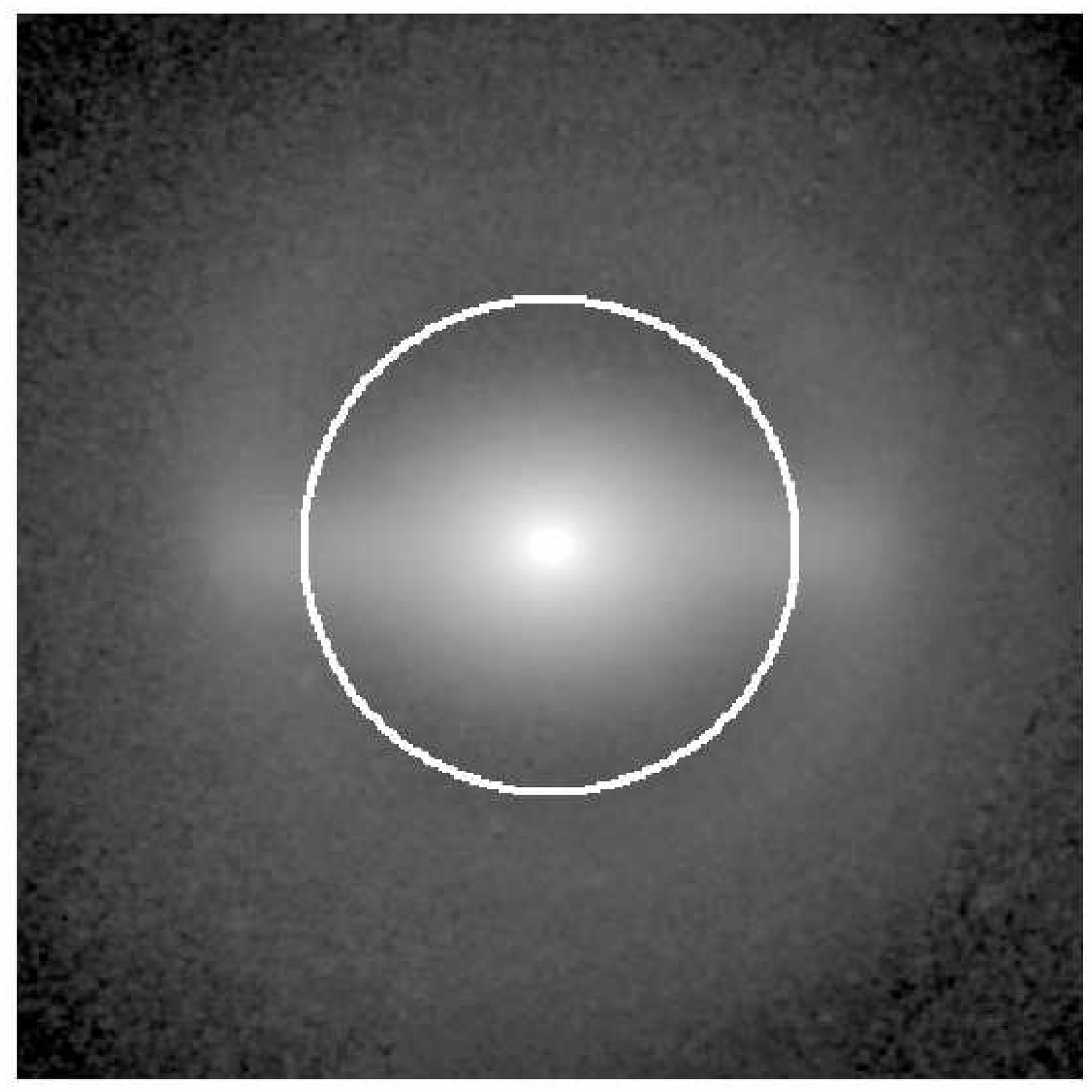}
 \hspace{0.1cm}
 \end{minipage}
 \begin{minipage}[t]{0.68\linewidth}
 \centering
\raisebox{0.5cm}{\includegraphics[width=\textwidth,trim=0 0 0 250,clip]{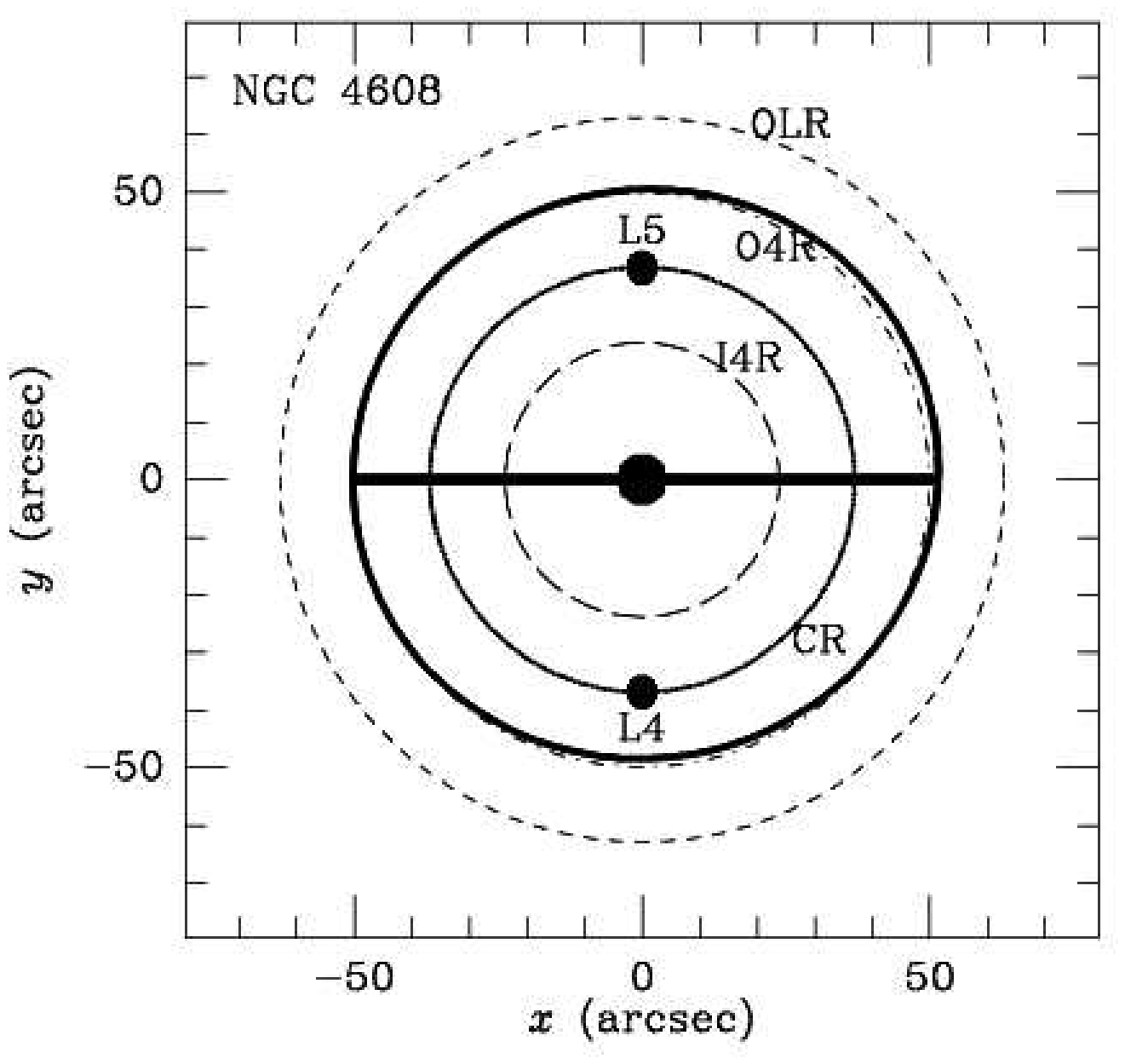}}
 \end{minipage}
 \begin{minipage}[b]{0.45\linewidth}
 \centering
\includegraphics[width=\textwidth]{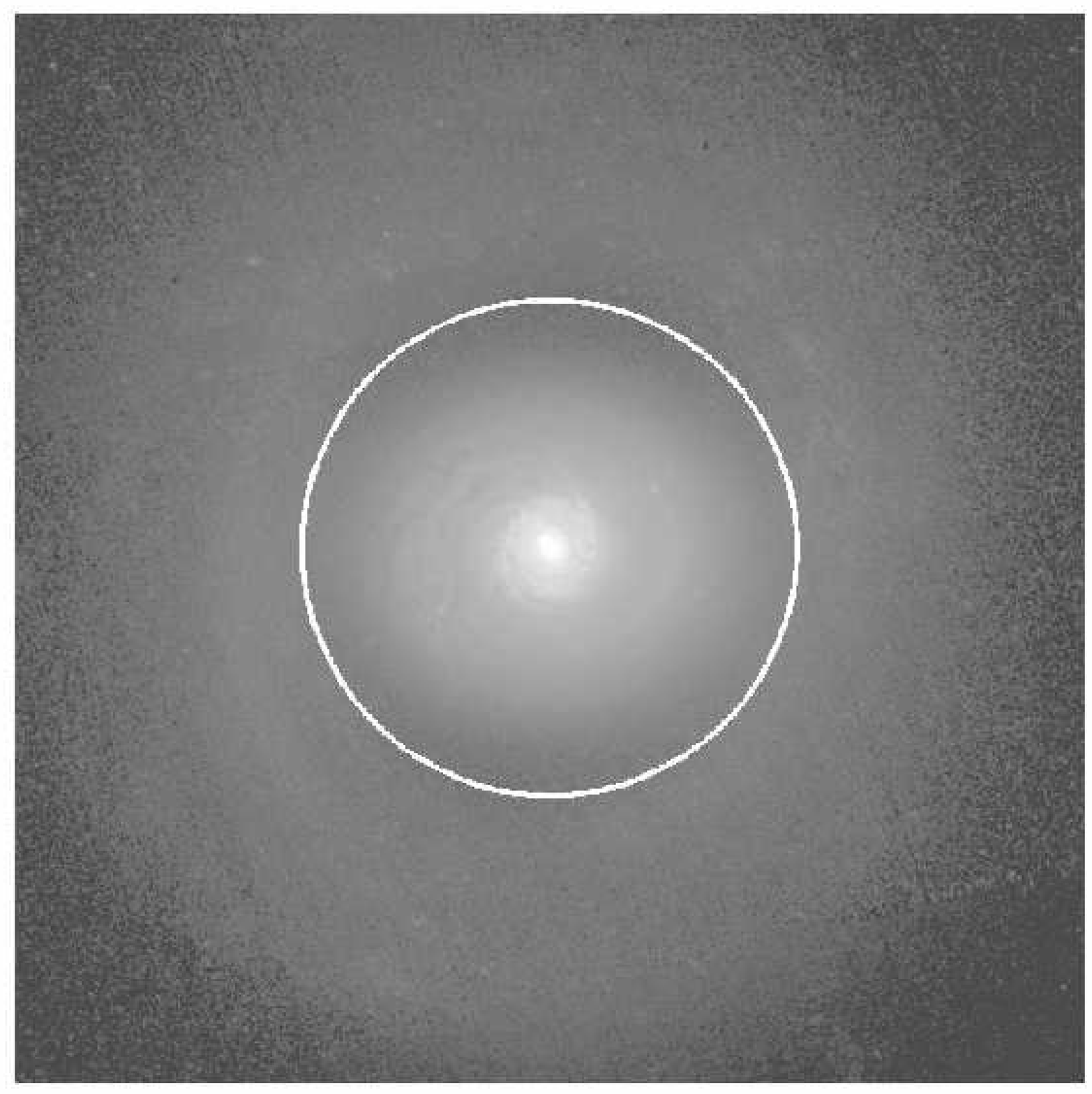}
 \hspace{0.1cm}
 \end{minipage}
 \begin{minipage}[t]{0.68\linewidth}
 \centering
\raisebox{0.5cm}{\includegraphics[width=\textwidth,trim=0 0 0 250,clip]{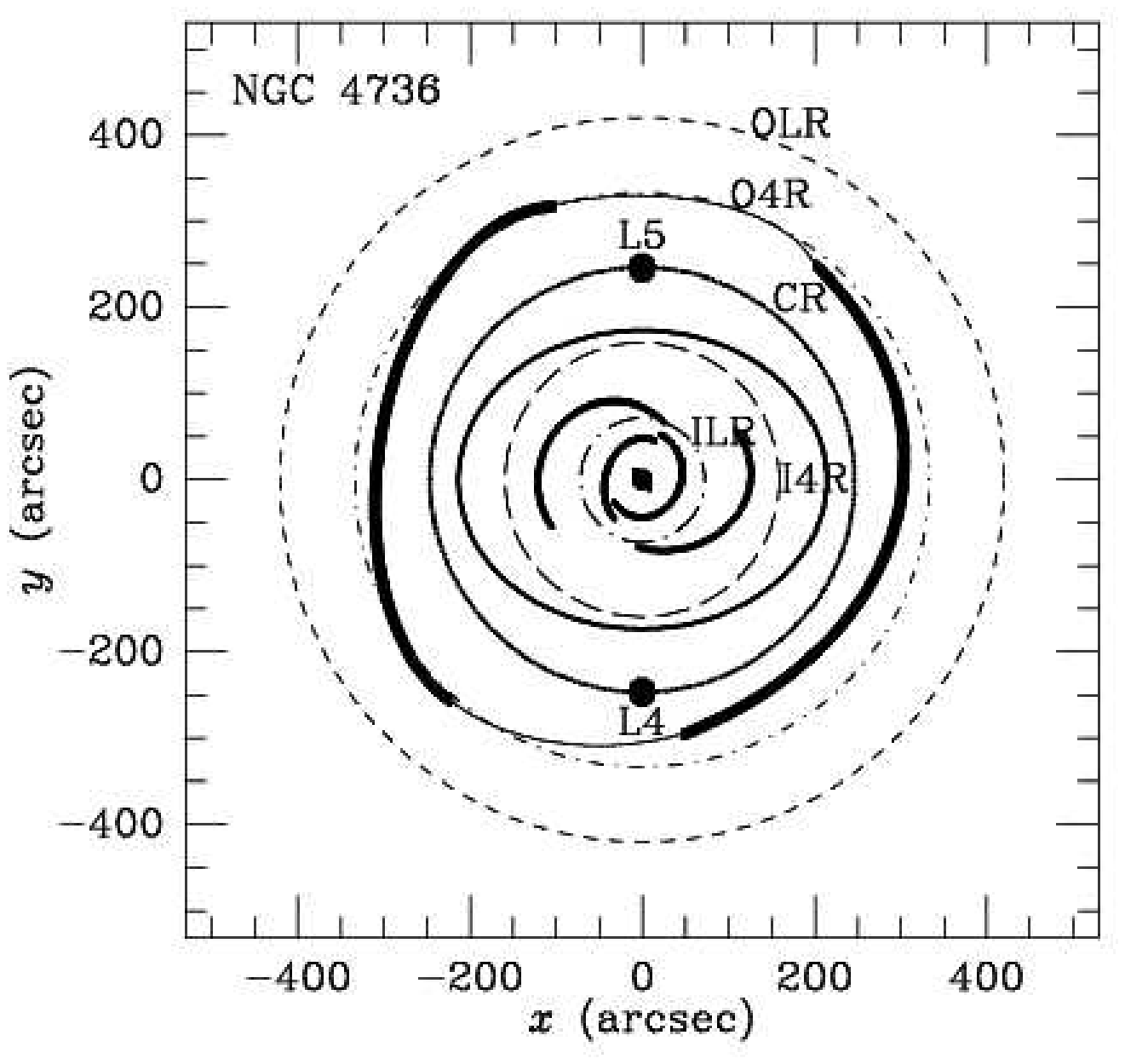}}
 \end{minipage}
 \begin{minipage}[b]{0.45\linewidth}
 \centering
\includegraphics[width=\textwidth]{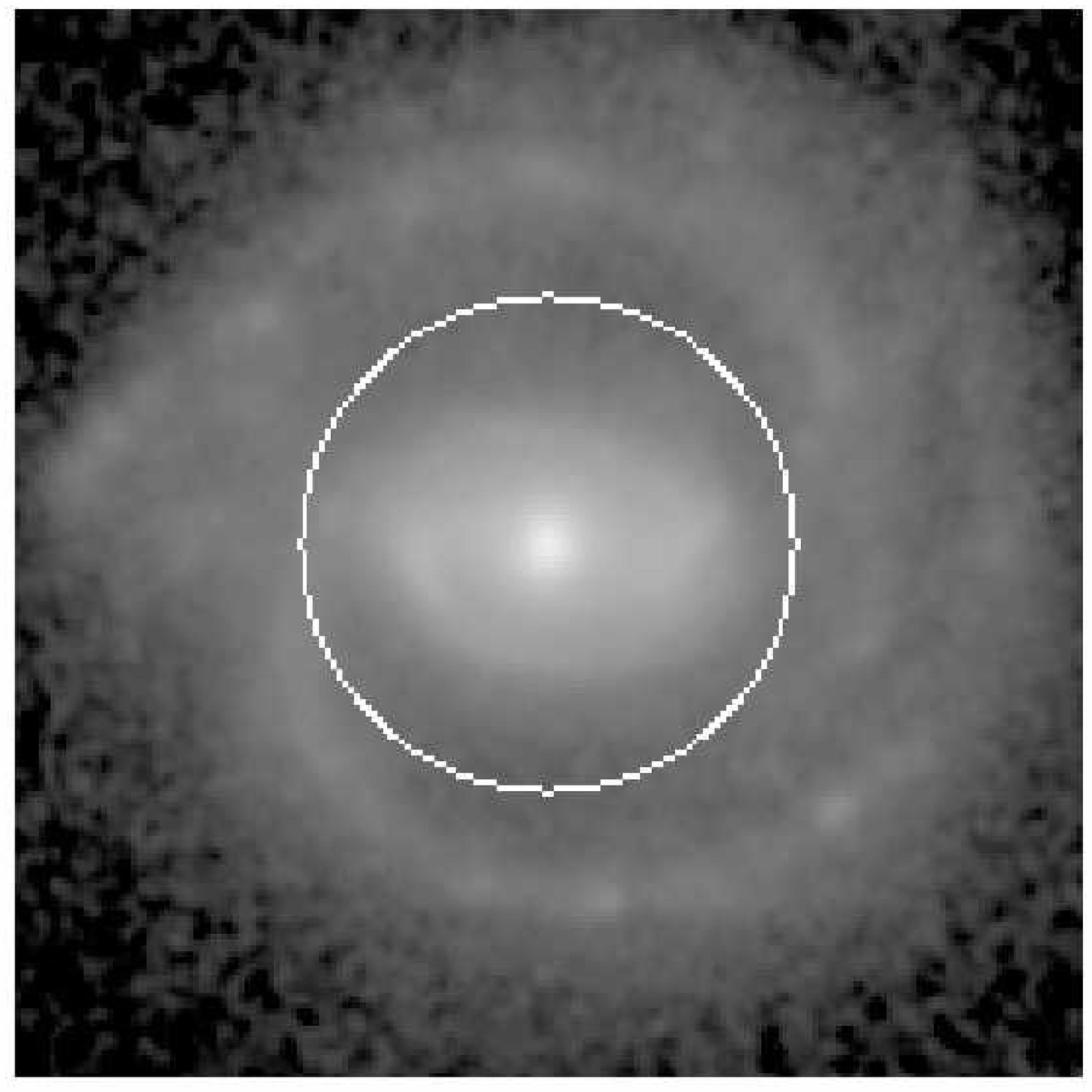}
 \hspace{0.1cm}
 \end{minipage}
 \begin{minipage}[t]{0.68\linewidth}
 \centering
\raisebox{0.5cm}{\includegraphics[width=\textwidth,trim=0 0 0 250,clip]{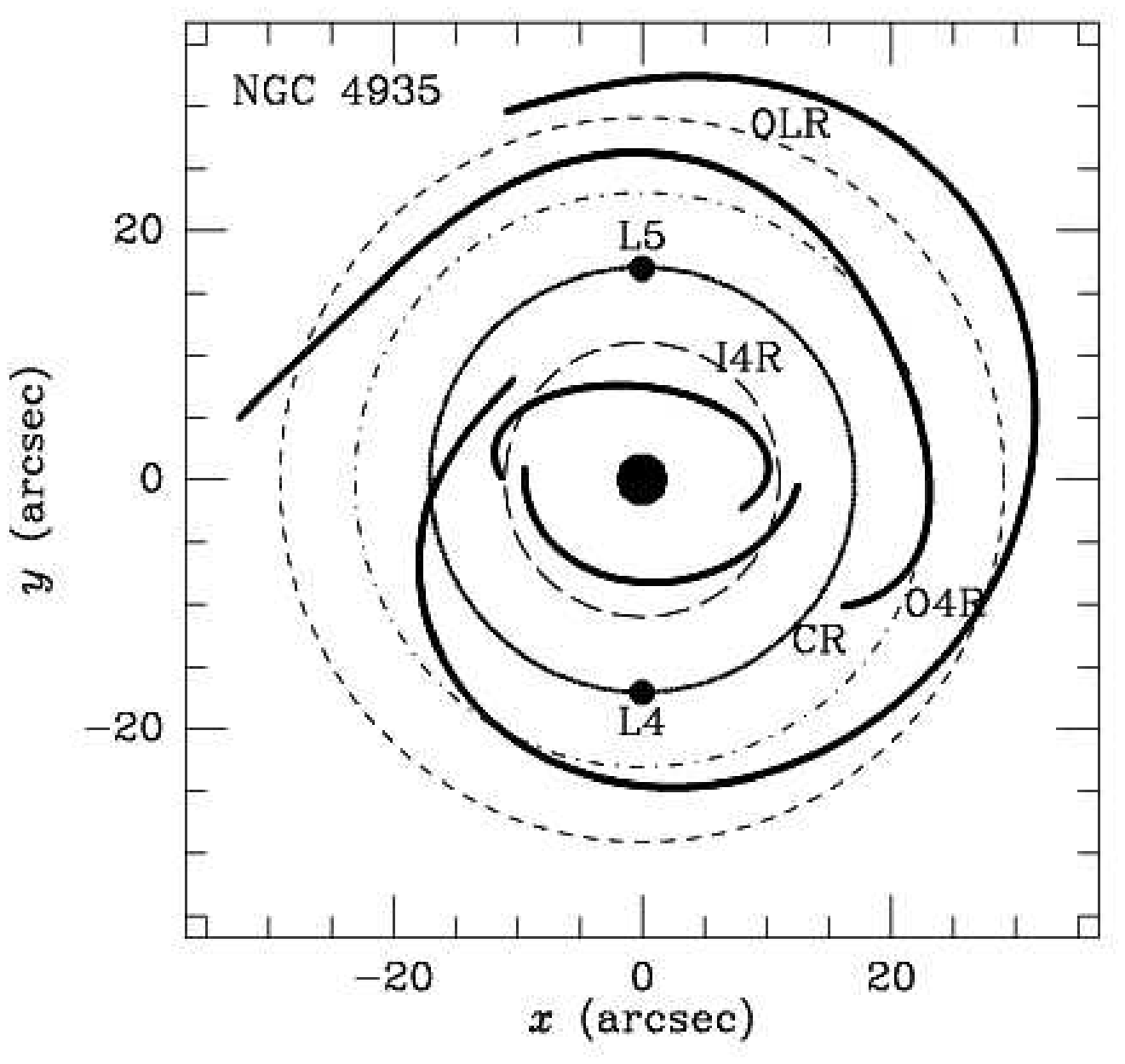}}
 \end{minipage}
\vspace{-1.0truecm}
\caption{(cont.)}
 \end{figure}
 \setcounter{figure}{12}
 \begin{figure}
\vspace{-1.27cm}
 \begin{minipage}[b]{0.45\linewidth}
 \centering
\includegraphics[width=\textwidth]{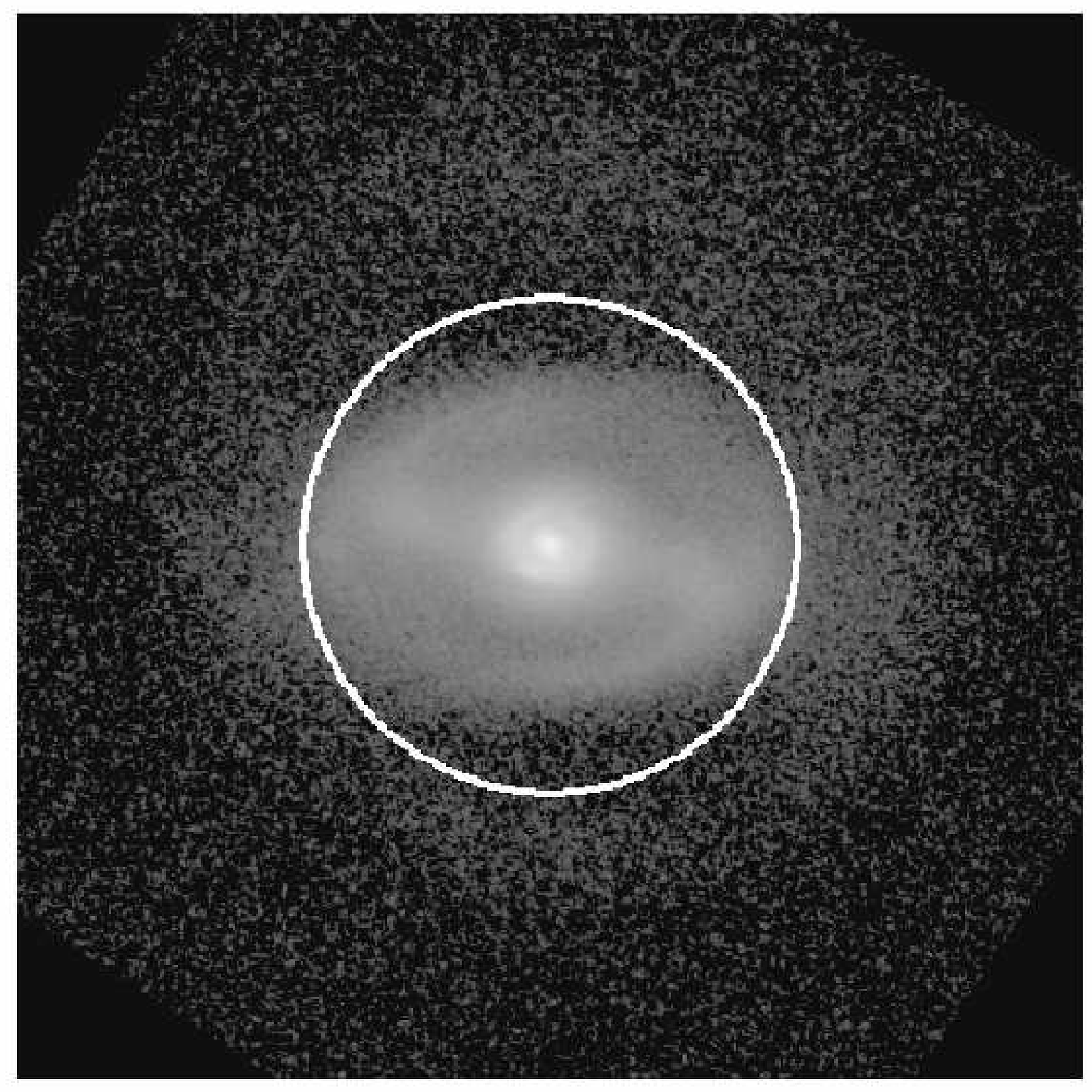}
 \hspace{0.1cm}
 \end{minipage}
 \begin{minipage}[t]{0.68\linewidth}
 \centering
\raisebox{0.5cm}{\includegraphics[width=\textwidth,trim=0 0 0 250,clip]{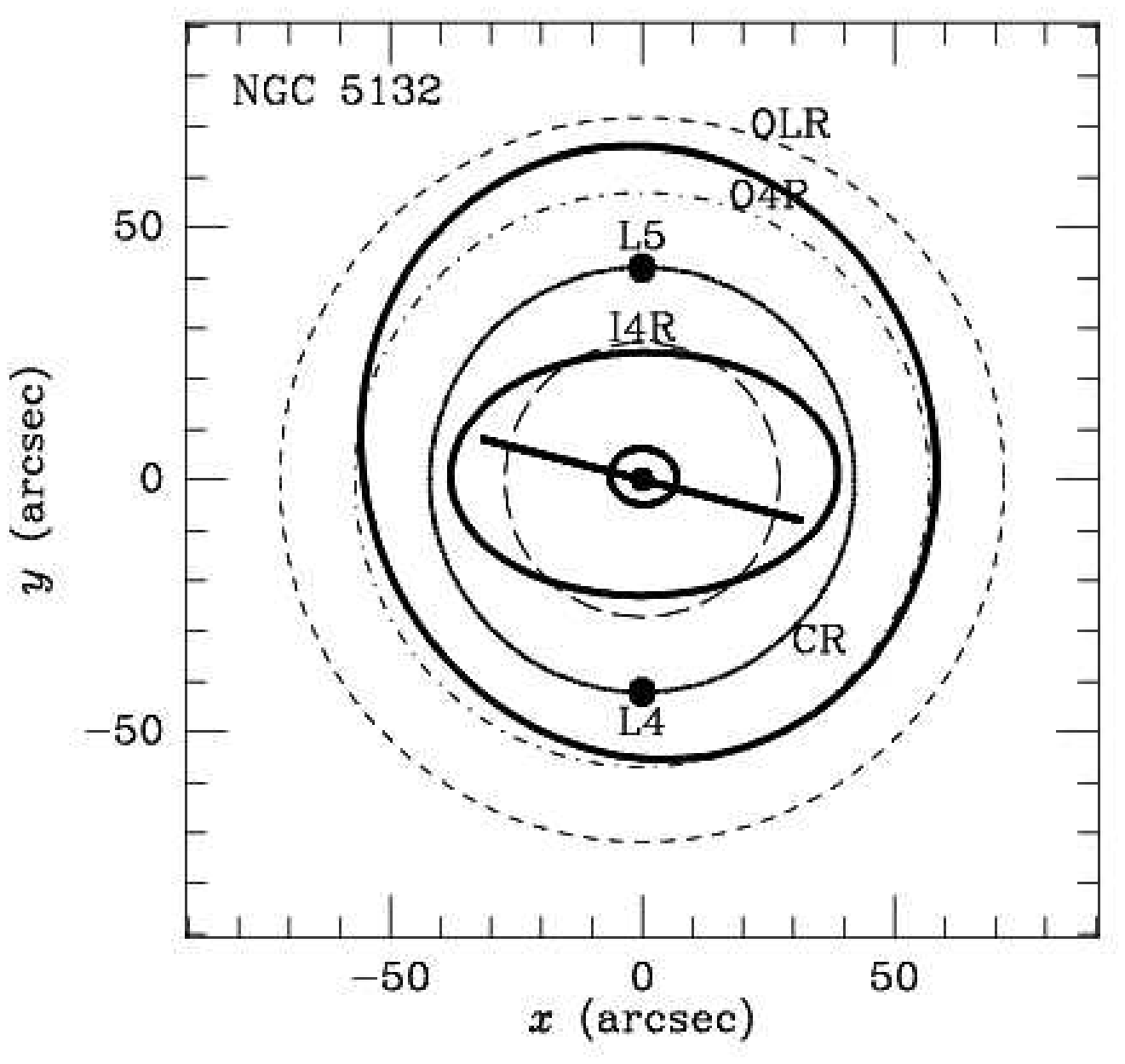}}
 \end{minipage}
 \begin{minipage}[b]{0.45\linewidth}
 \centering
\includegraphics[width=\textwidth]{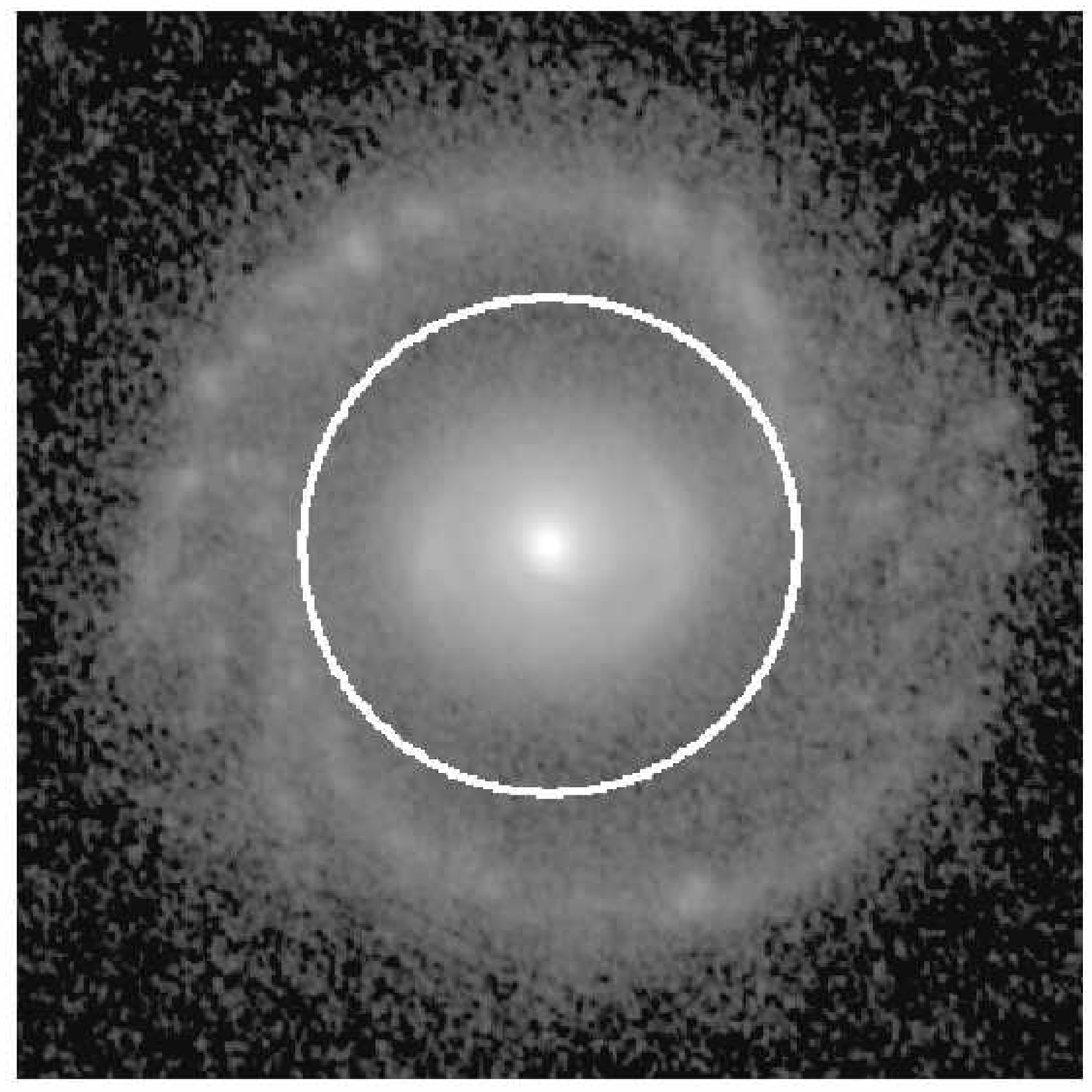}
 \hspace{0.1cm}
 \end{minipage}
 \begin{minipage}[t]{0.68\linewidth}
 \centering
\raisebox{0.5cm}{\includegraphics[width=\textwidth,trim=0 0 0 250,clip]{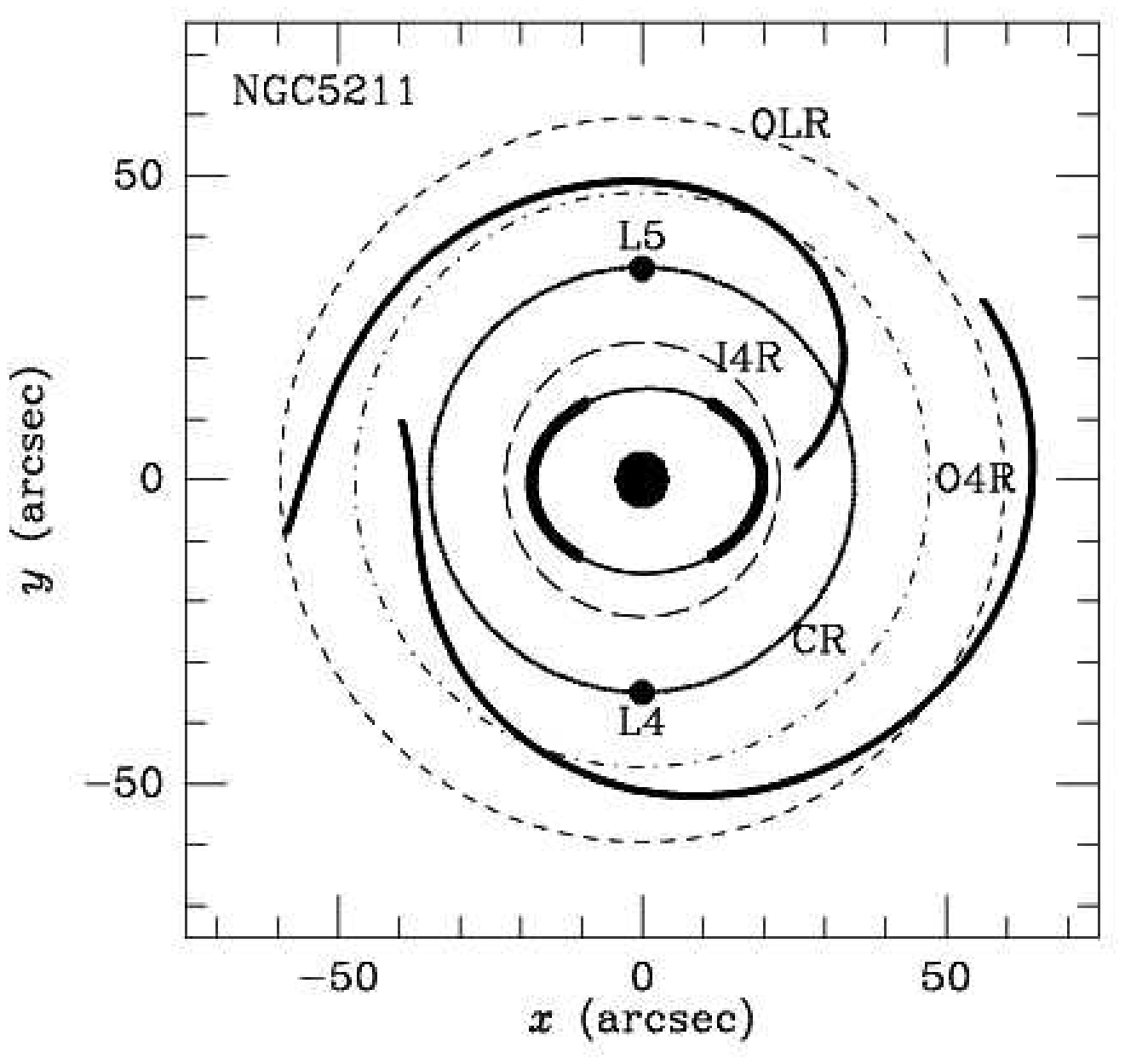}}
 \end{minipage}
 \begin{minipage}[b]{0.45\linewidth}
 \centering
\includegraphics[width=\textwidth]{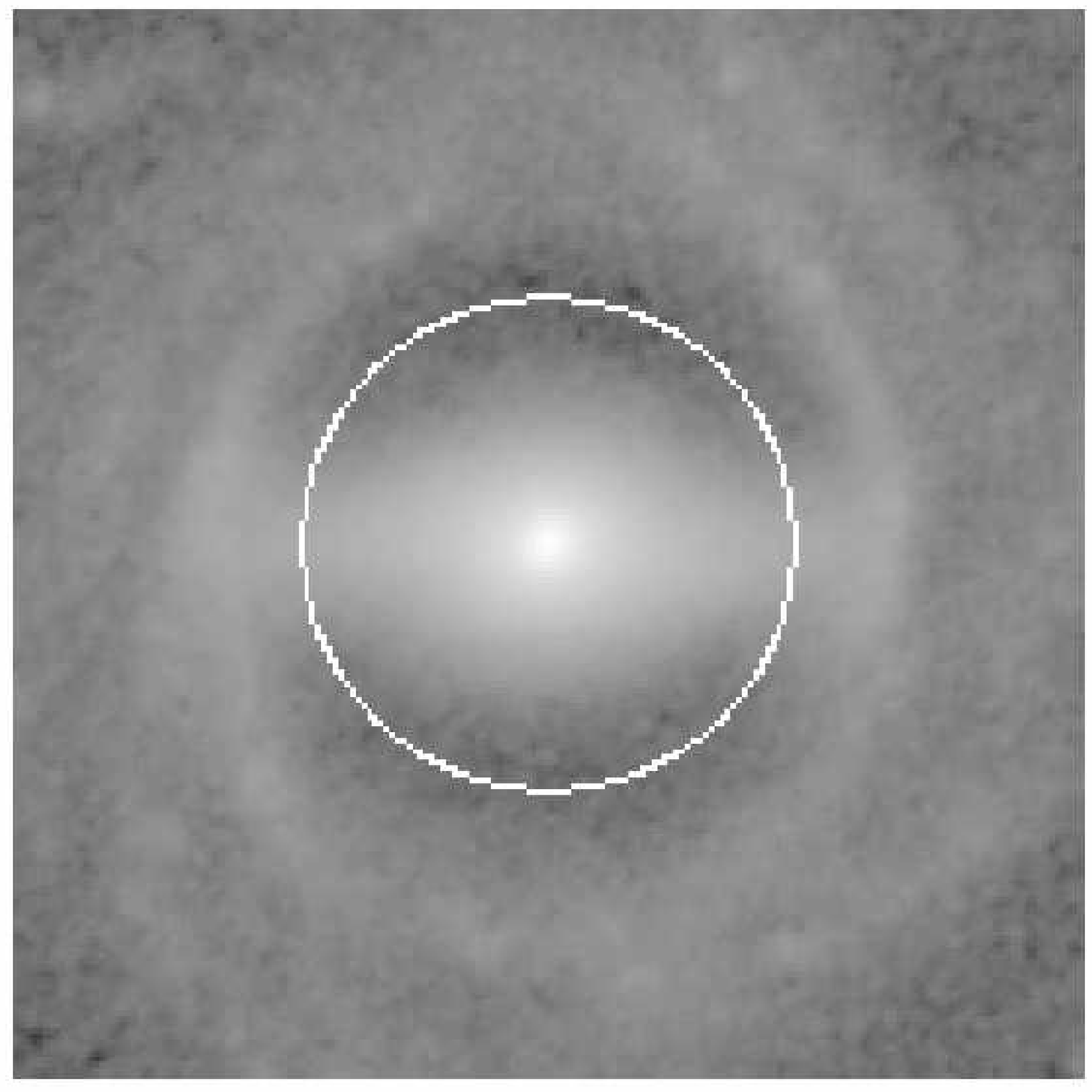}
 \hspace{0.1cm}
 \end{minipage}
 \begin{minipage}[t]{0.68\linewidth}
 \centering
\raisebox{0.5cm}{\includegraphics[width=\textwidth,trim=0 0 0 250,clip]{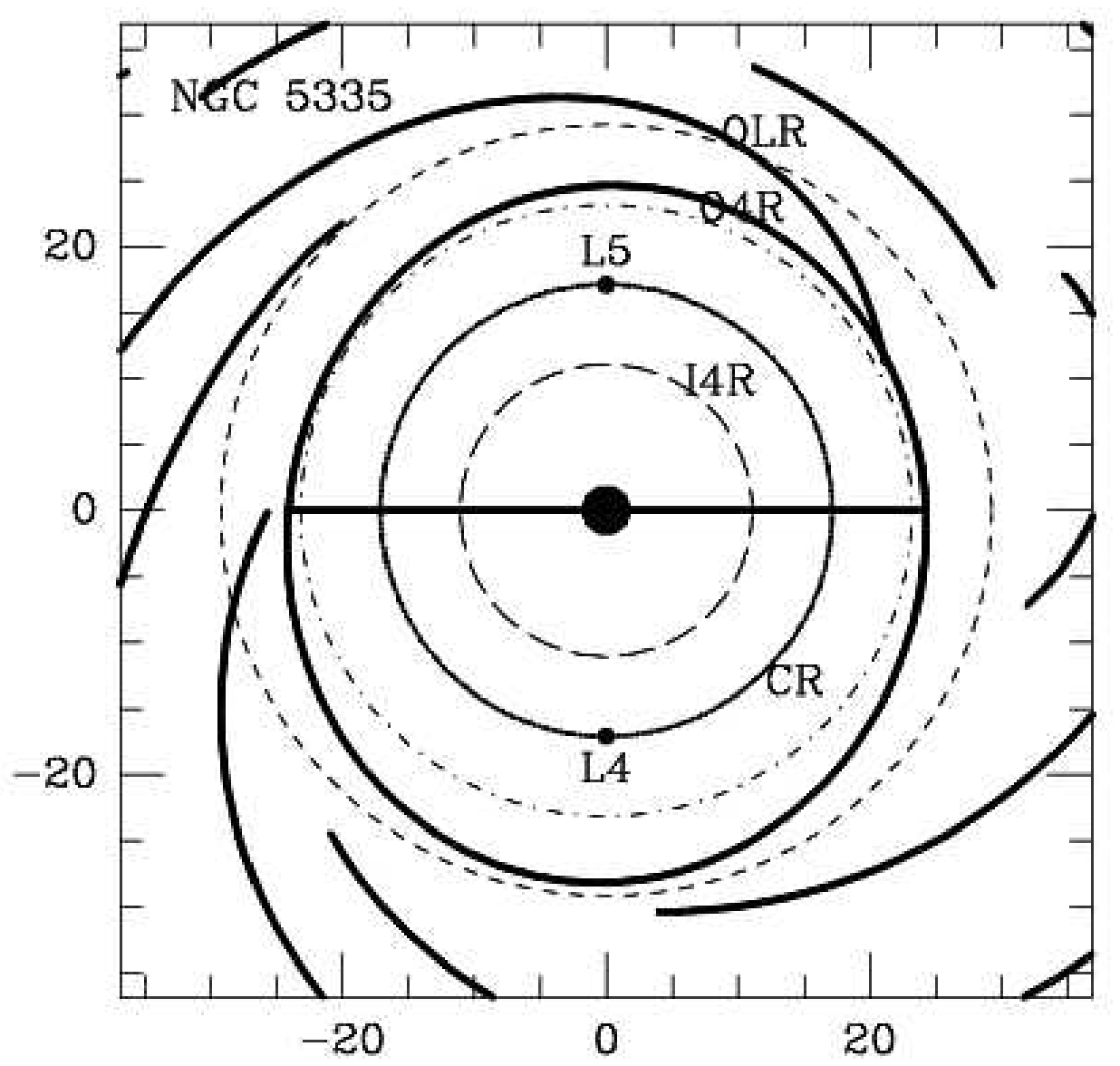}}
 \end{minipage}
 \begin{minipage}[b]{0.45\linewidth}
 \centering
\includegraphics[width=\textwidth]{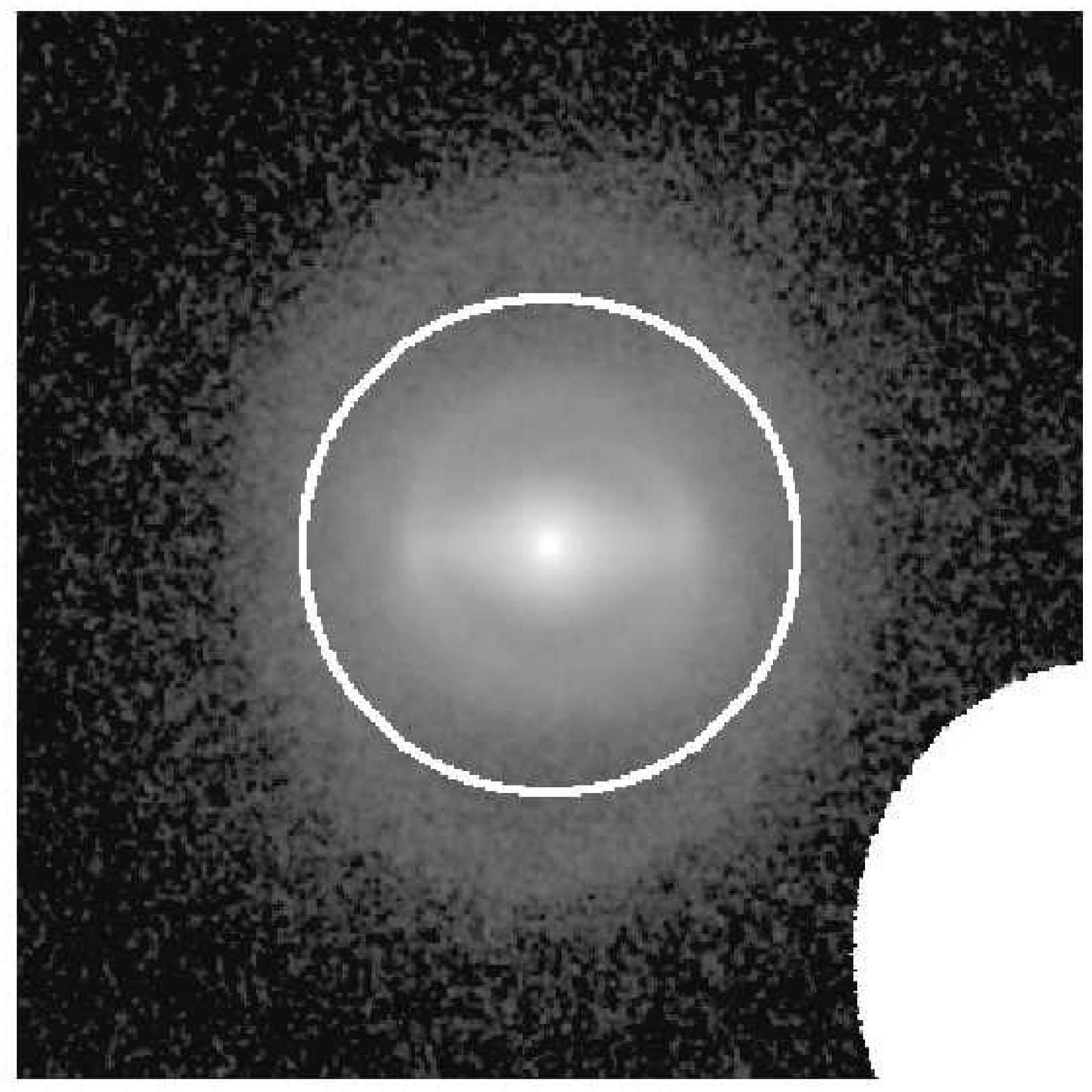}
 \hspace{0.1cm}
 \end{minipage}
 \begin{minipage}[t]{0.68\linewidth}
 \centering
\raisebox{0.5cm}{\includegraphics[width=\textwidth,trim=0 0 0 250,clip]{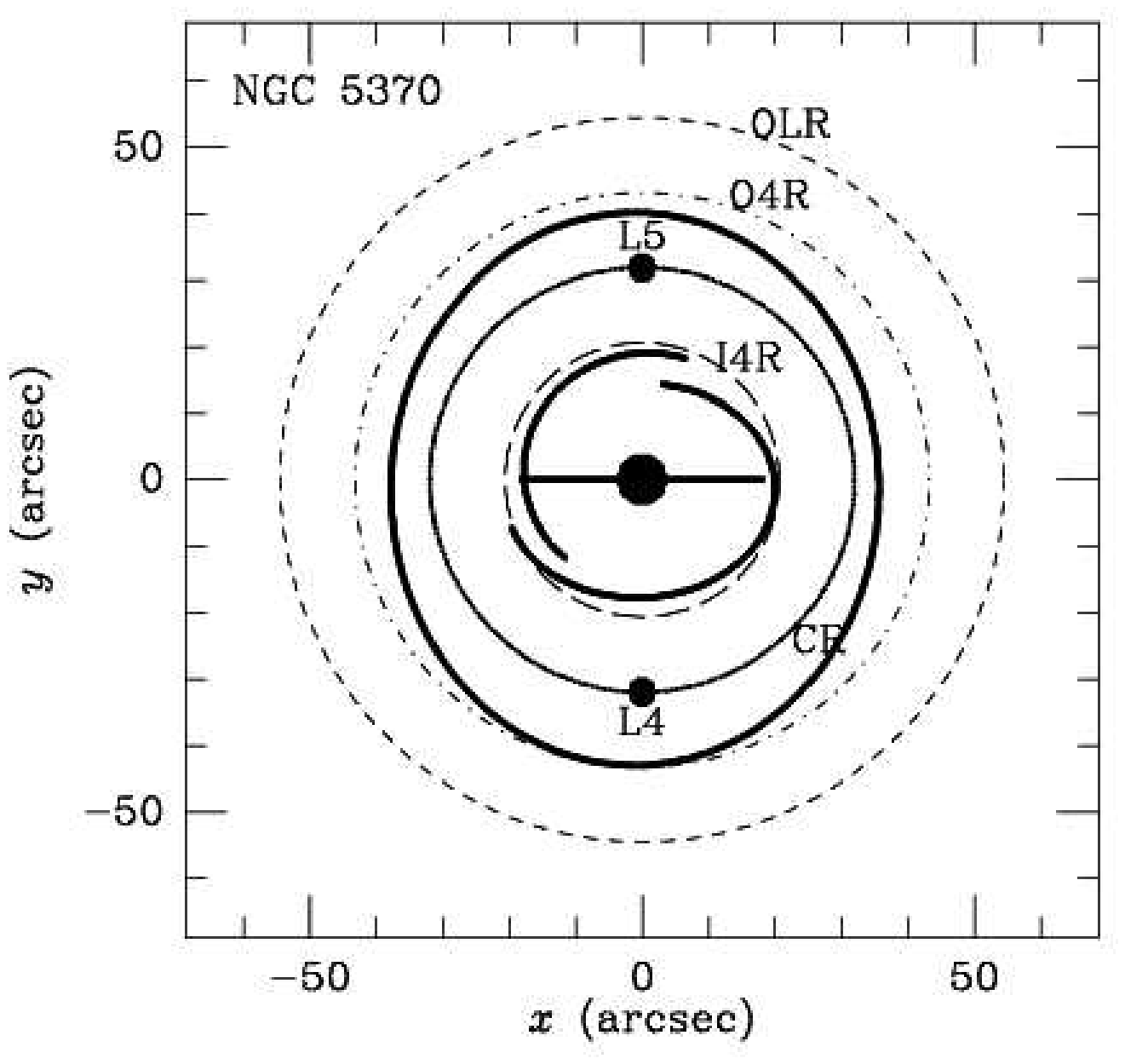}}
 \end{minipage}
 \begin{minipage}[b]{0.45\linewidth}
 \centering
\includegraphics[width=\textwidth]{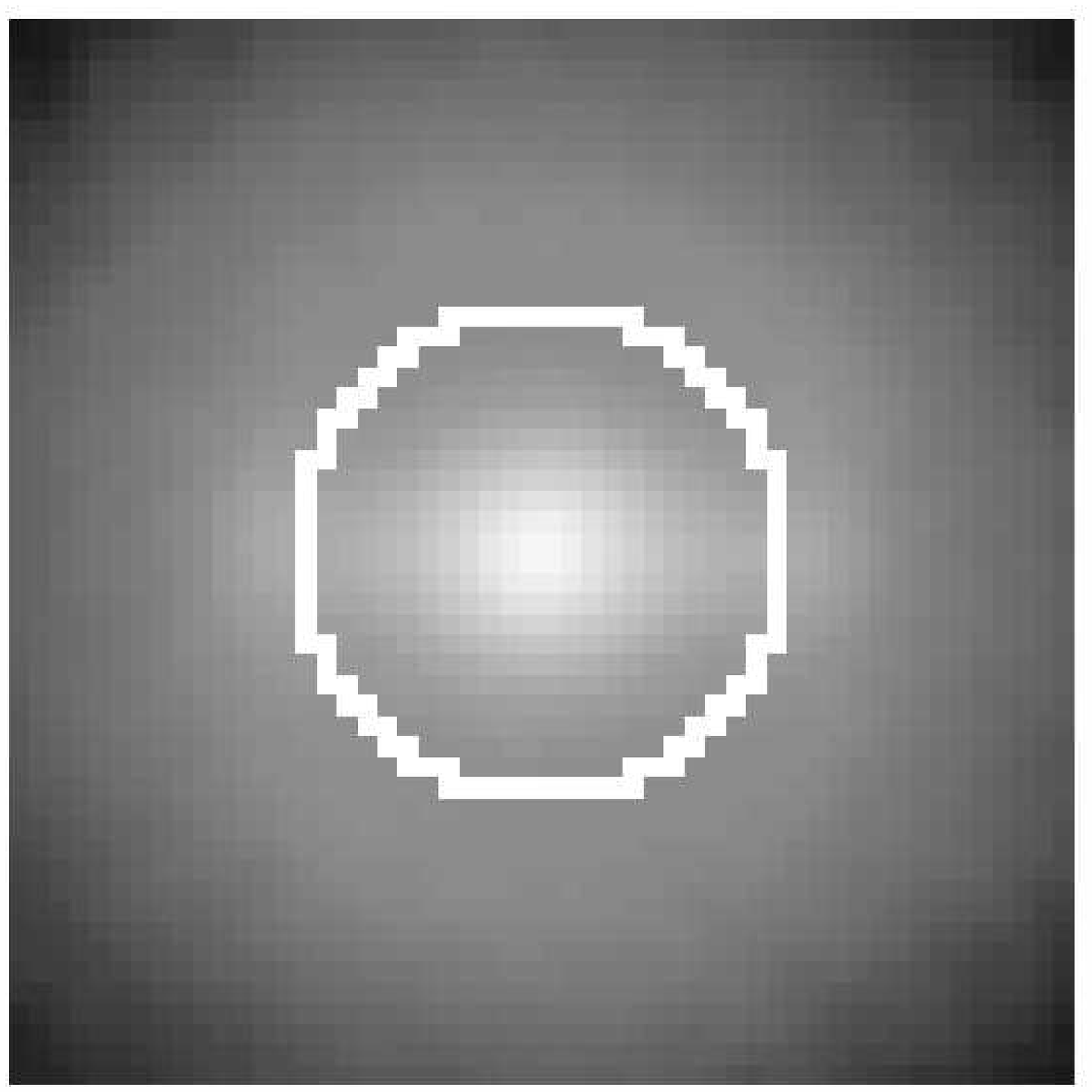}
 \hspace{0.1cm}
 \end{minipage}
 \begin{minipage}[t]{0.68\linewidth}
 \centering
\raisebox{0.5cm}{\includegraphics[width=\textwidth,trim=0 0 0 250,clip]{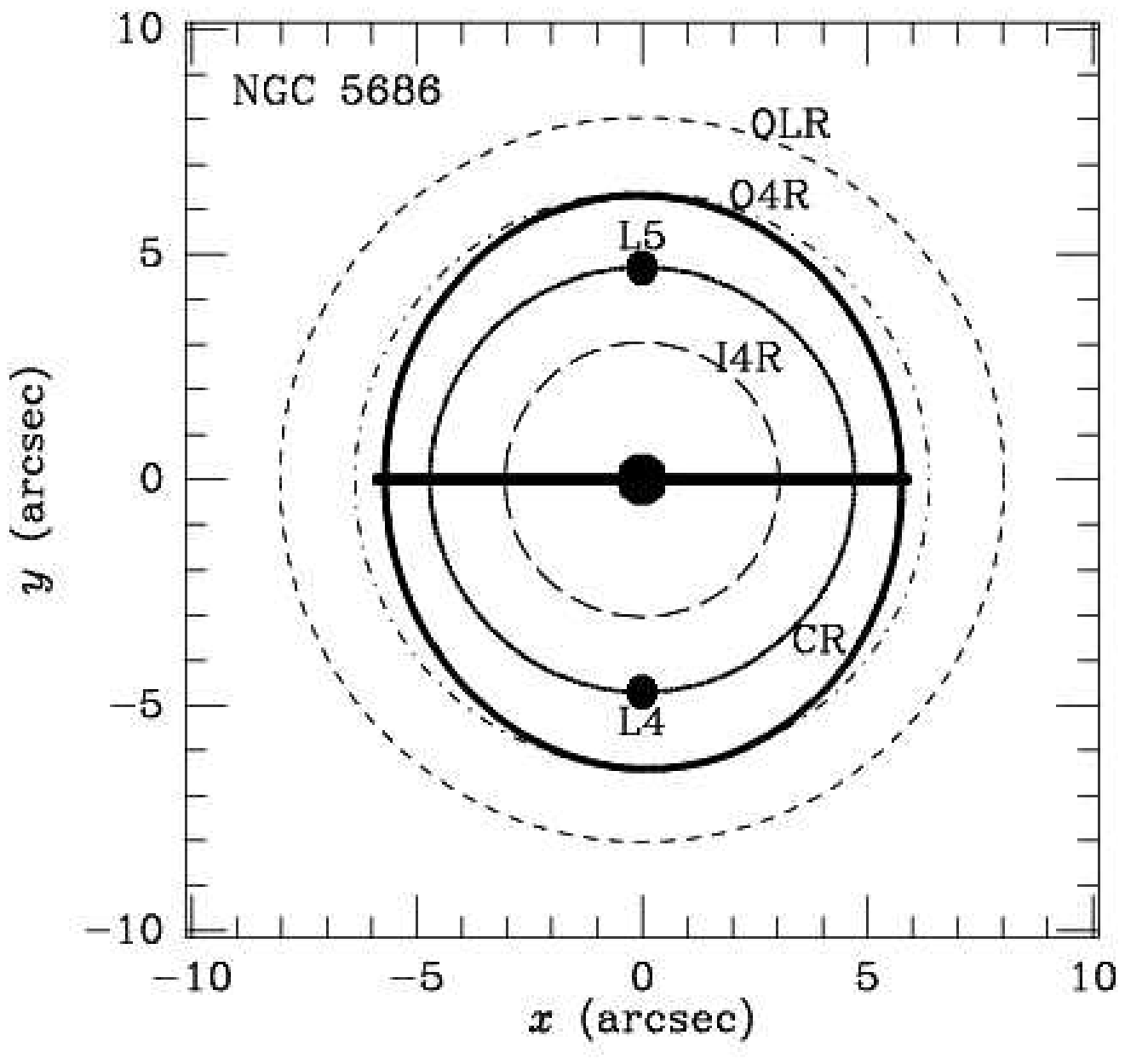}}
\vspace{-1.0truecm}
 \end{minipage}
\vspace{-1.0cm}
\caption{(cont.)}
 \end{figure}
 \setcounter{figure}{12}
 \begin{figure}
\vspace{-1.27cm}
 \begin{minipage}[b]{0.45\linewidth}
 \centering
\includegraphics[width=\textwidth]{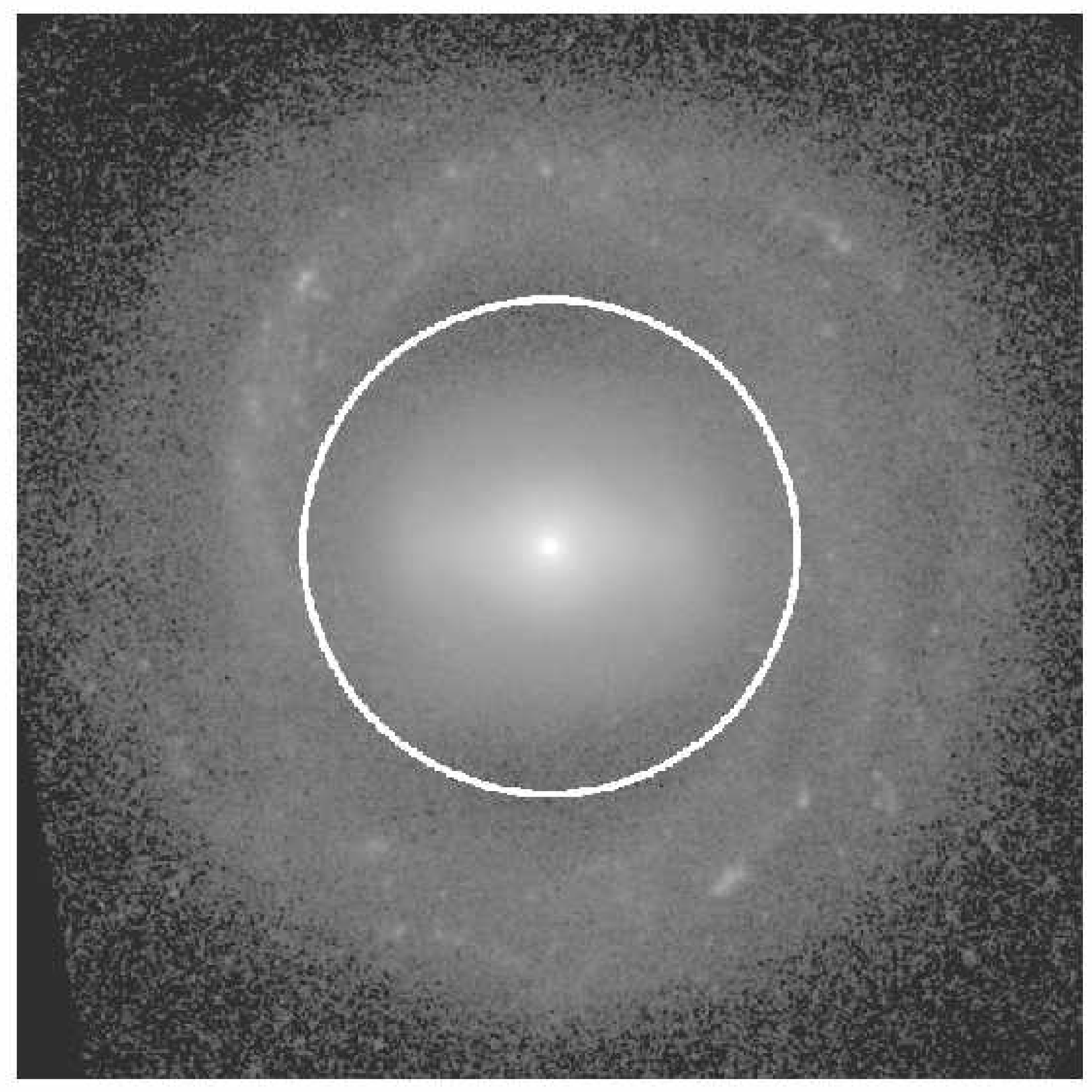}
 \hspace{0.1cm}
 \end{minipage}
 \begin{minipage}[t]{0.68\linewidth}
 \centering
\raisebox{0.5cm}{\includegraphics[width=\textwidth,trim=0 0 0 250,clip]{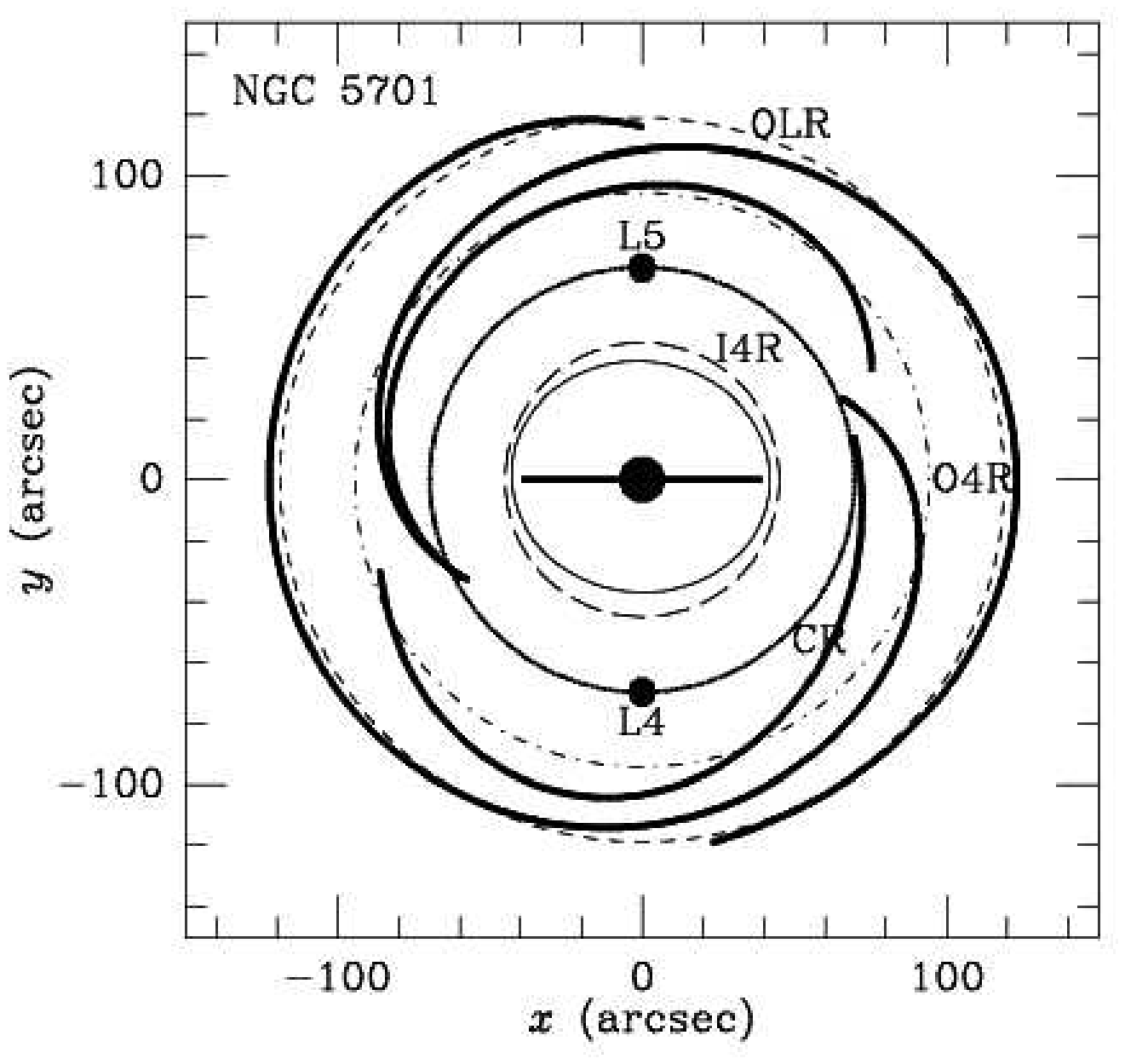}}
 \end{minipage}
 \begin{minipage}[b]{0.45\linewidth}
 \centering
\includegraphics[width=\textwidth]{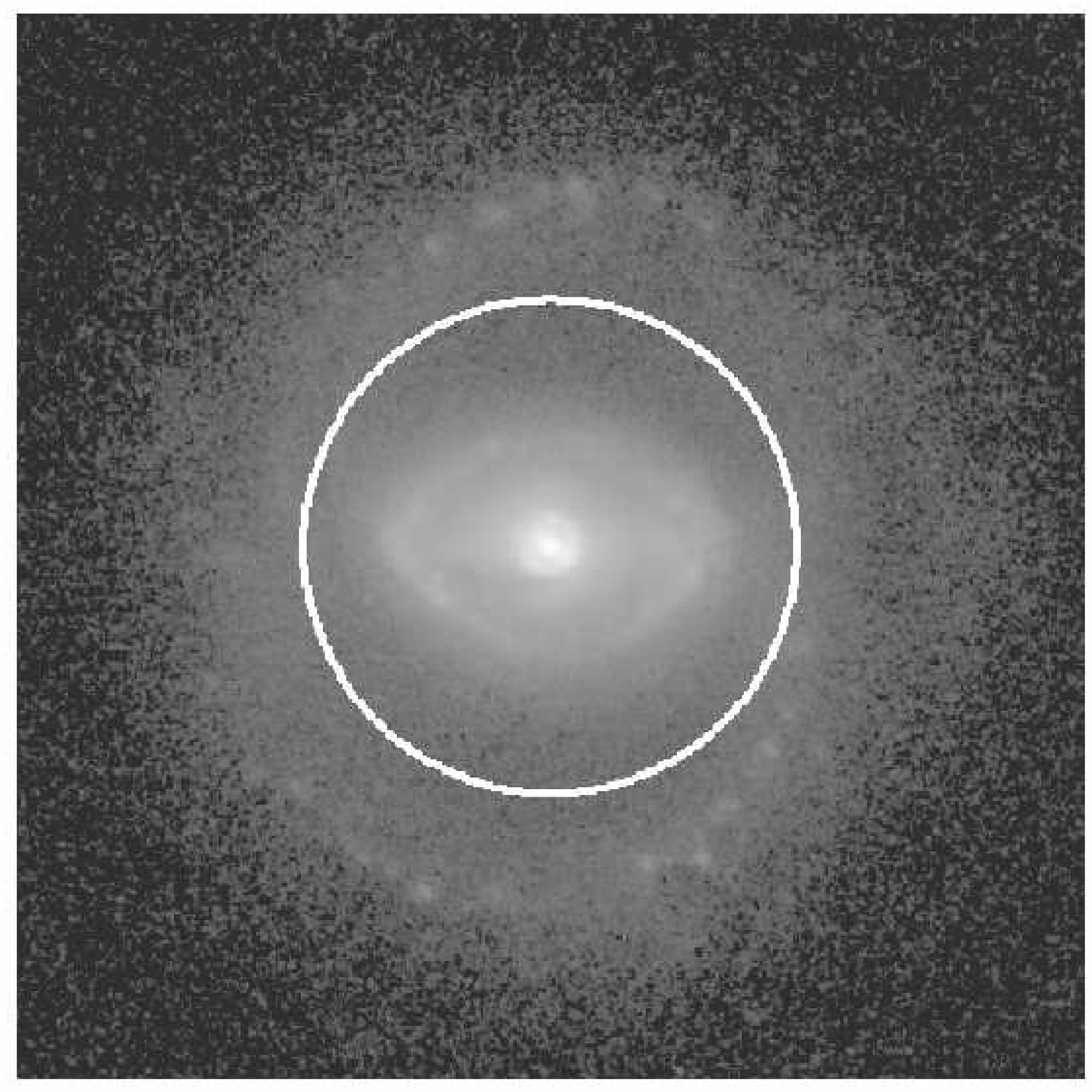}
 \hspace{0.1cm}
 \end{minipage}
 \begin{minipage}[t]{0.68\linewidth}
 \centering
\raisebox{0.5cm}{\includegraphics[width=\textwidth,trim=0 0 0 250,clip]{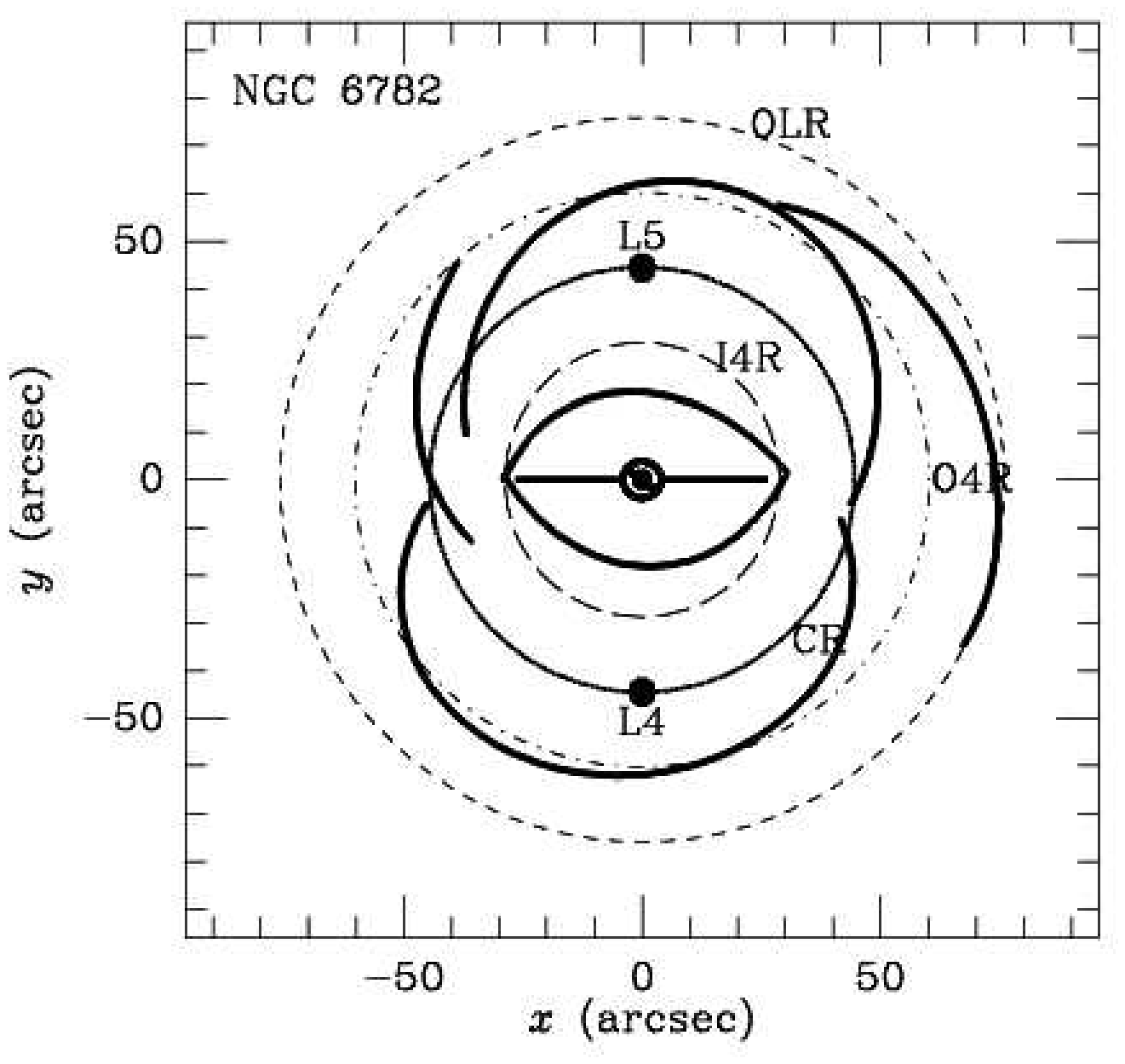}}
 \end{minipage}
 \begin{minipage}[b]{0.45\linewidth}
 \centering
\includegraphics[width=\textwidth]{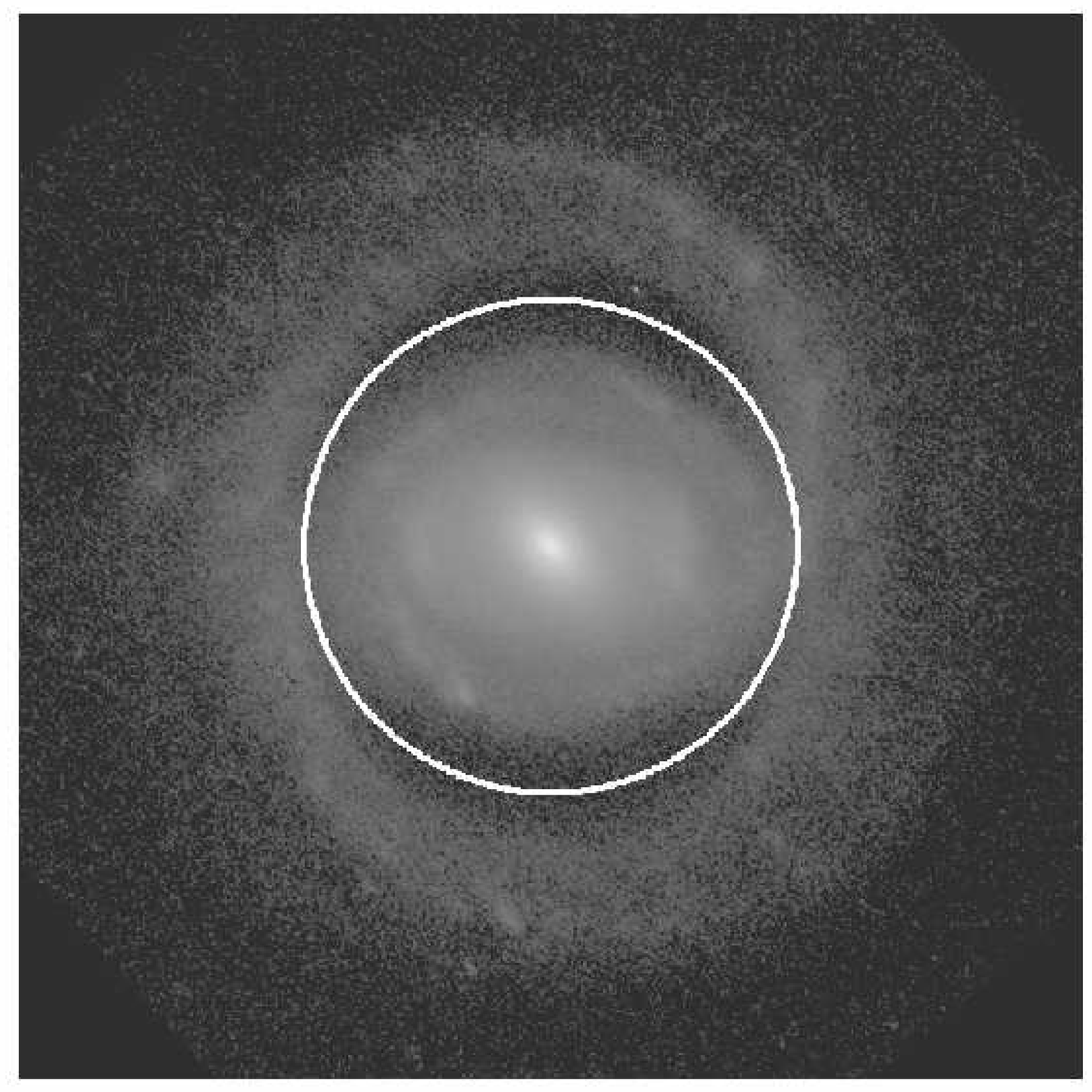}
 \hspace{0.1cm}
 \end{minipage}
 \begin{minipage}[t]{0.68\linewidth}
 \centering
\raisebox{0.5cm}{\includegraphics[width=\textwidth,trim=0 0 0 250,clip]{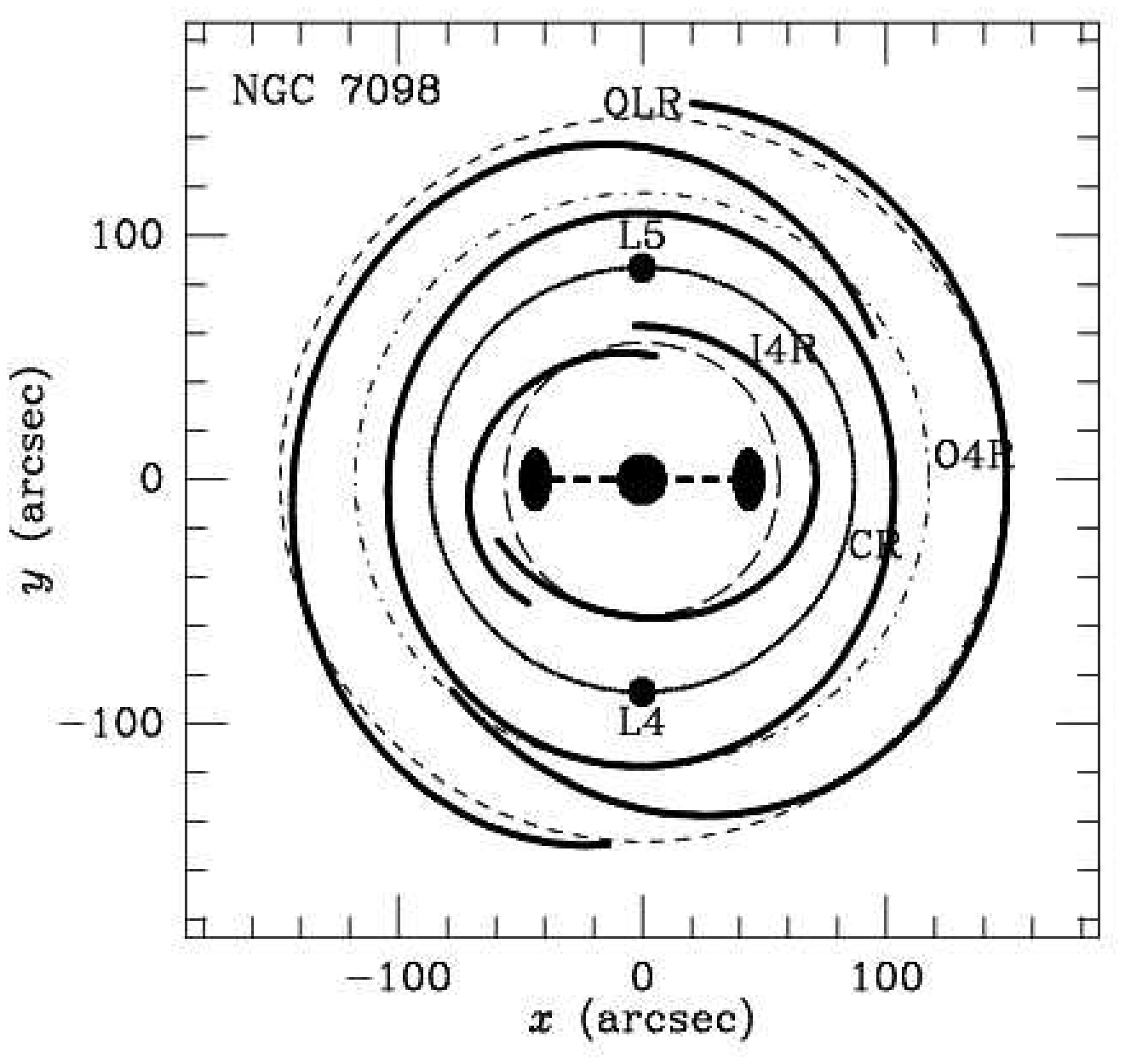}}
 \end{minipage}
 \begin{minipage}[b]{0.45\linewidth}
 \centering
\includegraphics[width=\textwidth]{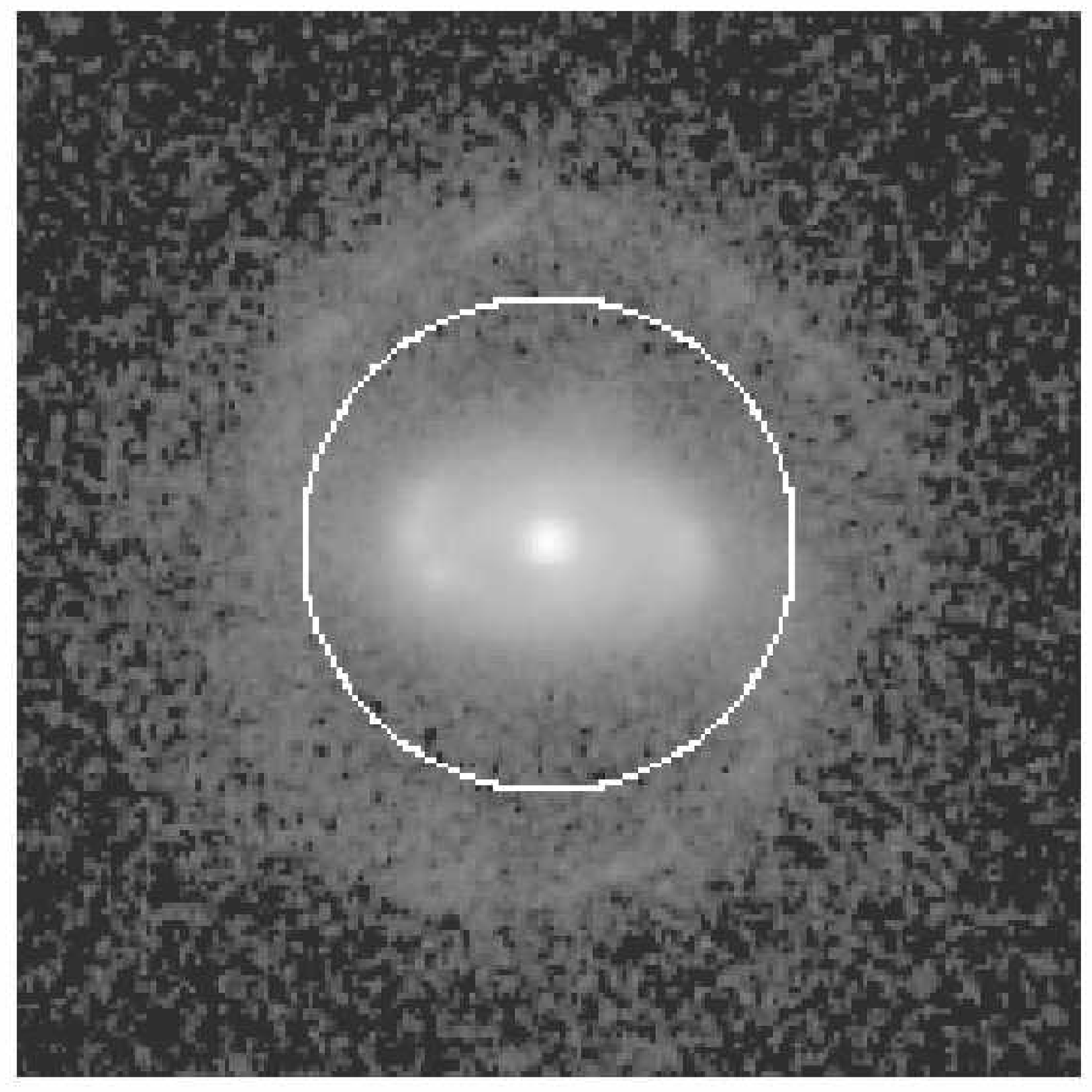}
 \hspace{0.1cm}
 \end{minipage}
 \begin{minipage}[t]{0.68\linewidth}
 \centering
\raisebox{0.5cm}{\includegraphics[width=\textwidth,trim=0 0 0 250,clip]{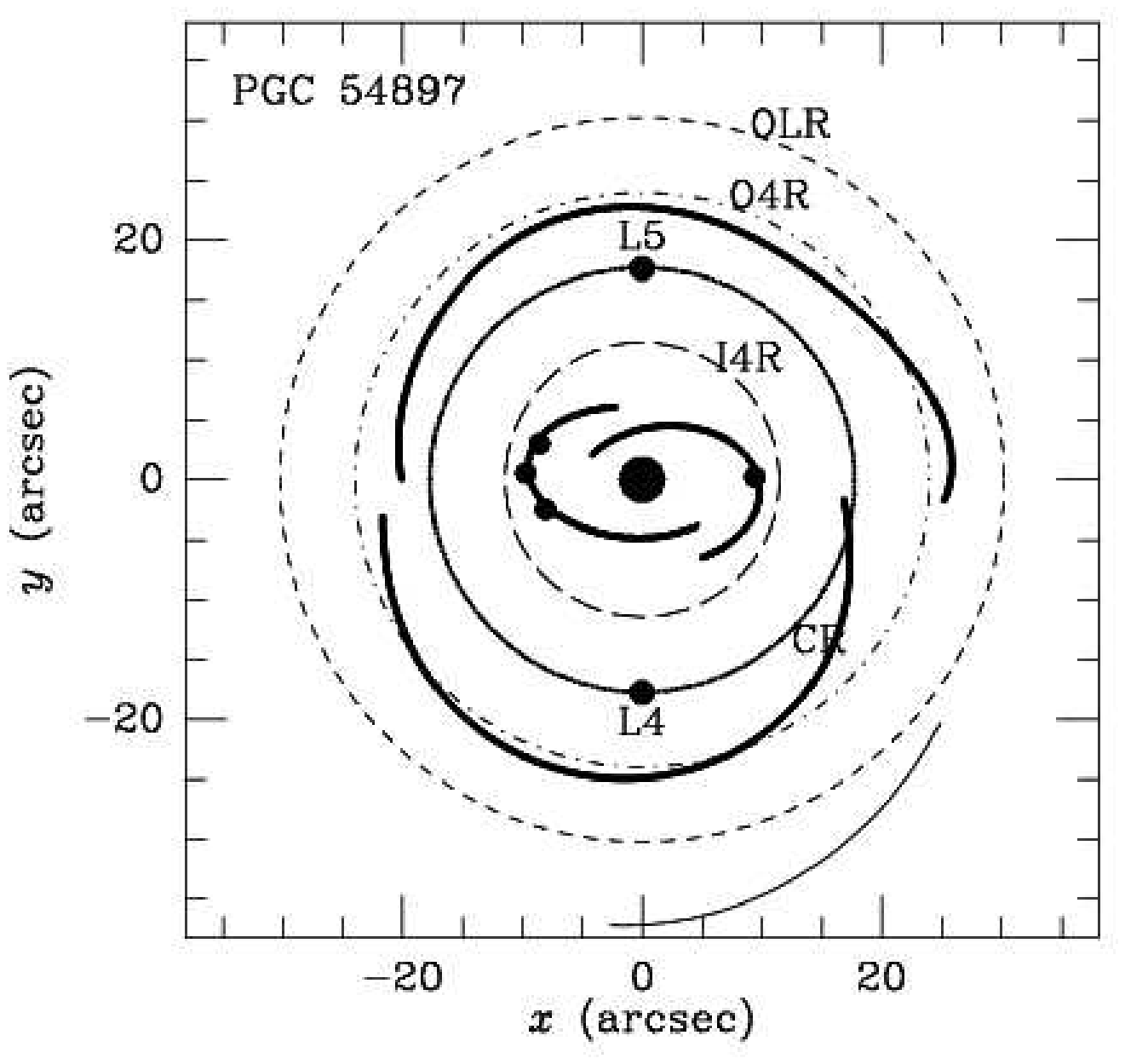}}
 \end{minipage}
 \begin{minipage}[b]{0.45\linewidth}
 \centering
\includegraphics[width=\textwidth]{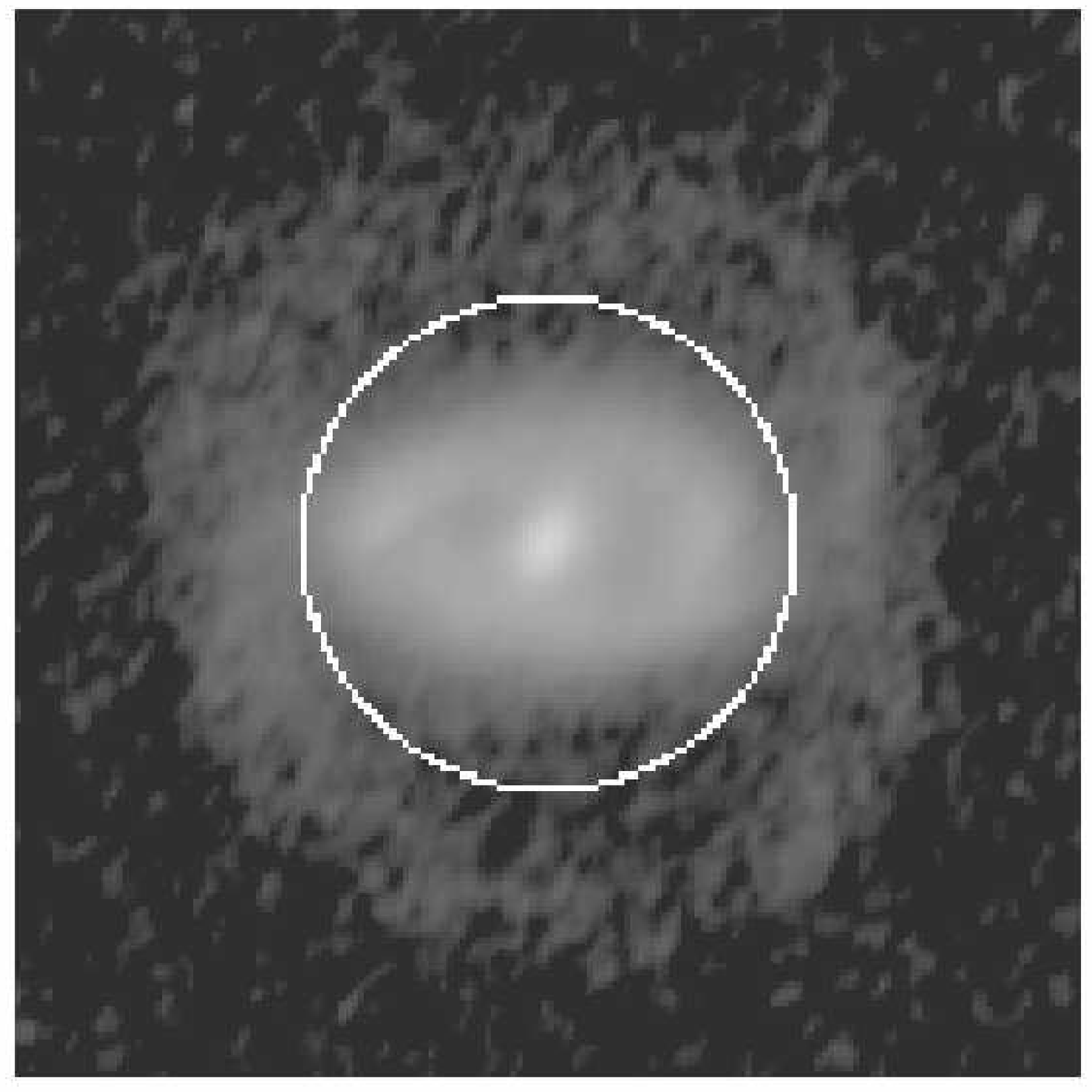}
 \hspace{0.1cm}
 \end{minipage}
 \begin{minipage}[t]{0.68\linewidth}
 \centering
\raisebox{0.5cm}{\includegraphics[width=\textwidth,trim=0 0 0 250,clip]{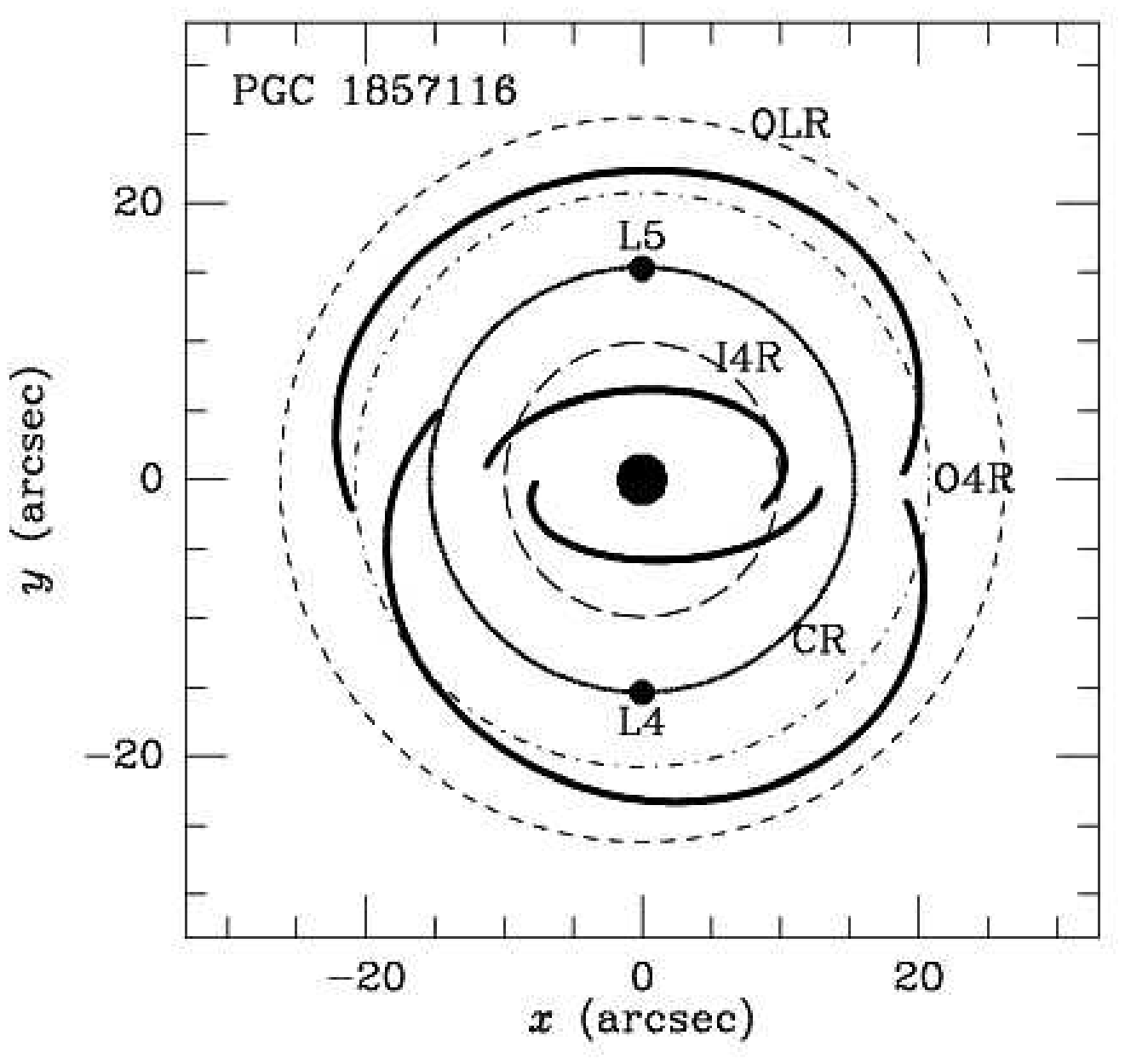}}
 \end{minipage}
\vspace{-1.0truecm}
\caption{(cont.)}
 \end{figure}
 \setcounter{figure}{12}
 \begin{figure}
\vspace{-1.27cm}
 \begin{minipage}[b]{0.45\linewidth}
 \centering
\includegraphics[width=\textwidth]{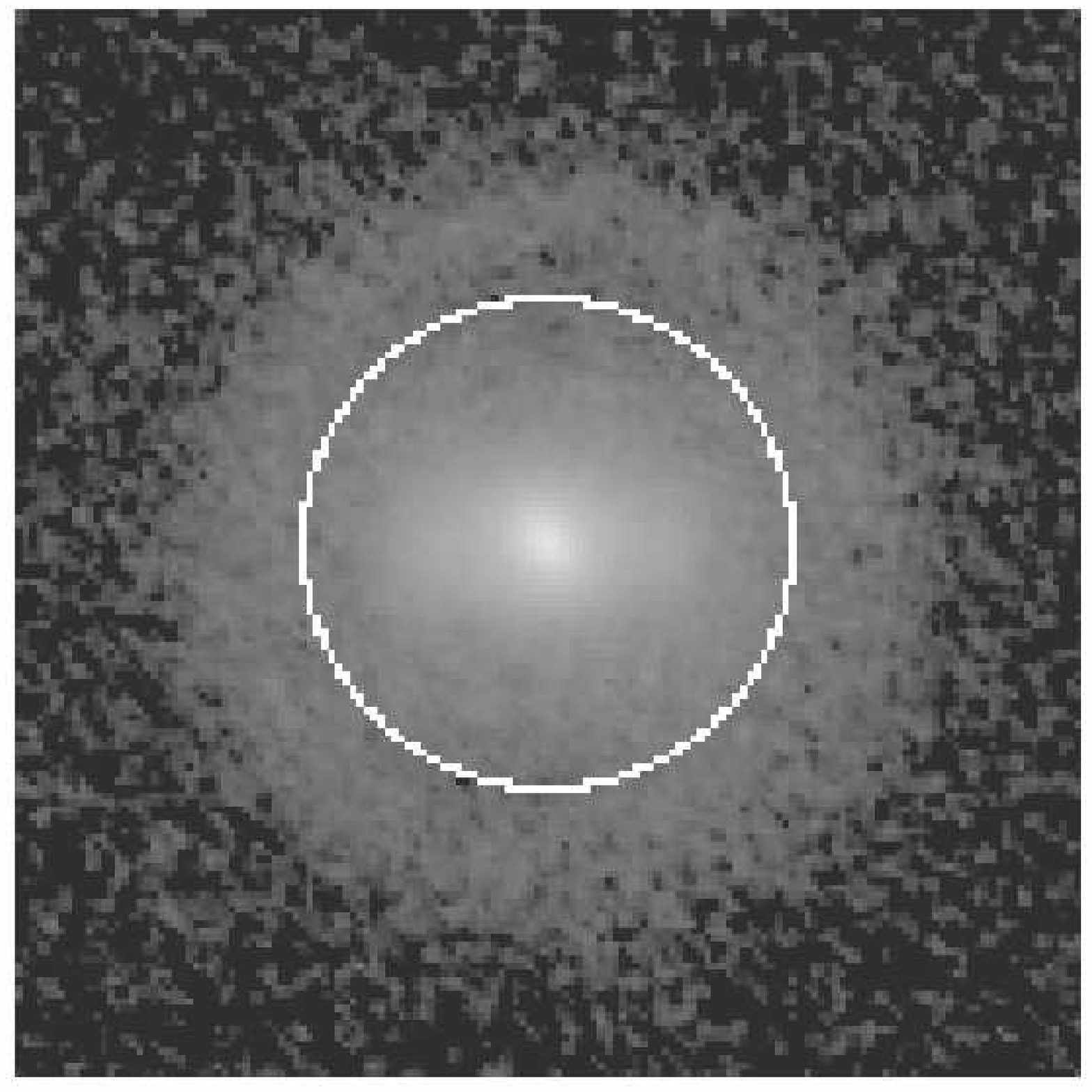}
 \hspace{0.1cm}
 \end{minipage}
 \begin{minipage}[t]{0.68\linewidth}
 \centering
\raisebox{0.5cm}{\includegraphics[width=\textwidth,trim=0 0 0 250,clip]{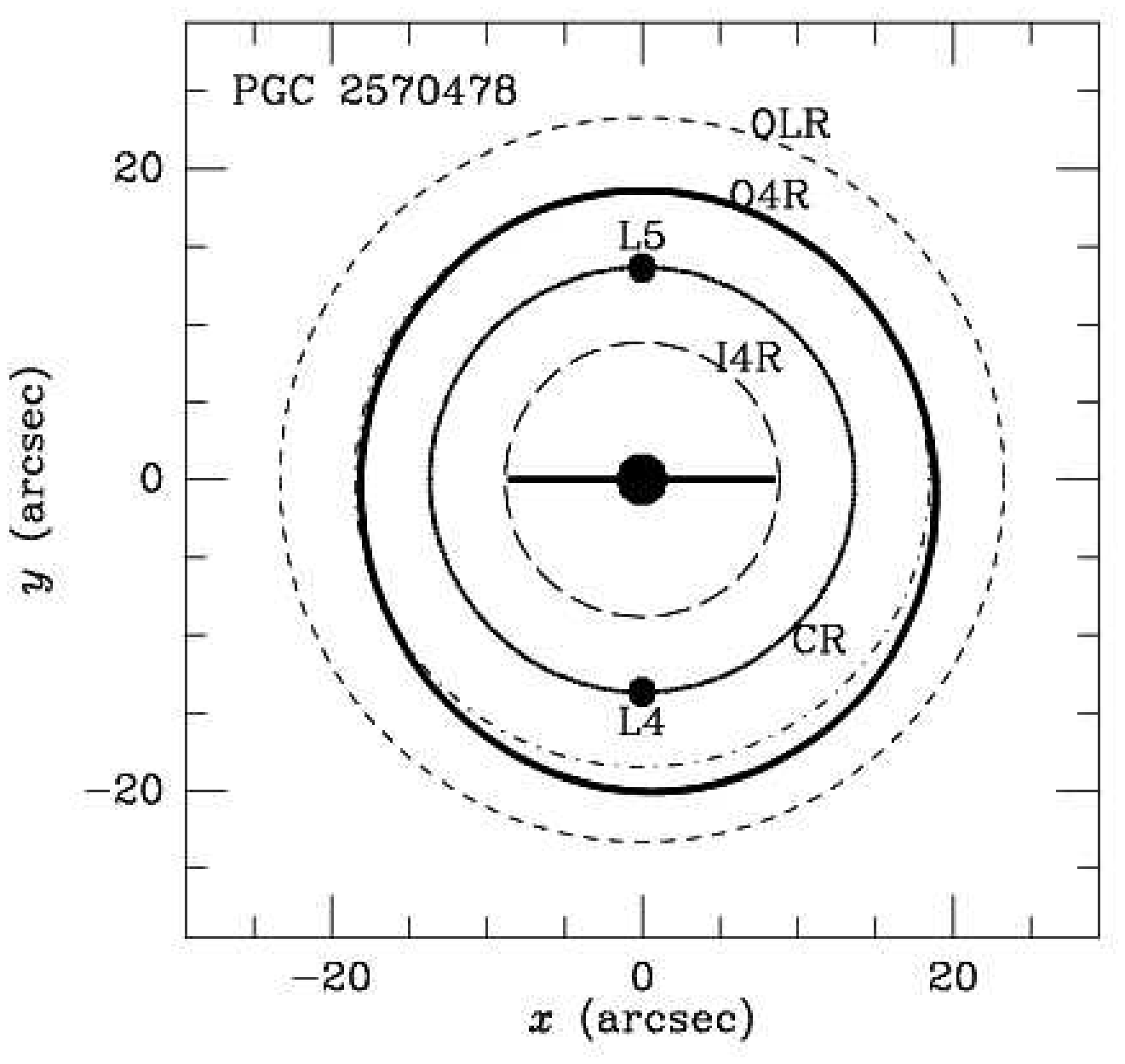}}
 \end{minipage}
 \begin{minipage}[b]{0.45\linewidth}
 \centering
\includegraphics[width=\textwidth]{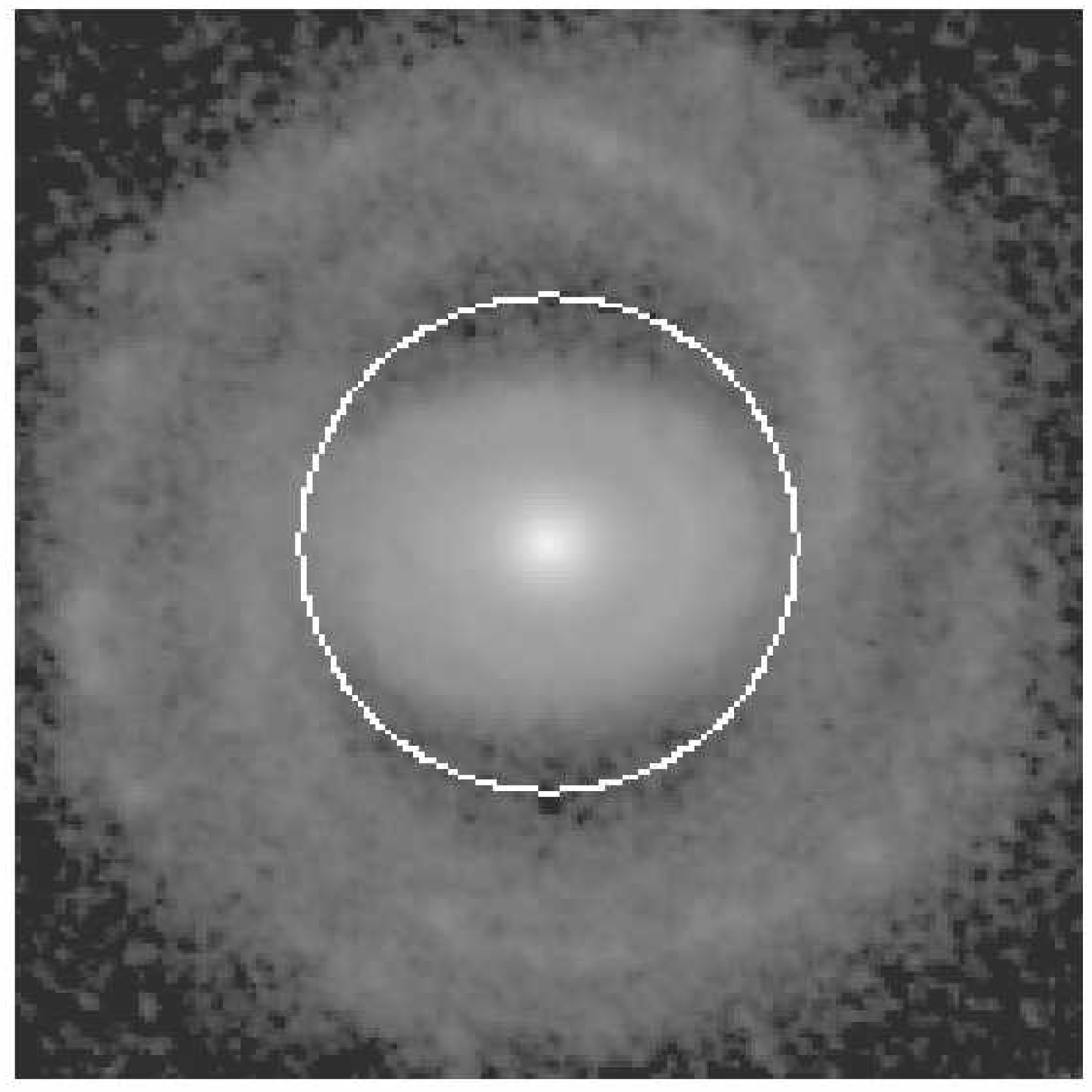}
 \hspace{0.1cm}
 \end{minipage}
 \begin{minipage}[t]{0.68\linewidth}
 \centering
\raisebox{0.5cm}{\includegraphics[width=\textwidth,trim=0 0 0 250,clip]{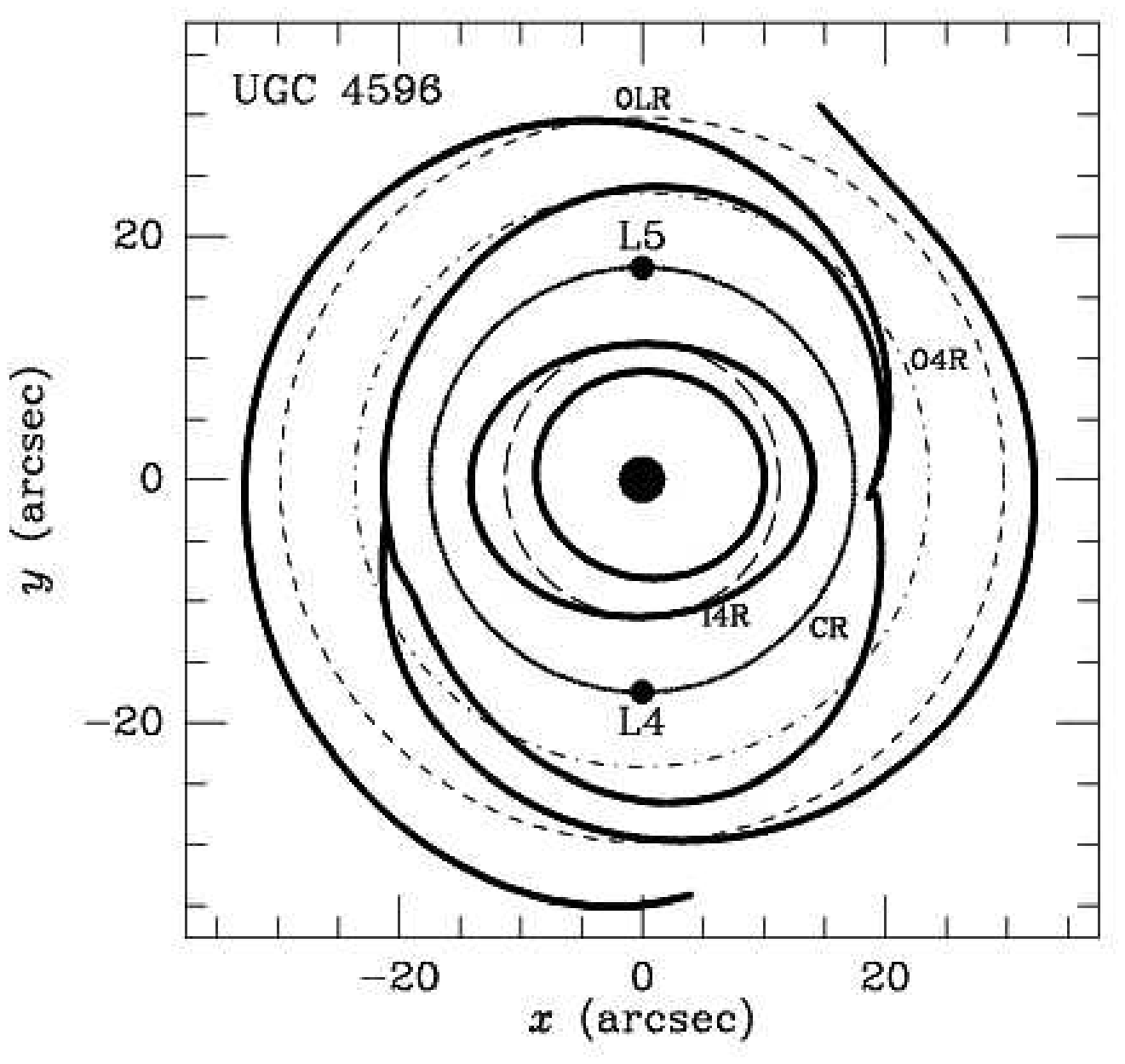}}
 \end{minipage}
 \begin{minipage}[b]{0.45\linewidth}
 \centering
\includegraphics[width=\textwidth]{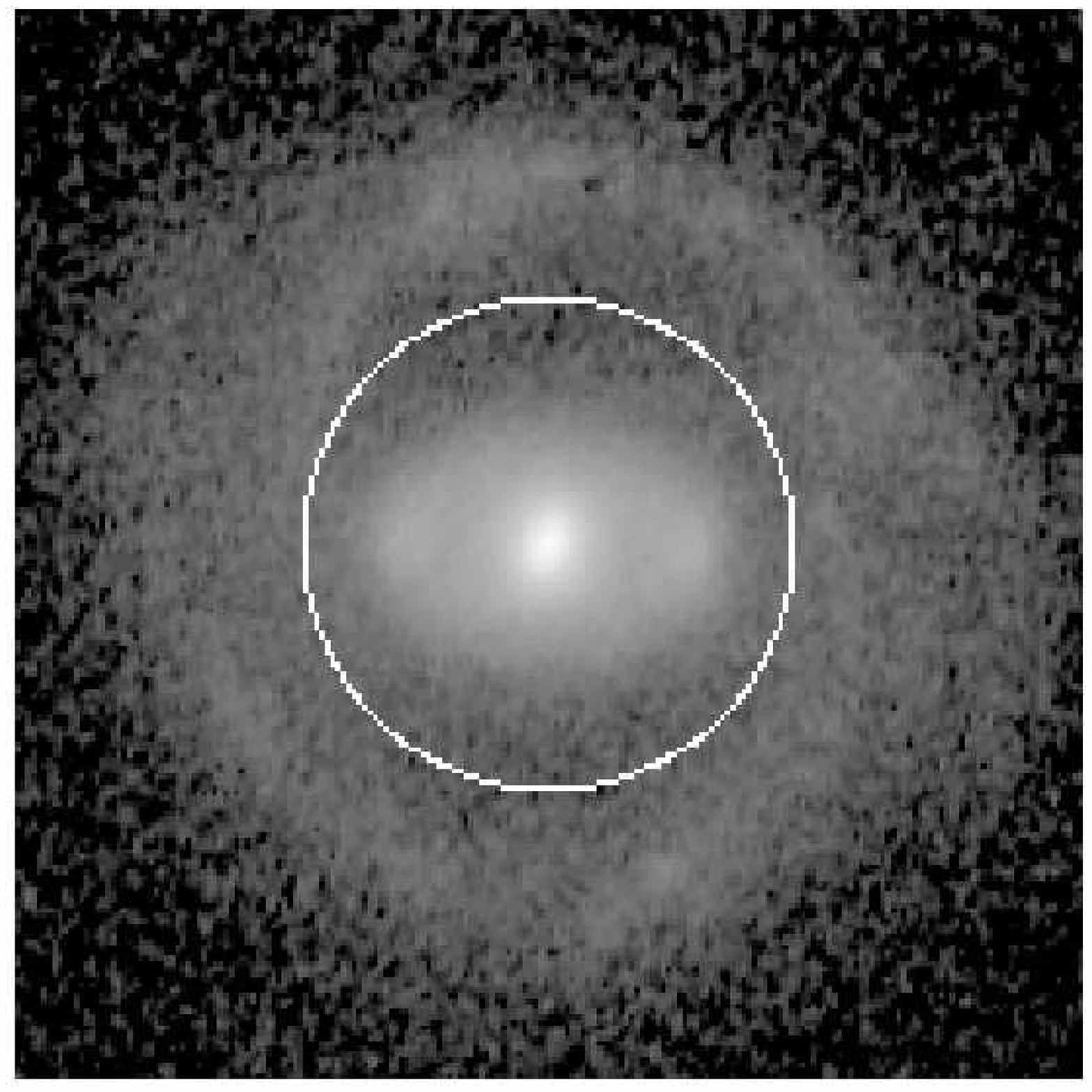}
 \hspace{0.1cm}
 \end{minipage}
 \begin{minipage}[t]{0.68\linewidth}
 \centering
\raisebox{0.5cm}{\includegraphics[width=\textwidth,trim=0 0 0 250,clip]{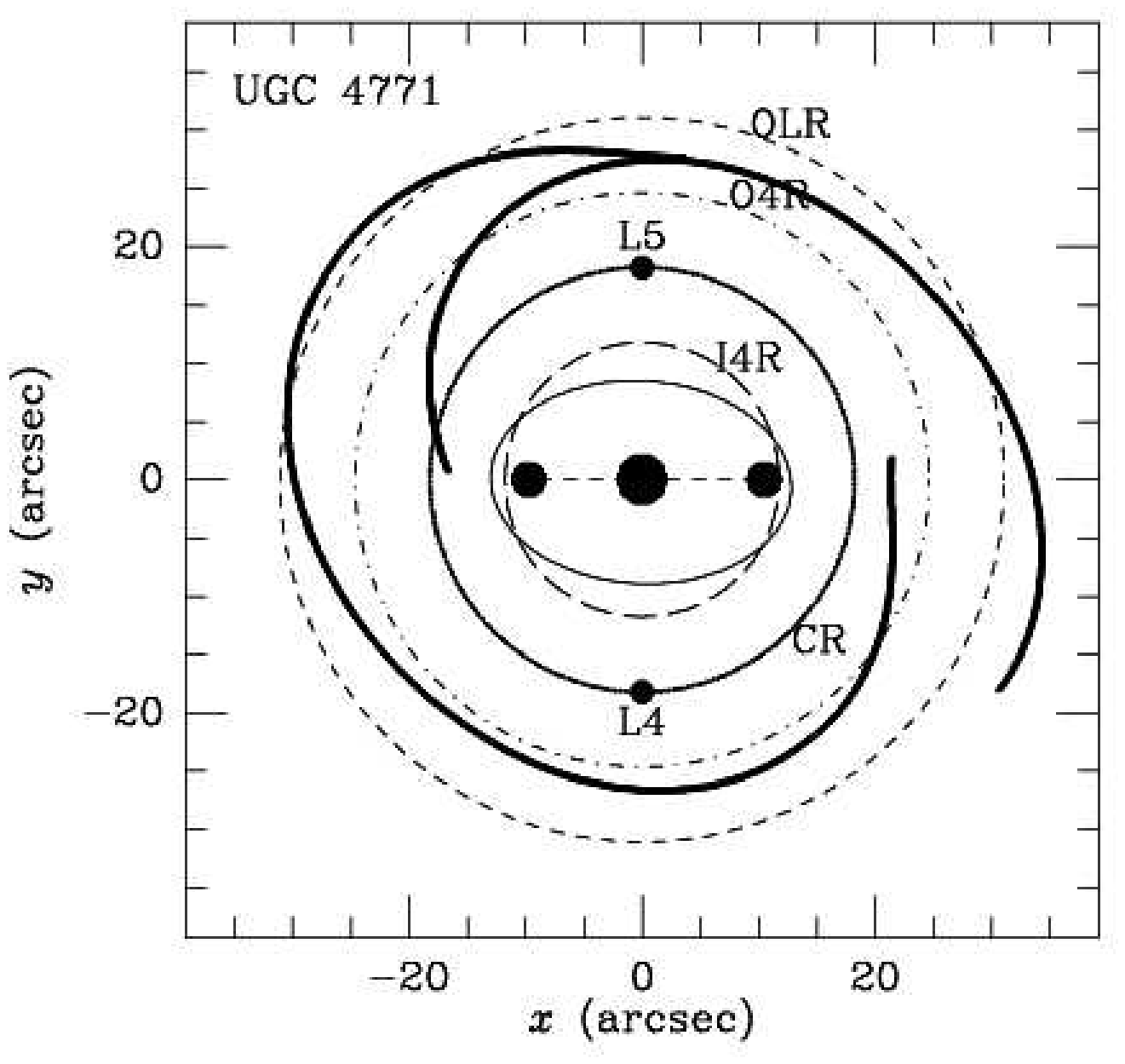}}
 \end{minipage}
 \begin{minipage}[b]{0.45\linewidth}
 \centering
\includegraphics[width=\textwidth]{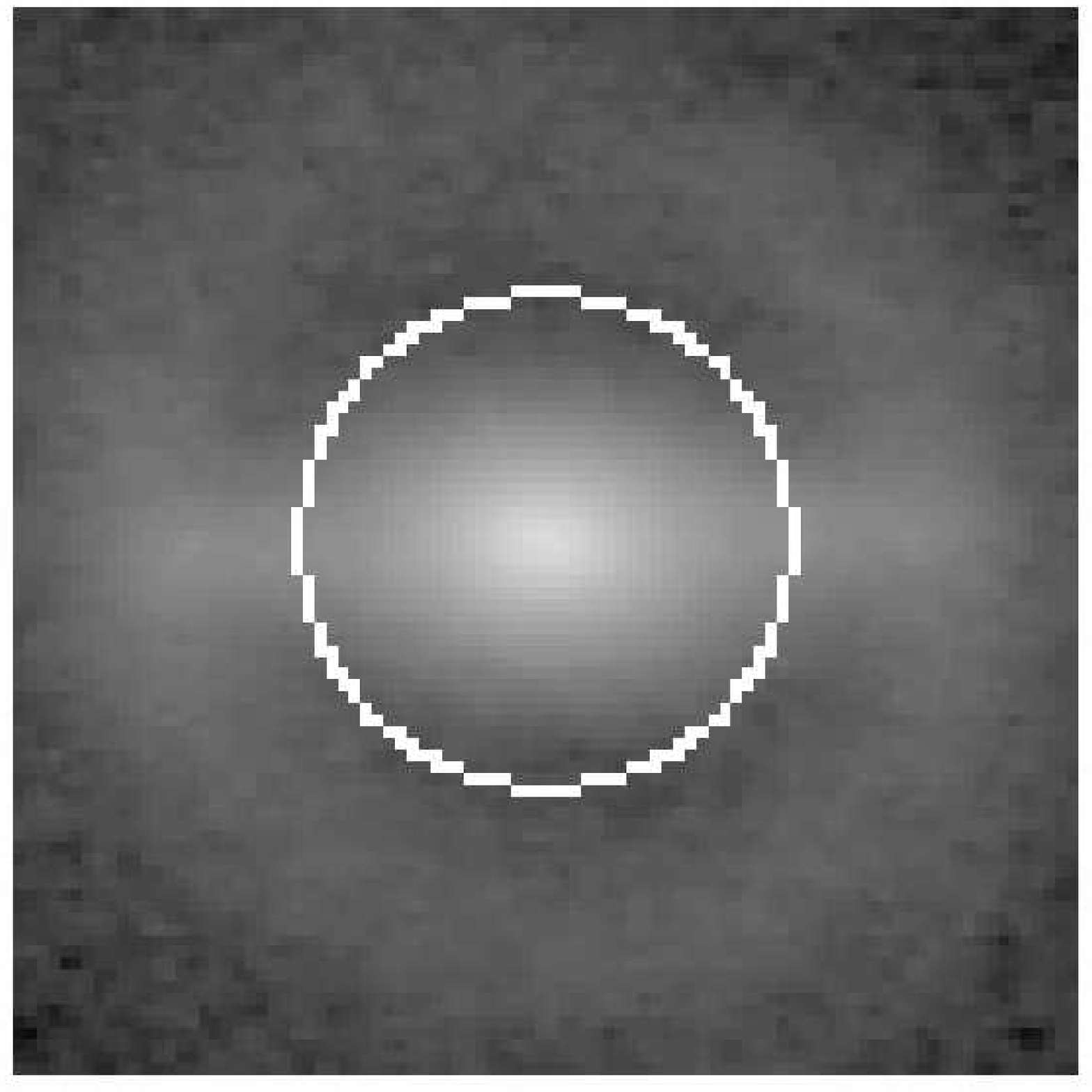}
 \hspace{0.1cm}
 \end{minipage}
 \begin{minipage}[t]{0.68\linewidth}
 \centering
\raisebox{0.5cm}{\includegraphics[width=\textwidth,trim=0 0 0 250,clip]{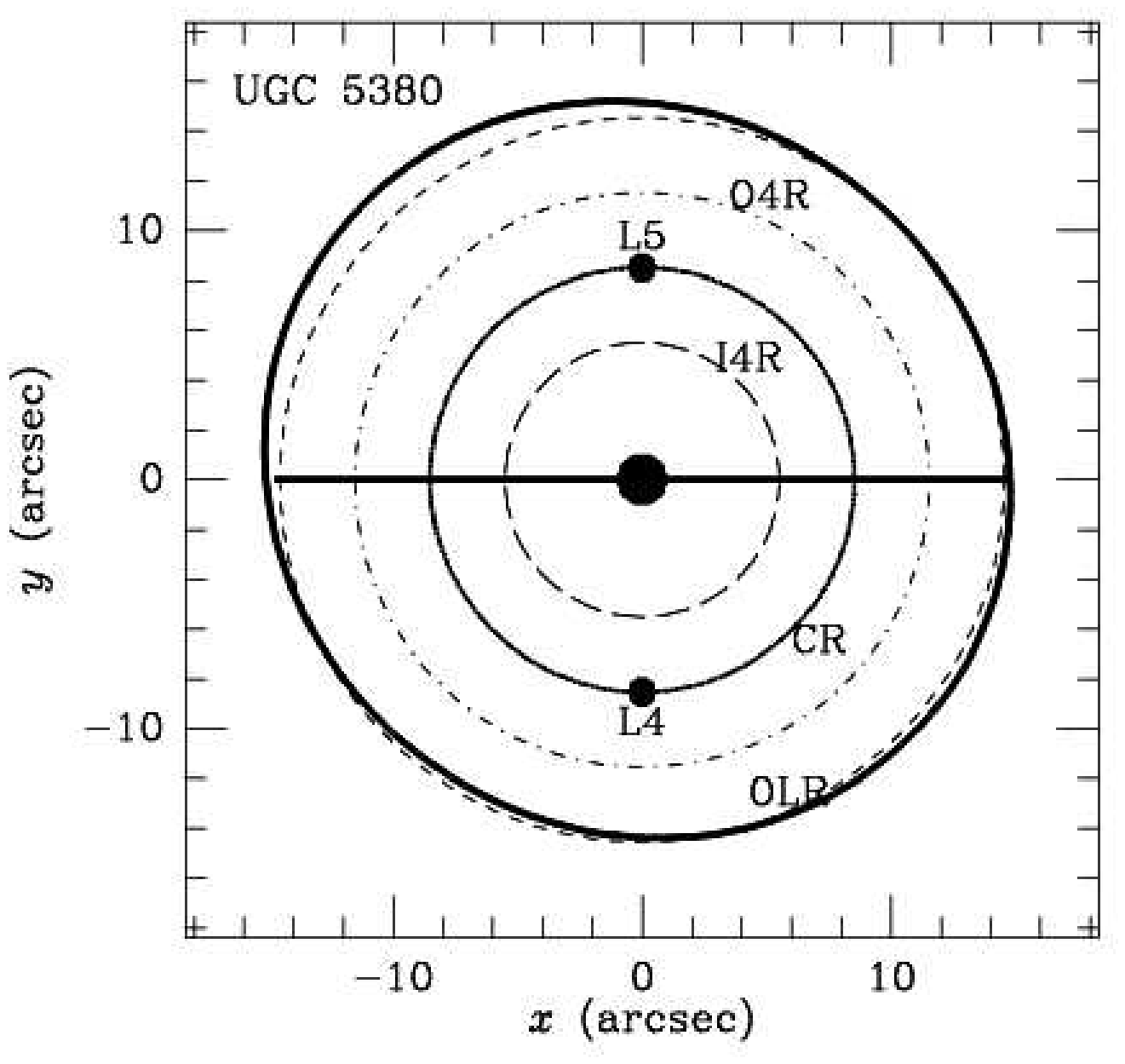}}
 \end{minipage}
 \begin{minipage}[b]{0.45\linewidth}
 \centering
\includegraphics[width=\textwidth]{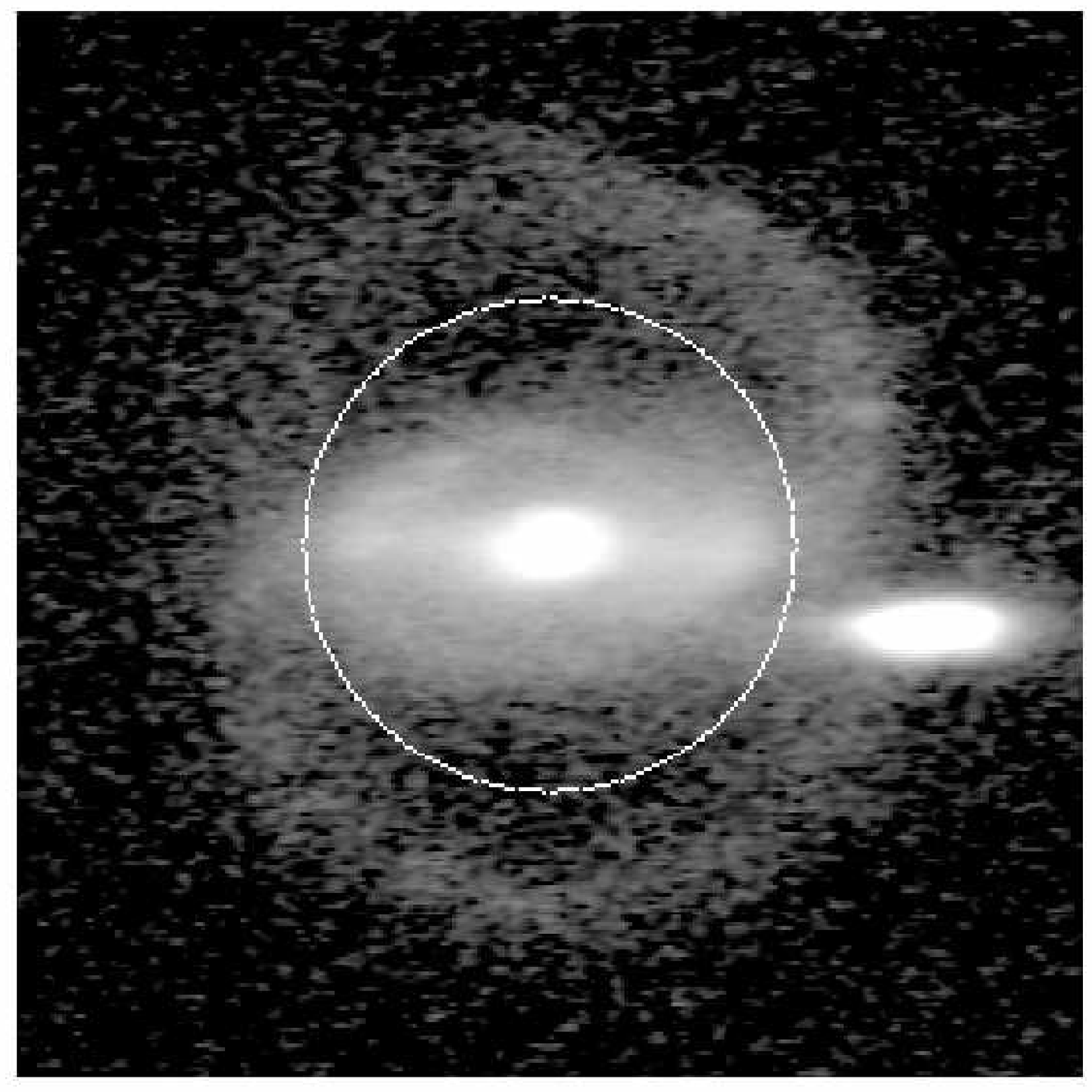}
 \hspace{0.1cm}
 \end{minipage}
 \begin{minipage}[t]{0.68\linewidth}
 \centering
\raisebox{0.5cm}{\includegraphics[width=\textwidth,trim=0 0 0 250,clip]{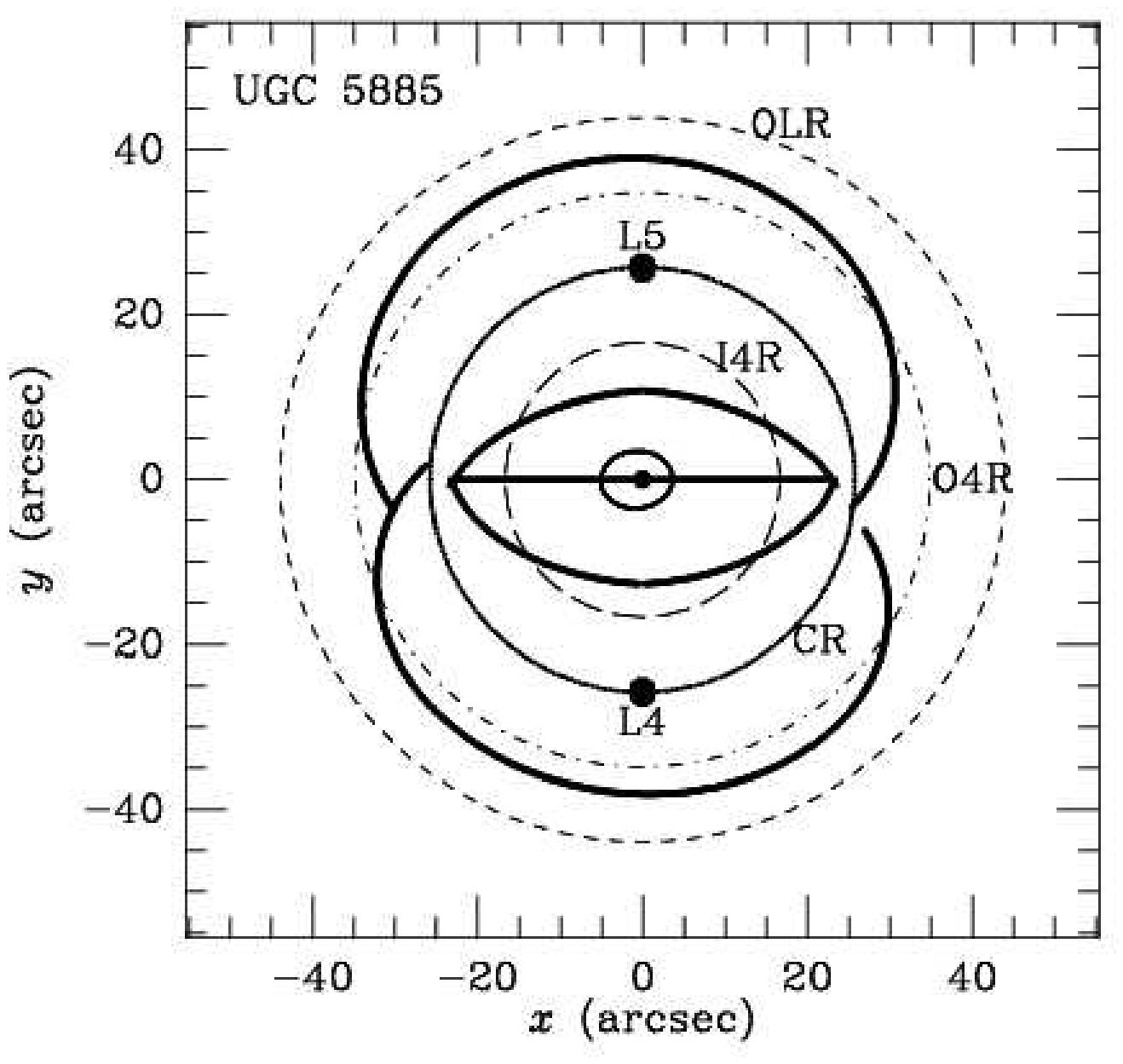}}
 \end{minipage}
\vspace{-1.0truecm}
\caption{(cont.)}
 \end{figure}
 \setcounter{figure}{12}
 \begin{figure}
\vspace{-1.27cm}
 \begin{minipage}[b]{0.45\linewidth}
 \centering
\includegraphics[width=\textwidth]{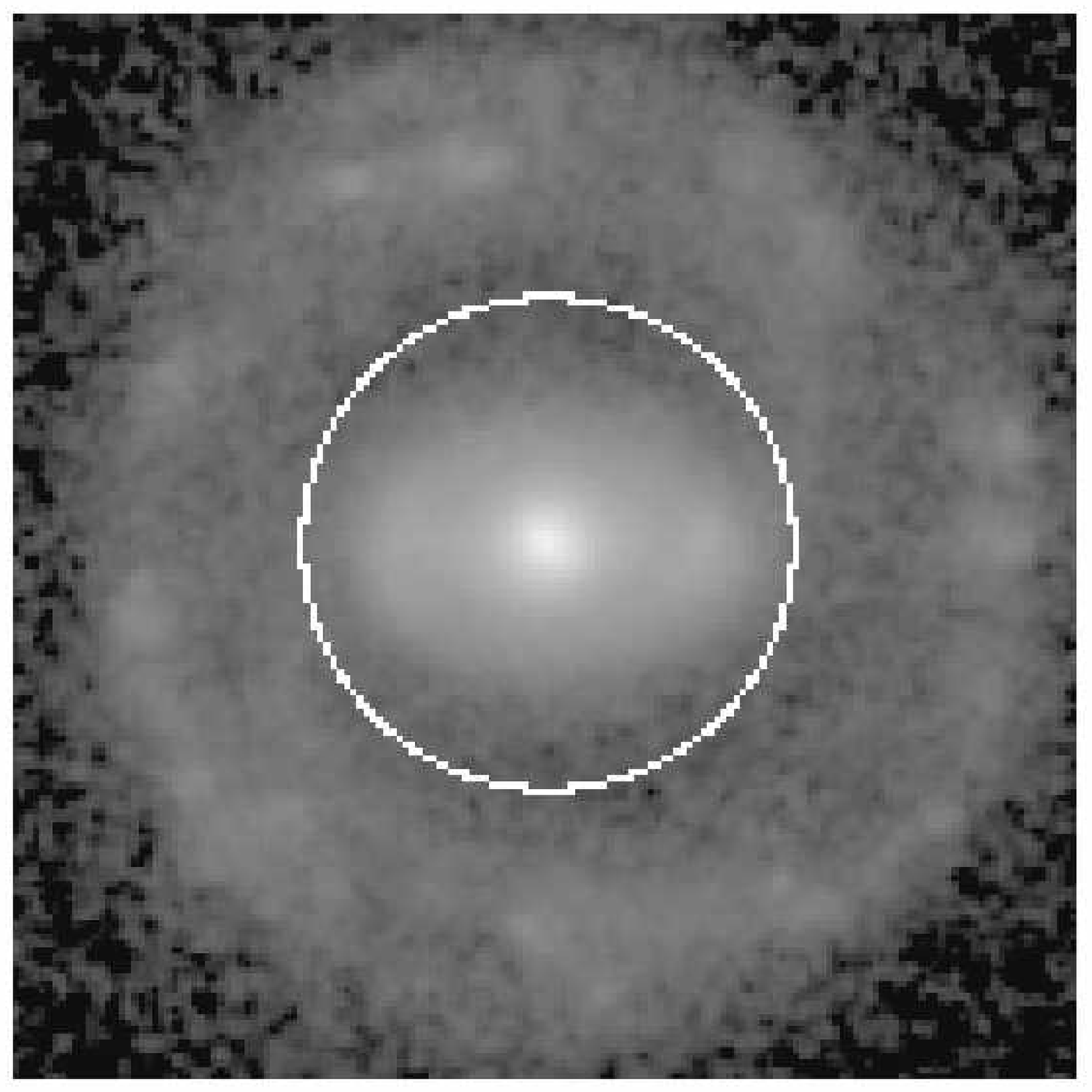}
 \hspace{0.1cm}
 \end{minipage}
 \begin{minipage}[t]{0.68\linewidth}
 \centering
\raisebox{0.5cm}{\includegraphics[width=\textwidth,trim=0 0 0 250,clip]{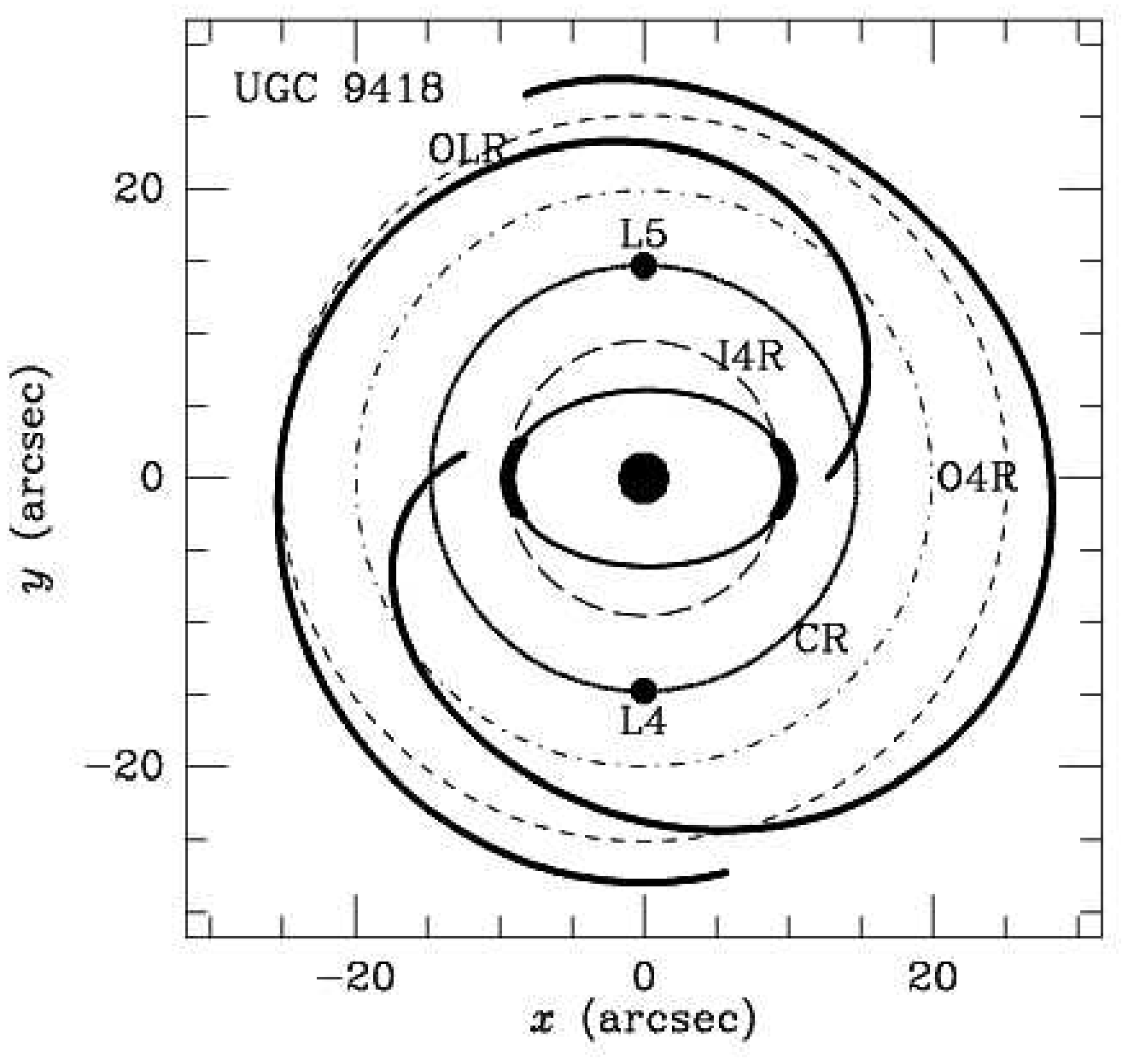}}
 \end{minipage}
 \begin{minipage}[b]{0.45\linewidth}
 \centering
\includegraphics[width=\textwidth]{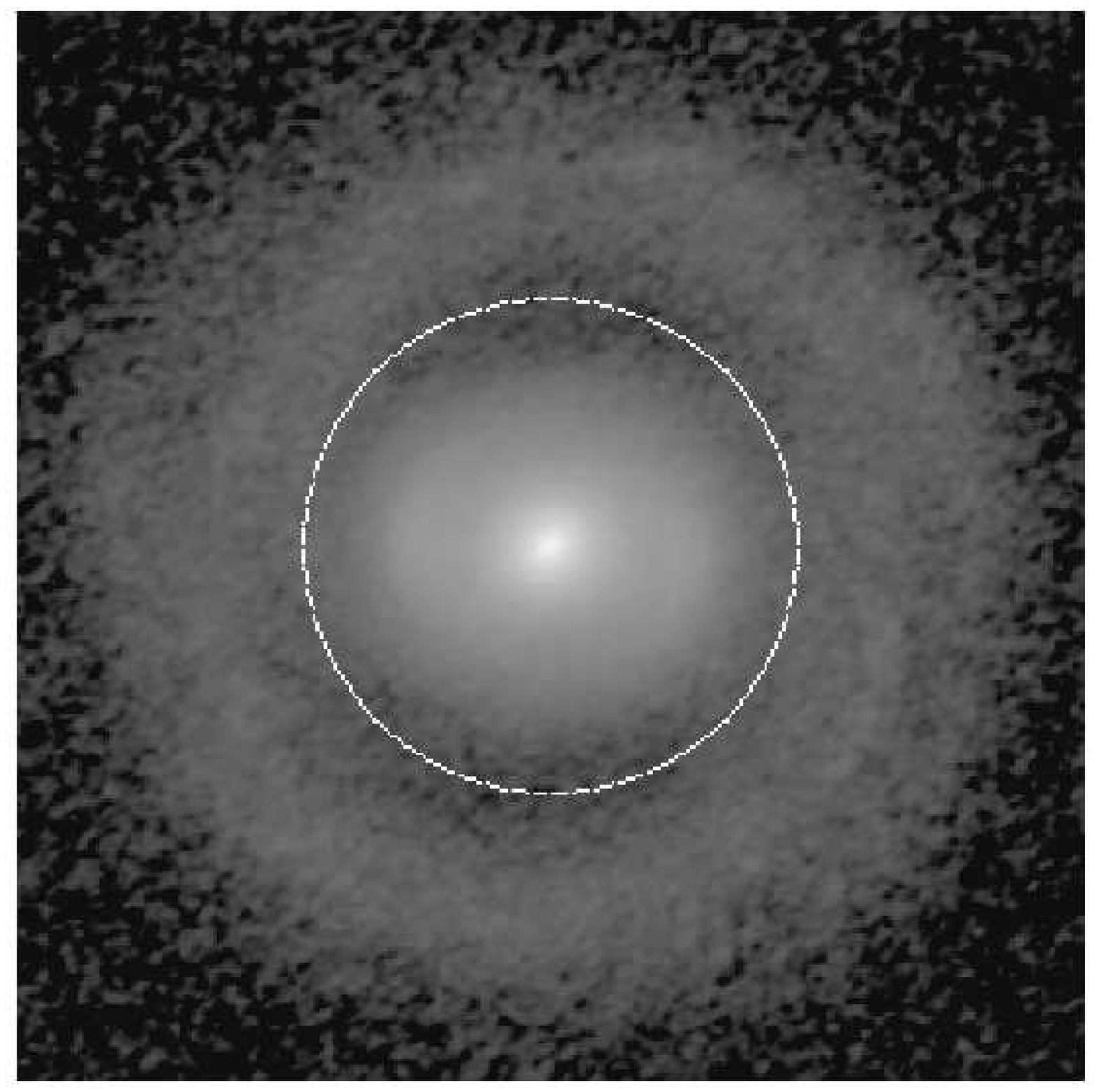}
 \hspace{0.1cm}
 \end{minipage}
 \begin{minipage}[t]{0.68\linewidth}
 \centering
\raisebox{0.5cm}{\includegraphics[width=\textwidth,trim=0 0 0 250,clip]{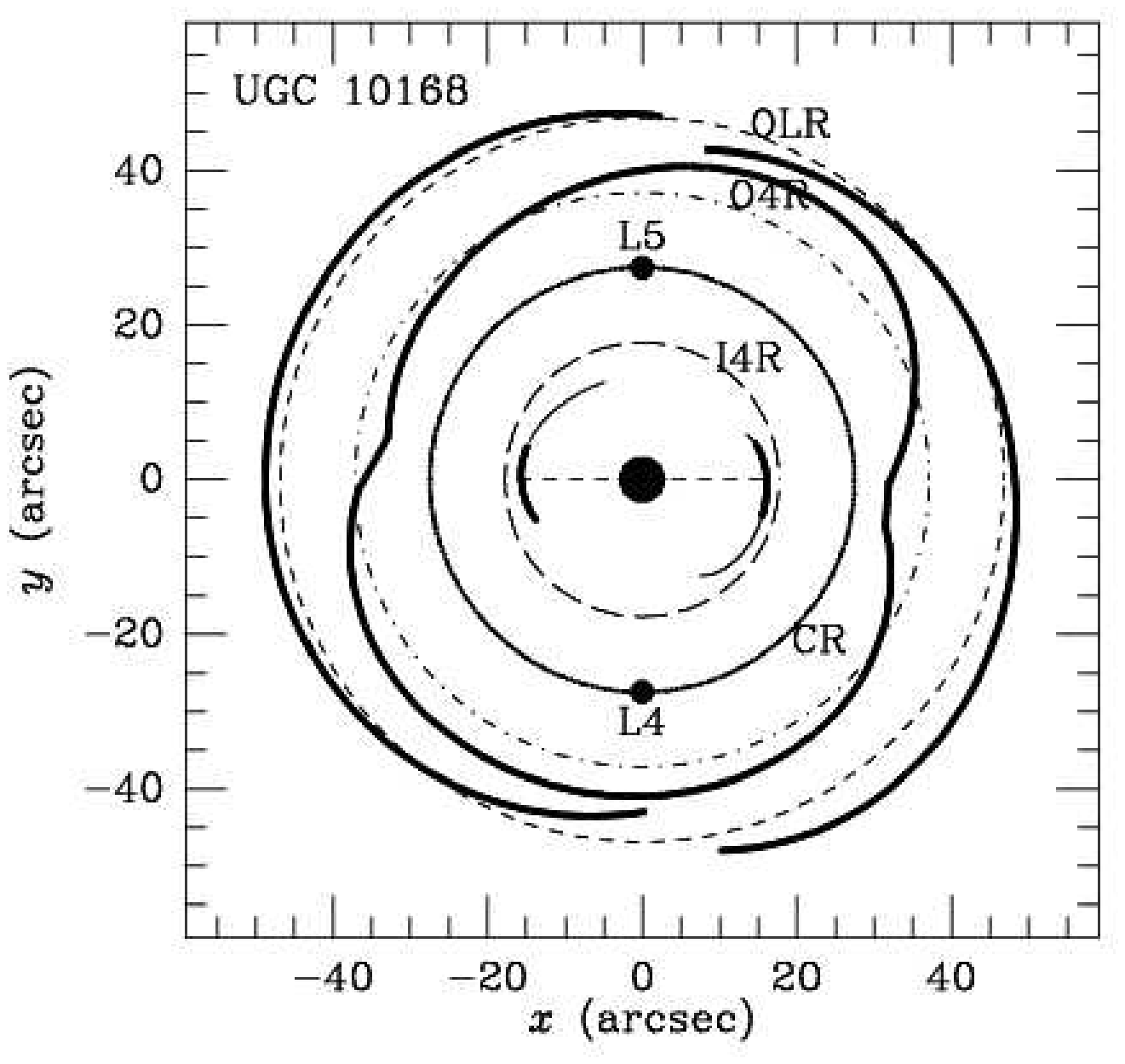}}
 \end{minipage}
 \begin{minipage}[b]{0.45\linewidth}
 \centering
\includegraphics[width=\textwidth]{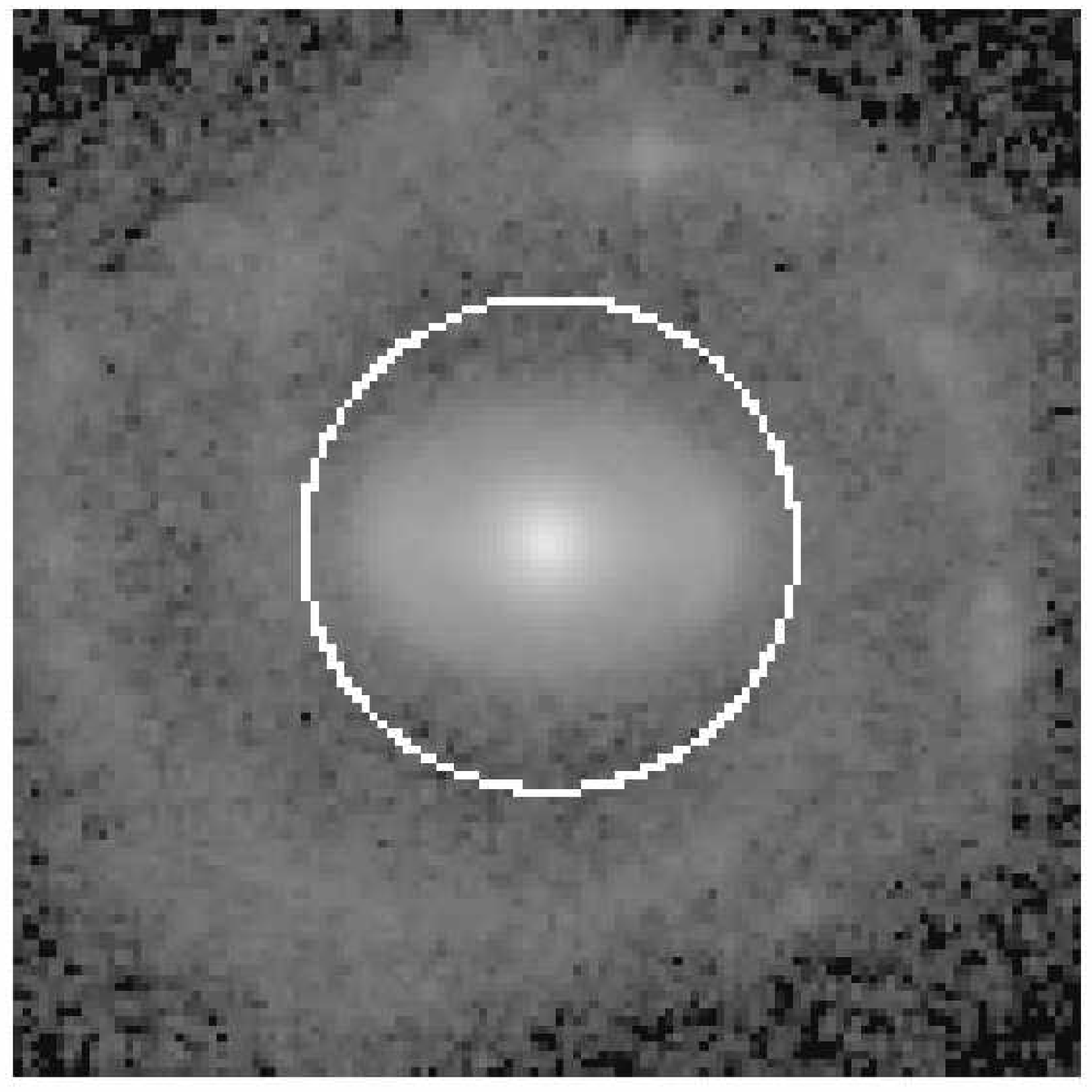}
 \hspace{0.1cm}
 \end{minipage}
 \begin{minipage}[t]{0.68\linewidth}
 \centering
\raisebox{0.5cm}{\includegraphics[width=\textwidth,trim=0 0 0 250,clip]{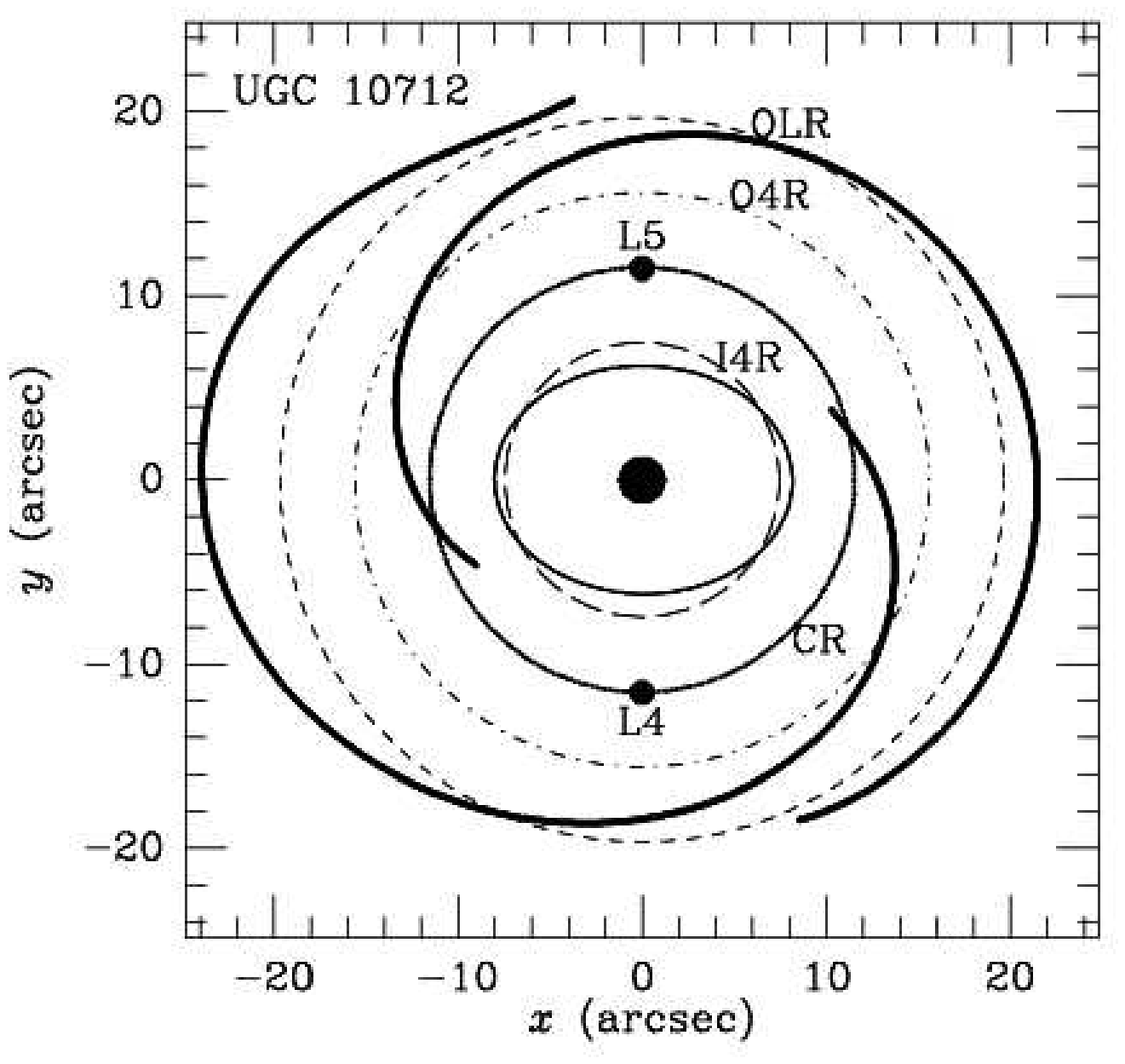}}
 \end{minipage}
 \begin{minipage}[b]{0.45\linewidth}
 \centering
\includegraphics[width=\textwidth]{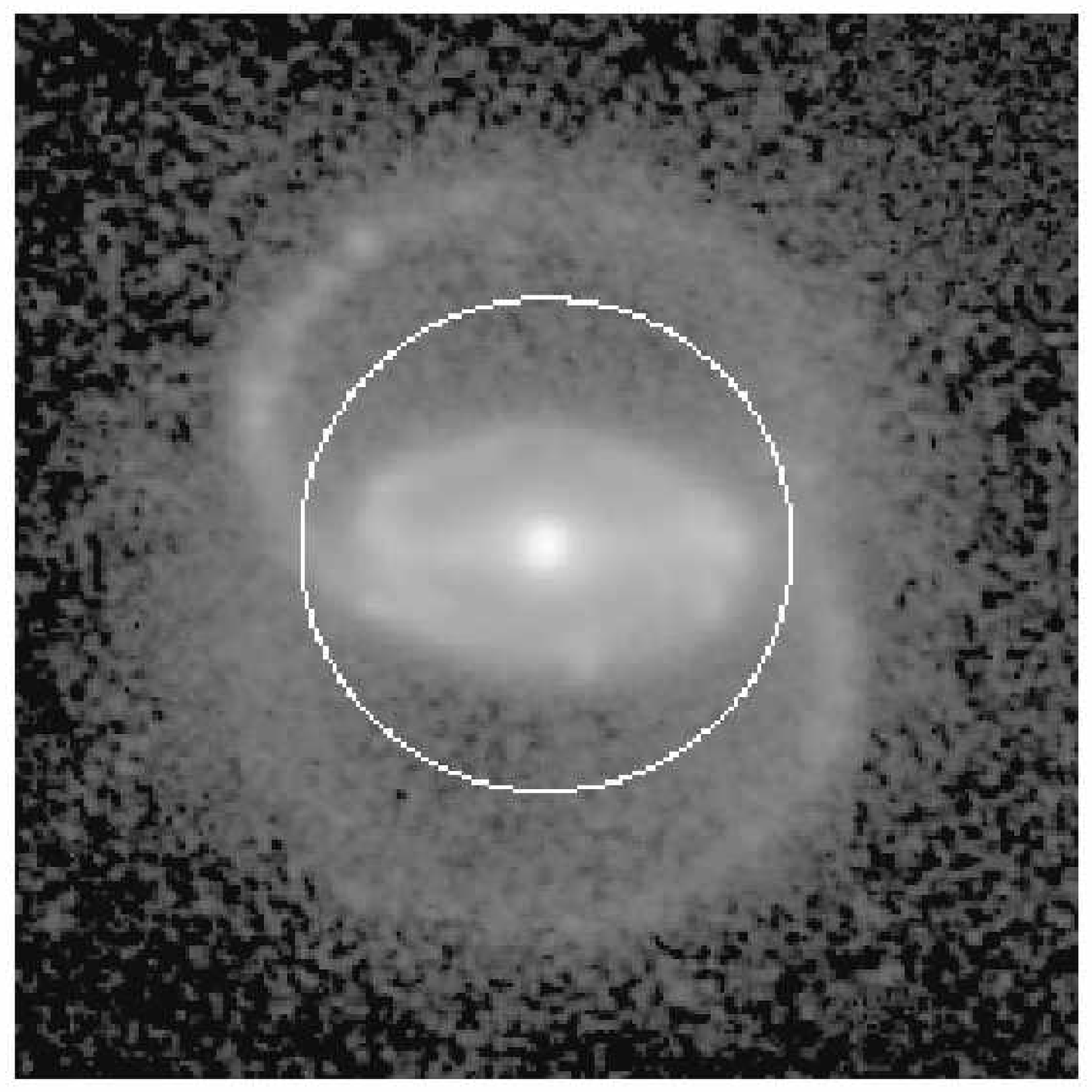}
 \hspace{0.1cm}
 \end{minipage}
 \begin{minipage}[t]{0.68\linewidth}
 \centering
\raisebox{0.5cm}{\includegraphics[width=\textwidth,trim=0 0 0 250,clip]{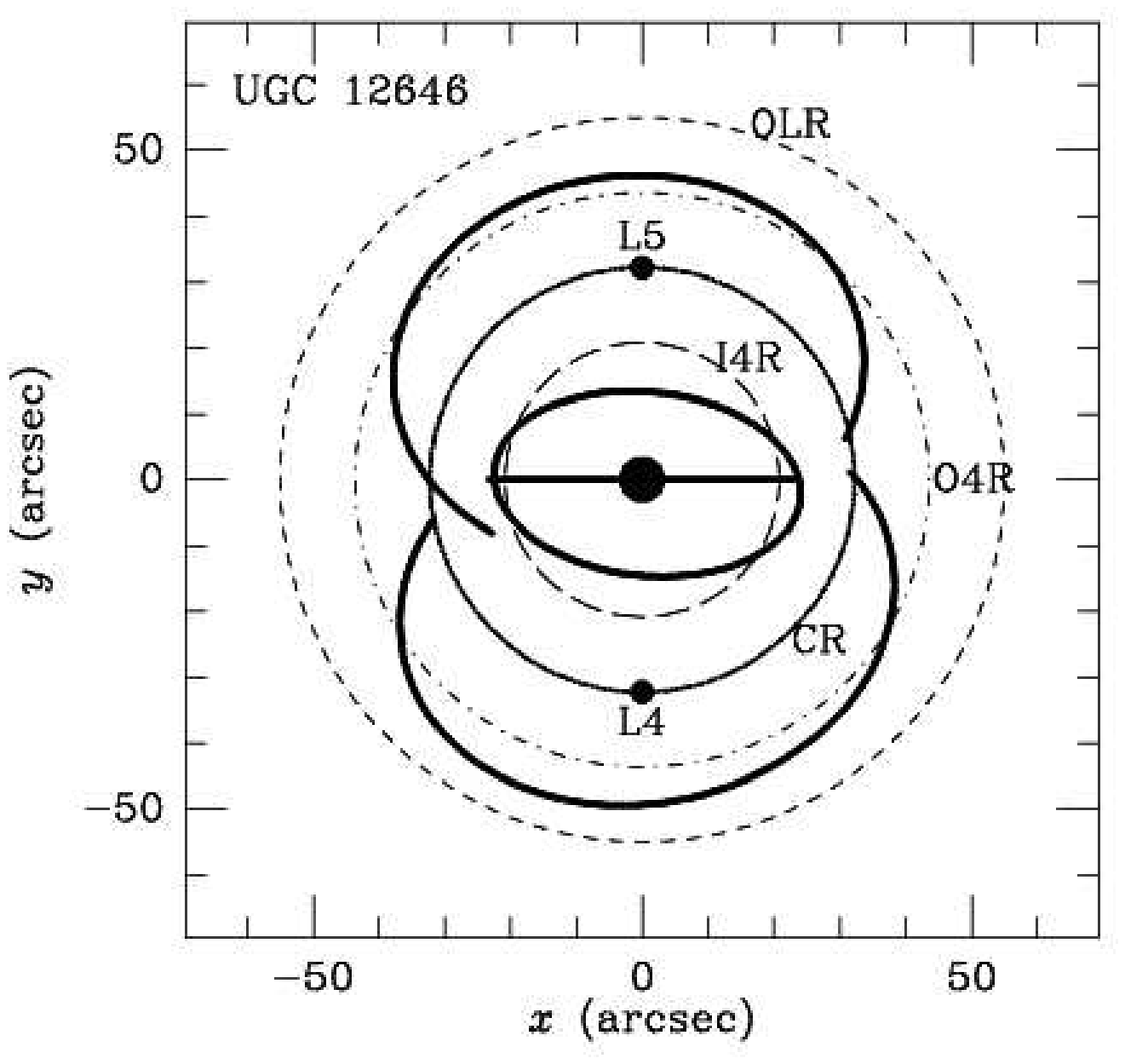}}
 \end{minipage}
\vspace{-1.0truecm}
\caption{Images and Schematics: (left) Deprojected $g$ or $B$-band
image with corotation circle superposed; (right) schematic showing
visually-mapped structures and their relation to corotation and other
labeled resonance radii.  Filled circles perpendicular to the bar axis
on the CR circle are the interpreted locations of the $L_4$, $L_5$
points.  In some cases, filled circles or ovals refer to bar ansae.}
\end{figure}

\noindent
CGCG 13-75 - The galaxy is of type
(\RtwoP)\S_AB(\rpl)ab. The gap method places the \RtwoP\ close to and
slightly outside \rolr, while both the \rpl\ and the bar (defined by
subtle ansae on the \rpl) just reach the \ri4r. The galaxy has some
slight asymmetry, and the colour index maps indicate the gaps on both
sides are red while the \RtwoP\ and the \rpl\ are slightly bluish.

\noindent
CGCG 65-2 - The galaxy is of type (\RoneP)\SB_a(\_rs,bl)ab and has a
clear bar. The bl, which stands for ``barlens," refers to the rounder
central zone (Laurikainen et al. 2011; see also paper 1). The galaxy
has a well-defined \RoneP\ outer pseudoring and clear dark gaps. The
gap CR method places the \RoneP\ entirely inside \rolr\ and very close
to the \ro4r. The schematic shows a four-part structure of the inner
pseudoring, and that the bar and the inner pseudoring extend beyond the
\ri4r\ but not beyond \rcr. In a colour index map, both the \_rs\ and
the \RoneP\ are slightly bluish. In the schematic, the \RoneP\ is not
displayed as a spiral because this aspect is very subtle.

\noindent
CGCG 67-4 - This interesting case has the type
(\RoneL)\SB_a(s,bl,nl)0/a and is noteworthy for its strong blue bar
ansae that do not appear to be merely the blue ends of an extremely
elongated inner ring. The ansae are not arcs but roughly circular
enhancements, and inside these features are both a barlens and a
nuclear lens. The outer structure includes an \RoneL\ with only subtle
gaps. The gap method places \rcr\ fully outside the bar ansae, which in
fact just reach \ri4r. The \RoneL\ also is fully within \rolr, and is
closer to \ro4r.

\noindent
CGCG 73-53 - This early-type galaxy is
classified as (\Rone\Rtwo)\S_AB\ans(l)0$^+$, an example where the outer
morphology is a subtle double ring where the \Rtwo\ may not be a
pseudoring. The gaps are still distinguishable, and the implied 
\rcr\ has the weak bar ending at \ri4r, the inner lens extending 
halfway between \ri4r\ and \rcr, the \Rone\ straddling the \ro4r, and
the \Rtwo\ almost exactly coinciding with \rolr. In spite of the
apparent early type, colour index maps indicate slightly bluer
colours in the (\Rone\Rtwo) compared to the gaps.

\noindent
CGCG 185-14 - This galaxy is
highly-inclined, leading to strong bulge deprojection stretch in the
bright central area (bright vertical oval). The type is
(\Rone\RtwoP)SAB\ans(rs)ab, and is a case where the two outer ring
features are separated by well-defined gaps both parallel and
perpendicular to the bar. If we take the gaps perpendicular to the line
joining the weak bar ansae to set \rcr, then the inner pseudoring and
the ansae extend to near \ri4r, the \Rone\ ring straddles \ro4r, and
the \RtwoP\ lies just beyond \rolr. All of the recognized ring features
show slightly enhanced blue colours in colour index maps.

\noindent
CGCG 263-22 - The morphology of this galaxy is
interesting in the sense that it has what appears to be an
\RoneP\ outer pseudoring surrounding a broad inner pseudoring/oval.
Within this pseudoring/oval, a second large pseudoring is seen. The
bright central region itself is slightly oval nearly perpendicular to
the main inner pseudoring/oval. The classification is
(\RoneP)\S_AB(\_rs,\_rs)a, accounting for all the ring features.
The well-defined gaps along the line perpendicular to the main
inner pseudoring/oval place \rcr\ almost exactly around the ends of the
oval. The \RoneP\ in this case just reaches the \ro4r\, well inside the
predicted \rolr. A $g-i$ colour index map shows that the rim of the large
inner pseudoring/oval is slightly bluer than the inside of the feature.
The smaller \_rs\ shows no clear colour contrast.

\noindent
ESO 325-28 - This CSRG galaxy has a type of (\RtwoP)SB(r)b (BC91) and
is basically face-on. The galaxy has strong, dark gaps in the zones
perpendicular to the obvious bar. The inner ring is extremely oval and
in blue light seems underfilled by the bar. This is less apparent in
the $I$-band. The gap method places \rcr\ such that the bar 
and the major axis of the inner ring slightly overextend
\ri4r.  The two main spiral arms lie largely outside \rcr, with the
prominent but slightly asymmetric \RtwoP\ lying  a little beyond \rolr.
The schematic shows also that the outer arms may begin between
$r_{I4R}$ and $r_{CR}$. A colour index map shown in BC91 reveals
enhanced blue colours in the central region.

\noindent
ESO 365-35 - This galaxy was classified by
BC91 as (\RtwoP)\S_AB(l)0/a. However, a subtle \Rone\
component is also recognizable, and we reclassify the galaxy as
(\Rone\RtwoP)\S_AB(l)0/a. The galaxy has strong gaps and only a very
weak bar. The gap method places the ends of the weak bar as well as the
major axis radius of the bright inner lens near \ri4r. The \Rone\
straddles \ro4r, while the \RtwoP\ just reaches \rolr. A colour index
map in BC91 shows that most of the recent star
formation in ESO 365-35 is confined to the \RtwoP\ outer pseudoring.

\noindent
ESO 426-2 - The strongest features of this galaxy are its inner ring
and bar. In blue light the latter appears to underfill the former, but in
the $I$-band they are matched in size. The type is
(\Rone\RtwoP)SB(r)0/a (BC91). From the well-defined gaps, \rcr\ is
placed such that the bar ends just beyond \ri4r, the inner ring
straddles \ri4r, the \Rone\ straddles \ro4r, and the \RtwoP\ extends
just  a little beyond \rolr. The colour index map of ESO 426-2 in BC91 shows the
inner ring to be a strong, smooth zone of recent star formation, with
some recent star formation also in the \Rone\ ring.

\noindent
ESO 437-33 - The CVRHS type is (\RoneP\RtwoP)SAB(rs,nr)ab. The gaps are
strong in this case, and the gap method places \rcr\ such that the bar
and the inner pseudoring extend to \ri4r, the strong \RoneP\ straddles
\ro4r, and the subtle \RtwoP\ straddles \rolr. 
BC91 show a colour index map of this galaxy that reveals the
\RtwoP\ outer pseudoring better than does the $B$-band image alone.
This map shows considerable star formation in this feature as well as
in the inner pseudoring and the nuclear ring.

\noindent
ESO 437-67 - Classified as (\RoneP)SB(\_rs,nr)\a_b\ by BC91, this
galaxy has well-defined features. The gap method places \rcr\ such that
the bar ends at \ri4r, while the major axis radius of the inner ring
extends a little outside \ri4r. The two arms of the strong and large
\RoneP\ lie mostly outside \rcr\ but do not extend much beyond \ro4r. A
subtle doubling of these arms in the regions just off the inner ring
major axis is more strongly emphasized in the schematic. This
doubling extends from between $r_{I4R}$ and $r_{CR}$ to between
$r_{O4R}$ and $r_{OLR}$. The galaxy strongly resembles NGC 1433.

\noindent
ESO 566-24 - Classified as (R)SB(r)b in the deVA, the galaxy is
remarkable for its strong inner four-armed spiral pattern. This zone is
surrounded by a subtle outer ring that the schematic shows lies just
outside $r_{OLR}$. The galaxy was studied in detail by Rautiainen et
al.  (2004) using numerical simulations. These authors suggested that
the four-armed spiral is linked to the O4R with the strong bar. The
gaps in ESO 566-24 are strong enough to re-evaluate this possible
connection. These give an \rcr\ such that the bar and the inner ring
extend to \ri4r. If the entire four-armed pattern is used to define a
pseudoring, the feature so defined is roughly circular and straddles
\ro4r. This supports the conclusion of Rautiainen et al.  (2004).  The
schematic shows how the four-armed pattern is largely confined between
$r_{I4R}$ and $r_{O4R}$.

\noindent
ESO 575-47 - Classified as (\RoneP)SB(\_rs)ab by BC91, this galaxy has
strong features and well-defined gaps. The gap method places \rcr\ such
that the bar extends to between \ri4r and \rcr, while
the (\_rs) just reaches $r_{CR}$. The \RoneP\ straddles
\ro4r\ closely, and lies well inside \rolr. The inner ring of this
galaxy bears a strong resemblance to that seen in NGC 1433.

\noindent
IC 1223 - The type is (\RP)SB(rs,rs,bl)a. The
most noteworthy feature is the doubled inner pseudoring.  The bar fills
the inner rs. The gap method \rcr\ places both the ends of the
strong apparent bar and the inner rs within \ri4r. The larger rs
overextends \ri4r\ but is entirely within \rcr. Surprisingly, the
outer feature lies just outside \rcr, mostly inside \ro4r. The outer rs
shows enhanced blue colours in a colour index map.

\noindent
IC 1438 - The CVRHS classification is (\Rone\RtwoP)SAB(r,nr)a. The gaps
are well-defined and give \rcr\ such that the inner ring and bar (the
latter shown in the schematic as arc ansae) extend to \ri4r\ and the
slightly asymmetric \Rone\ component straddles \ro4r.  The \RtwoP\ is
mostly outside \rolr. The schematic shows how the subtle dimples in the
R$_1$ component lie near \rcr.  Buta (1995) brought attention to the
\Rone, \RtwoP\ dichotomy in this galaxy, where the \Rone\ component is
most promiment in the $I$-band while the \RtwoP\ component is most
prominent in the $B$-band. A colour index map of the galaxy is shown in
Buta (1995).

\noindent
IC 2473 - This object [type (\RoneP)SB(r,bl)ab]
strongly resembles ESO 437-67, except that the bar appears twisted away
from the strong apparent dimples in the \RoneP\ outer pseudoring. The
displayed image rotates the dimples to the horizontal, instead of the
bar, since the dimples define the axis of a massive inner oval. The gap
method {\it relative to this oval} places both the bar ends and the
major axis of the inner ring just outside \ri4r, while the asymmetric
\RoneP\ is entirely within \rolr, mostly close to \ro4r.  Central star
formation, a circumnuclear dust ring, and enhanced blue colours in the
inner ring and \RoneP\ outer pseudoring are seen in a $g-i$ colour index
map.

The schematic shows how the $L_4$ and $L_5$ points would be twisted
towards the ends of the prominent gaps if the axis perpendicular to the
bar were used to get \rcr. The centre of the gaps is closer to the line
perpendicular to the dimple axis (horizontal as shown). Either
direction gives about the same value of $r_{CR}$.

\noindent
IC 2628 - The type is (\Rone\RtwoP)SAB(l)a. The
dark gaps are clear in this case, and using the gap method, \rcr\ is
placed such that the weak bar extends very close to \ri4r, the \Rone\
straddles \ro4r, and the \RtwoP\ lies just outside \rolr. The colour
index map shows enhanced blue colours in the zone of the \Rone\RtwoP\ 
feature.

\noindent
IC 4214 - This well-studied object (Buta et al 1999; Salo et al. 1999)
has a type of (\Rone)SA(\_rs,nr)a (BC91). It has no clear bar, although
the inner pseudoring defines a massive, bar-like oval. The gap method
places \rcr\ such that the inner pseudoring extends to \ri4r. The
\Rone\ outer ring just reaches the \ro4r, well inside \rolr.

\noindent
MCG 6-32-24 - This galaxy has strong
features.  The type is (\Rone)SAB(r)0/a, with the rings being much more
obvious than the bar. The latter is a short feature inside the inner
ring.  The gaps are well-defined, and the gap method places \rcr\ such
that the bar ends well inside \ri4r, the inner ring extends close to
\ri4r, and the \Rone\ outer ring extends just to \ro4r. A $g-i$ colour index
map shows strong star formation only in the highly-elliptical inner
ring.

\noindent
MCG 7-18-40 - This galaxy, type
(\RtwoP)SB\ans(l)b, shows four well-defined dark zones: two along the
line perpendicular to the bar, and two along the line parallel to the
bar. Each zone is distinct from each other zone, such that an \Rone\
component (dot-long dash curves in schematic) can be recognized. If the gaps
perpendicular to the bar (which has subtle ansae) are used to locate
\rcr, then the inner lens and bar extend close to the \ri4r, while the
\RtwoP\ lies almost exactly at \rolr. The subtle \Rone\ component also
reaches \rolr, which is different from most of the galaxies in the OLR
subclass sample. With \rcr\ defined by the perpendicular gaps, the
parallel gaps are at a larger radius, closer to \ro4r. The broad
short-dash/long-dash oval in the schematic maps the rough midsections
of the four dark areas, with a major axis extending between $r_{O4R}$
and $r_{OLR}$, and the minor axis at $r_{CR}$.

\noindent
NGC 210 - This galaxy is neither a GZ2 nor a CSRG case, but a large
example from the deVA. Our analysis is based on a red continuum image
(Crocker et al. 1996) rather than $BVI$ or $gri$. The CVRHS classification is
(\RoneP\RtwoP)SAB(l,nr)b. The dark gaps are well-defined and give
\rcr\ such that the bar extends very nearly to the \ri4r. The major
axis of the \RoneP\ extends nearly to \rolr, while the \RtwoP\ is close
to but mostly outside \rolr. The gap method places most of the spiral
structure of NGC 210 outside corotation, and the oval lens/bar inside
\ri4r.

\noindent
NGC 1079 -
This galaxy is an excellent, large and mostly symmetric example having
an \Rone\RtwoP\ morphology. Its full type is
(\Rone\RtwoP)SAB\ans(\rpl,bl)a. The dark gaps are clear, and the implied
\rcr\ places both the ends of the bar and the inner pseudoring-lens
just within \ri4r. The bright and virtually complete \Rone\ component
lies mostly at \ro4r\ and shows little or no dimpling. The \RtwoP\ is
predicted to lie outside but still near \rolr. The structure of the
galaxy is remarkably consistent with a single pattern speed model.

\noindent
NGC 1291 - This large example of an early-type outer-ringed galaxy (de
Vaucouleurs 1975) also falls neatly within the outer resonant subclass
morphologies. Its CVRHS type is (\Rone\RtwoP)SAB(l,nb)0/a The
well-defined gaps give \rcr\ such that the primary bar and inner lens
extend close to \ri4r. The \Rone\ component, which in this case is
not dimpled, appears to straddle \ro4r, while the \RtwoP\ component
closely straddles \rolr.  The \Rone\ components of NGC 1079 and 1291
are similar, but the \RtwoP\ component of NGC 1291 is more diffuse than
that in NGC 1079.  The schematic also shows the prominent secondary
bar.

\noindent
NGC 1326 - This galaxy has a strong, well-defined outer ring. The type
is (\Rone)SAB(r,nr)0/a. The gaps are diffuse but give \rcr\ such that
the bar extends to \ri4r, with the inner ring extending a little beyond
\ri4r. The bright \Rone\ outer ring straddles \ro4r, well inside
\rolr.  The schematic also highlights an interesting characteristic of
the R$_1$ outer ring in this galaxy: the ring is brighter in arcs
around the bar/oval major axis and weaker along the bar/oval minor
axis.

\noindent
NGC 1398 - This complex spiral, type
(R$^{\prime}$,R$_1$)SB($\underline{\rm r}$s)ab, has subtle gaps in an
outer \Rone\ ring feature that appears almost ``buried" within the
strong multi-armed spiral pattern. The two strong spiral arms just
outside the inner pseudoring form an unclosed outer pseudoring (the
R$^{\prime}$ in the classification) breaking fom the minor axis zones
of the \Rone\ component. The latter also looks much smoother than the
the former. When the gaps are taken to define \rcr, the strong bar and
inner pseudoring are found to extend almost exactly to \ri4r, while the
\Rone\ extends from \rcr\ to just under \ro4r. The inner two-armed
spiral interpreted as an unclosed \RtwoP\ just reaches \rolr. Much of the
pattern outside \rolr\ is multi-armed and lies at radii
where no bisymmetric resonances would exist if the system has only a
single pattern speed. 

\noindent
NGC 1433 - This classic ringed barred spiral, type
(\RoneP)SB(p,\_rs,nr,nb)ab has well-defined gaps. When these are used
to locate \rcr, the bar is found to extend exactly to \ri4r, while the
inner ring extends a little further and is asymmetric. The distinct
four-fold pattern of the inner ring (including the short arms off the
bar ends) lies almost entirely within \rcr. The detached secondary arcs
located just off the leading quadrants of the strong inner ring appear
to be confined between \rcr\ and \ro4r. These features, which Buta
(1984) called ``plumes," are recognized by the ``p" in the
classification. Also, the prominent \RoneP\ outer pseudoring straddles
\ro4r, well inside \rolr. The identification of the \RoneP\ feature in
NGC 1433 with the O4R was also proposed by Treuthardt et al. (2008).

\noindent
NGC 2665 - The BC91 classification is (\RoneP)\SA_B(rs)a. The gaps are
apparent but not deep. In this case, the gap method places both the
major axis of the inner ring and the bar ends about halfway between
\ri4r\ and \rcr. The bright, slightly asymmetric \RoneP\ extends 
to \ro4r. The schematic emphasizes the symmetric aspects of
this feature.

\noindent
NGC 2766 - This interesting case has a type of (\Rone\RtwoP)\S_AB(r)a.
The system is highly-inclined, however, and the deprojected image shows
significant bulge deprojection stretch. The galaxy is noteworthy for
its extremely cuspy oval inner ring, a feature which also has strong
azimuthal colour variations in the sense that the ring is bluer around
the cusps. The galaxy has no clear bar, although the inner ring is
bar-like. The gap method places \rcr\ such that the inner ring
straddles \ri4r. \rcr\ lies outside the ring ends, while the
\Rone\ straddles \ro4r\ and the \RtwoP\ lies close to but outside
\rolr. The inner ring cusps are close to but inside $r_{CR}$ in this
case.

\noindent
NGC 3081 - This well-studied object (Buta \& Purcell 1998; Buta et al.
2004) is a prototype of the (R$_1$R$_2^{\prime}$) morphology, and has a
CVRHS type of (R$_1$R$_2^{\prime}$)SAB(r,nr,nb)0/a.  The galaxy
strongly resembles models of barred galaxies (Schwarz 1981; 1984a; Buta
\& Purcell 1998). The gaps are subtle and best measured on residual
intensity images.  With this approach, the gap method places \rcr\ such
that the bar ends a little inside \ri4r, while the major axis of the
inner ring extends up to \ri4r. While in past studies both the \Rone\ and
\RtwoP\ components were considered linked to the OLR, the schematic
gives a different interpretation. The \Rone\ component extends to \ro4r\
while the \RtwoP\ extends to \rolr.

\noindent
NGC 3380 - Most of the star formation in this
galaxy peculiarly lies within the bar/oval zone. Little or no star
formation is detectable in the prominent outer ring/lens. The galaxy has a
type of (\RoneL)SAB(l,rs,bl)ab, where the inner pseudoring is one of the
most highly-elongated cases known. The gaps are subtle but definable
in a median difference image. The \rcr\ so implied places the inner ring
and the ends of the bar at \ri4r. The inner ring and bar are also
embedded within an oval lens-like zone that extends to \rcr. The \RoneL\
closely straddles \ro4r, and is well inside \rolr. The barlens (bl)
recognized in the classification is seen best in the $i$-band image.
A $g-i$ colour index map also reveals some central star formation.

\noindent
NGC 4113 - This galaxy, type
(\RoneP)\SA_B(\_rs,nr)ab, has an extremely elongated, cuspy-oval inner
ring. The ring and the bar are slightly misaligned. Although the
\RoneP\ is faint and the gaps are weak, the gap method is still
applicable and yields an \rcr\ such that the bar and inner ring both
extend between \rcr\ and \ri4r. The \RoneP\ extends between \ro4r\ and
\rolr. A $g-i$ colour index map reveals a small nuclear star-forming ring.
The schematic shows that the cusps in the inner ring extend close to but
just inside $r_{CR}$.

\noindent
NGC 4608 - This galaxy, type (RL)SB(rl,bl)0/a, is not the same as any of
the others described so far in the sense that there are no dark gaps
between an inner ring and outer ring. Instead, the dark zones are {\it
inside} an apparent inner ring. If these gaps are identified as the
$L_4$, $L_5$ Lagrangian points, then both the inner ring and bar in
this galaxy lie at \ro4r, outside \rcr\ (i. e., a region forbidden to
support a bar in standard barred galaxy orbit theory). The faint
structure outside the bar would then lie outside \rolr. A $g-i$ colour
index map shows little or no recent star formation.

\noindent
NGC 4736 - This large, nearly face-on object was selected from the deVA
for its strong inner ring of star formation and a very broad and
diffuse outer ring. The galaxy has no clear bar; instead, it has a
broad oval bar-like structure. Lindblad (1960) and Schommer \& Sullivan
(1976) proposed that corotation would likely be in the prominent gap
region between the massive oval (which is lens-like) and the outer
ring.  The CVRHS classification of NGC 4736 is (R)\S_AB(\rpl,\_rs)ab,
where the \rpl\ refers to the boundary of the massive oval. Although
the outer ring is very diffuse and extends well-beyond \rolr, the main
ridge of the feature straddles \ro4r. The best known feature of the
galaxy is the bright inner pseudoring, which is defined by active star
formation and a tightly-wrapped spiral character. This feature is an
analogue of a nuclear ring in a barred spiral and is probably linked
with an ILR. The schematic of this galaxy is the only one showing the
predicted location of $r_{ILR}$, and it highlights how the \_rs\ lies
just inside $r_{ILR}$ for a perfectly flat rotation curve. The
schematic also highlights how the outer ring is, like that in NGC 1326,
weak along the axis perpendicular to the oval.  The gap method gives
\rcr\ such that the massive oval extends to between \ri4r\ and
$r_{CR}$.

\noindent
NGC 4935 - This somewhat asymmetric object, type (\RtwoP)SAB\ans(rs)b,
has a partial \RtwoP\ and does not appear to also have a trace of an
\Rone. The bar is mostly an oval having the same extent as the inner
pseudoring. The gaps are not deep, but median difference images define
the gaps well enough to measure $r_{gp}$. The resulting value of
\rcr\ places the bar/oval/inner pseudoring as extending to \ri4r\ and
the \RtwoP\ as extending close to and outside \rolr. A $g-i$ colour
index map shows that both the inner pseudoring and the \RtwoP\ outer
pseudoring are zones of enhanced blue colours, with no clear central
star formation. 

\noindent
NGC 5132 - This galaxy, type
(\RoneP)SB(\_rs,bl,nr)a, has three well-defined ring features.  A
noteworthy characteristic of the galaxy is the slight misalignment
between the prominent bar and the major axis of the inner ring.  In
this case, the massive oval is used to define the locations of $L_4$
and $L_5$ in the schematic. In spite of the faintness of the \RoneP,
the gap method can still be used and gives \rcr\ such that the bar
extends just past \ri4r\ and the inner ring extends to just inside
\rcr. The \RoneP\ outer pseudoring is asymmetric but appears to be
confined between \ro4r\ and \rolr. A $g-i$ colour index map reveals star
formation in the nuclear and inner rings, but little or no star
formation in the \RoneP.

\noindent
NGC 5211 - This galaxy, type (\RtwoP)SA(rs)b, strongly resembles NGC
4935. The gaps are well-defined, and these give an \rcr\ such that the
inner pseudoring and weak bar (which is clear in the $i$-band but less
so in the $g$-band) extend to just short of \ri4r, and the two main
spiral arms lie mostly outside \rcr. As for NGC 4935, the \RtwoP\ is
slightly asymmetric, and extends to \rolr.  Although the galaxy has
little conventional bar in $g$-band light, the schematic highlights
broad arc ansae in the inner pseudoring.

\noindent
NGC 5335 - As noted in section 5.2, this galaxy, classified as
SB(\_rs,bl)ab, is the darkest (r) dark-spacer in the sample.  As for
NGC 4608, the gap method places the ends of the bar and the bright
inner ring close to \ro4r. Much of the multi-armed outer spiral pattern
would be located outside \rolr, in an area where no bisymmetric
resonances would exist, if the object has only a single pattern speed.

\noindent
NGC 5370 - This galaxy, type (\RoneL)SB\ans(\_rs,bl)0/a, is unusual in
having a strong bar and a relatively circular inner ring. The features
are well-defined but the gaps are subtle. Using the median difference
approach, the method gives \rcr\ such that the inner ring and bar extend
close to but just inside \ri4r. The \RoneL, which is non-dimpled, is
confined mostly between \rcr\ and \ro4r. A $g-i$ colour index map shows
that the inner pseudoring and the \RoneL\ have slightly enhanced blue
colours, with no central star formation. The bar of NGC 5370 is also
unusual in having a diamond shape and small ansae.

\noindent
NGC 5686 - This interesting case, type SB\ans(r,bl)0\^o, has an
excellent ansae-type bar and a faint, circular inner ring. The ring is
a subtle (r) dark-spacer, and if the weak gaps are taken to define
\rcr, then the inner ring and bar extend to \ro4r, with no further
structure in a diffuse outer disk. A $g-i$ colour index map reveals no
active star formation in the galaxy.

\noindent
NGC 5701  - This excellent large example, type
(\Rone\RtwoP)SB(\rpl,bl)a, is nearly face-on and has strong dark gaps
perpendicular to the bar. The gap method places \rcr\ such that the bar
and \rpl\ extend to just inside \ri4r.  The \Rone\ component straddles
\ro4r, while the \RtwoP\ is either at or slightly larger than \rolr. A
$g-i$ colour index map reveals star formation mainly in the \RtwoP, with
a trace of bluer colours also on one side of the \Rone. No star
formation is associated with the \rpl\ or in the central region.

\noindent
NGC 6782 - This CSRG galaxy, type
(\RoneP)SB(r,nr,nb)a, has an excellent cuspy-oval inner ring, a strong
nuclear ring and secondary bar, and a conspicuous knotty \RoneP. The
gaps are well-defined, and give an \rcr\ which places the bar ends and
the major axis cusp points of the inner ring at \ri4r.  The
\RoneP\ outer pseudoring straddles \ro4r, and seems well inside \rolr.
NGC 6782 is a well-studied object and has been the subject of
sophisticated numerical simulations (Lin et al. 2008).

\noindent
NGC 7098 - This exceptional large object, of
type (\Rone\RtwoP)SAB\ans(\_rs,nb)ab, has strong dark, banana-shaped gaps
between its inner and outer ring features. The gap method places \rcr\
such the bar extends to \ri4r, while the inner ring extends between
\ri4r\ and \rcr. The \Rone\ component, which is slightly dimpled,
extends only to \ro4r\ while the \RtwoP\ component straddles \rolr.  A
$B-I$ colour index map (Buta 1995) shows that all three ring features in
the galaxy have slightly enhanced blue colours and are zones of recent
star formation, while the prominent linear ansae have no colour
contrast. The nb in the classification refers to an inner secondary
oval. The schematic emphasizes the strong, nearly linear bar ansae.

\noindent
PGC 54897 - This galaxy, type
(\RoneP)\SA_B\ans(\rpl)ab, has an excellent \RoneP\ with subtle gaps.
The gap method places \rcr\ such that the inner pseudoring/bar-oval
extend to \ri4r\ while the \RoneP\ straddles \ro4r, well-inside \rolr.
A $g-i$ colour index map reveals active star formation in both the \rpl\ and
the \RoneP, in addition to some nuclear star formation (a partial
nuclear ring).

\noindent
PGC 1857116 - A small and distant object with strong features. The type
is (\RoneP)SAB(\_rs)a. The gaps are well-defined and give \rcr\ such
that the inner pseudoring/bar-oval (which is a subtle ansae-type)
extends to \ri4r. The prominent \RoneP\ extends to between
\ro4r\ and \rolr.

\noindent
PGC 2570478 - A small object, type (R$^{\prime}$)SB(s)a. The bar is
boxy, and the gaps are well-defined.  These give an \rcr\ such that the
bar extends exactly to \ri4r\ while the \RP\ straddles \ro4r. The
bar is surrounded by a broad diffuse zone that cannot be characterized
as an inner lens. A $g-i$ colour index map reveals some enhanced star
formation in the \RP.

\noindent
UGC 4596 - As described already in section 5, this galaxy, type
(\Rone\RtwoP)SA(rr)b, has strong dark gaps in spite of the lack of an
apparent bar. Instead the inner zone is a bar-like oval rimmed by two
inner rings. The \Rone\RtwoP\ morphology is well-defined, and the gap
method has the \Rone\ component straddling \ro4r\ while the
\RtwoP\ lies just outside \rolr. The doubled inner ring straddles
\ri4r.

\noindent
UGC 4771 - This excellent case, type
(\RtwoP)SAB\ans(l)b, strongly resembles NGC 4935 and 5211. The gaps are
well-defined and give \rcr\ such that the bar extends to \ri4r\ with
the inner lens straddling \ri4r. The \RtwoP\ straddles \rolr.
A $g-i$ colour index map indicates recent star formation in the \RtwoP\ and
possibly also near the rim on one side of the inner lens.

\noindent
UGC 5380 - A typical (r) dark-spacer, type
(L)SB(r,bl)0/a. The gap method applied to this galaxy places \rcr\ such
that the bar and inner ring straddle \rolr. The outer disk has a sharp
outer edge, but if the bar in UGC 5380 is in fact a superfast type, then
the outer lens would not have a resonant origin. A $g-i$ colour index map
reveals little or no recent star formation in the galaxy.

\noindent
UGC 5885 - This galaxy, type
(\RoneP)SAB\ans(\_rs,nr)ab, is highly-inclined although the centre is
likely to be a pseudobulge since it shows little deprojection stretch.
The \RoneP\ is an excellent example with strong dimpling. The gap
method places \rcr\ such that the inner ring and bar extend close to
but slightly inside corotation, one of the very few cases in the sample
like this. The cusps of the inner ring lie close to $r_{CR}$.  The
\RoneP\ mostly straddles \ro4r, although its major axis is halfway
between \ro4r\ and \rolr. All of the ring features in the galaxy show
some trace of recent star formation.

\noindent
UGC 9418 - This galaxy, type (\RtwoP)SAB\ans(\rpl)b,
has a strong \RtwoP\ outer pseudoring. The gaps are well-defined, and
there is little evidence for an \Rone\ component. The gap method places
\rcr\ such that the bar and inner pseudoring-lens straddle \ri4r, while
the \RtwoP\ lies close to but mostly outside \rolr. The colour index map
shows strong star formation in the \RtwoP\ and subtle blue colours along
the rim of the inner pseudoring-lens. There is little or no central star
formation.

\noindent
UGC 10168 - This galaxy has strong features and
is an excellent example of an \Rone\RtwoP\ morphology. The full
classification is (\Rone\RtwoP)SAB\ans(\rpl)a. The gaps are strong, and
give \rcr\ such that the bar and inner pseudoring-lens extend close to
but just inside \ri4r. The \Rone\ component straddles \ro4r, while the
\RtwoP\ straddles \rolr. A $g-i$ colour index map shows enhanced blue colours
in the \Rone\RtwoP, but no star formation in the inner pseudoring-lens.

\noindent
UGC 10712 - This excellent example, type
(\Rone\RtwoP)SAB(rl)b, has nearly circular isophotes at the faintest
detectable light level and has been assumed to be face-on. The gaps
are well-defined and give an \rcr\ such that the inner ring-lens and
the bar straddle \ri4r. The major axis of the \Rone\ component and the
minor axis of the \RtwoP\ component lie almost exactly at \rolr\ in the
manner of the two families of OLR orbits shown by Schwarz (1981).
A $g-i$ colour index map shows strong enhanced blue colours in the
\Rone\RtwoP\ feature and in the inner ring-lens. There appears to be little
or no central star formation. The schematic excludes the R$_1$ component,
which is very faint in the $g$-band.

\noindent
UGC 12646 - This galaxy was selected from Appendix 3 of RC3 as an
exceptional example of its type, (\RoneP)SB(r)ab. The obviously
dimpled \RoneP\ outer pseudoring is well-defined. The gaps are clear
and give an \rcr\ such that the bar extends to \ri4r, while the
highly-elongated inner ring extends just outside \ri4r.  The
\RoneP\ mostly straddles \ro4r, but on one side nearly reaches \rolr.
The \RoneP\ is nevertheless entirely confined within \rolr.

\begin{figure}
\includegraphics[width=\columnwidth,bb=14 14 600 600]{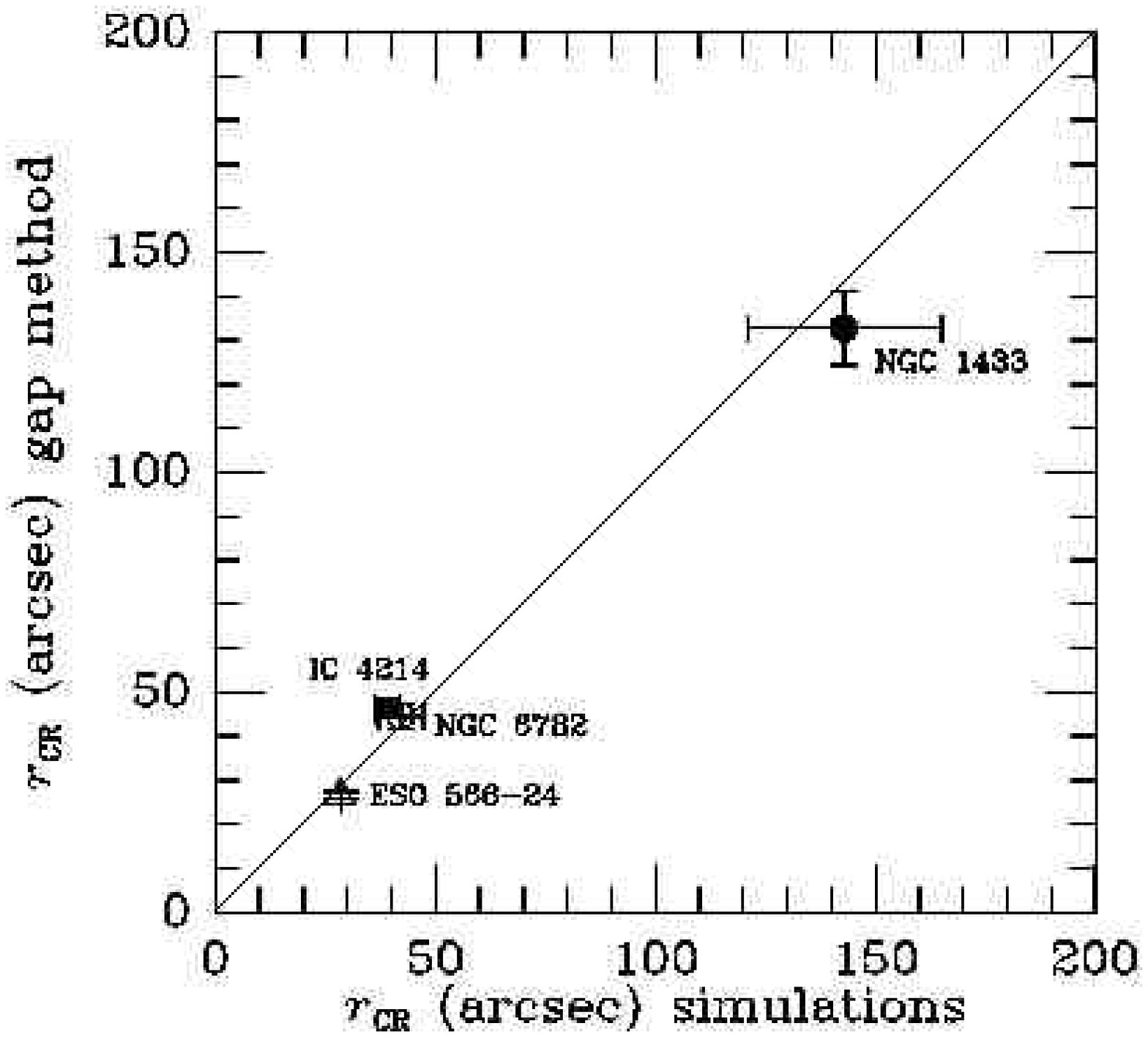}
\caption{Graph of CR radii estimated from numerical simulations
of four galaxies in our sample, and the CR radii estimated in this paper
from the gap method.}
\label{fig:cresrads}
\end{figure}

{\it Comparison of CR radii} - Numerical simulations have been used to
model the structure of four of our sample galaxies: NGC 1433, NGC 6782,
IC 4214, and ESO 566$-$24.  A comparison between estimates of $r_{CR}$
from these models and from the gap method is shown in
Figure~\ref{fig:cresrads}.  For NGC 1433, two estimates of $r_{CR}$ are
available from simulations:  121$^{\prime\prime}$ from Buta et al.
(2001) and 165$^{\prime\prime}$ from Treuthardt et al. (2008); these
have been averaged and the average deviation between them is plotted in
Figure~\ref{fig:cresrads} as the horizontal error bar. The same has
been done for NGC 6782, which has $r_{CR}$ estimates of
47\rlap{.}$^{\prime\prime}$2 from Lin et al.  (2008) and
37\rlap{.}$^{\prime\prime}$1 from Rautiainen et al. (2008).  The former
paper explicitly used a flat approximation to an observed rotation
curve, while latter used a rotation curve inferred from an $H$-band
image augmented by a halo contribution, which made the rotation curve
also nearly flat. In models similar to those used by Rautiainen et al.
(2008), Salo et al. (1999) estimated $r_{CR}$
=39$^{\prime\prime}$$\pm$3$^{\prime\prime}$ for IC 4214, while
Rautiainen et al. (2004) estimated $r_{CR}$ =
28\rlap{.}$^{\prime\prime}$5 for ESO 566-24. In these cases, the
authors used models of observed nearly flat rotation curves for their
analysis. Figure~\ref{fig:cresrads} shows that the agreement between
$r_{CR}$ from the gap method and the corresponding values from
simulations is fairly good.

\section{Analysis}

Our analysis is based on the data in
Tables~\ref{tab:resrads}--~\ref{tab:inner-features}. In these tables,
the mean of any column is the average of the values in that column and
$\sigma_1$ is the standard deviation. The mean error is
$\sigma_1/\sqrt{N}$, where $N$ is the number of values used for the
means.

\subsection{(rR) dark-spacers}

\subsubsection{Bar Extent}

Let us begin by examining where the gap method places the bars in the
sample galaxies relative to the main resonances. For the 50 (rR) cases
in the combined GZ2/CSRG samples, Table~\ref{tab:resrads} shows that
$<a_{bar}/r_{CR}>$ = 0.65$\pm$0.02 with a standard deviation of 0.12. Thus,
the bars are predicted to lie on average well inside the CR radius. If
we consider instead the ratio of the bar radius to the radius of the
inner 4:1 resonance, the sample gives $<a_{bar}/r_{I4R}>$ = 1.01$\pm$0.03
with  a standard deviation of 0.19. The gap method therefore has the
bars in the (rR) dark-spacer sample ending very close to the inner 4:1
resonance, on average.

The reciprocal of $<a_{bar}/r_{CR}>$ is the parameter $\cal{R}$, which was
studied by Debattista \& Sellwood (2000) who suggested that $\cal{R}$ =
1.4 divides bars into ``fast" and ``slow" types. For the sample
analyzed here, $<\cal{R}>$ = 1.58 $\pm$ 0.04 with a standard deviation
$\sigma_1$ = 0.28. Thus, the gap method as applied in this paper favours
the apparent bars of OLR subclass galaxies to be relatively slow. This
differs from other pattern speed studies (e.g., Rautiainen et al.
2008; Buta \& Zhang 2009) which favour fast bars over slow bars. The
only way for the gap method to give lower values of $\cal{R}$ is for
the bar radius to be defined as something larger than it appears to be,
e.g., by including the full extent of the oval the bars are typically
imbedded within.

Contopoulos (1980) used orbit theory of barred galaxies to conclude
that a bar cannot extend beyond its CR radius. This is because the main
periodic orbits supporting the elongated shape of the bar belong to the
$x_1$ family, all members of which lie inside the CR radius. Outside
the CR radius, the main periodic orbits are aligned perpendicular to
the bar (Contopoulos \& Grosbol 1989).  Also, Contopoulos (1985)
concluded that one of the main limiting factors for the extent of bars
is stochasticity of orbits of the $x_1$ family. He concluded that bars
could extend all the way to the CR radius until limited by such
effects.

Contopoulos (1988) examined the inner 4:1 resonance in more detail but
did not specifically link the ends of bars with the location of this
resonance, only noting that 4:1 resonant orbits become increasingly
more elongated as the resonance is approached. However, Contopoulos
(1985) concluded that strong spirals end at the inner 4:1 resonance,
with only weaker spirals extending to the CR radius.

\subsubsection{Rings}

For the OLR subclass rings, we have a few theoretical values to compare with
our observed ratios:

\noindent
1. R$_1$ outer rings are aligned perpendicular to the bar and have
$a/r_{OLR}$$\approx$ 1. Such rings should be mostly within
$r_{OLR}$ (Schwarz 1981).

\noindent
2. R$_2^{\prime}$ outer pseudorings are aligned parallel to the bar and
have $b/r_{OLR}$$\approx$ 1. Such rings should be mostly
outside $r_{OLR}$ (Schwarz 1981).

\noindent
3. The average radius of an inner ring $\approx$ $r_{I4R}$, but $a/r_{CR}$
$\leq$ 1 (Simkin, Su \& Schwarz 1980). Inner rings should be aligned
parallel to the bar.

Table~\ref{tab:R1gals} summarizes the results for 36 R$_1$ and R$_1^{\prime}$ rings.
These are found to have $<a/r_{OLR}>$ = 0.85 $\pm$ 0.01, with a
standard deviation of 0.05, significantly less than the expected value
of 1.0. The mean ratios in columns 6 and 7 show that R$_1$ and
R$_1^{\prime}$ rings may be linked instead to the outer 4:1 resonance:
the values ``straddle" the outer 4:1 resonance in the sense that the
major axis radii of these features are  9\% larger than $r_{O4R}$ and
the minor axis radii are 10\% smaller than $r_{O4R}$, on average.  If we
set $r^*$ = $\sqrt{ab}$, the geometric mean radius, then
$<r^*/r_{O4R}>$ = 0.99. Consistent with Schwarz (1981), R$_1$ and
R$_1^{\prime}$ rings lie entirely within the OLR radius, but their
association with the outer 4:1 resonance was not predicted by Schwarz.

\begin{table*}
\centering
\caption{R$_1$ and R$_1^{\prime}$ Outer Rings and Pseudorings.
Col. 1: galaxy name; col. 2: feature type; cols. 3,4:
deprojected major and minor axis ring dimensions, respectively, in
arcseconds; cols. 5-7: ratios of ring dimensions to outer 4:1 and outer
Lindblad resonance radii; col. 8: deprojected ring minor to major axis ratio; col. 9:
deprojected relative angle between the bar and the major axis of the ring.
}
\label{tab:R1gals}
\begin{tabular}{llrrrrrrr}
\hline
Name & Feature & $a$ & $b$ & $a/r_{OLR}$ & $a/r_{O4R}$ & $b/r_{O4R}$ & $q_0$ & $|\theta_{Bf}|$ \\
 1 & 2 & 3 & 4 & 5 & 6 & 7 & 8 & 9  \\
\hline
CGCG 8-10     & \Rone              &     32.5 &     25.7 &     0.87 &     1.10 &     0.87 &     0.79 &     84.4 \\
CGCG 65-2     & \Rone              &     17.6 &     14.5 &     0.80 &     1.01 &     0.83 &     0.82 &     89.6 \\
CGCG 73-53    & \Rone              &     15.6 &     11.7 &     0.88 &     1.11 &     0.83 &     0.75 &     89.3 \\
CGCG 185-14   & \Rone              &     16.9 &     13.7 &     0.87 &     1.09 &     0.88 &     0.81 &     82.1 \\
CGCG 263-22   & \RoneP             &     29.1 &     27.3 &     0.78 &     0.98 &     0.92 &     0.94 &     57.4 \\
ESO 365-35    & \Rone              &     31.2 &     23.8 &     0.88 &     1.11 &     0.84 &     0.76 &     85.3 \\
ESO 426-2     & \Rone              &     34.8 &     28.5 &     0.92 &     1.16 &     0.95 &     0.82 &     86.6 \\
ESO 437-33    & \RoneP             &     33.0 &     26.7 &     0.82 &     1.03 &     0.83 &     0.81 &     76.6 \\
ESO 437-67    & \RoneP             &     78.4 &     72.3 &     0.85 &     1.07 &     0.99 &     0.92 &     83.6 \\
ESO 575-47    & \RoneP             &     49.0 &     43.1 &     0.78 &     0.98 &     0.87 &     0.88 &     89.1 \\
IC 1438       & \Rone              &     50.9 &     41.6 &     0.89 &     1.13 &     0.92 &     0.82 &     82.7 \\
IC 2473       & \RoneP             &     52.2 &     47.8 &     0.87 &     1.09 &     1.00 &     0.92 &     65.8 \\
IC 2628       & \Rone              &     15.1 &     11.9 &     0.88 &     1.11 &     0.87 &     0.79 &     77.8 \\
IC 4214       & \Rone              &     63.4 &     54.8 &     0.80 &     1.01 &     0.88 &     0.86 &     88.5 \\
MCG 6-32-24   & \Rone              &     16.6 &     13.7 &     0.83 &     1.05 &     0.87 &     0.83 &     69.6 \\
MCG 7-18-40   & \RoneP             &     13.0 &      8.5 &     1.01 &     1.27 &     0.83 &     0.65 &     79.2 \\
NGC  210      & \Rone              &     93.6 &     73.6 &     0.91 &     1.15 &     0.90 &     0.79 &     88.3 \\
NGC 1079      & \Rone              &    114.1 &    104.5 &     0.80 &     1.01 &     0.92 &     0.92 &     86.8 \\
NGC 1291      & \Rone              &    249.3 &    218.9 &     0.83 &     1.05 &     0.92 &     0.88 &     84.4 \\
NGC 1326      & \Rone              &     86.5 &     76.4 &     0.81 &     1.02 &     0.90 &     0.88 &     79.0 \\
NGC 1398      & \Rone              &    110.6 &     90.1 &     0.73 &     0.92 &     0.75 &     0.81 &     83.2 \\
NGC 1433      & \RoneP             &    190.8 &    167.7 &     0.84 &     1.06 &     0.93 &     0.88 &     78.9 \\
NGC 2665      & \Rone              &     58.3 &     51.2 &     0.80 &     1.01 &     0.89 &     0.88 &     81.6 \\
NGC 2766      & \Rone              &     30.6 &     28.4 &     0.84 &     1.06 &     0.99 &     0.93 &     81.0 \\
NGC 3081      & \Rone              &     79.1 &     66.1 &     0.78 &     0.99 &     0.83 &     0.84 &     84.4 \\
NGC 4113      & \RoneP             &     35.1 &     28.6 &     0.85 &     1.08 &     0.88 &     0.82 &     80.4 \\
NGC 5132      & \Rone              &     61.4 &     56.4 &     0.86 &     1.08 &     0.99 &     0.92 &     69.4 \\
NGC 5701      & \Rone              &    104.6 &     88.0 &     0.88 &     1.11 &     0.93 &     0.84 &     87.6 \\
NGC 6782      & \RoneP             &     65.1 &     49.0 &     0.86 &     1.08 &     0.81 &     0.75 &     78.4 \\
NGC 7098      & \Rone              &    113.4 &    103.7 &     0.77 &     0.96 &     0.88 &     0.91 &     88.2 \\
PGC 54897     & \RoneP             &     24.4 &     20.1 &     0.81 &     1.02 &     0.84 &     0.82 &     83.1 \\
PGC 1857116   & \RoneP             &     23.8 &     20.7 &     0.91 &     1.14 &     1.00 &     0.87 &     73.2 \\
UGC 4596      & \Rone              &     25.4 &     20.7 &     0.85 &     1.07 &     0.88 &     0.82 &     84.7 \\
UGC 5885      & \RoneP             &     40.0 &     33.1 &     0.91 &     1.15 &     0.95 &     0.83 &     87.0 \\
UGC 10168     & \Rone              &     42.6 &     34.9 &     0.91 &     1.15 &     0.94 &     0.82 &     68.3 \\
UGC 12646     & \RoneP             &     48.6 &     41.8 &     0.88 &     1.11 &     0.96 &     0.86 &     89.6 \\
             &                     &          &          &          &          &          &          &          \\
mean         &                     &    ..... &    ..... &     0.85 &     1.07 &     0.90 &     0.84 &     81.3 \\ 
mean error   &                     &    ..... &    ..... &     0.01 &     0.01 &     0.01 &     0.01 &      1.3 \\
$\sigma_1$   &                     &    ..... &    ..... &     0.05 &     0.07 &     0.06 &     0.06 &      7.6 \\
$N$          &                     &    ..... &    ..... &       36 &       36 &       36 &       36        & 36 \\
\hline
\end{tabular}
\end{table*}

Table~\ref{tab:R2gals} summarizes the results for 25 R$_2^{\prime}$ outer
pseudorings.  The mean ratios are $<a/r_{OLR}>$ = 1.08$\pm$ 0.01 and
$<b/r_{OLR}>$ = 1.00$\pm$0.01, with standard deviations of 0.07 and
0.06, respectively. Consistent with Schwarz (1981), the minor axis
radii of these features are thus located closest to the OLR radius, and
the features on the whole are outside the OLR radius.

\begin{table*}
\centering
\caption{R$_2$ and R$_2^{\prime}$ Outer Rings and Pseudorings.
Col. 1: galaxy name; col. 2: feature type; cols.
3,4:  deprojected major and minor axis ring dimensions, respectively,
in arcseconds; cols. 5-6: ratios of ring dimensions to outer Lindblad
resonance radii; col. 7: deprojected ring minor to major axis ratio; col. 8:
deprojected relative angle between the bar and the major axis of the ring.
}
\label{tab:R2gals}
\begin{tabular}{llrrrrrr}
\hline
Name & Feature & $a$ & $b$ & $a/r_{OLR}$ & $b/r_{OLR}$ & $q_0$ & $|\theta_{Bf}|$ \\
 1 & 2 & 3 & 4 & 5 & 6 & 7 & 8 \\ 
\hline
CGCG 8-10     & \RtwoP             &     36.4 &     33.7 &     0.97 &     0.90 &     0.93 &     11.4 \\
CGCG 13-75    & \RtwoP             &     23.8 &     22.0 &     1.10 &     1.02 &     0.93 &     73.8 \\
CGCG 73-53    & \Rtwo              &     17.0 &     16.7 &     0.96 &     0.95 &     0.98 &      3.7 \\
CGCG 185-14   & \RtwoP             &     20.6 &     20.0 &     1.06 &     1.03 &     0.97 &     18.0 \\
ESO 325-28    & \RtwoP             &     33.1 &     32.2 &     1.10 &     1.07 &     0.97 &     26.8 \\
ESO 365-35    & \RtwoP             &     36.7 &     32.4 &     1.04 &     0.91 &     0.88 &     11.0 \\
ESO 426-2     & \RtwoP             &     43.4 &     40.9 &     1.15 &     1.08 &     0.94 &     19.1 \\
ESO 437-33    & \RtwoP             &     41.2 &     38.3 &     1.02 &     0.95 &     0.93 &     10.1 \\
IC 1438       & \RtwoP             &     65.2 &     61.3 &     1.15 &     1.08 &     0.94 &     30.8 \\
IC 2628       & \RtwoP             &     18.9 &     17.1 &     1.10 &     1.00 &     0.91 &     29.3 \\
MCG 7-18-40   & \RtwoP             &     13.8 &     13.1 &     1.07 &     1.01 &     0.94 &     47.6 \\
NGC 210       & \RtwoP             &    128.4 &    112.9 &     1.25 &     1.10 &     0.88 &      5.3 \\
NGC 1079      & \RtwoP             &    162.9 &    151.5 &     1.14 &     1.06 &     0.93 &     10.1 \\
NGC 1291      & \RtwoP             &    306.6 &    299.6 &     1.03 &     1.00 &     0.98 &     88.8 \\
NGC 2766      & \RtwoP             &     40.7 &     36.5 &     1.12 &     1.01 &     0.90 &      0.4 \\
NGC 3081      & \RtwoP             &    104.1 &     96.4 &     1.03 &     0.96 &     0.93 &      3.2 \\
NGC 4935      & \RtwoP             &     29.6 &     27.6 &     1.02 &     0.95 &     0.93 &      0.1 \\
NGC 5211      & \RtwoP             &     59.4 &     51.5 &     1.00 &     0.86 &     0.87 &     11.2 \\
NGC 5701      & \RtwoP             &    125.5 &    120.4 &     1.06 &     1.01 &     0.96 &     18.9 \\
NGC 7098      & \RtwoP             &    153.6 &    144.0 &     1.04 &     0.97 &     0.94 &     58.1 \\
UGC 4596      & \RtwoP             &     33.6 &     31.5 &     1.13 &     1.06 &     0.94 &     41.3 \\
UGC 4771      & \RtwoP             &     31.6 &     27.2 &     1.02 &     0.88 &     0.86 &     33.9 \\
UGC 9418      & \RtwoP             &     27.5 &     26.4 &     1.10 &     1.05 &     0.96 &     79.7 \\
UGC 10168     & \RtwoP             &     49.9 &     46.7 &     1.07 &     1.00 &     0.94 &     39.8 \\
UGC 10712     & \RtwoP             &     23.0 &     19.5 &     1.17 &     0.99 &     0.85 &      7.4 \\
             &                     &          &          &          &          &          &          \\
means        &                     &    ..... &    ......&     1.08 &     1.00 &     0.93 &     27.2 \\
mean error   &                     &    ..... &    ......&     0.01 &     0.01 &     0.01 &      5.1 \\
$\sigma_1$   &                     &    ..... &    ......&     0.07 &     0.06 &     0.04 &     25.4 \\
$N$          &                     &    ..... &    ......&       25 &       25 &       25 &       25 \\
\hline
\end{tabular}
\end{table*}

Table~\ref{tab:inner-features} summarizes the results for the various
inner features in the (rR) dark-spacer galaxies. The table lists 50
features, but only 47 galaxies are represented since three have double
inner features. Cols. 5 and 6 show that inner features straddle the
inner 4:1 resonance, an interpretation consistent with Schwarz (1984a)
and Simkin, Su, \& Schwarz (1980). The geometric mean radius of these
features has on average $<r^*/r_{I4R}>$ = 0.91.  However, the standard
deviations of the ratios, 0.19 for $a/r_{I4R}$ and 0.15 for
$b/r_{I4R}$, reflects in part the wide range of intrinsic shapes of
inner features.

\begin{table*}
\centering
\caption{Inner Rings, Pseudorings, Lenses, Ring-Lenses, and Pseudoring-Lenses.
Col. 1: galaxy name; col. 2: type of feature (r=inner
ring, rs=inner pseudoring, rl=inner ring-lens, r$^{\prime}$l = inner
pseudoring-lens, r$_i$=inner-most of two inner rings, r$_o$=outer-most
of two inner rings; l=inner lens); cols. 3,4: deprojected
major and minor axis ring dimensions, respectively, in arcseconds;
cols. 5-6:  ratios of ring dimensions to the inner 4:1 resonance
radius; col. 7:  feature minor-to-major axis ratio; col. 8: relative
deprojected angle between the bar major axis and the major axis
position angle of the feature.
}
\label{tab:inner-features}
\begin{tabular}{llrrrrrr}
\hline
 Name & Feature & $a$ & $b$ & $a/r_{I4R}$ & $b/r_{I4R}$ & $q_0$ & $|\theta_{Bf}|$ \\
 1 & 2 & 3 & 4 & 5 & 6 & 7 & 8  \\ 
\hline
CGCG 8-10     & r               &    13.0 &    10.1 &    0.92 &    0.71 &    0.77 &     8.7 \\
CGCG 13-75    & \rpl            &     8.0 &     6.1 &    0.98 &    0.75 &    0.77 &     0.0 \\
CGCG 65-2     & rs              &    11.5 &     6.9 &    1.38 &    0.83 &    0.60 &     5.8 \\
CGCG 73-53    & l               &     8.1 &     6.8 &    1.22 &    1.01 &    0.83 &     3.4 \\
CGCG 185-14   & rs              &     7.8 &     6.8 &    1.05 &    0.93 &    0.88 &    11.1 \\
CGCG 263-22   & rs$_i$          &    14.5 &    11.1 &    1.02 &    0.78 &    0.77 &    87.0 \\
CGCG 263-22   & rs$_o$          &    20.8 &    14.5 &    1.47 &    1.02 &    0.70 &     0.8 \\
ESO 325-28    & r               &    13.3 &     8.3 &    1.17 &    0.73 &    0.62 &     3.7 \\
ESO 365-35    & l               &    14.0 &    11.9 &    1.04 &    0.89 &    0.85 &     2.5 \\
ESO 426-2     & r               &    17.4 &    11.4 &    1.22 &    0.80 &    0.65 &     3.1 \\
ESO 437-33    & rs              &    14.2 &     9.7 &    0.93 &    0.64 &    0.68 &     0.7 \\
ESO 437-67    & rs              &    41.6 &    26.0 &    1.19 &    0.75 &    0.62 &     2.3 \\
ESO 566-24    & r               &    18.6 &    14.4 &    1.08 &    0.84 &    0.78 &     6.7 \\
ESO 575-47    & rs              &    29.3 &    19.6 &    1.23 &    0.83 &    0.67 &     2.5 \\
IC 1223       & rs$_i$          &    13.7 &    11.5 &    0.94 &    0.79 &    0.84 &    16.2 \\
IC 1223       & rs$_o$          &    18.9 &    16.0 &    1.30 &    1.10 &    0.84 &     5.7 \\
IC 1438       & r               &    22.3 &    15.0 &    1.03 &    0.69 &    0.67 &     1.3 \\
IC 2473       & r               &    26.7 &    17.8 &    1.17 &    0.78 &    0.67 &     9.4 \\
IC 4214       & rs              &    30.6 &    19.9 &    1.03 &    0.67 &    0.65 &     4.2 \\
MCG 6-32-24   & r               &     8.4 &     5.3 &    1.11 &    0.70 &    0.63 &     0.0 \\
MCG 7-18-40   & rl              &     5.3 &     4.2 &    1.09 &    0.86 &    0.79 &     1.9 \\
NGC  210      & rs              &    32.4 &    20.4 &    0.83 &    0.52 &    0.63 &     1.4 \\
NGC 1079      & \rpl            &    45.4 &    37.9 &    0.84 &    0.70 &    0.84 &    14.2 \\
NGC 1291      & l               &   140.1 &   111.4 &    1.24 &    0.98 &    0.80 &     3.5 \\
NGC 1326      & r               &    43.3 &    30.7 &    1.07 &    0.76 &    0.71 &     1.8 \\
NGC 1398      & rs              &    58.7 &    54.4 &    1.02 &    0.95 &    0.93 &    18.4 \\
NGC 1433      & r               &   103.8 &    71.0 &    1.21 &    0.83 &    0.68 &     1.9 \\
NGC 2665      & rs              &    34.9 &    18.0 &    1.27 &    0.66 &    0.52 &     4.9 \\
NGC 2766      & r               &    16.5 &    12.3 &    1.20 &    0.89 &    0.74 &     2.9 \\
NGC 3081      & r               &    35.3 &    24.8 &    0.92 &    0.65 &    0.70 &     1.1 \\
NGC 3380      & rs              &    20.4 &    11.1 &    1.00 &    0.54 &    0.54 &     2.6 \\
NGC 4113      & rs              &    20.9 &     9.9 &    1.34 &    0.64 &    0.48 &     8.3 \\
NGC 4736      & \_rs            &    50.6 &    42.7 &    0.32 &    0.27 &    0.84 &    49.9 \\
NGC 4935      & rs              &    11.2 &     7.7 &    1.02 &    0.70 &    0.69 &    13.3 \\
NGC 5132      & rs              &    38.4 &    24.1 &    1.41 &    0.89 &    0.63 &     0.7 \\
NGC 5211      & rs              &    18.9 &    15.2 &    0.84 &    0.67 &    0.80 &     2.1 \\
NGC 5370      & r               &    19.3 &    17.9 &    0.94 &    0.87 &    0.93 &    16.2 \\
NGC 5701      & l               &    42.5 &    37.9 &    0.94 &    0.84 &    0.89 &     5.4 \\
NGC 6782      & r               &    27.6 &    17.6 &    0.96 &    0.61 &    0.64 &     4.1 \\
NGC 7098      & \_rs            &    69.1 &    58.6 &    1.23 &    1.04 &    0.85 &     1.9 \\
PGC 54897     & \rpl            &    10.1 &     6.0 &    0.88 &    0.52 &    0.59 &     6.0 \\
PGC 1857116   & rs              &    11.5 &     6.5 &    1.16 &    0.65 &    0.56 &     2.4 \\
UGC 4596      & r$_i$           &     9.4 &     8.5 &    0.83 &    0.75 &    0.90 &     9.6 \\
UGC 4596      & r$_o$           &    14.1 &    11.2 &    1.25 &    0.99 &    0.80 &     2.6 \\
UGC 4771      & l               &    12.9 &     8.7 &    1.10 &    0.74 &    0.67 &     3.1 \\
UGC 5885      & r               &    21.8 &    11.3 &    1.31 &    0.68 &    0.52 &     0.0 \\
UGC 9418      & rl              &     9.8 &     6.1 &    1.03 &    0.64 &    0.62 &     0.3 \\
UGC 10168     & \rpl            &    16.0 &    13.1 &    0.90 &    0.74 &    0.82 &    11.3 \\
UGC 10712     & rl              &     8.1 &     6.2 &    1.09 &    0.83 &    0.76 &     0.7 \\
UGC 12646     & r               &    23.4 &    13.9 &    1.12 &    0.67 &    0.60 &     5.5 \\
              &                 &         &          &        &         &         &         \\
means         &                 &   ..... &    ..... &   1.08 &    0.77 &    0.72 &     7.5 \\
mean error    &                 &   ..... &    ..... &   0.03 &    0.02 &    0.02 &     2.0 \\
$\sigma_1$    &                 &   ..... &    ..... &   0.19 &    0.15 &    0.12 &    13.9 \\
$N$           &                 &   ..... &    ..... &     50 &      50 &      50 &      50 \\
\hline
\end{tabular}
\end{table*}

Two other points of comparison between the (rR) dark-spacer sample
galaxies and previous studies are the average minor-to-major axis
ratios and relative bar/ring position angles of the different features.
The deprojected values of these parameters are listed in the last two
columns of Tables~\ref{tab:R1gals}-~\ref{tab:inner-features}. It is
important to note that the projected shapes of the rings themselves
were not used to deproject the galaxies, only faint outer isophotes.
The results show that R$_2^{\prime}$ outer pseudorings in the sample
have a mean axis ratio of 0.93$\pm$0.01 and a mean bar-ring position
angle of 27\rlap{.}$^o$2$\pm$5\rlap{.}$^o$1. The standard deviation of
the angle is large, 25\rlap{.}$^o$4, but some of this is simply due to
the difficulty of measuring major axis position angles when the feature
is nearly circular.

For R$_1$ and R$_1^{\prime}$ outer rings, the mean axis ratio is 0.84$\pm$0.01
and the mean bar/ring position angle is
81\rlap{.}$^o$3$\pm$1\rlap{.}$^o$3, the latter with a standard deviation
of 7\rlap{.}$^o$6. Thus, these features deproject to a large angle
to the bar, nearly perpendicular on average, and are more elongated 
intrinsically than are R$_2^{\prime}$ rings.

For the inner features in the sample, the average axis ratio and
bar/ring position angle are 0.72$\pm$0.02 and
7\rlap{.}$^o$5$\pm$2\rlap{.}$^o$0, respectively. Both parameters have
large standard deviations: 0.12 for the axis ratio and 13\rlap{.}$^o$9
for the bar/ring position angle. Thus, inner features have the most
elongated intrinsic shapes of all the rings in the sample. Subtle
misalignments between bars and inner features, such as that seen in NGC
5132, are likely to be significant as a result.

These shape and alignment results are consistent with the predictions
of Schwarz (1981, 1984a), and with previous studies of the shapes and
orientations of galactic rings (Buta 1995; Comer\'on et al. 2014).

\subsection{(r) dark-spacers}

In contrast to the (rR) dark-spacers, Table~\ref{tab:rdkresrads} shows
that under the same assumption about the nature of the dark gaps, the
four (r) dark-spacers in the sample have $<a_{bar}/r_{CR}>$ =
1.44$\pm$0.10 and $<a_{bar}/r_{I4R}>$ = 2.23$\pm$0.16, values that are
higher by more than a factor of 2 than the means for (rR)
dark-spacers.  This shows that if the dark spaces in galaxies having
R$_1$ and R$_2$ features and those in galaxies like NGC 5335, which
lack such features, are viewed in the same manner, then the bars in (r)
dark-spacers extend well beyond their CR radius, violating the rules
established by Contopoulos (1980, 1985) which state that bars cannot
extend beyond their CR radius. The fact that there is no outer ring in
the outer disks of the four (r) dark-spacers studied here could
indicate that no major even-order resonances exist in those regions.
For example, most of the spiral structure in NGC 5335 would lie beyond
its implied OLR radius from the gap method. Other aspects of (r)
dark-spacers are discussed in section 10.2.

\section{Gap Colours}

The gap method assumes that dark spaces like those seen in (rR)
dark-spacers are dynamical phenomena, not artifacts of an unusual dust
distribution. To examine further the nature of dark gaps, the
integrated colours of each sample galaxy were derived at four points,
two within the dark gaps at the presumed $L_4$ and $L_5$ points, and
two along the bar axis at the same radius. The fluxes were integrated
at these points within small circular apertures having sizes chosen to
be approximately the apparent width of the gap. Each pair was combined
to give average colours. The results, corrected for galactic extinction
values from NED, are listed in Tables~\ref{tab:grri} - ~\ref{tab:bvvi}.
These tables also give the radius of the circular aperture used for the
integrations. The mean gap and bar axis point colours are summarized in
Table~\ref{tab:meancols}. The combination of the GZ2 and CSRG samples
allows us to derive this information in both the SDSS $ugri$ system and
in the Johnson-Cousins $BVI$ system.

\begin{table*}
\centering
\caption{
Corrected $g-r$ and $r-i$ colours of Gap and Bar Axis Points.  Col. 1:
galaxy name; cols. 2-3: combined colours of the two gap points at $r_{gp}$
within a circular aperture of radius $A$; cols. 4-5: combined colours of
the two points also at $r_{gp}$ but along the axis defined by the bar;
col. 6: radius of integrating aperture in arcseconds. All colours are
corrected for galactic extinction values from NED (Schlafly \&
Finkbeiner 2011).
}
\label{tab:grri}
\begin{tabular}{lccccc}
\hline
Name & $(g-r)_o$ & $(r-i)_o$ & $(g-r)_o$ & $(r-i)_o$ & $A$ \\
 & gap & gap & bar axis & bar axis & (arcsec) \\
1 & 2 & 3 & 4 & 5 & 6 \\ 
\hline
CGCG 8-10        &  0.91$\pm$0.03 &  0.43$\pm$0.03 &  0.71$\pm$0.02 &  0.41$\pm$0.02 &  5.94\\
CGCG 13-75       &  0.68$\pm$0.04 &  0.53$\pm$0.03 &  0.58$\pm$0.02 &  0.39$\pm$0.02 &  1.98\\
CGCG 65-2        &  0.64$\pm$0.04 &  0.43$\pm$0.03 &  0.44$\pm$0.02 &  0.27$\pm$0.02 &  1.98\\
CGCG 67-4        &  0.88$\pm$0.05 &  0.36$\pm$0.05 &  0.55$\pm$0.03 &  0.28$\pm$0.03 &  1.98\\
CGCG 73-53       &  0.74$\pm$0.04 &  0.41$\pm$0.03 &  0.68$\pm$0.03 &  0.34$\pm$0.02 &  1.98\\
CGCG 185-14      &  0.67$\pm$0.03 &  0.38$\pm$0.02 &  0.58$\pm$0.02 &  0.34$\pm$0.02 &  1.98\\
CGCG 263-22      &  0.75$\pm$0.05 &  0.70$\pm$0.04 &  0.66$\pm$0.02 &  0.42$\pm$0.02 &  2.97\\
IC 1223          &  0.87$\pm$0.03 &  0.51$\pm$0.02 &  0.73$\pm$0.02 &  0.41$\pm$0.02 &  3.96\\
IC 2473          &  0.71$\pm$0.02 &  0.34$\pm$0.02 &  0.46$\pm$0.01 &  0.26$\pm$0.01 &  7.92\\
IC 2628          &  0.70$\pm$0.03 &  0.38$\pm$0.03 &  0.68$\pm$0.03 &  0.33$\pm$0.02 &  1.98\\
MCG 6-32-24      &  0.66$\pm$0.04 &  0.42$\pm$0.03 &  0.64$\pm$0.02 &  0.28$\pm$0.02 &  1.98\\
MCG 7-18-40      &  0.68$\pm$0.03 &  0.38$\pm$0.03 &  0.57$\pm$0.02 &  0.34$\pm$0.02 &  1.98\\
NGC 2766         &  0.70$\pm$0.02 &  0.40$\pm$0.02 &  0.61$\pm$0.02 &  0.32$\pm$0.01 &  2.97\\
NGC 3380         &  0.62$\pm$0.01 &  0.46$\pm$0.01 &  0.63$\pm$0.01 &  0.36$\pm$0.01 &  5.94\\
NGC 4113         &  0.48$\pm$0.10 &  0.25$\pm$0.10 &  0.43$\pm$0.03 &  0.24$\pm$0.03 &  1.98\\
NGC 4608         &  0.73$\pm$0.01 &  0.41$\pm$0.01 &  0.77$\pm$0.01 &  0.42$\pm$0.01 &  3.96\\
NGC 4935         &  0.62$\pm$0.02 &  0.40$\pm$0.02 &  0.52$\pm$0.01 &  0.32$\pm$0.01 &  2.97\\
NGC 5132         &  0.79$\pm$0.02 &  0.47$\pm$0.02 &  0.70$\pm$0.01 &  0.35$\pm$0.01 &  7.92\\
NGC 5211         &  0.67$\pm$0.02 &  0.39$\pm$0.02 &  0.62$\pm$0.02 &  0.32$\pm$0.01 &  3.96\\
NGC 5335         &  0.81$\pm$0.04 &  0.54$\pm$0.03 &  0.79$\pm$0.01 &  0.41$\pm$0.01 &  1.98\\
NGC 5370         &  0.76$\pm$0.02 &  0.44$\pm$0.02 &  0.68$\pm$0.02 &  0.42$\pm$0.02 &  3.96\\
NGC 5686         &  0.72$\pm$0.01 &  0.35$\pm$0.01 &  0.73$\pm$0.01 &  0.34$\pm$0.01 &  1.58\\
NGC 5701         &  0.64$\pm$0.01 &  0.47$\pm$0.01 &  0.60$\pm$0.01 &  0.42$\pm$0.01 &  7.92\\
PGC 54897        &  0.51$\pm$0.04 &  0.53$\pm$0.03 &  0.59$\pm$0.02 &  0.35$\pm$0.02 &  2.97\\
PGC 1857116      &  0.87$\pm$0.04 &  0.34$\pm$0.04 &  0.57$\pm$0.02 &  0.34$\pm$0.02 &  3.96\\
PGC 2570478      &  0.89$\pm$0.05 &  0.42$\pm$0.04 &  0.88$\pm$0.04 &  0.40$\pm$0.03 &  1.98\\
UGC 4596         &  0.80$\pm$0.04 &  0.59$\pm$0.04 &  0.59$\pm$0.02 &  0.34$\pm$0.02 &  2.97\\
UGC 4771         &  0.69$\pm$0.02 &  0.38$\pm$0.02 &  0.62$\pm$0.02 &  0.41$\pm$0.02 &  3.96\\
UGC 5380         &  0.74$\pm$0.02 &  0.41$\pm$0.02 &  0.80$\pm$0.01 &  0.39$\pm$0.01 &  1.98\\
UGC 5885         &  0.73$\pm$0.05 &  0.38$\pm$0.04 &  0.58$\pm$0.02 &  0.35$\pm$0.02 &  3.96\\
UGC 9418         &  0.65$\pm$0.05 &  0.49$\pm$0.04 &  0.67$\pm$0.03 &  0.36$\pm$0.03 &  1.98\\
UGC 10168        &  0.75$\pm$0.04 &  0.46$\pm$0.04 &  0.69$\pm$0.03 &  0.40$\pm$0.02 &  2.97\\
UGC 10712        &  0.79$\pm$0.04 &  0.37$\pm$0.04 &  0.64$\pm$0.03 &  0.34$\pm$0.03 &  1.98\\
\hline
\end{tabular}
\end{table*}

\begin{figure} 
\includegraphics[width=\columnwidth,bb=14 14 650 350]{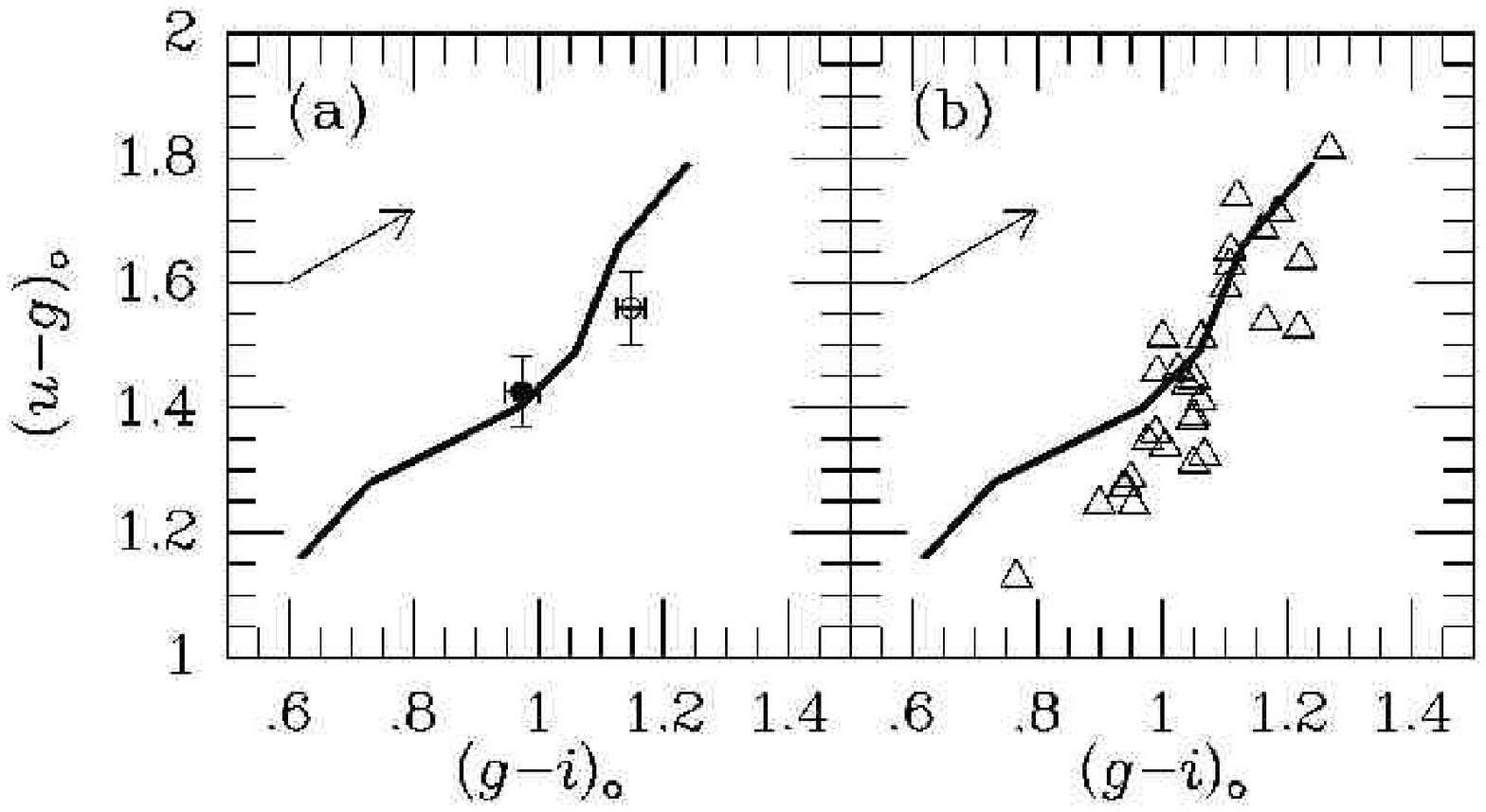}
\includegraphics[width=\columnwidth,bb=14 14 650 350]{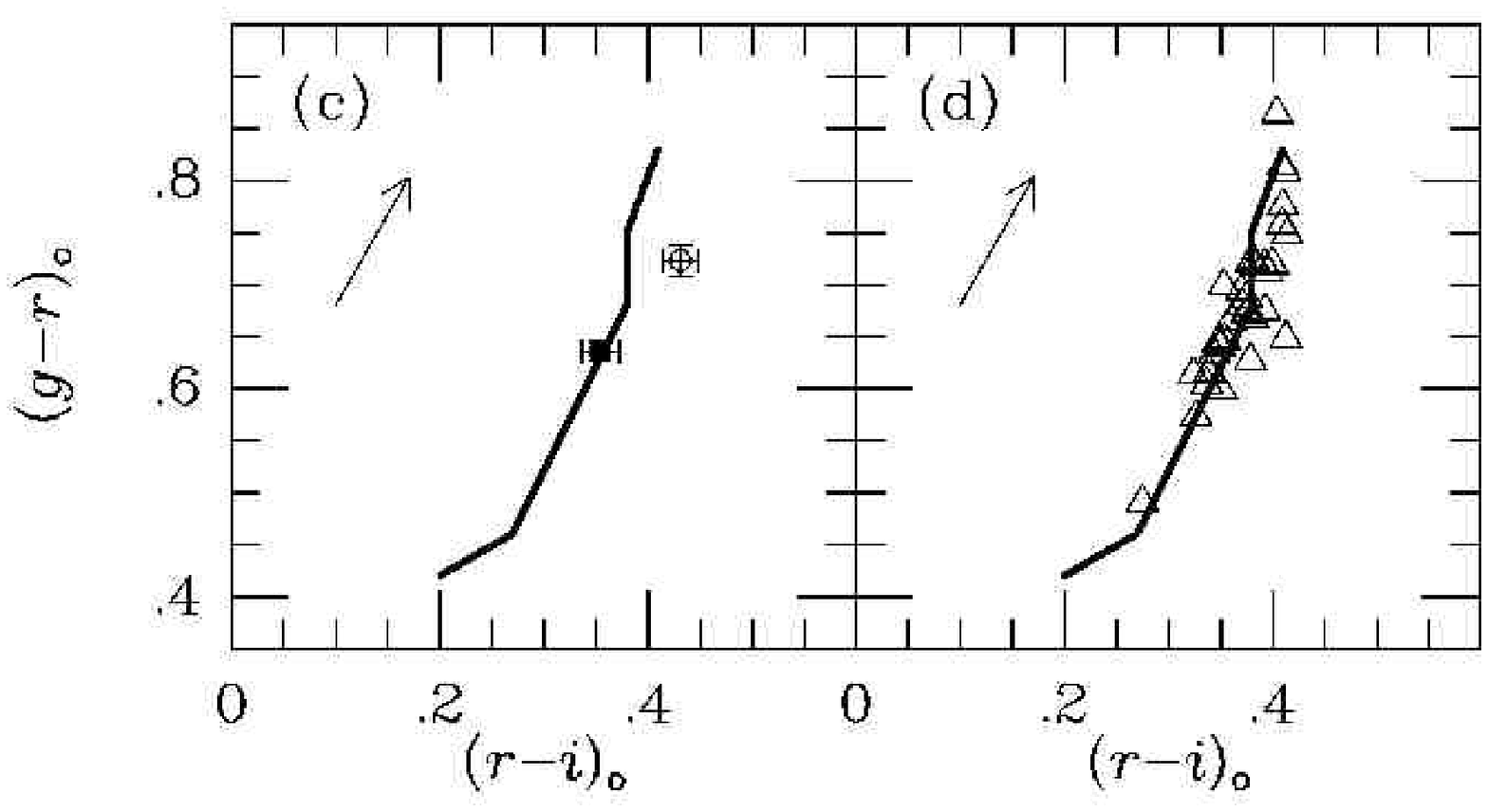}
\caption{(a) and (c) show the colour-colour plots for the average
colours of the gap (open circles) and bar axis points (filled circles)
from Table~\ref{tab:meancols}, while (b) and (d) show the same plots
for the total colours of the galaxies (open triangles) from
Table~\ref{tab:totmags}. The solid curves are mean colour-colour
relations for normal galaxies of different types in the SDSS filter
system from Shimasaku et al. (2001). All colours are corrected for
galactic extinction values from NED. The arrows indicate the reddening
vectors.}

\label{fig:colours} 
\end{figure}

\begin{figure} 
\includegraphics[width=\columnwidth,bb=14 14 600 400]{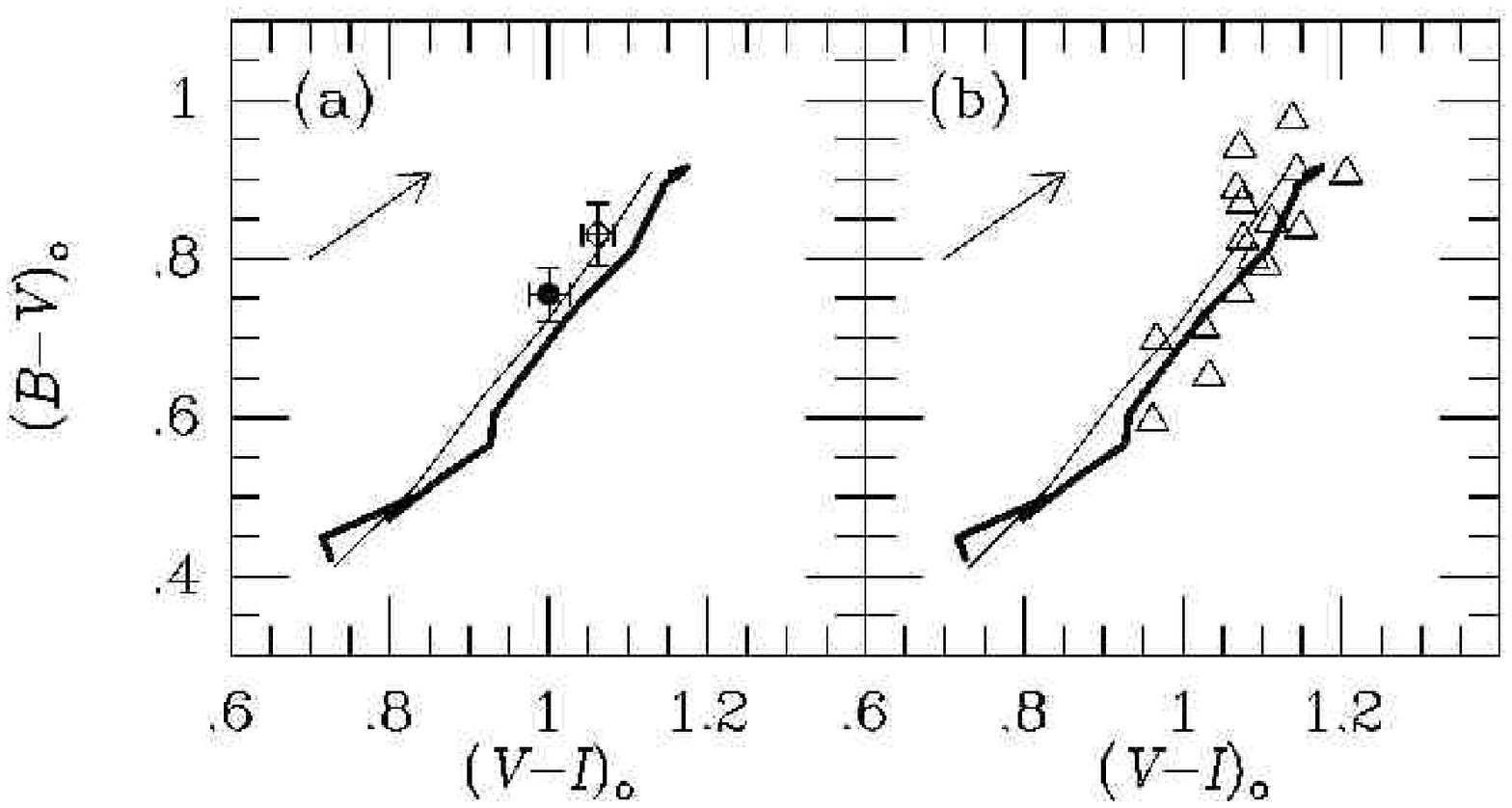}
\caption{(a) Colour-colour plot for the gap (open circle) and the bar
axis points (filled circle) in the non-GZ2 subsample. For comparison,
the thin curve shows model galactic colours for a Salpeter initial mass
function from Kennicutt et al. (1994), while the thick solid curve
shows the colour-colour correlations for bright RC3 galaxies from Buta
al. (1994) and Buta \& Williams (1995). (b) Same as (a) for the total
colours of the non-GZ2 galaxies (open triangles).  The arrows in both
graphs indicate the reddening vectors.}

\label{fig:ubvri} 
\end{figure}

Figure~\ref{fig:colours} shows the resulting colour-colour plots
$(g-r)_o$ versus $(r-i)_o$ and $(u-g)_o$ versus $(g-i)_o$ for the GZ2
galaxies, while Figure~\ref{fig:ubvri} shows $(B-V)_o$ versus $(V-I)_o$
for the non-GZ2 cases. The left panel in each pair shows the mean
gap/bar colour points from Table~\ref{tab:meancols} while the right
panel shows the same plot for the integrated colours of the galaxies
from Tables~\ref{tab:totmags} and ~\ref{tab:totmagsBVI} (open
triangles). For comparison, each plot shows the mean colour-colour
relation for normal galaxies of different types. For SDSS filters, the
mean relation shown is from Shimasaku et al. (2001). For $BVI$ filters,
the relation shown is from Buta et al. (1994) and Buta \& Williams
(1995). All colours are corrected for Galactic extinction.  The graphs
show first that the total colours of the sample galaxies cluster around
the redder half of the normal galaxy sequence, consistent with spirals
of early-to-intermediate types. The graphs also show that the
distribution of the points is nearly parallel to the reddening vector
in each case, making it difficult to separate the effects of stellar
population differences from reddening effects.

\begin{table*}
\centering
\caption{
Corrected $u-g$ and $g-i$ colours of Gap and Bar Axis Points.  Col. 1:
galaxy name; cols. 2-3: combined colours of the two gap points at $r_{gp}$
within a circular aperture of radius $A$; cols. 4-5: combined colours of
the two points also at $r_{gp}$ but along the axis defined by the bar;
col. 6: radius of integrating aperture in arcseconds. All colours are
corrected for galactic extinction values from NED (Schlafly \&
Finkbeiner 2011).
}
\label{tab:uggi}
\begin{tabular}{lccccc}
\hline
Name & $(u-g)_o$ & $(g-i)_o$ & $(u-g)_o$ & $(g-i)_o$ & $A$ \\
 & gap & gap & bar axis & bar axis & (arcsec) \\
1 & 2 & 3 & 4 & 5 & 6 \\ 
\hline
CGCG 13-75       &  1.78$\pm$0.11 &  1.21$\pm$0.04 &  1.39$\pm$0.06 &  0.97$\pm$0.03 &  1.98\\
CGCG 65-2        &  1.34$\pm$0.10 &  1.08$\pm$0.04 &  0.84$\pm$0.05 &  0.71$\pm$0.02 &  1.98\\
CGCG 67-4        &  1.32$\pm$0.13 &  1.24$\pm$0.05 &  1.25$\pm$0.06 &  0.83$\pm$0.03 &  1.98\\
CGCG 185-14      &  1.44$\pm$0.08 &  1.05$\pm$0.03 &  1.23$\pm$0.06 &  0.91$\pm$0.02 &  1.98\\
CGCG 263-22      &  1.62$\pm$0.12 &  1.44$\pm$0.04 &  1.36$\pm$0.06 &  1.08$\pm$0.02 &  2.97\\
IC 2473          &  1.70$\pm$0.05 &  1.05$\pm$0.02 &  1.11$\pm$0.02 &  0.72$\pm$0.01 &  7.92\\
IC 2628          &  1.87$\pm$0.10 &  1.08$\pm$0.03 &  1.34$\pm$0.06 &  1.02$\pm$0.02 &  1.98\\
MCG 6-32-24      &  2.36$\pm$0.13 &  1.09$\pm$0.04 &  1.49$\pm$0.06 &  0.92$\pm$0.03 &  1.98\\
MCG 7-18-40      &  1.38$\pm$0.08 &  1.06$\pm$0.03 &  1.26$\pm$0.05 &  0.91$\pm$0.02 &  1.98\\
NGC 2766         &  1.44$\pm$0.06 &  1.11$\pm$0.02 &  1.34$\pm$0.04 &  0.93$\pm$0.02 &  2.97\\
NGC 3380         &  1.57$\pm$0.03 &  1.08$\pm$0.01 &  1.53$\pm$0.02 &  0.99$\pm$0.01 &  5.94\\
NGC 4608         &  1.67$\pm$0.04 &  1.13$\pm$0.01 &  1.82$\pm$0.02 &  1.18$\pm$0.01 &  3.96\\
NGC 4935         &  1.64$\pm$0.05 &  1.02$\pm$0.02 &  1.28$\pm$0.03 &  0.84$\pm$0.01 &  2.97\\
NGC 5211         &  1.49$\pm$0.05 &  1.05$\pm$0.02 &  1.44$\pm$0.04 &  0.95$\pm$0.02 &  3.96\\
NGC 5335         &  1.70$\pm$0.11 &  1.35$\pm$0.04 &  1.87$\pm$0.04 &  1.19$\pm$0.01 &  1.98\\
NGC 5370         &  1.72$\pm$0.06 &  1.20$\pm$0.02 &  2.05$\pm$0.06 &  1.10$\pm$0.02 &  3.96\\
NGC 5686         &  1.58$\pm$0.03 &  1.07$\pm$0.01 &  1.73$\pm$0.03 &  1.07$\pm$0.01 &  1.58\\
NGC 5701         &  1.66$\pm$0.03 &  1.10$\pm$0.01 &  1.27$\pm$0.02 &  1.02$\pm$0.01 &  7.92\\
PGC 54897        &  0.91$\pm$0.07 &  1.04$\pm$0.03 &  1.50$\pm$0.06 &  0.94$\pm$0.02 &  2.97\\
PGC 1857116      &  1.10$\pm$0.09 &  1.21$\pm$0.04 &  1.27$\pm$0.03 &  0.90$\pm$0.02 &  3.96\\
PGC 2570478      &  1.43$\pm$0.13 &  1.31$\pm$0.04 &  2.02$\pm$0.13 &  1.28$\pm$0.03 &  1.98\\
UGC 4596         &  1.51$\pm$0.11 &  1.39$\pm$0.04 &  1.29$\pm$0.04 &  0.93$\pm$0.02 &  2.97\\
UGC 4771         &  1.26$\pm$0.06 &  1.07$\pm$0.02 &  1.42$\pm$0.05 &  1.03$\pm$0.02 &  3.96\\
UGC 5885         &  1.43$\pm$0.13 &  1.11$\pm$0.05 &  1.25$\pm$0.04 &  0.93$\pm$0.02 &  3.96\\
UGC 10712        &  2.05$\pm$0.13 &  1.16$\pm$0.04 &  1.30$\pm$0.06 &  0.98$\pm$0.03 &  1.98\\
\hline
\end{tabular}
\end{table*}

\begin{table*}
\centering
\caption{
Corrected $B-V$ and $V-I$ colours of Gap and Bar Axis Points.  Col. 1:
galaxy name; cols. 2-3: combined colours of the two gap points at $r_{gp}$
within a circular aperture of radius $A$; cols. 4-5: combined colours of
the two points also at $r_{gp}$ but along the axis defined by the bar;
col. 6: radius of integrating aperture in arcseconds. All colours are
corrected for galactic extinction values from NED (Schlafly \& Finkbeiner 2011).
}
\label{tab:bvvi}
\begin{tabular}{lccccc}
\hline
Name & $(B-V)_o$ & $(V-I)_o$ & $(B-V)_o$ & $(V-I)_o$ & $A$ \\
 & gap & gap & bar axis & bar axis & (arcsec) \\
1 & 2 & 3 & 4 & 5 & 6 \\
\hline
ESO 325-28      & 0.76$\pm$0.07 & 1.01$\pm$0.05 & 0.60$\pm$0.03 & 0.88$\pm$0.02 &  4.10\\
ESO 365-35      & 1.00$\pm$0.15 & 1.09$\pm$0.03 & 0.98$\pm$0.03 & 1.04$\pm$0.03 &  4.10\\
ESO 426-2       & 0.95$\pm$0.05 & 1.13$\pm$0.04 & 0.81$\pm$0.03 & 1.07$\pm$0.07 &  4.10\\
ESO 437-33      & 0.70$\pm$0.02 & 0.96$\pm$0.05 & 0.69$\pm$0.03 & 0.95$\pm$0.04 &  4.10\\
ESO 437-67      & 0.55$\pm$0.23 & 1.00$\pm$0.21 & 0.65$\pm$0.05 & 0.95$\pm$0.03 & 10.94\\
ESO 566-24      & 0.66$\pm$0.02 & 1.03$\pm$0.07 & 0.59$\pm$0.01 & 0.92$\pm$0.02 &  4.10\\
ESO 575-47      & 0.79$\pm$0.10 & 1.16$\pm$0.24 & 0.65$\pm$0.04 & 0.89$\pm$0.05 &  5.47\\
IC 1438         & 0.84$\pm$0.04 & 1.15$\pm$0.06 & 0.76$\pm$0.04 & 1.06$\pm$0.06 &  8.70\\
IC 4214         & 0.79$\pm$0.04 & 1.02$\pm$0.02 & 0.74$\pm$0.03 & 1.02$\pm$0.02 &  8.21\\
NGC 1079        & 0.92$\pm$0.12 & 1.03$\pm$0.04 & 0.87$\pm$0.03 & 1.02$\pm$0.03 & 17.40\\
NGC 1291        & 0.81$\pm$0.02 & 1.01$\pm$0.04 & 0.83$\pm$0.02 & 1.02$\pm$0.05 & 30.00\\
NGC 1326        & 0.84$\pm$0.01 & 1.11$\pm$0.01 & 0.86$\pm$0.01 & 1.10$\pm$0.04 & 10.94\\
NGC 1398        & 0.93$\pm$0.02 & 1.20$\pm$0.02 & 0.89$\pm$0.02 & 1.18$\pm$0.02 &  8.70\\
NGC 2665        & 0.68$\pm$0.05 & 1.05$\pm$0.19 & 0.47$\pm$0.07 & 0.81$\pm$0.09 &  5.47\\
NGC 6782        & 0.86$\pm$0.05 & 1.14$\pm$0.06 & 0.87$\pm$0.04 & 1.12$\pm$0.06 &  8.70\\
NGC 7098        & 1.20$\pm$0.07 & 0.91$\pm$0.06 & 0.85$\pm$0.02 & 0.97$\pm$0.05 & 10.87\\
\hline
\end{tabular}
\end{table*}

\begin{table*}
\centering
\caption{
Mean gap and bar axis colours.  Eight of the GZ2 sample galaxies did
not have a measureable $u$-band total magnitude.
}
\label{tab:meancols}
\begin{tabular}{lcccccc}
\hline
Points & $<(u-g)_o>$ & $<(g-i)_o>$ & $<(g-r)_o>$ & $<(r-i)_o>$ & $<(B-V)_o>$ & $<(V-I)_o>$\\
1 & 2 & 3 & 4 & 5 & 6 & 7 \\
\hline
gap &  1.56  &  1.15 &  0.72  &  0.43  &  0.83  &  1.06 \\
mean error & 0.06  & 0.02 & 0.02 & 0.01 & 0.04 & 0.02 \\
$\sigma_1$ & 0.29  & 0.12 & 0.10 & 0.08 & 0.15 & 0.08 \\
           &       &      &      &      &      &      \\
bar axis  &  1.43 &  0.97 & 0.64 &  0.35 &  0.75 &  1.00\\
mean error & 0.06 & 0.03 & 0.02 & 0.01 & 0.03 & 0.02 \\
$\sigma_1$ & 0.28 & 0.13 & 0.10 & 0.05 & 0.14 & 0.10 \\
           &       &      &      &      &      &      \\
$<$gap$-$bar$>$ & 0.13 & 0.17 & 0.09 & 0.08 & 0.08 & 0.06 \\
mean error & 0.07 & 0.03 & 0.02 & 0.01 & 0.03 & 0.02 \\
$\sigma_1$ & 0.37 & 0.13 & 0.10 & 0.07 & 0.11 & 0.09 \\
           &       &      &      &      &      &      \\
$N$        & 25    &   25 & 33   &   33 &   16 &   16 \\
\hline
\end{tabular}
\end{table*}

Figures~\ref{fig:colours} and ~\ref{fig:ubvri} show that for all of the
colours measured, the gap points are redder than the bar axis points.
However, the difference amounts to only $\approx$0.1 mag on average
(Table~\ref{tab:meancols}). To evaluate what such a difference might
mean, the CSRG $BVI$ colours are compared in Figure~\ref{fig:ubvri} with
the colours of synthetic galaxies (thin solid curves) calculated by
Kennicutt et al. (1994). The latter are based on models having an age
of 10 Gyr and a star formation rate that declines exponentially with an
e-folding time $\tau$.  The models are also characterized by a
birthrate parameter, which is the current star formation rate relative
to the past average, and an inital mass function (IMF), of which three
different versions were used.

Irrespective of the IMF used, the comparison with the synthetic
galaxies suggests that dark gaps are made of a stellar population with
an e-folding time $\tau$ approximately 1 Gyr less than for the bar axis
points, implying a more rapidly declining star formation rate for the
gap regions. This basically means that recent star formation is not a
characteristic of dark gaps. Even if the rings and pseudorings have
considerable star formation, the gaps tend to be devoid of such
activity. This supports the idea that dark gaps are a purely stellar
dynamical phenomenon, and that the colour difference between the gaps
and the bar axis points is due mainly to stellar population
differences, not to extinction. Examination of colour index maps of all
of the sample galaxies do not show any evidence that recent star
formation is ever found in the gaps of (rR)- or (r) dark-spacer
galaxies. BC96 note: ``When the bar is strong, most orbits
perpendicular to the bar between CR and OLR become unstable. That is
why this region is often depopulated and, in real galaxies, never
includes recent star formation."

\section{Discussion}

\subsection{Comparison with Manifold Theory}

The outer ring/pseudoring subclasses R$_1$, R$_1^{\prime}$, and
R$_2^{\prime}$ are remarkable not only because of their distinctive
shapes, but also because they were predicted to exist before they were
actually found.  The early models of the gas flow in barred galaxies by
Schwarz (1981, 1984a) showed that features resembling these
morphologies can develop naturally in response to a bar perturbation.
The implication of these models is that gas in a barred galaxy
accumulates near resonances, under the continuous action of gravity
torques from the bar pattern (BC96). On the basis of these models, the
link of the R$_1$, R$_1^{\prime}$, and R$_2^{\prime}$ features to outer
resonances like the O4R and the OLR seems strong, especially when the
combined morphology, R$_1$R$_2^{\prime}$, is added to the mix of types.
Being able to recognize these features in real galaxies provides a
direct link between morphology and dynamics, which makes further
studies of these galaxies imperative as a stepping stone to
understanding other galactic morphologies.

The analysis in this paper has shown that there is considerable
homogeneity in the structure of outer resonant subclass galaxies. Many
show characteristic double-humped azimuthally-averaged surface
brightness profiles (Figure~\ref{fig:azim}), and the ring features have
characteristic intrinsic shapes and orientations with respect to bars
and ovals (highlighted by the schematics in Figure~\ref{fig:results}).
Dark gaps are a common theme in these galaxies, and this study has
shown that linking these gaps to the corotation resonance provides
consistent interpretations of the structure of these galaxies in terms
of characteristic low-order bisymmetric resonances. This consistency
suggests that the bars and the ring patterns by and large share the
same pattern speed. This does not, however, mean that multiple pattern
speeds are not present in these galaxies. Our analysis, for example, does not
consider the role of the ILR in the structure of these galaxies.

Although the resonant idea seems to explain many of the features of the
galaxies studied in this paper, the alternative ``manifold theory"
(Romero-G\'omez et al. 2006, 2007; Athanassoula et al. 2009a,b; 2010)
also can account for the \Rone, \RoneP, \RtwoP, and
\Rone\RtwoP\ morphologies. Rather than linking these features to
specific bisymmetric resonances, manifold theory is based around the
behavior of particles in the vicinity of the $L_1$ and $L_2$ Lagrangian
points of the bar gravitational potential. These points are located
near the ends of the bar, and like the $L_4$ and $L_5$ points, have
characteristic periodic orbits around them. Because $L_1$ and $L_2$ are
saddle points in the effective potential, the periodic orbits around
the points are generally unstable and chaotic, which would lead to a
deficiency of light in these regions were it not for the existence of
manifold tubes that confine the chaotic motions along specific
trajectories called homoclinic and heteroclinic orbits.  These orbits
can account not only for the \Rone, \RoneP, \RtwoP, and
\Rone\RtwoP\ morphologies, but also \rRone\ morphologies consisting
of a cuspy, elongated inner ring oriented perpendicular to a dimpled
\Rone\ outer ring, as well as pure spiral morphologies.

Of the sample of galaxies examined here, several with strong R$_1$
components resemble the homoclinic manifold models, where particles
emanate from the vicinity of $L_1$ or $L_2$, and approach the opposite
point at the other end of the bar.  For example, the schematics of NGC
2665, ESO 437-67, NGC 6782, UGC 5885, and UGC 12646 in
Figure~\ref{fig:results} strongly resemble the manifold morphology
\rRone, described in detail by Romero-G\'omez et al. (2006). To this
category we could also add NGC 1326 and PGC 1857116. A possible point
of disagreement is that the \rRone\ models place the cusps of the
manifold inner ring as coincident with the dimples of the manifold
R$_1$ outer ring. This is not necessarily seen in all of these
examples. In NGC 6782, for example, there is considerable distance
between the strong inner ring cusps and the R$_1$ dimples. Such a
separation would be expected if the R$_1$ outer ring begins outside
\rcr\ and the inner ring ends near \ri4r, as is suggested by the
resonance idea. Athanassoula et al. (2010) argue that if the cusps of
the inner ring do not overlap the dimples of the R$_1$ outer ring, then
this just means that the $L_1$ and $L_2$ points lie between the cusps
and dimples. Examination of the schematics in Figure~\ref{fig:results}
shows that the gap method in general places R$_1$-component dimples
either close to or just outside \rcr.

Romero-G\'omez et al. (2006) also make the interesting note that the
R$_1$ ring in the manifold \rRone\ morphology is not necessarily
associated with the OLR, but instead the outer branches of the
manifolds extend to 0.86\rolr. This actually agrees with our estimate
of $<a/r_{OLR}>$ = 0.85 for 36 galaxies with R$_1$ components in
Table~\ref{tab:R1gals}, i.e., that R$_1$ rings lie well inside \rolr.
Another issue is that inner rings like those seen in NGC 1433 and NGC
7098 have a tightly-wrapped spiral structure that suggests a link to
the I4R. This is implied, for example, by the test-particle model
illustrated by Simkin et al. (1980), which shows the development of a
strong, cuspy inner ring made of at least two and possibly four
separate sections of spiral arms, developing in a relatively weak bar
model.

Manifolds can also well explain the \Rone\RtwoP\ as well as
R$_1$R$_2$ closed double outer ring morphologies. Athanassoula et al.
(2009a) show that a different type of manifold orbit can account for
these types: the heteroclinic type where particles emanate from $L_1$
or $L_2$ and return to the same point. Such manifolds do resemble
objects like NGC 1079, 2766, 3081, and UGC 10168 in the present sample.
Pure R$_2^{\prime}$ rings are interpreted differently. In the manifold
theory, these are instead associated with ``escaping" manifold orbits,
where particles emanating from $L_1$ or $L_2$ do not go to either the
same or the opposite Lagrangian point, but spiral outward. It is
possible for these escaping manifolds to intersect on opposite sides of
a galaxy roughly after winding about 270$^o$. This would then form an
R$_2^{\prime}$ outer pseudoring. In the present sample, the purest
R$_2^{\prime}$ rings are found in CGCG 13-75, ESO 325-28, NGC 4935,
NGC 5211, UGC 4771, and UGC 10712.

The case of ESO 566-24 and its strong four-armed spiral pattern is
interesting from the manifold point of view. As shown in
Figure~\ref{fig:results}, the gap method places the pattern largely
between \ri4r\ and \ro4r, consistent  with the numerical study of
Rautiainen et al. (2004). Athanassoula et al. (2009b) state that the
manifold theory most easily explains two-armed barred spirals, but to
get a regular $m$=4 manifold pattern in a barred galaxy would require
more than two saddle points and a potential which is bar-like inside
\rcr\ and spiral-like outside \rcr.

\subsection{Dark Spaces and Bar Evolution}

The recent study by Kim et al. (2016) touches on issues of dark spaces
in barred galaxies from a somewhat different point of view. In this
paper, the authors quantify light deficits in barred galaxies by
comparing surface brightness profiles along the deprojected bar axis
with those along an axis perpendicular to the bar. They find that the
maximum difference between the profiles, Max($\Delta\mu$) =
Max[$\mu$(bar minor axis)$-$$\mu$(bar major axis)], correlates
with bar length and bar-to-total luminosity ratio. Using numerical
simulations, they show that Max($\Delta\mu$) should increase with time as
a bar grows stronger and longer by capturing disk stars in the vicinity
of the bar. 

The profiles shown in Figures~\ref{fig:ugc4596b} and
~\ref{fig:NGC5335b} are the same kinds as those used by Kim et al.
(2016; see also Buta et al. 2006).  These figures give Max($\Delta\mu$)
in $i$-band light as 1.32 mag for UGC 4596 and 2.47 mag for NGC 5335.
The Kim et al. study does not distinguish (r) dark-spacers from
(rR) dark-spacers but nevertheless their Figure 1 shows an example of
both: NGC 1015 and IC 1438. NGC 1015 would be interpreted here as an
(r) dark-spacer, while IC 1438 is an (rR) dark-spacer in common with
the present study.

As discussed in section 8, if the gaps in surface
brightness or residual intensity between inner and outer ring features
in (rR) dark-spacers are linked to the $L_4$ and $L_5$ Lagrangian
points in a bar field, then a very consistent picture
of the structure of galaxies having \Rone, \RoneP, \RtwoP, and
\Rone\RtwoP\ outer rings/pseudoring in terms of basic galactic
resonances emerges. The analysis has suggested that, in general, the
bars in (rR) dark-spacers extend to \ri4r, not \rcr. The inner rings in
such galaxies also have a wide range of shapes, from nearly circular to
intrinsic axis ratios of $\approx$0.5 (see also Figure 5 of Buta 2014).

A very different interpretation of (r) dark-spacers follows from the
Lagrangian points assumption. The bars of such galaxies would extend
beyond their implied CR radii as shown in Table~\ref{tab:rdkresrads}.
While this may only signify a failure of the assumption in these cases,
(r) dark-spacers have additional differences compared to (rR)
dark-spacers that may have dynamical significance.  First,
Table~\ref{tab:rdksp} shows that the four examples of (r) dark-spacers
in the GZ2 subsample have inner rings whose deprojected axis ratio
averages much rounder than those in (rR) dark-spacer galaxies:  $<q_0>$
= 0.92$\pm$0.02 versus 0.72$\pm$0.02. This difference would be
unchanged if NGC 1015 from Kim et al. (2016) were added to the subset.
While this could be a selection effect if (r) dark-spacers are easier
to recognize when a ring is intrinsically circular, it is interesting
that Kim et al.  (2016) show a model of an evolved barred galaxy which
strongly resembles an (r) dark-spacer with a nearly circular inner
ring.

Another difference is highlighted in Table~\ref{tab:fouriermeans},
which lists the average $m$ = 2, 4, and 6 relative Fourier intensity
amplitudes for (rR) dark-spacers as compared to (r) dark-spacers, using
the information from Table~\ref{tab:fourier}. Although the two types
have about the same $<A_2>$, (r) dark-spacers tend to have bars with
larger $<A_4>$ and $<A_6>$ amplitudes. This reflects how many of the
(rR) dark-spacers in our sample have massive ovals as their main
perturbation rather than conventional bars. Ovals would tend to have
weaker higher-order Fourier terms than would conventional bars. From
Figure 1 of Kim et al. (2016), adding NGC 1015 to our sample would not
change this result. Also, their model after 9Gyr clearly shows a bar
with significant higher-order Fourier terms.

\begin{figure}
\centering
\includegraphics[width=\columnwidth,bb=14 14 600 600]{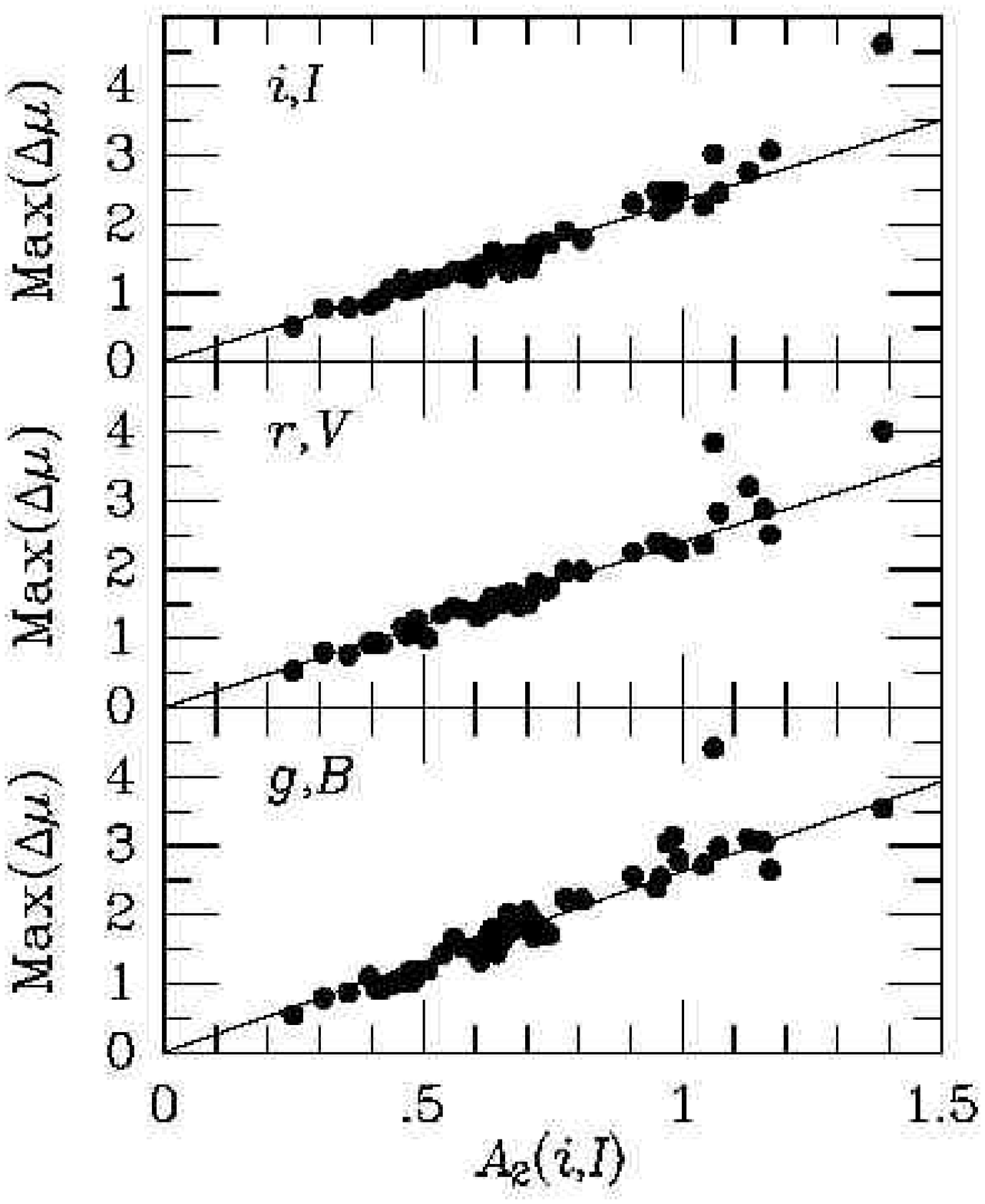}
\caption{Graphs showing how well the parameter Max($\Delta\mu$)
(mag arcsec$^{-2}$) correlates with the maximum relative
$m$=2 Fourier intensity amplitude $A_2$ which is based only on the
infrared band images. The lines are least squares fits that have been
forced to pass through the origin. These lines are after two cycles of
2$\sigma$ rejection.}
\label{fig:maxdmuA2}
\end{figure}

Kim et al. (2016) further examine correlations between Max($\Delta\mu$)
and parameters such as the bar radius relative to the isophotal radius
of the galaxy, the bar-to-total luminosity ratio, the bar S\'ersic
index $n_{bar}$, and the bar ellipticity $\epsilon_{bar}$. These
parameters are based partly on decompositions. Table~\ref{tab:maxdmu}
lists Max($\Delta\mu$) values for all 54 of the galaxies in the present
sample, as well as the radii in arcseconds where these maxima occur.
The values for filters $g$ and $B$, $r$ and $V$, and $i$ and $I$ are
listed in each separate column.  Figure~\ref{fig:maxdmuA2} first shows
that Max($\Delta\mu$) for the GZ2/CSRG sample correlates very well with
the near-infrared relative $m$ = 2 Fourier amplitude $A_2(i,I)$
(Table~\ref{tab:fourier}). The lines in the graph are least squares
fits that force the lines to pass through the origin. The relations
after two cycles of 2$\sigma$ rejection are

$$Max[\Delta\mu(g,B)] = (2.618 \pm 0.029)A_2(i,I)$$
$$Max[\Delta\mu(r,V)] = (2.394 \pm 0.021)A_2(i,I)$$
$$Max[\Delta\mu(i,I)] = (2.337 \pm 0.020)A_2(i,I)$$

\begin{figure}
\centering
\includegraphics[width=\columnwidth,bb=14 14 600 600]{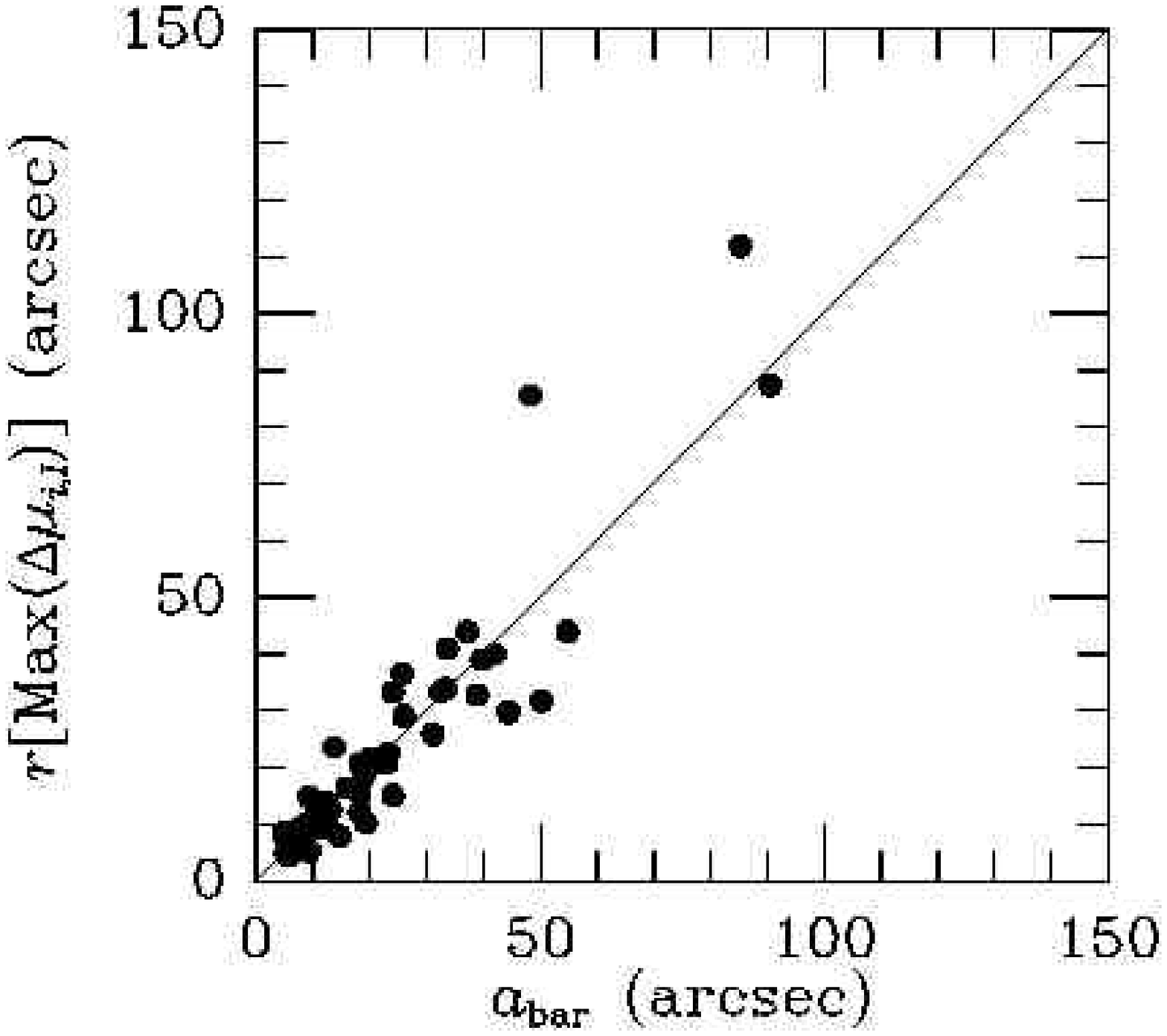}
\caption{Graph showing that the radius of Max[$\Delta\mu$($i$,$I$)]
correlates well with the estimated bar radius from
Tables~\ref{tab:resrads} and ~\ref{tab:rdkresrads}. The line is not
based on a least squares fit but is shown to guide the eye.  }

\label{fig:radsdmu}
\end{figure}

\noindent
These relations and Table~\ref{tab:maxdmu} in addition show that
$\Delta\mu$ can vary with wavelength, being somewhat larger in $g$ and
$B$ compared to $i$ and $I$, and that the radii, $r$[Max($\Delta\mu$)],
where these maxima occur are not necessarily the same for each filter,
although many are similar.  Figure~\ref{fig:radsdmu} shows that
$r$[Max($\Delta\mu(i,I$)] correlates well with the bar radii listed in
Tables~\ref{tab:resrads} and ~\ref{tab:rdkresrads}.

\begin{table*}
\centering
\caption{Inner rings in (r) dark-spacer galaxies.
Col. 1: galaxy name; cl. 2: type of feature (r=inner
ring, \_rs=inner pseudoring; cols. 3,4: deprojected
major and minor axis ring dimensions, respectively, in arcseconds;
cols. 5-6:  ratios of ring dimensions to the inner 4:1 resonance
radius; col. 7:  feature minor-to-major axis ratio; col. 8: relative
deprojected angle between the bar major axis and the deprojected major axis
position angle of the feature.
}
\label{tab:rdksp}
\begin{tabular}{llrrrrrr}
\hline
 Name & Feature & $a$ & $b$ & $a/r_{I4R}$ & $b/r_{I4R}$ & $q_0$ & $|\theta_{Bf}|$ \\
 1 & 2 & 3 & 4 & 5 & 6 & 7 & 8  \\ 
\hline
NGC4608      & r               &    51.0 &    49.3 &    2.14 &    2.07 &    0.97 &    17.1 \\
NGC5335      & \_rs            &    27.2 &    24.3 &    2.46 &    2.20 &    0.89 &    81.7 \\
NGC5686      & r               &     6.4 &     5.7 &    2.09 &    1.88 &    0.90 &    85.9 \\
UGC05380     & r               &    15.3 &    14.4 &    2.79 &    2.62 &    0.94 &    39.2 \\
             &                 &         &         &         &         &         &         \\
means        &                 &    .... &    .... &    2.37 &    2.19 &    0.92 &    56.0 \\
mean error   &                 &    .... &    .... &    0.16 &    0.16 &    0.02 &    16.7 \\
$\sigma_1$   &                 &    .... &    .... &    0.32 &    0.31 &    0.04 &    33.4 \\
$N$          &                 &    .... &    .... &     4   &     4   &     4   &      4  \\
\hline
\end{tabular}
\end{table*}

\begin{table*}
\centering
\caption{Mean relative Fourier intensity amplitudes for (rR) dark-spacers as compared with
(r) dark-spacers, from data in Table~\ref{tab:fourier}.
Col. 1: Type of dark-spacer; cols. 2-4: mean $m$ = 2, 4, and 6 amplitudes (with mean error
followed by standard deviation in parentheses); col. 5: number of galaxies in averages.}
\label{tab:fouriermeans}
\begin{tabular}{lcccr}
\hline
Morphology & $<A_2>$ & $<A_4>$ & $<A_6>$ & $N$ \\
 1 & 2 & 3 & 4 & 5 \\
\hline
(rR) & 0.69$\pm$0.04 (0.25) & 0.29$\pm$0.02 (0.17) & 0.15$\pm$0.02 (0.16) & 50 \\
(r)  & 0.68$\pm$0.13 (0.27) & 0.39$\pm$0.08 (0.16) & 0.24$\pm$0.05 (0.10) &  4 \\
\hline
\end{tabular}
\end{table*}

\begin{table*}
\centering
\caption{Maximum values of $\Delta\mu$=$\mu$(bar minor axis)$-$$\mu$(bar major axis)
for full sample.
(a) Max($\Delta\mu$) values in mag arcsec$^{-2}$; (b): $r[Max(\Delta\mu)]$ in arcseconds.
Col. 1: galaxy name; col. 2: values in $g$-band (GZ2) or $B$-band (non-GZ2);
col. 3: values in $r$-band (GZ2) or $V$-band (non-GZ2);
col. 4: values in $i$-band (GZ2) or $I$-band (non-GZ2)
}
\label{tab:maxdmu}
\begin{tabular}{lrrrlrrrlrrr}
\hline
Galaxy & $g,B$ & $r,V$ & $i,I$ & Galaxy & $g,B$ & $r,V$ & $i,I$ & Galaxy & $g,B$ & $r,V$ & $i,I$ \\ 
\hline
 1 & 2 & 3 & 4 & 1 & 2 & 3 & 4 & 1 & 2 & 3 & 4 \\
\hline
            &        &        &        &            &        &        &        &             &        &        &       \\ 
\noalign{\centerline{(a) Max($\Delta\mu$)}}
            &        &        &        &            &        &        &        &             &        &        &       \\ 
CGCG 8-10   &   2.04 &   1.50 &   1.37 & IC 4214    &   1.74 &   1.53 &   1.45 & NGC 5211    &   0.94 &   0.95 &   0.91 \\
CGCG 13-75  &   1.43 &   1.35 &   1.22 & MCG 6-32-24 &   1.54 &   1.32 &   1.22 & NGC 5335  &   2.39 &   2.38 &   2.46 \\
CGCG 65-2   &   1.67 &   1.40 &   1.39 & MCG 7-18-40 &   1.17 &   1.09 &   1.09 & NGC 5370  &   1.44 &   1.51 &   1.45 \\
CGCG 67-4   &   3.14 &   2.30 &   2.33 & NGC  210    &   1.18 &   .... &   .... & NGC 5686  &   0.79 &   0.81 &   0.79 \\
CGCG 73-53  &   0.92 &   0.92 &   0.92 & NGC 1079    &   1.62 &   1.52 &   1.52 & NGC 5701  &   1.21 &   1.23 &   1.16 \\
CGCG 185-14 &   0.56 &   0.54 &   0.53 & NGC 1291    &   1.05 &   1.12 &   1.10 & NGC 6782  &   1.80 &   1.61 &   1.60 \\
CGCG 263-22 &   1.67 &   1.47 &   1.33 & NGC 1326    &   1.47 &   1.44 &   1.42 & NGC 7098  &   1.91 &   1.48 &   1.57 \\
ESO 325-28  &   2.23 &   1.98 &   1.90 & NGC 1398    &   1.09 &   1.17 &   1.19 & PGC 54897 &   2.54 &   2.38 &   2.19 \\
ESO 365-35  &   1.20 &   1.01 &   1.19 & NGC 1433    &   3.04 &   .... &   2.48 & PGC 1857116 &   2.79 &   2.26 &   2.46 \\
ESO 426-2   &   1.75 &   1.71 &   1.76 & NGC 2665    &   4.42 &   3.84 &   3.02 & PGC 2570478 &   1.02 &   1.05 &   1.05 \\
ESO 437-33  &   1.52 &   1.39 &   1.35 & NGC 2766    &   1.10 &   0.93 &   0.82 & UGC  4596   &   2.03 &   1.65 &   1.32 \\
ESO 437-67  &   2.66 &   2.51 &   3.07 & NGC 3081    &   1.91 &   .... &   1.57 & UGC  4771   &   1.79 &   1.66 &   1.53 \\
ESO 566-24  &   1.33 &   1.34 &   1.42 & NGC 3380    &   2.22 &   1.98 &   1.79 & UGC  5380   &   1.72 &   1.75 &   1.72 \\
ESO 575-47  &   2.56 &   2.25 &   2.31 & NGC 4113    &   3.54 &   4.01 &   4.61 & UGC  5885   &   3.10 &   3.21 &   2.75 \\
IC 1223     &   1.19 &   1.07 &   1.08 & NGC 4608    &   1.69 &   1.70 &   1.70 & UGC  9418   &   1.89 &   1.81 &   1.65 \\
IC 1438     &   1.93 &   1.64 &   1.55 & NGC 4736    &   0.98 &   1.07 &   0.00 & UGC 10168   &   1.08 &   1.27 &   1.09 \\
IC 2473     &   2.72 &   2.36 &   2.28 & NGC 4935    &   1.53 &   1.43 &   1.30 & UGC 10712   &   1.85 &   1.63 &   1.51 \\
IC 2628     &   0.88 &   0.76 &   0.80 & NGC 5132    &   2.98 &   2.81 &   2.45 & UGC 12646   &   3.05 &   2.88 &   .... \\
            &        &        &        &            &        &        &        &             &        &        &       \\ 
\noalign{\centerline{(b) $r$[Max($\Delta\mu$)]}}
            &        &        &        &            &        &        &        &             &        &        &       \\ 
CGCG 8-10   &   21.2 &   12.8 &   12.8 & IC 4214    &   29.1 &   29.7 &   29.7 & NGC 5211    &   17.2 &   18.0 &   18.4 \\
CGCG 13-75  &    8.4 &    8.4 &    9.2 & MCG 6-32-24 &    8.8 &    8.4 &    8.4 & NGC 5335  &   16.0 &   15.2 &   15.2 \\
CGCG 65-2   &   10.8 &   10.0 &    5.2 & MCG 7-18-40 &    6.4 &    5.2 &    4.8 & NGC 5370  &   11.6 &   11.2 &   12.0 \\
CGCG 67-4   &    8.8 &    9.2 &   10.0 & NGC  210    &   40.9 &   17.0 &   17.0 & NGC 5686  &    4.4 &    4.8 &    4.4 \\
CGCG 73-53  &    6.8 &    7.2 &    6.8 & NGC 1079    &   42.7 &   42.2 &   40.0 & NGC 5701  &   33.2 &   33.2 &   32.8 \\
CGCG 185-14 &    7.6 &    7.2 &    7.6 & NGC 1291    &   88.9 &   95.6 &   87.4 & NGC 6782  &   26.5 &   27.2 &   29.1 \\
CGCG 263-22 &   21.2 &   22.4 &   21.0 & NGC 1326    &   38.0 &   38.2 &   39.0 & NGC 7098  &   86.8 &   87.6 &   85.6 \\
ESO 325-28  &   13.2 &   12.7 &   12.7 & NGC 1398    &   45.7 &   45.5 &   43.9 & PGC 54897 &   10.0 &    9.6 &    9.2 \\
ESO 365-35  &   15.9 &   14.3 &   13.8 & NGC 1433    &  110.5 &   .... &  111.8 & PGC 1857116 &   12.6 &   11.4 &   11.8 \\
ESO 426-2   &   18.7 &   15.9 &   15.9 & NGC 2665    &   33.4 &   33.9 &   33.9 & PGC 2570478 &    7.6 &    6.8 &    6.8 \\
ESO 437-33  &   15.4 &   14.8 &   23.6 & NGC 2766    &   16.0 &   15.4 &   15.4 & UGC  4596   &   15.6 &   16.0 &   14.8 \\
ESO 437-67  &   39.0 &   39.6 &   44.0 & NGC 3081    &   36.4 &   .... &   36.4 & UGC  4771   &   11.6 &   11.2 &   10.8 \\
ESO 566-24  &   22.0 &    9.9 &   10.4 & NGC 3380    &   20.0 &   19.6 &   21.6 & UGC  5380   &    7.6 &    7.6 &    8.0 \\
ESO 575-47  &   31.4 &   29.7 &   25.9 & NGC 4113    &   19.2 &   20.0 &   20.8 & UGC  5885   &   21.2 &   21.6 &   21.2 \\
IC 1223     &   20.4 &    9.0 &    9.2 & NGC 4608    &   34.0 &   30.4 &   31.6 & UGC  9418   &   10.4 &    9.6 &    9.6 \\
IC 1438     &   22.6 &   21.8 &   21.0 & NGC 4736    &  216.8 &  234.0 &   71.2 & UGC 10168   &   16.8 &   23.2 &   16.4 \\
IC 2473     &   29.2 &   29.2 &   33.2 & NGC 4935    &   10.4 &   10.0 &   10.0 & UGC 10712   &    8.0 &    7.6 &    7.6 \\
IC 2628     &    8.0 &    5.2 &    5.2 & NGC 5132    &   35.2 &   36.4 &   33.2 & UGC 12646   &   23.2 &   22.5 &   .... \\
\hline
\end{tabular}
\end{table*}

\begin{table*}
\centering
\caption{Effective radii for full sample. Col. 1: galaxy name; col. 2: effective radii in $g$-band (GZ2) or $B$-band (non-GZ2);
col. 3: effective radii in $r$-band (GZ2) or $V$-band (non-GZ2); col. 4: effective radii
in $i$-band (GZ2) or $I$-band (non-GZ2). All values in arcseconds.
}
\label{tab:effrads}
\begin{tabular}{lrrrlrrrlrrr}
\hline
Galaxy & $a_e(g,B)$ & $a_e(r,V)$ & $a_e(i,I)$ &
Galaxy & $a_e(g,B)$ & $a_e(r,V)$ & $a_e(i,I)$ &
Galaxy & $a_e(g,B)$ & $a_e(r,V)$ & $a_e(i,I)$ \\
 1 & 2 & 3 & 4 & 
 1 & 2 & 3 & 4 & 
 1 & 2 & 3 & 4  \\
\hline
CGCG 8-10        &    8.5 &    7.2 &    7.4 & IC 4214          &   26.9 &   25.8 &   22.7 & NGC 5211         &   18.3 &   15.5 &   14.2 \\
CGCG 13-75       &    7.8 &    6.4 &    6.2 & MCG 6-32-24      &    6.6 &    6.4 &    6.5 & NGC 5335         &   30.5 &   26.6 &   24.9 \\
CGCG 65-2        &    6.9 &    6.2 &    6.0 & MCG 7-18-40      &    8.8 &    8.0 &    7.8 & NGC 5370         &   11.8 &   10.9 &   10.9 \\
CGCG 67-4        &    6.0 &    5.8 &    5.8 & NGC 210          &   .... &   37.1 &   .... & NGC 5686         &    6.1 &    6.0 &    6.2 \\
CGCG 73-53       &    5.2 &    4.8 &    5.0 & NGC 1079         &   34.8 &   33.6 &   31.1 & NGC 5701         &   39.3 &   33.6 &   32.9 \\
CGCG 185-14      &    8.9 &    7.8 &    7.6 & NGC 1291         &   89.0 &   78.8 &   71.2 & NGC 6782         &   19.7 &   19.3 &   18.2 \\
CGCG 263-22      &   11.4 &   10.7 &   10.6 & NGC 1326         &   30.8 &   30.2 &   29.2 & NGC 7098         &   49.1 &   45.4 &   38.2 \\
ESO 325-28       &   12.6 &   11.8 &    9.9 & NGC 1398         &   63.0 &   58.4 &   60.2 & PGC 54897        &    7.0 &    6.5 &    6.4 \\
ESO 365-35       &   10.4 &    9.5 &    8.2 & NGC 1433         &   82.1 &   .... &   71.1 & PGC 1857116      &    8.0 &    7.5 &    7.2 \\
ESO 426-2        &   14.3 &   12.1 &   10.8 & NGC 2665         &   22.6 &   21.0 &   19.7 & PGC 2570478      &    7.5 &    7.3 &    7.2 \\
ESO 437-33       &   10.1 &    9.3 &    7.7 & NGC 2766         &   13.9 &   12.7 &   12.4 & UGC 4596         &   12.1 &   10.3 &    9.8 \\
ESO 437-67       &   30.1 &   26.5 &   22.7 & NGC 3081         &   27.0 &   .... &   24.5 & UGC 4771         &   14.3 &   12.2 &   11.9 \\
ESO 566-24       &   19.6 &   17.6 &   16.4 & NGC 3380         &   15.0 &   14.8 &   15.2 & UGC 5380         &   13.2 &   12.1 &   11.9 \\
ESO 575-47       &   23.2 &   21.3 &   19.4 & NGC 4113         &    8.2 &    7.4 &    7.3 & UGC 5885         &   14.6 &   12.7 &   12.4 \\
IC 1223          &    9.1 &    8.8 &    8.9 & NGC 4608         &   33.8 &   31.9 &   30.4 & UGC 9418         &   10.8 &    8.8 &    8.4 \\
IC 1438          &   19.8 &   18.5 &   16.8 & NGC 4736         &   61.2 &   .... &   58.7 & UGC 10168        &   13.4 &   12.1 &   12.2 \\
IC 2473          &   17.9 &   16.2 &   15.6 & NGC 4935         &   12.7 &   10.7 &   10.3 & UGC 10712        &   10.9 &    7.8 &    7.6 \\
IC 2628          &   10.3 &    7.8 &    7.3 & NGC 5132         &   16.9 &   16.9 &   17.2 & UGC 12646        &   17.6 &   16.0 &   .... \\
\hline
\end{tabular}
\end{table*}

\begin{figure}
\centering
\includegraphics[width=\columnwidth,bb=14 14 581 326]{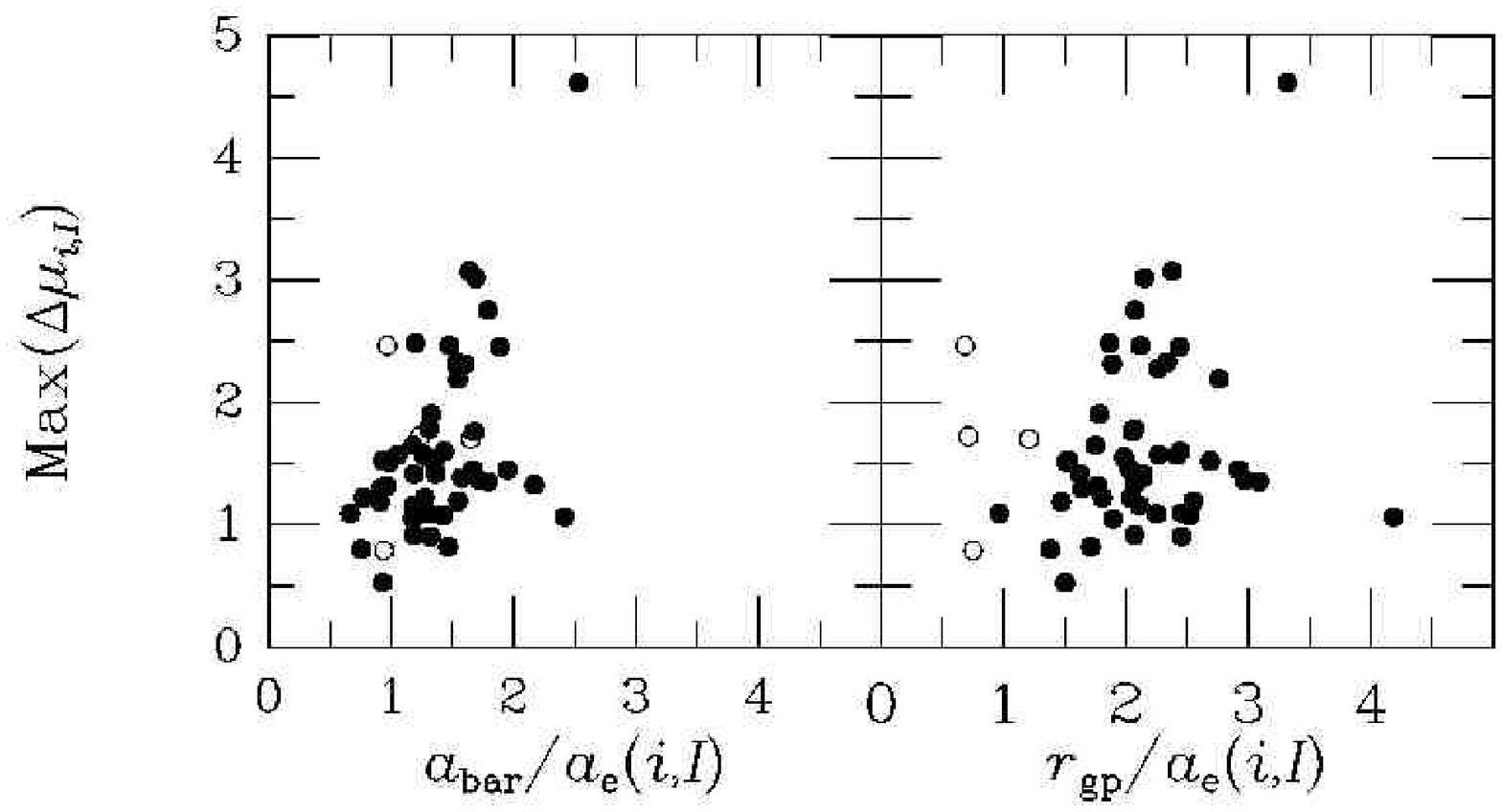}
\caption{Graph showing Max[$\Delta\mu$($i$,$I$)] (mag arcsec$^{-2}$)
versus (left) the bar radius and (right) the gap radius relative to
the effective radius $a_e$ in infrared light.
}
\label{fig:aeff}
\end{figure}

Figure~\ref{fig:aeff}, left, shows a graph similar to Figure 6 of Kim
et al. (2016), except that here the bar radius is normalized to the
effective (half-power) radius $a_e(i,I)$ in the infrared bands, rather
than an isophotal radius. These radii were derived from the integrated
luminosity distributions and the total magnitudes and colours listed in
Tables~\ref{tab:totmags} and ~\ref{tab:totmagsBVI}.  The values of
$a_e(i,I)$ as well as $a_e(g,B)$ and $a_e(r,V)$ in arcseconds are
listed in Table~\ref{tab:effrads}. The points for (rR) dark-spacers in
Figure~\ref{fig:aeff} are indicated by filled circles while those for
the four (r) dark-spacers are indicated by open circles. In this graph,
a clear correlation is seen, although it is not as strong as that found
by Kim et al. The Spearman rank correlation coefficient is $\rho$ =0.41
with a statistical significance of 0.001. In this graph, the (r)
dark-spacers blend with the the (rR) dark-spacers and do not stand out.
In Figure~\ref{fig:aeff}, right, however, which plots
Max[$\Delta\mu(i,I)$] versus the mean gap radius $r_{gp}$ (listed as
$r_{CR}$ in Table~\ref{tab:resrads}) normalized to $a_e(i,I)$ shows a
poorer correlation and that the (r) dark-spacers stand out relative to
the (rR) dark-spacers.

\begin{figure}
\centering
\includegraphics[width=\columnwidth,bb=14 14 575 383]{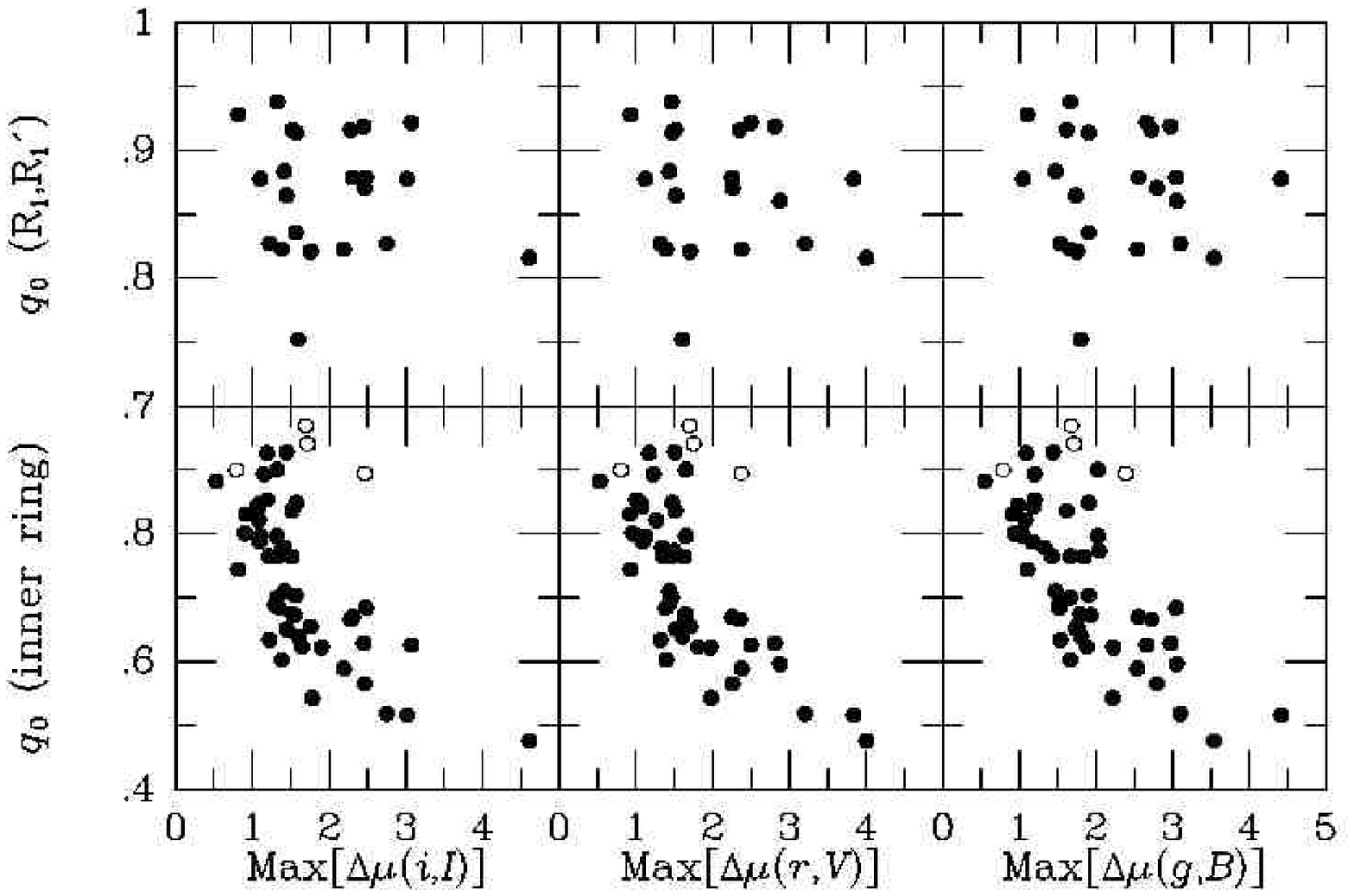}
\caption{Graphs showing intrinsic axis ratios versus
Max[$\Delta\mu$($i$,$I$)] (mag arcsec$^{-2}$) for inner rings and
related features (lower frames) and outer \Rone\ and \RoneP\ outer
rings (upper frames). Filled circles are for rings in (rR)
dark-spacers, open circles are for rings in (r) dark-spacers.}
\label{fig:q0rings}
\end{figure}

In the context of ring shapes, both manifold theory (Athanassoula et
al. 2009b) and resonance theory (Schwarz 1984b; Salo et al. 1999)
predict that the intrinsic axis ratios of inner rings and outer \Rone,
\RoneP\ rings in barred and oval galaxies should depend significantly
on bar strength. Figure~\ref{fig:q0rings} shows graphs of the deprojected
axis ratios $q_0$ versus Max($\Delta\mu$) for these ring types in the
GZ2/CSRG sample. In the inner ring graphs, (r) dark-spacers are
distinguished from (rR) dark-spacers using open circles. If
Max($\Delta\mu$) is taken as an indicator of bar strength, then
inner ring shapes show a significant correlation with bar strength (see
also Grouchy et al. 2010). The Spearman rank correlation coefficients
are $\rho(g,B)$=$-$0.646 ($P<$10$^{-6}$), $\rho(r,V)$=$-$0.588
($P=$3$\times$10$^{-6}$), $\rho(i,I)$=$-$0.588
($P=$2.5$\times$10$^{-6}$), for $N$= 53, 51, and 52 inner rings,
respectively, from Tables~\ref{tab:inner-features} and
~\ref{tab:rdksp}.

In contrast, the present sample shows little dependence of \Rone,
\RoneP\ intrinsic ring shapes on Max($\Delta\mu)$. Restricting to the
best-defined cases as depicted in Figure~\ref{fig:results}, the
Spearman rank correlation coefficients are $\rho(g,B)$=$-$0.166
($P$=0.225), $\rho(r,V)$=$-$0.220 ($P$=0.170), and $\rho(i,I)$=$-$0.139
($P$=0.268), for $N$= 23, 21, and 22 rings, respectively, from
Table~\ref{tab:R1gals}.

\section{Summary}

This paper has developed the ``gap method" of locating the radius of
the corotation resonance in mainly early-to-intermediate type barred
and oval disk galaxies. The study was motivated by the extreme dark
spaces noticed in some galaxies in the GZ2 ring sample (paper I),
specifically UGC 4596 and NGC 5335. These represent two types of
``dark-spacer" morphologies:

\noindent
- (rR) dark-spacers, where gaps of low surface brightness are found between
inner and outer ring features, the latter often including features classified
as \Rone, \RoneP, \RtwoP, and \Rone\RtwoP, outer rings and pseudorings; and\br
\noindent
- (r) dark-spacers, where dark zones lie inside an inner ring crossed by
a conspicuous bar.\br

The paper makes the assumption that the gaps in these galaxies are
linked to the $L_4$ and $L_5$ Lagrangian points that would exist in a
bar field. Two approaches have been used to derive the gap radius $r_{gp}$:
direct parabolic fits to the gap surface brightnesses (when these are
deep) and parabolic fits to residual intensities after subtraction of a
heavily median-smoothed background (when the gap surface brightnesses
are shallow). Taking the corotation radius $r_{CR}$ = $r_{gp}$, usually
based on an average of three filters, this paper has shown that

\noindent
- Most R$_1$ and R$_1^{\prime}$ outer rings and pseudorings are likely
associated with the outer 4:1 resonance, not the OLR as was originally
suggested by Schwarz (1981). Only 1 out of the 36 cases analyzed have
$a/r_{OLR}$$\approx$1 as predicted by Schwarz (1981), Some support for
an association with $r_{O4R}$ in general comes from numerical
simulation studies of NGC 1433 (Treuthardt et al. 2008), ESO 566$-$24
(Rautiainen et al. 2004) and NGC 6782 (Lin et al. 2008), where the
best-fitting pattern speeds place an R$_1^{\prime}$ ring closer to
$r_{O4R}$ than to $r_{OLR}$. R$_1$ and R$_1^{\prime}$ rings are
approximately aligned perpendicular to the bar in our sample.

\noindent
- The gap method links R$_2^{\prime}$ outer pseudorings to the
OLR. The predictions from Schwarz (1981) are confirmed for these
features. 

The revised interpretation of R$_1$ and R$_1^{\prime}$ rings means that
the combined category, R$_1$R$_2^{\prime}$, is actually a two-resonance
ring morphology rather than both being associated with the OLR. Since all of
these ring types are completely outside the CR radius (no visual
mapping led to any of these features significantly crossing this
radius), then a better way to describe them is as ``outer resonant
subclasses."

\noindent
- Virtually all of the inner features in the (rR) dark-spacer
sample are placed very close to or straddling the inner
4:1 resonance. In no case does an inner feature significantly cross the
CR radius. This association is consistent with Schwarz (1984a) and
Simkin, Su, \& Schwarz (1980).

\noindent
- With few exceptions, the bars identified in the (rR) sample appear to
extend to the I4R; some come close to \rcr. The average value
$<\cal{R}>$ = 1.58 $\pm$ 0.04 (standard deviation 0.28) is large
compared to previous studies, but cannot be dismissed completely
because many of the bars in the sample of galaxies studied here lie
inside an inner ring or related feature. If inner rings are linked to
the inner 4:1 resonance as has been strongly suggested by the models of
Schwarz (1984a) and Simkin, Su, \& Schwarz (1980), then the bars within
such rings cannot extend much beyond the I4R. Debattista \& Sellwood
(2000) classified a bar as slow if $\cal{R}$ $>$ 1.4 and ``fast" if $\cal{R}$
$<$ 1.4. The analysis in this paper places these galaxies into the slow
bar domain for the estimated bar radii.

The situation for (r) dark-spacers is less certain. If the dark regions
in these galaxies also trace the $L_4$ and $L_5$ points, then the bars
would exceed their corotation resonance radii, violating the
Contopoulos (1980) rule. This paper has shown that (r) dark-spacers
have rounder inner rings, stronger high-order relative Fourier
intensity amplitudes, and lower values of $r_{gp}$ relative to the
near-infrared effective radii of the galaxies as compared to (rR)
dark-spacers. Because the azimuthally-averaged colours are uniform
across the bar-gap region in NGC 5335, it seems most likely that bar
evolution is responsible for (r) dark-spacers in the sense that the bar
depletes the old disk stars and grows stronger with time, probably by
the mechanism highlighted by Kim et al. (2016). Indeed, on the basis of
the uniform average colours across the gap, Buta (2014) had already
noted: ``Dust may not be an issue, because the spaces [in NGC 5335] are
still dark even in the 2.1$\mu$m $K_s$-band (Buta et al. 2009). Could
this imply that the bar in NGC 5335 has recently assembled itself from
the old disk stellar population?"

\noindent
- Analysis of (rR) gap colours and the colours of points along the bar
axis at the same radius shows that dark gaps are redder than the bar
axis points by $\approx$0.1 mag on average in all colours used.
Although the trends are roughly parallel to the reddening lines in
colour-colour diagrams, it is most likely that the colour differences
are due to a stellar population difference, rather than higher internal
extinction in the gaps. Gaps have average $BVI$ and $ugri$ colours
comparable to the integrated colours of Sa galaxies, while the bar axis
points have average colours similar to the integrated colours of Sb
galaxies.  Because (rR) dark-spacer gaps are not necessarily the same
colour as the bar axis points, owing to the frequent presence of
star-formation around the ends of the bars or ovals of these galaxies,
it is likely that these gaps are not simply a question of evolutionary
state of the bar and therefore just part of a continuum with (r)
dark-spacers. Instead, the $L_4$, $L_5$ points and their stability
provide the most reasonable interpretation of the dark gaps seen in
(rR) dark-spacers.

\noindent
Finally, several important issues remain:

\noindent
(1) Although many galaxies in the GZ2/CSRG sample show the characteristic
(rR) dark spaces, many, if not more, early-type disk galaxies do not
show these features, even cases with an apparently strong bar. Do such
cases imply that the $L_4$, $L_5$ Lagrangian points are stable, thus
preventing the expected depletion of material around the points, or
that these points are located so far out that the surface brightness is
too low to see them? Why should these points be consequential in some
barred and oval galaxies, and largely inconsequential in others?

\noindent
(2) Even though a single pattern speed can account for much of the
structure of galaxies showing R$_1$, R$_1^{\prime}$, R$_2^{\prime}$,
and R$_1$R$_2^{\prime}$ outer rings and pseudorings, and that numerical
simulation surveys (e.g., Rautiainen et al. 2008; Treuthardt et al.
2012) have shown that single pattern speed models can effectively
reproduce structure even in galaxies without these features, other
methods used to locate corotation resonances in galaxies have
nevertheless strongly favoured the idea of multiple pattern speeds
(e.g., Corsini et al. 2003; Buta \& Zhang 2009; Font et al. 2011,
2014). This idea has also found some support from theory (e.g.,
Sellwood \& Sparke 1988; Pfenniger \& Norman 1990; Rautiainen \& Salo
2000). These issues will be furthered examined in later papers in this
series.

\noindent
(3) Although flat rotation curves have been assumed for the analysis in
this paper, the shape of the rotation curve could have a bearing on
whether a galaxy might be seen to be an (rR) dark-spacer or an (r)
dark-spacer. The referee has noted that some galaxies in the sample
(e.g., NGC 1433 and NGC 3081) could be interpreted as having both types
of dark spaces. The referee suggests that a slowly rising rotation
curve might place the outer resonances so far out that only an inner
ring would be seen. However, it is noteworthy that cases like NGC 1433
and NGC 3081 have extremely oval inner rings unlike the typical (r)
dark-spacers highlighted in this paper.

I thank the anonymous referee for many helpful comments that
substantially improved this paper. This work was supported in part by
a grant from the Research Grants Committee, University of Alabama. The
development of Galaxy Zoo was supported in part by the Alfred P. Sloan
Foundation and by the Leverhulme Trust. I thank K. L. Masters and Arfon
Smith for sending the links to the images of the sample of GZ2 ringed
galaxies.  This research has made use of the NASA/IPAC Extragalactic
Database (NED) which is operated by the Jet Propulsion Laboratory,
California Institute of Technology, under contract with the National
Aeronautics and Space Administration.  Funding for the creation and
distribution of the SDSS Archive has been provided by the Alfred P.
Sloan Foundation, the Participating Institutions, NASA, NSF, the U. S.
Department of Energy, the Japanese Monbukagakusho, and the Max Planck
Society.  IRAF is written and supported by the National Optical
Astronomy Observatories (NOAO) in Tucson, Arizona. NOAO is operated by
the Association of Universities for Research in Astronomy (AURA), Inc.,
under cooperative agreement with the National Science Foundation.

\section{Appendix: Azimuthally-averaged, Relative Fourier Intensity,
and Bar Minor Axis Profiles of 52 Ringed Galaxies}

The azimuthally-averaged profiles of the 31 GZ2 and 21 non-GZ2 ringed
galaxies in the sample (excluding UGC 4596 and NGC 5335 which were
described separately in section 5) are shown in Figure~\ref{fig:azim}.
These were obtained in each case by averaging the luminosity
distribution over fixed ellipses having a shape and position angle
equal to the disk axis ratio and major axis position angle listed in
Table~\ref{tab:orient}. These profiles were extrapolated to get total
magnitudes and colours by fitting exponentials to the outer points. In
most cases the slope was freely-fitted, but for the $u$-band it was
generally necessary to force the slope to be the same as found for the
$g$-band.

The relative Fourier intensity profiles for the same galaxies are
shown in Figure~\ref{fig:allfours}. (The profiles for UGC 4596 and NGC
5335 are shown in Figures ~\ref{fig:ugc4596c} and ~\ref{fig:NGC5335c},
respectively.) These are based only on the deprojected images in the
$i$-band. Since no decompositions were performed, a few of the galaxies
(e.g., NGC 2766 and CGCG 185-14) show inner amplitudes due to
significant bulge deprojection stretch. The relative intensity profiles
are shown only for the $m$ = 2, 4, and 6 components, while the phase is
shown only for the $m$ = 2 component in each case.

Figure~\ref{fig:allrgp} shows the bar minor profiles and how they were
fitted to derive $<r_{gp}>$. Those cases where the surface brightness
profiles were fitted directly show only one panel analogous to
Figures~\ref{fig:ugc4596-proc} and ~\ref{fig:NGC5335-proc} for UGC 4596
and NGC 5335, respectively.  These cases are indicated as ``gri" or
``BVI" in Table~\ref{tab:orient}. For cases where residual intensities
were fitted instead, the left panel shows three graphs of the residual
intensity profiles and the parabolic fits used to get $r_{gp}$ in each
filter, while the right panel shows the minor axis bar profiles and the
adopted value of $<r_{gp}>$ superposed as vertical lines.  These cases
are indicated as, for example, ``intgri" or ``intBVI" in
Table~\ref{tab:orient} and are based on fits to images after
subtraction of a heavily median-smoothed version of the same image. The
smoothing box ranged from 51$\times$51 to 161$\times$161 pixels,
depending roughly on the size of the galaxy.

 \setcounter{figure}{20}
 \begin{figure}
 \begin{minipage}[b]{0.49\linewidth}
 \centering
\includegraphics[width=\textwidth,trim=0 0 275 400,clip]{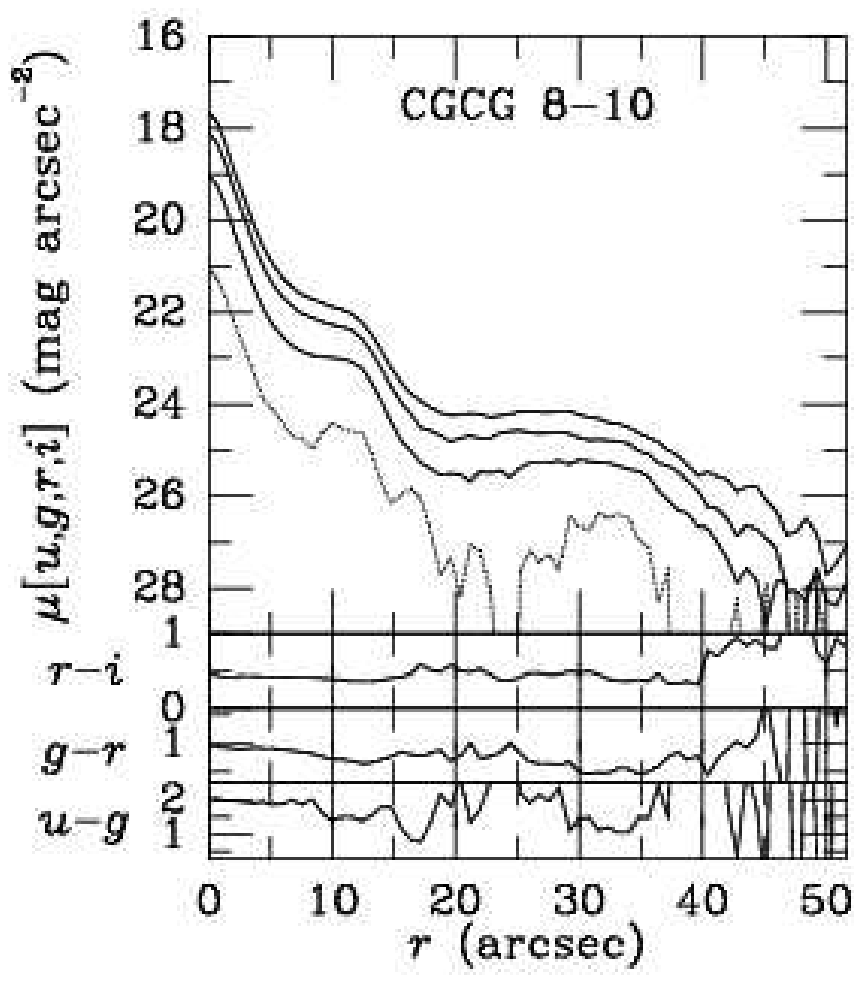}
 \hspace{0.1cm}
 \end{minipage}
 \begin{minipage}[t]{0.49\linewidth}
 \centering
\raisebox{0.35cm}{\includegraphics[width=\textwidth,trim=0 0 275 400,clip]{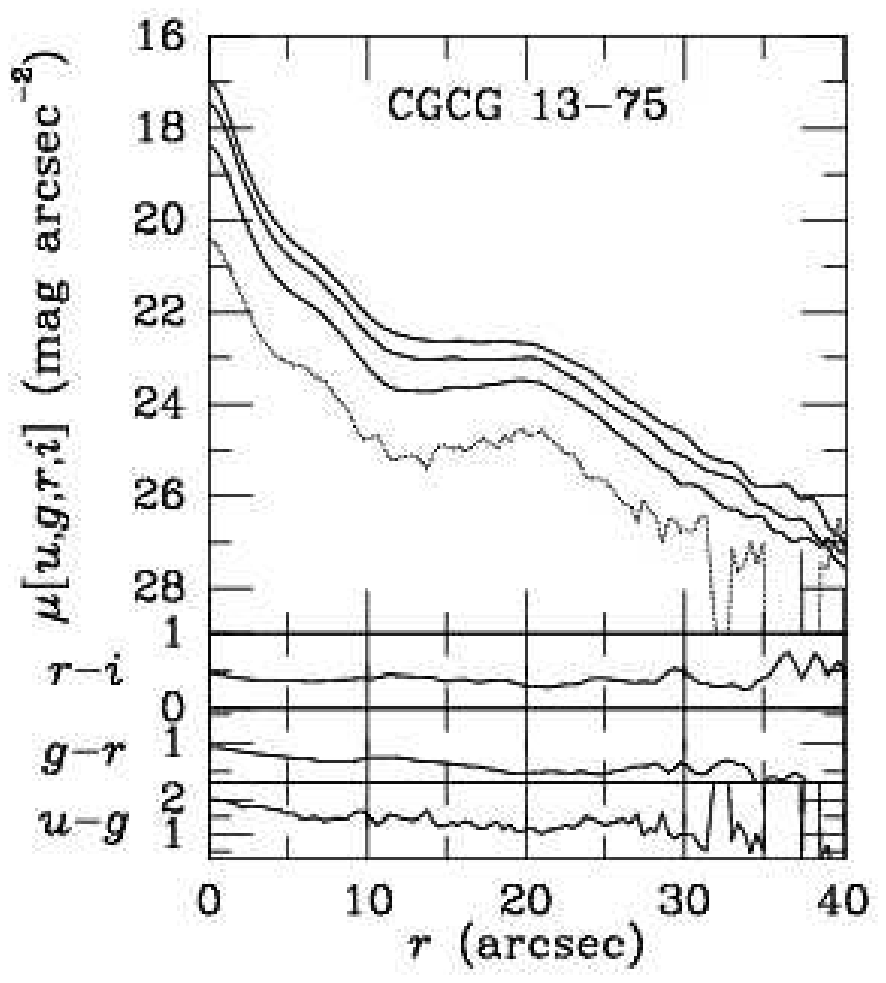}}
 \end{minipage}
 \begin{minipage}[b]{0.49\linewidth}
 \centering
\includegraphics[width=\textwidth,trim=0 0 275 400,clip]{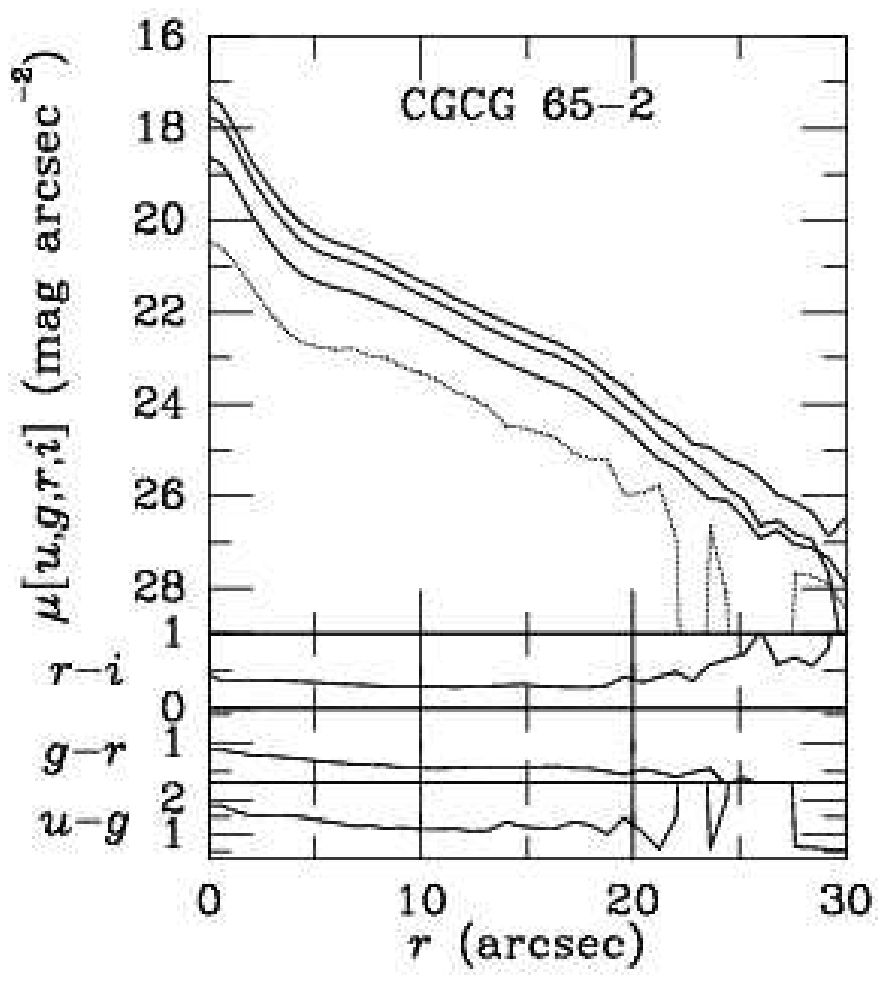}
 \hspace{0.1cm}
 \end{minipage}
 \begin{minipage}[t]{0.49\linewidth}
 \centering
\raisebox{0.35cm}{\includegraphics[width=\textwidth,trim=0 0 275 400,clip]{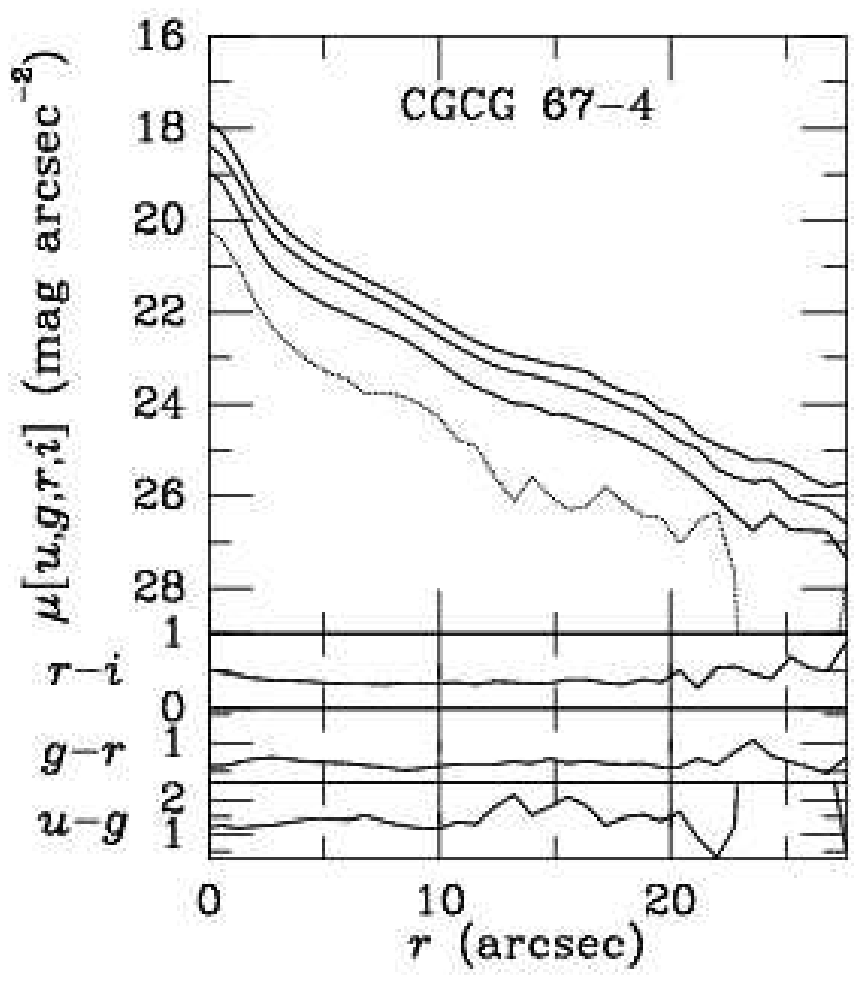}}
 \end{minipage}
 \begin{minipage}[b]{0.49\linewidth}
 \centering
\includegraphics[width=\textwidth,trim=0 0 275 400,clip]{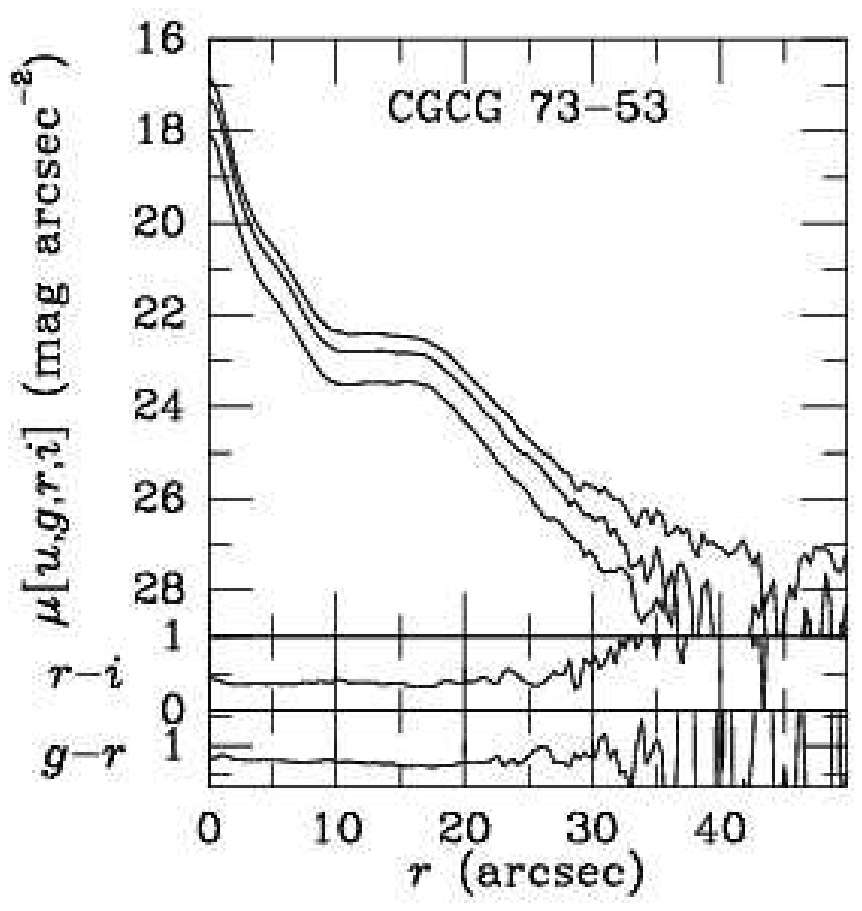}
 \hspace{0.1cm}
 \end{minipage}
 \begin{minipage}[t]{0.49\linewidth}
 \centering
\raisebox{0.35cm}{\includegraphics[width=\textwidth,trim=0 0 275 400,clip]{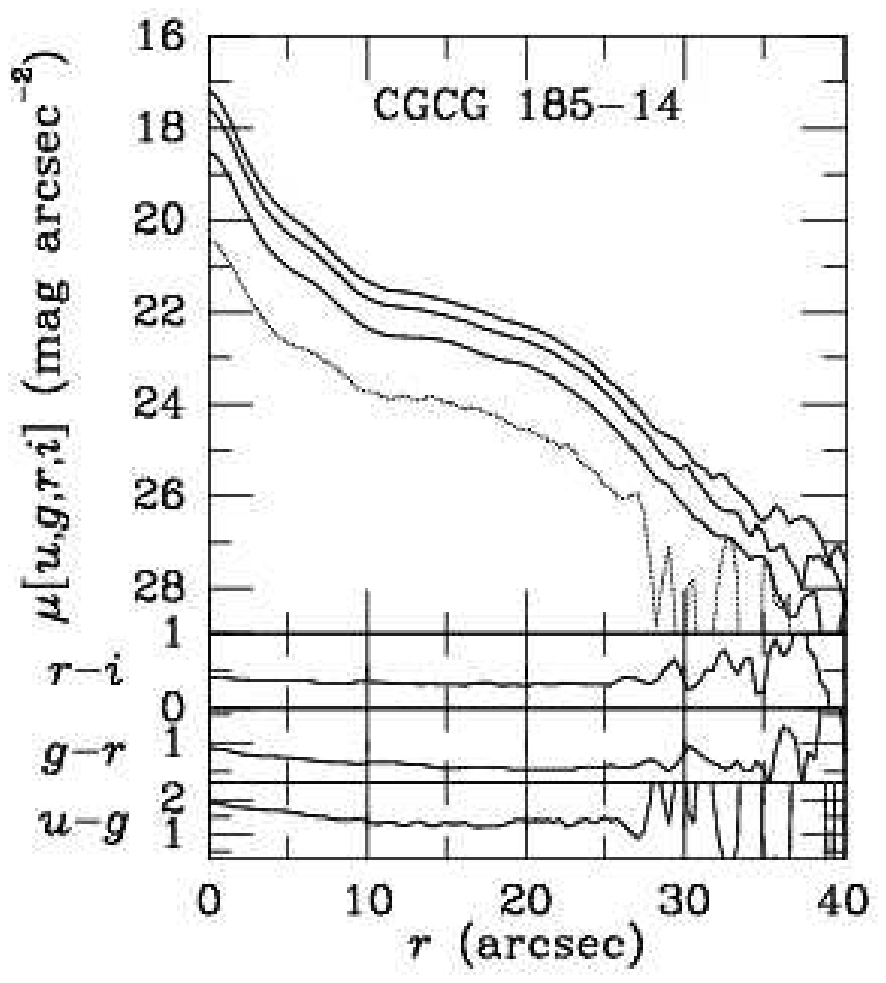}}
 \end{minipage}
 \begin{minipage}[b]{0.49\linewidth}
 \centering
\includegraphics[width=\textwidth,trim=0 0 275 400,clip]{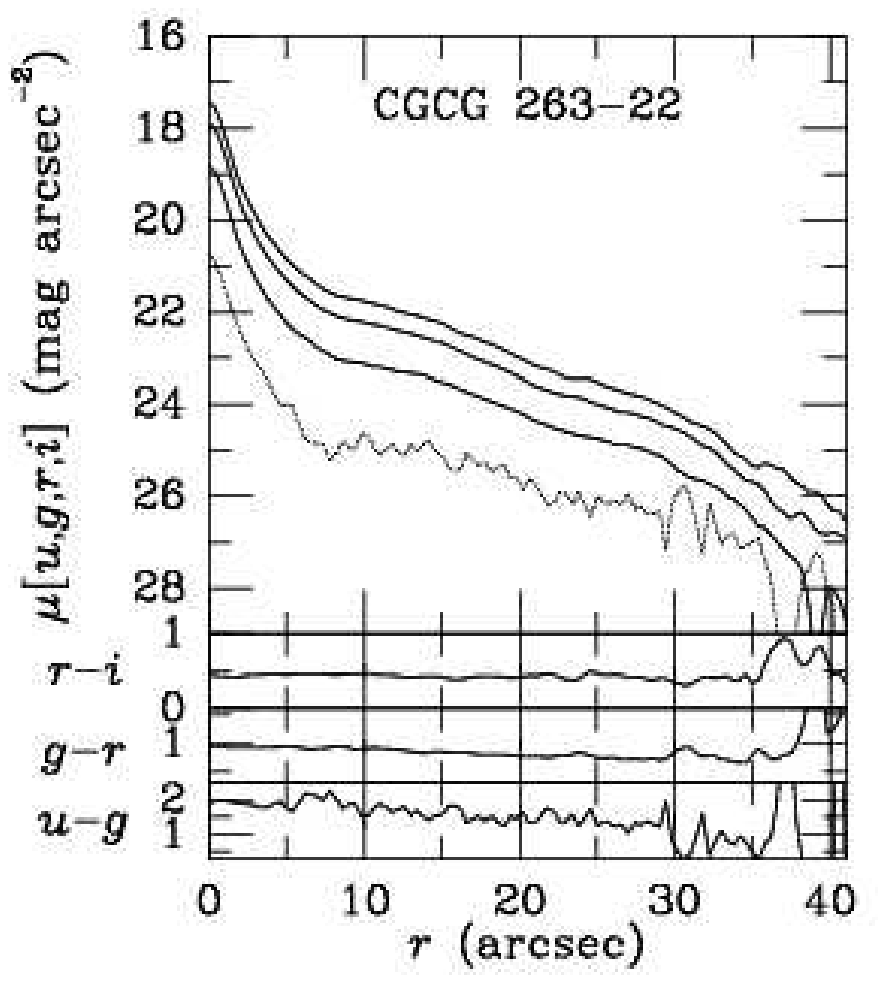}
 \hspace{0.1cm}
 \end{minipage}
 \begin{minipage}[t]{0.49\linewidth}
 \centering
\raisebox{0.35cm}{\includegraphics[width=\textwidth,trim=0 0 275 400,clip]{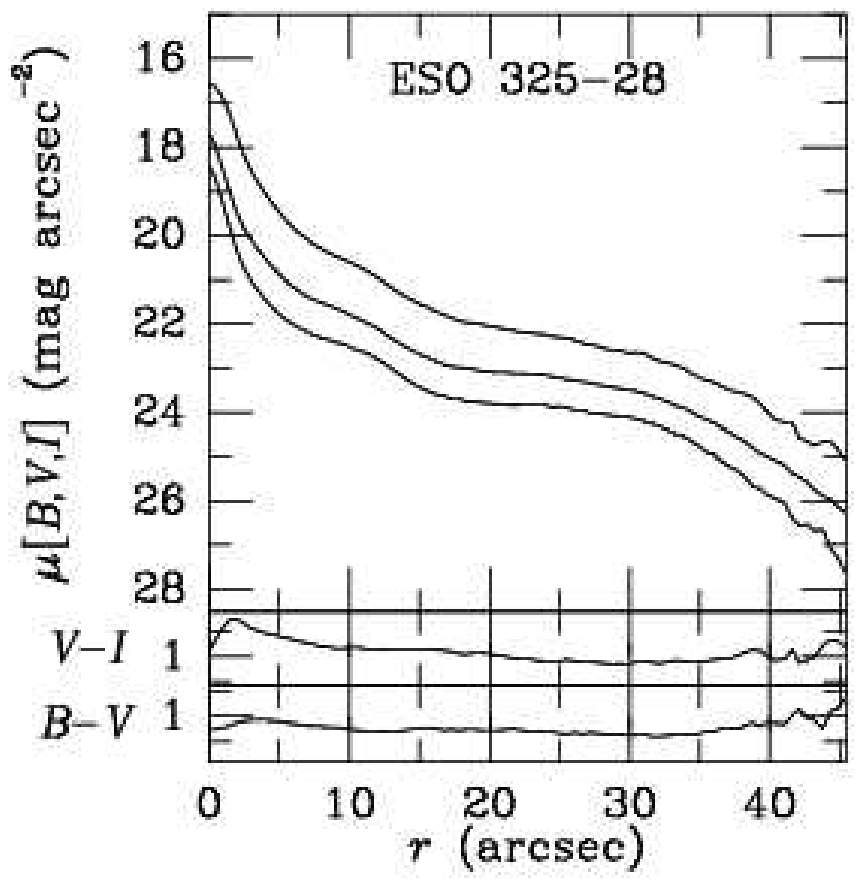}}
 \end{minipage}
 \begin{minipage}[b]{0.49\linewidth}
 \centering
\includegraphics[width=\textwidth,trim=0 0 275 400,clip]{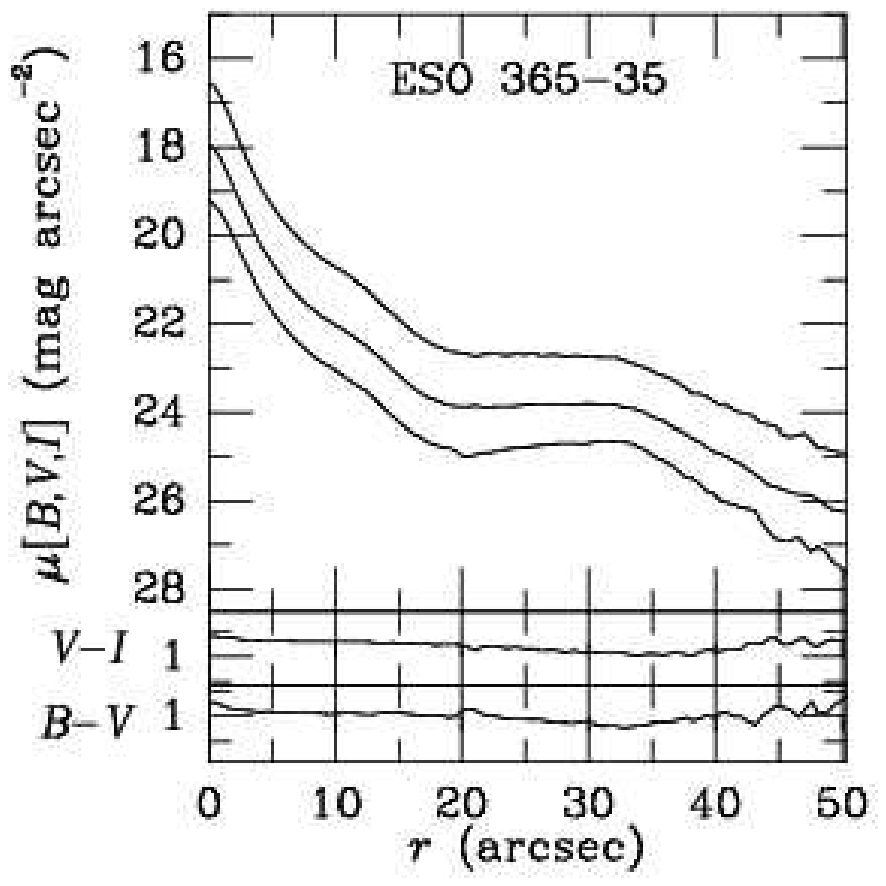}
 \hspace{0.1cm}
 \end{minipage}
 \begin{minipage}[t]{0.49\linewidth}
 \centering
\raisebox{0.35cm}{\includegraphics[width=\textwidth,trim=0 0 275 400,clip]{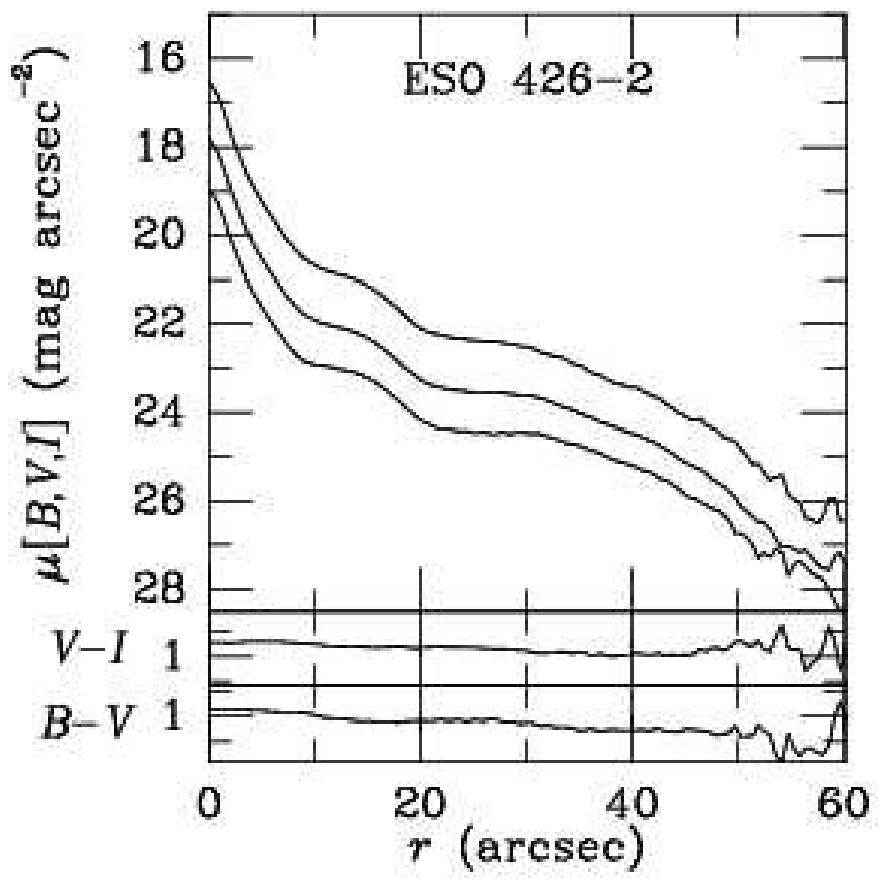}}
 \end{minipage}
\vspace{-1.0truecm}
\caption{
}
\label{fig:azim}
\end{figure}
 \setcounter{figure}{20}
 \begin{figure}
 \begin{minipage}[b]{0.49\linewidth}
 \centering
\includegraphics[width=\textwidth,trim=0 0 275 400,clip]{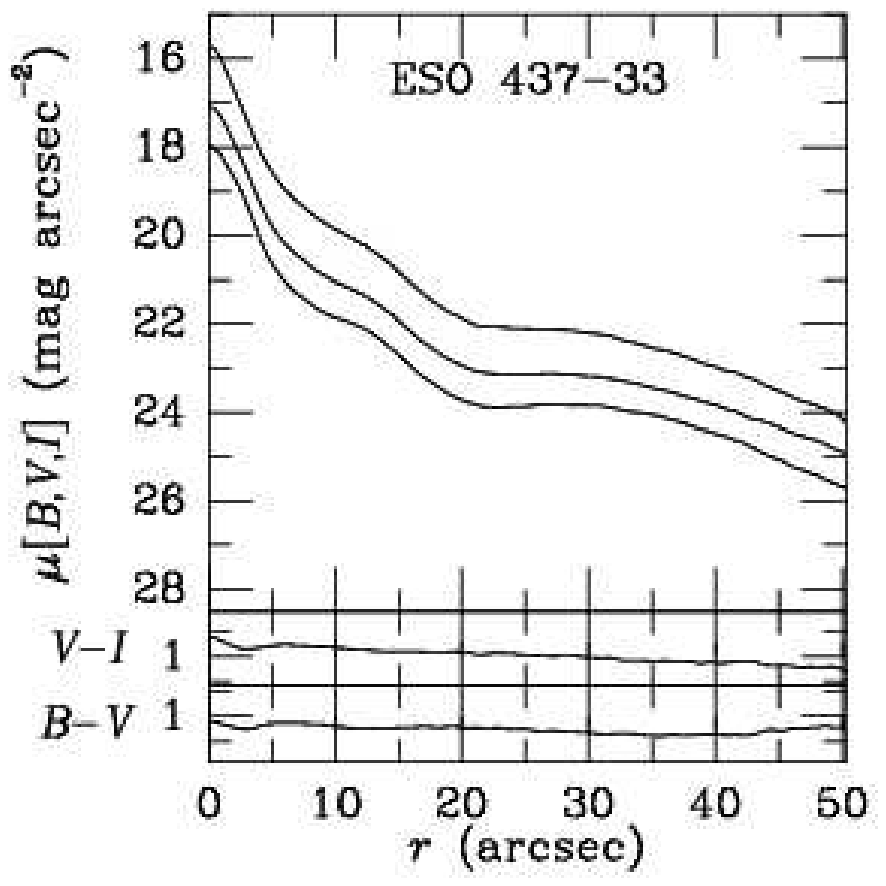}
 \hspace{0.1cm}
 \end{minipage}
 \begin{minipage}[t]{0.49\linewidth}
 \centering
\raisebox{0.35cm}{\includegraphics[width=\textwidth,trim=0 0 275 400,clip]{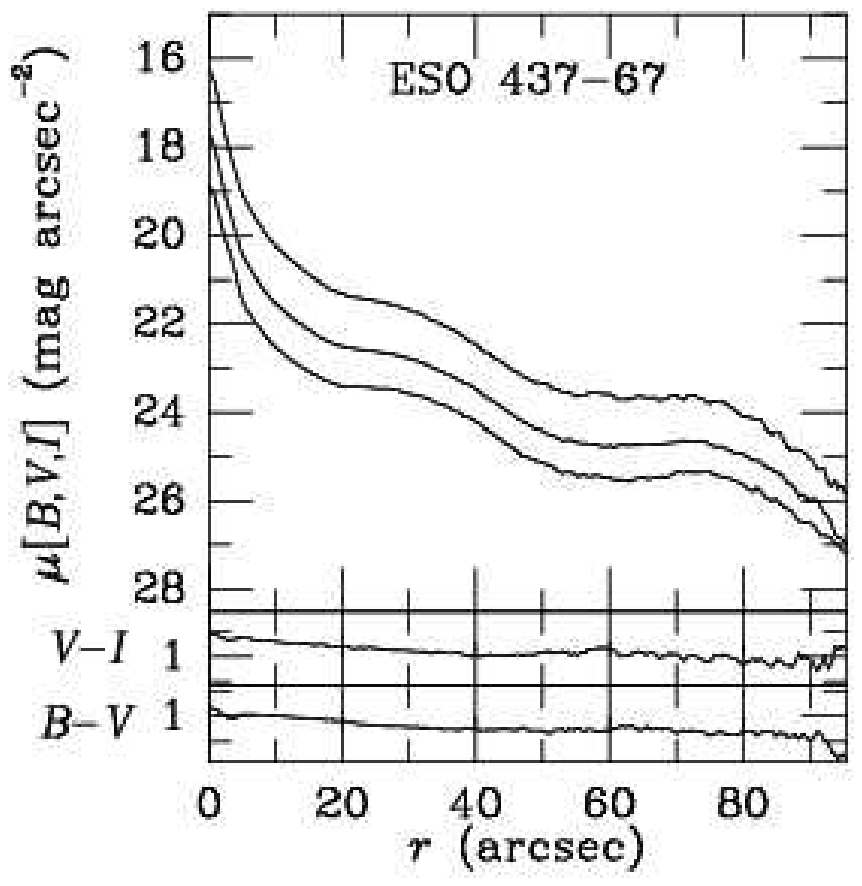}}
 \end{minipage}
 \begin{minipage}[b]{0.49\linewidth}
 \centering
\includegraphics[width=\textwidth,trim=0 0 275 400,clip]{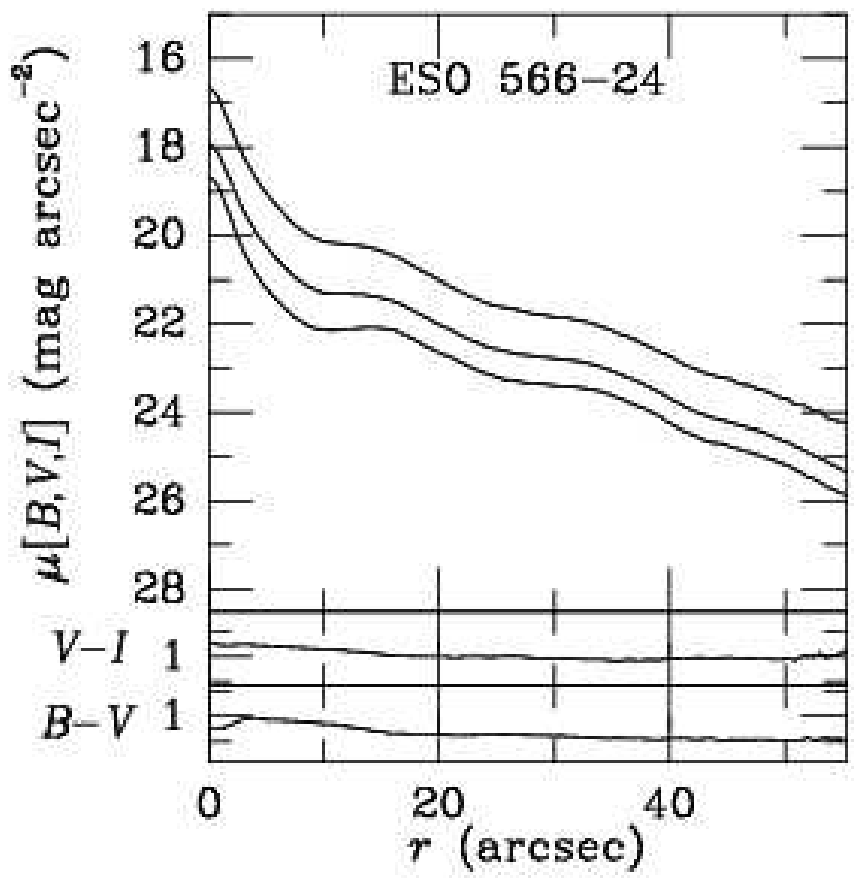}
 \hspace{0.1cm}
 \end{minipage}
 \begin{minipage}[t]{0.49\linewidth}
 \centering
\raisebox{0.35cm}{\includegraphics[width=\textwidth,trim=0 0 275 400,clip]{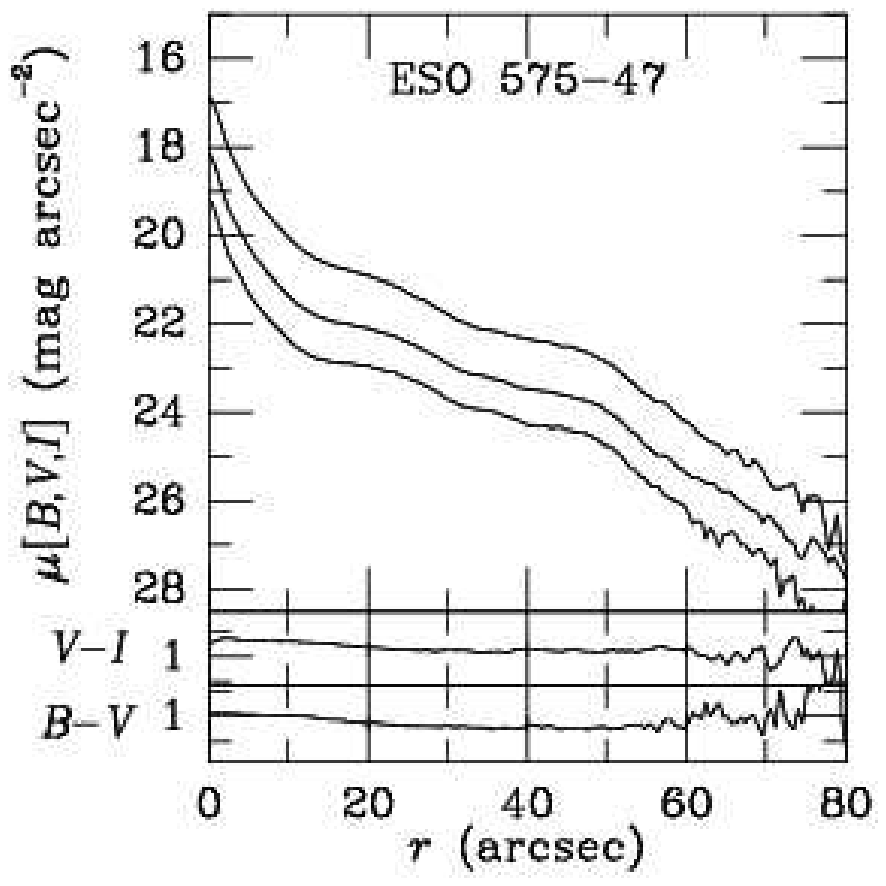}}
 \end{minipage}
 \begin{minipage}[b]{0.49\linewidth}
 \centering
\includegraphics[width=\textwidth,trim=0 0 275 400,clip]{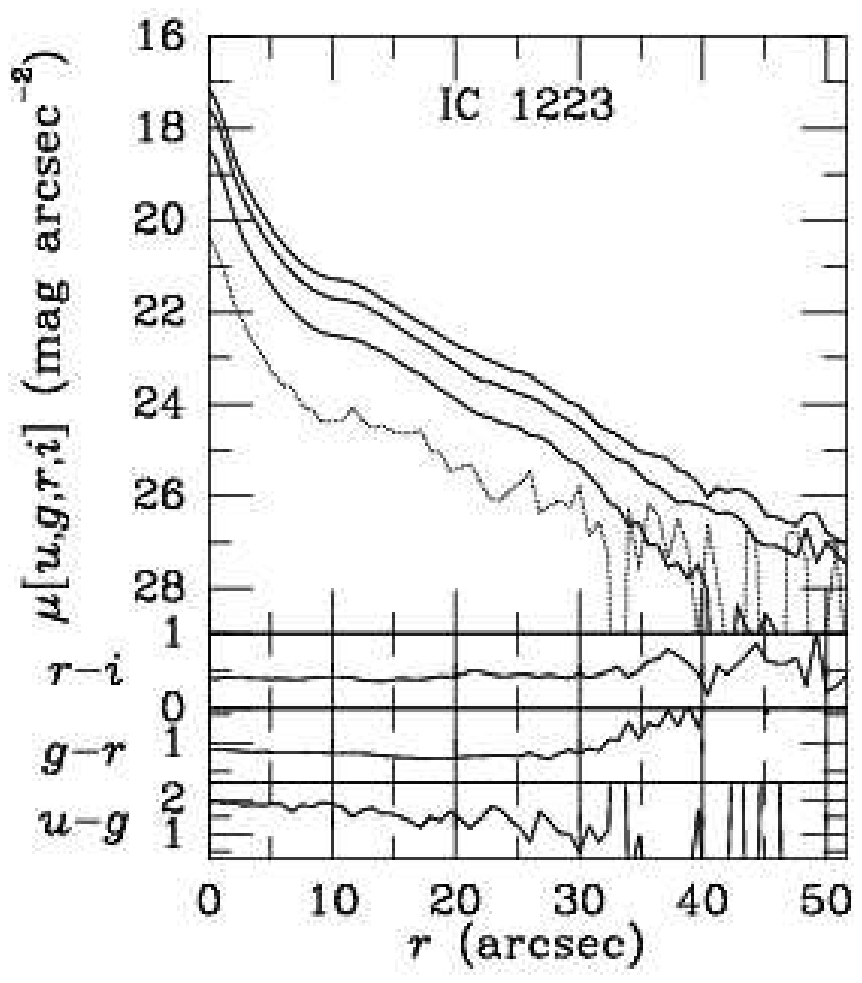}
 \hspace{0.1cm}
 \end{minipage}
 \begin{minipage}[t]{0.49\linewidth}
 \centering
\raisebox{0.35cm}{\includegraphics[width=\textwidth,trim=0 0 275 400,clip]{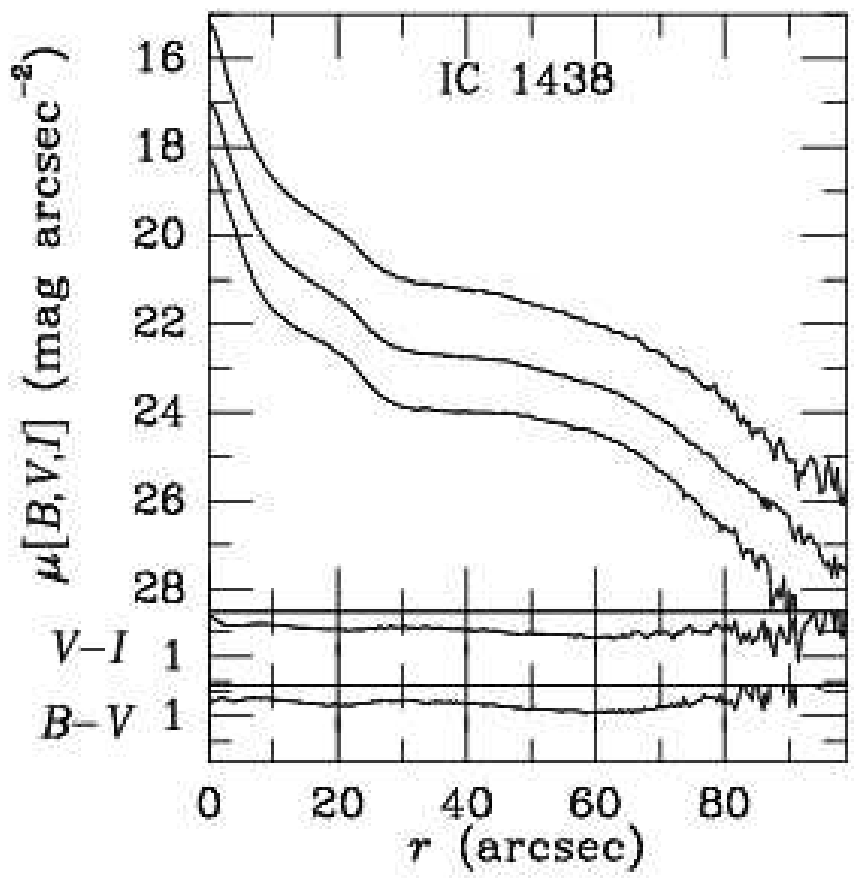}}
 \end{minipage}
 \begin{minipage}[b]{0.49\linewidth}
 \centering
\includegraphics[width=\textwidth,trim=0 0 275 400,clip]{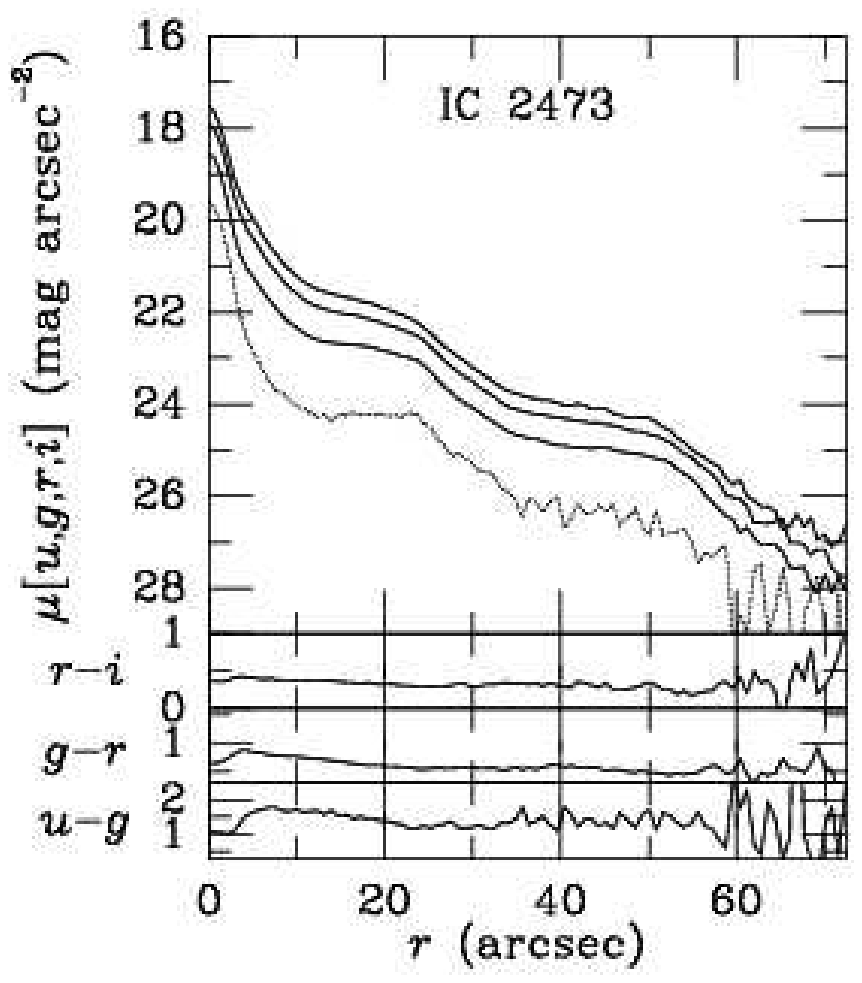}
 \hspace{0.1cm}
 \end{minipage}
 \begin{minipage}[t]{0.49\linewidth}
 \centering
\raisebox{0.35cm}{\includegraphics[width=\textwidth,trim=0 0 275 400,clip]{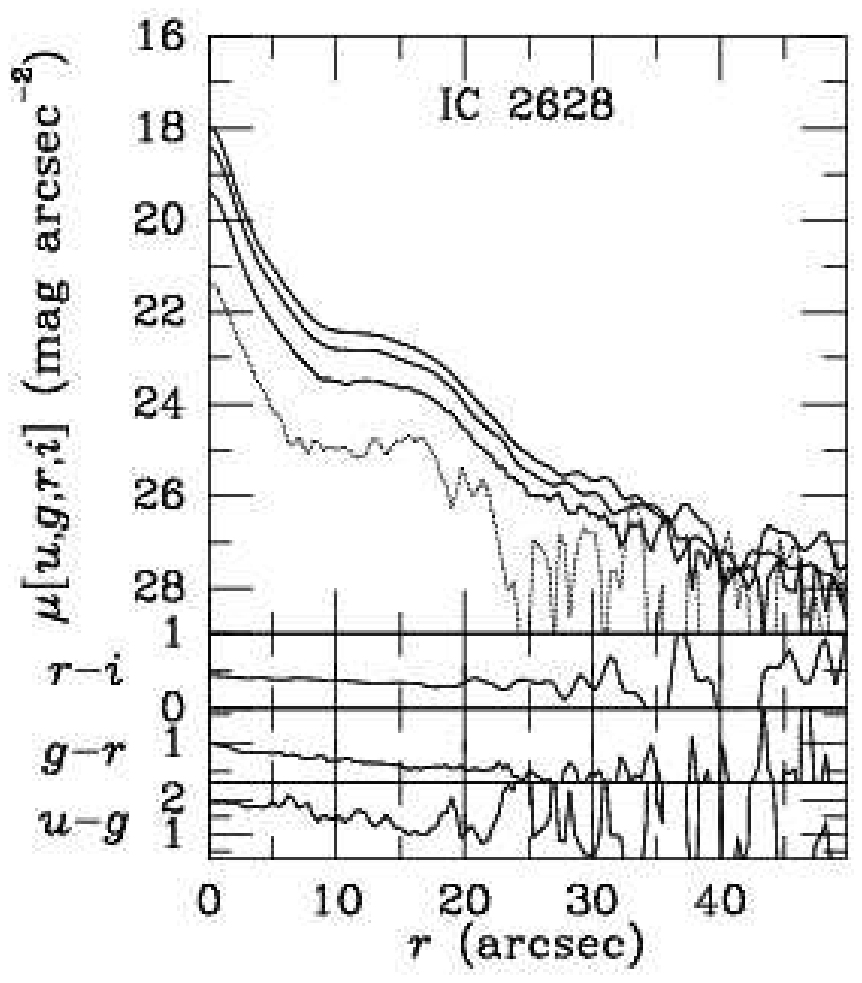}}
 \end{minipage}
 \begin{minipage}[b]{0.49\linewidth}
 \centering
\includegraphics[width=\textwidth,trim=0 0 275 400,clip]{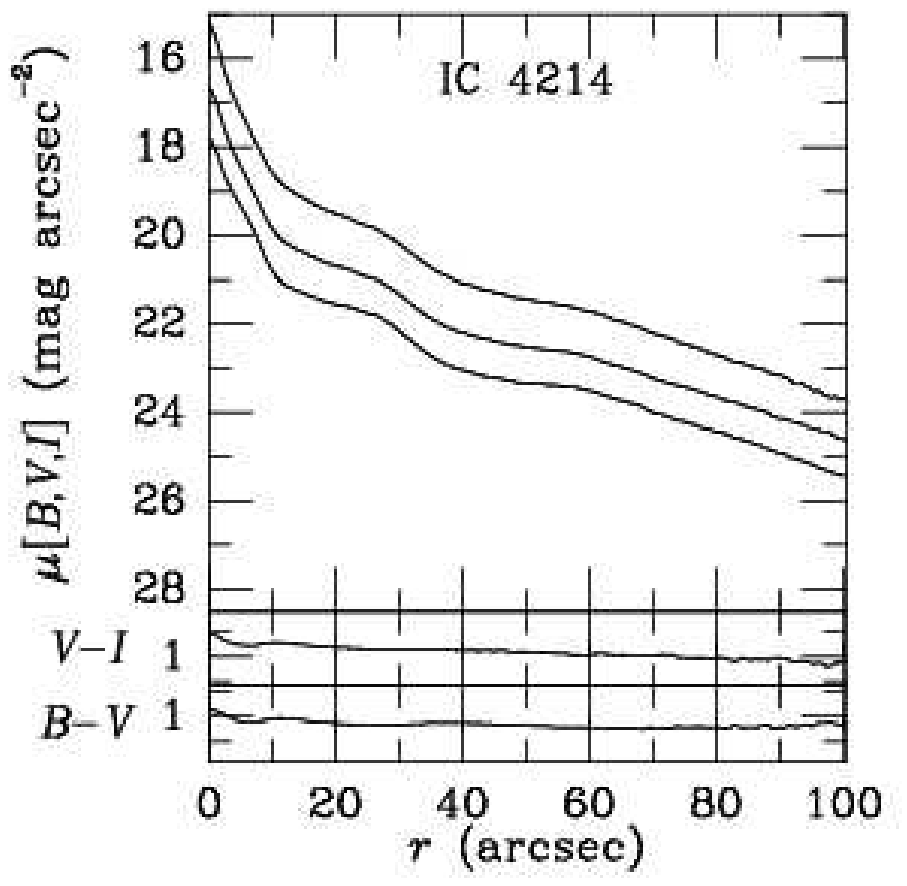}
 \hspace{0.1cm}
 \end{minipage}
 \begin{minipage}[t]{0.49\linewidth}
 \centering
\raisebox{0.35cm}{\includegraphics[width=\textwidth,trim=0 0 275 400,clip]{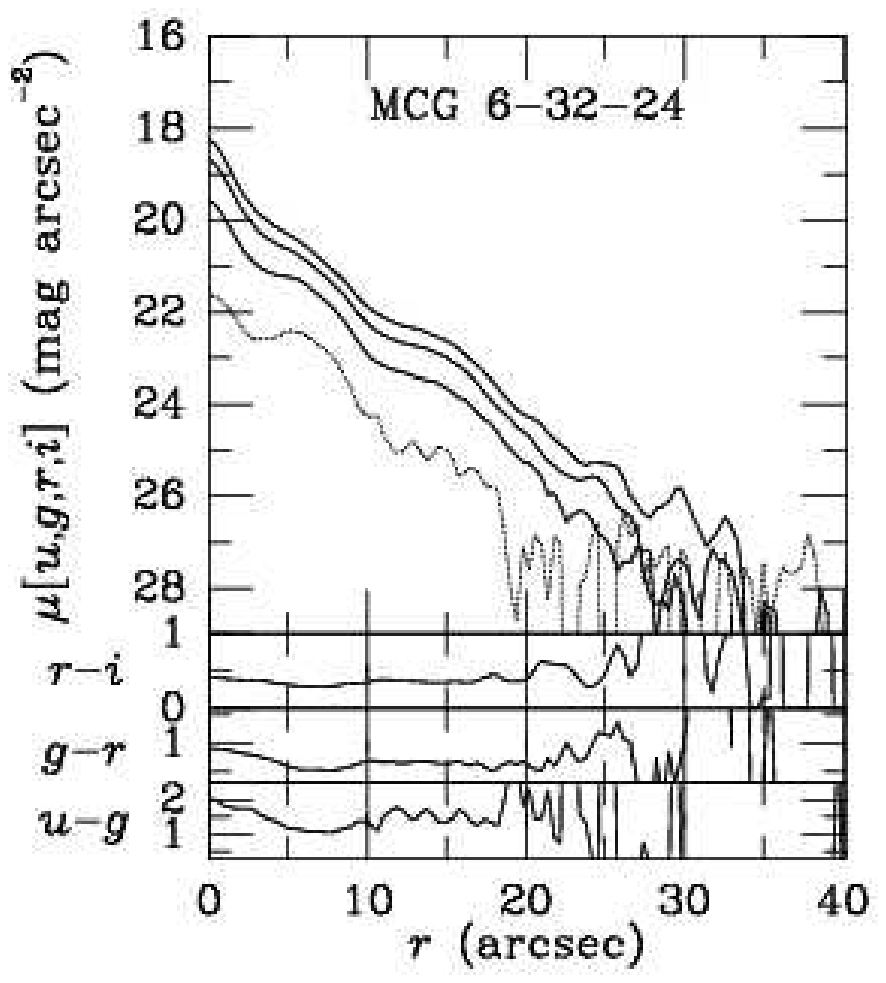}}
 \end{minipage}
\vspace{-1.0truecm}
\caption{(cont.)}
\end{figure}
 \setcounter{figure}{20}
 \begin{figure}
 \begin{minipage}[b]{0.49\linewidth}
 \centering
\includegraphics[width=\textwidth,trim=0 0 275 400,clip]{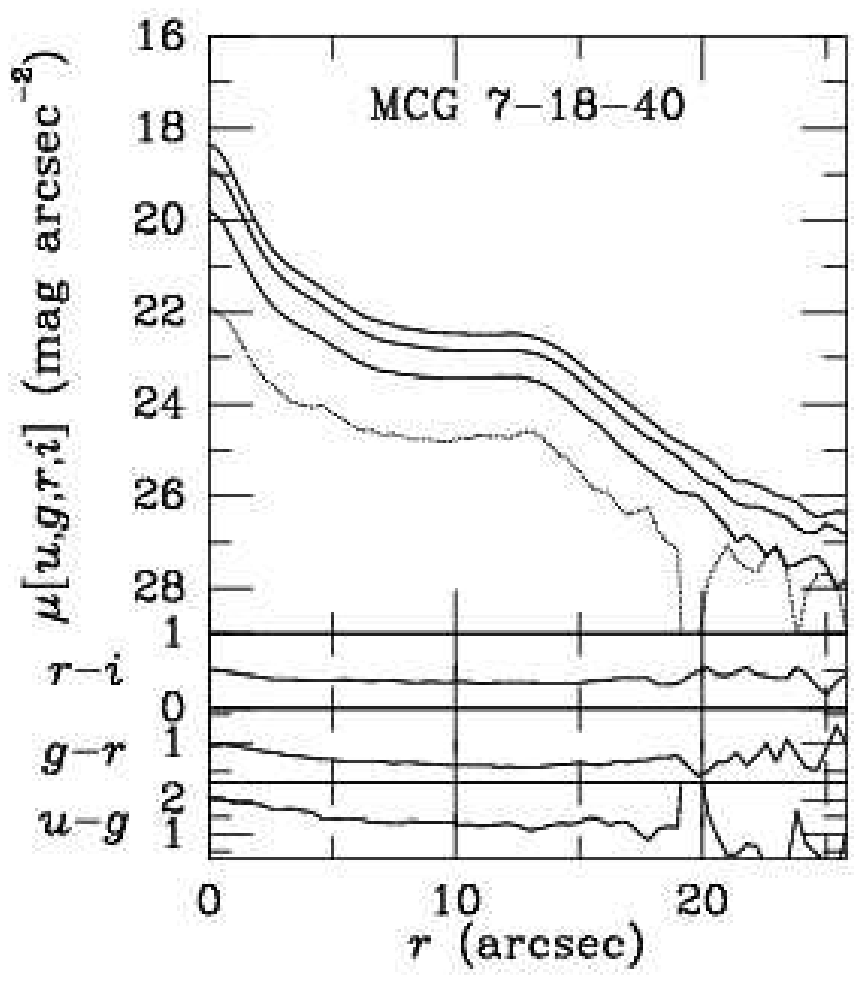}
 \hspace{0.1cm}
 \end{minipage}
 \begin{minipage}[t]{0.49\linewidth}
 \centering
\raisebox{0.35cm}{\includegraphics[width=\textwidth,trim=0 0 275 400,clip]{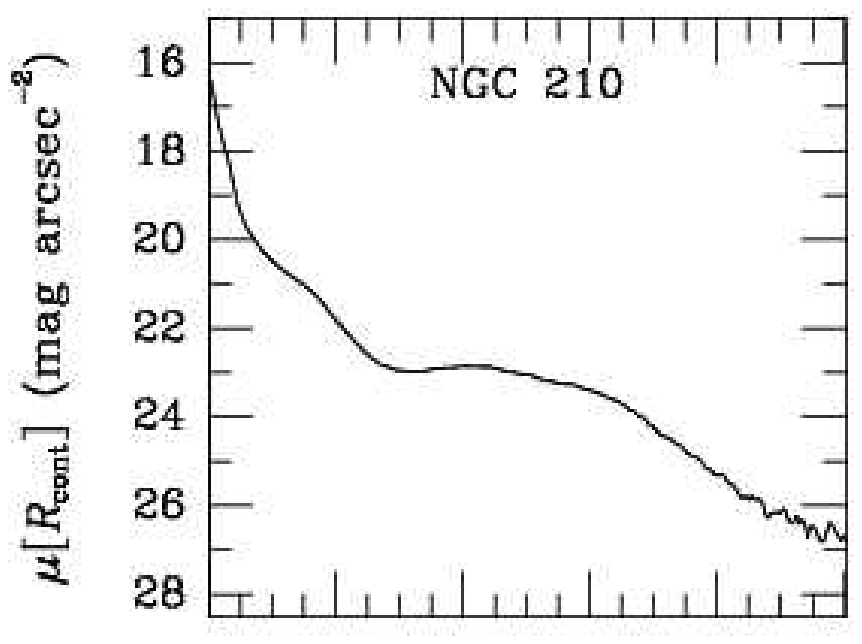}}
 \end{minipage}
 \begin{minipage}[b]{0.49\linewidth}
 \centering
\includegraphics[width=\textwidth,trim=0 0 275 400,clip]{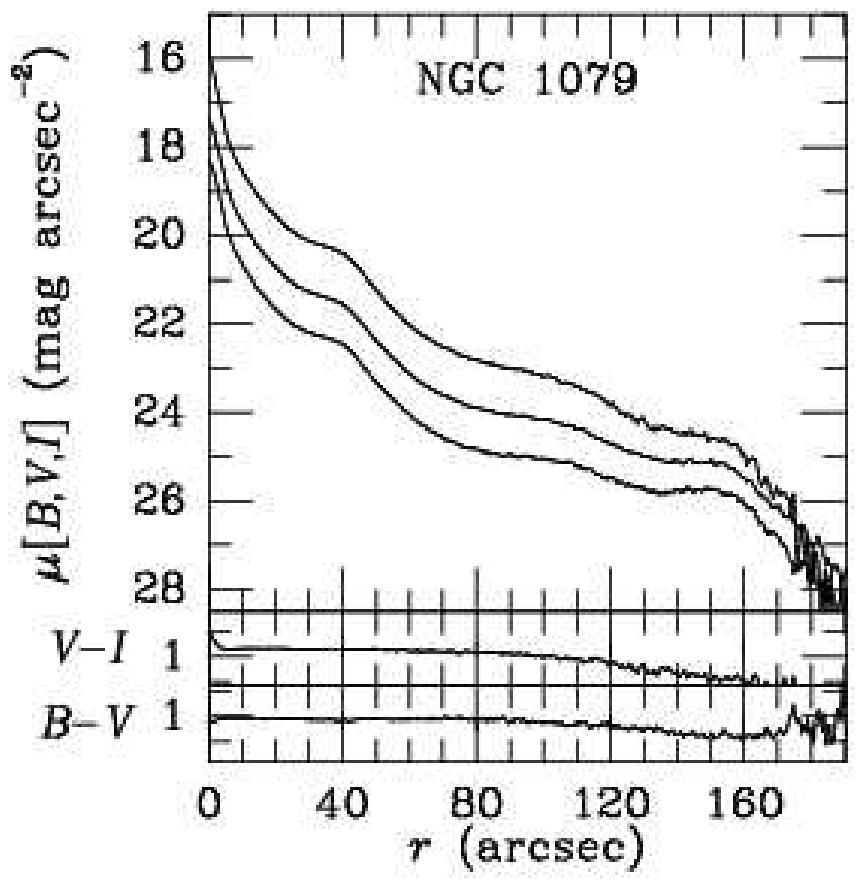}
 \hspace{0.1cm}
 \end{minipage}
 \begin{minipage}[t]{0.49\linewidth}
 \centering
\raisebox{0.35cm}{\includegraphics[width=\textwidth,trim=0 0 275 400,clip]{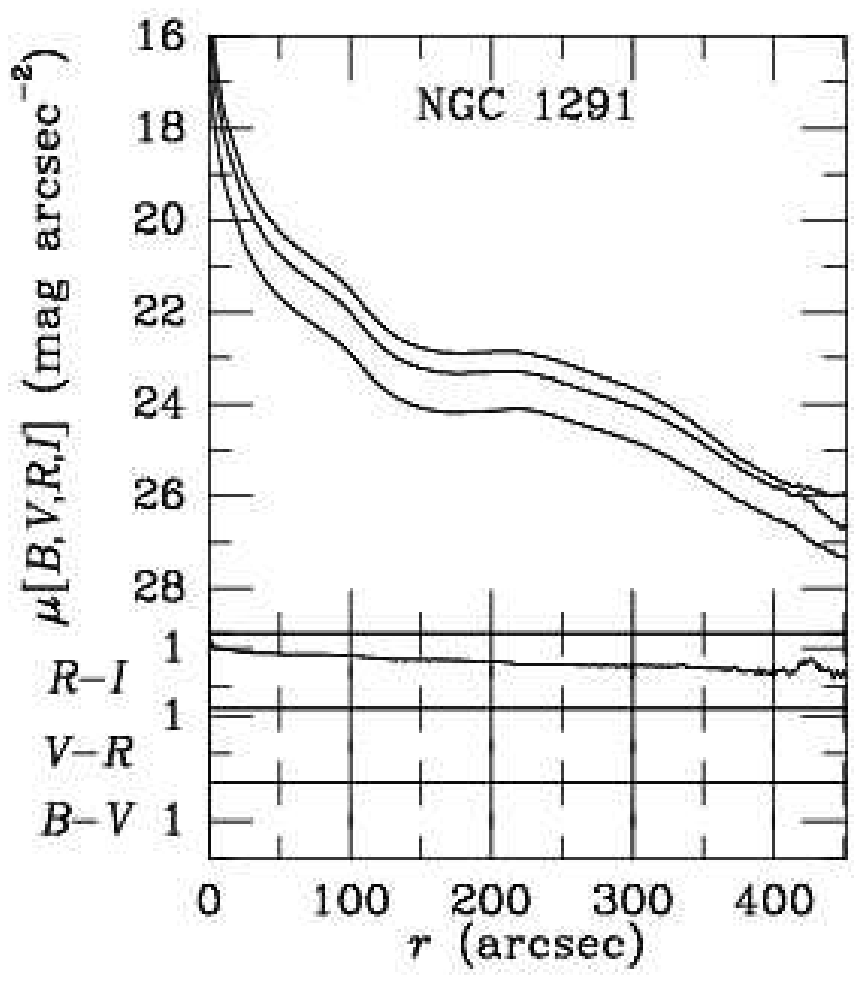}}
 \end{minipage}
 \begin{minipage}[b]{0.49\linewidth}
 \centering
\includegraphics[width=\textwidth,trim=0 0 275 400,clip]{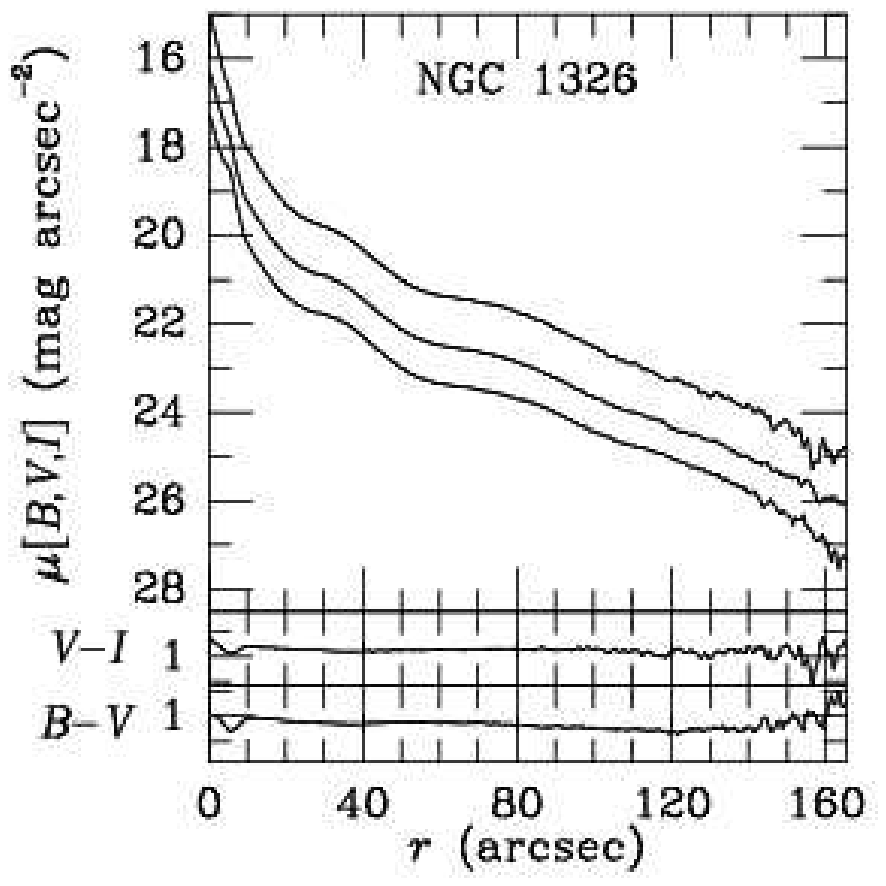}
 \hspace{0.1cm}
 \end{minipage}
 \begin{minipage}[t]{0.49\linewidth}
 \centering
\raisebox{0.35cm}{\includegraphics[width=\textwidth,trim=0 0 275 400,clip]{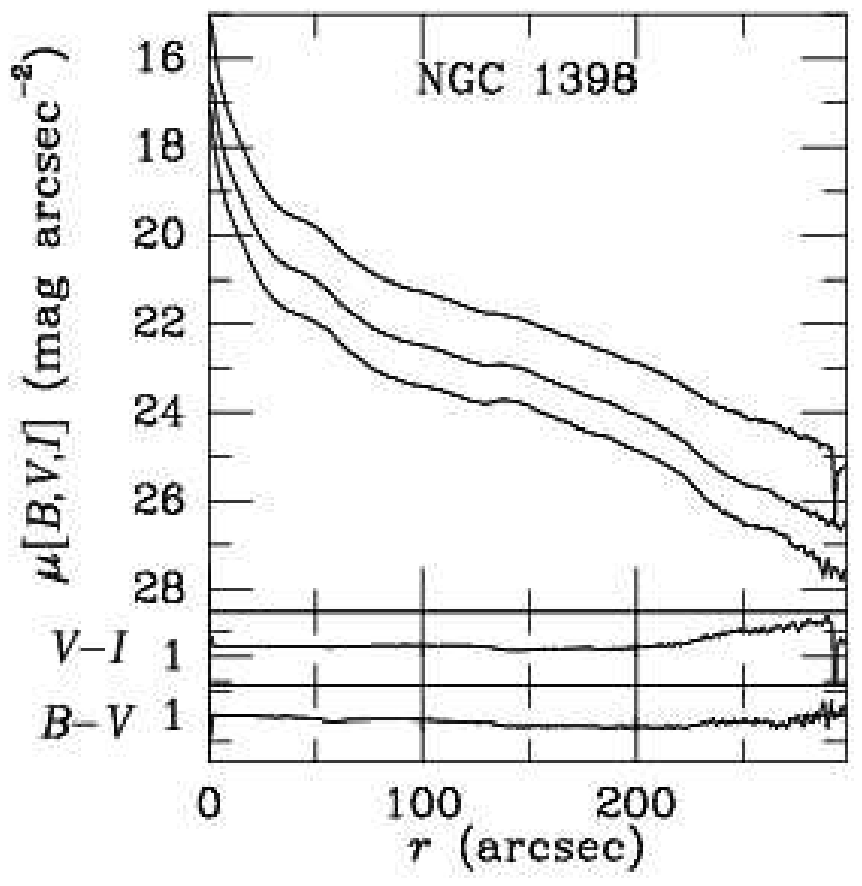}}
 \end{minipage}
 \begin{minipage}[b]{0.49\linewidth}
 \centering
\includegraphics[width=\textwidth,trim=0 0 275 400,clip]{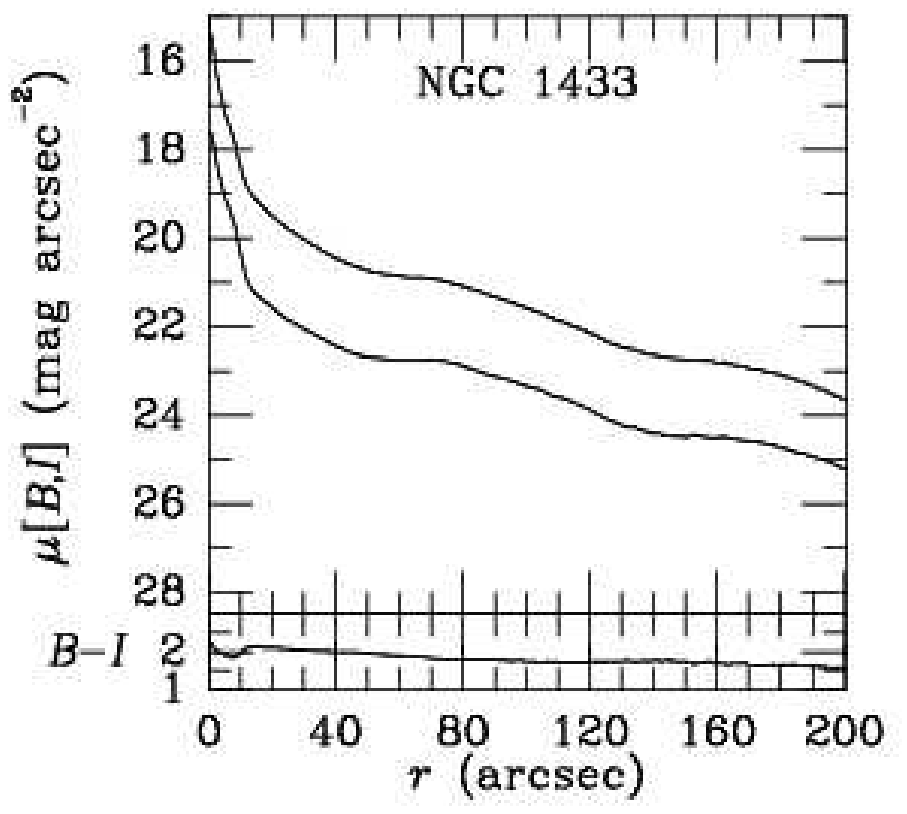}
 \hspace{0.1cm}
 \end{minipage}
 \begin{minipage}[t]{0.49\linewidth}
 \centering
\raisebox{0.35cm}{\includegraphics[width=\textwidth,trim=0 0 275 400,clip]{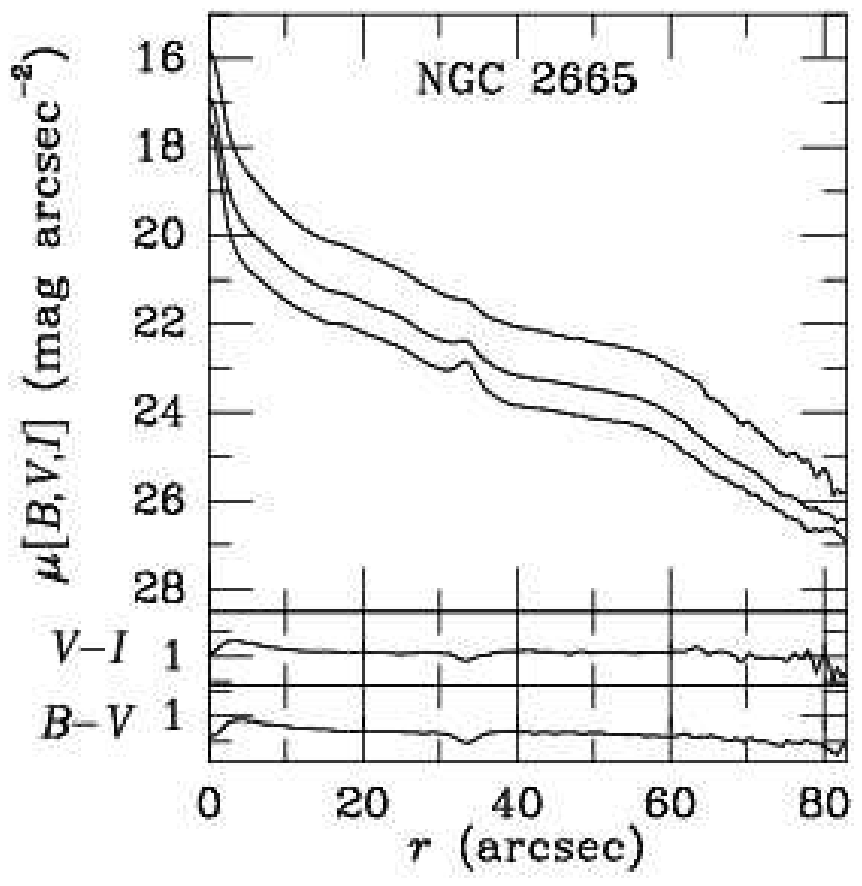}}
 \end{minipage}
 \begin{minipage}[b]{0.49\linewidth}
 \centering
\includegraphics[width=\textwidth,trim=0 0 275 400,clip]{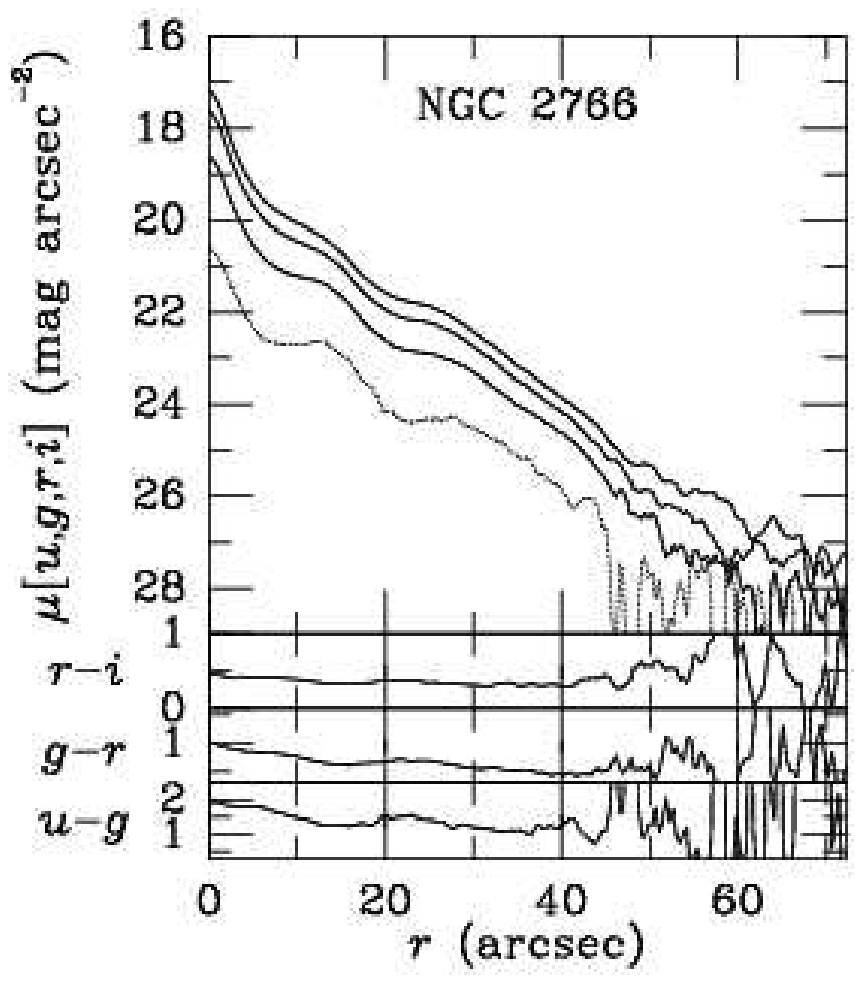}
 \hspace{0.1cm}
 \end{minipage}
 \begin{minipage}[t]{0.49\linewidth}
 \centering
\raisebox{0.35cm}{\includegraphics[width=\textwidth,trim=0 0 275 400,clip]{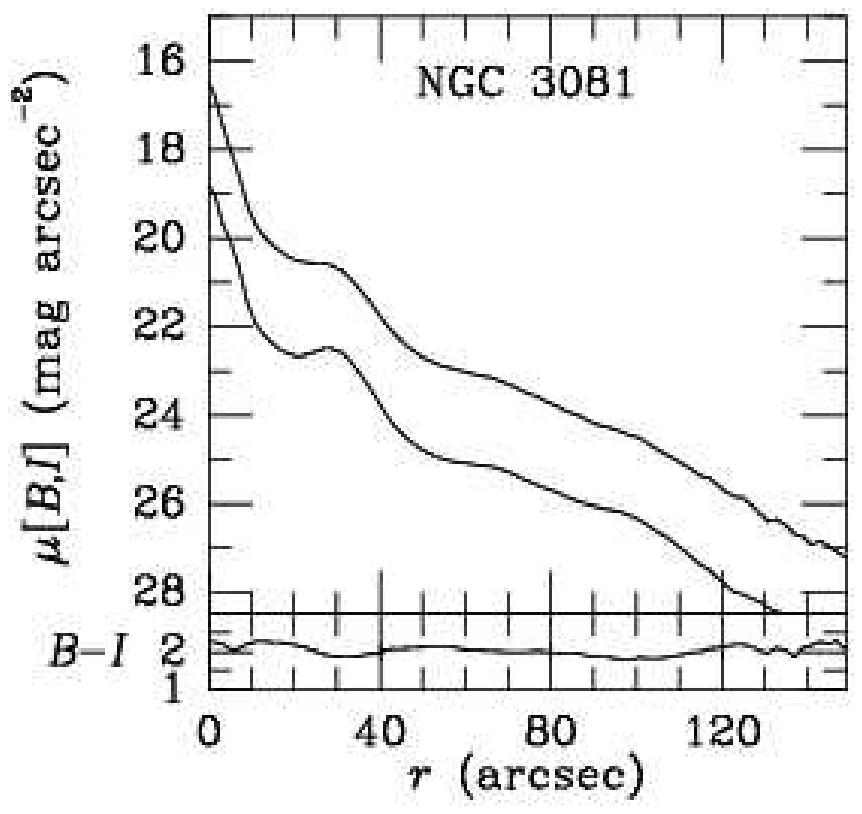}}
 \end{minipage}
\vspace{-0.5truecm}
\caption{(cont.)}
\end{figure}
 \setcounter{figure}{20}
 \begin{figure}
 \begin{minipage}[b]{0.49\linewidth}
 \centering
\includegraphics[width=\textwidth,trim=0 0 275 400,clip]{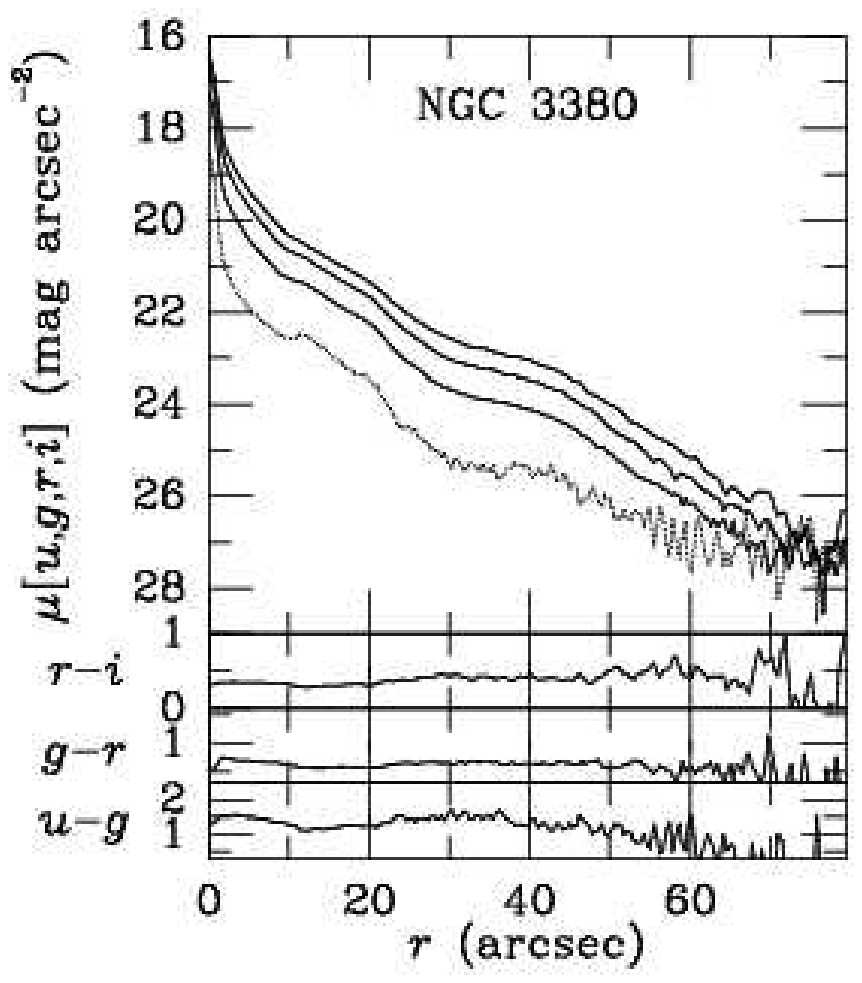}
 \hspace{0.1cm}
 \end{minipage}
 \begin{minipage}[t]{0.49\linewidth}
 \centering
\raisebox{0.35cm}{\includegraphics[width=\textwidth,trim=0 0 275 400,clip]{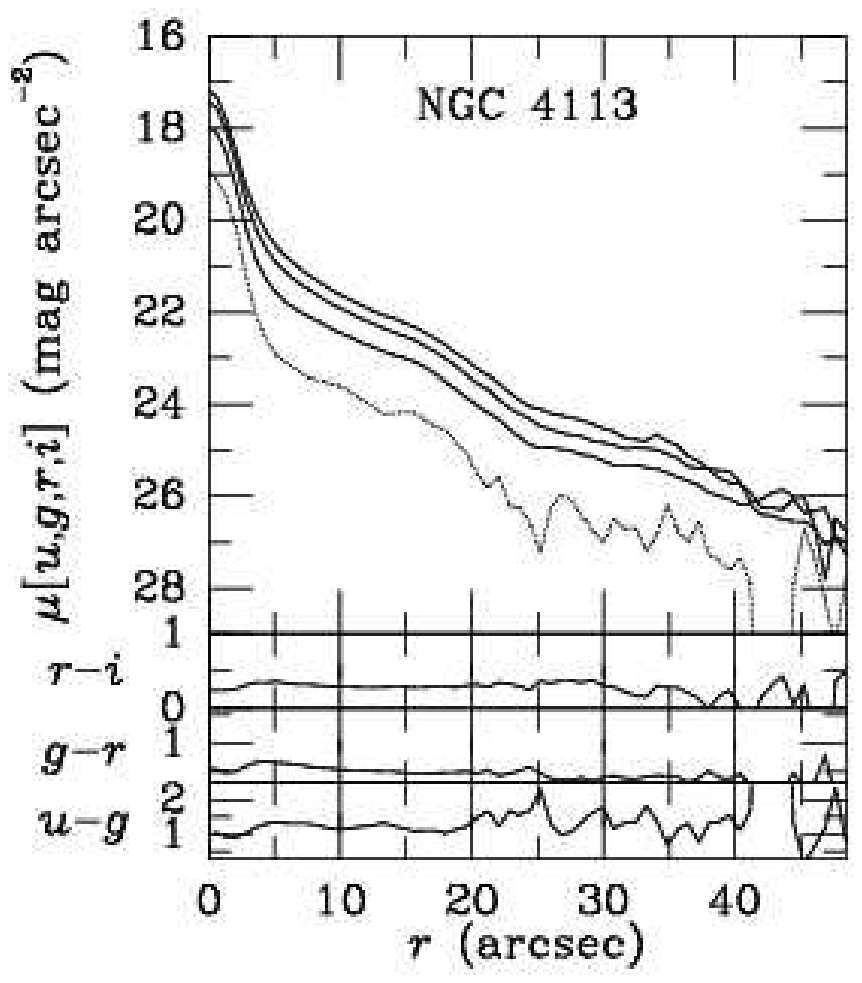}}
 \end{minipage}
 \begin{minipage}[b]{0.49\linewidth}
 \centering
\includegraphics[width=\textwidth,trim=0 0 275 400,clip]{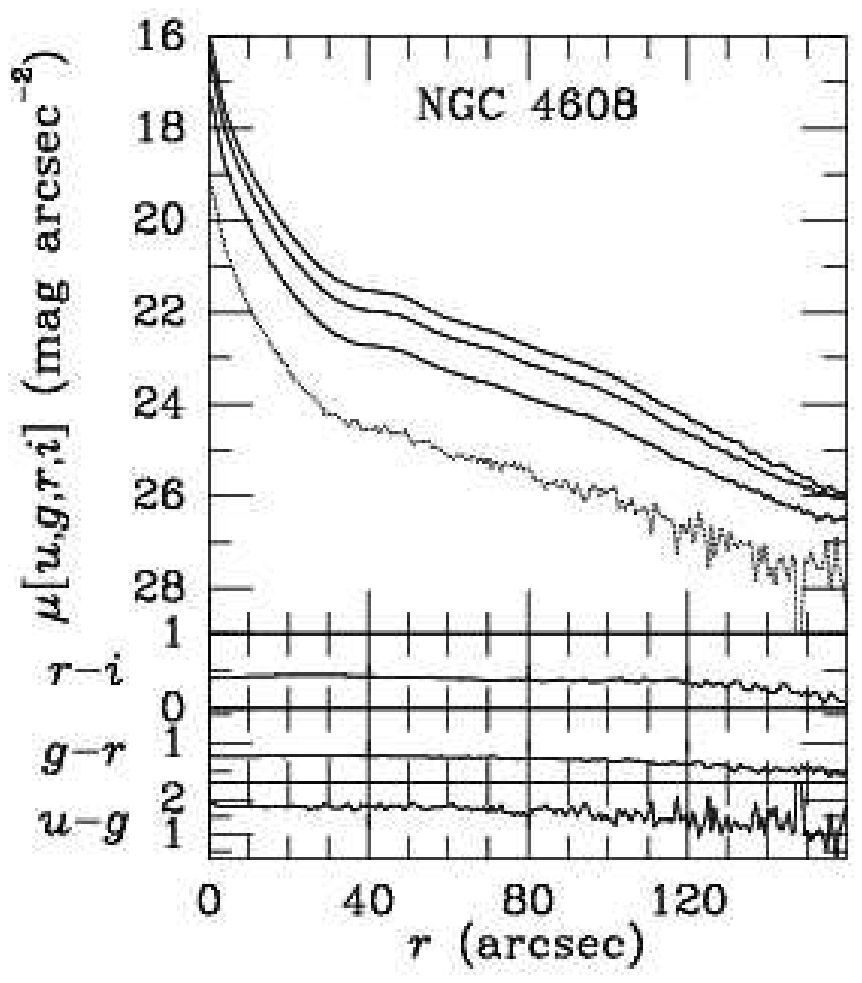}
 \hspace{0.1cm}
 \end{minipage}
 \begin{minipage}[t]{0.49\linewidth}
 \centering
\raisebox{0.35cm}{\includegraphics[width=\textwidth,trim=0 0 275 400,clip]{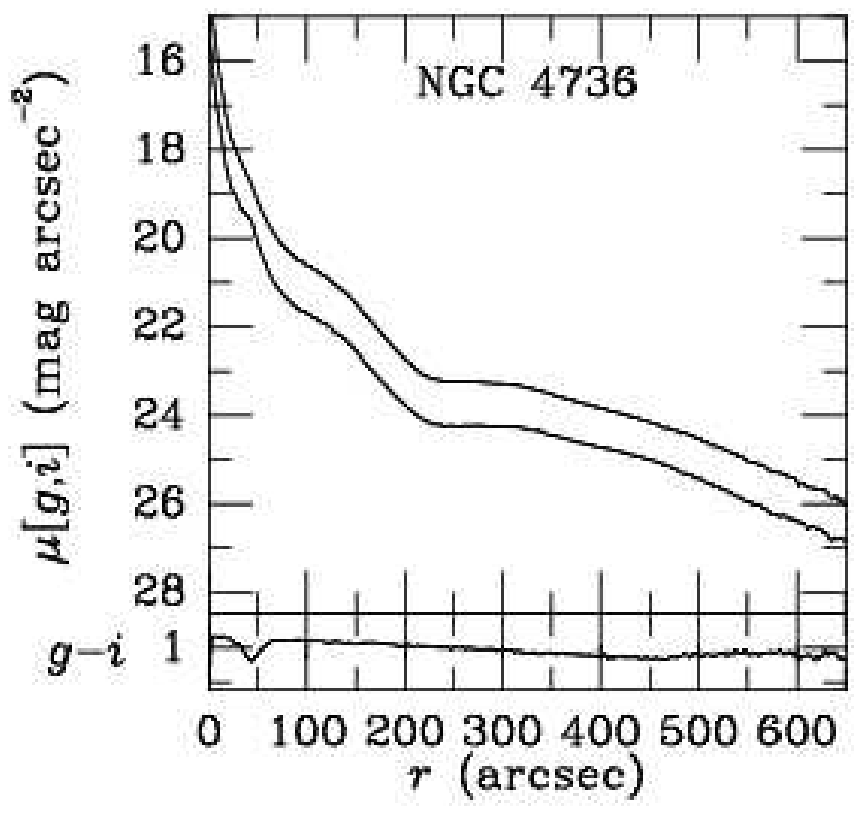}}
 \end{minipage}
 \begin{minipage}[b]{0.49\linewidth}
 \centering
\includegraphics[width=\textwidth,trim=0 0 275 400,clip]{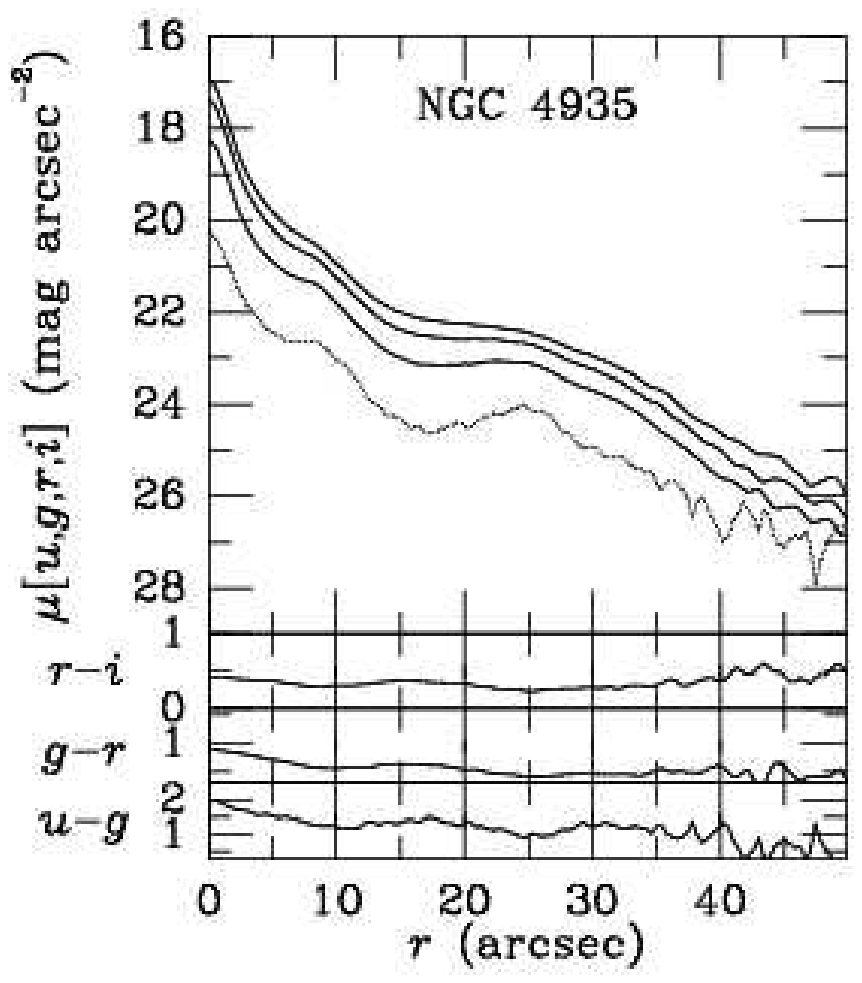}
 \hspace{0.1cm}
 \end{minipage}
 \begin{minipage}[t]{0.49\linewidth}
 \centering
\raisebox{0.35cm}{\includegraphics[width=\textwidth,trim=0 0 275 400,clip]{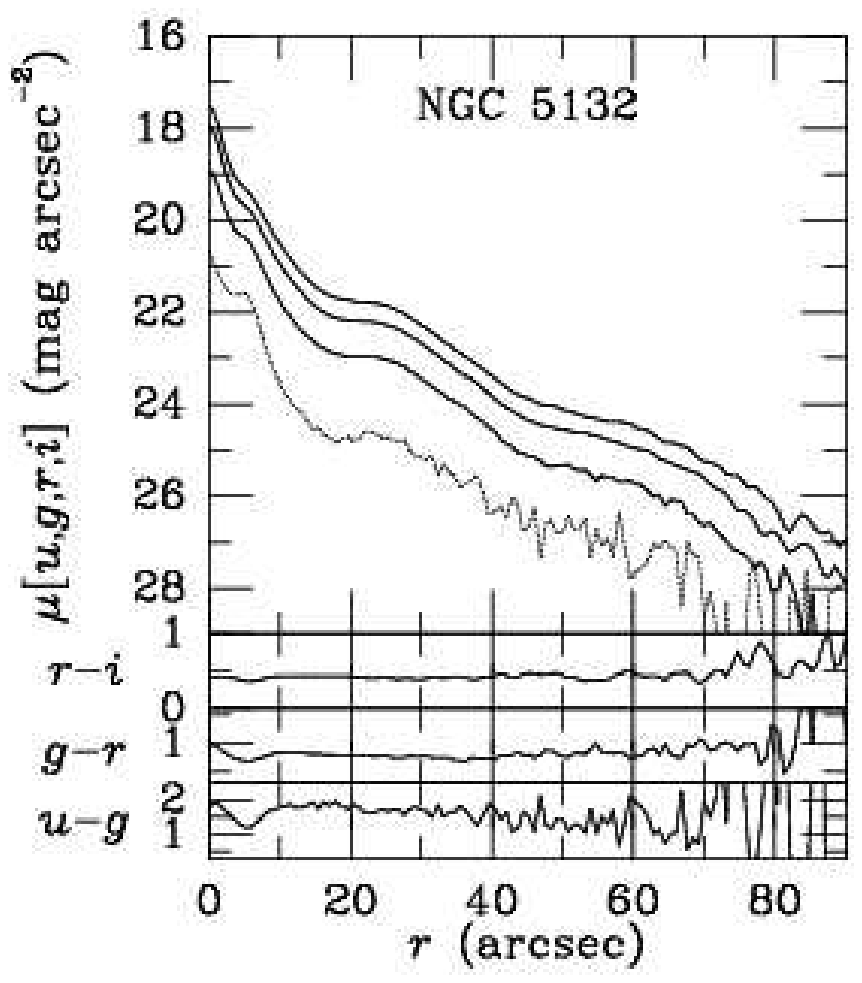}}
 \end{minipage}
 \begin{minipage}[b]{0.49\linewidth}
 \centering
\includegraphics[width=\textwidth,trim=0 0 275 400,clip]{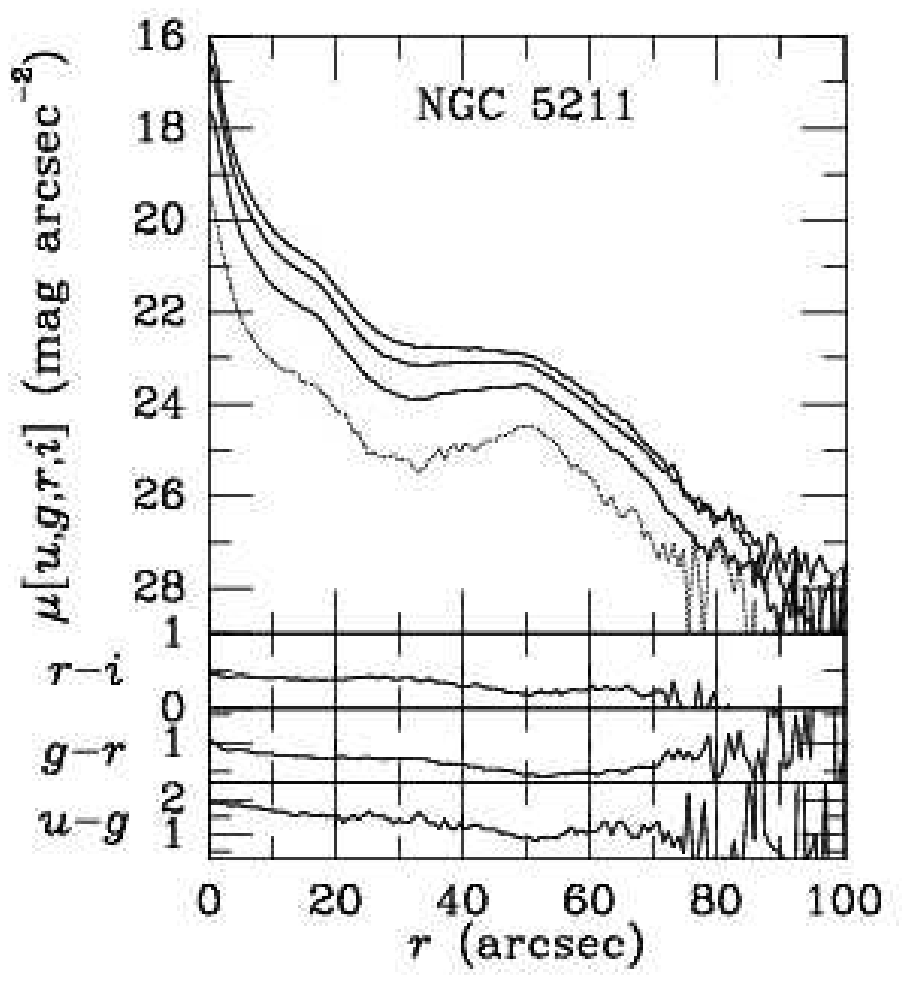}
 \hspace{0.1cm}
 \end{minipage}
 \begin{minipage}[t]{0.49\linewidth}
 \centering
\raisebox{0.35cm}{\includegraphics[width=\textwidth,trim=0 0 275 400,clip]{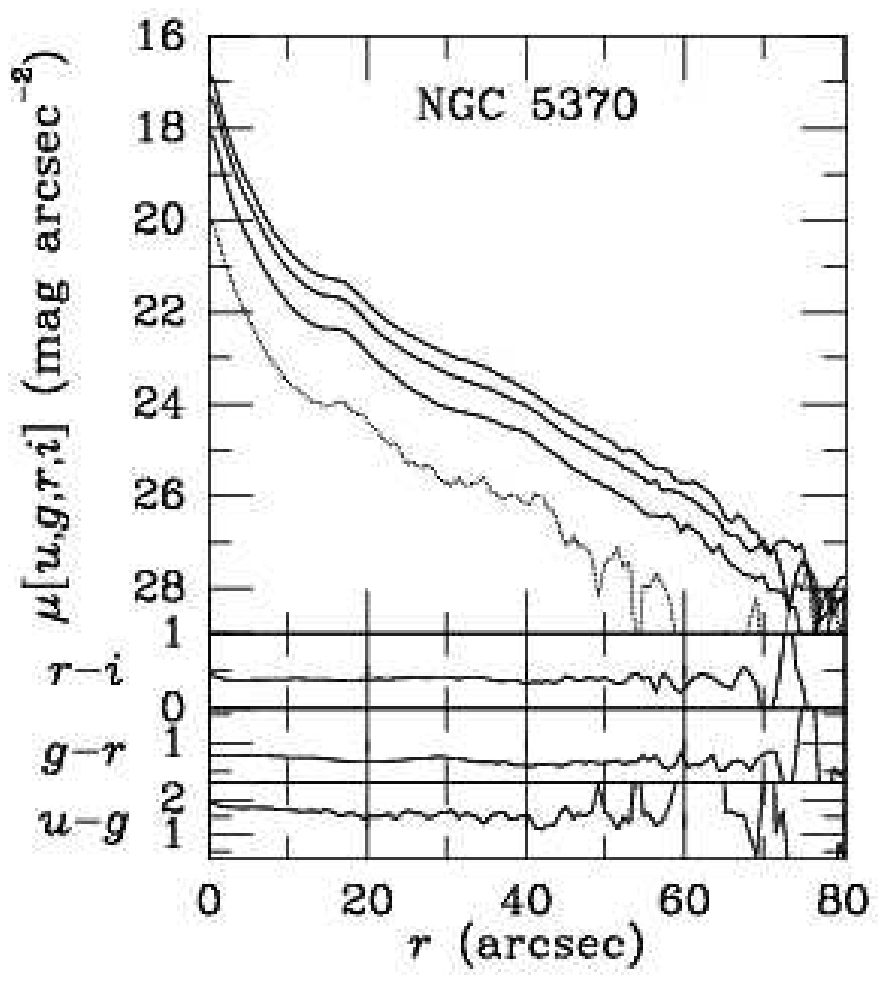}}
 \end{minipage}
 \begin{minipage}[b]{0.49\linewidth}
 \centering
\includegraphics[width=\textwidth,trim=0 0 275 400,clip]{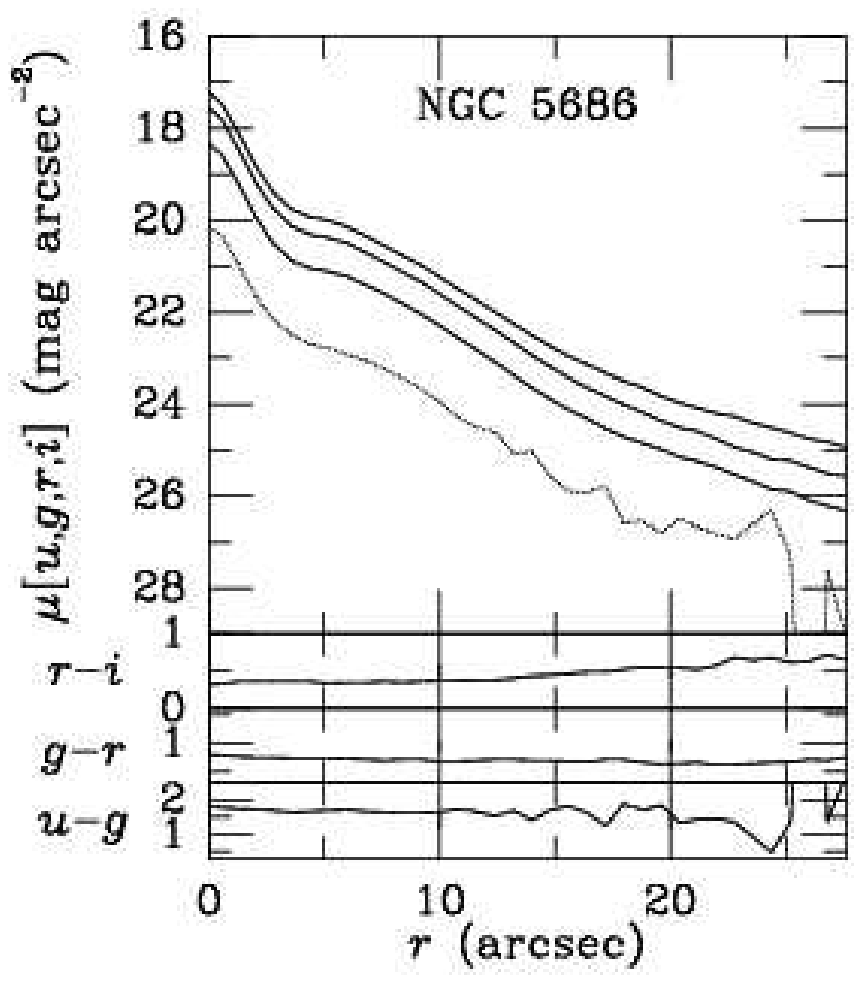}
 \hspace{0.1cm}
 \end{minipage}
 \begin{minipage}[t]{0.49\linewidth}
 \centering
\raisebox{0.35cm}{\includegraphics[width=\textwidth,trim=0 0 275 400,clip]{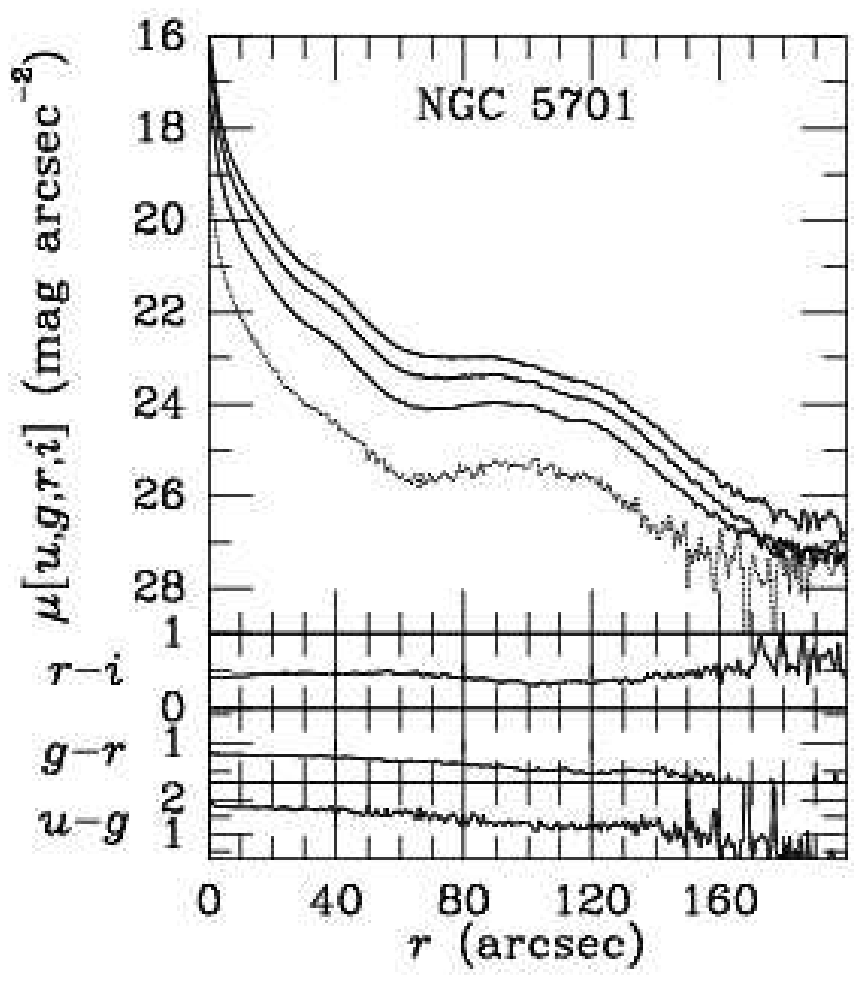}}
 \end{minipage}
\vspace{-0.5truecm}
\caption{(cont.)}
\end{figure}
 \setcounter{figure}{20}
 \begin{figure}
 \begin{minipage}[b]{0.49\linewidth}
 \centering
\includegraphics[width=\textwidth,trim=0 0 275 400,clip]{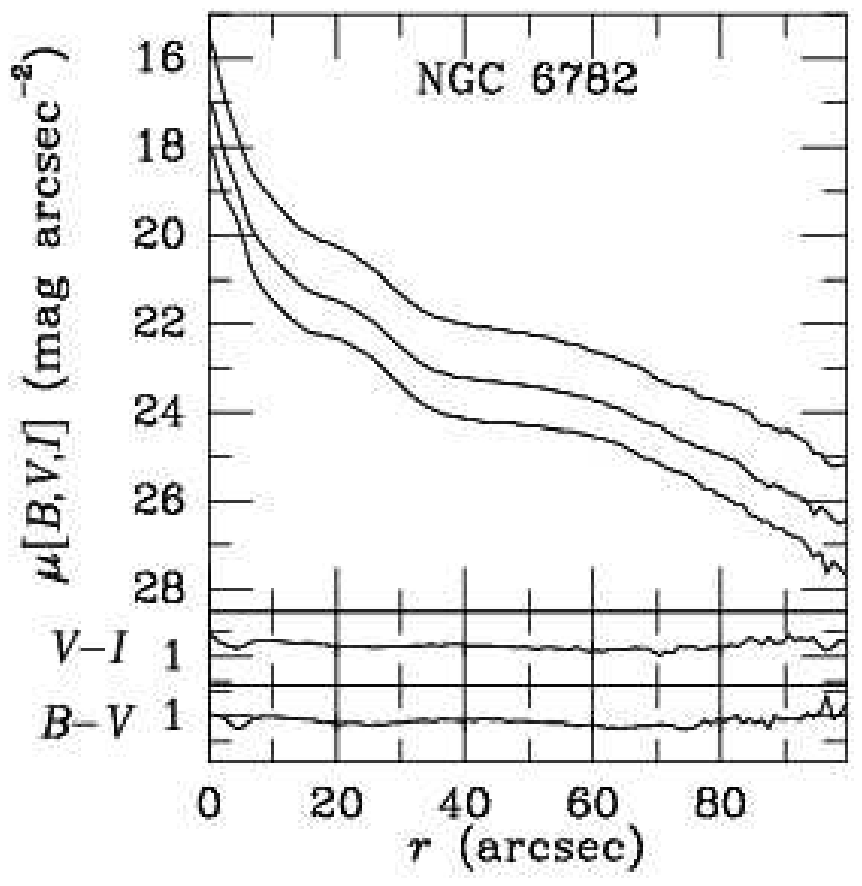}
 \hspace{0.1cm}
 \end{minipage}
 \begin{minipage}[t]{0.49\linewidth}
 \centering
\raisebox{0.35cm}{\includegraphics[width=\textwidth,trim=0 0 275 400,clip]{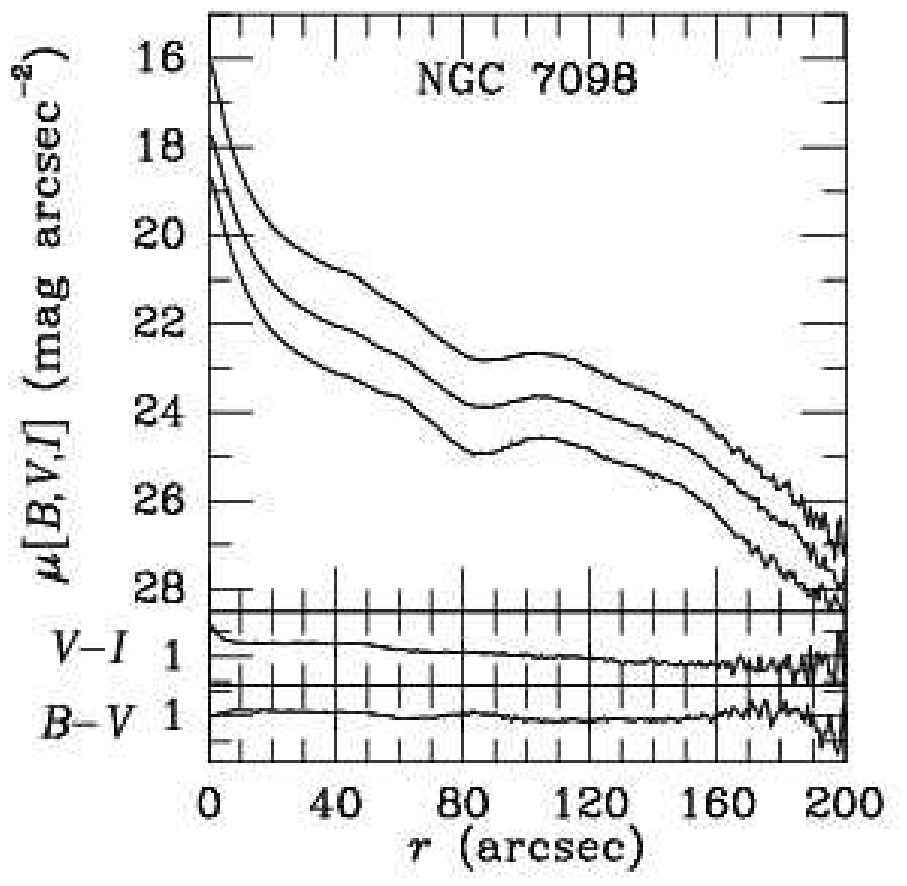}}
 \end{minipage}
 \begin{minipage}[b]{0.49\linewidth}
 \centering
\includegraphics[width=\textwidth,trim=0 0 275 400,clip]{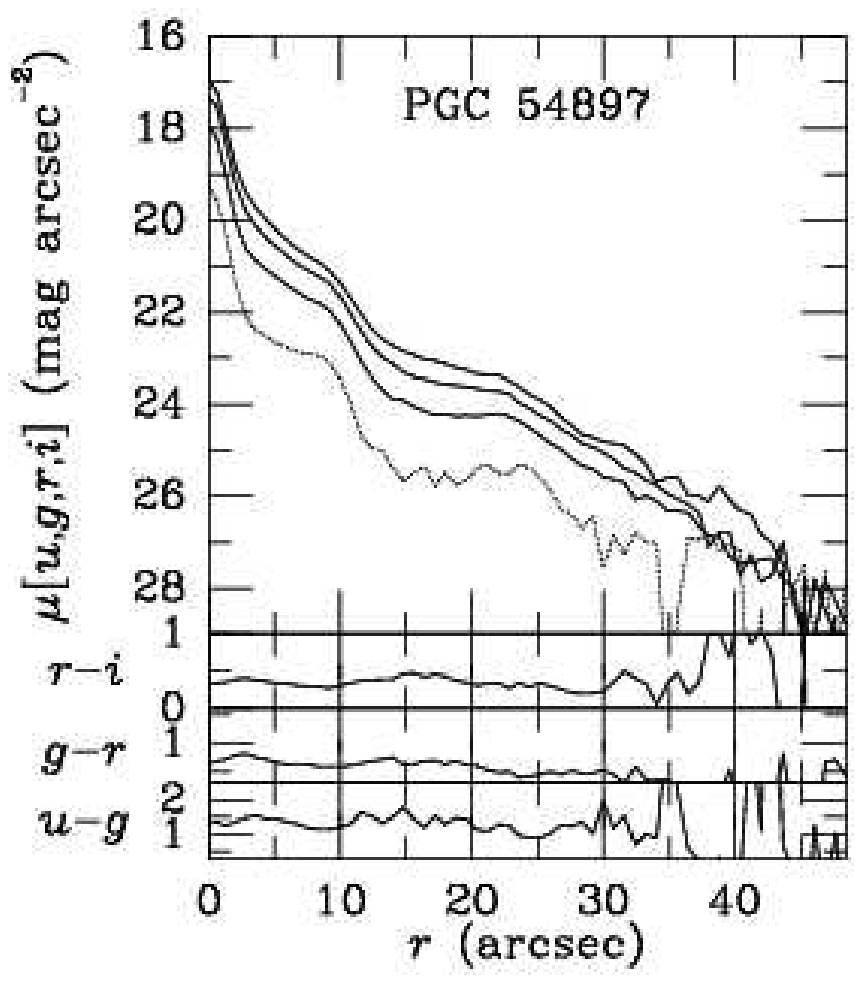}
 \hspace{0.1cm}
 \end{minipage}
 \begin{minipage}[t]{0.49\linewidth}
 \centering
\raisebox{0.35cm}{\includegraphics[width=\textwidth,trim=0 0 275 400,clip]{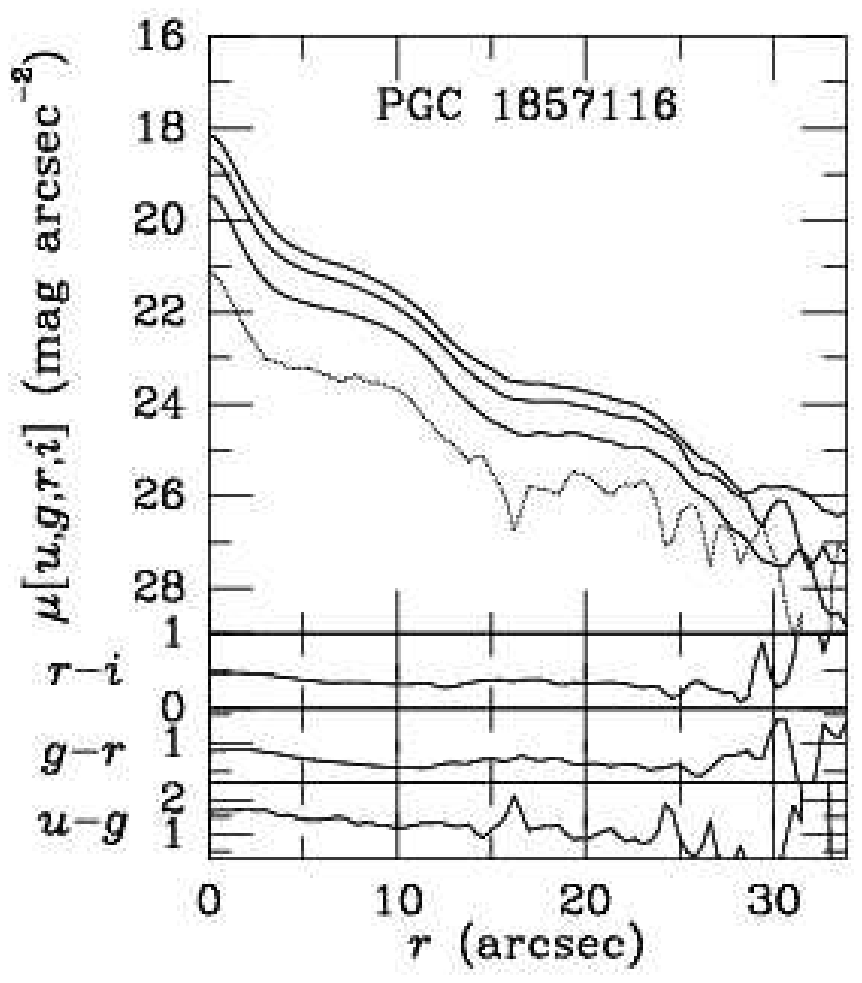}}
 \end{minipage}
 \begin{minipage}[b]{0.49\linewidth}
 \centering
\includegraphics[width=\textwidth,trim=0 0 275 400,clip]{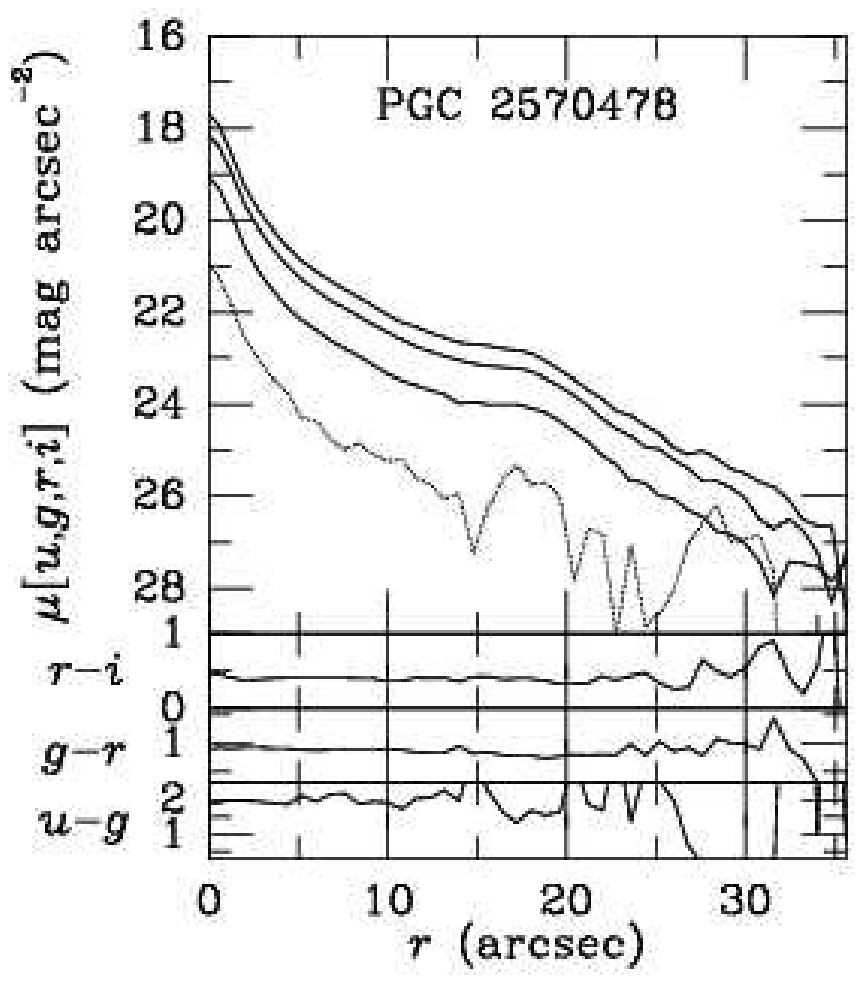}
 \hspace{0.1cm}
 \end{minipage}
 \begin{minipage}[t]{0.49\linewidth}
 \centering
\raisebox{0.35cm}{\includegraphics[width=\textwidth,trim=0 0 275 400,clip]{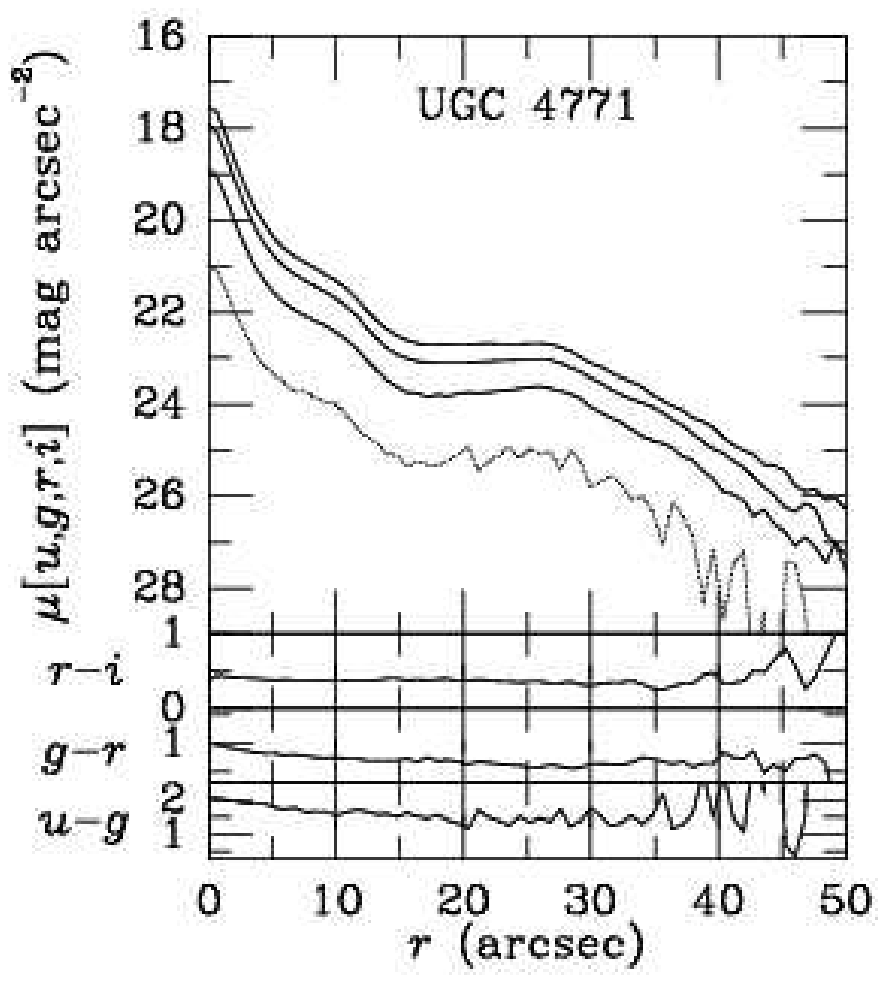}}
 \end{minipage}
 \begin{minipage}[b]{0.49\linewidth}
 \centering
\includegraphics[width=\textwidth,trim=0 0 275 400,clip]{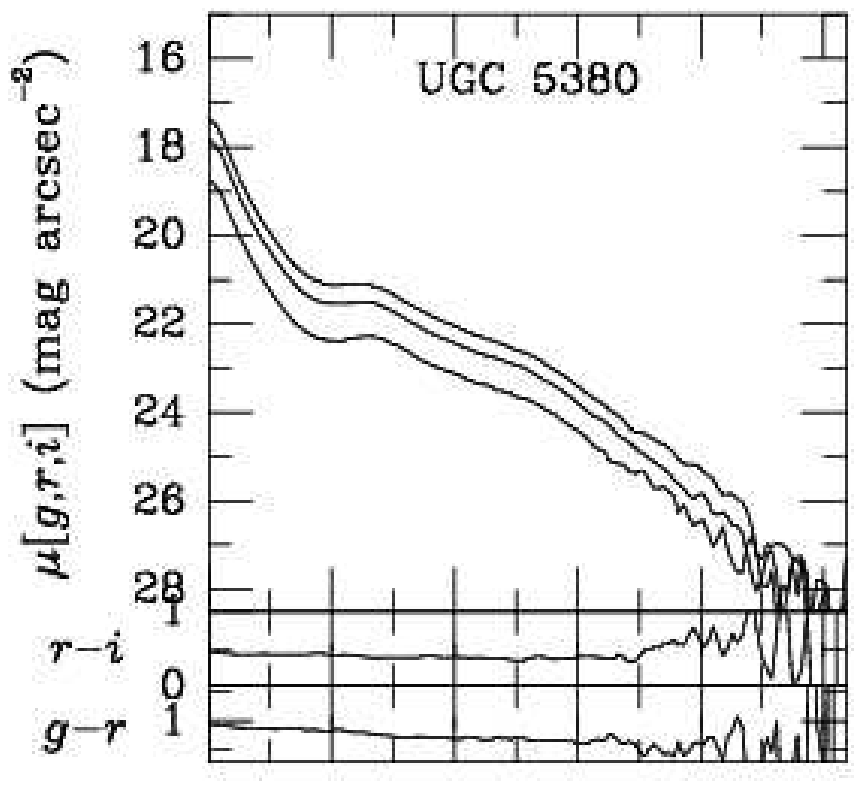}
 \hspace{0.1cm}
 \end{minipage}
 \begin{minipage}[t]{0.49\linewidth}
 \centering
\raisebox{0.35cm}{\includegraphics[width=\textwidth,trim=0 0 275 400,clip]{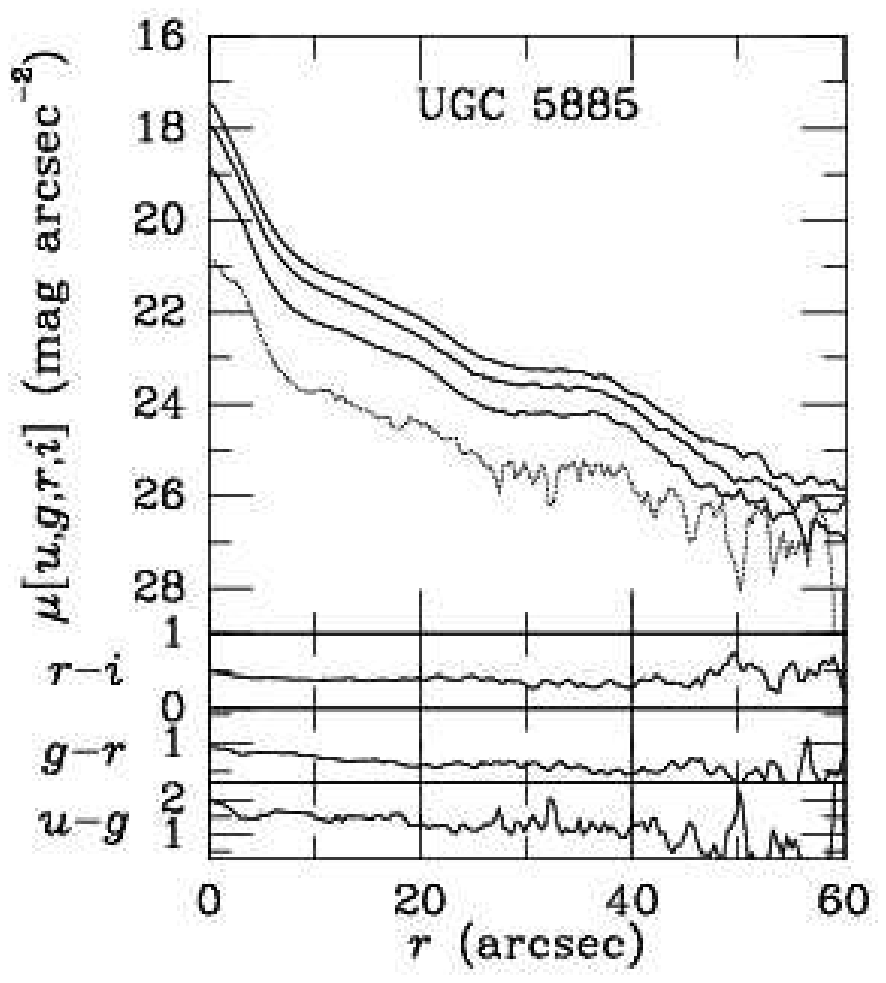}}
 \end{minipage}
 \begin{minipage}[b]{0.49\linewidth}
 \centering
\includegraphics[width=\textwidth,trim=0 0 275 400,clip]{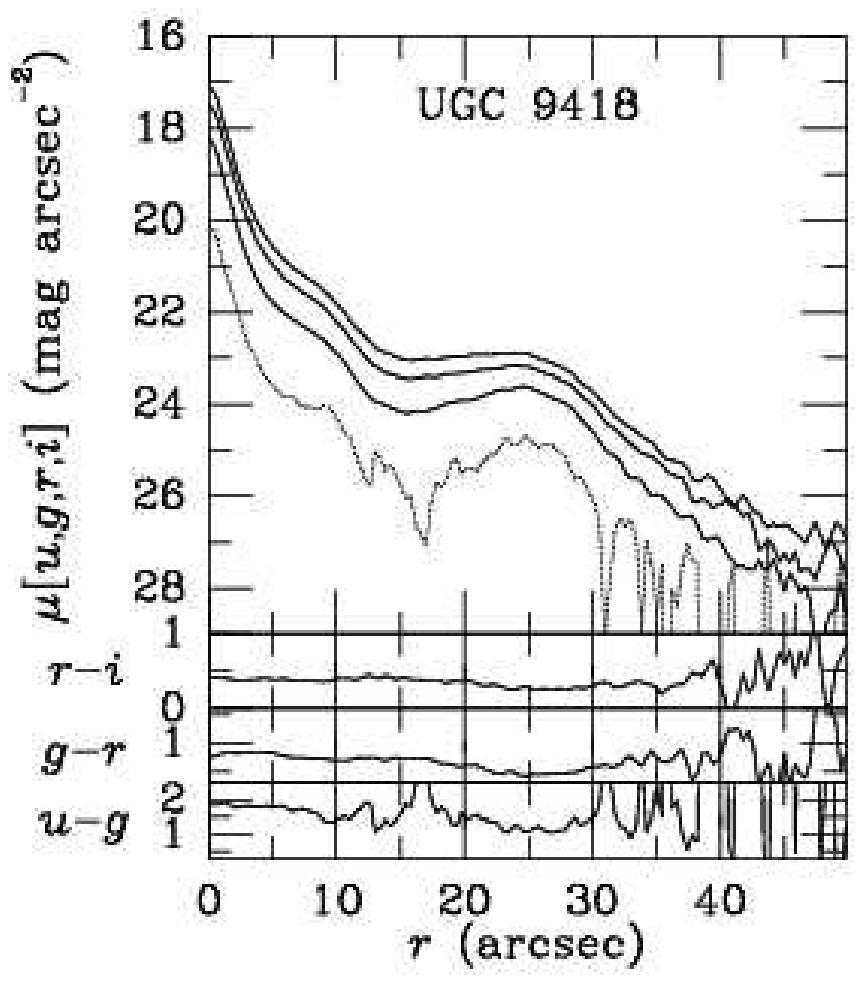}
 \hspace{0.1cm}
 \end{minipage}
 \begin{minipage}[t]{0.49\linewidth}
 \centering
\raisebox{0.35cm}{\includegraphics[width=\textwidth,trim=0 0 275 400,clip]{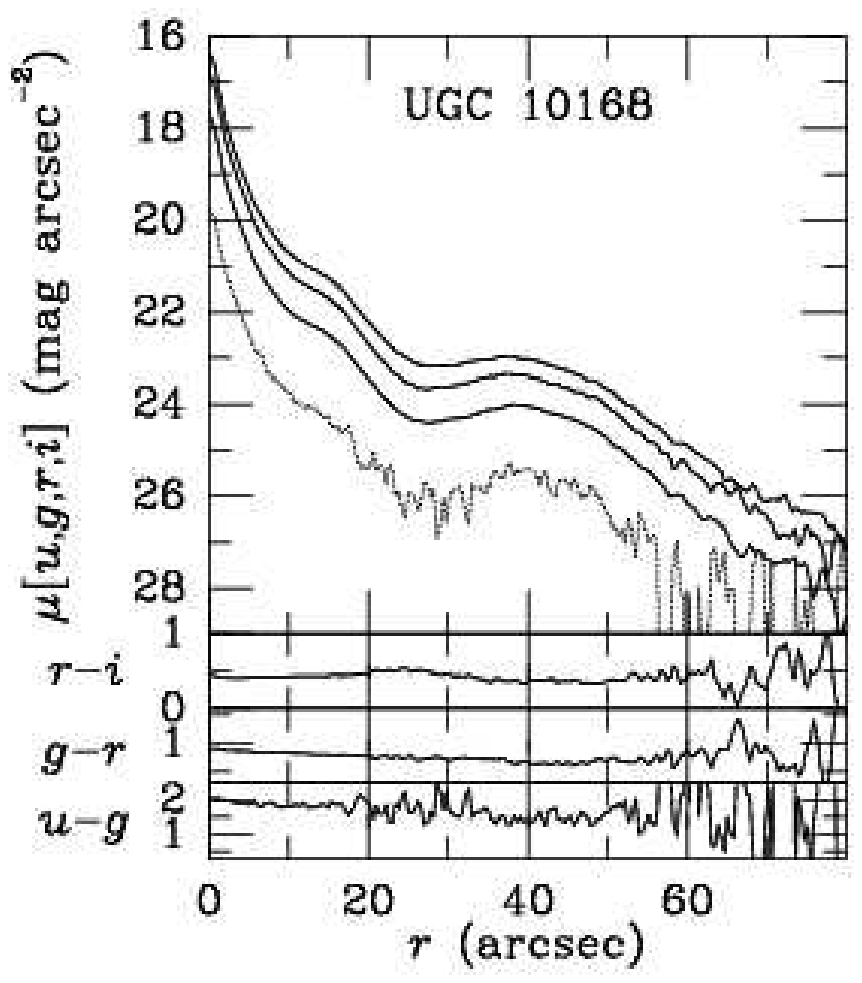}}
 \end{minipage}
\vspace{-0.5truecm}
\caption{(cont.)}
\end{figure}
 \setcounter{figure}{20}
 \begin{figure}
 \begin{minipage}[b]{0.49\linewidth}
 \centering
\includegraphics[width=\textwidth,trim=0 0 275 400,clip]{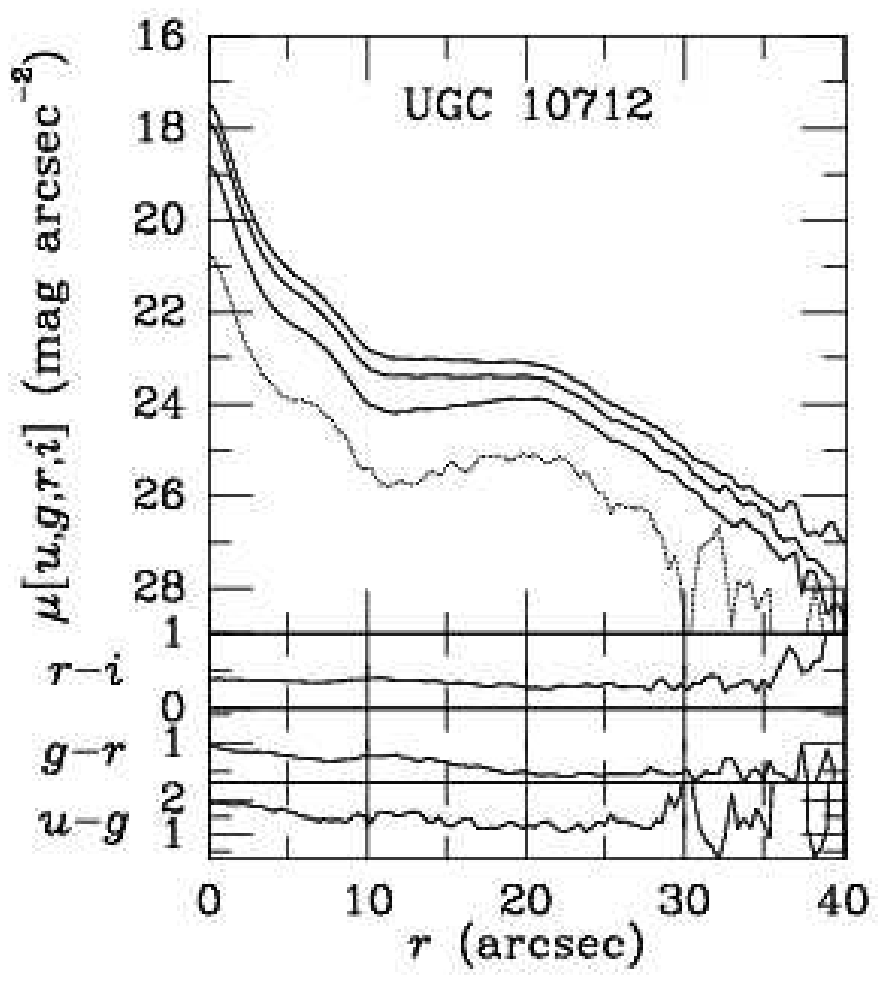}
 \hspace{0.1cm}
 \end{minipage}
 \begin{minipage}[t]{0.49\linewidth}
 \centering
\raisebox{0.35cm}{\includegraphics[width=\textwidth,trim=0 0 275 400,clip]{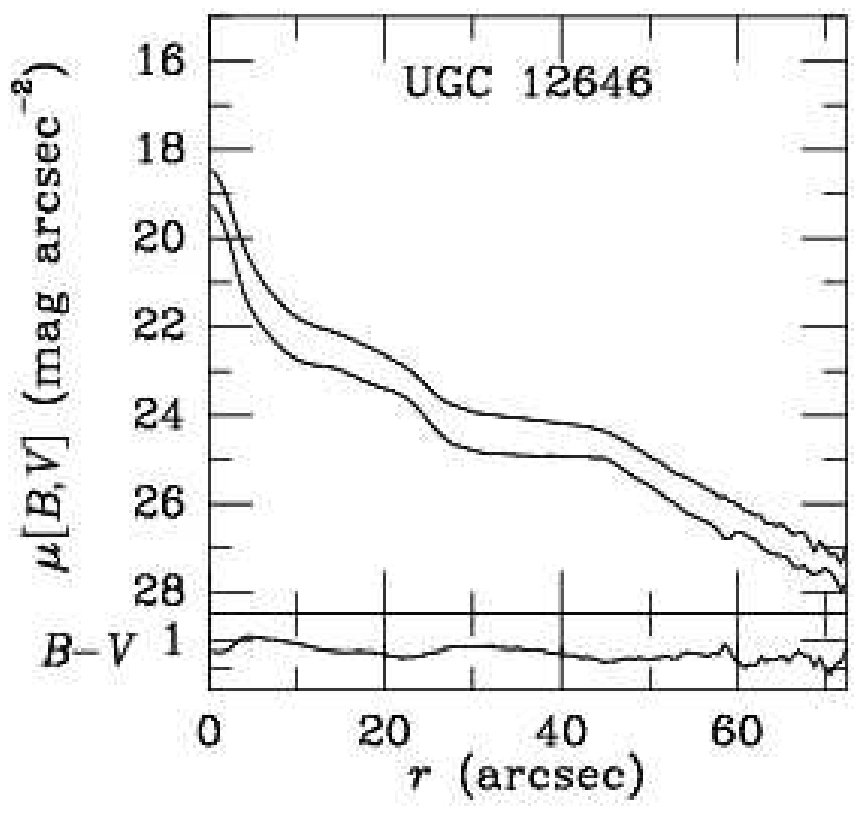}}
 \end{minipage}
\vspace{-0.5truecm}
\caption{Azimuthally-averaged surface brightness profiles of 31 GZ2 and
21 non-GZ2 ringed galaxies. The profiles use the orientation parameters
listed in Table~\ref{tab:orient}. For GZ2 galaxies, the profiles from
bottom to top are $u$, $g$, $r$, and $i$. For non-GZ2 galaxies, the
profiles from bottom to top are $B$, $V$, and $I$. A few non-GZ2 sample
galaxies have only $B$ and $V$, $B$ and $I$, $R$, and (in one case) $g$
and $i$.}
\end{figure}

\begin{figure}
\includegraphics[width=\columnwidth,trim=0 -10 0 50,clip]{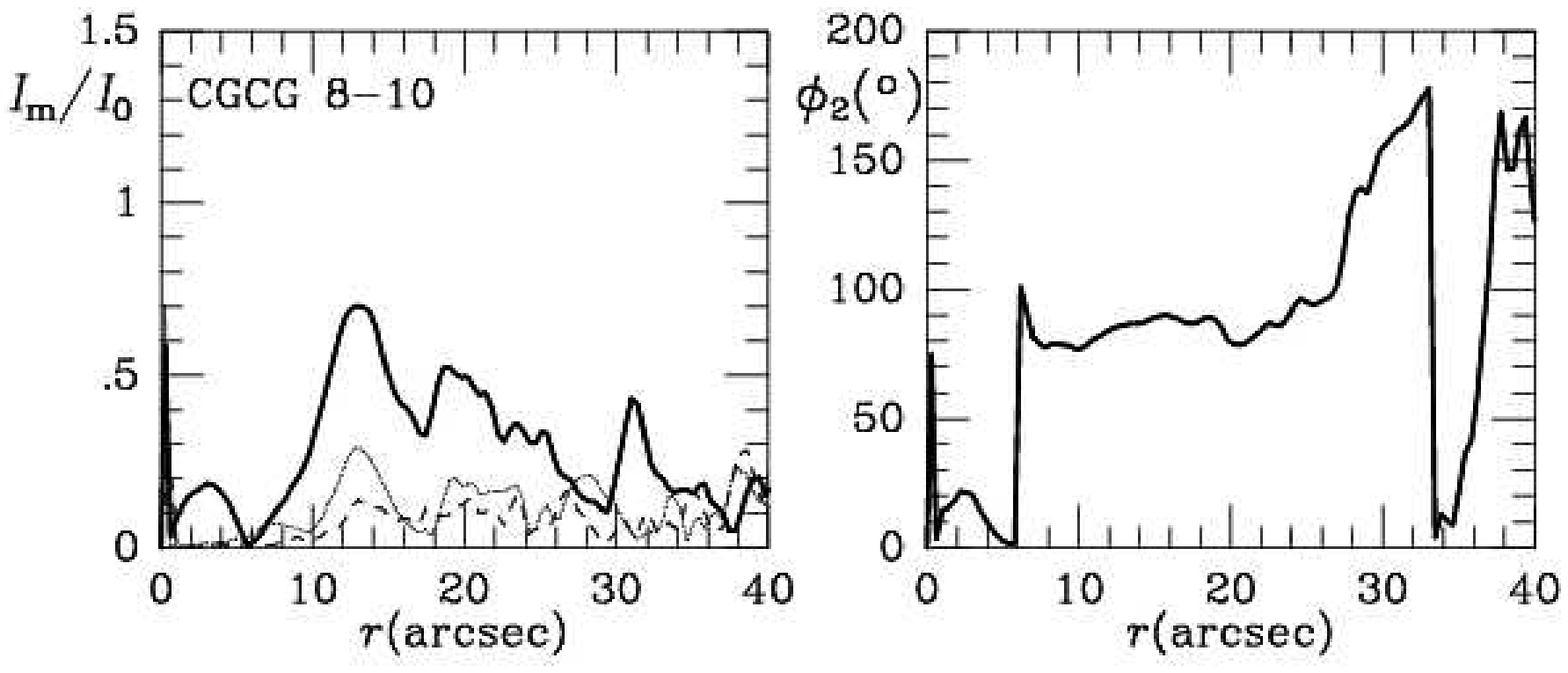}
\vskip -6.5cm
\includegraphics[width=\columnwidth,trim=0 -10 0 50,clip]{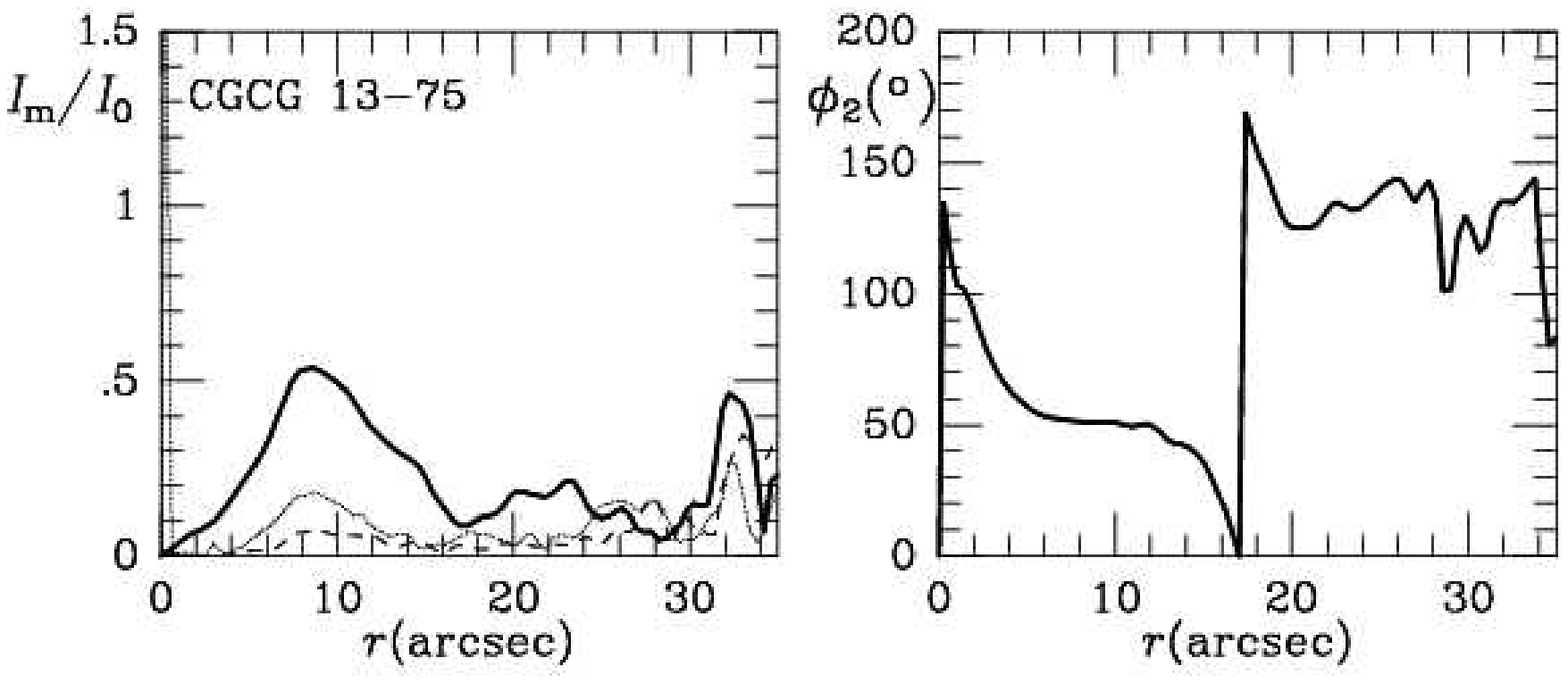}
\vskip -6.5cm
\includegraphics[width=\columnwidth,trim=0 -10 0 50,clip]{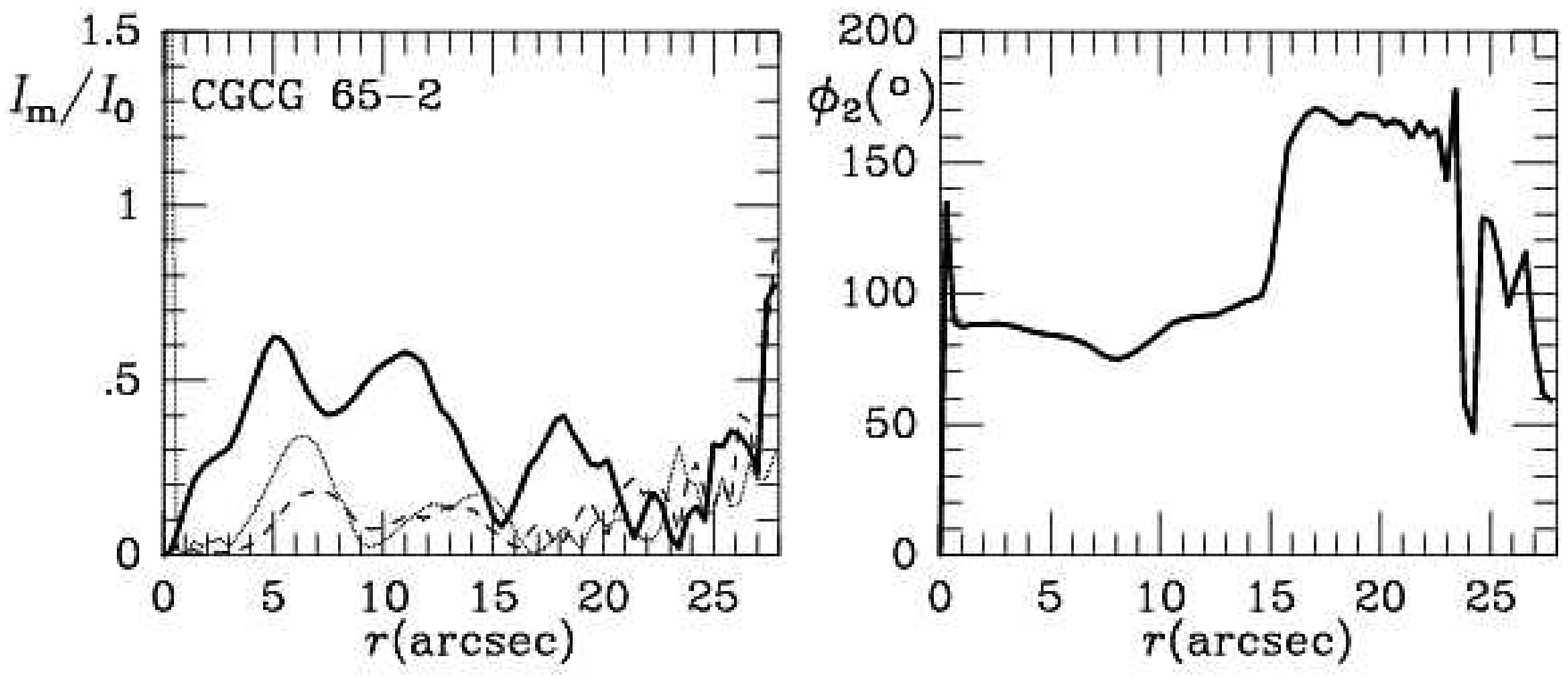}
\vskip -6.5cm
\includegraphics[width=\columnwidth,trim=0 -10 0 50,clip]{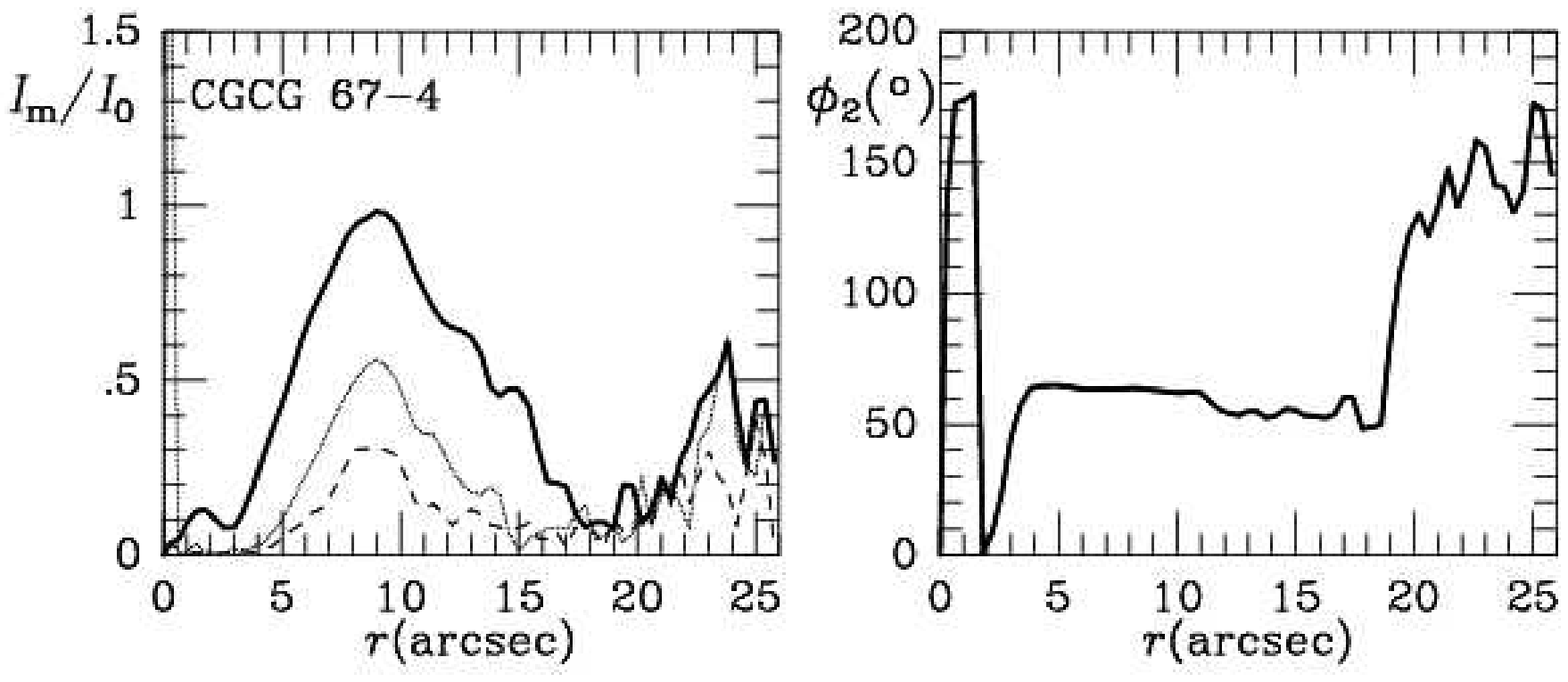}
\vskip -6.5cm
\includegraphics[width=\columnwidth,trim=0 -10 0 50,clip]{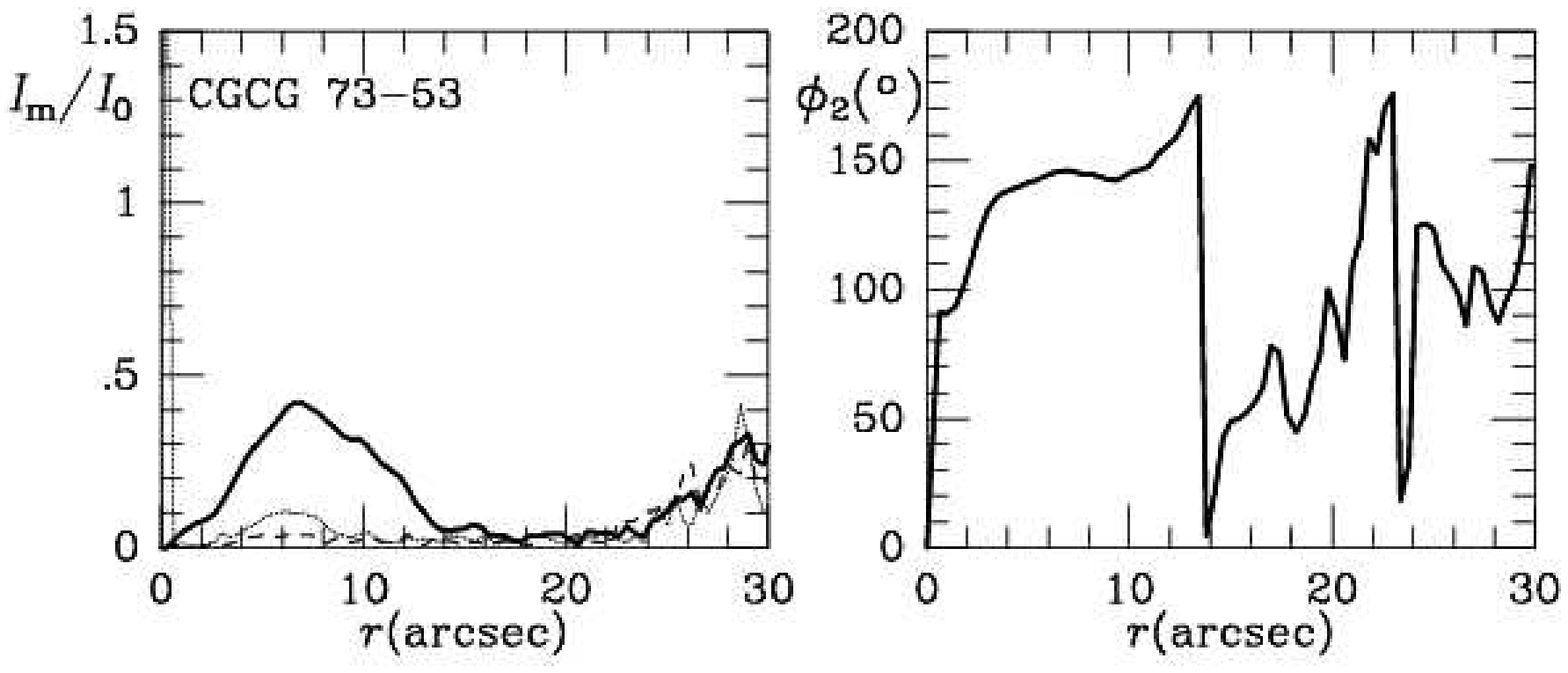}
\vskip -6.5cm
\includegraphics[width=\columnwidth,trim=0 -10 0 50,clip]{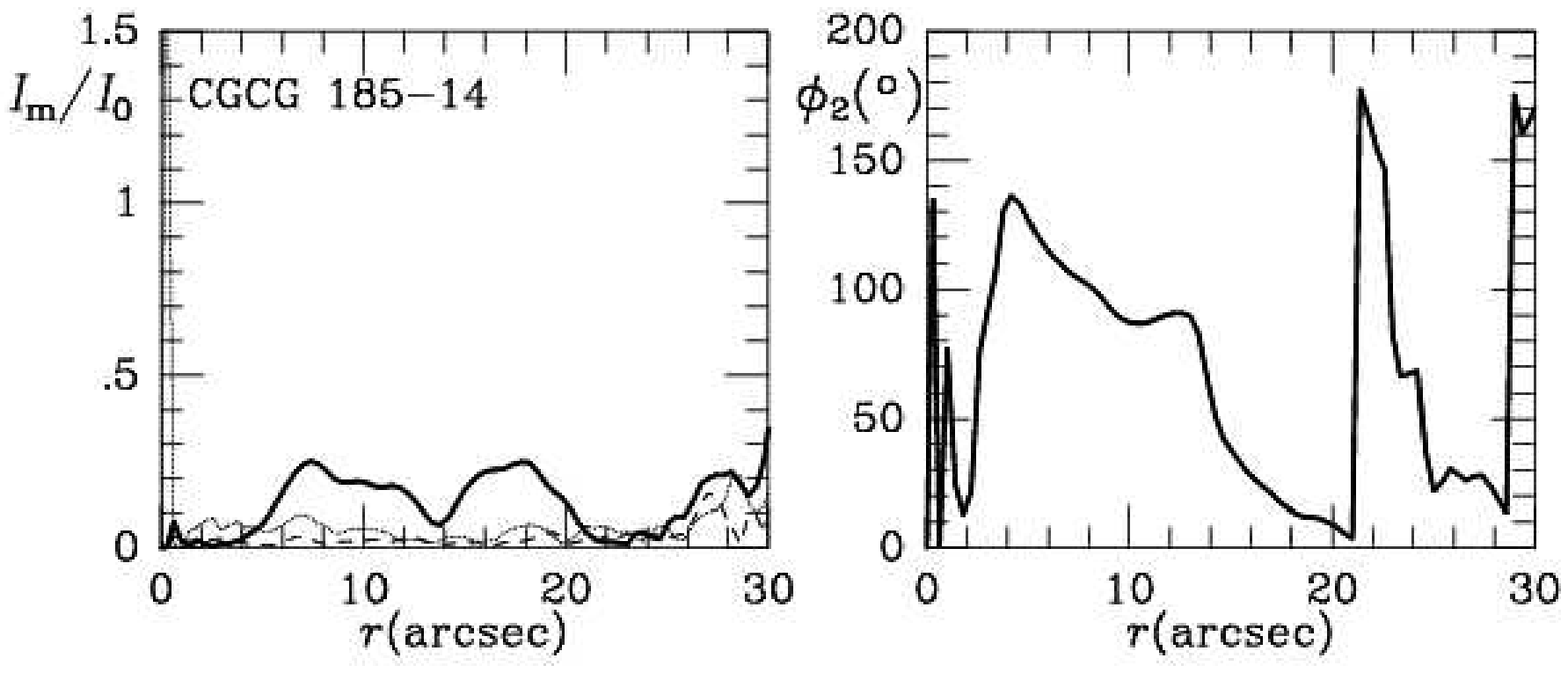}
\vskip -6.5cm
\caption{
}
\label{fig:allfours}
\end{figure}
\setcounter{figure}{21}
\begin{figure}
\includegraphics[width=\columnwidth,trim=0 -10 0 50,clip]{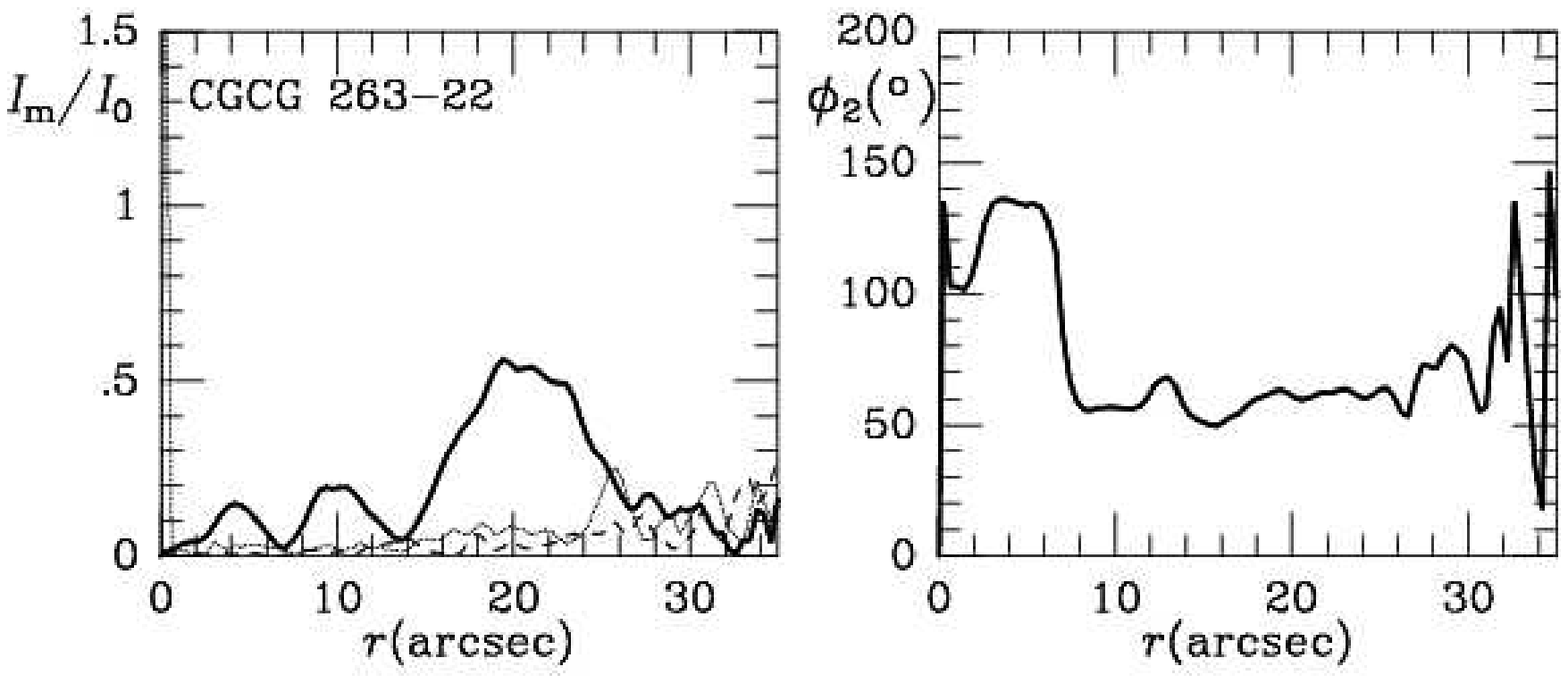}
\vskip -6.5cm
\includegraphics[width=\columnwidth,trim=0 -10 0 50,clip]{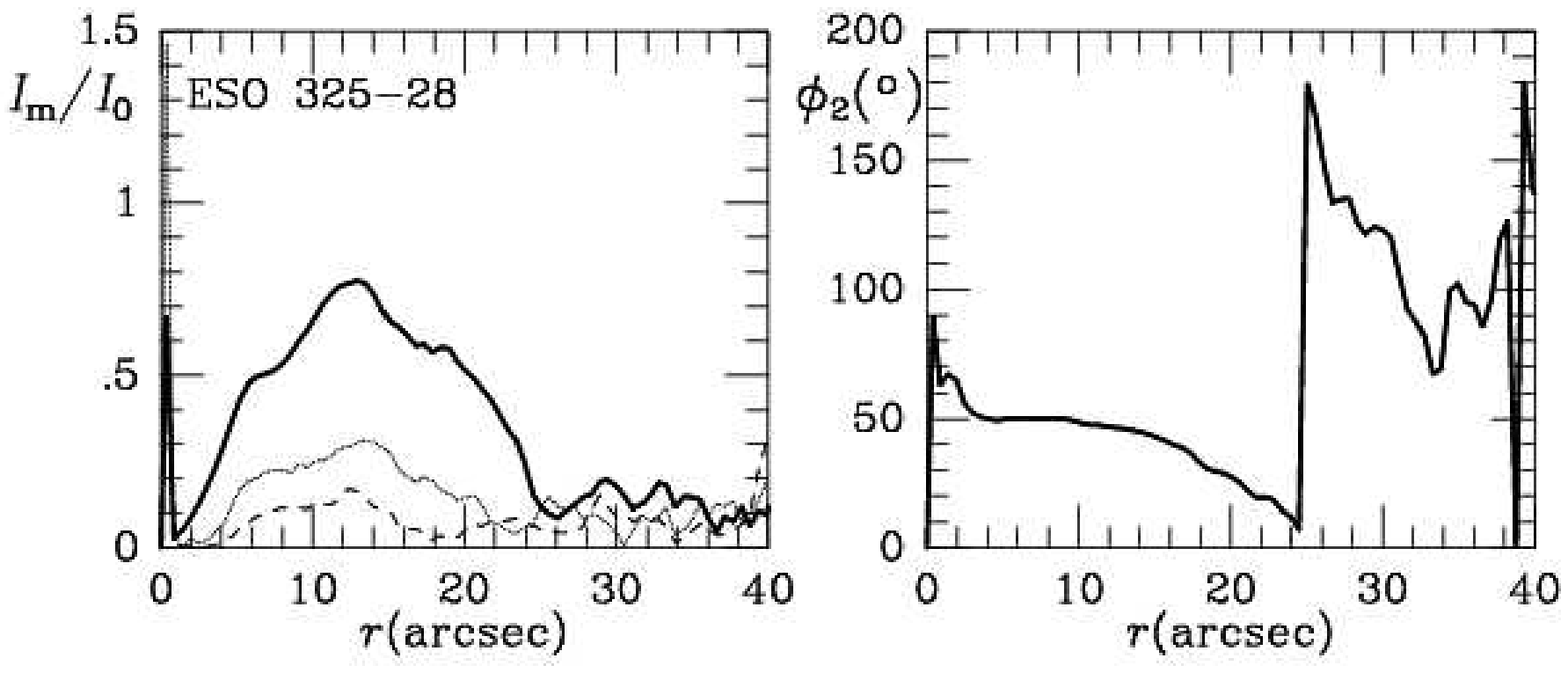}
\vskip -6.5cm
\includegraphics[width=\columnwidth,trim=0 -10 0 50,clip]{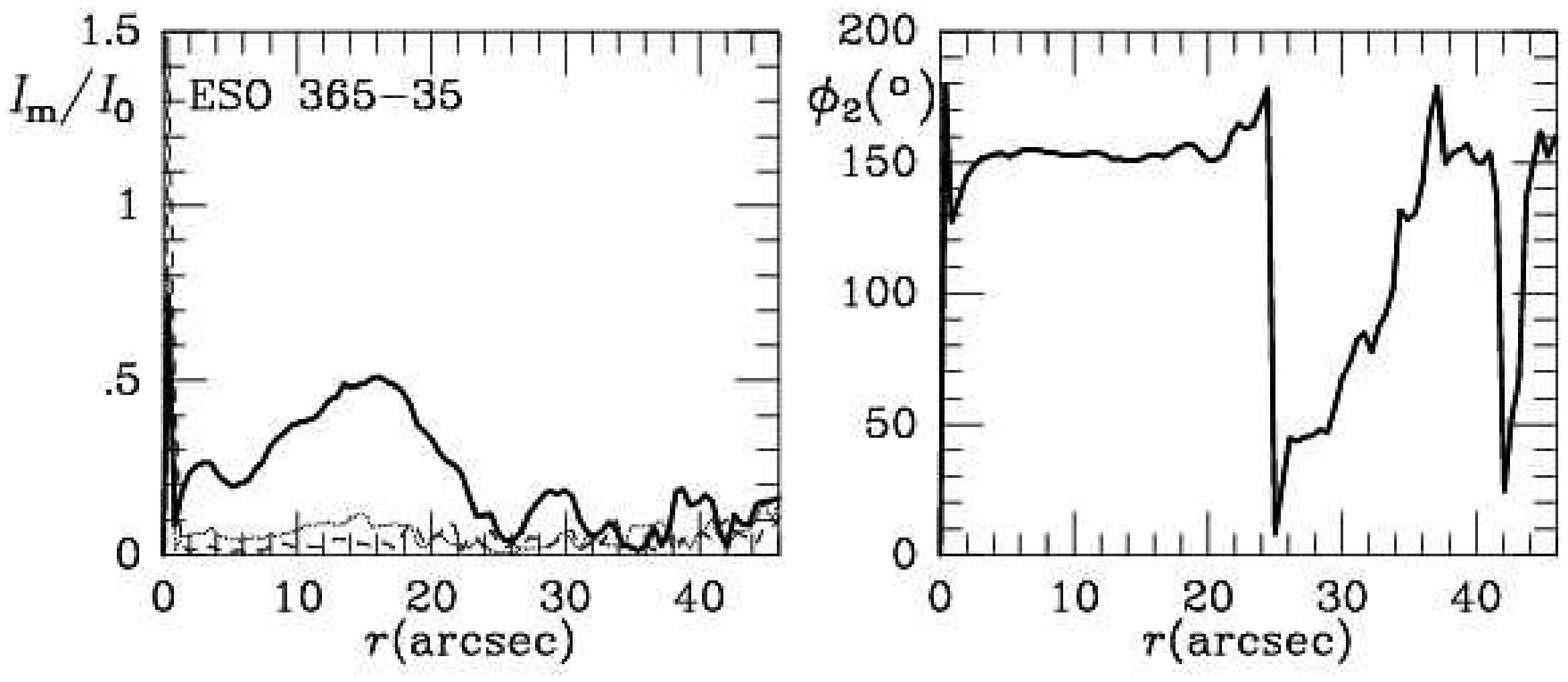}
\vskip -6.5cm
\includegraphics[width=\columnwidth,trim=0 -10 0 50,clip]{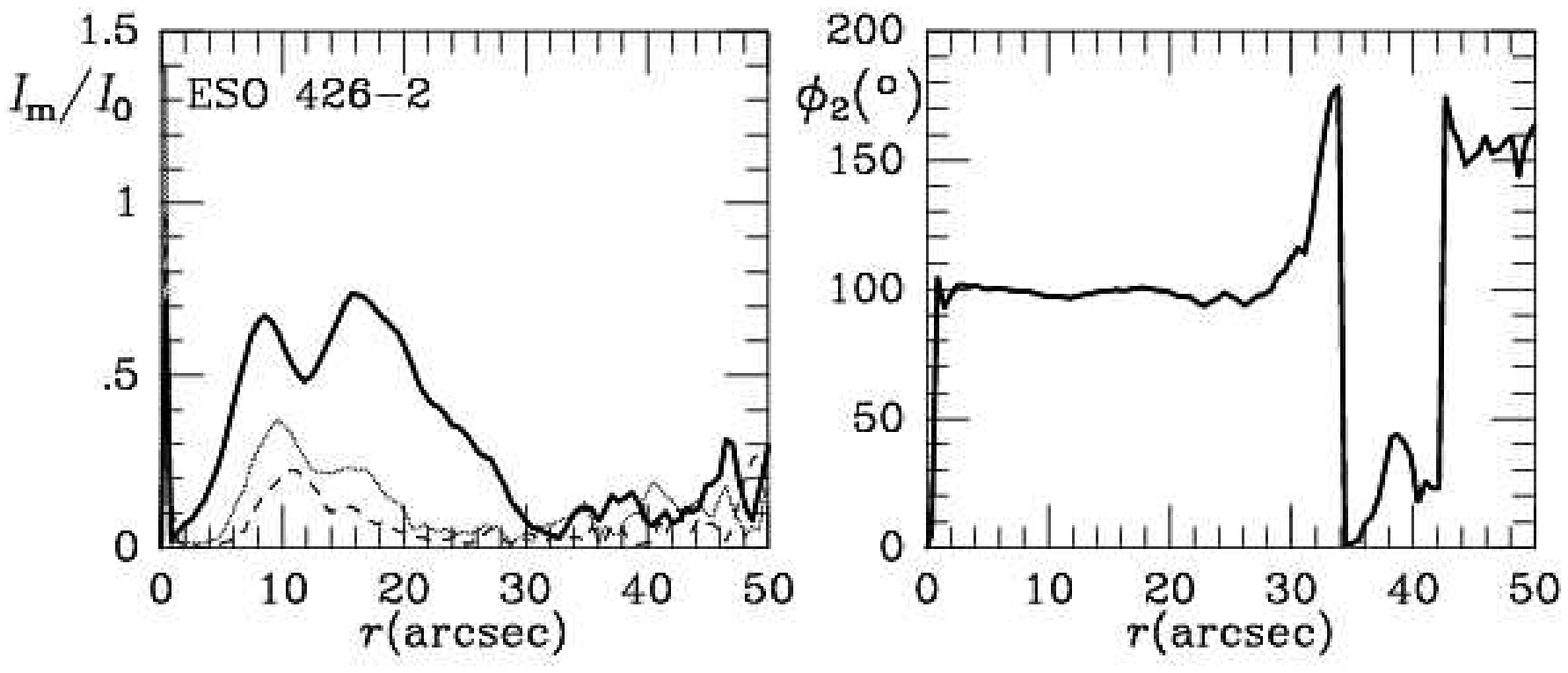}
\vskip -6.5cm
\includegraphics[width=\columnwidth,trim=0 -10 0 50,clip]{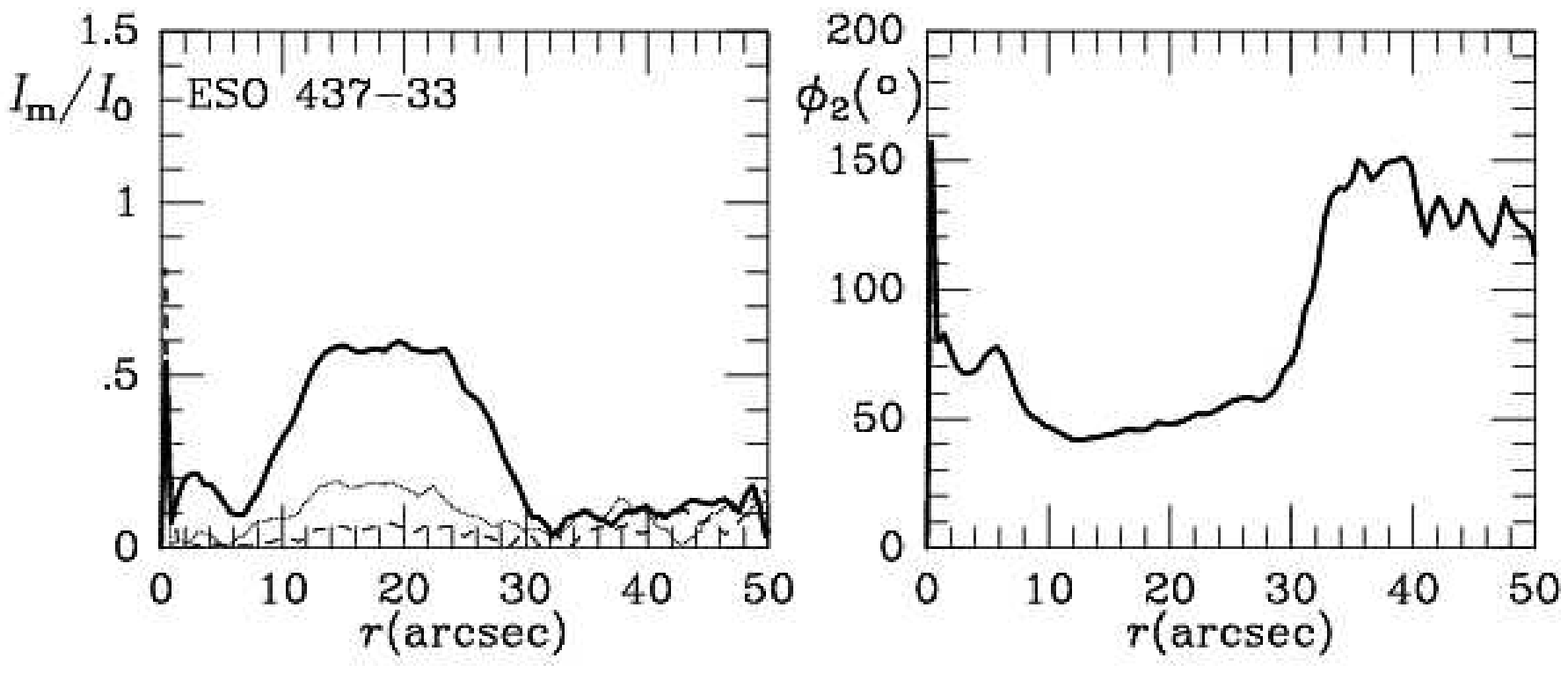}
\vskip -6.5cm
\includegraphics[width=\columnwidth,trim=0 -10 0 50,clip]{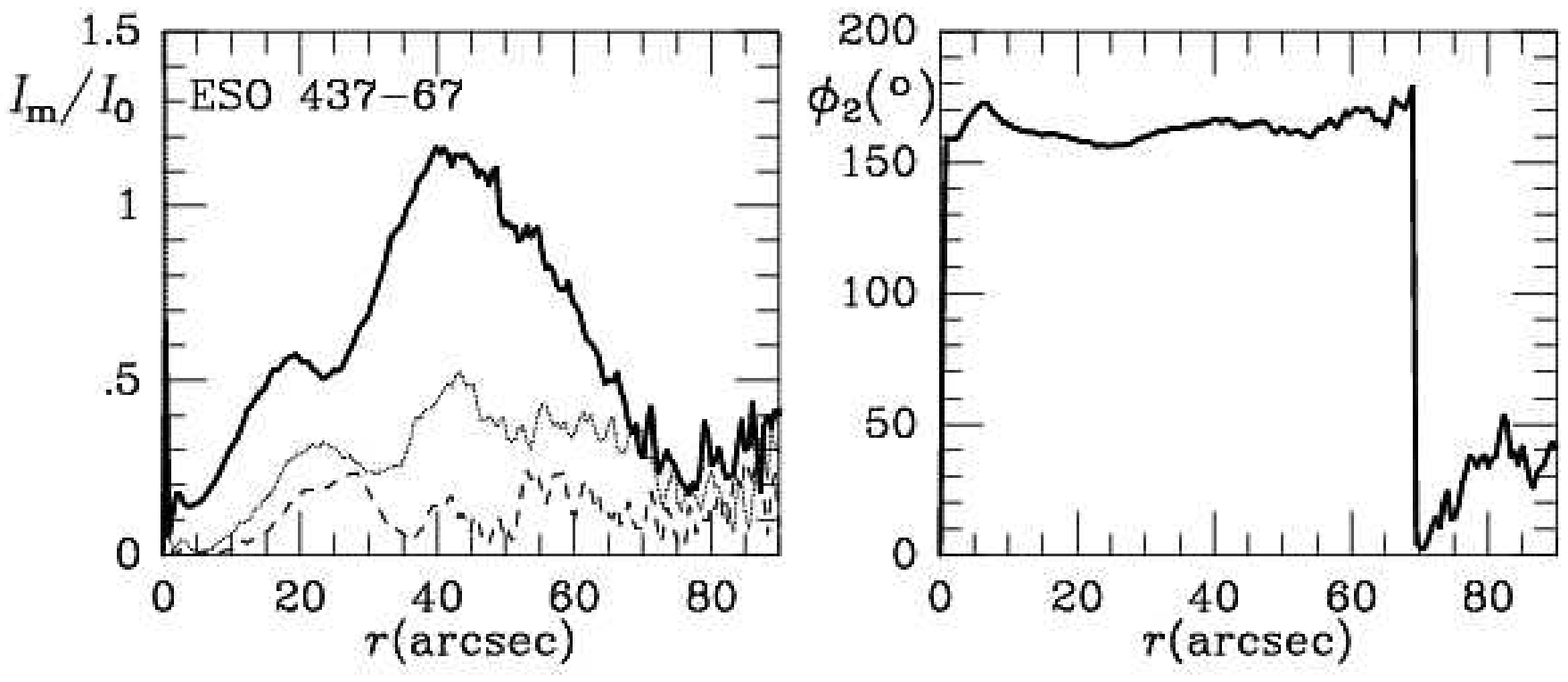}
\vskip -6.5cm
\caption{(cont.)}
\end{figure}
\setcounter{figure}{21}
\begin{figure}
\includegraphics[width=\columnwidth,trim=0 -10 0 50,clip]{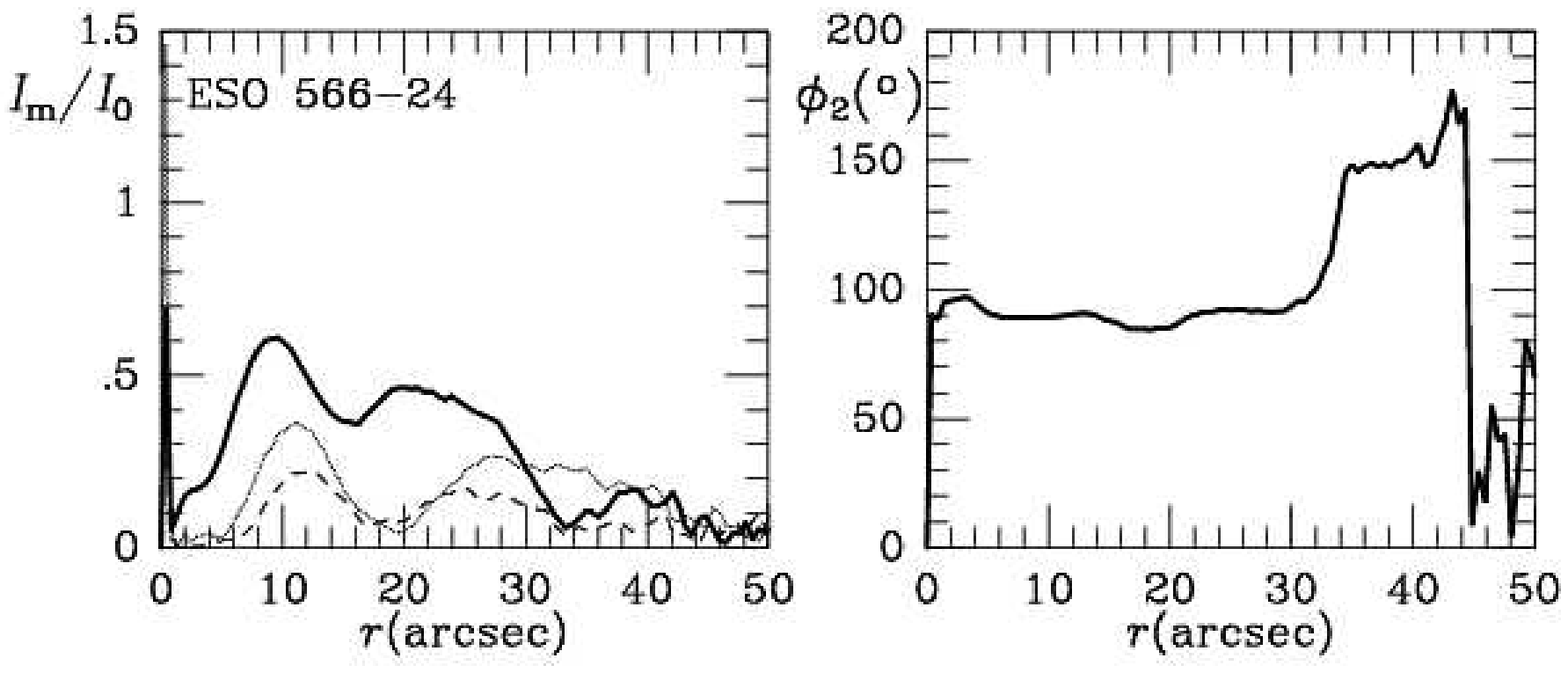}
\vskip -6.5cm
\includegraphics[width=\columnwidth,trim=0 -10 0 50,clip]{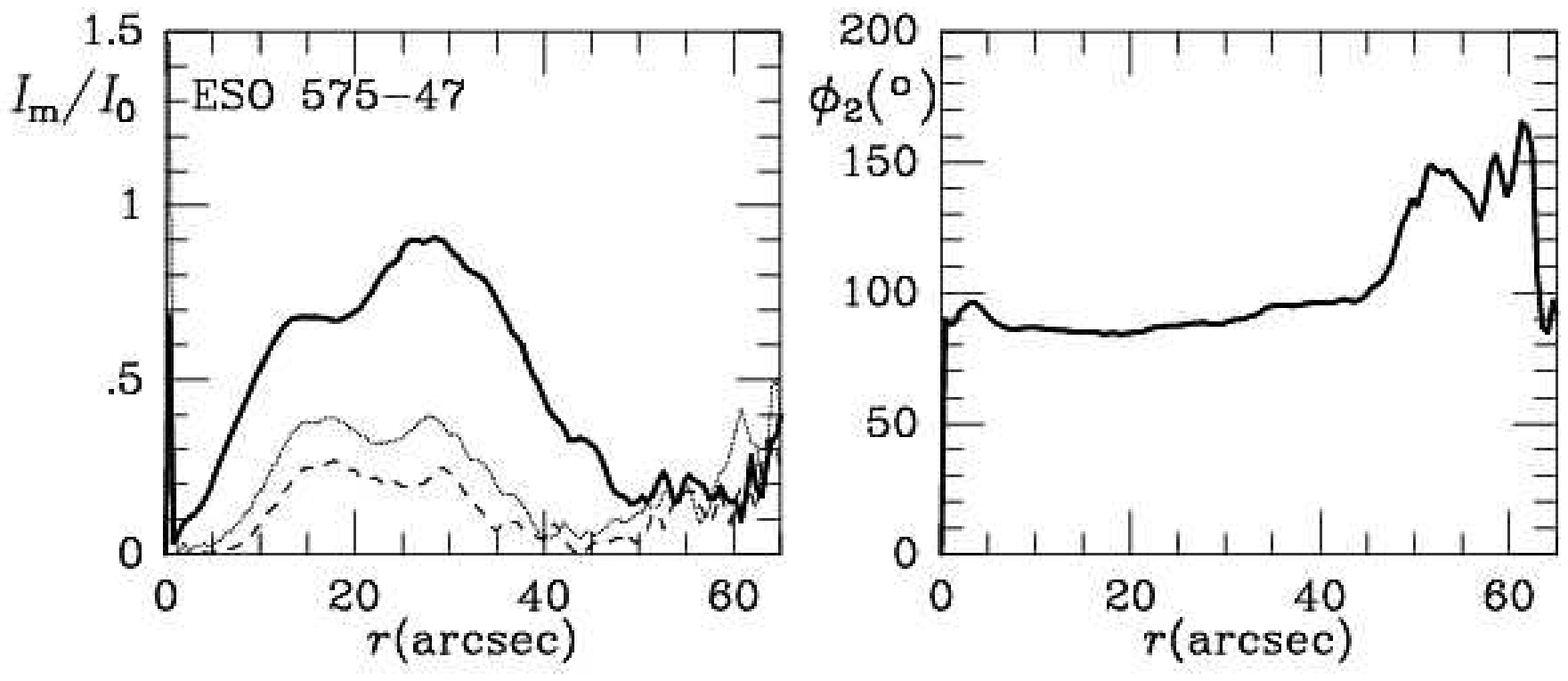}
\vskip -6.5cm
\includegraphics[width=\columnwidth,trim=0 -10 0 50,clip]{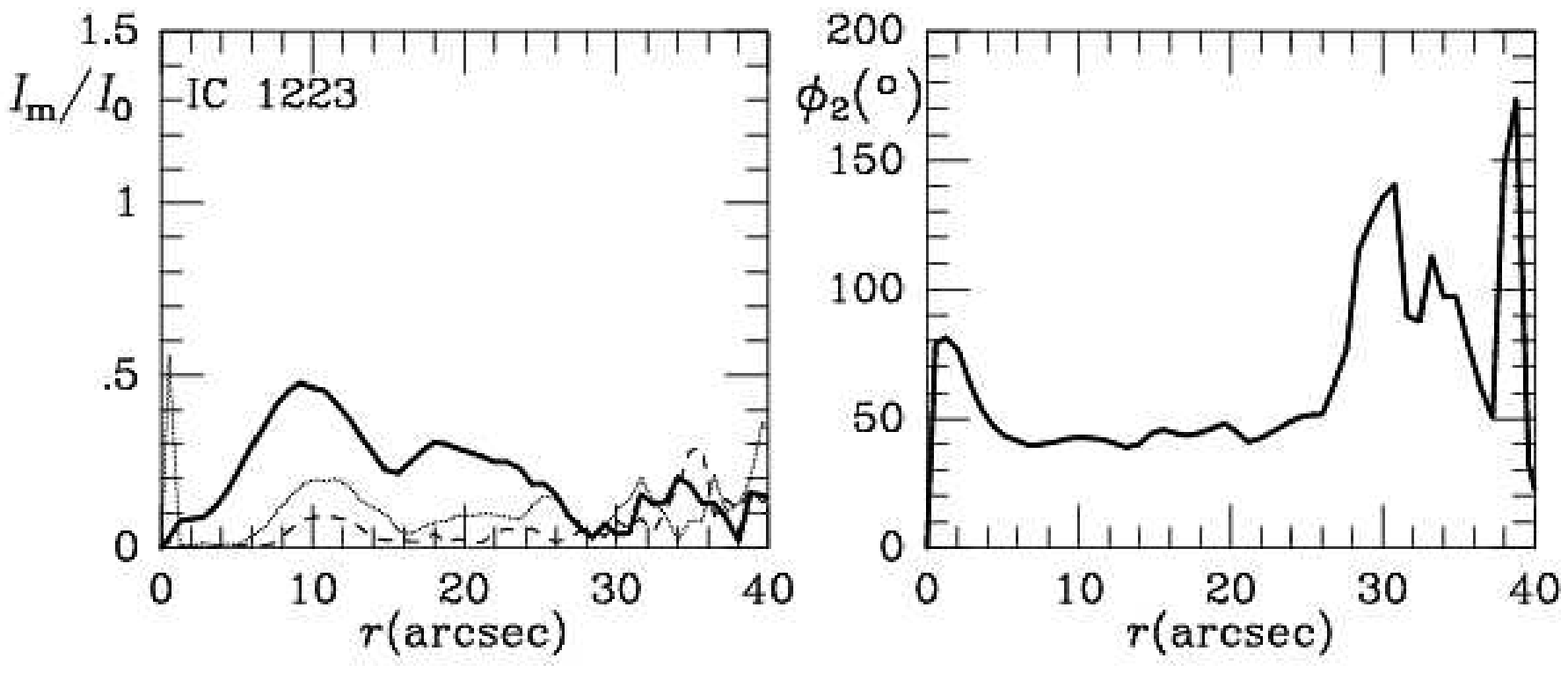}
\vskip -6.5cm
\includegraphics[width=\columnwidth,trim=0 -10 0 50,clip]{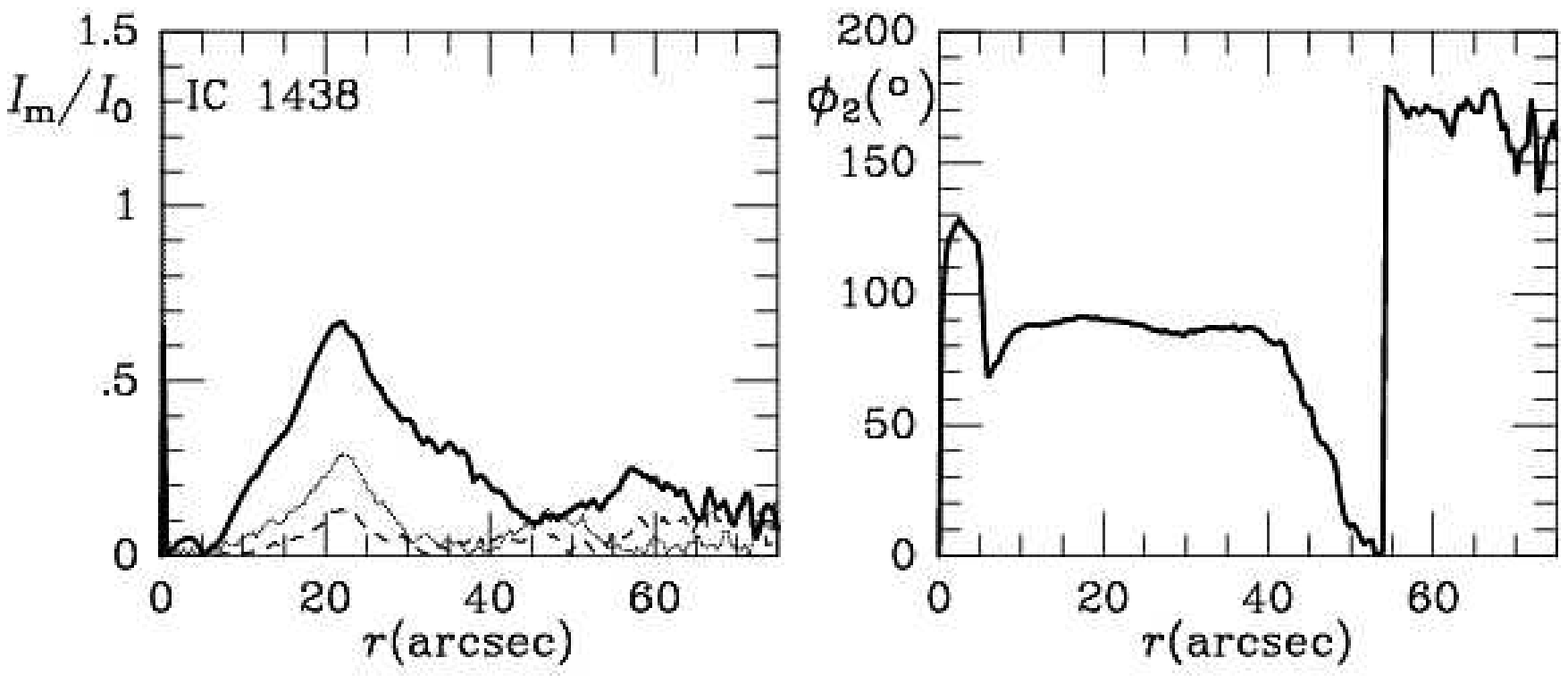}
\vskip -6.5cm
\includegraphics[width=\columnwidth,trim=0 -10 0 50,clip]{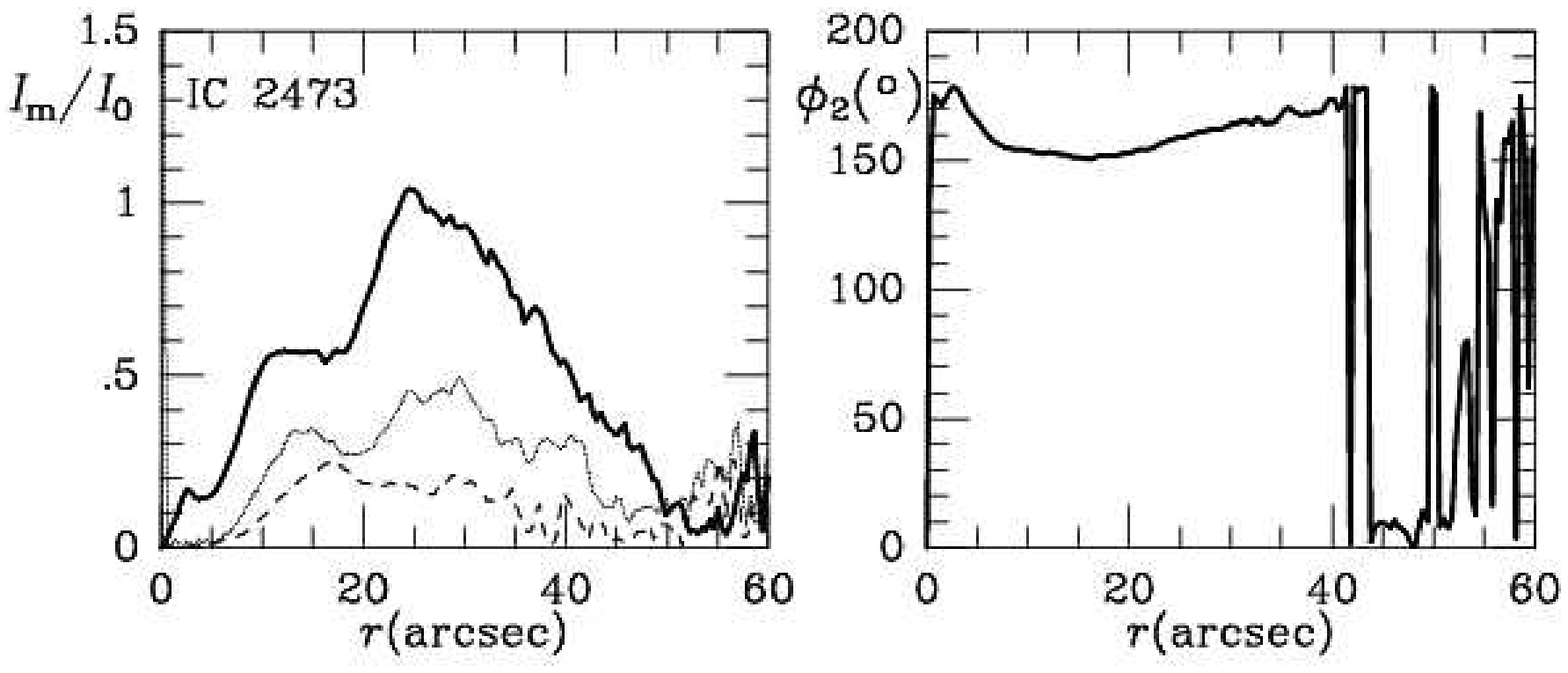}
\vskip -6.5cm
\includegraphics[width=\columnwidth,trim=0 -10 0 50,clip]{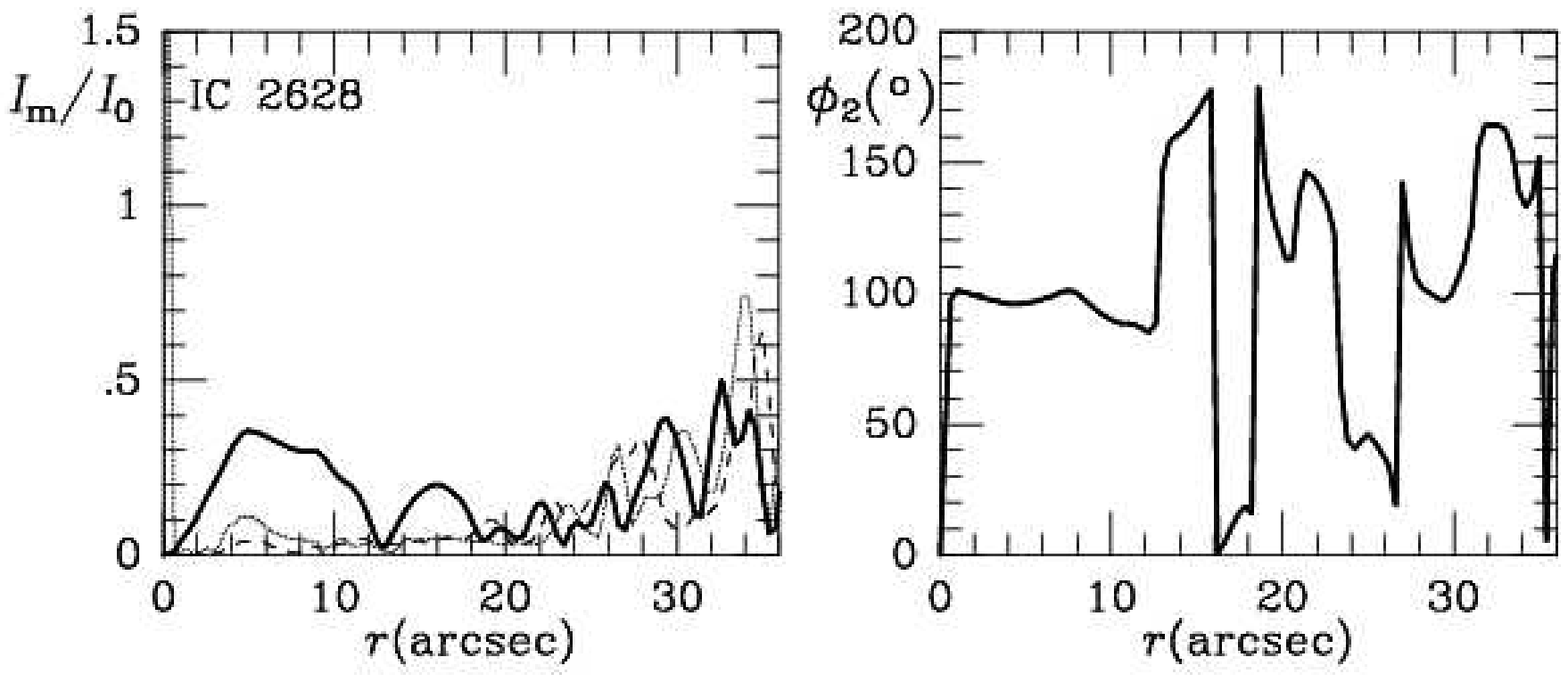}
\vskip -6.5cm
\caption{(cont.)}
\end{figure}
\setcounter{figure}{21}
\begin{figure}
\includegraphics[width=\columnwidth,trim=0 -10 0 50,clip]{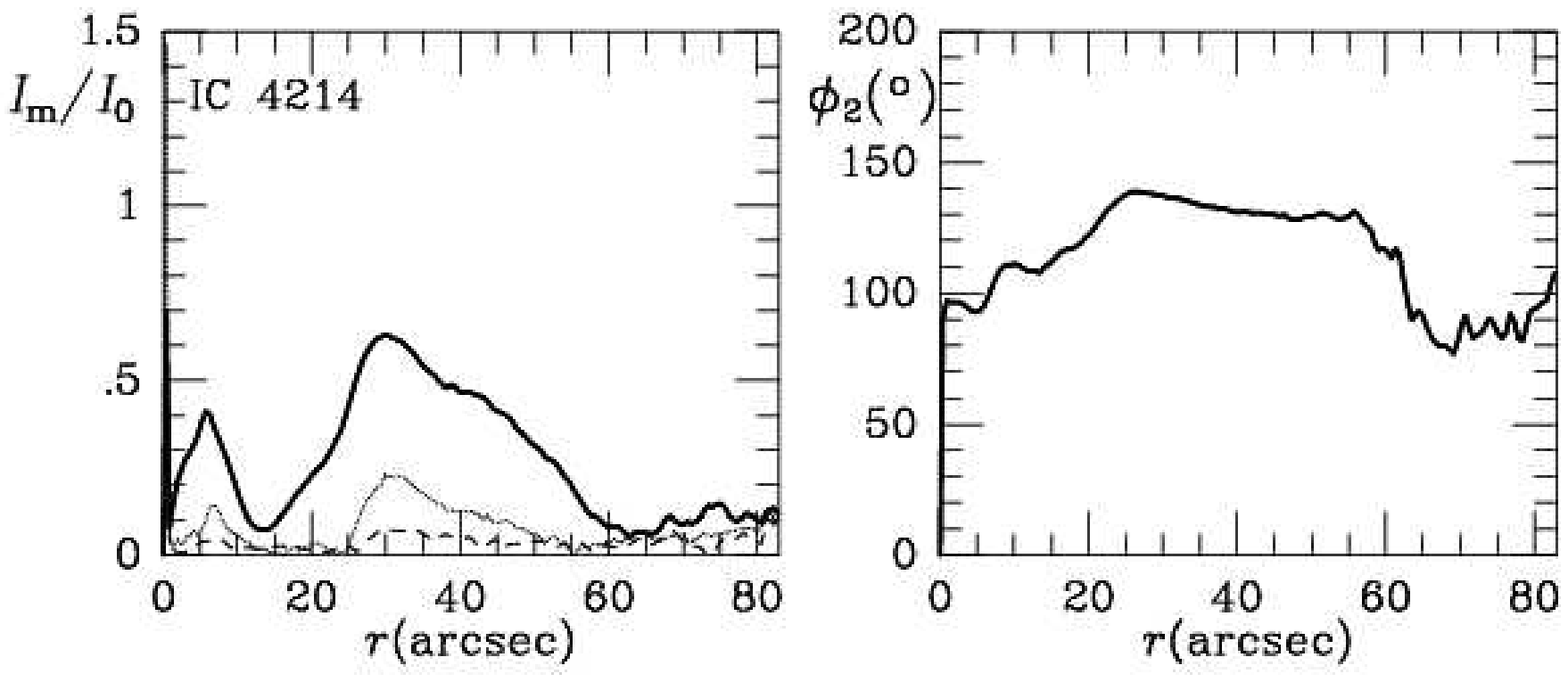}
\vskip -6.5cm
\includegraphics[width=\columnwidth,trim=0 -10 0 50,clip]{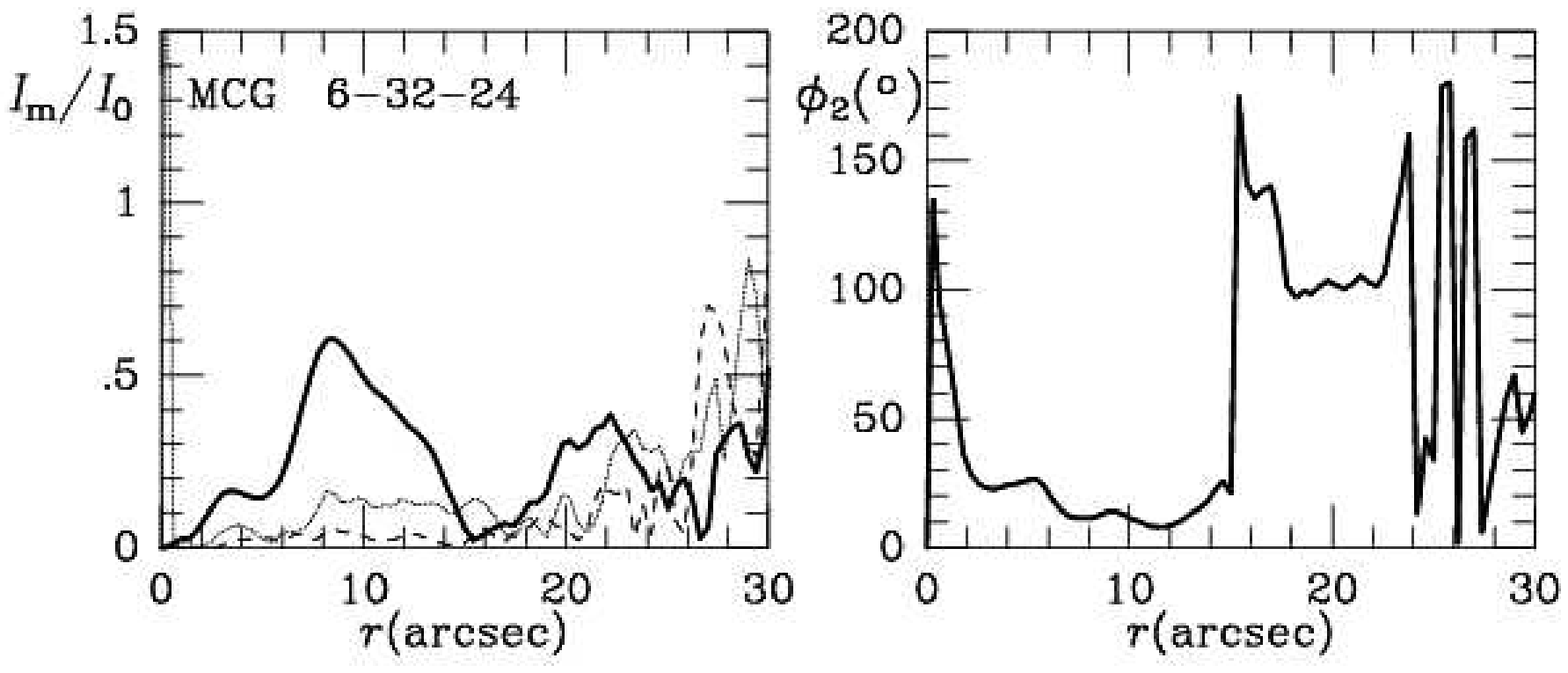}
\vskip -6.5cm
\includegraphics[width=\columnwidth,trim=0 -10 0 50,clip]{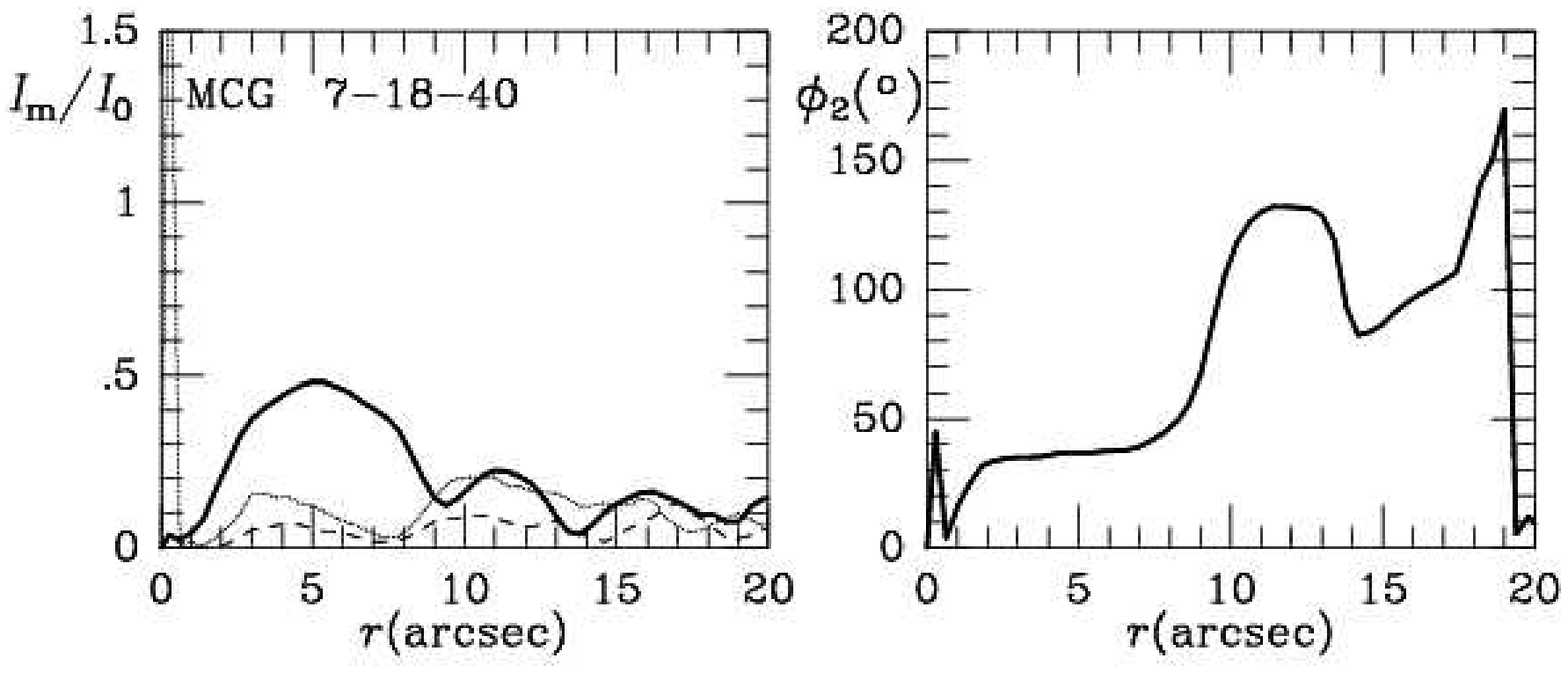}
\vskip -6.5cm
\includegraphics[width=\columnwidth,trim=0 -10 0 50,clip]{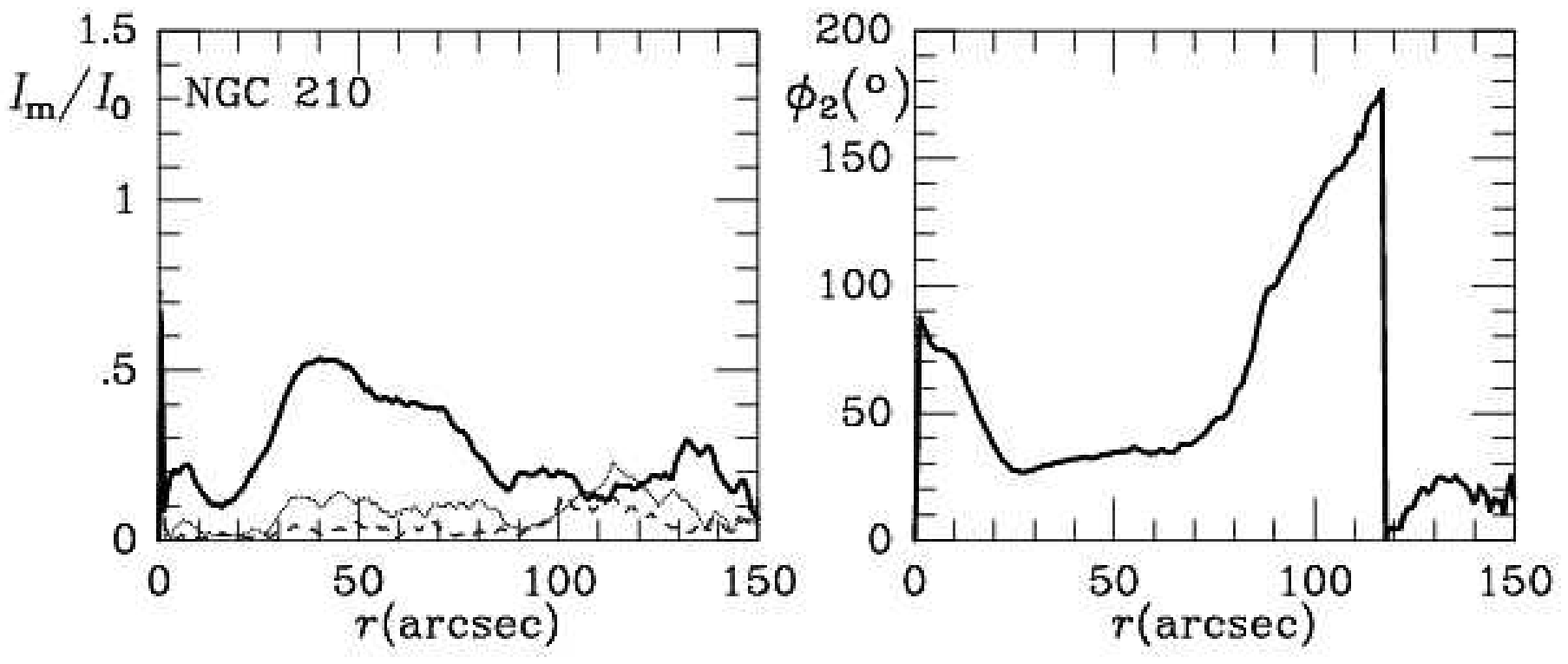}
\vskip -6.5cm
\includegraphics[width=\columnwidth,trim=0 -10 0 50,clip]{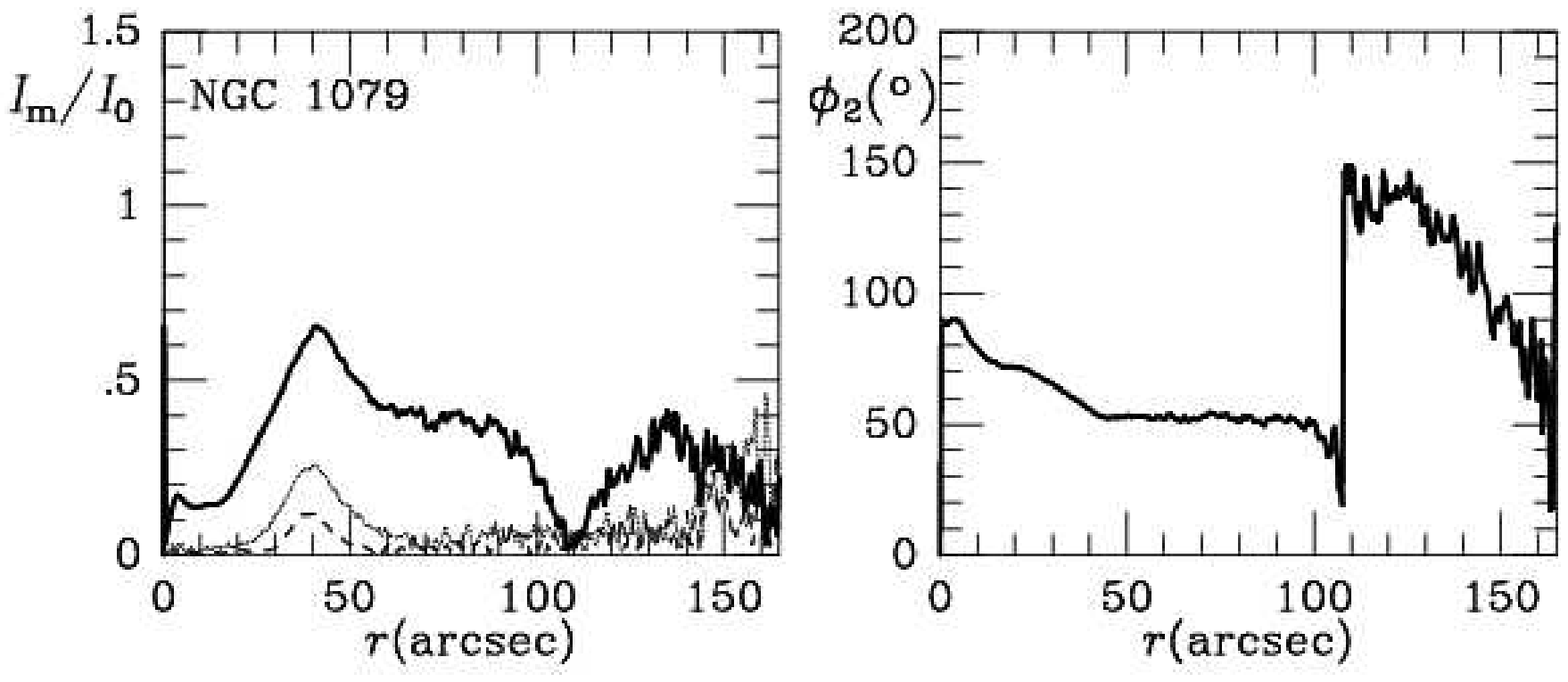}
\vskip -6.5cm
\includegraphics[width=\columnwidth,trim=0 -10 0 50,clip]{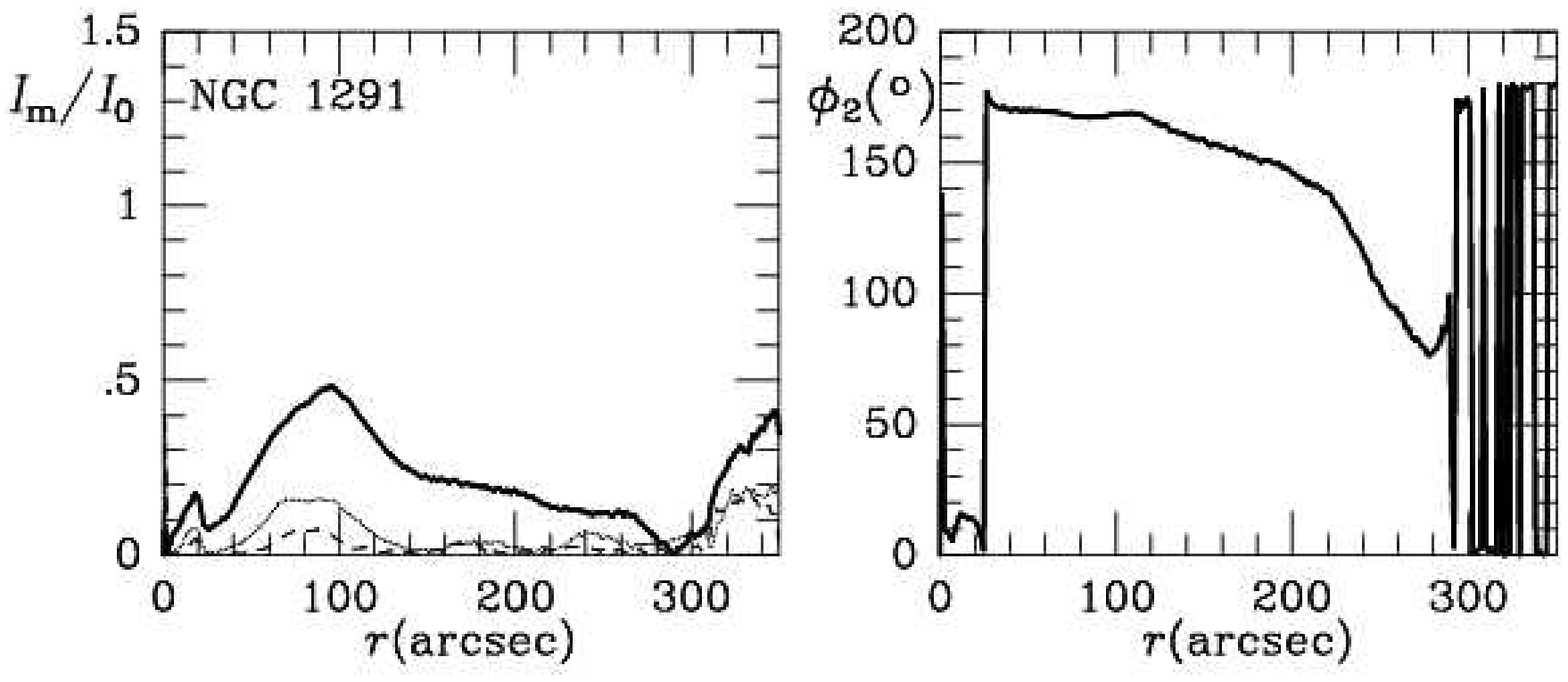}
\vskip -6.5cm
\caption{(cont.)}
\end{figure}
\setcounter{figure}{21}
\begin{figure}
\includegraphics[width=\columnwidth,trim=0 -10 0 50,clip]{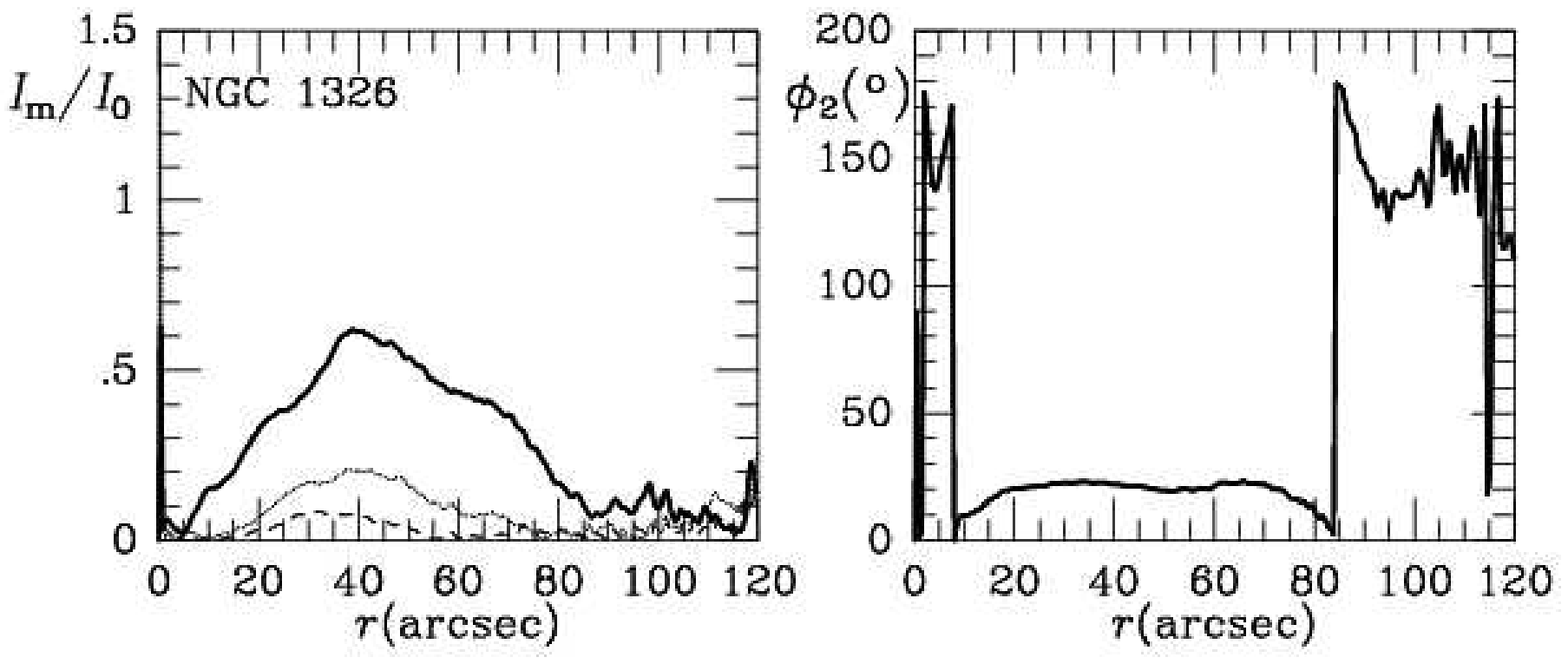}
\vskip -6.5cm
\includegraphics[width=\columnwidth,trim=0 -10 0 50,clip]{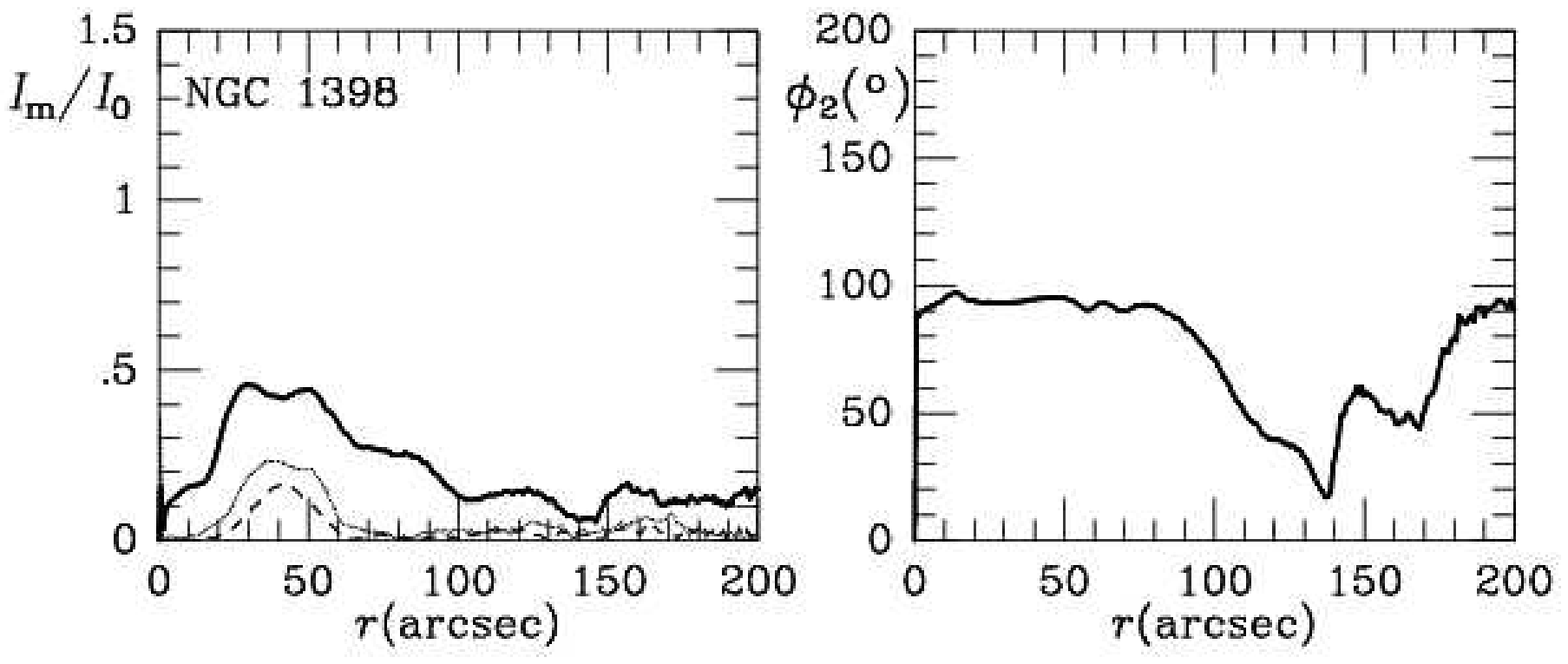}
\vskip -6.5cm
\includegraphics[width=\columnwidth,trim=0 -10 0 50,clip]{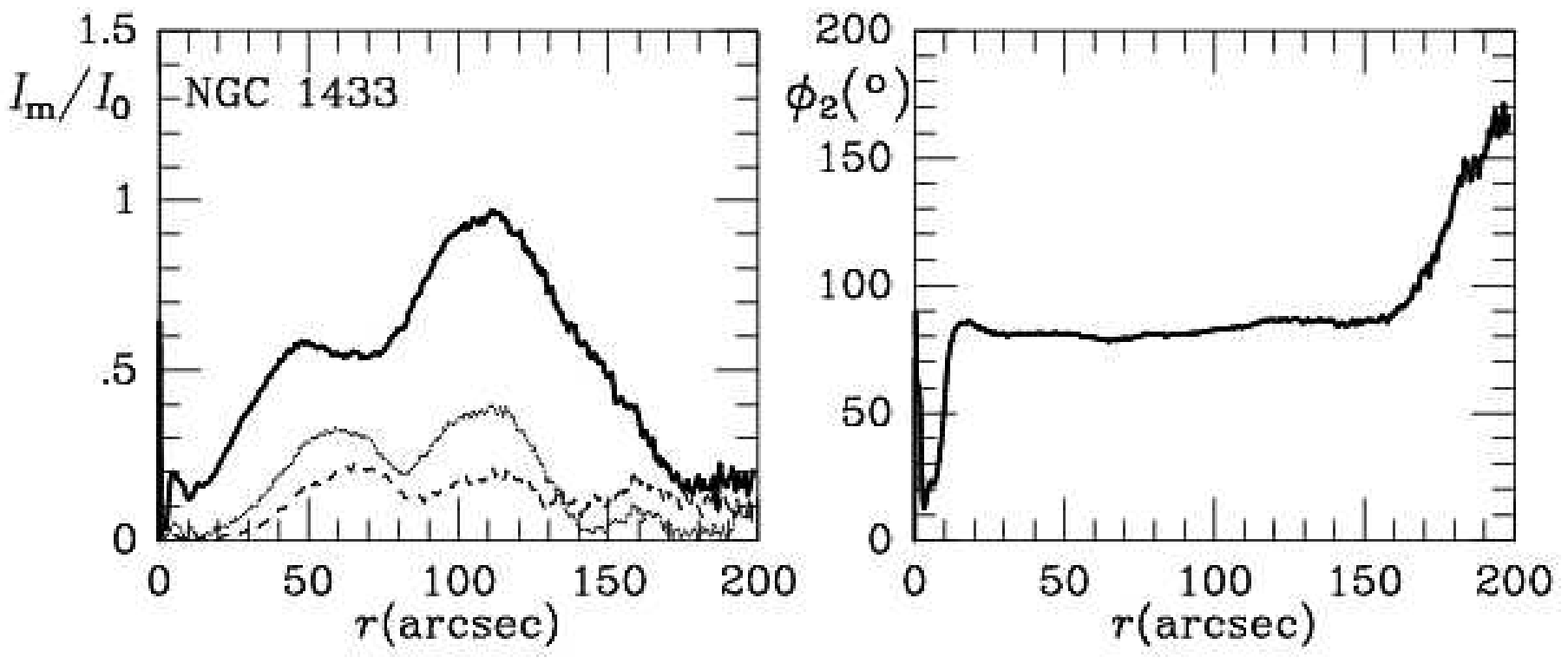}
\vskip -6.5cm
\includegraphics[width=\columnwidth,trim=0 -10 0 50,clip]{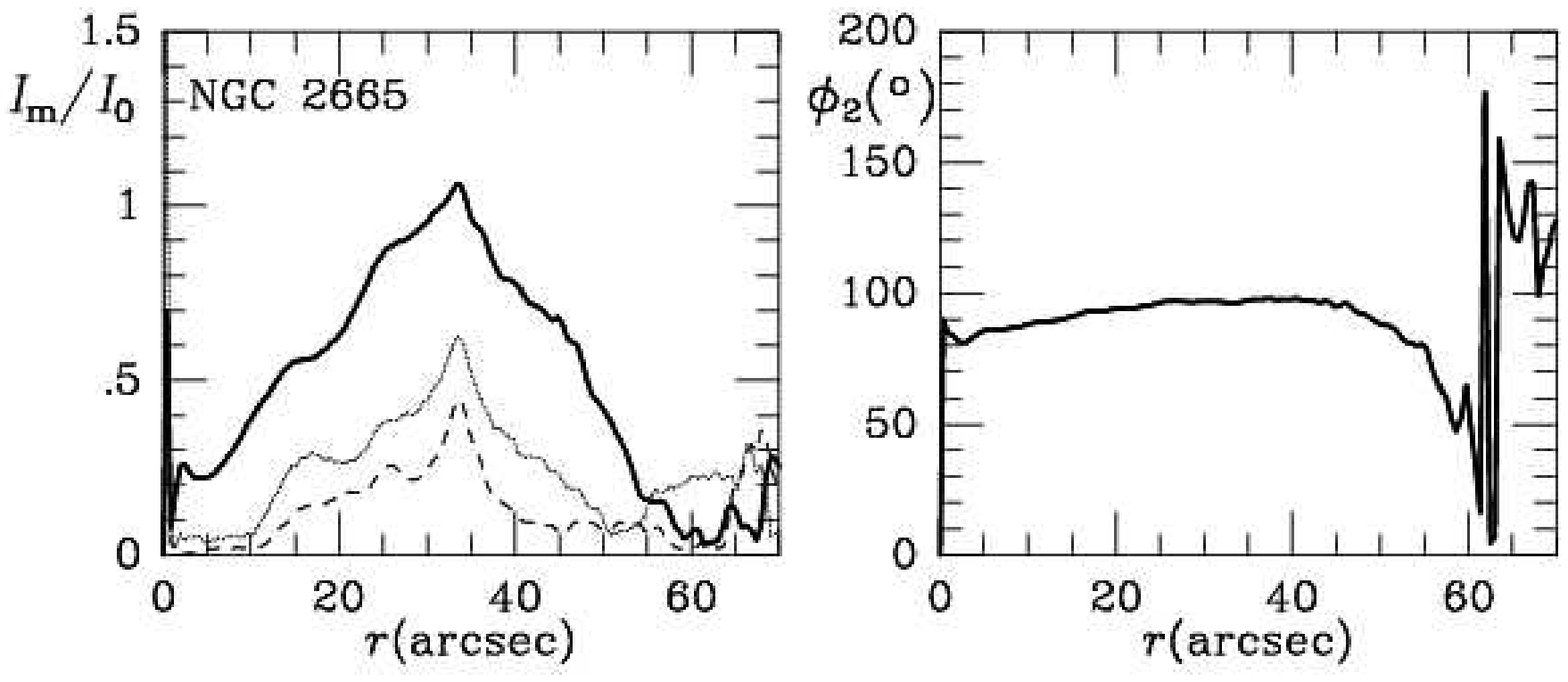}
\vskip -6.5cm
\includegraphics[width=\columnwidth,trim=0 -10 0 50,clip]{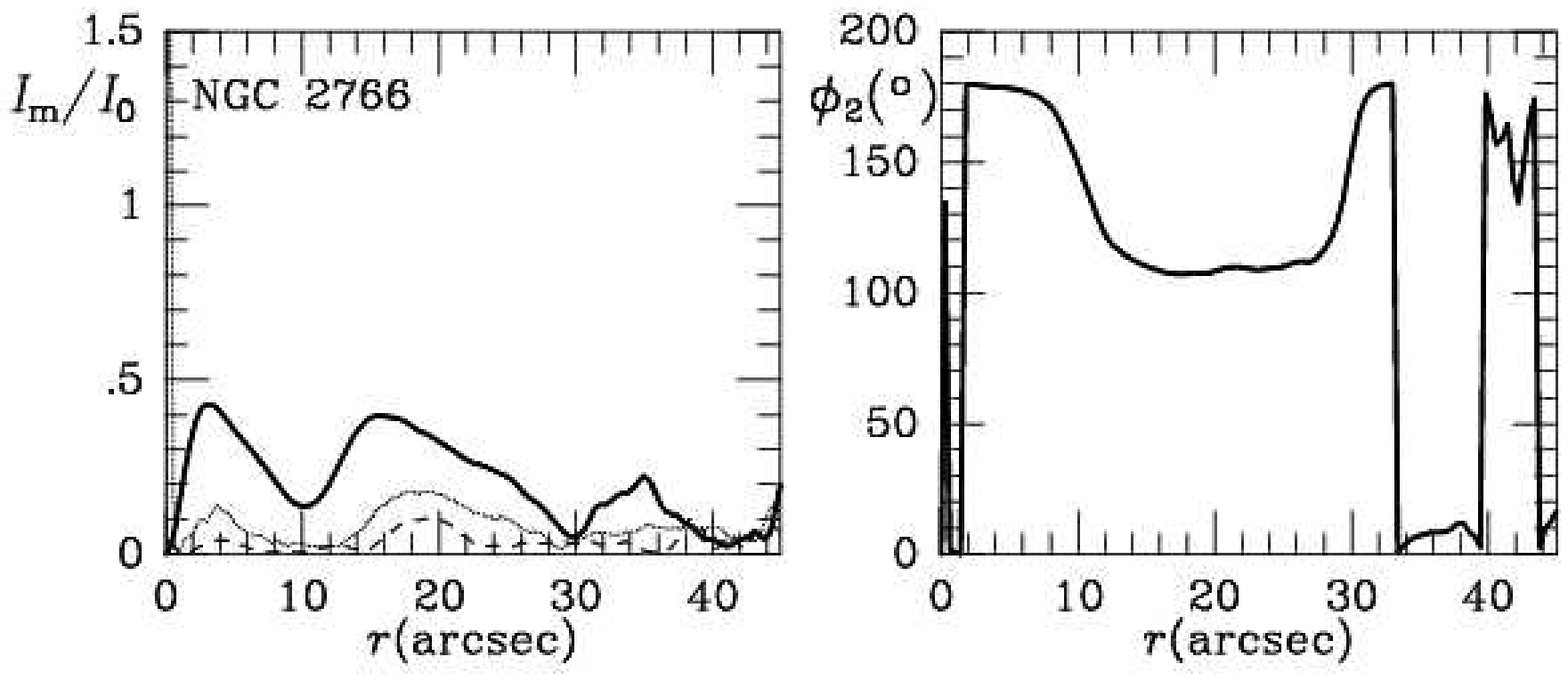}
\vskip -6.5cm
\includegraphics[width=\columnwidth,trim=0 -10 0 50,clip]{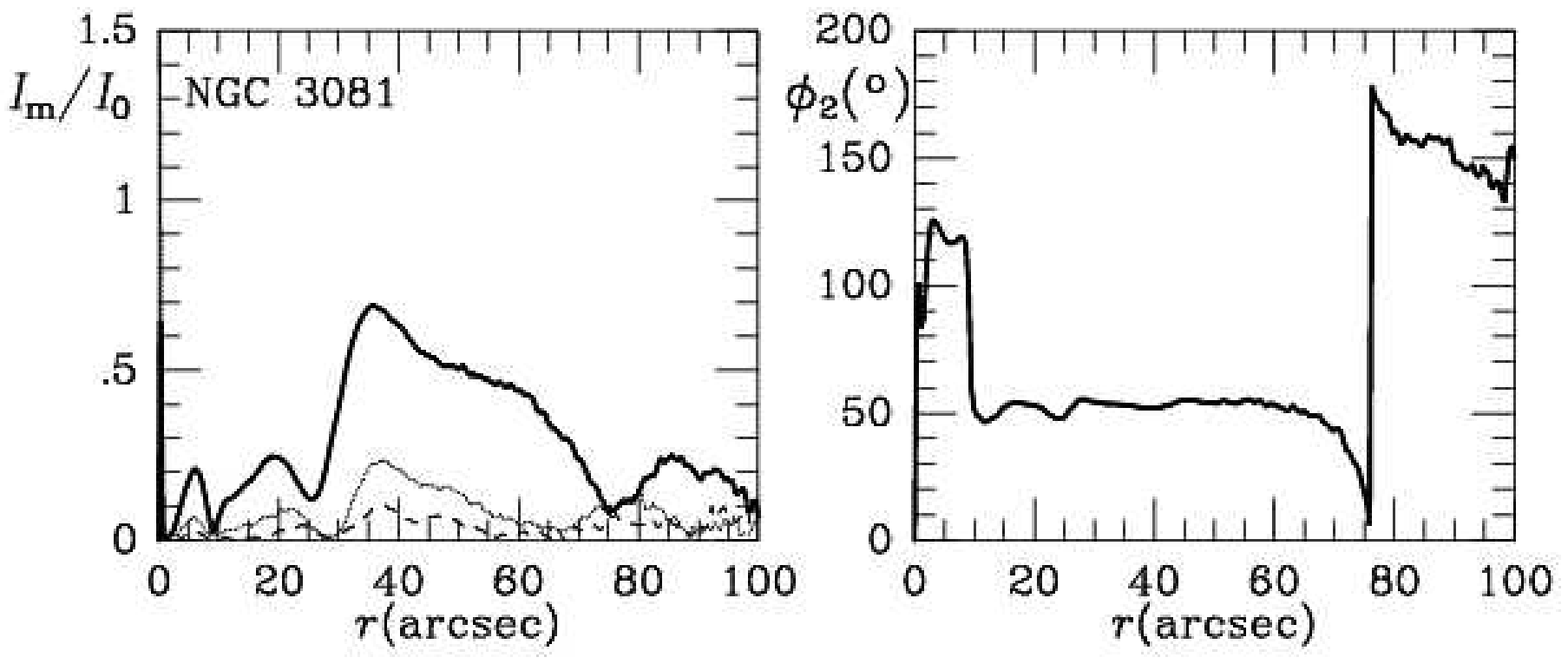}
\vskip -6.5cm
\caption{(cont.)}
\end{figure}
\setcounter{figure}{21}
\begin{figure}
\includegraphics[width=\columnwidth,trim=0 -10 0 50,clip]{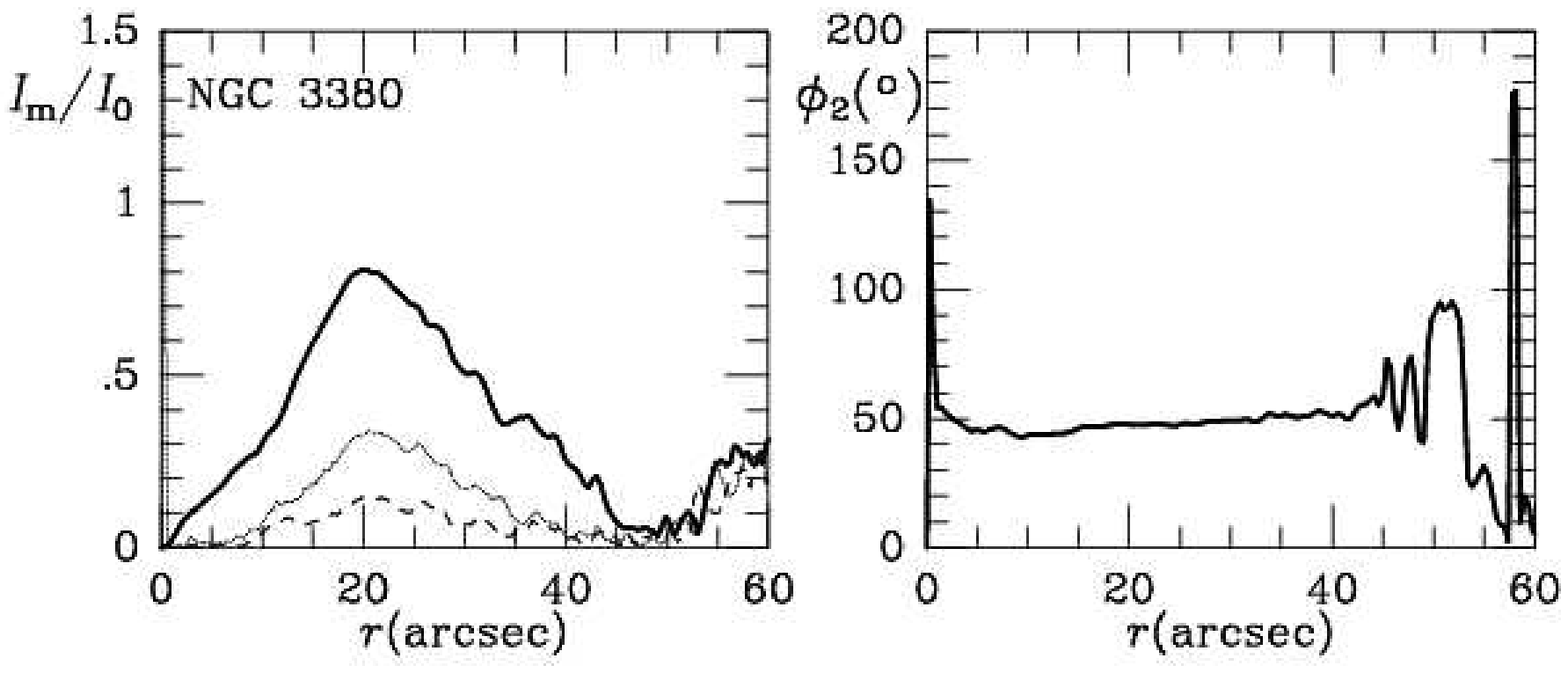}
\vskip -6.5cm
\includegraphics[width=\columnwidth,trim=0 -10 0 50,clip]{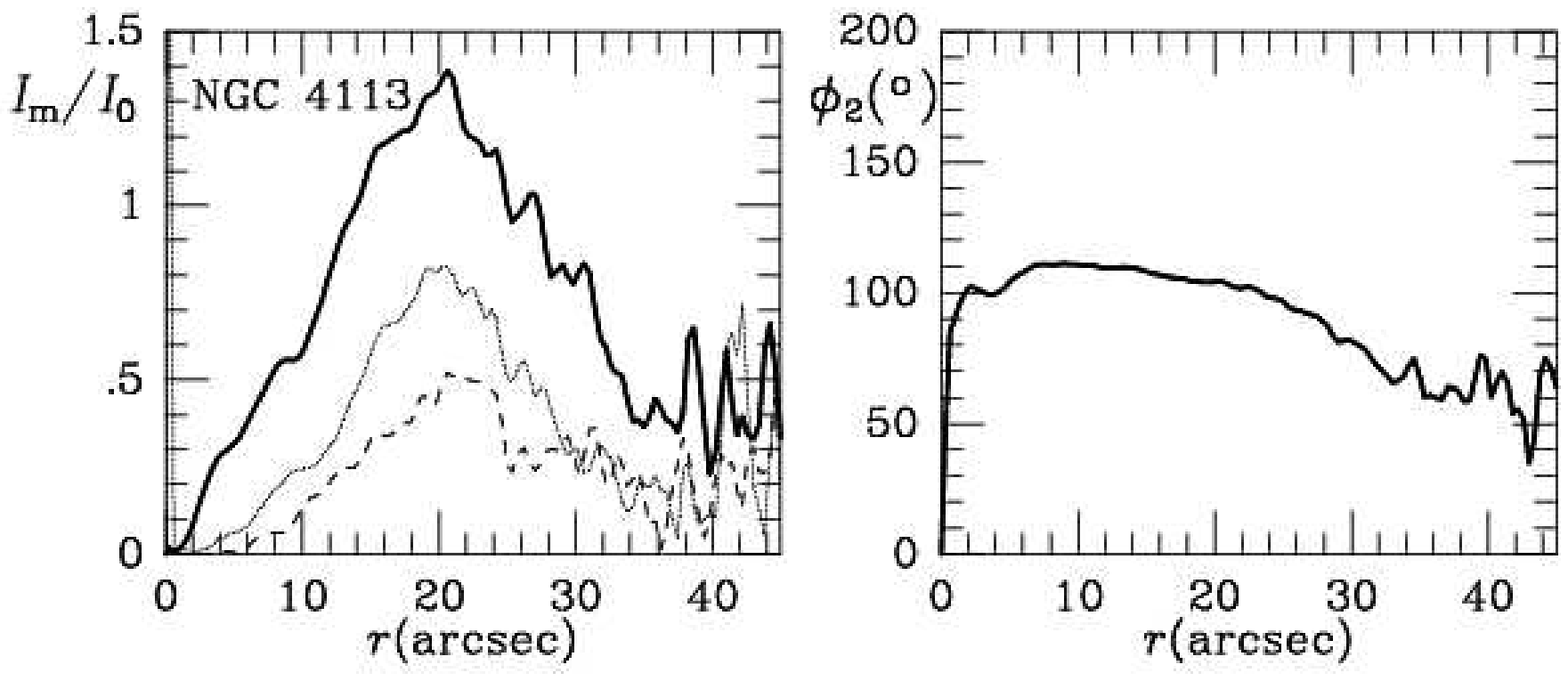}
\vskip -6.5cm
\includegraphics[width=\columnwidth,trim=0 -10 0 50,clip]{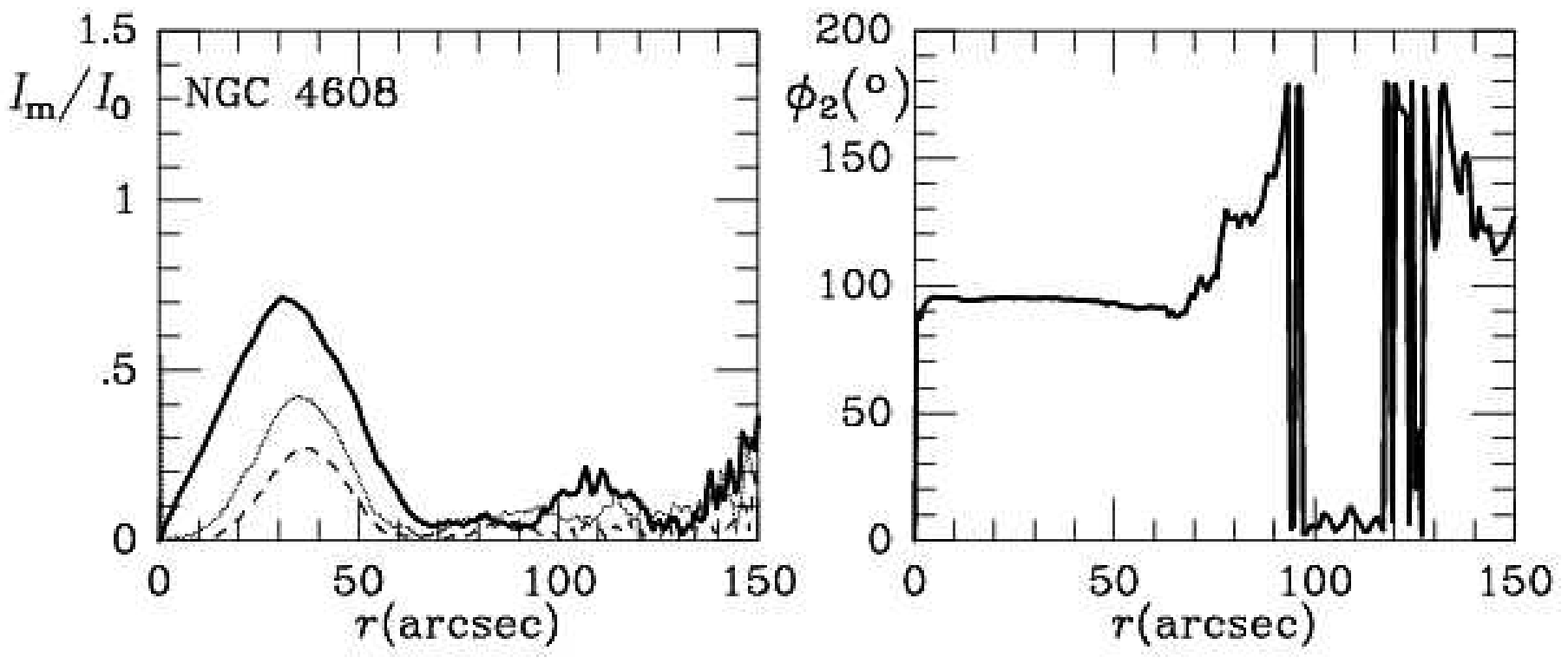}
\vskip -6.5cm
\includegraphics[width=\columnwidth,trim=0 -10 0 50,clip]{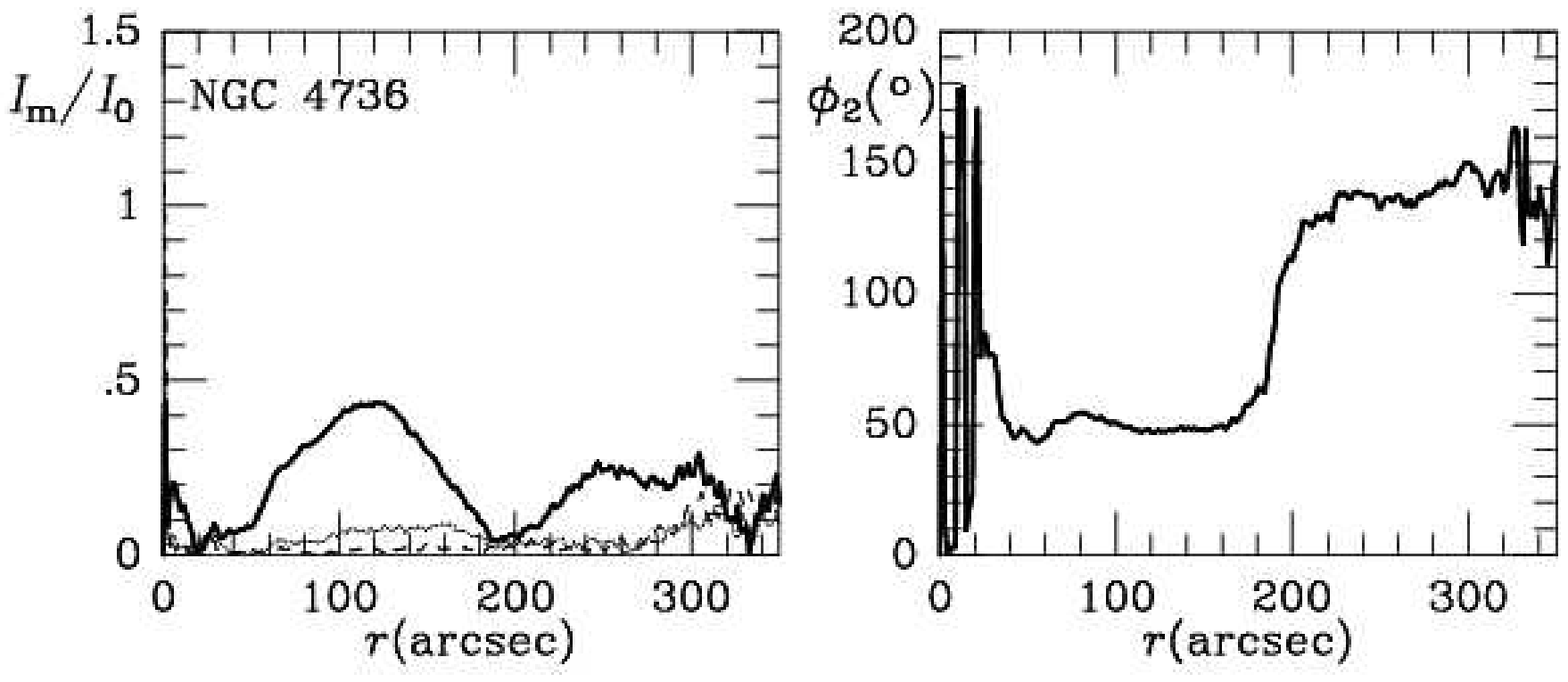}
\vskip -6.5cm
\includegraphics[width=\columnwidth,trim=0 -10 0 50,clip]{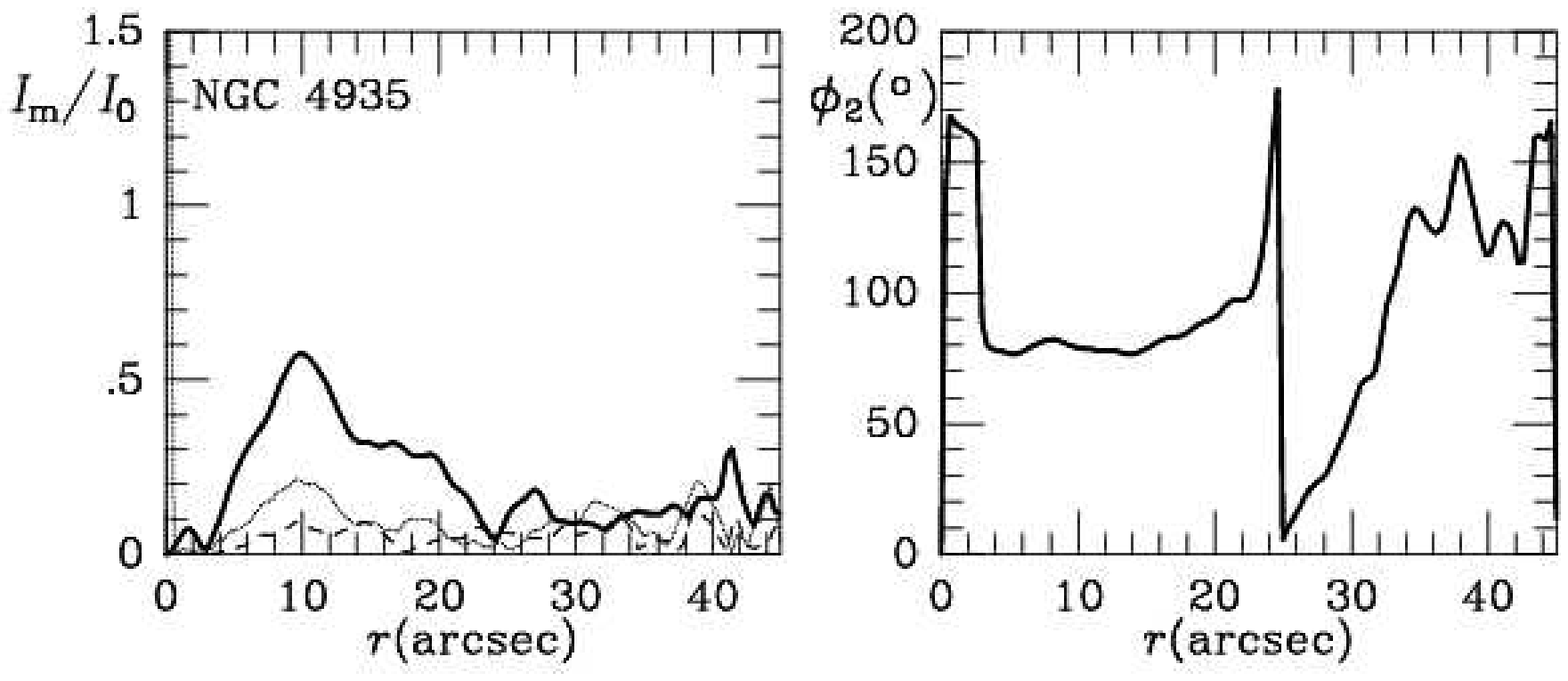}
\vskip -6.5cm
\includegraphics[width=\columnwidth,trim=0 -10 0 50,clip]{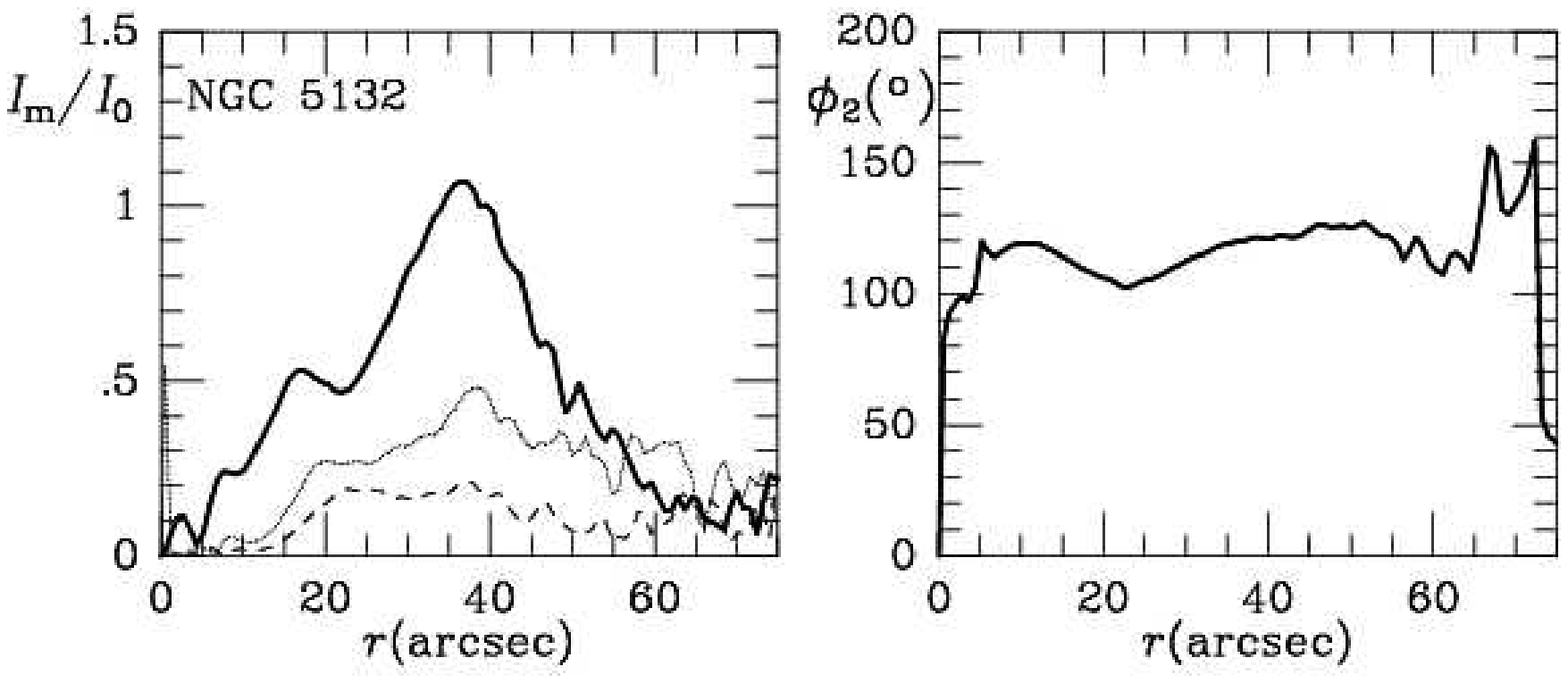}
\vskip -6.5cm
\caption{(cont.)}
\end{figure}
\setcounter{figure}{21}
\begin{figure}
\includegraphics[width=\columnwidth,trim=0 -10 0 50,clip]{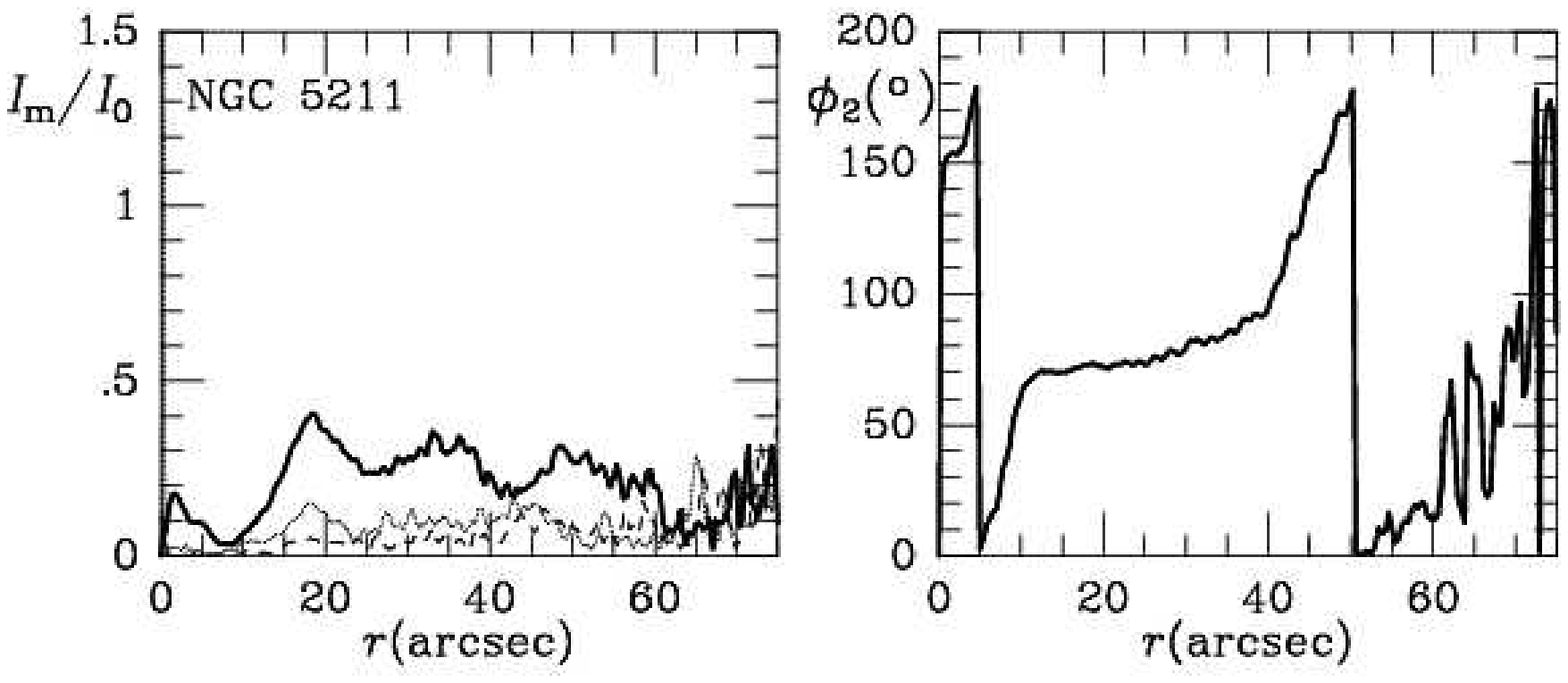}
\vskip -6.5cm
\includegraphics[width=\columnwidth,trim=0 -10 0 50,clip]{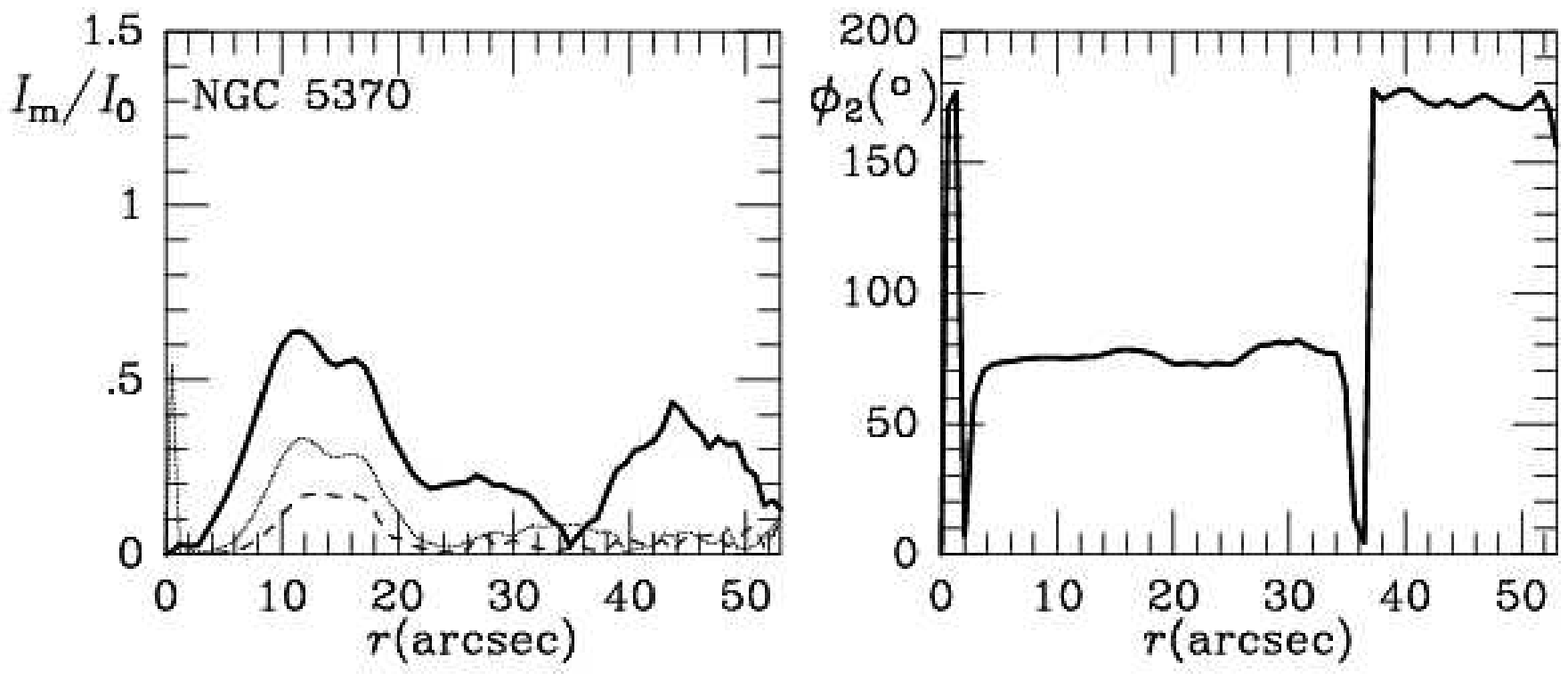}
\vskip -6.5cm
\includegraphics[width=\columnwidth,trim=0 -10 0 50,clip]{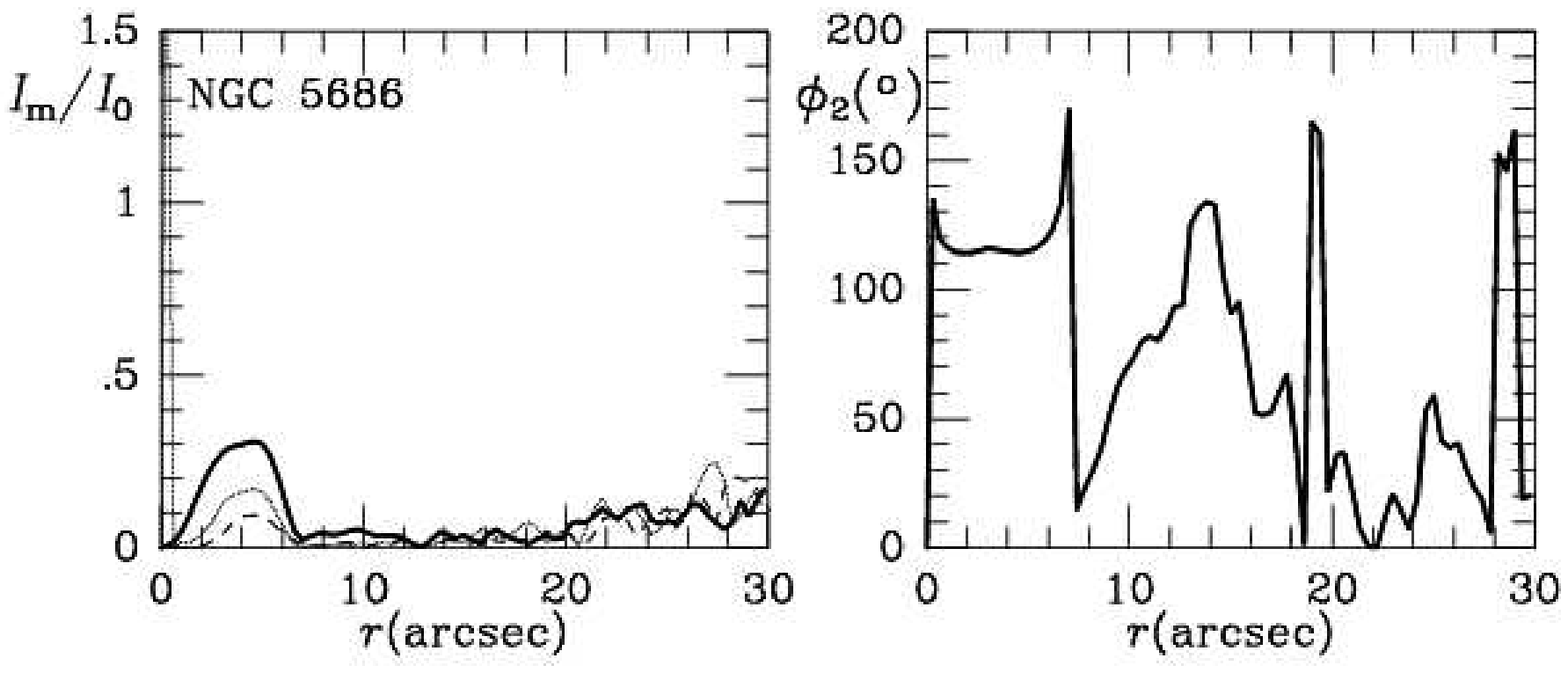}
\vskip -6.5cm
\includegraphics[width=\columnwidth,trim=0 -10 0 50,clip]{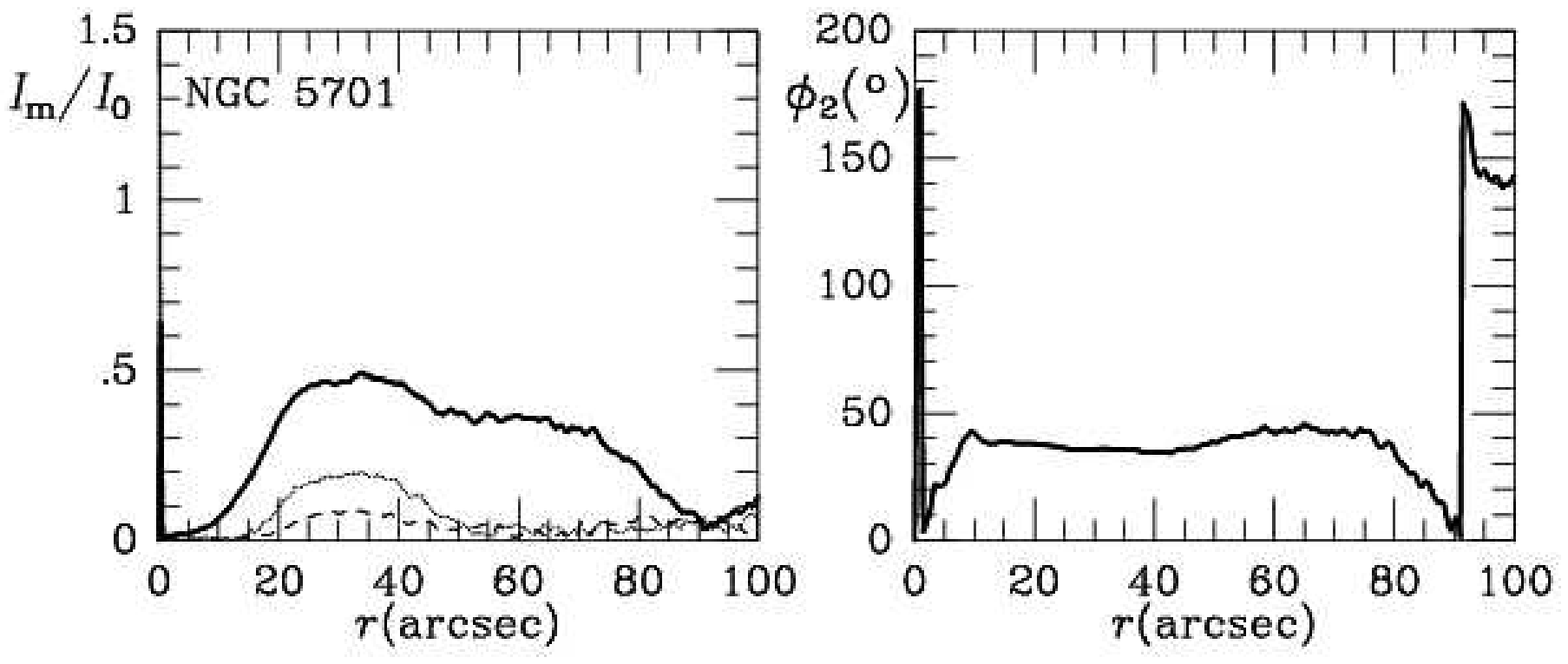}
\vskip -6.5cm
\includegraphics[width=\columnwidth,trim=0 -10 0 50,clip]{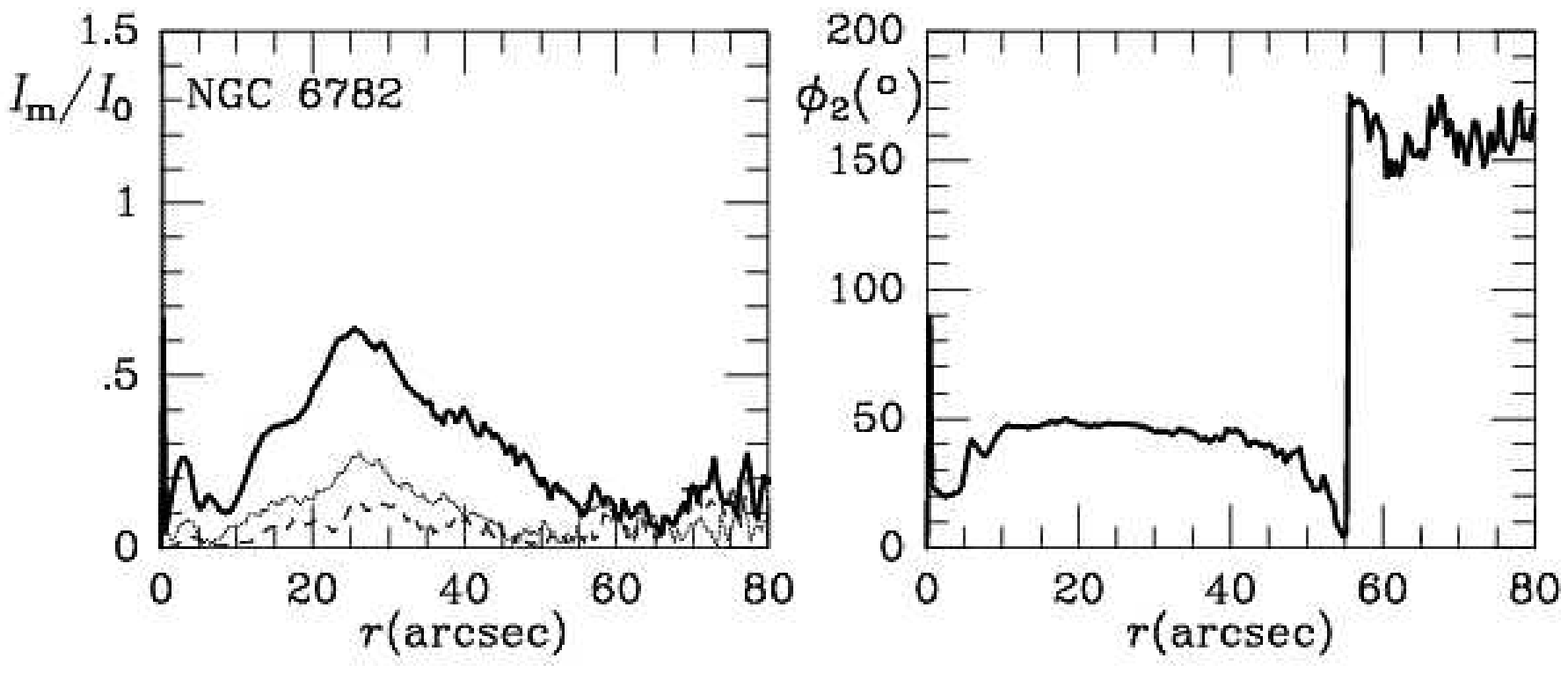}
\vskip -6.5cm
\includegraphics[width=\columnwidth,trim=0 -10 0 50,clip]{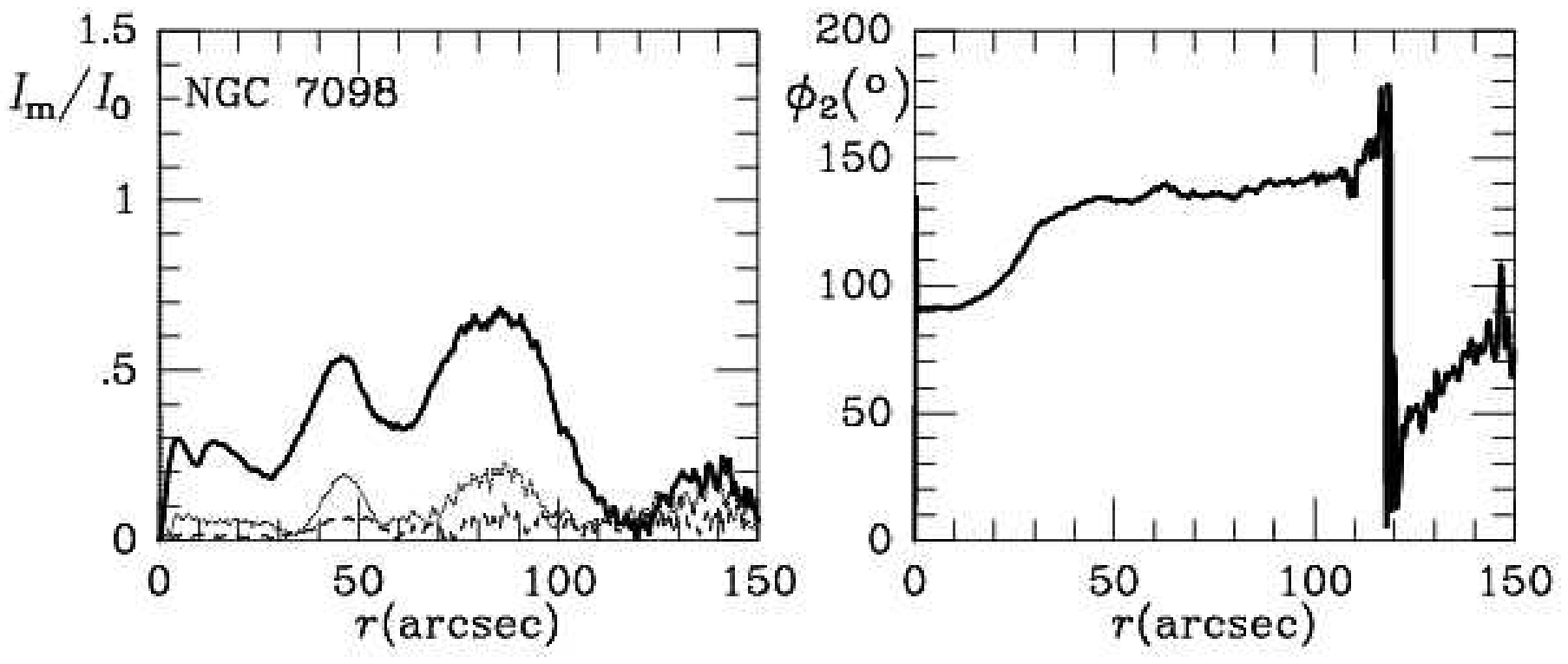}
\vskip -6.5cm
\caption{(cont.)}
\end{figure}
\setcounter{figure}{21}
\begin{figure}
\includegraphics[width=\columnwidth,trim=0 -10 0 50,clip]{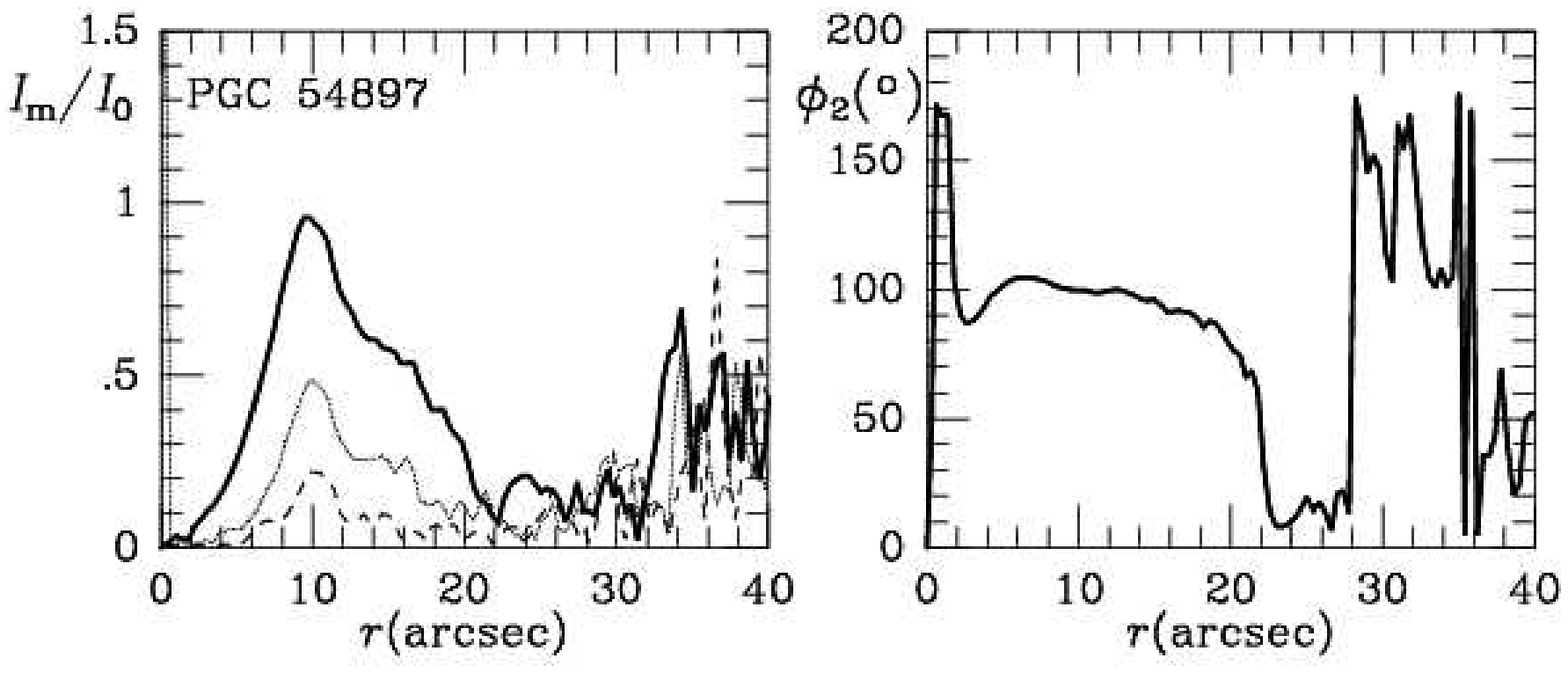}
\vskip -6.5cm
\includegraphics[width=\columnwidth,trim=0 -10 0 50,clip]{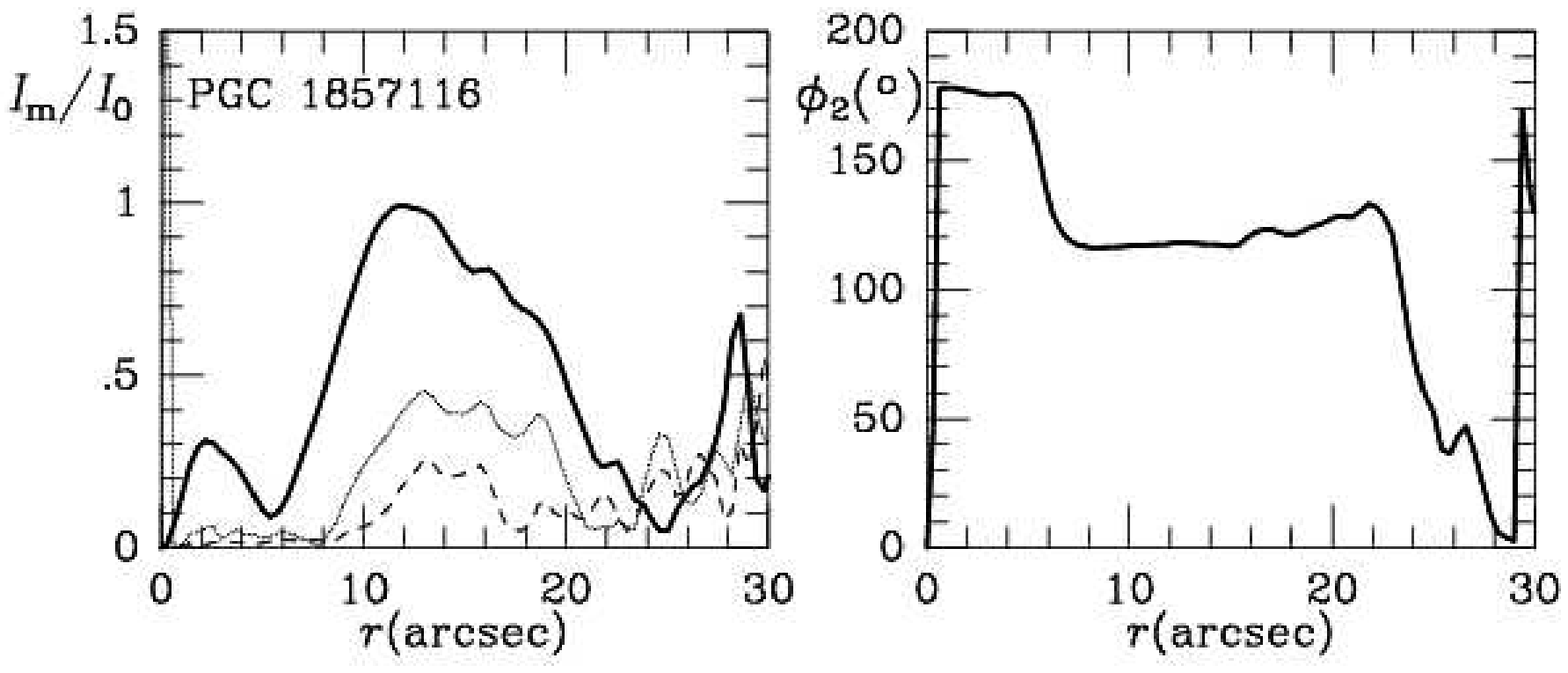}
\vskip -6.5cm
\includegraphics[width=\columnwidth,trim=0 -10 0 50,clip]{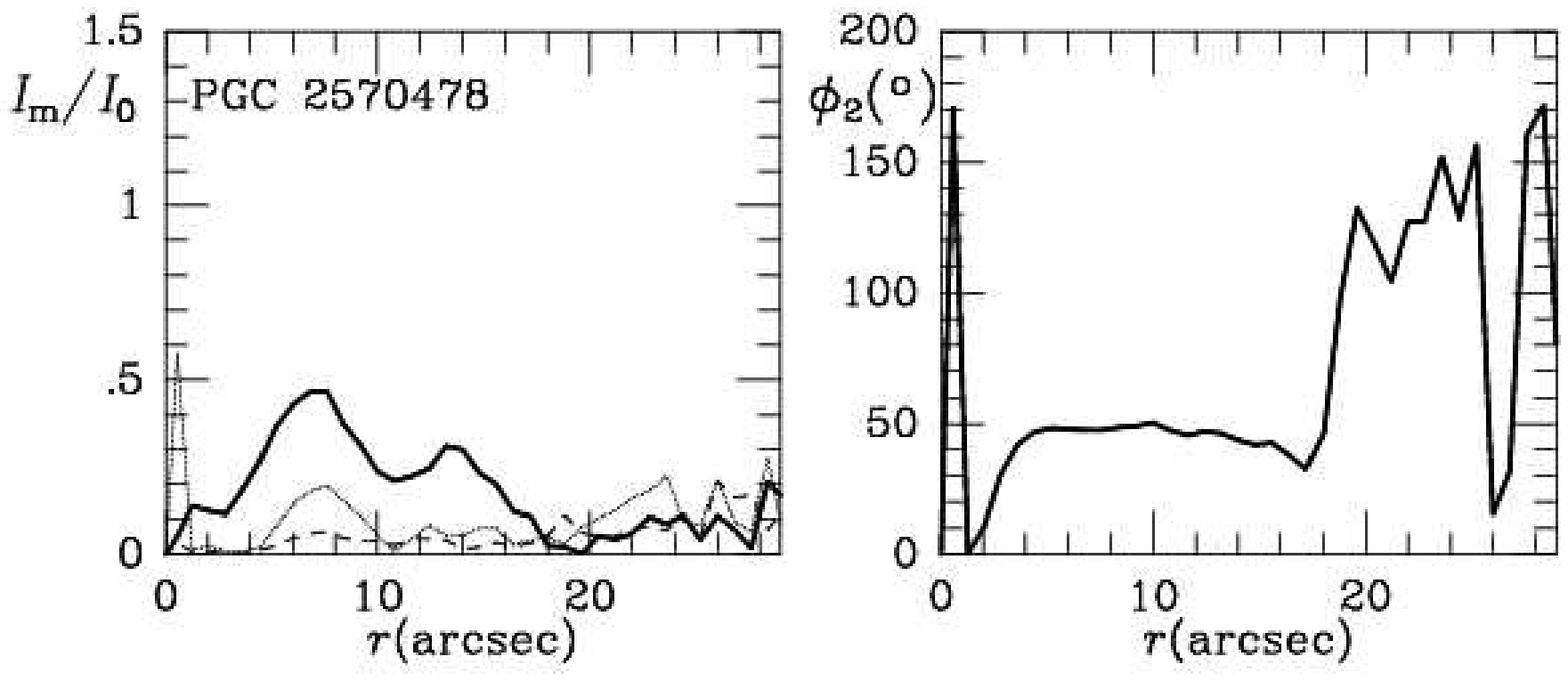}
\vskip -6.5cm
\includegraphics[width=\columnwidth,trim=0 -10 0 50,clip]{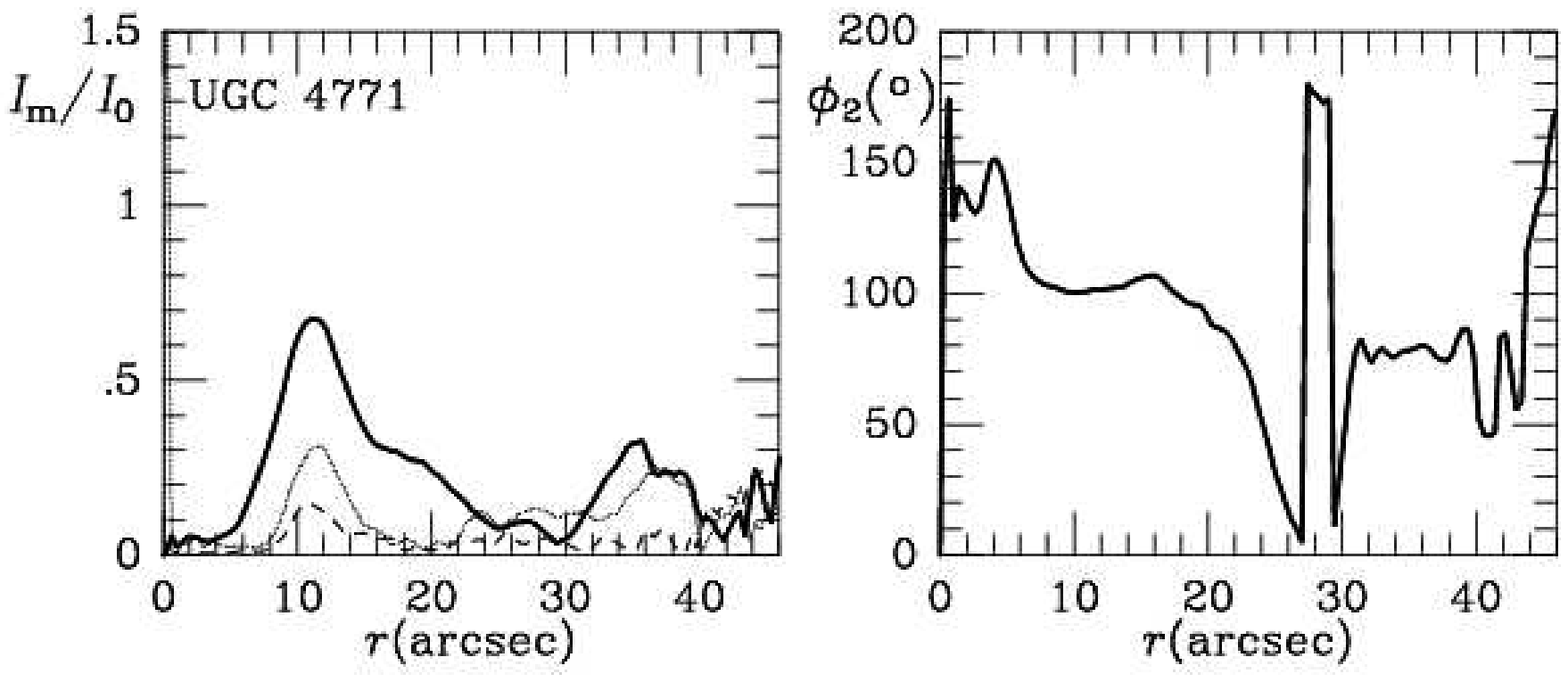}
\vskip -6.5cm
\includegraphics[width=\columnwidth,trim=0 -10 0 50,clip]{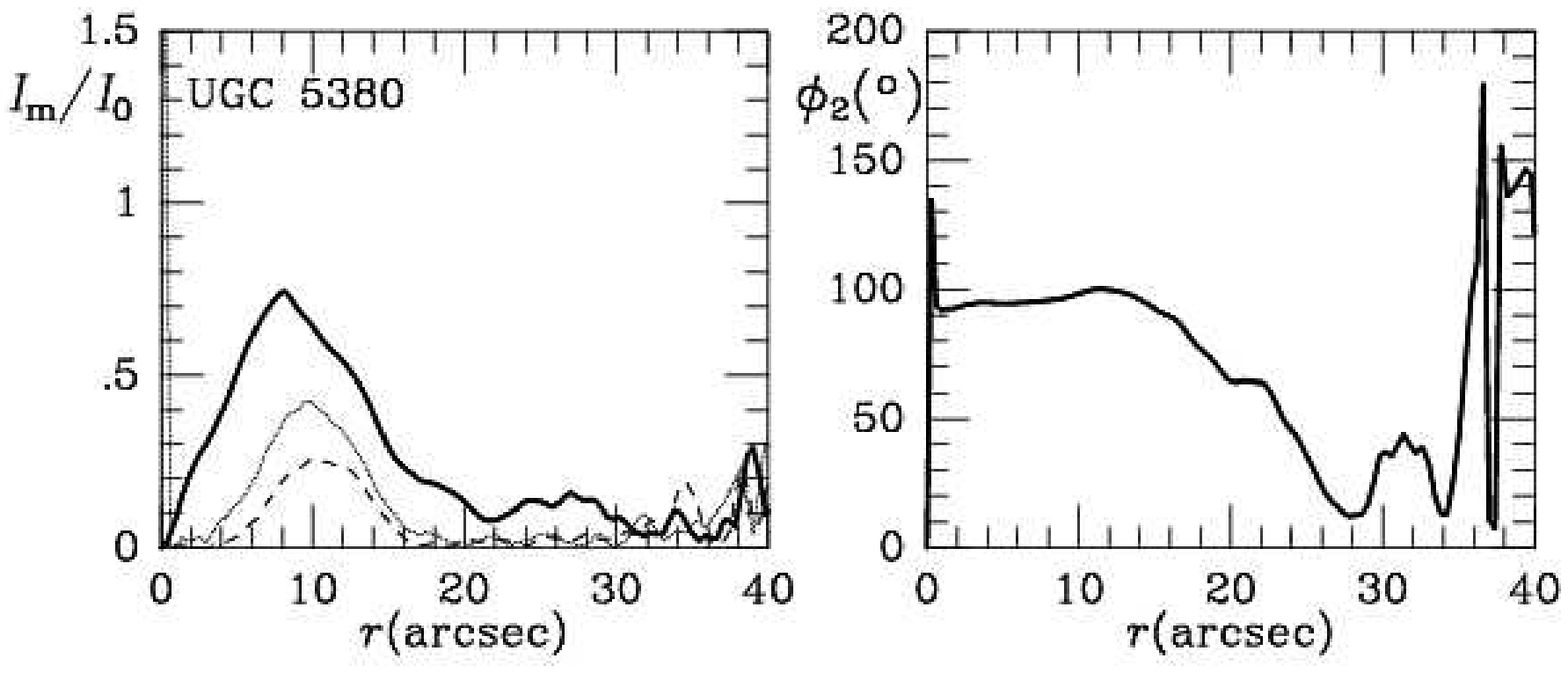}
\vskip -6.5cm
\includegraphics[width=\columnwidth,trim=0 -10 0 50,clip]{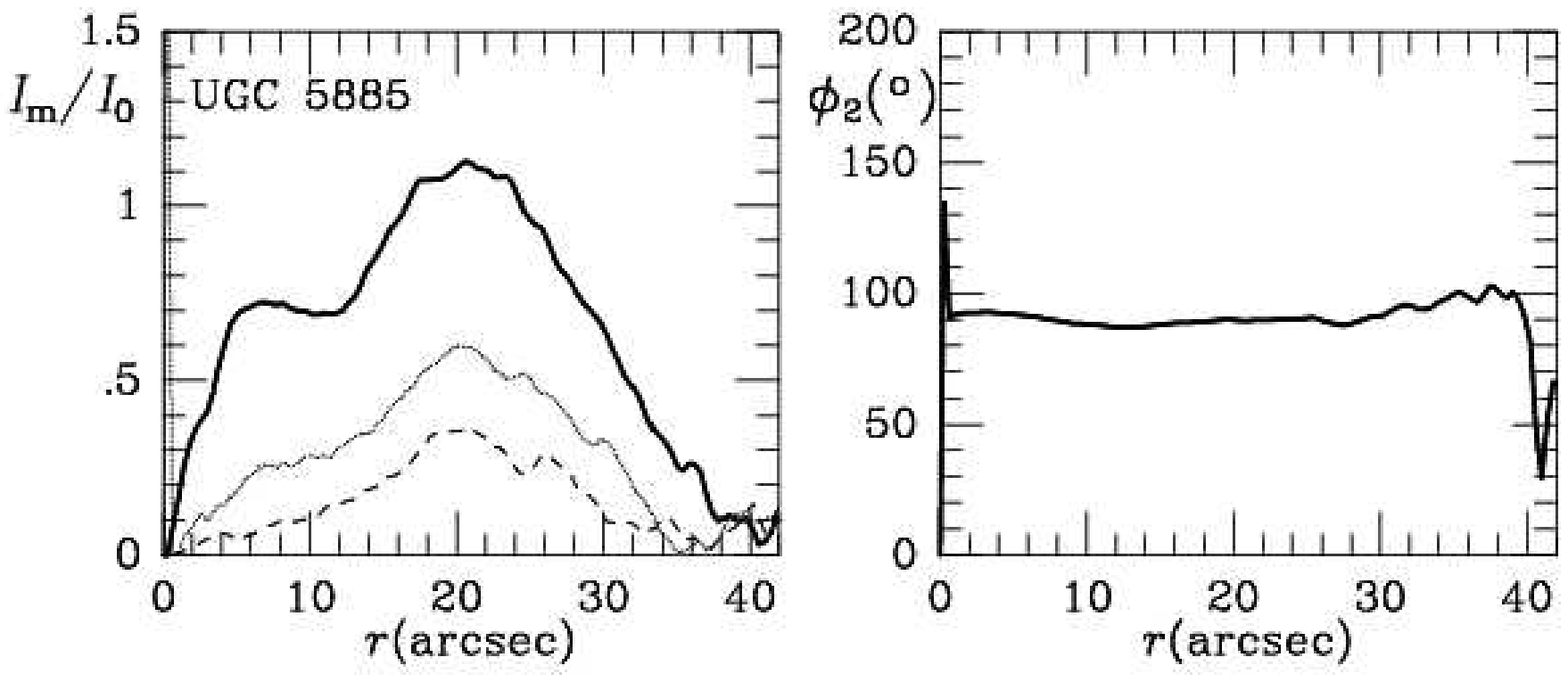}
\vskip -6.5cm
\caption{(cont.)}
\end{figure}
\setcounter{figure}{21}
\begin{figure}
\includegraphics[width=\columnwidth,trim=0 -10 0 50,clip]{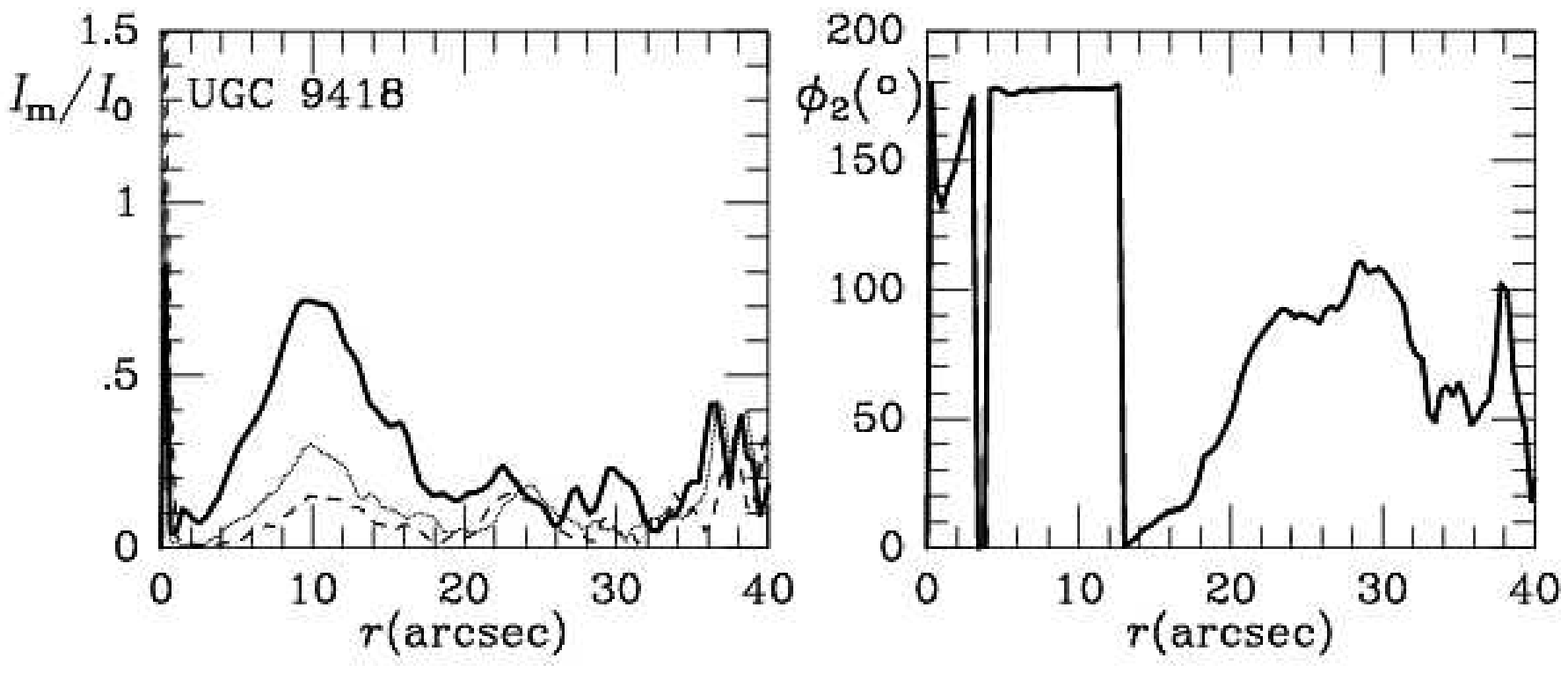}
\vskip -6.5cm
\includegraphics[width=\columnwidth,trim=0 -10 0 50,clip]{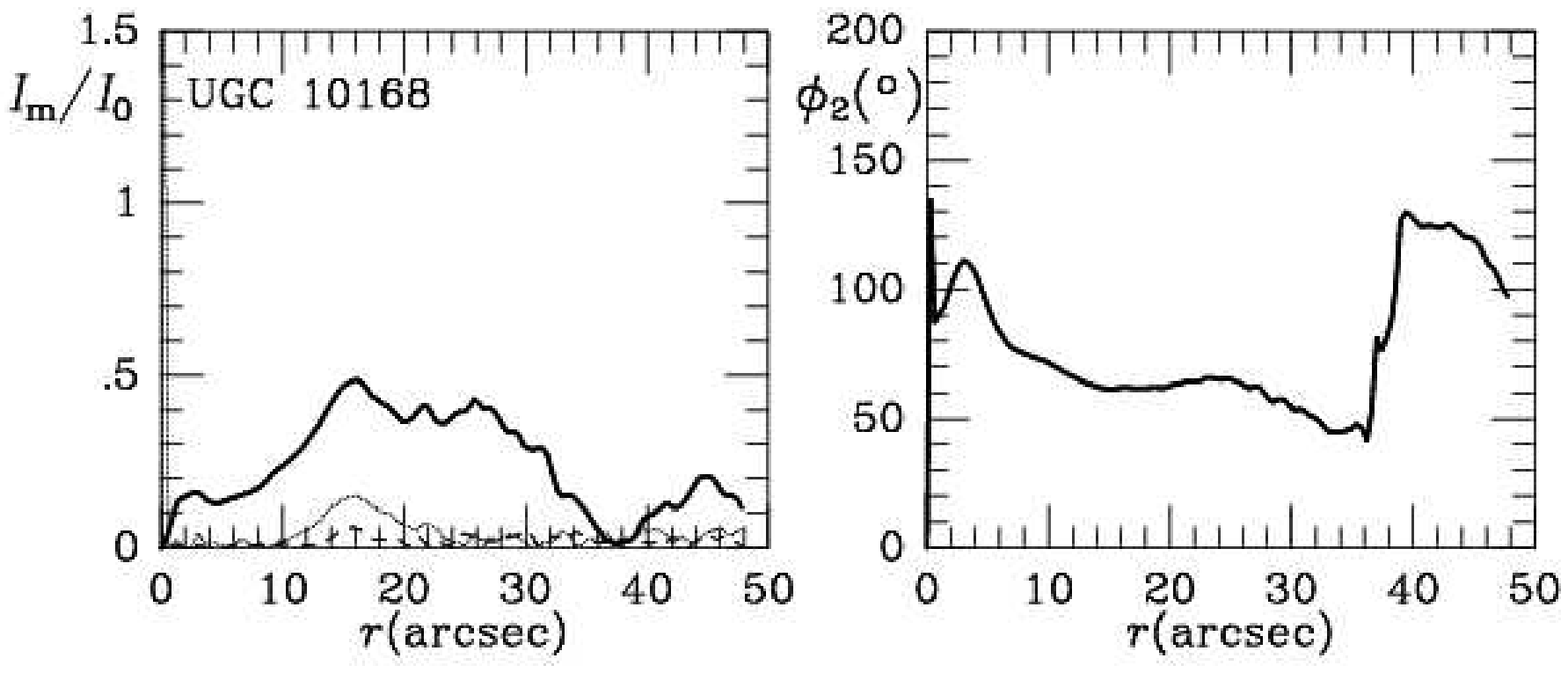}
\vskip -6.5cm
\includegraphics[width=\columnwidth,trim=0 -10 0 50,clip]{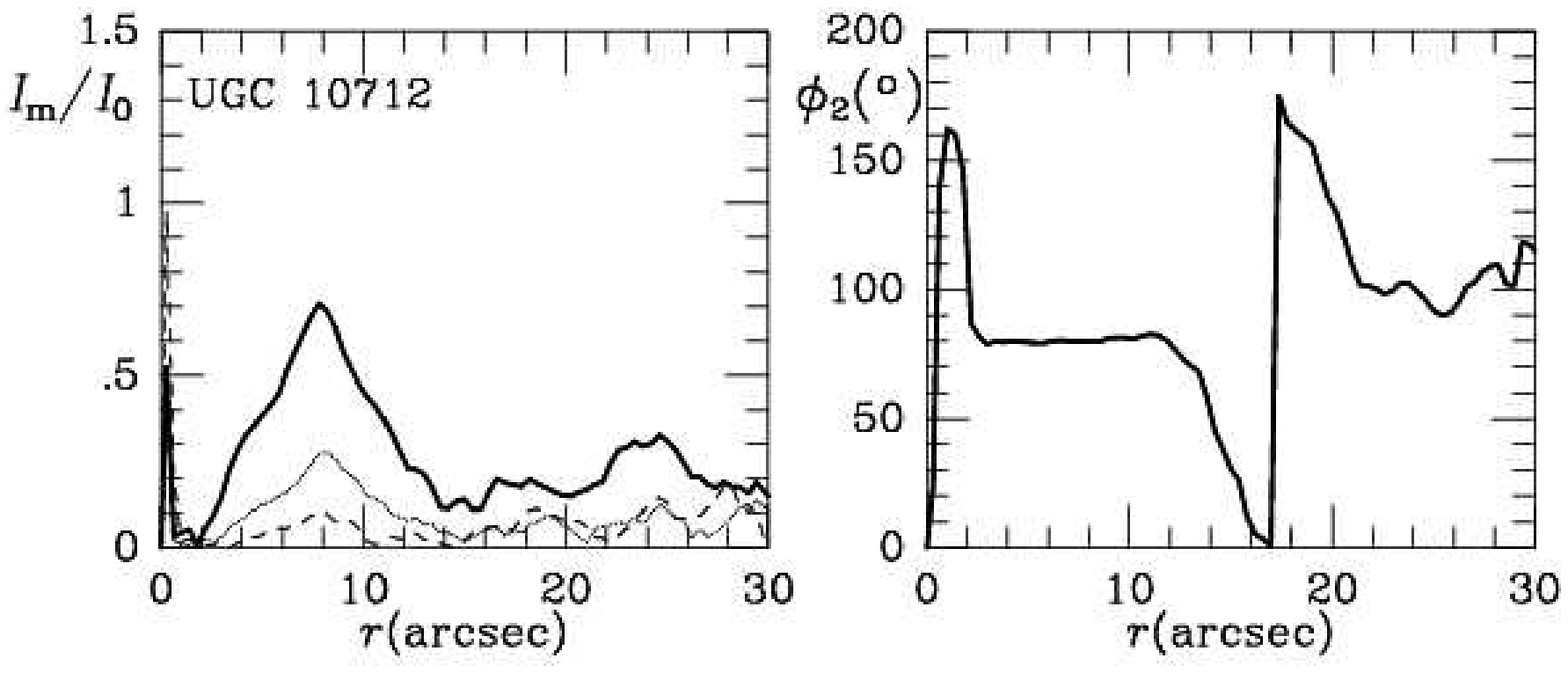}
\vskip -6.5cm
\includegraphics[width=\columnwidth,trim=0 -10 0 50,clip]{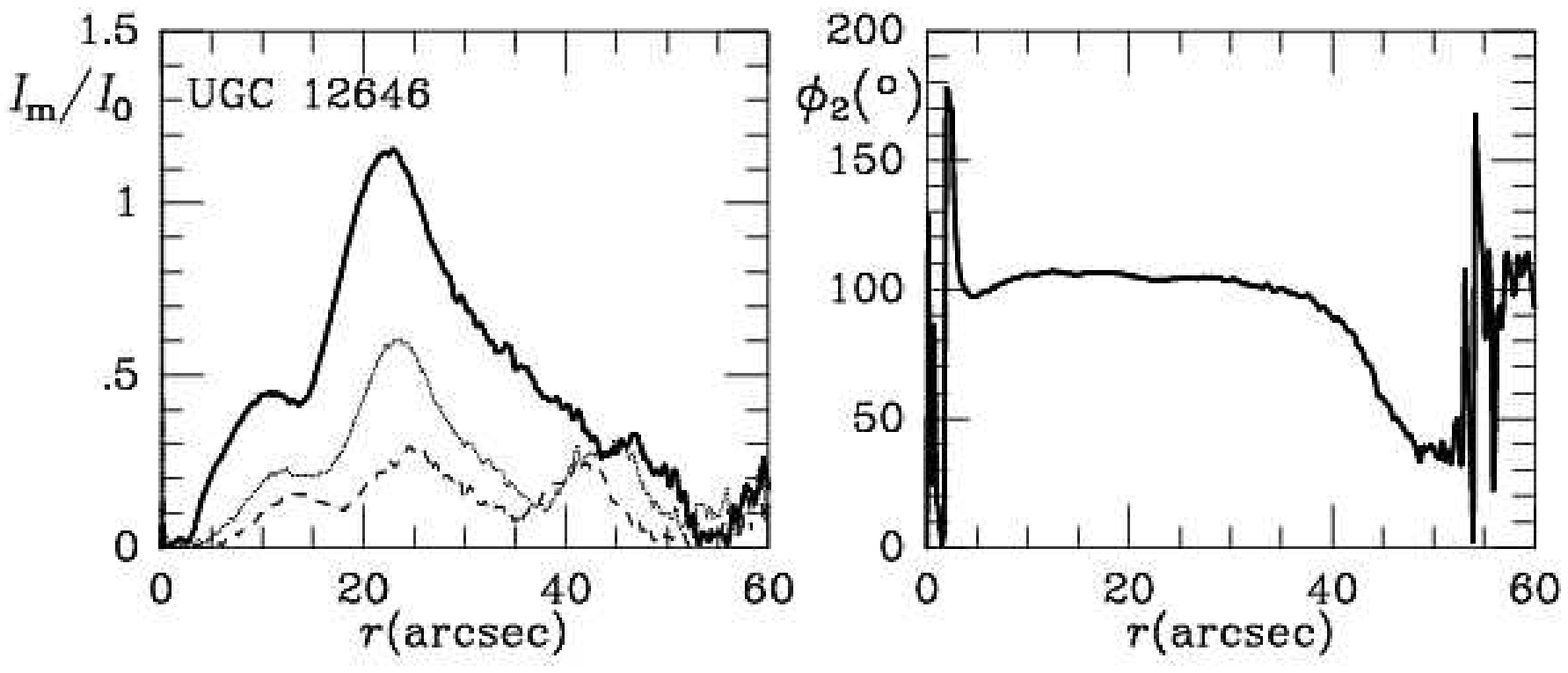}
\vskip -6.5cm
\caption{(Left): Relative Fourier intensity amplitudes for $m$ = 2
(solid curves), 4 (dotted curves), and 6 (dashed curves) for 31 GZ2 and 21
non-GZ2 ringed galaxies. (Right):  Phase of the $m$ = 2 component.}
\end{figure}


\begin{figure}
\includegraphics[width=\columnwidth,trim=0 25 0 400,clip]{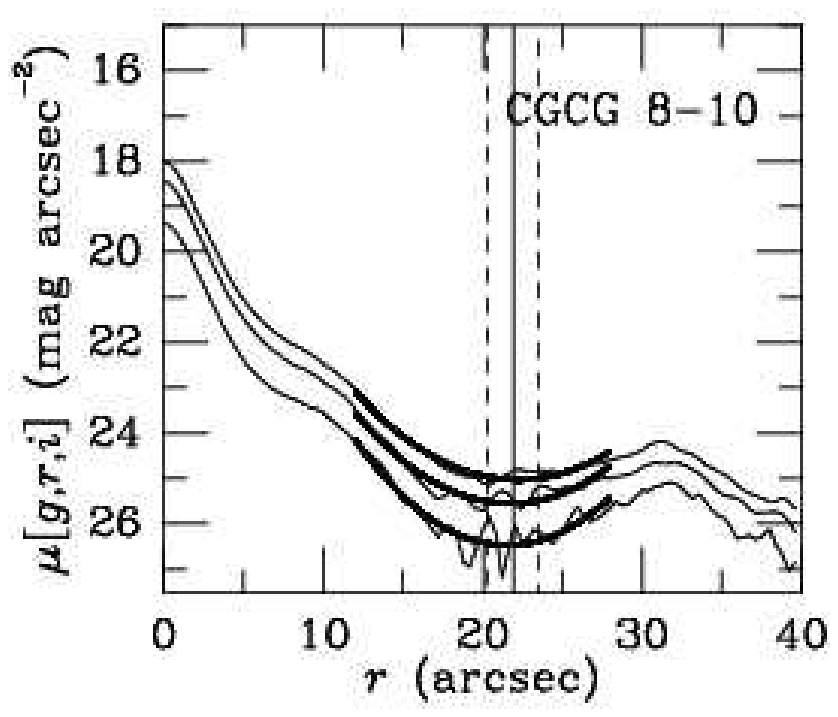}
\includegraphics[width=\columnwidth,trim=0 25 0 400,clip]{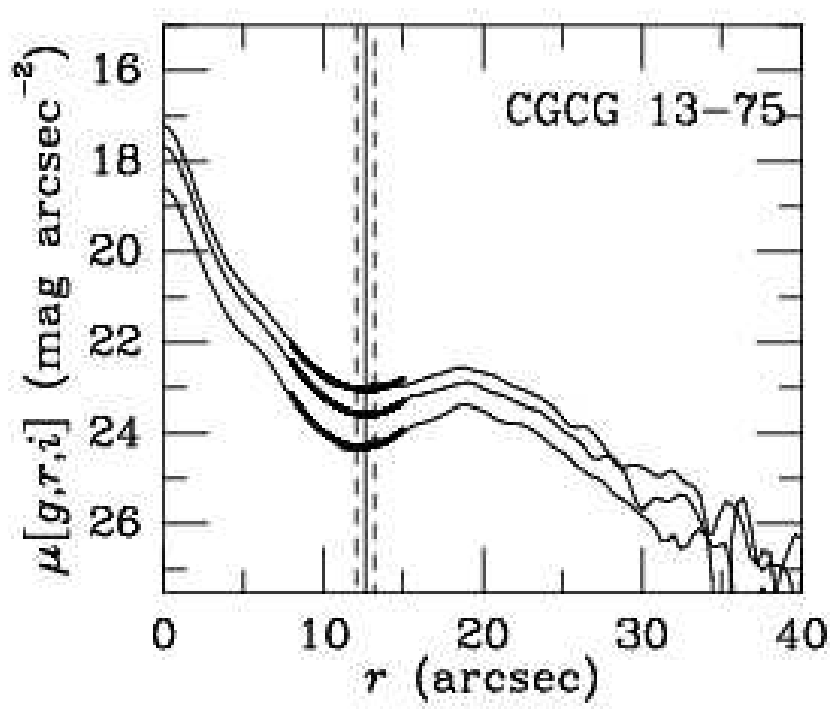}
\includegraphics[width=\columnwidth,trim=0 25 0 400,clip]{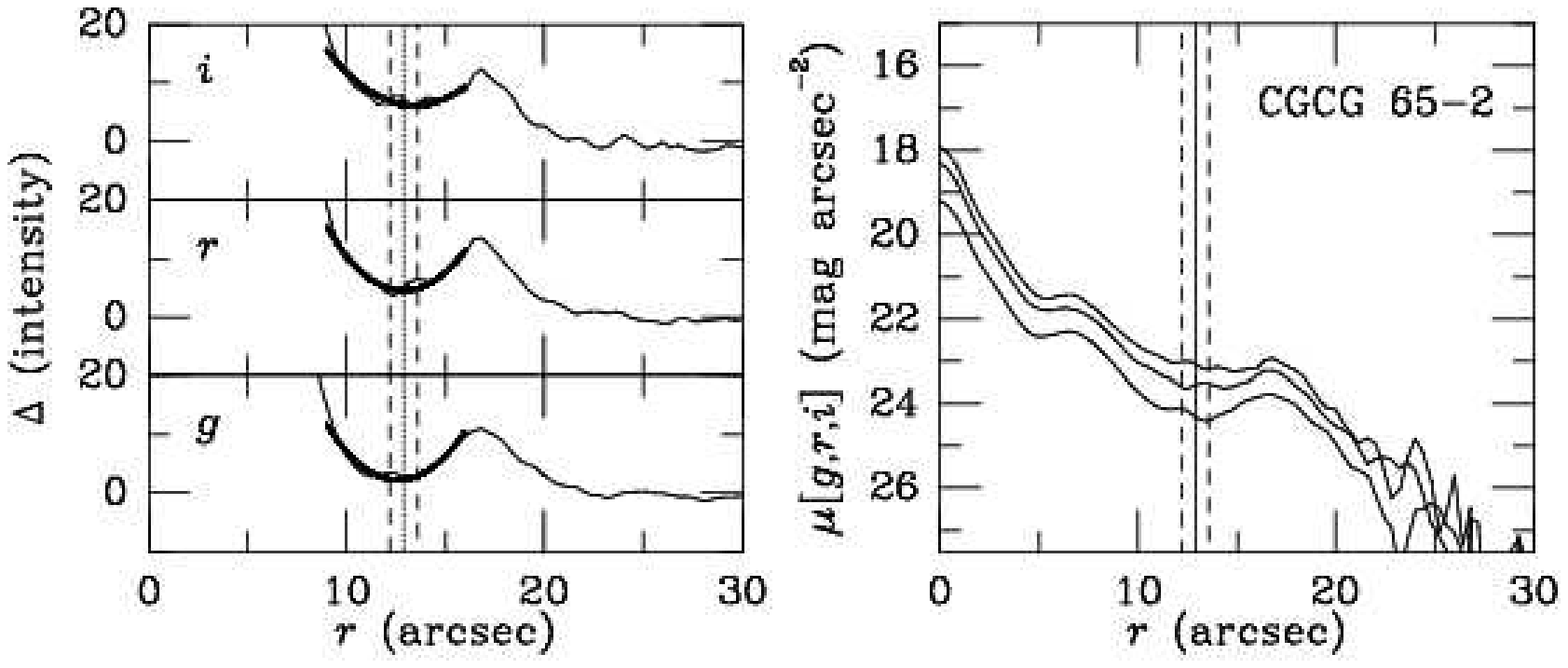}
\includegraphics[width=\columnwidth,trim=0 25 0 400,clip]{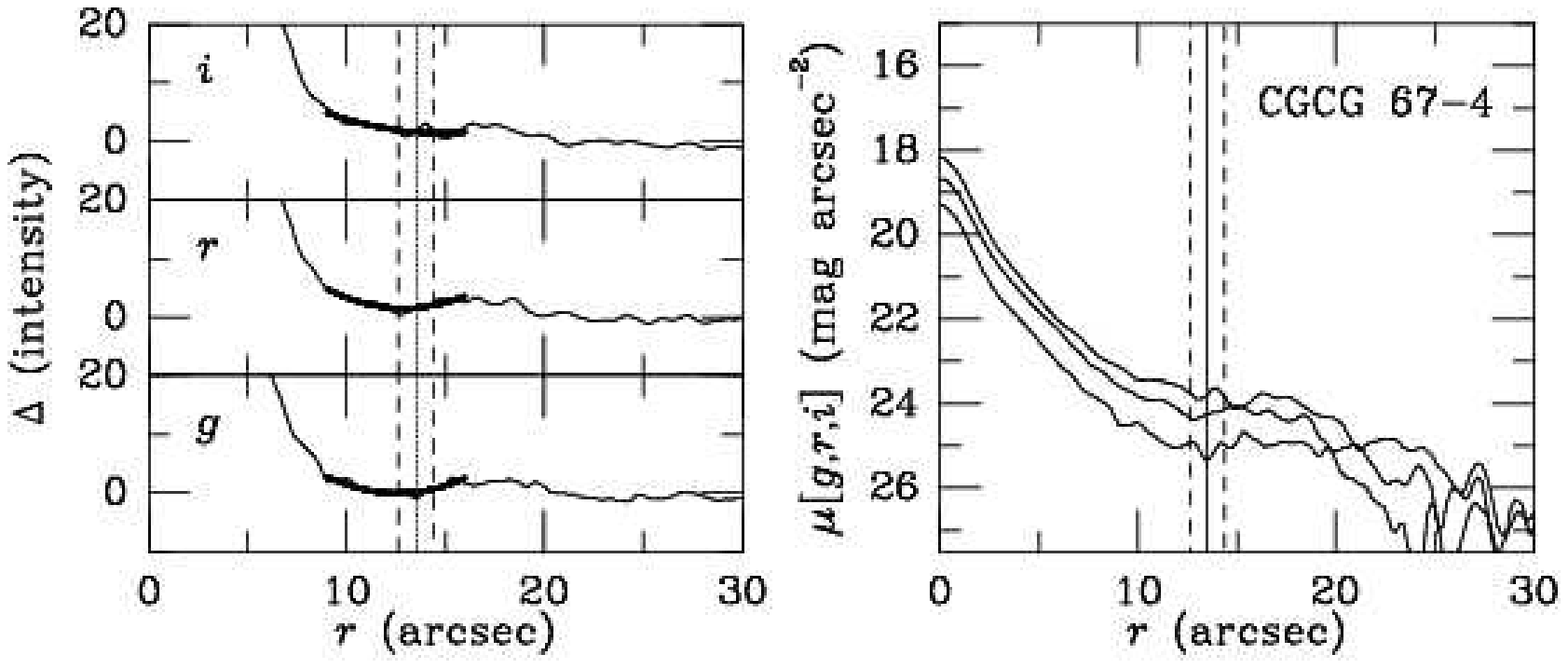}
\includegraphics[width=\columnwidth,trim=0 25 0 400,clip]{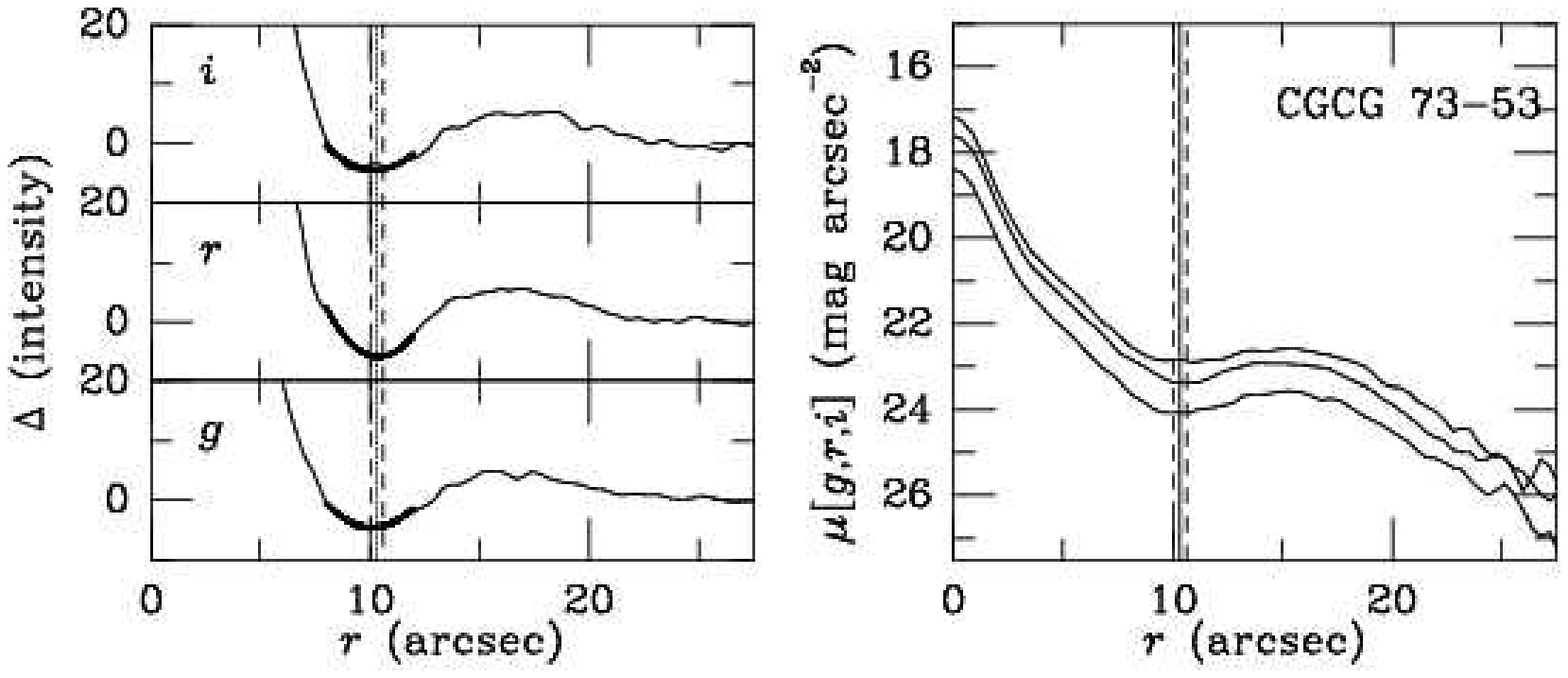}
\caption{
}
\label{fig:allrgp}
\end{figure}
\setcounter{figure}{22}
\begin{figure}
\includegraphics[width=\columnwidth,trim=0 25 0 400,clip]{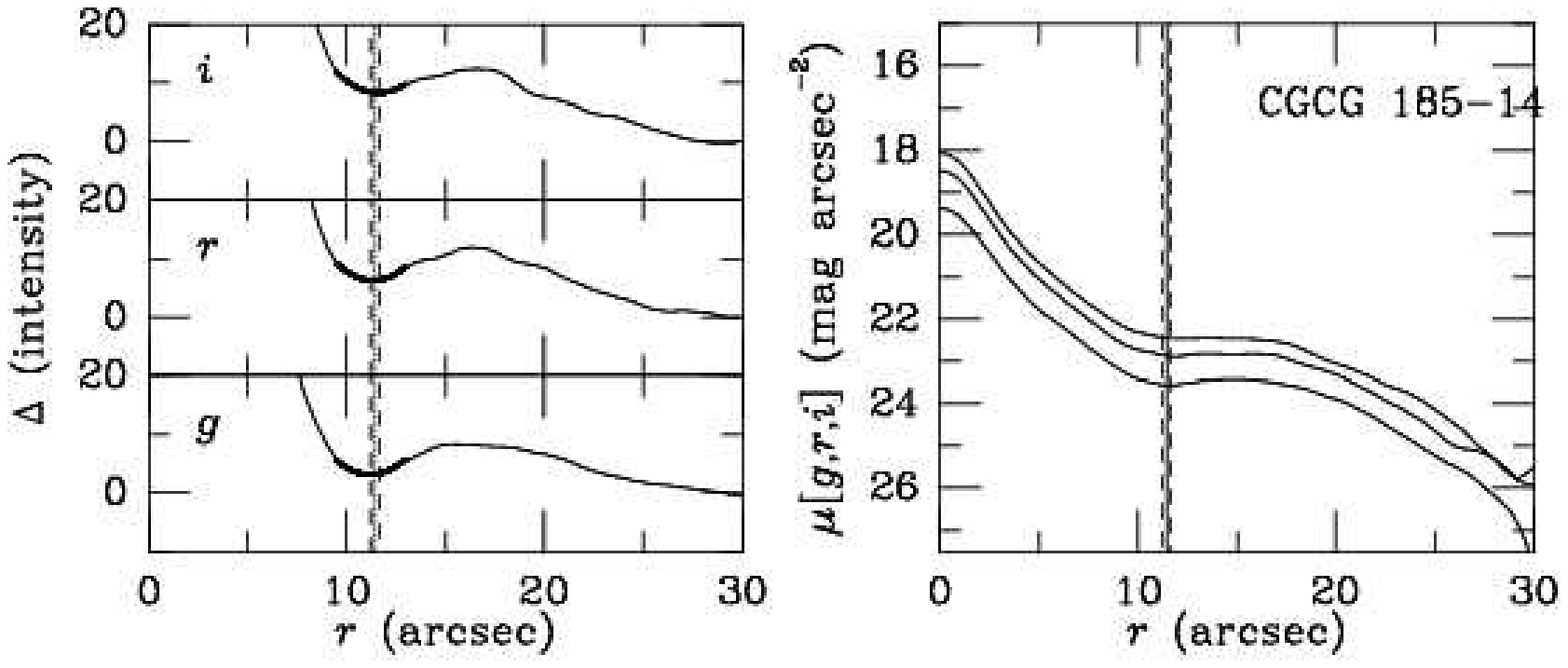}
\includegraphics[width=\columnwidth,trim=0 25 0 400,clip]{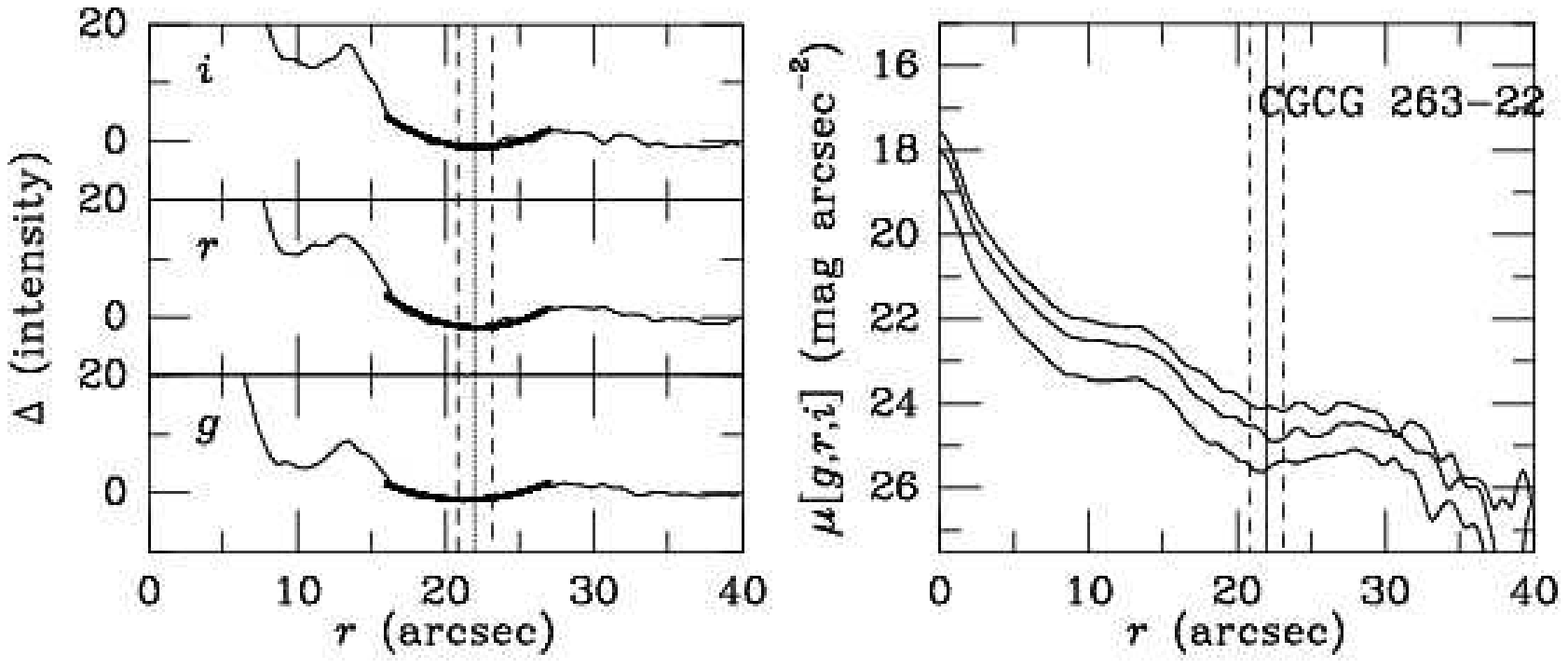}
\includegraphics[width=\columnwidth,trim=0 25 0 400,clip]{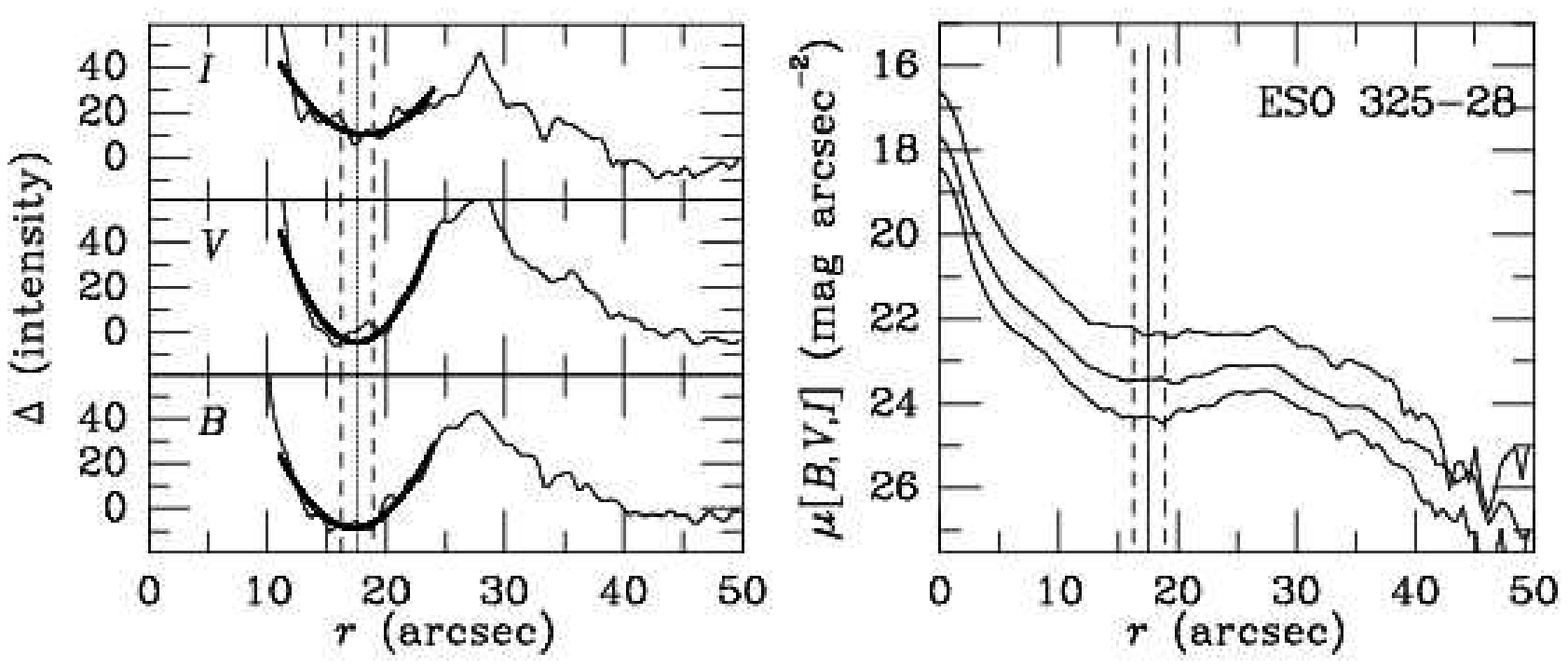}
\includegraphics[width=\columnwidth,trim=0 25 0 400,clip]{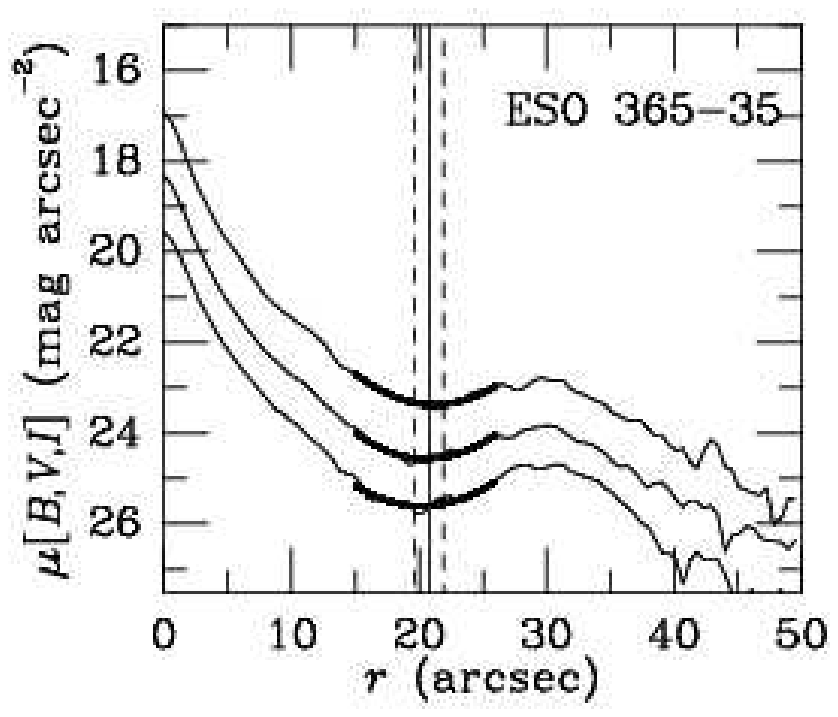}
\includegraphics[width=\columnwidth,trim=0 25 0 400,clip]{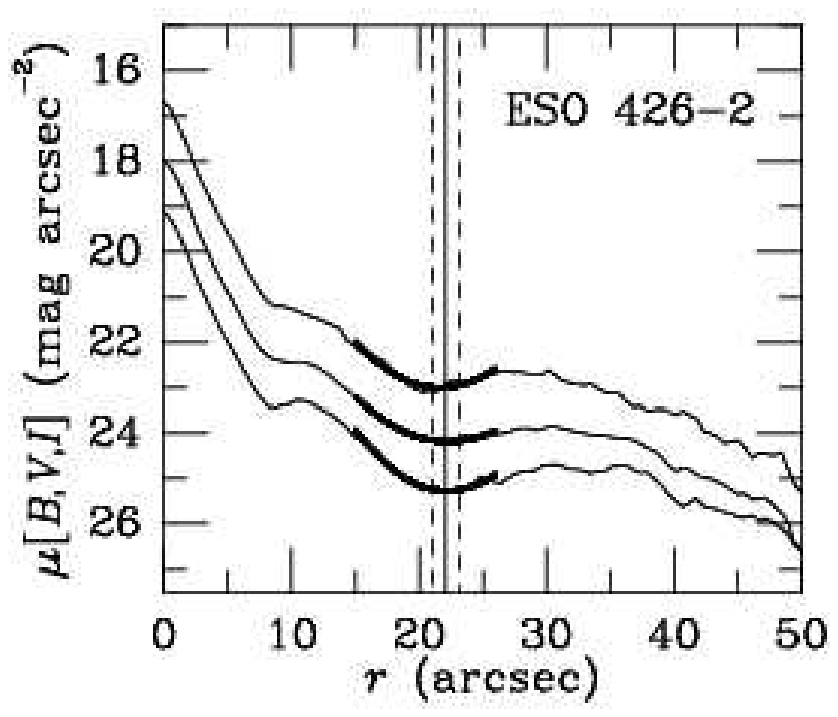}
\caption{(cont.)}
\end{figure}
\setcounter{figure}{22}
\begin{figure}
\includegraphics[width=\columnwidth,trim=0 25 0 400,clip]{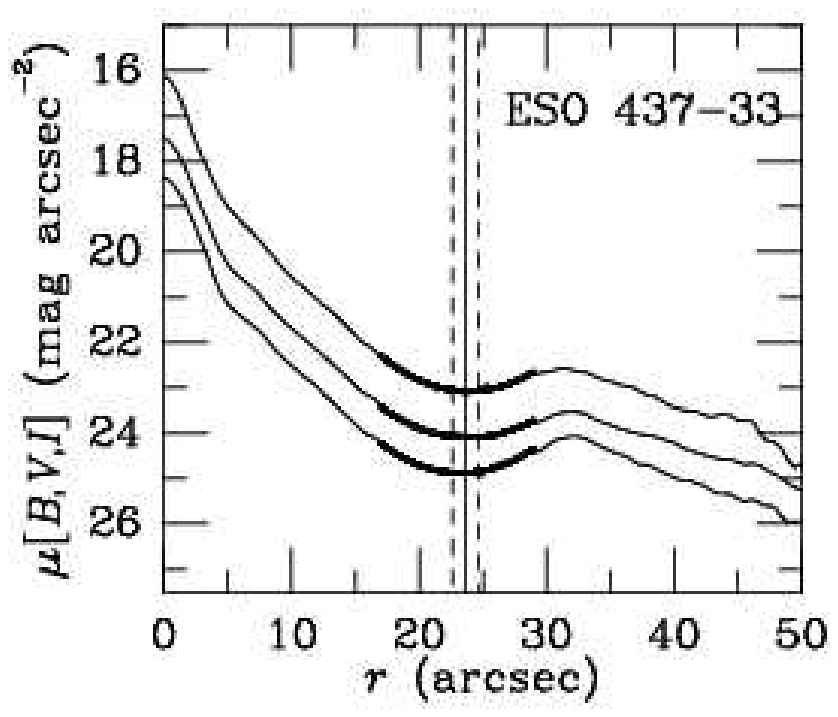}
\includegraphics[width=\columnwidth,trim=0 25 0 400,clip]{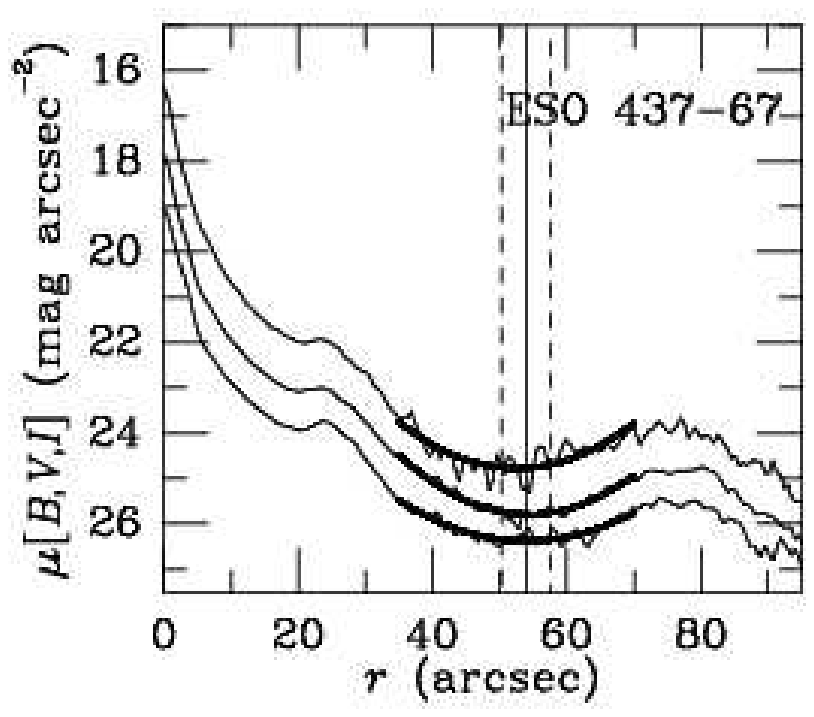}
\includegraphics[width=\columnwidth,trim=0 25 0 400,clip]{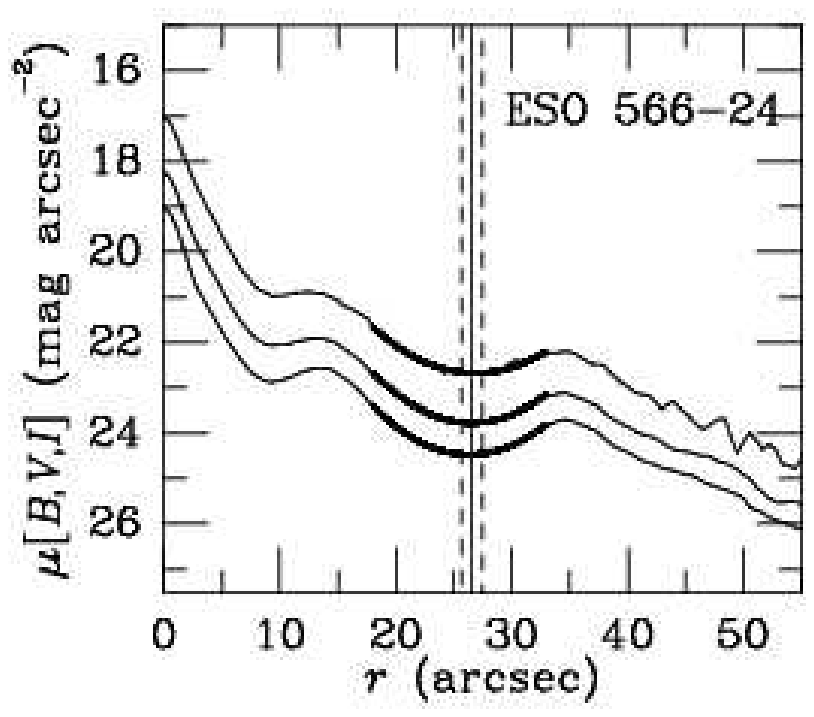}
\includegraphics[width=\columnwidth,trim=0 25 0 400,clip]{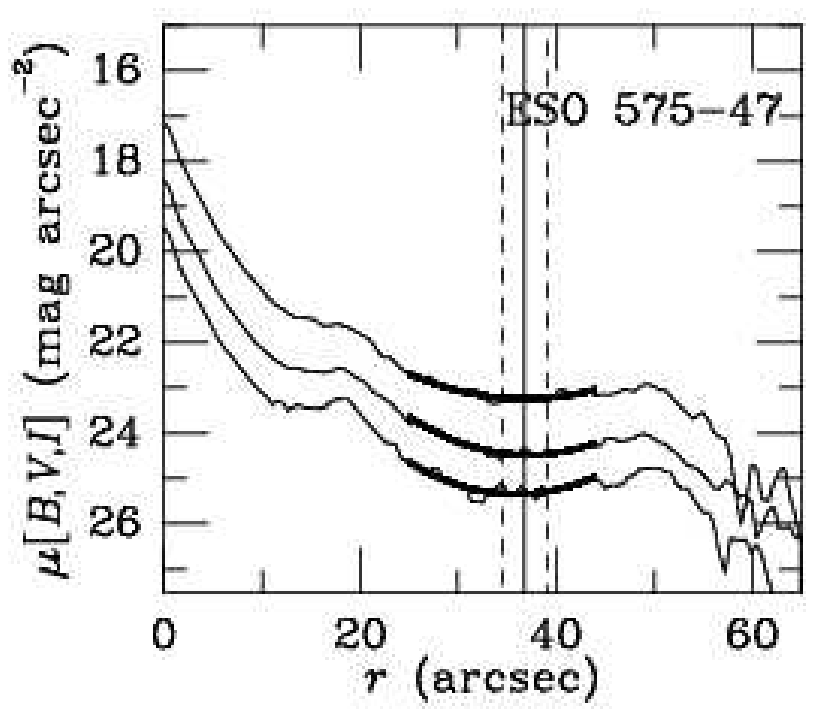}
\includegraphics[width=\columnwidth,trim=0 25 0 400,clip]{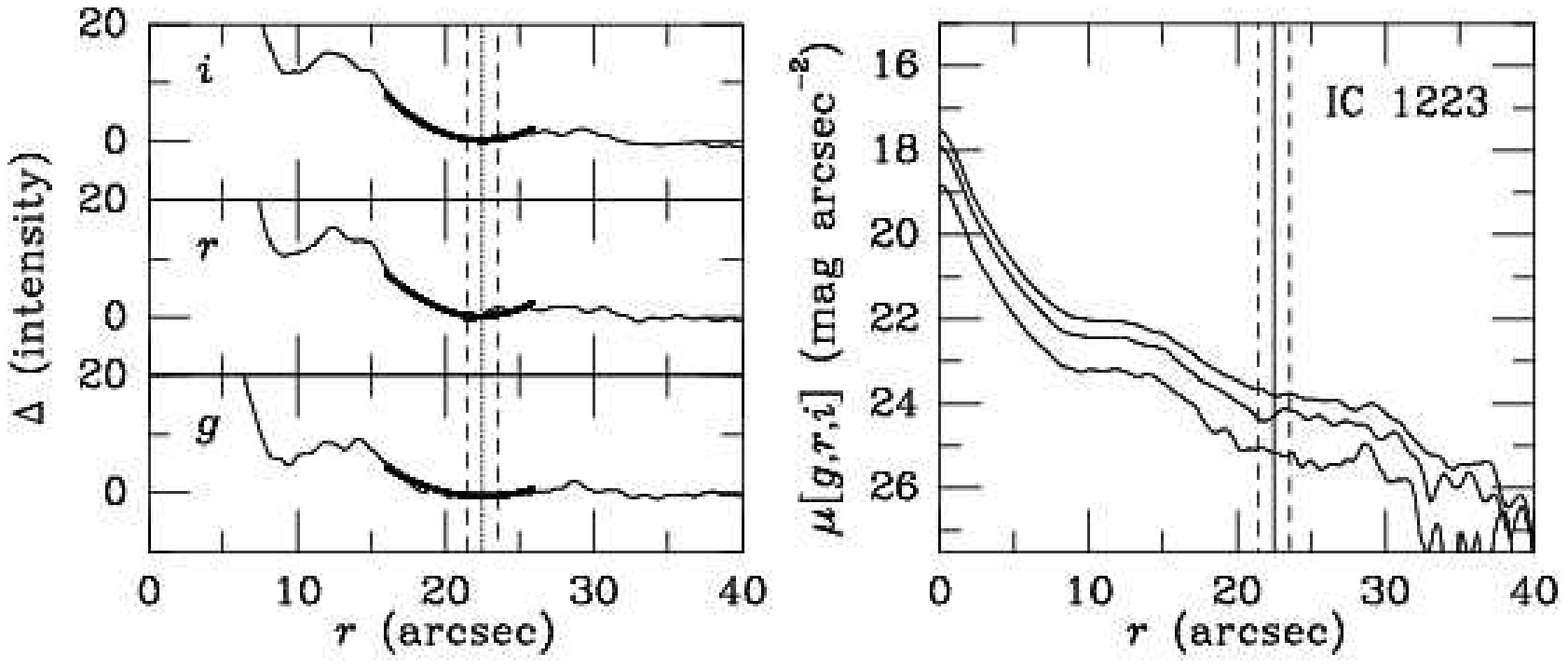}
\caption{(cont.)}
\end{figure}
\setcounter{figure}{22}
\begin{figure}
\includegraphics[width=\columnwidth,trim=0 25 0 400,clip]{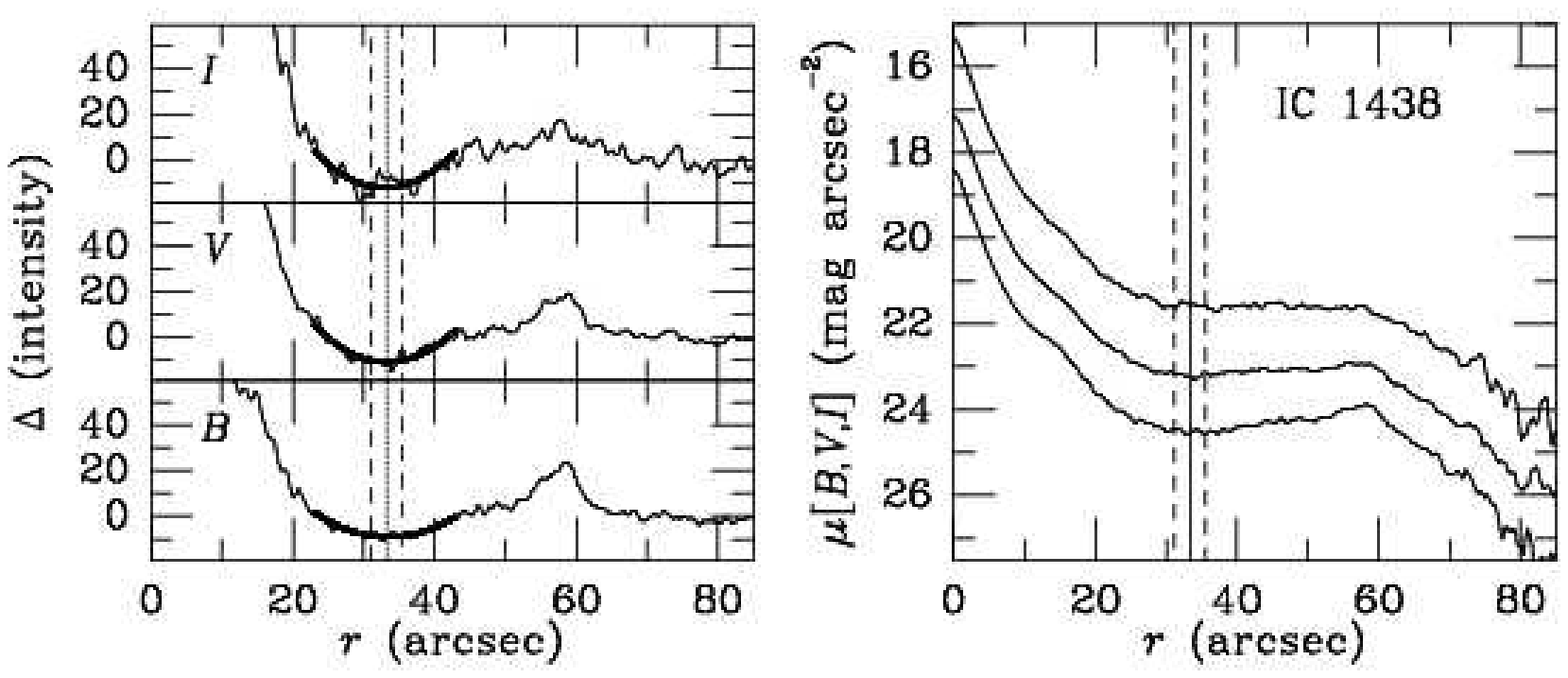}
\includegraphics[width=\columnwidth,trim=0 25 0 400,clip]{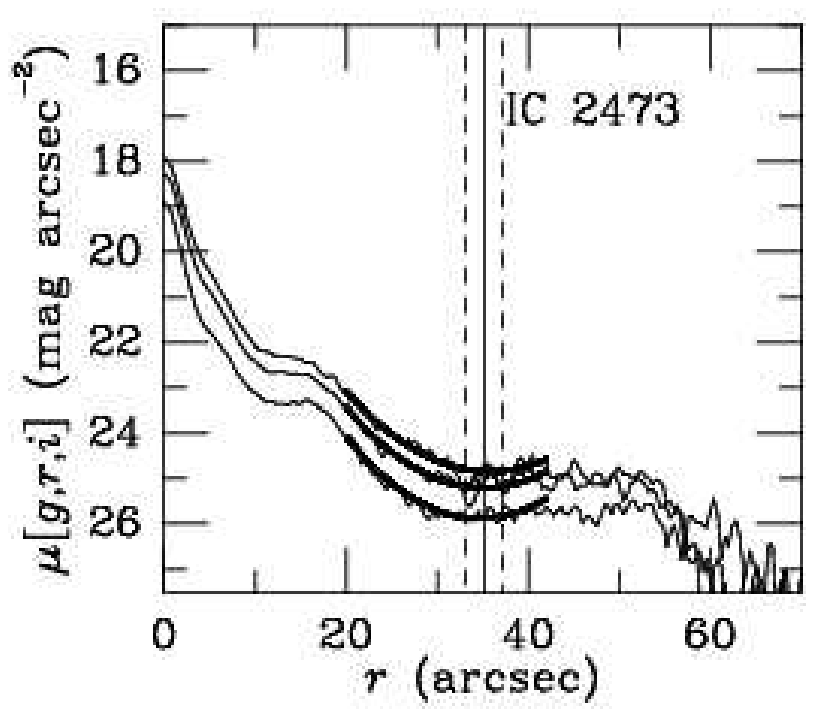}
\includegraphics[width=\columnwidth,trim=0 25 0 400,clip]{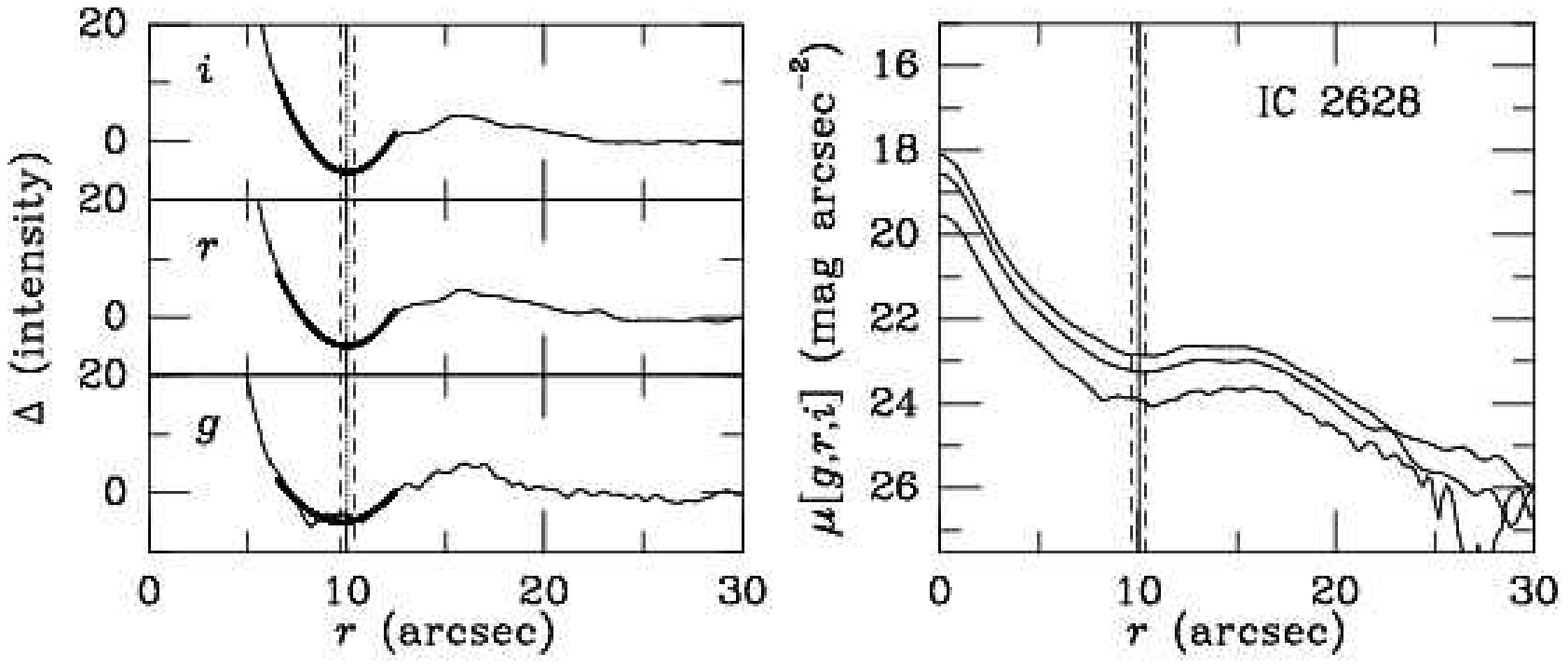}
\includegraphics[width=\columnwidth,trim=0 25 0 400,clip]{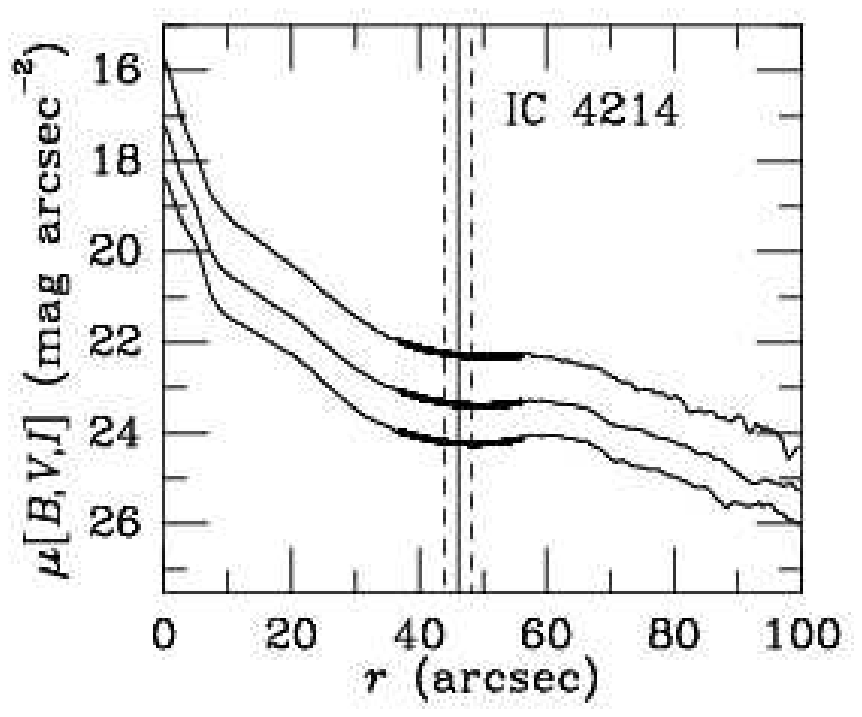}
\includegraphics[width=\columnwidth,trim=0 25 0 400,clip]{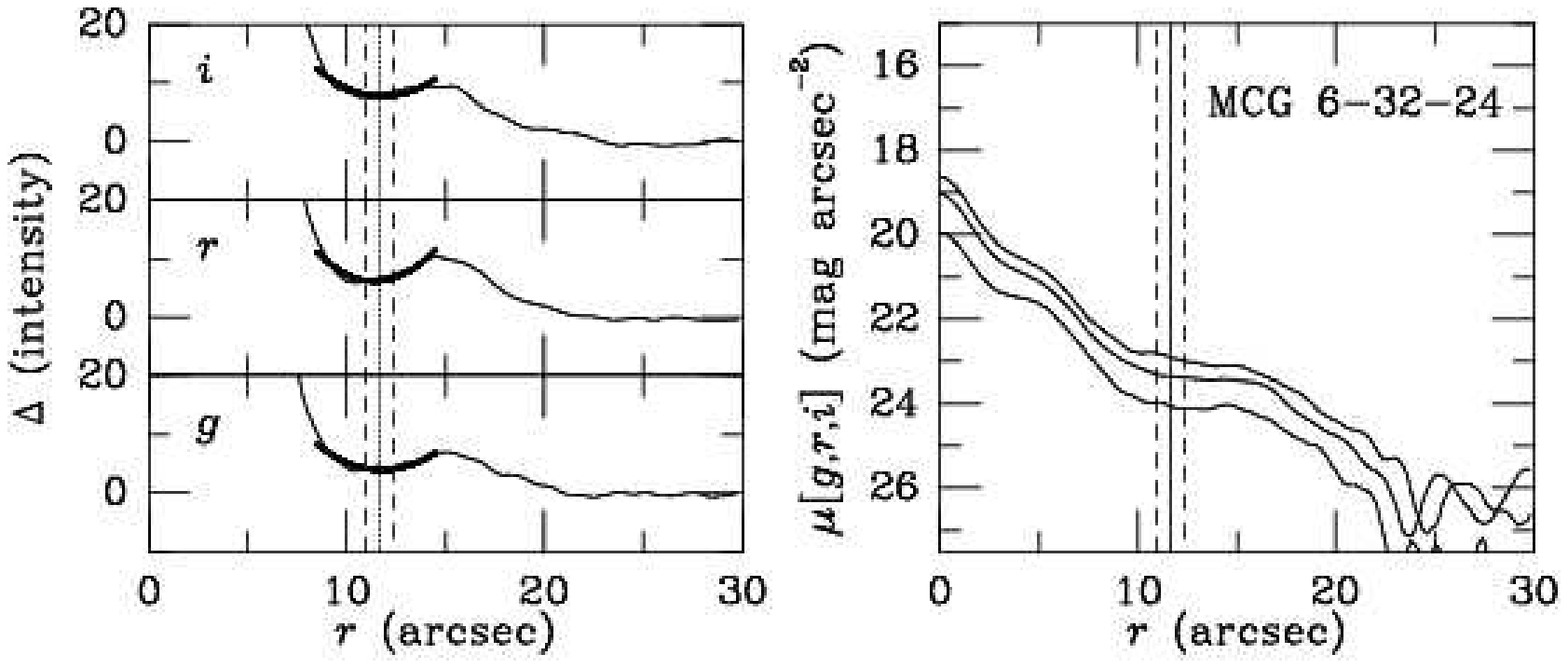}
\caption{(cont.)}
\end{figure}
\setcounter{figure}{22}
\begin{figure}
\includegraphics[width=\columnwidth,trim=0 25 0 400,clip]{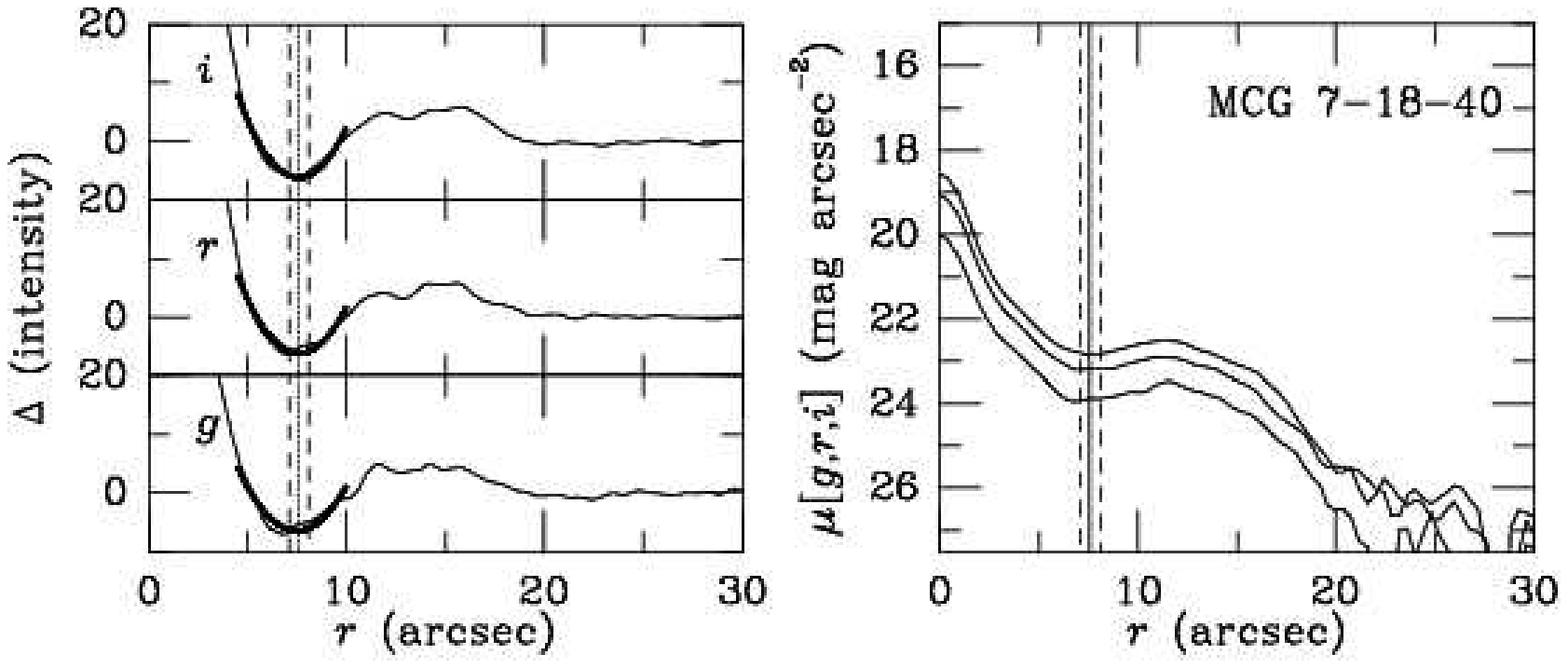}
\includegraphics[width=\columnwidth,trim=0 25 0 400,clip]{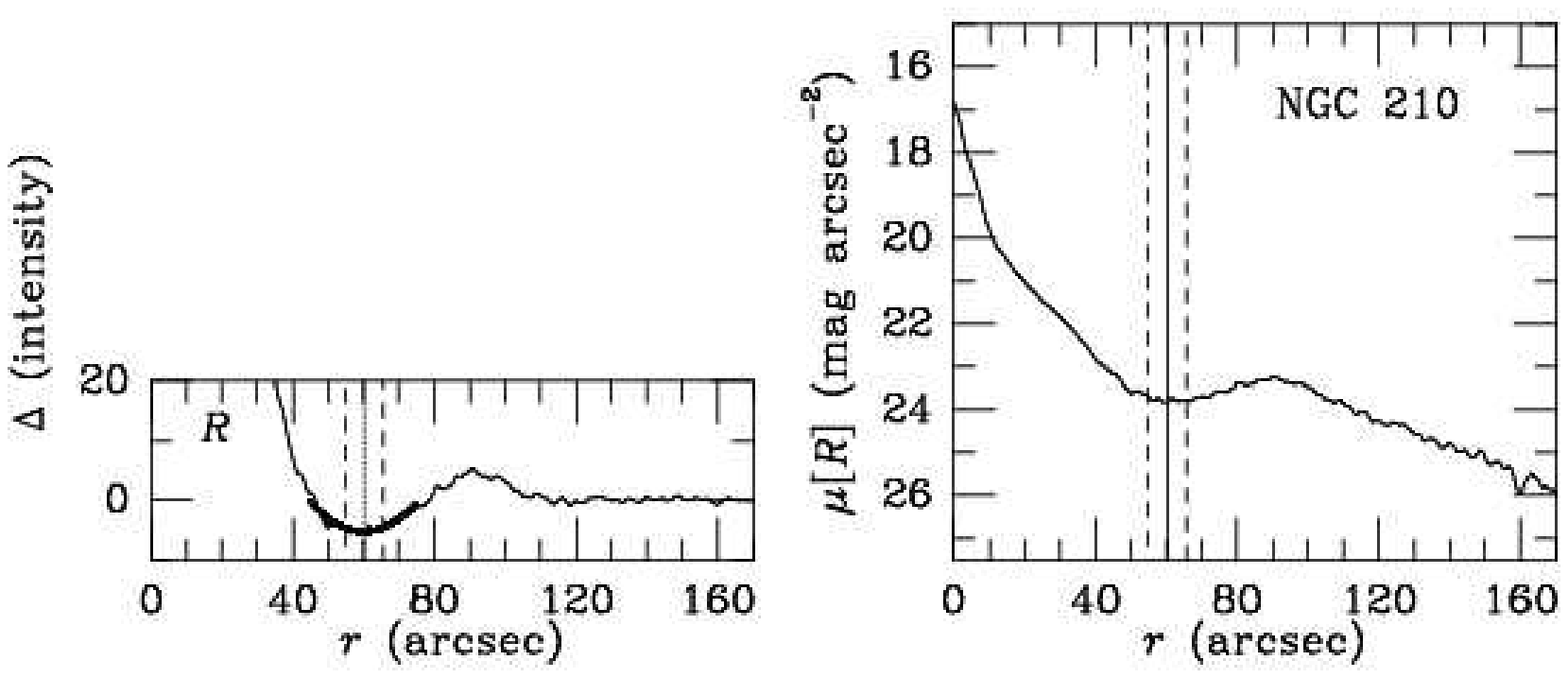}
\includegraphics[width=\columnwidth,trim=0 25 0 400,clip]{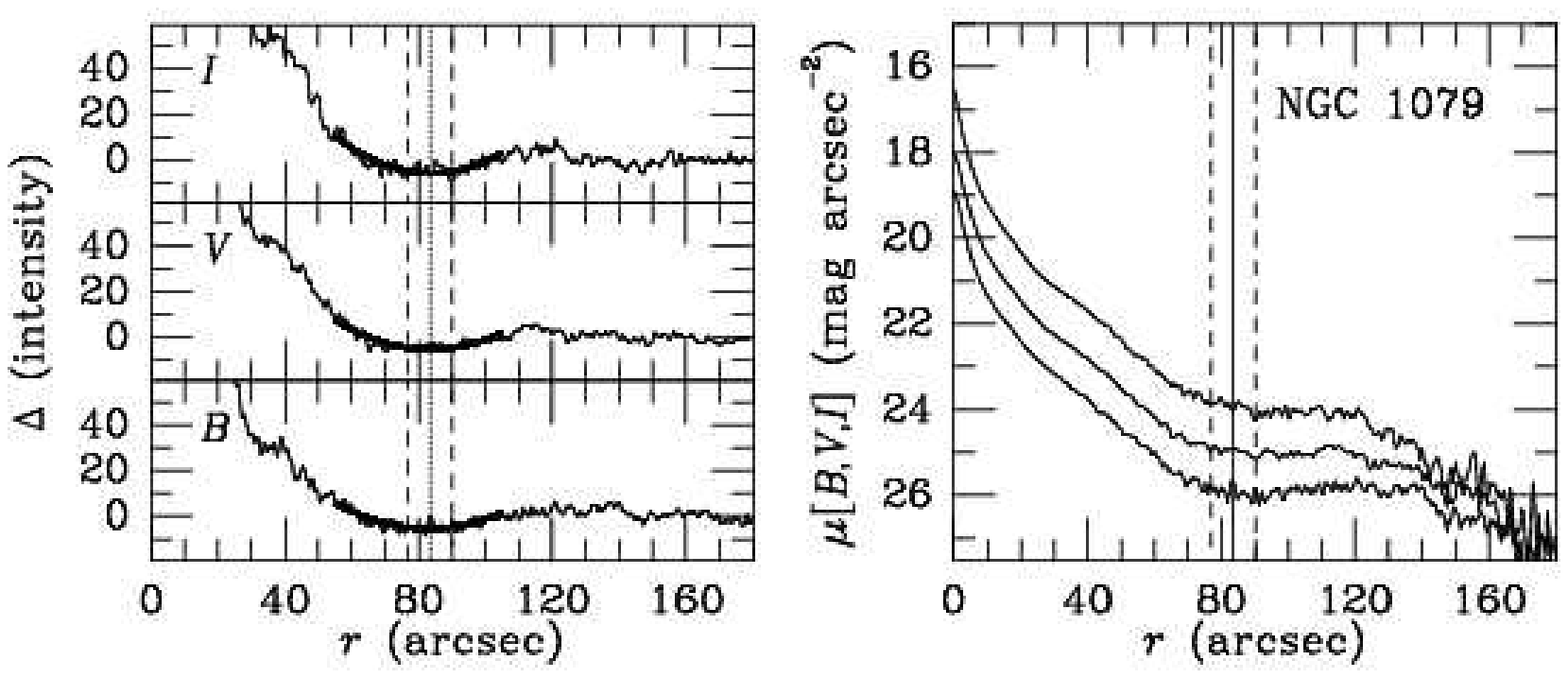}
\includegraphics[width=\columnwidth,trim=0 25 0 400,clip]{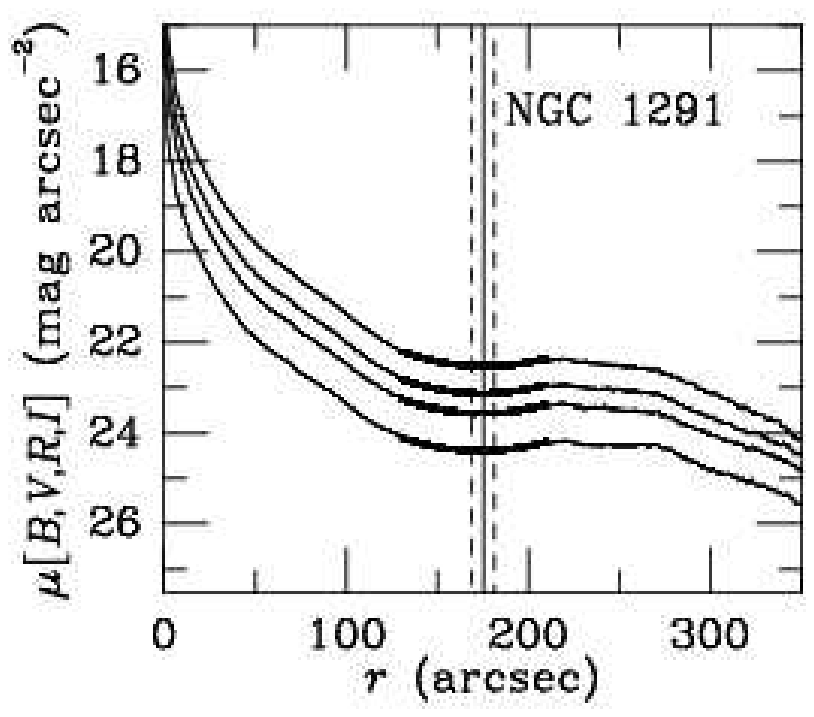}
\includegraphics[width=\columnwidth,trim=0 25 0 400,clip]{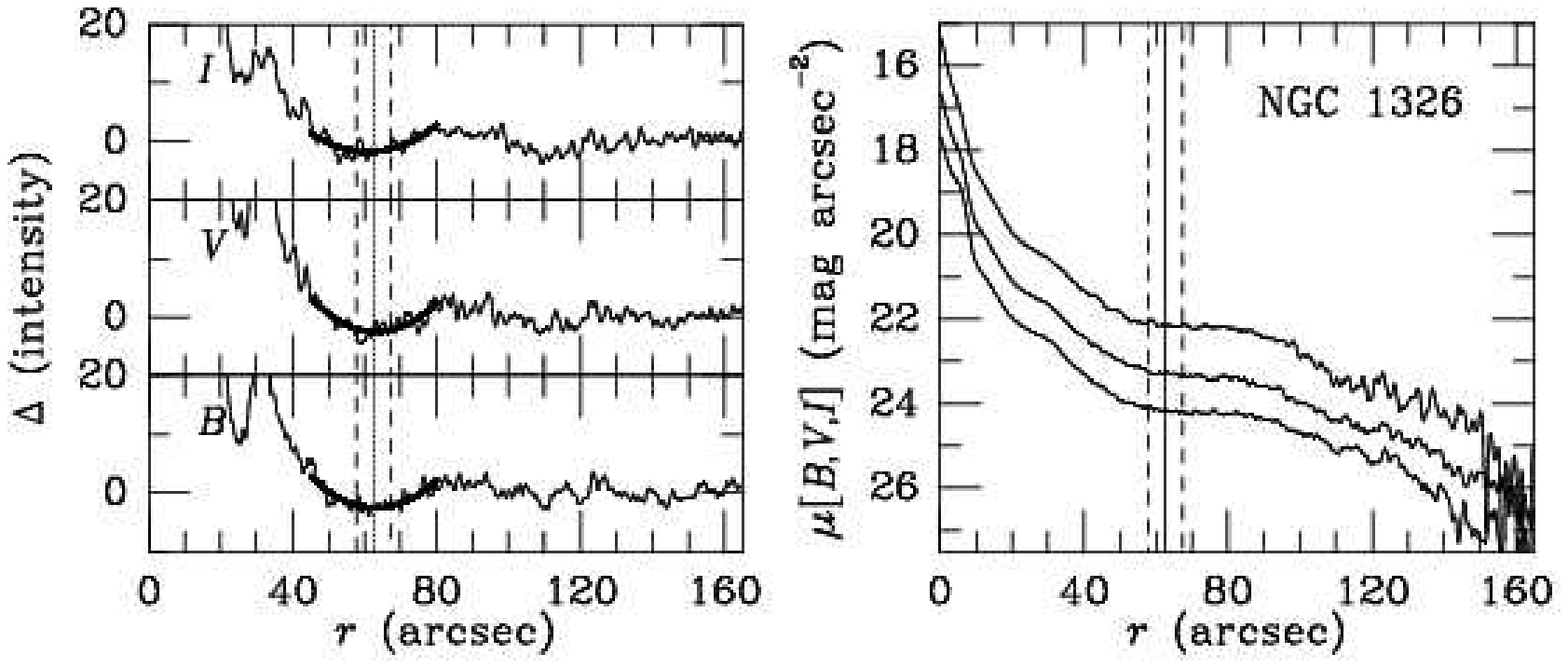}
\caption{(cont.)}
\end{figure}
\setcounter{figure}{22}
\begin{figure}
\includegraphics[width=\columnwidth,trim=0 25 0 400,clip]{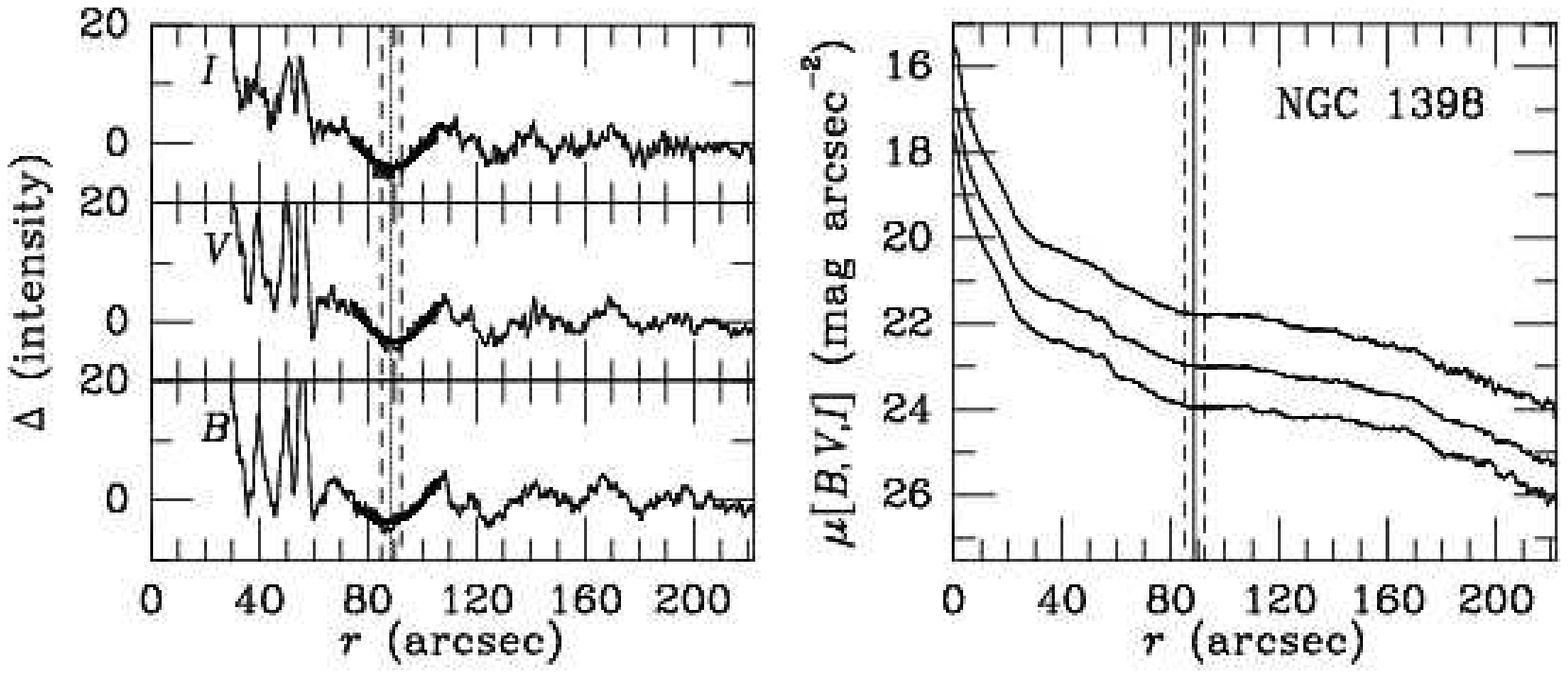}
\includegraphics[width=\columnwidth,trim=0 25 0 400,clip]{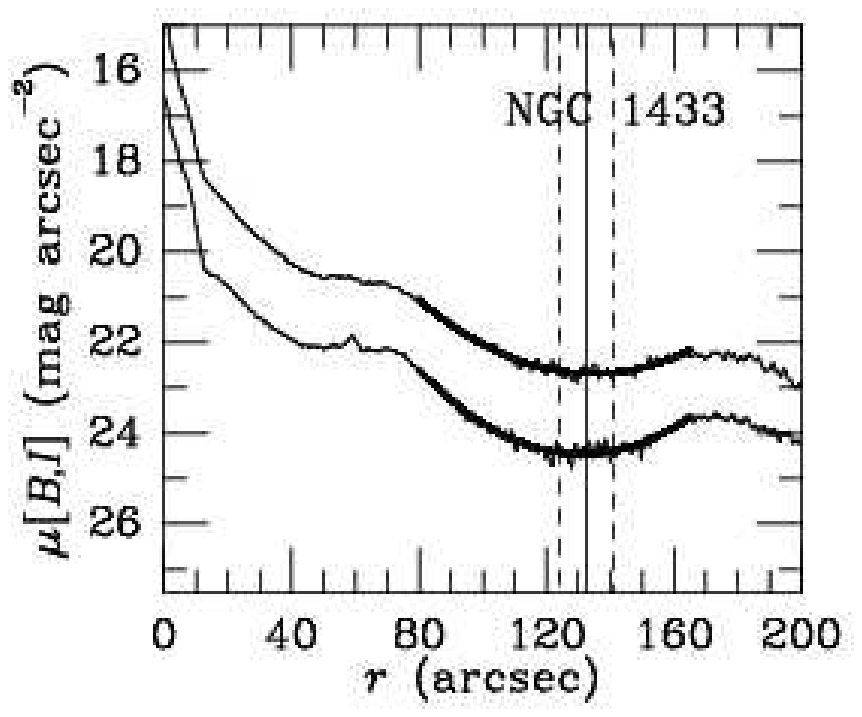}
\includegraphics[width=\columnwidth,trim=0 25 0 400,clip]{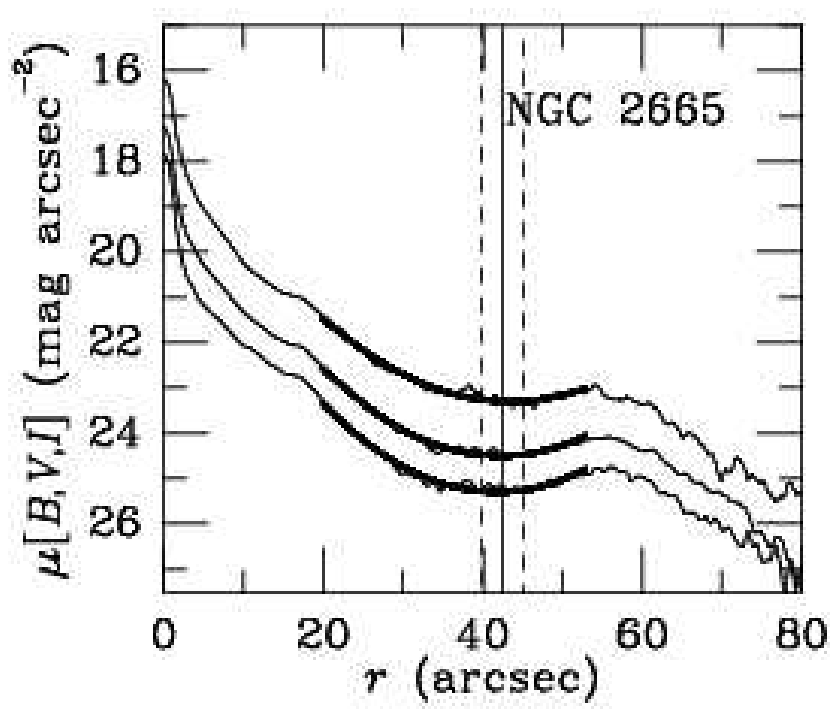}
\includegraphics[width=\columnwidth,trim=0 25 0 400,clip]{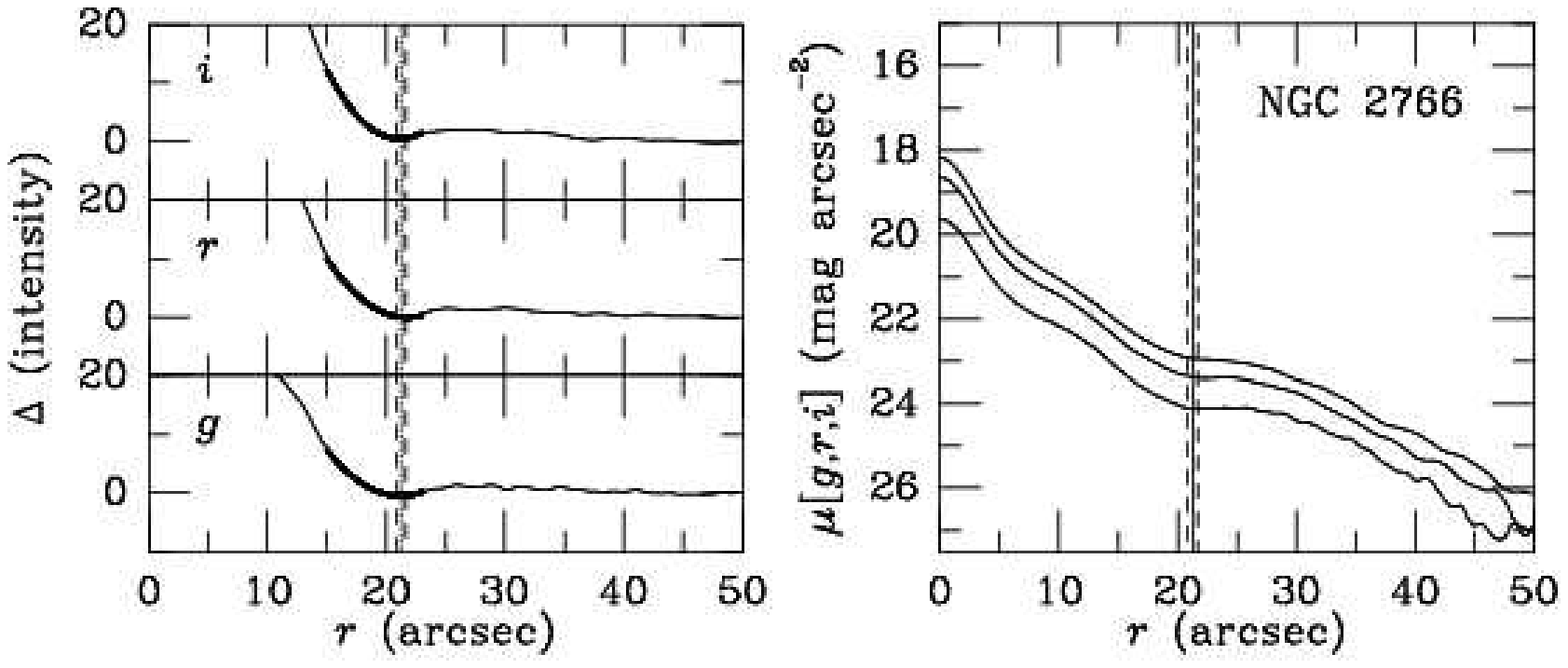}
\includegraphics[width=\columnwidth,trim=0 25 0 400,clip]{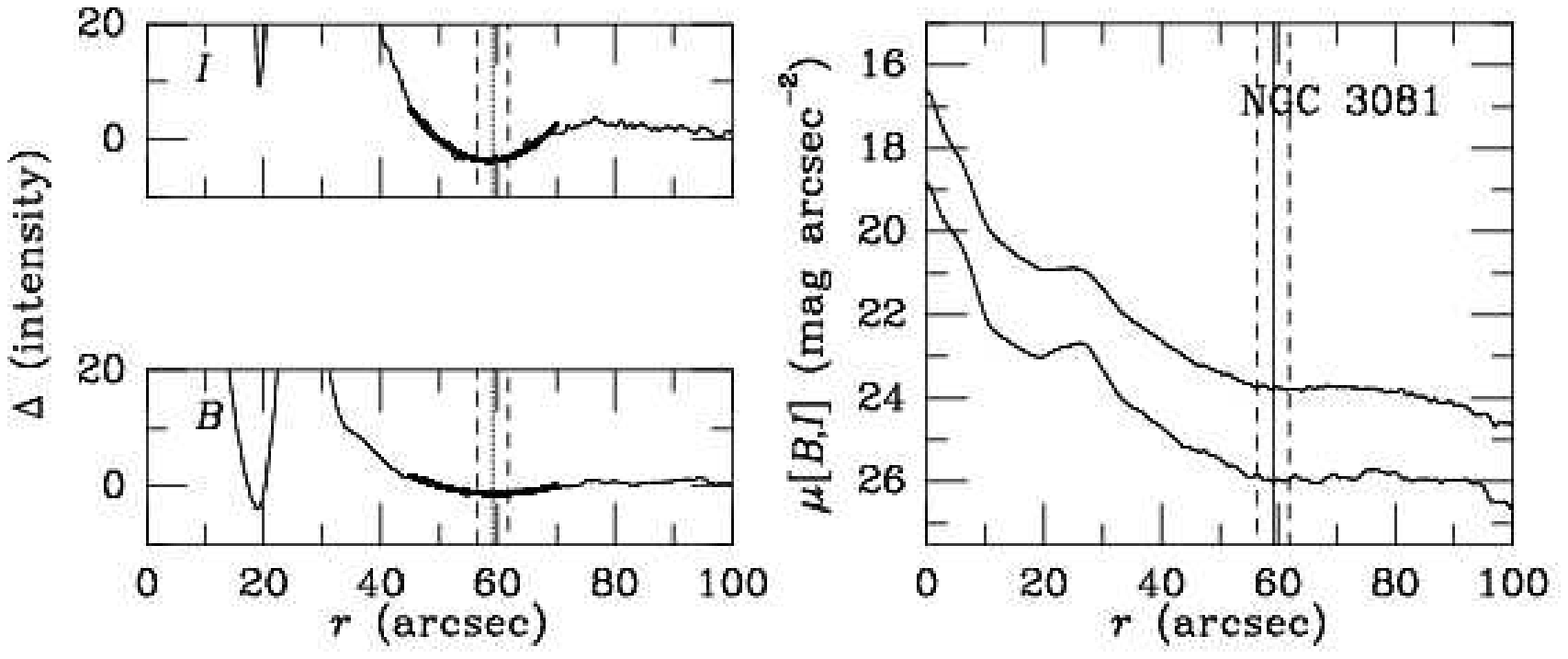}
\caption{(cont.)}
\end{figure}
\setcounter{figure}{22}
\begin{figure}
\includegraphics[width=\columnwidth,trim=0 25 0 400,clip]{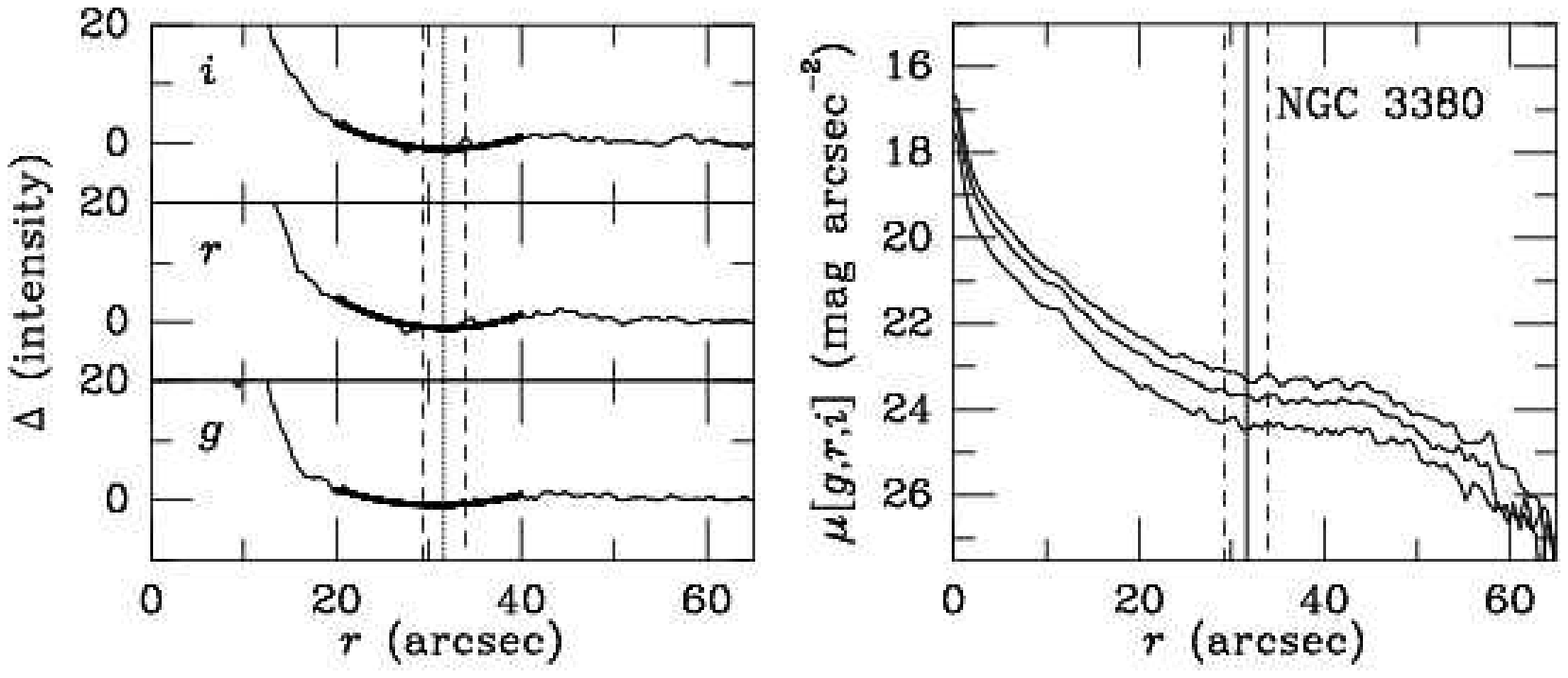}
\includegraphics[width=\columnwidth,trim=0 25 0 400,clip]{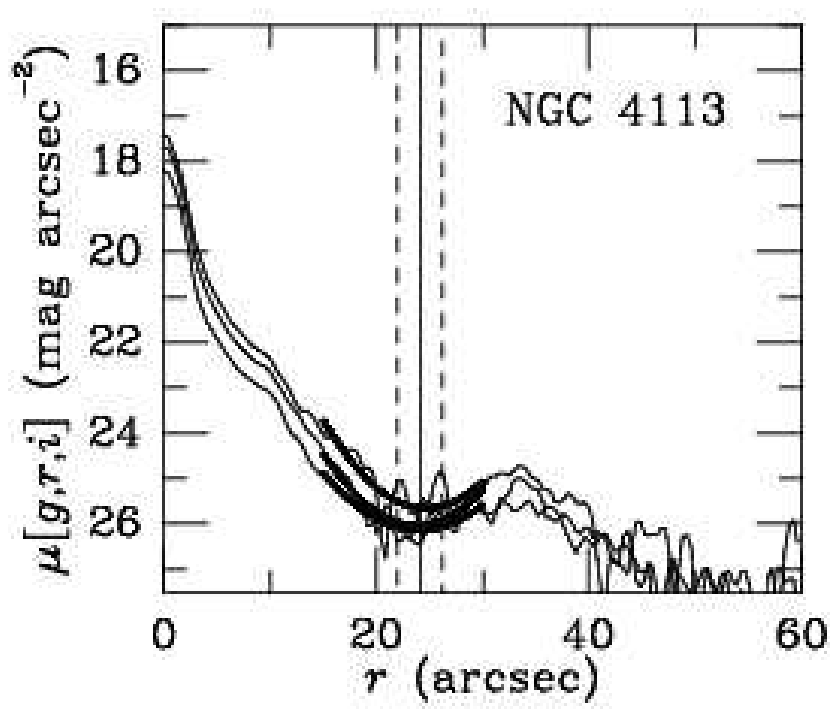}
\includegraphics[width=\columnwidth,trim=0 25 0 400,clip]{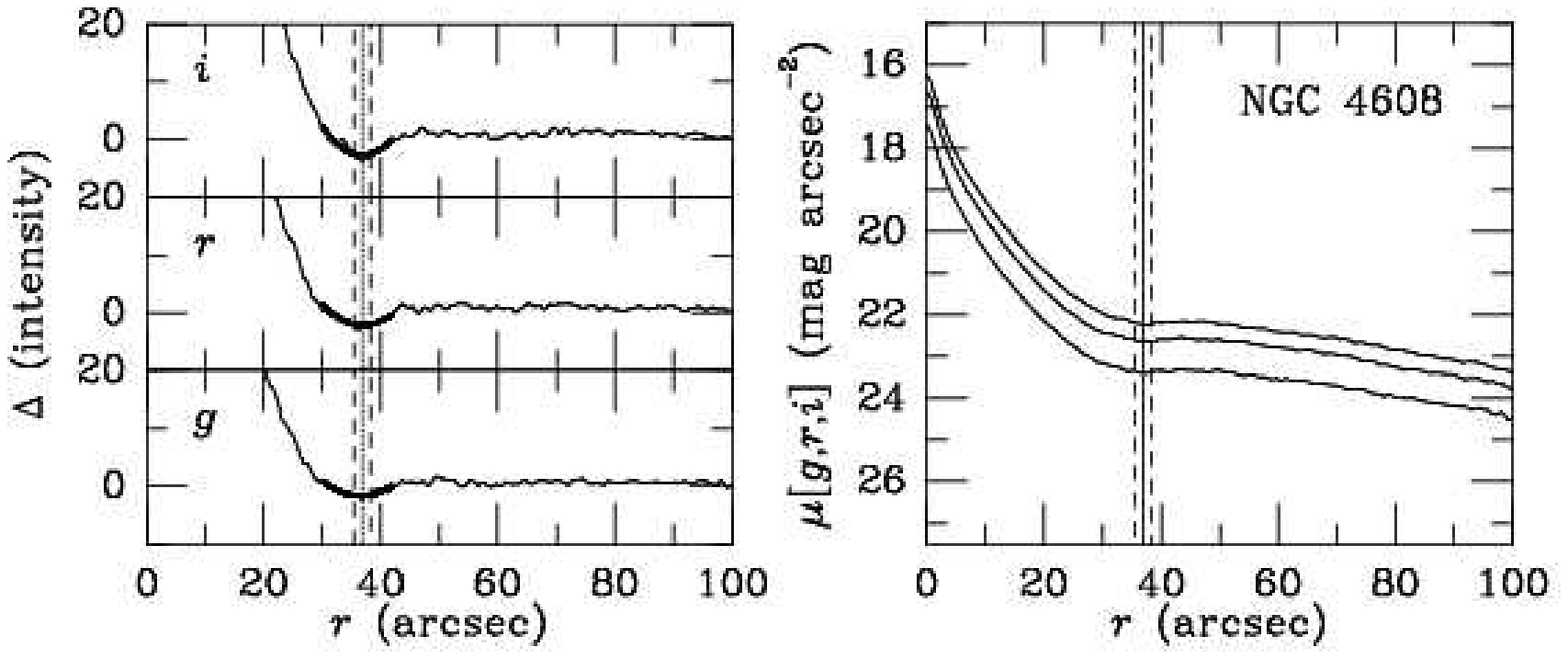}
\includegraphics[width=\columnwidth,trim=0 25 0 400,clip]{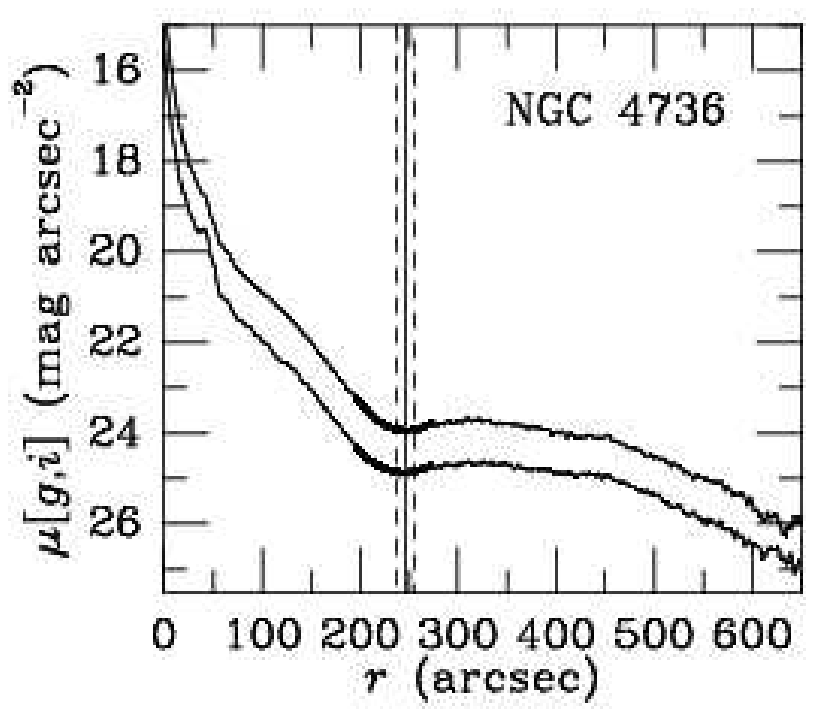}
\includegraphics[width=\columnwidth,trim=0 25 0 400,clip]{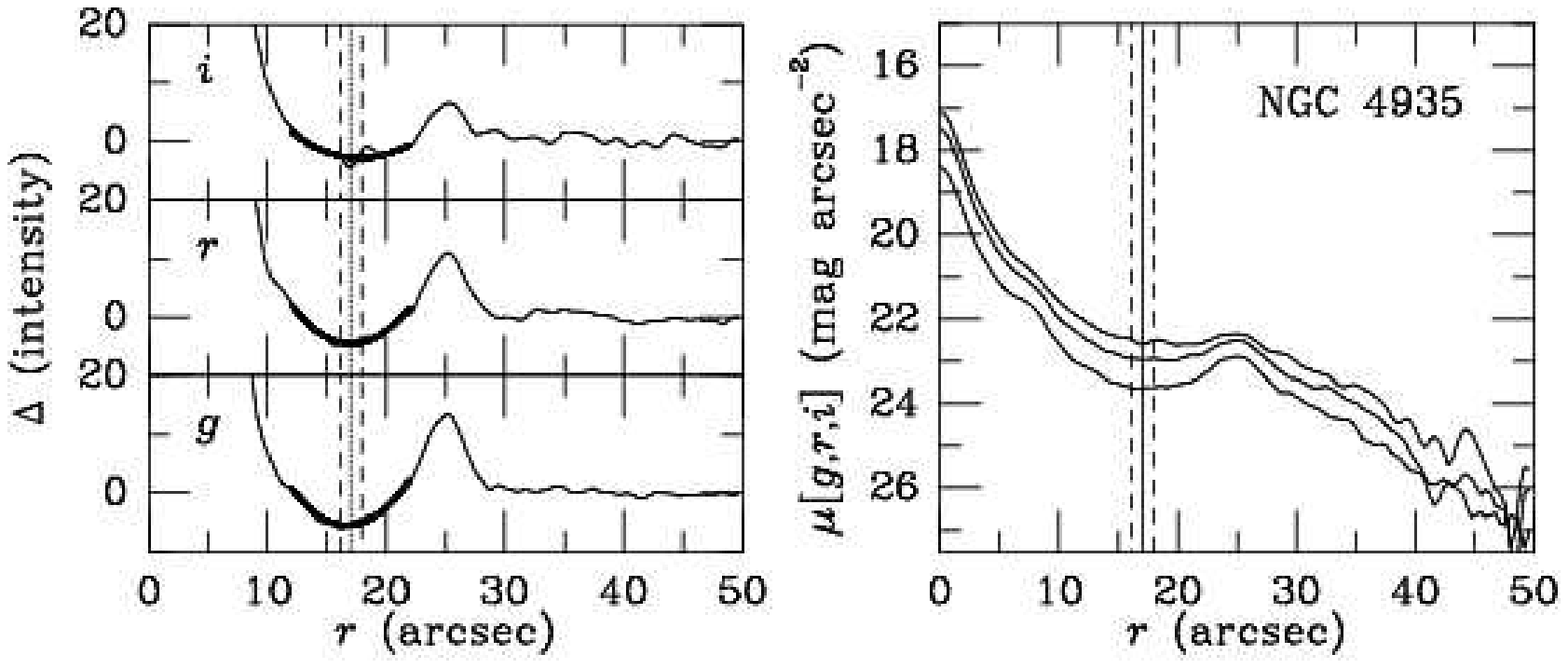}
\caption{(cont.)}
\end{figure}
\setcounter{figure}{22}
\begin{figure}
\includegraphics[width=\columnwidth,trim=0 25 0 400,clip]{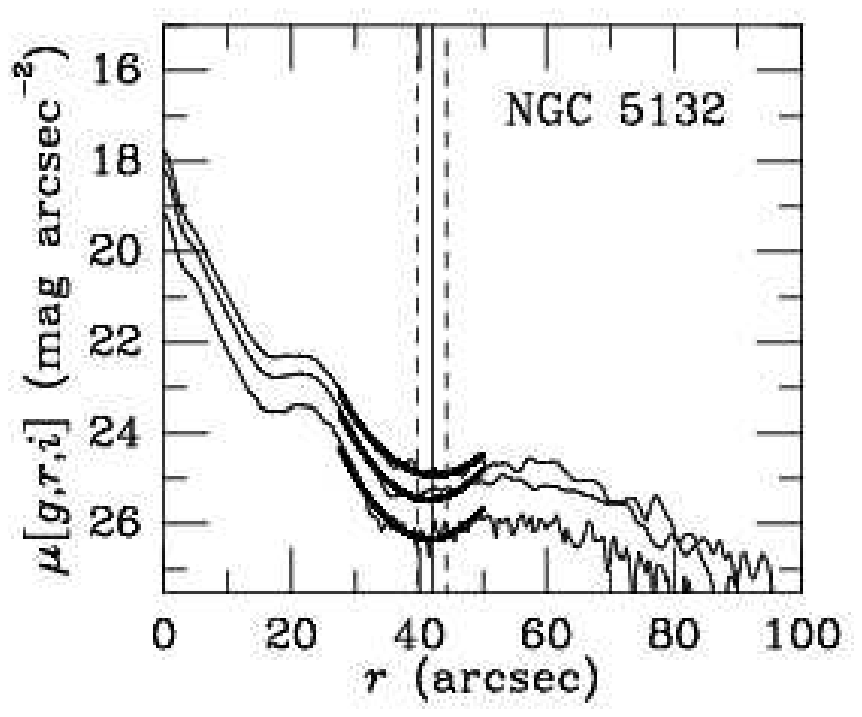}
\includegraphics[width=\columnwidth,trim=0 25 0 400,clip]{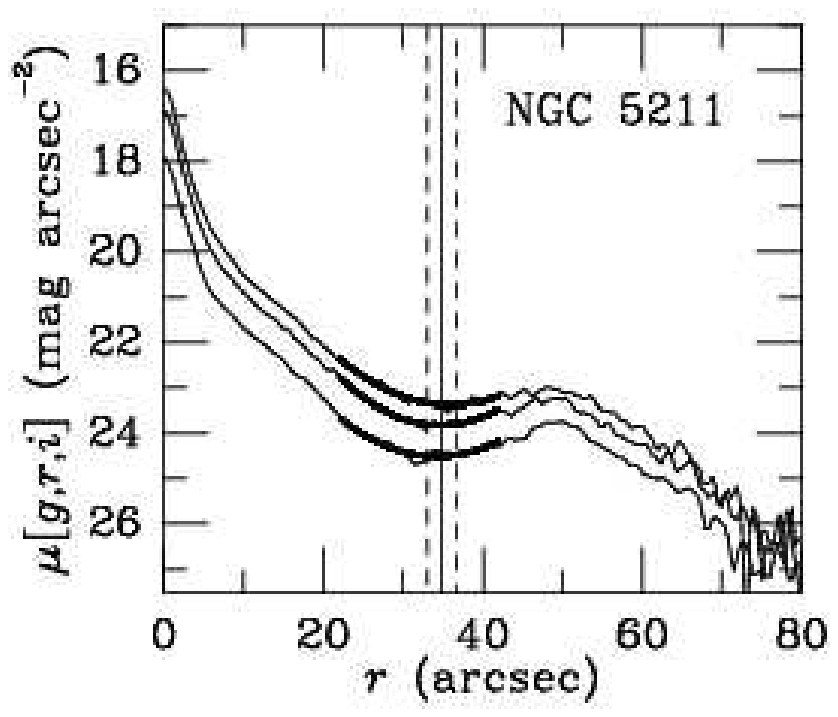}
\includegraphics[width=\columnwidth,trim=0 25 0 400,clip]{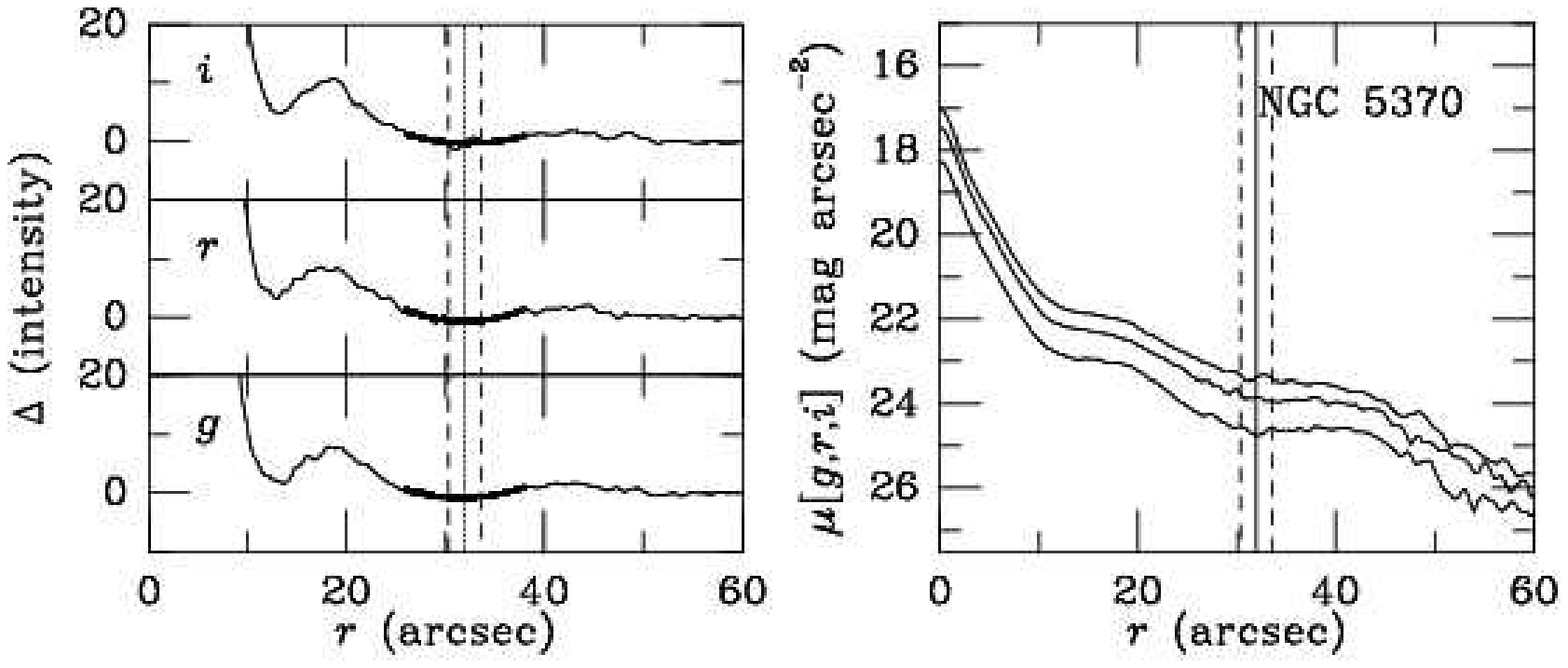}
\includegraphics[width=\columnwidth,trim=0 25 0 400,clip]{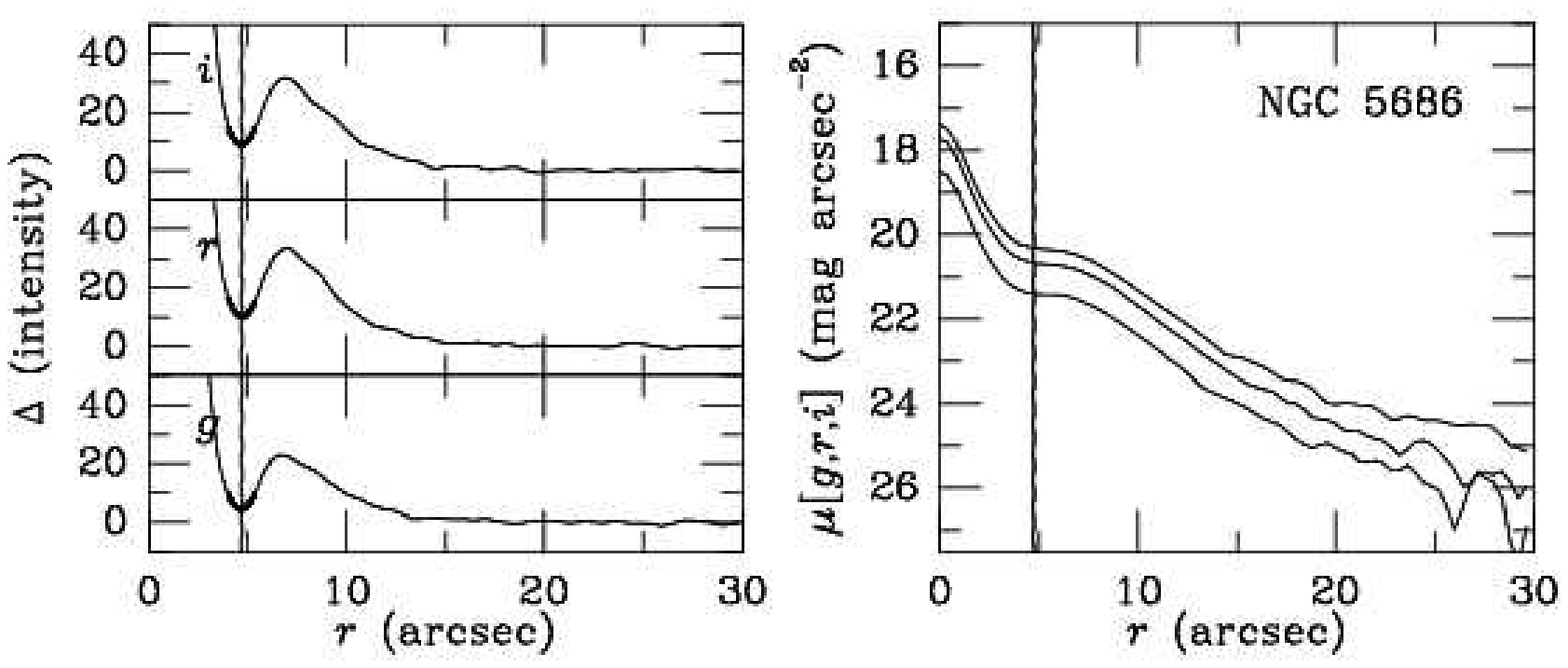}
\includegraphics[width=\columnwidth,trim=0 25 0 400,clip]{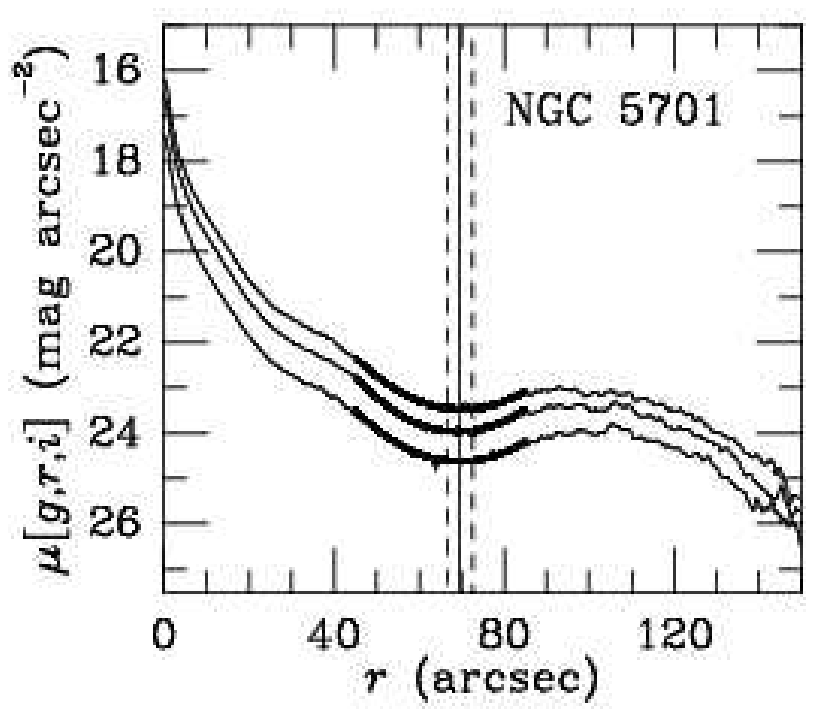}
\caption{(cont.)}
\end{figure}
\setcounter{figure}{22}
\begin{figure}
\includegraphics[width=\columnwidth,trim=0 25 0 400,clip]{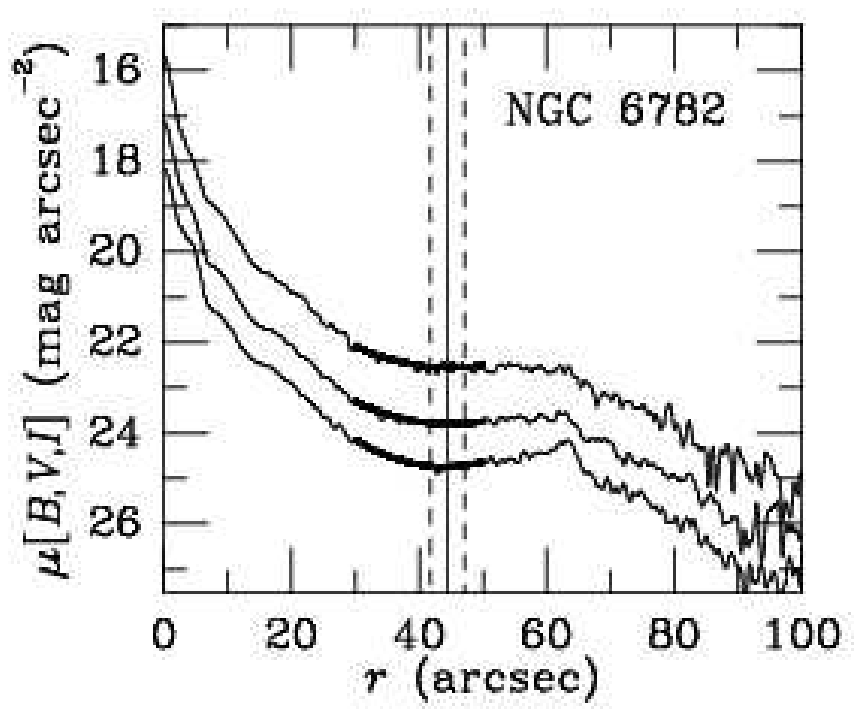}
\includegraphics[width=\columnwidth,trim=0 25 0 400,clip]{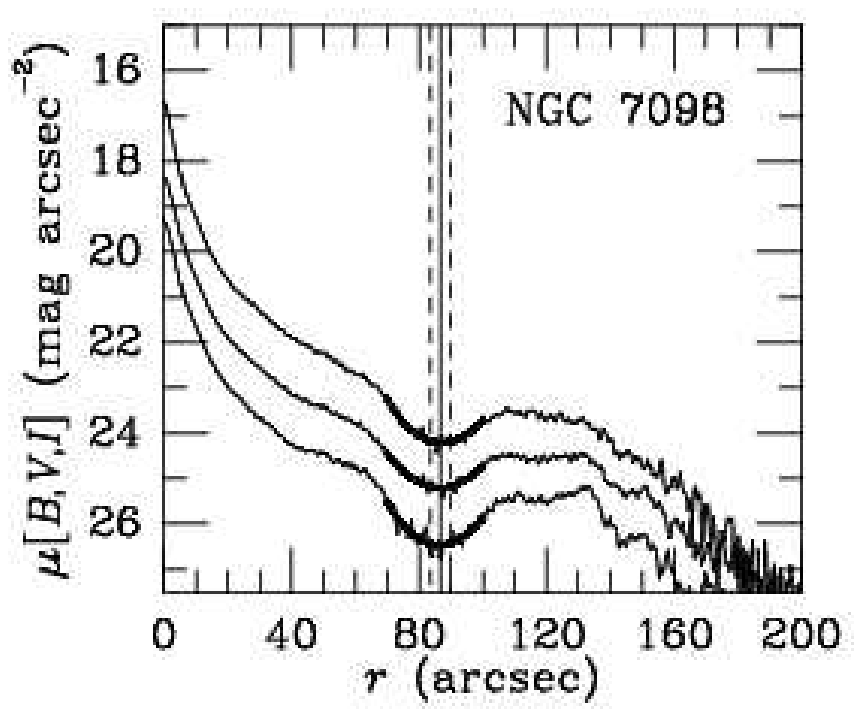}
\includegraphics[width=\columnwidth,trim=0 25 0 400,clip]{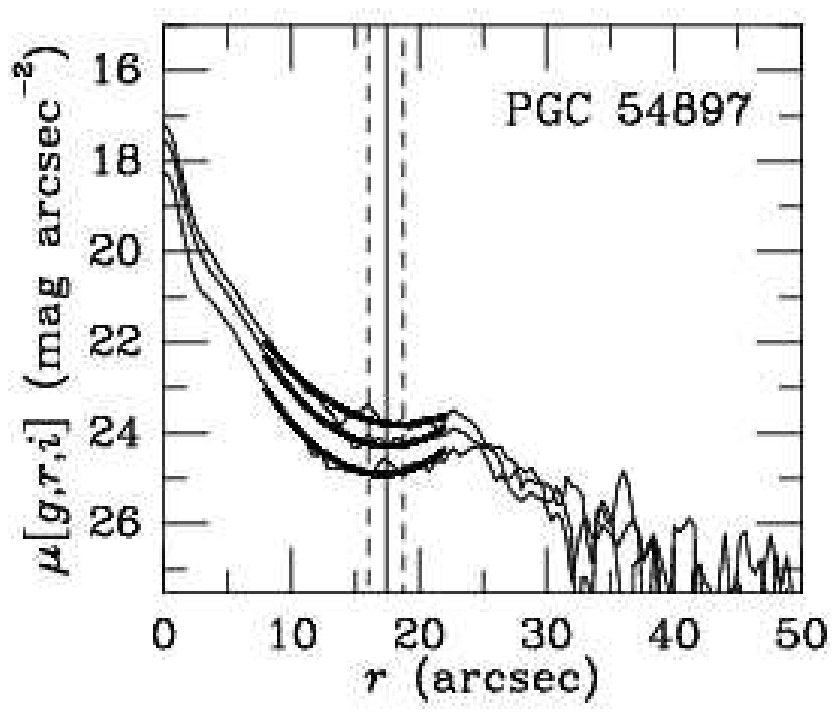}
\includegraphics[width=\columnwidth,trim=0 25 0 400,clip]{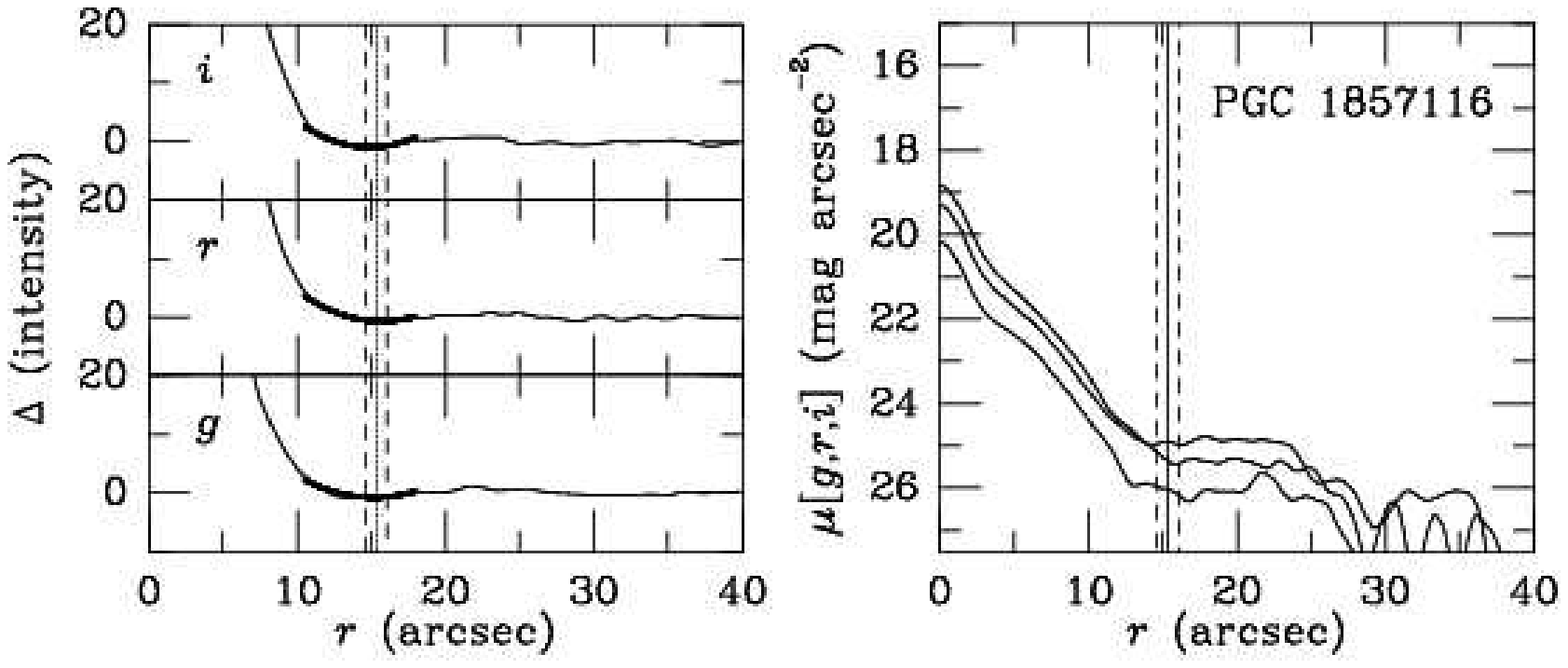}
\includegraphics[width=\columnwidth,trim=0 25 0 400,clip]{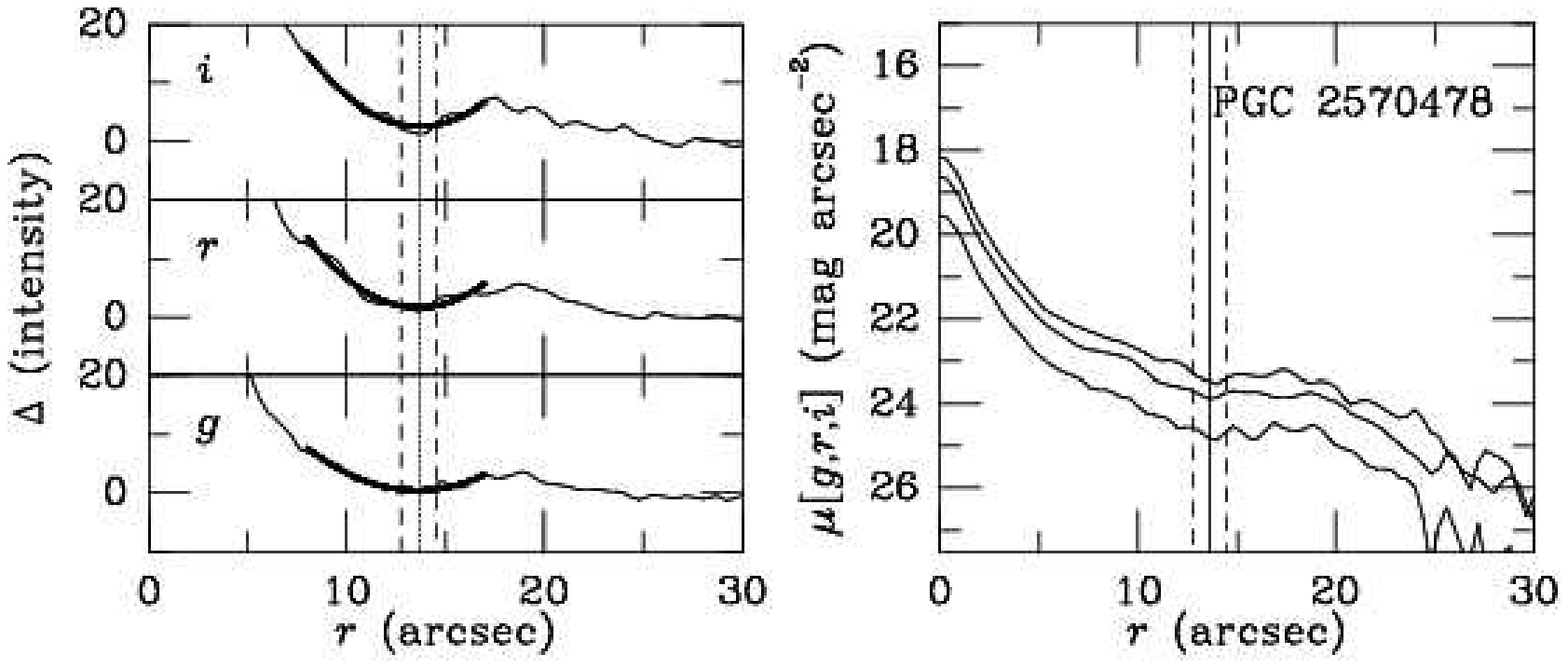}
\caption{(cont.)}
\end{figure}
\setcounter{figure}{22}
\begin{figure}
\includegraphics[width=\columnwidth,trim=0 25 0 400,clip]{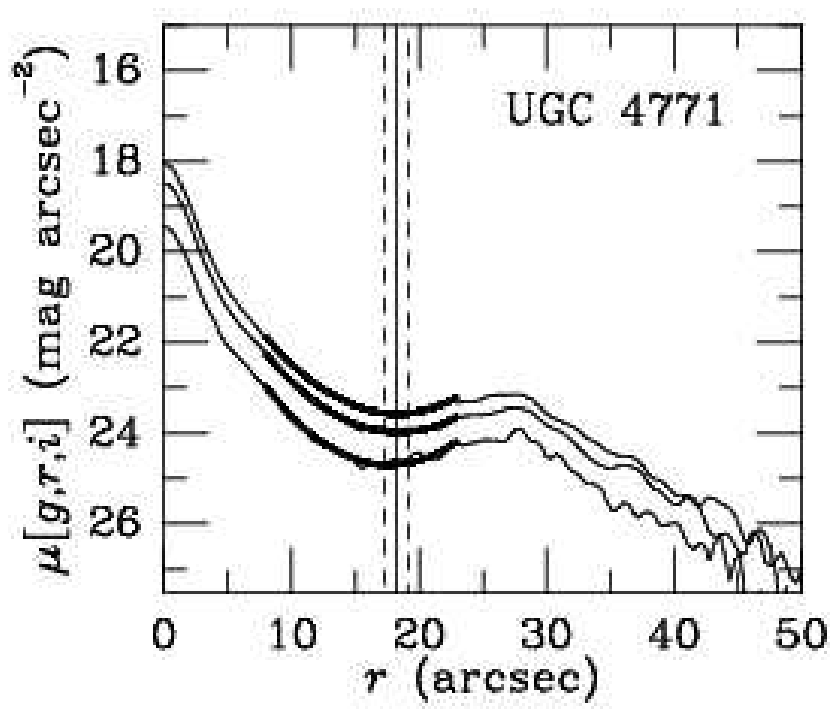}
\includegraphics[width=\columnwidth,trim=0 25 0 400,clip]{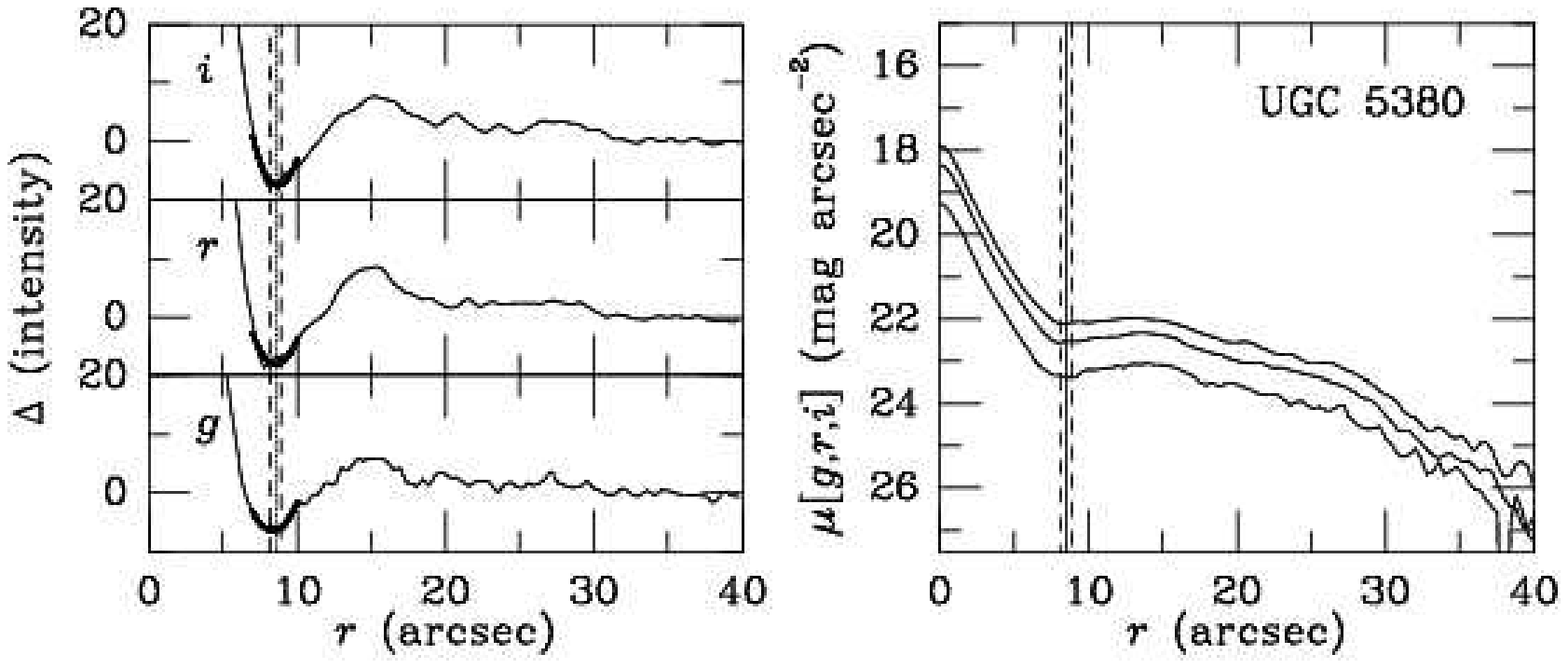}
\includegraphics[width=\columnwidth,trim=0 25 0 400,clip]{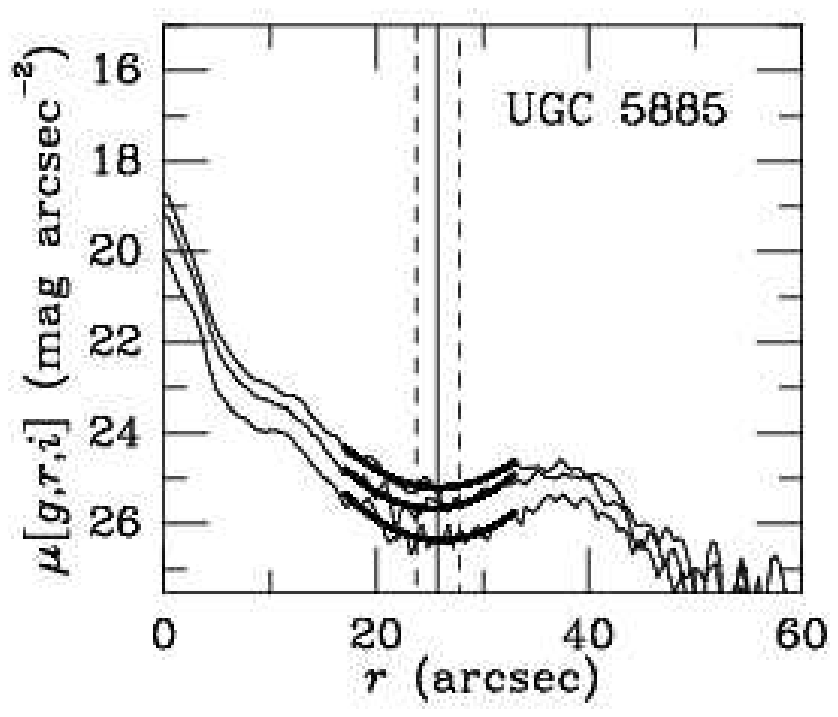}
\includegraphics[width=\columnwidth,trim=0 25 0 400,clip]{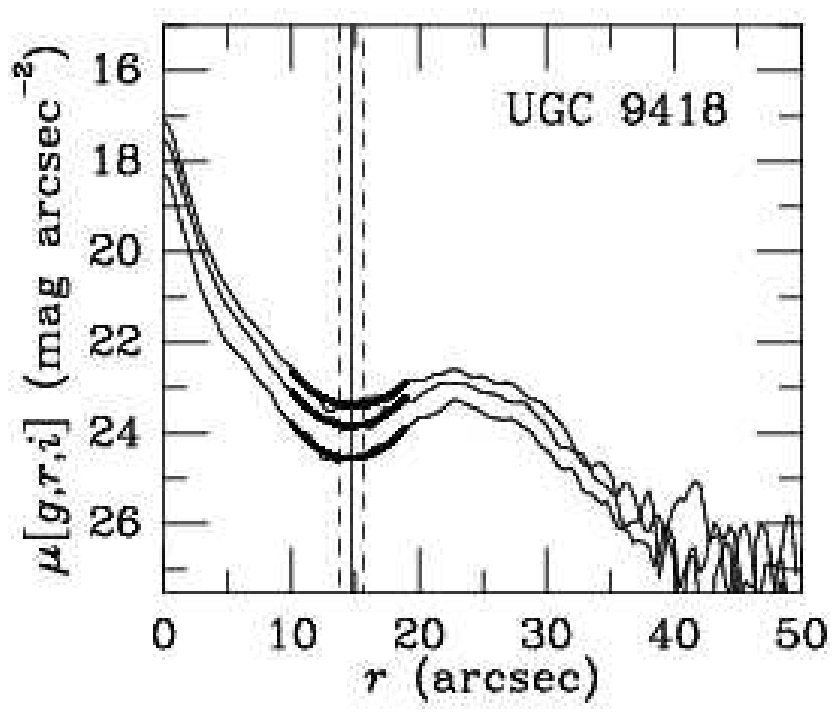}
\includegraphics[width=\columnwidth,trim=0 25 0 400,clip]{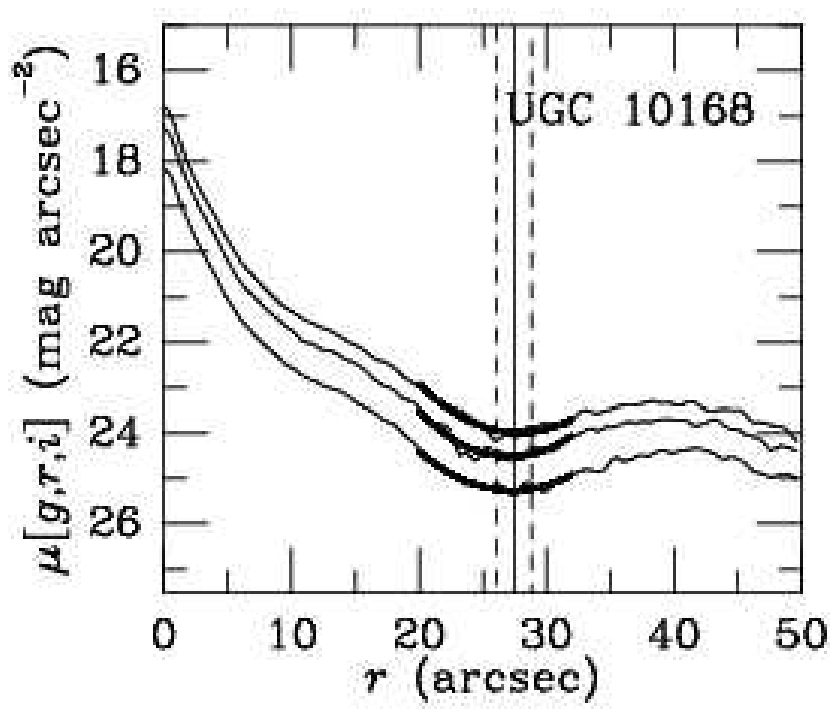}
\caption{(cont.)}
\end{figure}
\setcounter{figure}{22}
\begin{figure}
\includegraphics[width=\columnwidth,trim=0 25 0 400,clip]{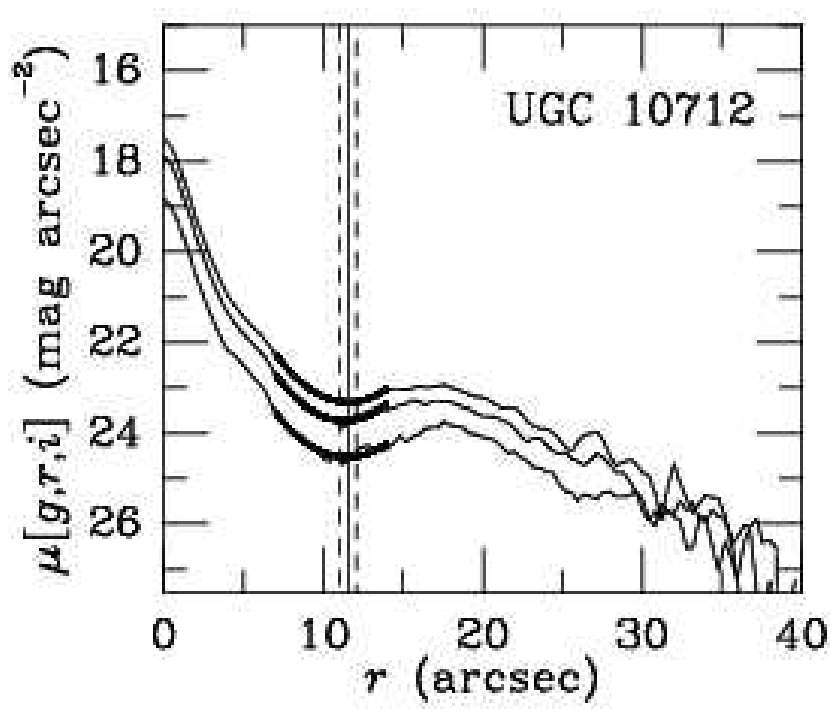}
\includegraphics[width=\columnwidth,trim=0 25 0 400,clip]{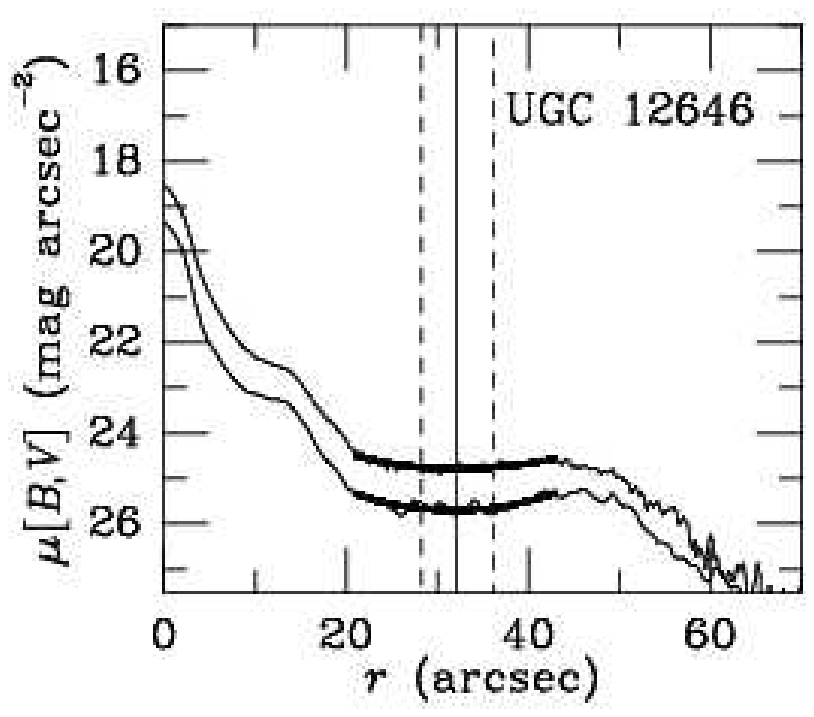}
\caption{The derivation of $r_{gp}$ for each sample galaxy. Those cases
where $r_{gp}$ is based on parabolic fits to residual intensities
$\Delta$ (as for CGCG 65-2) include up to three plots in a left panel
with a darker line showing the fits. In these cases, the right panel
shows the bar minor axis profiles and the vertical lines indicate the
mean value of $r_{gp}$ from the three filters and its estimated
uncertainty. Those cases where $r_{gp}$ is instead derived from
parabolic fits to the suface brightness ($\mu$) profiles (as for UGC
4596 and NGC 5335 in section 5) have only a single panel with a darker
line showing the fits and vertical lines showing $<r_{gp}>$ and its
estimated uncertainty. 
}
\end{figure}

\end{document}